\documentclass[a4paper,12pt]{book}
\usepackage{hyperref}
\hypersetup{dvips,dvipdfm,colorlinks=true,urlcolor=magenta,filecolor=magenta,linktoc=page,citecolor=red,linkcolor=blue,bookmarks=true}
\usepackage[usenames,dvipsnames,svgnames]{xcolor}
\usepackage{graphicx,epstopdf}
\usepackage{wrapfig}
\usepackage{subfig}
\usepackage{multirow}
\usepackage{amsfonts, amsmath, amssymb, latexsym}
\usepackage{setspace}
\usepackage{cancel}
\usepackage[Conny]{fncychap}
\usepackage{braket}
\usepackage{bigints}
\usepackage{titlesec}

\titleformat*{\section}{\normalfont\LARGE\bfseries}
\titleformat*{\subsection}{\normalfont\large\bfseries}
\titleformat*{\subsubsection}{\normalfont\normalsize\mdseries}

\usepackage{fancyhdr}
\usepackage{etoolbox}
\apptocmd{\thebibliography}{\setlength{\itemsep}{-2pt}}{}{}

\title{\huge{Dynamics: A different outlook}}
\author{Subenoy Chakraborty}

\begin{document}

%
\begin{titlepage}
 \maketitle
\end{titlepage}
%
%




\tableofcontents



\pagestyle{fancy}

\chapter{Motion in 2-dimension}

\section{Velocity and Acceleration of a particle moving in a plane curve in different frames of references: a vector treatment}

We shall now derive the components of velocity and acceleration in different frames of references namely (a) Cartesian frame of reference, (b) Polar co-ordinates, (c) Intrinsic co-ordinate system.\\

\subsection{Cartesian frame of references}

\subsubsection{Case I: When the axes are fixed in space:}

Let $P(x,y)$ be the position of the particle at time $t$. So the position vector of the particle is given by

\begin{wrapfigure}[7]{r}{0.34\textwidth}\vspace{-1.2\intextsep}
	\includegraphics[height=4.5 cm , width=5.5 cm ]{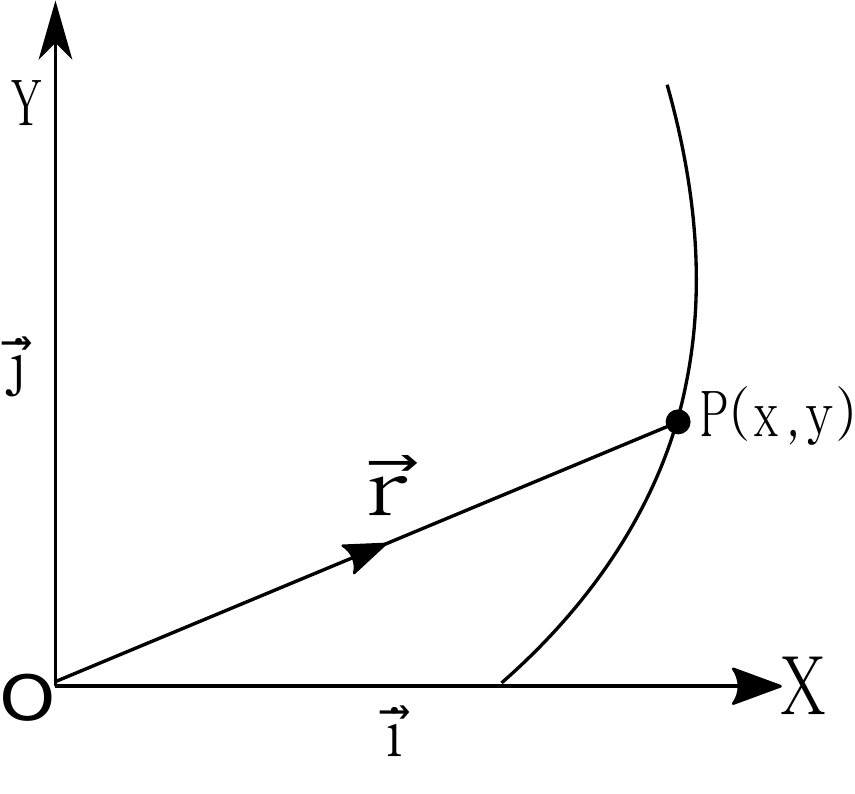}
	\begin{center}\vspace{-\intextsep}
		Fig. 2.1
	\end{center}
\end{wrapfigure}

$\vec{r}=x\vec{i}+y\vec{j}$

where $\vec{i}$ and $\vec{j}$ are constant unit vectors parallel to the axes $OX$ and $OY$ respectively.

$\therefore \vec{v}=\dfrac{\mathrm{d}\vec{r}}{\mathrm{d}t}=\dfrac{\mathrm{d}x}{\mathrm{d}t}\vec{i}+\dfrac{\mathrm{d}y}{\mathrm{d}t}\vec{j}=\dot{x}\vec{i}+\dot{y}\vec{j}$

$\vec{f}=\dfrac{\mathrm{d}\vec{v}}{\mathrm{d}t}=\ddot{x}\vec{i}+\ddot{y}\vec{j}$

\subsubsection{Case II: When the axes are not fixed but are rotating:}

In this case $\vec{i}$ and $\vec{j}$ are not constant vectors, so

$\dfrac{\mathrm{d}\vec{i}}{\mathrm{d}t}=\dfrac{\mathrm{d}\vec{i}}{\mathrm{d}\theta}\dot{\theta}=\dot{\theta}\vec{j}$ ($\because \dfrac{\mathrm{d}\vec{i}}{\mathrm{d}\theta}$ is perpendicular to $\vec{i}$ and is a unit vector, see the appendix I)

Similarly, $\dfrac{\mathrm{d}\vec{j}}{\mathrm{d}t}=-\dot{\theta}\vec{i}$

Therefore $\vec{r}=x\vec{i}+y\vec{j}$
\begin{eqnarray}
\vec{v}=\dfrac{\mathrm{d}\vec{r}}{\mathrm{d}t}&=&\dot{x}\vec{i}+\dot{y}\vec{j}+x\dot{\theta}\vec{j}-y\dot{\theta}\vec{j}\nonumber\\&=&(\dot{x}-y\dot{\theta})\vec{i}+(\dot{y}+x\dot{\theta})\vec{j}\nonumber
\end{eqnarray}

Differentiating again
$\vec{f}=\dfrac{\mathrm{d}\vec{v}}{\mathrm{d}t}=(\ddot{x}-x\dot{\theta}^2-2\dot{y}\dot{\theta}-y\ddot{\theta})\vec{i}+(\ddot{y}-y\dot{\theta}^2+2\dot{x}\dot{\theta}+x\ddot{\theta})\vec{j}$

In particular, if the axes are rotating with constant angular velocity $\omega$, then $\dot{\theta}=\omega$, $\ddot{\theta}=0$, so the velocity and acceleration simplifies to
$$v_x=\dot{x}-y\omega,~~ v_y=\dot{y}+x\omega$$and
$$f_x=\ddot{x}-2\omega\dot{y}-x\omega^2,~~f_y=\ddot{y}+2\dot{x}\omega-y\omega^2$$

\subsection{Polar Co-ordinates}

In polar co-ordinate system let $P(r,\theta)$ be the position of the particle at time $t$. Suppose $\hat{r}$ and $\hat{\theta}$ be the unit vectors along and perpendicular to the radius vector.

Now,  $\vec{r}=r\hat{r}$,  $r=OP$, $\hat{r}$ is the unit vector along the radial direction.

\begin{wrapfigure}[7]{r}{0.34\textwidth}\vspace{-1.9\intextsep}
	\includegraphics[height=4.5 cm , width=5.5 cm ]{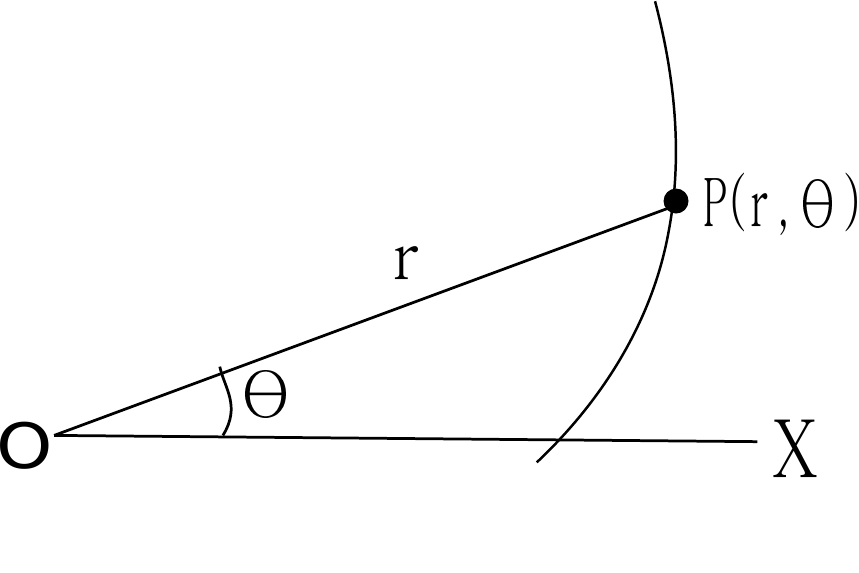}
	\begin{center}\vspace{-1.9\intextsep}
		Fig. 2.2
	\end{center}
\end{wrapfigure}

$\therefore ~\vec{v}=\dfrac{\mathrm{d}\vec{r}}{\mathrm{d}t}=\dfrac{\mathrm{d}r}{\mathrm{d}t}\hat{r}+r\dfrac{\mathrm{d}\hat{r}}{\mathrm{d}t}=\dot{r}\hat{r}+r\dfrac{\mathrm{d}\theta}{\mathrm{d}t}\dfrac{\mathrm{d}\hat{r}}{\mathrm{d}\theta}=\dot{r}\hat{r}+r\dot{\theta}\hat{\theta} $

i.e., radial velocity = $\dot{r}$,

Cross-radial velocity = $r\dot{\theta}$

Differentiating once more, we have, 
\begin{eqnarray}
	\vec{f}=\dfrac{\mathrm{d}\vec{v}}{\mathrm{d}t}&=&
	\ddot{r}\hat{r}+\dot{r}\dfrac{\mathrm{d}\hat{r}}{\mathrm{d}t}+\dfrac{\mathrm{d}(r\dot{\theta})}{\mathrm{d}t}\hat{\theta}+r\dot{\theta}\dfrac{\mathrm{d}\hat{\theta}}{\mathrm{d}t}\nonumber\\&=&
	\ddot{r}\hat{r}+\dot{r}\dot{\theta}\hat{\theta}+\dfrac{\mathrm{d}(r\dot{\theta})}{\mathrm{d}t}\hat{\theta}-r\dot{\theta}^2\hat{r}~~~\left(\because \dfrac{\mathrm{d}\hat{\theta}}{\mathrm{d}t}=\dfrac{\mathrm{d}\theta}{\mathrm{d}t}\dfrac{\mathrm{d}\hat{\theta}}{\mathrm{d}\theta}=-\dot{\theta}\hat{r}\right)\nonumber\\&=&
	(\ddot{r}-r\dot{\theta}^2)\hat{r}+\dfrac{1}{r}\dfrac{\mathrm{d}(r^2\dot{\theta})}{\mathrm{d}t}\hat{\theta}\nonumber
\end{eqnarray}

\subsection{Intrinsic co-ordinate system}

We shall first show the following:

``If a particle moves in a plane curve then its velocity will always be along the tangent to the curve"

\begin{wrapfigure}[7]{r}{0.34\textwidth}\vspace{-1.2\intextsep}
	\includegraphics[height=4.5 cm , width=5.5 cm ]{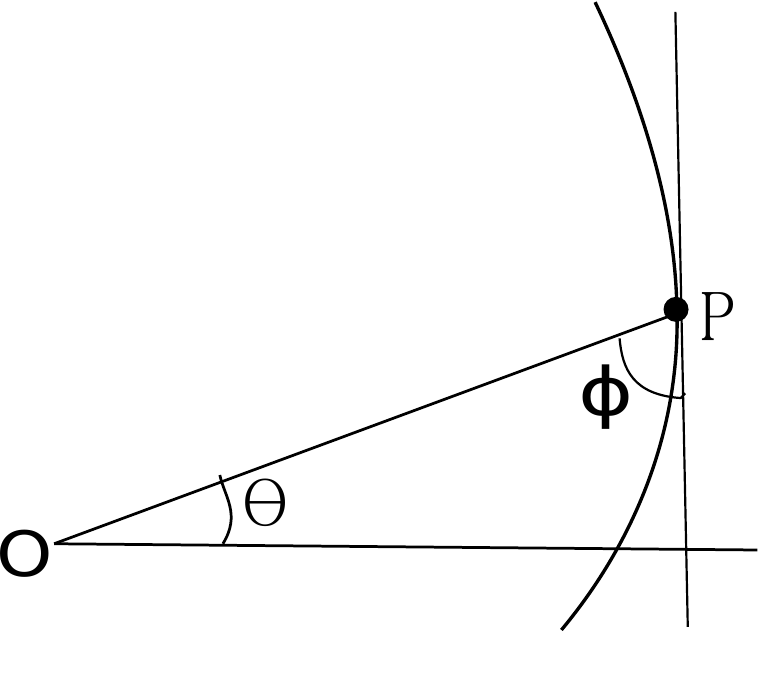}
	\begin{center}\vspace{-\intextsep}
		Fig. 2.3
	\end{center}
\end{wrapfigure}

Let $s$ be the arc length along the curve.
\begin{eqnarray}
	\vec{v}&=&\dfrac{\mathrm{d}\vec{r}}{\mathrm{d}t}\nonumber\\&=&\dfrac{\mathrm{d}\vec{r}}{\mathrm{d}s}\dot{s}=\dot{s}\vec{t}\nonumber
\end{eqnarray}

Now,\begin{eqnarray}
	\vec{t}=\dfrac{\mathrm{d}\vec{r}}{\mathrm{d}s}&=&\dfrac{\mathrm{d}}{\mathrm{d}s}(r\hat{r})=\dfrac{\mathrm{d}r}{\mathrm{d}s}\hat{r}+r\dfrac{\mathrm{d}\theta}{\mathrm{d}s}\dfrac{\mathrm{d}\hat{r}}{\mathrm{d}\theta}\nonumber\\&=&\dfrac{\mathrm{d}r}{\mathrm{d}s}\hat{r}+r\dfrac{\mathrm{d}\theta}{\mathrm{d}s}\hat{\theta}\nonumber\\&=&\cos\phi~\hat{r}+\sin\phi~\hat{\theta}\nonumber
\end{eqnarray}
which shows that $\vec{t}$ is a unit vector along the tangent to the curve. Hence velocity is always tangent to the curve.

\begin{eqnarray}
	\therefore~\vec{f}=\dfrac{\mathrm{d}\vec{v}}{\mathrm{d}t}&=&\dfrac{\mathrm{d}}{\mathrm{d}t}(\dot{\vec{t}})=\ddot{s}\vec{t}+\dot{s}\dot{s}\dfrac{\mathrm{d}\vec{t}}{\mathrm{d}s}\nonumber\\&=&\ddot{s}\vec{t}+\dot{s}^2\cdot\kappa\vec{n}~~~\mbox{(by Frenet Formula)}\nonumber\\&=& \ddot{s}\vec{t}+\frac{\dot{s}^2}{\rho}\vec{n}\nonumber
\end{eqnarray}
$\kappa$ is the curvature of the curve at the given point, $\rho$ is the radius of curvature of the curve and $\vec{n}$ is the unit vector along the normal.\\

\section{A particle is acted on by a given force. Determine the differential equation of the path of a particle.}

\subsection{Path of the particle in the Cartesian form}

Let $F_x$ and $F_y$ be the components of the force (per unit mass) along the two Cartesian axes $x$ and $y$. So the equation of motion along the axes are 
$$\ddot{x}=F_x\mbox{~~and~~}\ddot{y}=F_y$$

Now, $\dot{y}=\dfrac{\mathrm{d}y}{\mathrm{d}t}=\dfrac{\mathrm{d}y}{\mathrm{d}x}\dot{x}$

$\therefore~\ddot{y}=\dfrac{\mathrm{d}y}{\mathrm{d}x}\ddot{x}+\dfrac{\mathrm{d}^2y}{\mathrm{d}x^2}\dot{x}^2$

$\therefore~F_y=F_x\dfrac{\mathrm{d}y}{\mathrm{d}x}+\dfrac{\mathrm{d}^2y}{\mathrm{d}x^2}\dot{x}^2$

Also from the equation of motion $\ddot{x}=F_x$, if we multiply both sides by $2\dot{x}$ and integrate then we obtain
$$\dot{x}^2=2\int F_x \mathrm{d}x=\chi(x)~~\mbox{(say)}$$

$\therefore$ The differential equation of the path of the particle be
$$\dfrac{\mathrm{d}^2y}{\mathrm{d}x^2}=\frac{F_y-F_x\dfrac{\mathrm{d}y}{\mathrm{d}x}}{\chi(x)}$$\\

\subsection{Path of the particle in Polar form}

Suppose $F_r$ and $F_\theta$ be the component of the force (per unit mass) along the radial and cross-radial direction. So the equations of motion along these directions are
\begin{eqnarray}
		\ddot{r}-r\dot{\theta}^2&=&-F_r\label{eq2.1}\\\mbox{and}~~~\frac{1}{r}\frac{\mathrm{d}}{\mathrm{d}t}(r^2\dot{\theta})&=&F_\theta
\end{eqnarray}

Let us denote $h=r^2\dot{\theta}$, $u=\dfrac{1}{r}$.

$\therefore F_\theta=u\dfrac{\mathrm{d}h}{\mathrm{d}t}=u\dfrac{\mathrm{d}h}{\mathrm{d}\theta}\dfrac{\mathrm{d}\theta}{\mathrm{d}t}=hu^3\dfrac{\mathrm{d}h}{\mathrm{d}\theta}$

Now, $\dot{r}=-\dfrac{1}{u^2}\dfrac{\mathrm{d}u}{\mathrm{d}t}=-\dfrac{1}{u^2}\dfrac{\mathrm{d}u}{\mathrm{d}\theta}\dot{\theta}=-h\dfrac{\mathrm{d}u}{\mathrm{d}\theta}$
\begin{eqnarray}
	\ddot{r}=\dfrac{\mathrm{d}}{\mathrm{d}\theta}\left(-h\dfrac{\mathrm{d}u}{\mathrm{d}\theta}\right)\dot{\theta}=-\left[h\dfrac{\mathrm{d}^2u}{\mathrm{d}\theta^2}+\dfrac{\mathrm{d}h}{\mathrm{d}\theta}\dfrac{\mathrm{d}u}{\mathrm{d}\theta}\right]hu^2\nonumber\\=-hu^2\left[h\dfrac{\mathrm{d}^2u}{\mathrm{d}\theta^2}+\dfrac{F_\theta}{hu^3}\dfrac{\mathrm{d}u}{\mathrm{d}\theta}\right]\nonumber
\end{eqnarray}

So equation (\ref{eq2.1}) can be written as
\begin{eqnarray}
&&-h^2u^2\dfrac{\mathrm{d}^2u}{\mathrm{d}\theta^2}-\dfrac{F_\theta}{u}\dfrac{\mathrm{d}u}{\mathrm{d}\theta}	-h^2u^3=-F_r\nonumber\\
&\therefore& h^2u^2\left[\dfrac{\mathrm{d}^2u}{\mathrm{d}\theta^2}+u\right]=F_r-\dfrac{F_\theta}{u}\dfrac{\mathrm{d}u}{\mathrm{d}\theta}\nonumber
\end{eqnarray}

This is the differential equation of the path of a particle moving under a force.

In particular, if $F_\theta=0$ and $F_r=F(r)$, a function of the radial co-ordinate  (i.e., central force) then the differential equation of the path of the particle is 
$$h^2u^2\left[\dfrac{\mathrm{d}^2u}{\mathrm{d}\theta^2}+u\right]=F$$

\subsection{Differential equation of the path of a particle in intrinsic co-ordinates}

Let $F_T$ and $F_N$ be the components of the force (per unit mass) along the tangential and normal directions. So the equations of motion be
$$\ddot{s}=F_T~,~~\frac{v^2}{\rho}=F_N$$

$v^2=\rho F_N=F_N\dfrac{\mathrm{d}s}{\mathrm{d}\psi}$

$\dot{s}=\left(F_N\dfrac{\mathrm{d}s}{\mathrm{d}\psi}\right)^{\frac{1}{2}}$

$\implies \dfrac{\mathrm{d}s}{\mathrm{d}\psi}\dot{\psi}=\left(F_N\dfrac{\mathrm{d}s}{\mathrm{d}\psi}\right)^{\frac{1}{2}}$

$\implies \dot{\psi}=\left[\dfrac{F_N}{\frac{\mathrm{d}s}{\mathrm{d}\psi}}\right]^{\frac{1}{2}}$

Now, 
\begin{eqnarray}
	F_T&=&\ddot{s}=\dfrac{\mathrm{d}}{\mathrm{d}t}\left[F_N\dfrac{\mathrm{d}s}{\mathrm{d}\psi}\right]^{\frac{1}{2}}=\dfrac{\mathrm{d}}{\mathrm{d}\psi}\left[F_N\dfrac{\mathrm{d}s}{\mathrm{d}\psi}\right]^{\frac{1}{2}}\dot{\psi}\nonumber\\F_T&=&\frac{1}{2}\left[F_N\dfrac{\mathrm{d}s}{\mathrm{d}\psi}\right]^{-\frac{1}{2}}\left\{\dfrac{\mathrm{d}F_N}{\mathrm{d}\psi}\dfrac{\mathrm{d}s}{\mathrm{d}\psi}+F_N\dfrac{\mathrm{d}^2s}{\mathrm{d}\psi^2}\right\}\left[\dfrac{F_N}{\frac{\mathrm{d}s}{\mathrm{d}\psi}}\right]^{\frac{1}{2}}\nonumber\\&=&\frac{1}{2\frac{\mathrm{d}s}{\mathrm{d}\psi}}\left\{\dfrac{\mathrm{d}F_N}{\mathrm{d}\psi}\dfrac{\mathrm{d}s}{\mathrm{d}\psi}+F_N\dfrac{\mathrm{d}^2s}{\mathrm{d}\psi^2}\right\}\nonumber
\end{eqnarray}

Thus the differential equation of the path of the particle in intrinsic co-ordinates can be written as
$$\dfrac{\mathrm{d}^2s}{\mathrm{d}\psi^2}=\frac{1}{F_N}\left(2F_T-\dfrac{\mathrm{d}F_N}{\mathrm{d}\psi}\right)\dfrac{\mathrm{d}s}{\mathrm{d}\psi}$$

\chapter{Motion in 3-dimensions}

\section{Components of velocity and acceleration in spherical polar coordinates}

\vspace{0.6cm}

\begin{wrapfigure}[10]{r}{0.4\textwidth}\vspace{-3.5\intextsep}
\centering	\includegraphics[height=5.5 cm , width=5.5 cm ]{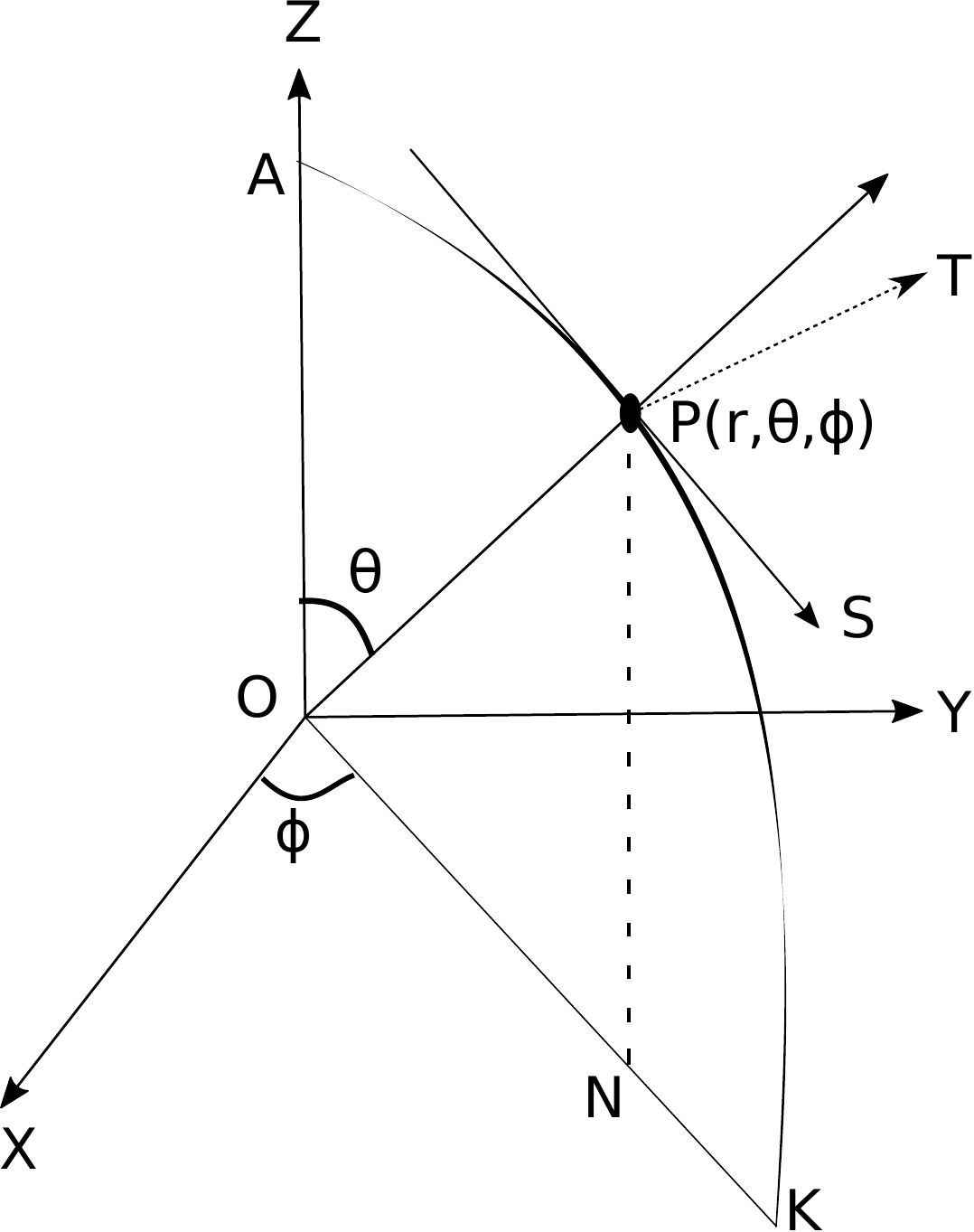}
	\begin{center}\vspace{-\intextsep}
		Fig. 3.1
	\end{center}
\end{wrapfigure}

~~Let $P(r,\theta,\phi)$ be the position of a particle. In the figure $APK$ be the meridian plane and $PT$ be the normal to the meridian plane and PS be in the meridian plane but perpendicular to OP. From the figure $r=OP$, $\angle AOP=\theta$, and $\angle XON=\phi$.\\

Now,
\begin{eqnarray}
	\mbox{velocity of }P&=&\mbox{velocity of }P\mbox{ in the meridian plane}\nonumber\\&&+\mbox{ velocity of }P\perp\mbox{to the meridian plane}\nonumber\\
	&=&\dot{r}\mbox{ along }OP+r\dot{\theta}\mbox{ along }PS+\mbox{velocity of }N\perp\mbox{to the meridian plane}\nonumber\\&=&\dot{r}\mbox{ along }OP+r\dot{\theta}\mbox{ along }PS+r\sin\theta\dot{\phi}\mbox{ along }PT\nonumber	
\end{eqnarray}

Thus $v_r=\dot{r}$, $v_\theta=r\dot{\theta}$ and $v_\phi=r\sin\theta\dot{\phi}$ are the components of the velocity along $OP$, $PS$ and $PT$ respectively.
\begin{eqnarray}
	\mbox{The acceleration of }P&=&\mbox{acceleration of }P\parallel\mbox{ to the }XOY\mbox{plane}\nonumber\\&&+\mbox{ acceleration of }P\perp\mbox{ to the }XOY\mbox{plane}\nonumber\\&=&\mbox{acceleration of }N\mbox{ in }XOY\mbox{plane}+\frac{\mathrm{d}^2}{\mathrm{d}t^2}(NP)\mbox{ along }NP\nonumber\\&=&\left[\frac{\mathrm{d}^2}{\mathrm{d}t^2}(ON)-(ON)\dot{\phi}^2\right]\mbox{ along }ON\nonumber\\&&+\left[\frac{1}{ON}\frac{\mathrm{d}}{\mathrm{d}t}(ON^2\cdot\dot{\phi})\right]\perp\mbox{ to the meridian plane}+\frac{\mathrm{d}^2}{\mathrm{d}t^2}(NP)\mbox{ along }NP\nonumber
\end{eqnarray}

AS $ON=r\sin\theta$, $NP=r\cos\theta$, so,
\begin{eqnarray}
	f_r&=&\sin\theta\left[\frac{\mathrm{d}^2}{\mathrm{d}t^2}(ON)-(ON)\dot{\phi}^2\right]+\cos\theta\frac{\mathrm{d}^2}{\mathrm{d}t^2}(NP)\nonumber\\&=&\sin\theta\left[\frac{\mathrm{d}}{\mathrm{d}t}(\dot{r}\sin\theta+r\cos\theta\dot{\theta})-r\sin\theta\dot{\phi}^2\right]+\cos\theta\frac{\mathrm{d}}{\mathrm{d}t}(\dot{r}\cos\theta-r\sin\theta\dot{\theta})\nonumber\\&=&\ddot{r}-r\dot{\theta}^2-r\sin^2\theta\dot{\phi}^2\nonumber
\end{eqnarray}
\begin{eqnarray}
	f_\theta&=&\cos\theta\left[\frac{\mathrm{d}^2}{\mathrm{d}t^2}(ON)-(ON)\dot{\phi}^2\right]-\sin\theta\frac{\mathrm{d}^2}{\mathrm{d}t^2}(NP)\nonumber\\&=&\cos\theta\left[\frac{\mathrm{d}}{\mathrm{d}t}(\dot{r}\sin\theta+r\cos\theta\dot{\theta})-r\sin\theta\dot{\phi}^2\right]-\sin\theta\frac{\mathrm{d}}{\mathrm{d}t}(\dot{r}\cos\theta-r\sin\theta\dot{\theta})\nonumber\\&=&2\dot{r}\dot{\theta}+r\ddot{\theta}-r\sin\theta\cos\theta\dot{\phi}^2\nonumber
\end{eqnarray}
\begin{eqnarray}
	f_\phi&=&\frac{1}{ON}\frac{\mathrm{d}}{\mathrm{d}t}(ON^2\cdot\dot{\phi})\nonumber\\&=&r\sin\theta\ddot{\phi}+2\dot{r}\sin\theta\dot{\phi}+2r\cos\theta\dot{\theta}\dot{\phi}\nonumber
\end{eqnarray}
\subsection{Vector Method}

\begin{wrapfigure}[20]{r}{0.4\textwidth}
	\centering	\includegraphics[height=5.5 cm , width=5.5 cm ]{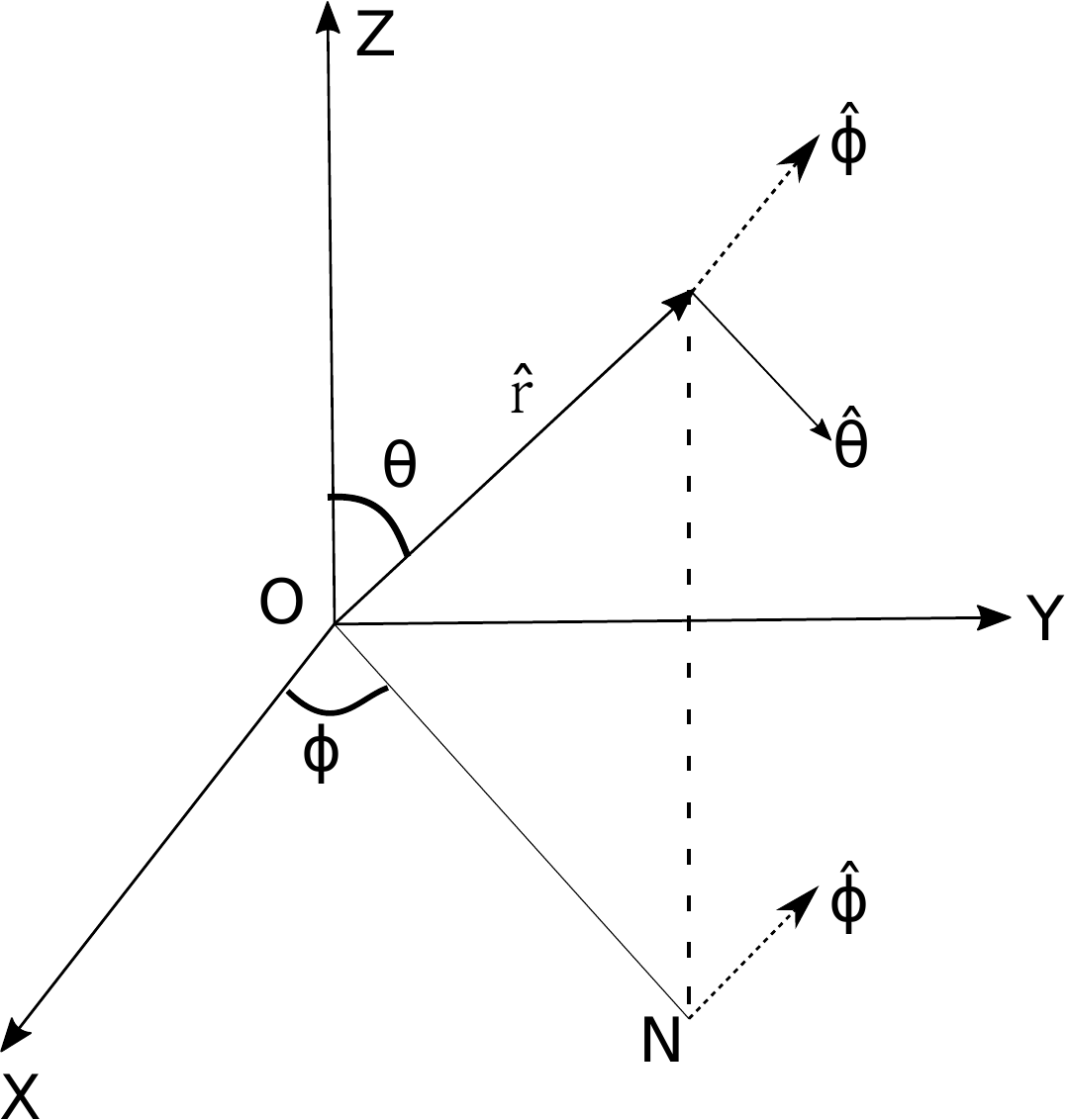}
	\begin{center}
		Fig. 3.2
	\end{center}
\end{wrapfigure}

Let $\hat{r}$ be the unit vector along the radial direction $OP$, $\hat{\theta}$ be the unit vector $\perp$ to $OP$ but in the meridian plane and $\hat{\phi}$ be the unit vector perpendicular to the meridian plane as shown in the figure.
$$\vec{r}=r~\hat{r}$$
$$\therefore~\vec{v}=\frac{\mathrm{d}\vec{r}}{\mathrm{d}t}=\dot{r}\hat{r}+r\frac{\mathrm{d}\hat{r}}{\mathrm{d}t}$$.

Here $\dfrac{\mathrm{d}\hat{r}}{\mathrm{d}t}$ is orthogonal to $\hat{r}$, so it will be in the plane of $\hat{\theta}$ and $\hat{\phi}$. Hence we have
$$\frac{\mathrm{d}\hat{r}}{\mathrm{d}t}=\dot{\theta}\hat{\theta}+\sin\theta\dot{\phi}\hat{\phi}$$
$$\therefore~\vec{v}=\frac{\mathrm{d}\vec{r}}{\mathrm{d}t}=\dot{r}\hat{r}+r\dot{\theta}\hat{\theta}+r\sin\theta\dot{\phi}\hat{\phi}$$
\begin{equation}
	\vec{f}=\frac{\mathrm{d}\vec{v}}{\mathrm{d}t}=\ddot{r}\hat{r}+\dot{r}\frac{\mathrm{d}\hat{r}}{\mathrm{d}t}+\left(\dot{r}\dot{\theta}+r\ddot{\theta}\right)\hat{\theta}+r\dot{\theta}\frac{\mathrm{d}\hat{\theta}}{\mathrm{d}t}+\left(\dot{r}\sin\theta\dot{\phi}+r\cos\theta\dot{\theta}\dot{\phi}+r\sin\theta\ddot{\phi}\right)\hat{\phi}+r\sin\theta\dot{\phi}\frac{\mathrm{d}\hat{\phi}}{\mathrm{d}t}\nonumber
\end{equation}

As before, $\dfrac{\mathrm{d}\hat{\theta}}{\mathrm{d}t}$ and $\dfrac{\mathrm{d}\hat{\phi}}{\mathrm{d}t}$ are respectively perpendicular to $\hat{\theta}$ and $\hat{\phi}$. So they are respectively the linear combination of ($\hat{r}$, $\hat{\phi}$) and ($\hat{r}$, $\hat{\theta}$) as follows:
$$\frac{\mathrm{d}\hat{\theta}}{\mathrm{d}t}=-\dot{\theta}\hat{r}+\cos\theta\dot{\phi}\hat{\phi}$$
$$\frac{\mathrm{d}\hat{\phi}}{\mathrm{d}t}=-\sin\theta\dot{\phi}\hat{r}-\cos\theta\dot{\phi}\hat{\theta}$$
\begin{equation}
	\therefore~\vec{f}=\left[\ddot{r}-r\dot{\theta}^2-r\sin^2\theta\dot{\phi}^2\right]\hat{r}+\left[\frac{1}{r}\frac{\mathrm{d}}{\mathrm{d}t}\left(r^2\dot{\theta}\right)-r\sin\theta\cos\theta\dot{\phi}^2\right]\hat{\theta}+\frac{1}{r\sin\theta}\frac{\mathrm{d}}{\mathrm{d}t}\left(r^2\sin^2\theta\dot{\phi}\right)\hat{\phi}\nonumber
\end{equation}

\begin{wrapfigure}[20]{r}{0.32\textwidth}
	\centering	\includegraphics[height=5.5 cm , width=4.5 cm ]{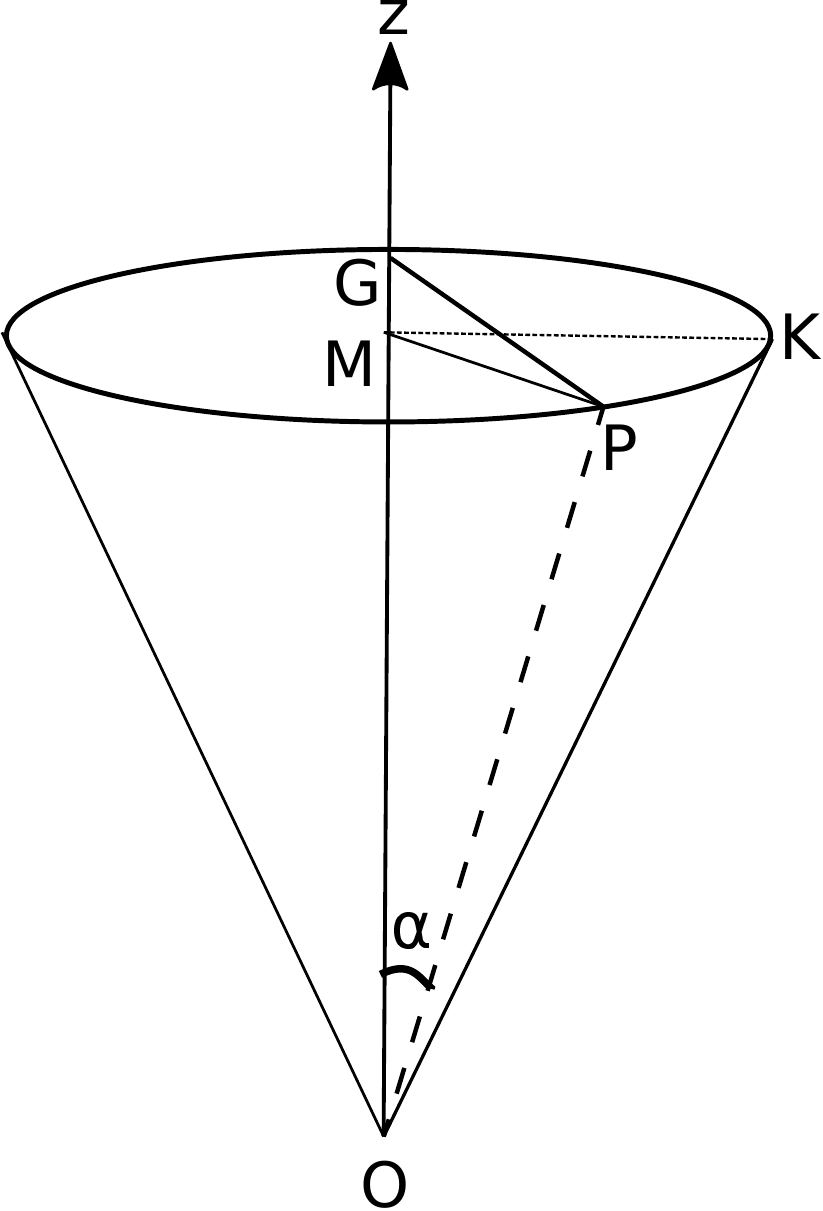}
	\begin{center}
		Fig. 3.3
	\end{center}
\end{wrapfigure}

$\bullet$ \textbf{Particular cases:}\\

\textbf{I. Motion of a particle on the surface of a cone}\\

In the figure $PG$ is orthogonal to $OP$. Let $OP=r$, $\theta=\alpha$, a constant

$$\dot{\theta}=0$$

$$v_r=\dot{r},~v_\theta=0,~v_\phi=r\sin\alpha\dot{\phi}$$

$$f_r=\ddot{r}-r\sin^2\alpha\dot{\phi}^2,~f_\theta=-r\sin\theta\cos\theta\dot{\phi}^2,~f_\phi=r\sin\alpha\ddot{\phi}+2\dot{r}\dot{\phi}\sin\alpha$$

\textbf{II. Motion of a particle on  the surface of a sphere}\\

Here $r=a$, a constant.

$$v_r=a,~v_\theta=a\dot{\theta},~v_\phi=a\sin\theta\dot{\phi}$$

$$f_r=-a\dot{\theta}^2-a\sin^2\theta\dot{\phi}^2~f_\theta=a\ddot{\theta}-a\sin\theta\cos\theta\dot{\phi}^2,~f_\phi=a\sin\theta\ddot{\phi}+2a\cos\theta\dot{\theta}\dot{\phi}$$

\section{Components of velocity and acceleration in cylindrical polar co-ordinates}

\begin{wrapfigure}[10]{r}{0.32\textwidth}
	\centering	\includegraphics[height=5 cm , width=4.5 cm ]{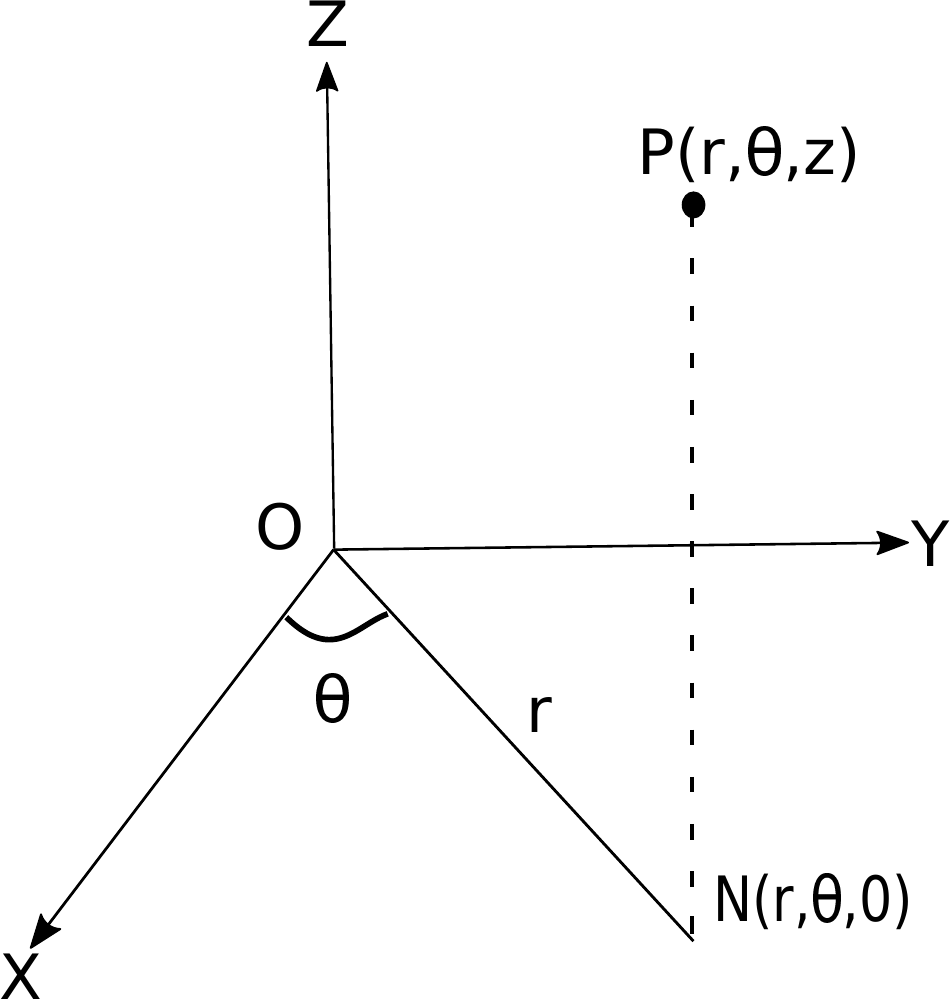}
	\begin{center}
		Fig. 3.4
	\end{center}
\end{wrapfigure}

Let $ON=r$, $\angle XON=\theta$.
\begin{eqnarray}
		\mbox{Velocity of }P&=&\mbox{velocity of }P~\parallel\mbox{to } XOY \mbox{ plane}\nonumber\\&&+\mbox{velocity of }P\perp\mbox{to the } XOY \mbox{ plane}\nonumber\\&=&\mbox{velocity of }N\mbox{ in the } XOY \mbox{ plane}\nonumber\\&&+\mbox{velocity of }P~\parallel\mbox{to } Z \mbox{ axis}\nonumber\\&=&\dot{r}\mbox{ along }ON+r\dot{\theta}\perp \mbox{ to }ON+\dot{z}\parallel \mbox{ to }Z\mbox{ axis}\nonumber
\end{eqnarray}
$$\therefore~~v_r=\dot{r},~v_\theta=r\dot{\theta}, v_z=\dot{z}$$

Similarly $f_r=\ddot{r}-r\dot{\theta}^2$, $f_\theta=\dfrac{1}{r}\dfrac{\mathrm{d}}{\mathrm{d}t}\left(r^2\dot{\theta}\right)$, $f_z=\ddot{z}$.

\textbf{Vector method}\\

$\vec{R}=\overrightarrow{OP}=\overrightarrow{ON}+\overrightarrow{NP}=r\hat{r}+z\hat{z}$
\begin{eqnarray}
\therefore~~\vec{v}&=&\frac{\mathrm{d}\vec{R}}{\mathrm{d}t}=\dot{r}\hat{r}+r\frac{\mathrm{d}\hat{r}}{\mathrm{d}t}+\dot{z}\hat{z}~~(\because~\hat{z}\mbox{ is a fixed direction})\nonumber\\&=&\dot{r}\hat{r}+r\dot{\theta}\hat{\theta}+\dot{z}\hat{z}\nonumber
	\end{eqnarray}
\begin{eqnarray}
	\vec{f}=\frac{\mathrm{d^2}\vec{R}}{\mathrm{d}t^2}&=&\ddot{r}\hat{r}+\dot{r}\frac{\mathrm{d}\hat{r}}{\mathrm{d}t}+\left(\dot{r}\dot{\theta}+r\ddot{\theta}\right)\hat{\theta}+r\dot{\theta}\frac{\mathrm{d}\hat{\theta}}{\mathrm{d}t}+\ddot{z}\hat{z}\nonumber\\&=&\ddot{r}\hat{r}+\dot{r}\dot{\theta}\hat{\theta}+\left(\dot{r}\dot{\theta}+r\ddot{\theta}\right)\hat{\theta}-r\dot{\theta}^2\hat{r}+\ddot{z}\hat{z}\nonumber\\&=&\left(\ddot{r}-r\dot{\theta}^2\right)\hat{r}+\left(2\dot{r}\dot{\theta}+r\ddot{\theta}\right)\hat{\theta}+\ddot{z}\hat{z}\nonumber
\end{eqnarray}

\section{Components of velocity and acceleration along tangent, principal normal and binormal of a particle moving along a space curve}

Let $\vec{r}=\vec{r}(s)$ be the space curve where $s$, the arc length is chosen as parameter
$$\vec{v}=\frac{\mathrm{d}\vec{r}}{\mathrm{d}t}=\frac{\mathrm{d}\vec{r}}{\mathrm{d}s}\dot{s}=\dot{s}\hat{t}$$
where $\hat{t}$ is the unit tangent vector to the space curve. The above result shows that the velocity of the particle is always along the tangent to the curve.
\begin{eqnarray}
	\vec{f}=\frac{\mathrm{d}\vec{v}}{\mathrm{d}t}&=&\ddot{s}\hat{t}+\dot{s}\frac{\mathrm{d}\hat{t}}{\mathrm{d}t}\nonumber\\&=&\ddot{s}\hat{t}+\dot{s}^2\frac{\mathrm{d}\hat{t}}{\mathrm{d}s}\nonumber\\&=&\ddot{s}\hat{t}+\dot{s}^2\kappa\vec{n}\mbox{~~~ (by Serret-Frenet formula)}\nonumber\\&=&\ddot{s}\hat{t}+\frac{\dot{s}^2}{\rho}\vec{n}\nonumber
\end{eqnarray}

Hence the acceleration vector has no component along the binormal direction. Similar to a plane curve the acceleration vector is always in the osculating plane of the curve.\\

\section{Angular velocity vector}

Let arc$PP'=\Delta s$, $\overrightarrow{OP}=\vec{r}$ and $\vec{\epsilon}$ is the unit vector along the axis of rotation. Suppose in small time $\Delta t$, $P$ goes to $P'$ such that $PP'=\Delta s$. Then the angular velocity
$$\Omega=\lim\limits_{\Delta t\rightarrow0}\frac{\Delta\theta}{\Delta t}=\frac{\mathrm{d}\theta}{\mathrm{d}t}=\dot{\theta}$$

Let $\vec{\omega}=\vec{\epsilon}\Omega$, is termed as the angular velocity vector at P about the axis of rotation, extending along the direction of the advanced right handed screw.\\

\begin{wrapfigure}[10]{r}{0.22\textwidth}
	\centering	\includegraphics[height=5 cm , width=3 cm ]{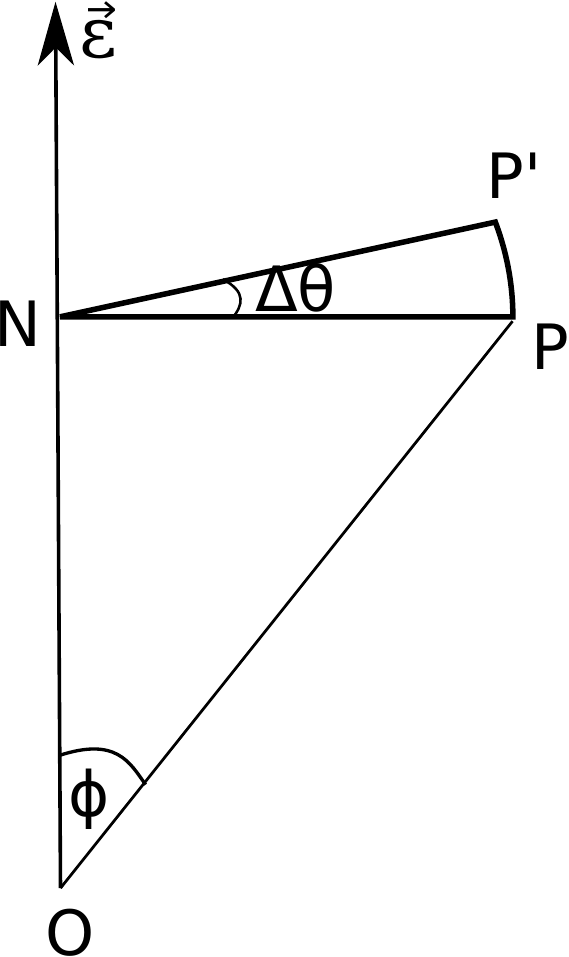}
	\begin{center}
		Fig. 3.5
	\end{center}
\end{wrapfigure}

Let $\vec{u}$ be the linear velocity at P due to rotation, then
\begin{minipage}{.3\textheight}
\begin{eqnarray}
	\vec{u}&=& NP~\dot{\theta}\frac{\vec{\epsilon}\times\vec{r}}{|\vec{\epsilon}\times\vec{r}|}\nonumber\\&=&\dot{\theta}\left(\vec{\epsilon}\times\vec{r}\right)\nonumber\\&=&\Omega\left(\vec{\epsilon}\times\vec{r}\right)\nonumber\\&=&\vec{\omega}\times\vec{r}\nonumber
\end{eqnarray}
\end{minipage}
\begin{minipage}{.2\textheight}
	\begin{eqnarray}
		|\vec{\epsilon}\times\vec{r}|&=&|\vec{\epsilon}||\vec{r}|\sin\phi\nonumber\\&=&r\sin\phi~=~NP\nonumber
	\end{eqnarray}
\end{minipage}

\bigskip

\section{Moving axes in three dimensions}

Let $P$ be a point whose co-ordinates are $(x,y,z)$ referred to $OX$, $OY$, $OZ$ as axes. Consider $\vec{i}$, $\vec{j}$ and $\vec{k}$ be the unit vectors along the $x$, $y$ and $z$ axes 
$$\overrightarrow{OP}=\vec{r}=\xi\vec{i}+\eta\vec{j}+\zeta\vec{k}$$
\begin{wrapfigure}[7]{r}{0.35\textwidth}
	\centering	\includegraphics[height=5 cm , width=5.3 cm ]{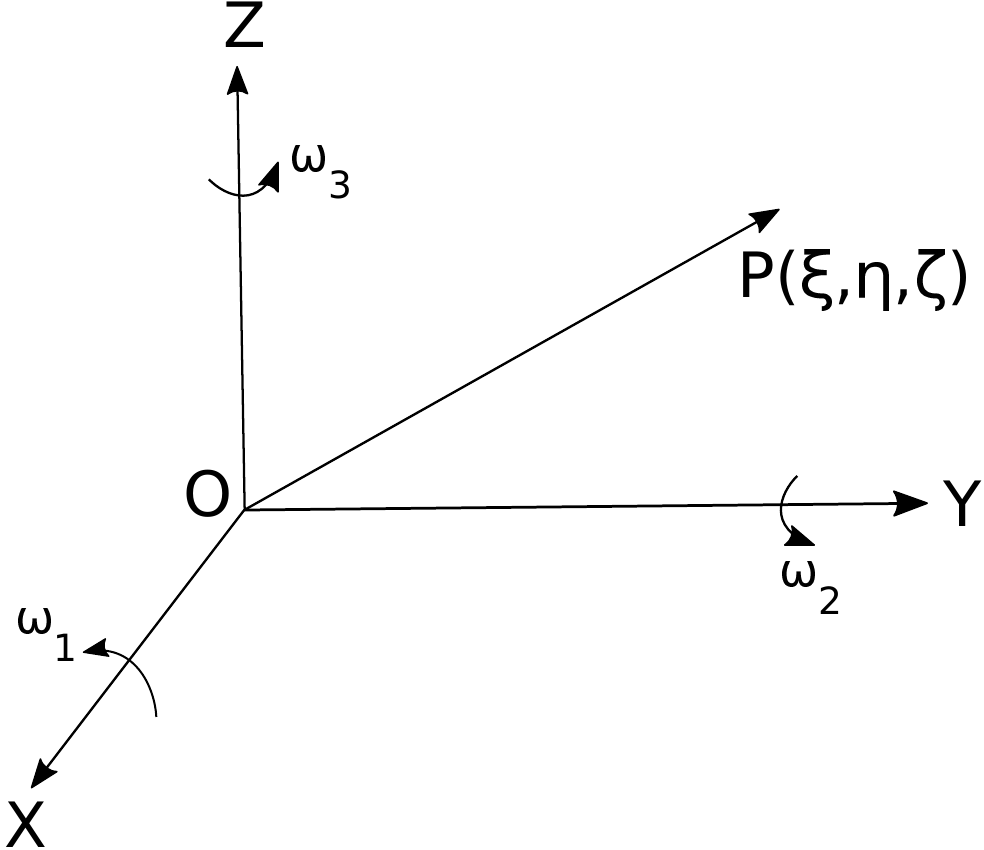}
	\begin{center}
		Fig. 3.6
	\end{center}
\end{wrapfigure}
\begin{equation}
\therefore~~	\vec{v}=\frac{\mathrm{d}\vec{r}}{\mathrm{d}t}=\left(\dot{\xi}\vec{i}+\dot{\eta}\vec{j}+\dot{\zeta}\vec{k}\right)+\left(\xi\frac{\mathrm{d}\vec{i}}{\mathrm{d}t}+\eta\frac{\mathrm{d}\vec{j}}{\mathrm{d}t}+\zeta\frac{\mathrm{d}\vec{k}}{\mathrm{d}t}\right)\nonumber
\end{equation}

Note that $\dfrac{\mathrm{d}\vec{i}}{\mathrm{d}t}$ is a vector whose components are the components of the velocity of the extremity of the unit vector $\vec{i}$ drawn from $O$. Suppose $\omega_1$, $\omega_2$ and $\omega_3$ be the components of the angular velocity about $OX$, $OY$ and $OZ$ respectively. Then $\dfrac{\mathrm{d}\vec{i}}{\mathrm{d}t}$ is given by
\begin{equation}
	\frac{\mathrm{d}\vec{i}}{\mathrm{d}t}=\begin{array}{|ccc|}
		\vec{i}&\vec{j}&\vec{k}\\\omega_1&\omega_2&\omega_3\\1&0&0
	\end{array}=\omega_3\vec{j}-\omega_2\vec{k}\nonumber
\end{equation}

Similarly $\dfrac{\mathrm{d}\vec{j}}{\mathrm{d}t}=\omega_1\vec{k}-\omega_3\vec{j}$ and $\dfrac{\mathrm{d}\vec{j}}{\mathrm{d}t}=\omega_2\vec{i}-\omega_1\vec{j}$.\\

$\therefore~~v_x=\dot{x}-\omega_3y+\omega_2z$, $v_y=\dot{y}-\omega_1z+\omega_3x$, $v_z=\dot{z}-\omega_2x+\omega_1y$.\\

(Note that we replace $\left(\xi,\eta,\zeta\right)$ by $\left(x,y,z\right)$).\\

$f_x=\dot{v}_x-\omega_3v_y+\omega_2v_z$, $f_y=\dot{v}_y-\omega_1v_z+\omega_3v_x$, $f_z=\dot{v}_z-\omega_2v_x+\omega_1v_y$.\\
\newpage
\section{Applications}
\subsection{Spherical Polar Co-ordinates}

\begin{wrapfigure}[13]{r}{0.35\textwidth}
	\centering	\includegraphics[height=5 cm , width=5.3 cm ]{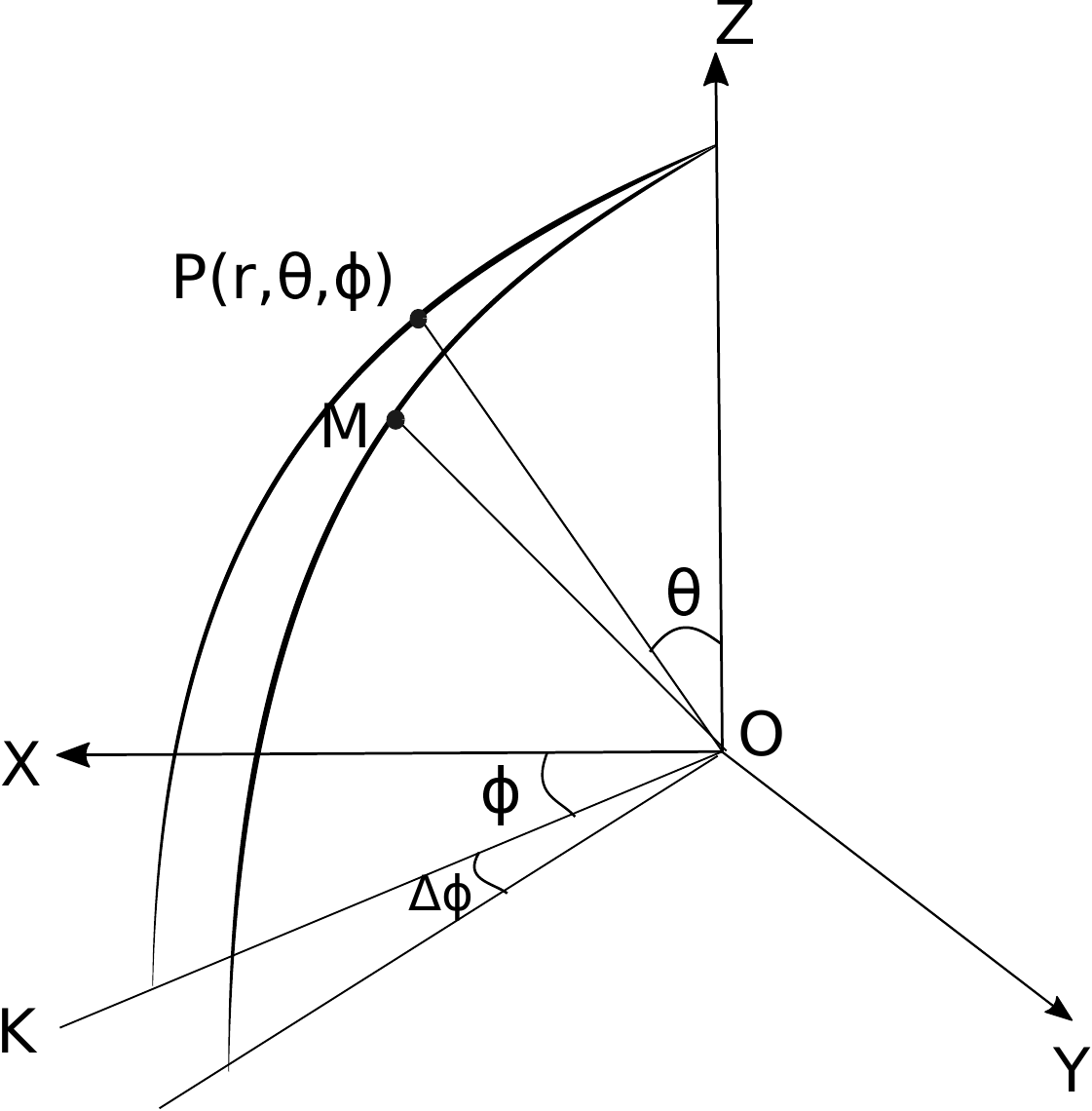}
	\begin{center}
		Fig. 3.7
	\end{center}
\end{wrapfigure}

Let the origin $O$ be fixed and the axes are rotating. Let $OP=r$, where $P\left(r,\theta,\phi\right)$ be the spherical polar co-ordinates. Suppose $M$ be the position of $P$ after a rotation in the increasing direction of $\theta$ and $\phi$. So the co-ordinates of $M$ be $\left(r,\theta+\mathrm{d}\theta,\phi+\mathrm{d}\phi\right)$. The angular velocity of the system is equivalent to $\dot{\phi}$ about $OZ$ together with $\dot{\theta}$ about an axis in the direction of $\phi$.. If $\left(\omega_1,\omega_2,\omega_3\right)$ be the component of angular velocity about $r$, $\theta$, $\phi$ direction, then we have
$$\omega_r=\dot{\phi}\cos\theta,~\omega_\theta=\dot{\phi}\sin\theta,~\omega_\phi=\dot{\theta}$$

So referred to $\left(r,\theta,\phi\right)$ as instantaneous axes of rotation, the co-ordinates of P are $(r,0,0)$.Thus if $\left(u_r,u_\theta,u_\phi\right)$ be the component of velocity along $r$, $\theta$ and $\phi$ direction, then
\begin{eqnarray}
	&&\left(u_r,u_\theta,u_\phi\right)=\left(\dot{r},0,0\right)+\begin{array}{|ccc|}
		\hat{r}&\hat{\theta}&\hat{\phi}\\\omega_r&\omega_\theta&\omega_\phi\\r&0&0
	\end{array}\nonumber\\\implies&& u_r=\dot{r},~u_\theta=r\dot{\theta},~u_\phi=r\sin\theta\dot{\phi}\nonumber
\end{eqnarray}

Similarly, for the acceleration vector,
\begin{eqnarray}
	&&\vec{f}=\left(f_r,f_\theta,f_\phi\right)=\left(\dot{u}_r,\dot{u}_\theta,\dot{u}_\phi\right)+\begin{array}{|ccc|}
		\hat{r}&\hat{\theta}&\hat{\phi}\\\omega_r&\omega_\theta&\omega_\phi\\u_r&u_\theta&u_\phi
	\end{array}\nonumber\\\implies&& f_r=\ddot{r}-r\dot{\theta}^2-r\sin^2\theta\dot{\phi}^2,~f_\theta=\frac{1}{r}\frac{\mathrm{d}}{\mathrm{d}t}\left(r^2\dot{\theta}\right)-r\sin\theta\cos\theta\dot{\phi}^2,~f_\phi=\frac{1}{r\sin\theta}\frac{\mathrm{d}}{\mathrm{d}t}\left(r^2\sin^2\theta\dot{\phi}\right)\nonumber
\end{eqnarray}

\subsection{Cylindrical co-ordinates}

\begin{wrapfigure}[10]{r}{0.35\textwidth}
	\centering	\includegraphics[height=5 cm , width=5.3 cm ]{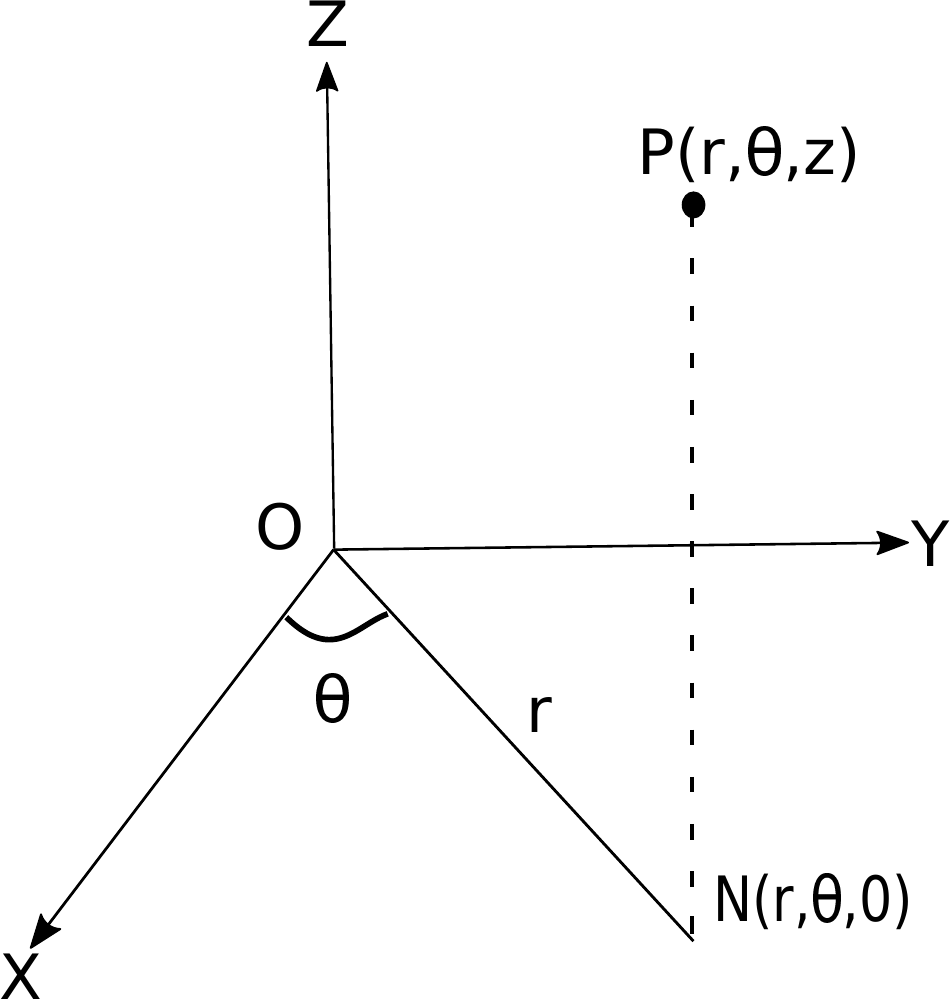}
	\begin{center}
		Fig. 3.8
	\end{center}
\end{wrapfigure}

In this case if $\hat{r}$, $\hat{\theta}$ and $\hat{z}$ be the unit vectors and $\omega_r$, $\omega_\theta$ and $\omega_z$ be the angular velocities along the axes $r$, $\theta$ and $z$, then
$$\omega_r=0=\omega_\theta,~\omega_z=\dot{\theta}$$

Now referred to $\left(r,\theta,z\right)$ as the instantaneous set of axes, co-ordinates of $P$ are $(r,0,z)$, the components of velocity are
\begin{eqnarray}
	&&\vec{v}=\left(u_r,u_\theta,u_z\right)=\left(\dot{r},0,\dot{z}\right)+\begin{array}{|ccc|}
		\hat{r}&\hat{\theta}&\hat{z}\\\omega_r&\omega_\theta&\omega_z\\r&0&z
	\end{array}\nonumber\\\implies&& u_r=\dot{r},~u_\theta=r\dot{\theta},~u_\phi=\dot{z}\nonumber
\end{eqnarray}

Similarly for acceleration vector,
\begin{eqnarray}
	&&\vec{f}=\left(f_r,f_\theta,f_z\right)=\left(\dot{u}_r,\dot{u}_\theta,\dot{u}_z\right)+\begin{array}{|ccc|}
		\hat{r}&\hat{\theta}&\hat{z}\\\omega_r&\omega_\theta&\omega_z\\u_r&u_\theta&u_z
	\end{array}\nonumber\\\implies&& f_r=\ddot{r}-r\dot{\theta}^2,~f_\theta=\frac{1}{r}\frac{\mathrm{d}}{\mathrm{d}t}\left(r^2\dot{\theta}\right),~f_z=\ddot{z}\nonumber
\end{eqnarray}

\section{Osculating  plane of a space curve}

Let $x=\phi_1(t)$, $y=\phi_2(t)$, $z=\phi_3(t)$ be the parametric equation of a space curve. We assume that $\phi_i(t),i=1,2,3$ are continuous in a certain range of $t$. Further, it is assumed that all points of the curve (within the given range) are arbitrary points so that $\phi'_i(t)$ and $\phi''_i(t)$ exist and continuous.\\

Consider a point $P_0$ on the curve. Suppose $P_1$, $P_2$ are two neighbouring points of $P_0$. Draw a plane through the points $P_0$, $P_1$ and $P_2$. The limiting position of the plane as $P_1$, $P_2\rightarrow P_0$ along the curve is called the osculating plane of the curve at $P_0$.\\

\subsection{Analytic form of the osculating plane}

As points on the space curve are ordinary points so $\phi'_1(t)=0=\phi'_2(t)=\phi'_3(t)$ will not hold simultaneously. Let $t_0$, $t_1$ and $t_2$ be the value of the parameter at the points $P_0$ and the neighbouring points $P_1$ and $P_2$ respectively. Suppose
\begin{equation}\label{eq3.1}
	ax+by+cz+d=0
\end{equation}
represent a plane in the three dimensional space, where $a$, $b$, $c$ and $d$ are constants. Let us define
\begin{equation}
	F(t)=a\phi_1(t)+b\phi_2(t)+c\phi_3(t)+d
\end{equation}

If we suppose that the plane (\ref{eq3.1}) passes through the points $P_0$, $P_1$ and $P_2$ then we have
\begin{equation}
	F(t_0)=0=F(t_1)=F(t_2)
\end{equation}

Note that $F(t)$ is a continuous function of $t$ having second order continuous derivatives. Now applying Role's theorem on $\left[t_0,t_1\right]$ and $\left[t_2,t_0\right]$ (assuming $t_2<t_0<t_1$) we have
\begin{equation}
	F'(\tau_1)=0 \mbox{ ~and ~}F'(\tau_2)=0
\end{equation}
where $t_0<\tau_1<t_1$ and $t_2<\tau_2<t_0$. Further, application  of Rolle's theorem on $F'(t)$ in $\left[\tau_2,\tau_1\right]$, we obtain $F''(\tau_3)=0$, $\tau_2<\tau_3<\tau_1$. Now due to limiting process $(P_1,P_2)\rightarrow P_)$ i.e. $(t_1,t_2)\rightarrow t_0$, we have $(\tau_1,\tau_2)\rightarrow t_0$ and also $\tau_3\rightarrow t_0$. Hence due to the conditioning of $F'(t)$ and $F''(t)$ we must have
\begin{equation}\label{eq3.5}
	F(t_0)=0=F'(t_0)=F''(t_0)
\end{equation}

As $P_0$ is any point on the plane (\ref{eq3.1}) so the result in equation (\ref{eq3.5}) holds for all points on the plane (\ref{eq3.1}) i.e.
\begin{equation}\label{eq3.6}
	F(t)=0=F'(t)=F''(t)
\end{equation}

Hence we have
\begin{eqnarray}
	ax+by+cz+d=0\nonumber\\ax'+by'+cz'~~~~=0\nonumber\\ax''+by''+cz''~~~=0
\end{eqnarray}

Thus solving the above relations one can determine $a$, $b$ and $c$  uniquely. Therefore, the osculating plane through a given point is always unique. We shall now determine the analytic form of the osculating plane through the point $P_0(x_0,y_0,z_0)$ as follows:\\

By notation, $x_0=x(t_0)$, $y_0=y(t_0)$, $z_0=z(t_0)$. As the plane (\ref{eq3.1})  passes through $P_0$ so we have
\begin{equation}\label{eq3.8}
	ax_0+by_0+cz_0+d=0
\end{equation}

Now, eliminating $d$  between  (\ref{eq3.1}) and (\ref{eq3.8})  we have
\begin{equation}\label{eq3.9}
	a(x-x_0)+b(y-y_0)+c(z-z_0)=0
\end{equation}

Also from relations (\ref{eq3.5}) (or (\ref{eq3.6}) ) we write
\begin{eqnarray}
	ax_0'+by_0'+cz_0'=0\\ax_0''+by_0''+cz_0''=0\label{eq3.11}
\end{eqnarray}

So eliminating $a$, $b$ and $c$ from equations (\ref{eq3.9}) - (\ref{eq3.11}) we obtain the equation of the osculating plane as
\begin{equation}
	\begin{array}{|ccc|}
		x-x_0&y-y_0&z-z_0\\x_0'&y_0'&z_0'\\x_0''&y_0''&z_0''
	\end{array}=0
\end{equation}

\subsection{Properties of the osculating plane}

The d.c. of the normal to the plane are proportional to $y_0'z_0''-z_0'y_0''$, $z_0'x_0''-x_0'z_0''$, $x_0'y_0''-y_0'x_0''$.

Now the d.c. of the tangent to the space curve at $P_0(x,y,z)$ are proportional to $\left(x_0',y_0',z_0'\right)$. Now, from the identity
$$x_0'\left(y_0'z_0''-z_0'y_0''\right)+y_0'\left(z_0'x_0''-x_0'z_0''\right)+z_0'\left(x_0'y_0''-y_0'x_0''\right)=0$$
it is evident that the tangent at $P_0$ to the space-curve lies on the osculating plane.\\

Let us now consider a line $L$ passing through $P_0$ having d.c. proportional to $\left(x_0'',y_0'',z_0''\right)$. Again from the identity 
$$x_0''\left(y_0'z_0''-z_0'y_0''\right)+y_0''\left(z_0'x_0''-x_0'z_0''\right)+z_0''\left(x_0'y_0''-y_0'x_0''\right)=0$$
we conclude that the line $L$ also lies on the osculating plane. Now to  find a relation between the tangent to the curve and the line $L$ we shall assume without any loss of generality that the arc length (measured from some fixed point) is taken as the parameter. So now $\left(x_0',y_0',z_0'\right)$ is the d.c. of the tangent (here $'$ stands for differentiation with respect to the arc length). Hence we write
\begin{eqnarray}
	\left(x_0'\right)^2+\left(y_0'\right)^2+\left(z_0'\right)^2=1\nonumber\\\mbox{i.e., }x_0'x_0''+y_0'y_0''+z_0'z_0''=0\nonumber
\end{eqnarray}
which shows that the line $L$ is orthogonal to the tangent line at $P_0$ but both of them lies on the osculating plane. Hence the line $L$ is directed along the principal normal to  the curve at $P_0$. Thus the osculating plane contains both the tangent line and the principal normal to the space curve at $P_0$.\\

\subsection{Interpretation of curvature}

We shall now give an interpretation of the curvature of the space-curve in analogy to a plane curve. Suppose with some fixed origin let $\vec{r}_0$ be the position vector of $P_0$ i.e.,
$$\vec{r}_0=x_0\vec{i}+y_0\vec{j}+z_0\vec{k}$$
where $\left(\vec{i},\vec{j},\vec{k}\right)$ are the unit vectors along the co-ordinate axes fixed in space. Choosing arc length as the parameter 
$$\vec{r}_0'=x_0'\vec{i}+y_0'\vec{j}+z_0'\vec{k}$$

Here $\vec{t}_1$ is the unit vector along the tangent to the space curve at $P_0$. So,
$$\vec{r}_0''=x_0''\vec{i}+y_0''\vec{j}+z_0''\vec{k}$$
is directed along the principal normal.\\

On the other hand, for the plane curve, 
$$\vec{r}_0''=x_0''\vec{i}+y_0''\vec{j}$$

Also, $	\left(x_0'\right)^2+\left(y_0'\right)^2=1$

 i.e., $x_0'x_0''+y_0'y_0'''=0$ (differentiating with respect to $s$)\\
 
 So we write
 \begin{eqnarray}
 	&&\frac{x_0'}{y_0''}=\frac{y_0'}{-x_0'}=\frac{x_0'y_0''-y_0'x_0''}{x_0''^2+y_0''^2}=\frac{\sqrt{x_0'^2+y_0'^2}}{\sqrt{x_0''^2+y_0''^2}}=\frac{1}{\sqrt{x_0''^2+y_0''^2}}\nonumber\\&\implies& x_0'y_0''-y_0'x_0''=\sqrt{x_0''^2+y_0''^2}=|\vec{r}_0''|\nonumber
 \end{eqnarray}

However, for a plane curve in parametric form, the curvature is expressed as
$$\frac{x'y''-y'x''}{\left(x'^2+y'^2\right)^{\frac{3}{2}}}=x'y''-y'x''=|\vec{r}_0''|$$

Hence the magnitude of $\vec{r}_0^{''}$ gives the curvature of a plane curve. Now generalizing this idea to space curve, we define the curvature at a point of a curve in space as the magnitude of the vector
$$\vec{r}_0^{''}=x_0''\vec{i}+y_0''\vec{j}+z_0''\vec{k}$$
i.e., $\kappa=\dfrac{1}{\rho}=|\vec{r}_0^{''}|$ (at $P_0$).

Thus the unit normal vector $\vec{n}$ can be defined as
$$\vec{n}=\rho\vec{r}_0^{''}=\frac{\vec{r}_0^{''}}{|\vec{r}_0^{''}|}$$
Consequently, the unit vector perpendicular to the osculating plane at $P_0$ is $$\vec{b}=\vec{t}_1\times\vec{n}$$
and is along the binormal vector of the curve at $P_0$.\\

\subsection{Curvature of a surface}

We now introduce the notion of curvature of a surface. Let us consider the section of aa surface by a plane parallel to an indefinitely near to tangent plane at any point (say $O$) on it. Clearly, the section will be a conic whose centre lies on the normal to the surface at $O$. This conic is called the \textit{indicatrix}.

\begin{wrapfigure}[13]{r}{0.35\textwidth}
	\centering	\includegraphics[height=5 cm , width=5.3 cm ]{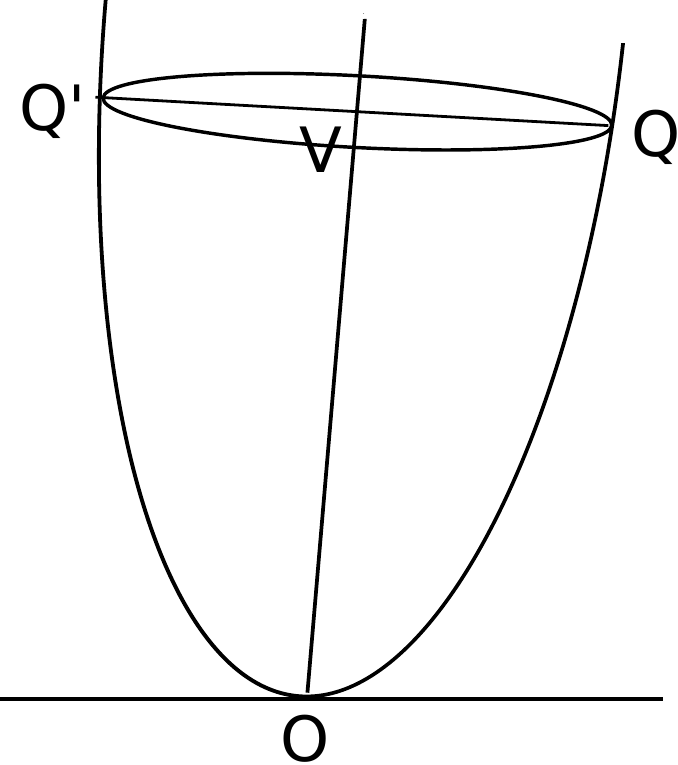}
	\begin{center}
		Fig. 3.9
	\end{center}
\end{wrapfigure}

Let any plane through the normal $OV$ at $O$ cut the indicatrix along the diameter $QVQ'$. Let $\rho$ be the radius of curvature at $O$ of the normal section. Then $$\rho=\lim\limits_{V\to O} \frac{{QV}^2}{2 OV}$$

Thus the radius of curvature of different normal sections varies as the square  of the diameter of the conic through that section.

Moreover, it is well known for a conic that the sum of the squares of the reciprocal of two perpendicular semi diameters is constant. This implies the sum of the reciprocal of radii of curvature of two perpendicular normal sections is constant. Also the diameter of a conic section has a maximum and a minimum value i.e. radius of curvature for different normal sections has a maximum and a minimum value. The sections having maximum and minimum value of the radius of curvature are called the principal sections and the corresponding radius of curvatures are the principal radius of curvature.\\

\subsection{Euler's Theorem}

\textbf{Statement:}\\
Suppose M be a surface in 3D Euclidian space and P is a point on M. A normal plane through P is a plane passing through P containing the normal vector to M. For each tangent vector $\vec{S}$ to the manifold at P, there exists a normal plane $M_X$ which intersects M in a curved having non-constant curvature $\kappa_X$. Let $\kappa_1=\kappa_{X_1}$ and $\kappa_2=\kappa_{X_2}$ are the largest and smallest possible curvatures then Euler's theorem state that corresponding  $\vec{X}_1$ and $\vec{X}_2$ are orthogonal to each other and if $\theta$ be the angle between $\vec{X}_1$ and $\vec{X}_2$ then $\kappa_X= \kappa_1 cos^2 \theta + \kappa_2~ sin^2\theta $ $i.e.,$ $1/\rho=cos^2 \theta/\rho_1 +sin^2 \theta/\rho_2$,
with $\rho,~ \rho_1$ and $\rho_2$ are the corresponding radii of the curvature.

\textbf{Proof:} We choose $XY$ plane as the tangent plane to the surface at $O$ and $z$ axis is along the normal to the surface at $O$. Suppose $x$ and $y$ axes are taken along the axes of the indicatrix. So upto second order the equation of the surface (i.e. neglecting 3rd and higher order terms) takes the form
$$2z=ax^2+by^2$$
Suppose $\rho_1$ and $\rho_2$ be the principal radii of curvature at $O$ i.e.
$$\rho_1=\lim\limits_{\substack{x\to0\\y\to0}}\frac{x^2}{2z}=\frac{1}{a},~\rho_2=\lim\limits_{\substack{x\to0\\y\to0}}\frac{y^2}{2z}=\frac{1}{b}$$
\begin{wrapfigure}[13]{r}{0.35\textwidth}
	\centering	\includegraphics[height=5 cm , width=5.3 cm ]{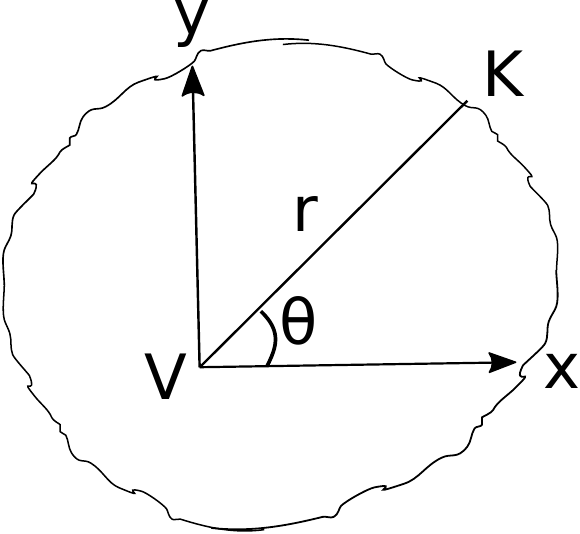}
	\begin{center}
		Fig. 3.10
	\end{center}
\end{wrapfigure}
So the above equation of the surface becomes
$$2z=\frac{x^2}{\rho_1}+\frac{y^2}{\rho_2}$$

If the diameter $VK$ makes an angle $\theta$ with $x$ direction, then 
\begin{eqnarray}
	2z&=&\frac{r^2\cos^2\theta}{\rho_1}+\frac{r^2\sin^2\theta}{\rho_2}\nonumber\\\mbox{i.e. }\frac{2z}{r^2}&=&\frac{\cos^2\theta}{\rho_1}+\frac{\sin^2\theta}{\rho_2}\nonumber
\end{eqnarray}

Hence if $\rho$ be the radius of curvature of the normal section of the surface through the diameter $VK$, then
$$\rho=\lim \frac{VK^2}{2OY}$$
i.e. in the limit $2\rho OV=VK^2=r^2$

i.e. $\dfrac{2z}{r^2}=\dfrac{1}{\rho}$ in the limit.

$\therefore$ $\dfrac{1}{\rho}=\dfrac{\cos^2\theta}{\rho_1}+\dfrac{\sin^2\theta}{\rho_2}$, which is the Euler's result.\\

\subsection{Meusnier's Theorem}

If $P_0$ and $P$ be the radii of curvature of a normal section and an oblique section of a surface through the same tangent line then $\rho=\rho_0\cos\theta$, $\theta$ is the angle between the sections.
\begin{wrapfigure}[16]{r}{0.35\textwidth}
	\centering	\includegraphics[height=6 cm , width=5.3 cm ]{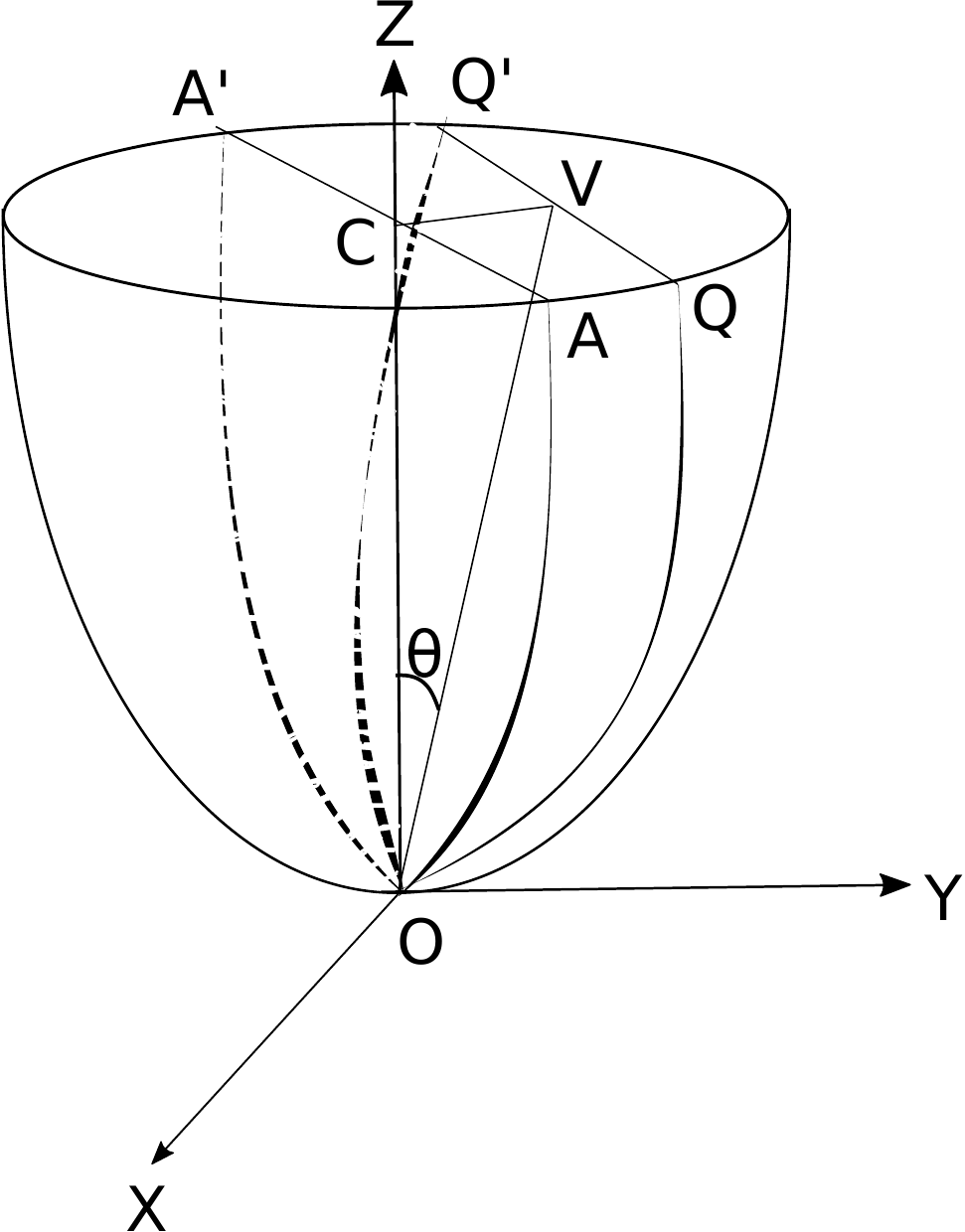}
	\begin{center}
		Fig. 3.11
	\end{center}
\end{wrapfigure}

\textbf{Proof:} Let $XOY$ plane be taken as the tangent plane to the surface at $O$, $z$ axis is along the normal to the surface at $O$. The $x$ axis is taken along the common tangent.\\

Let $OC=h$. The equation of the indicatrix, when third and higher order terms are neglected is given by
$$z=h,~2h=rx^2+2sxy+ty^2,$$ so that
$\rho_0=\lim\dfrac{CA^2}{2OC}=\dfrac{1}{r}$ (as equation of the normal section is $y=0$.

Equation of $QVQ'$ is $z=h$, $y=h\tan\theta$, so for points of intersection of $QVQ'$ with the surface  we have
$$2h=rx^2+2sxh\tan\theta+th^2\tan^2\theta$$

Note that if $x$, $y$ are 1st order quantities then $h$ (and also $z$) is a second order quantity. So neglecting $h^2$, $hx$ terms we get (upto 2nd order of smallness)
$$2h=rx^2\mbox{~ i.e. ~} QV^2=\frac{2h}{r}$$
$\therefore$ $\rho=\lim\limits_{h\to 0}\dfrac{QV^2}{2OV}=\lim\limits_{h\to0}\dfrac{\frac{2h}{r}}{2h\sec\theta}=\dfrac{\cos\theta}{r}=\rho_0\cos\theta$

\section{Motion of a particle on a fixed smooth surface}

\begin{wrapfigure}[13]{r}{0.35\textwidth}
	\centering	\includegraphics[height=5 cm , width=5.3 cm ]{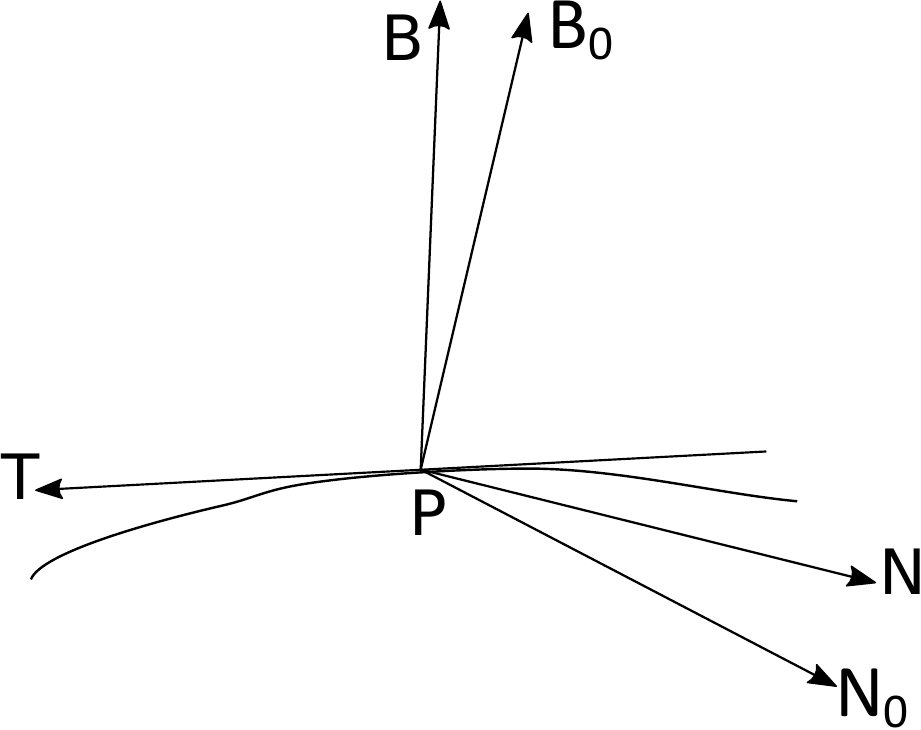}
	\begin{center}
		Fig. 3.12
	\end{center}
\end{wrapfigure}

Let $m$ be the mass of the particle $P$ moving on a fixed smooth surface with velocity $v$ at any time $t$. Suppose $PT$, $PN$, $PB$ be the tangent, principal normal and binormal to the path of the particle (on the surface) at $P$. Let $PN_0$ is the normal to the surface at $P$ and $PB_0$ is a tangent line on the surface at $P$ and is perpendicular to $PT$.\\

As $PT$  is perpendicular to $PN_0$, $PN$, $PB_0$, and $PB$ so they are all coplanar. Let $\chi$ be the angle which the osculating plane at $P$ makes with the plane normal to the surface and passing through $PT$. So we have $\angle N_0PN=\chi$.\\

Let $\rho$ be the radius of curvature of the path at $P$. Then the acceleration of the particle at $P$ has components $v\dfrac{\mathrm{d}v}{\mathrm{d}s}$ along $PT$ and $\dfrac{v^2}{\rho}$ along the principal normal $PN$.
$$\frac{v^2}{\rho}\mbox{ along }PN\equiv\frac{v^2}{\rho}\cos\chi\mbox{ along }PN_0+\frac{v^2}{\rho}\sin\chi\mbox{ along }PB_0$$

If $F$, $G$ and $H$ are the components of the external force acting on the particle $P$ along $PT$, $PN_0$ and $PB_0$ and $R$ be the reaction of the surface on the particle along $PN_0$, then the equation of motion of the particle are
$$mv\dfrac{\mathrm{d}v}{\mathrm{d}s}=F~,~~m\frac{v^2}{\rho}\cos\chi=G+R~,~~m\frac{v^2}{\rho}\sin\chi=H$$

Further, if $\rho_0$ be the radius of curvature of the normal section through $PT$, then by Meunier's theorem $\rho=\rho_0\cos\chi$, so that the equations of motion become
$$mv\dfrac{\mathrm{d}v}{\mathrm{d}s}=F~,~~m\frac{v^2}{\rho_0}=G+R~,~~m\frac{v^2}{\rho_0}\tan\chi=H$$

The first equation gives the velocity of the particle at any instant. Knowing $v$, the normal reation of the surface can be determined from the second equation.  The third equation gives the position of the osculating plane and also the differential equation of the path of the particle.\\

\textbf{Corollary:} If the external forces are absent then the equation of motion simplifies to
\begin{eqnarray}
	mv\dfrac{\mathrm{d}v}{\mathrm{d}s}=0\label{eq3.13}\\m\frac{v^2}{\rho_0}=R\label{eq3.14}\\m\frac{v^2}{\rho_0}\tan\chi=0\label{eq3.15}
\end{eqnarray}

The first equation shows that the particle moves on the surface with constant velocity. Equation (\ref{eq3.14}) gives the normal reaction of the surface on the particle. Equation (\ref{eq3.15}) implies $\chi=0$ (assuming $\dfrac{1}{\rho_0}\neq0$ i.e. the surface is not a plane surface) i.e. the osculating plane of the path of the particle contains the normal to the surface. So the path of the particle is a geodesic on the surface. Thus if a particle moves freely on a smooth surface then it describes a geodesic on the surface with constant velocity.\\

\textbf{Note:} The path of a particle on a smooth surface may be geodesic even in the presence of non-zero external forces. In particular, if the external forces be such that $H=0$ then the path of the particle will be a geodesic on the surface. Further, if a particle moves on a rough surface under no other external forces then the path of the particle will also be a geodesic and the velocity of the particle will gradually diminish until it comes to rest.\\

\subsection{Motion of a particle on a smooth surface of revolution}

\begin{wrapfigure}[13]{r}{0.35\textwidth}
	\centering	\includegraphics[height=5 cm , width=4 cm ]{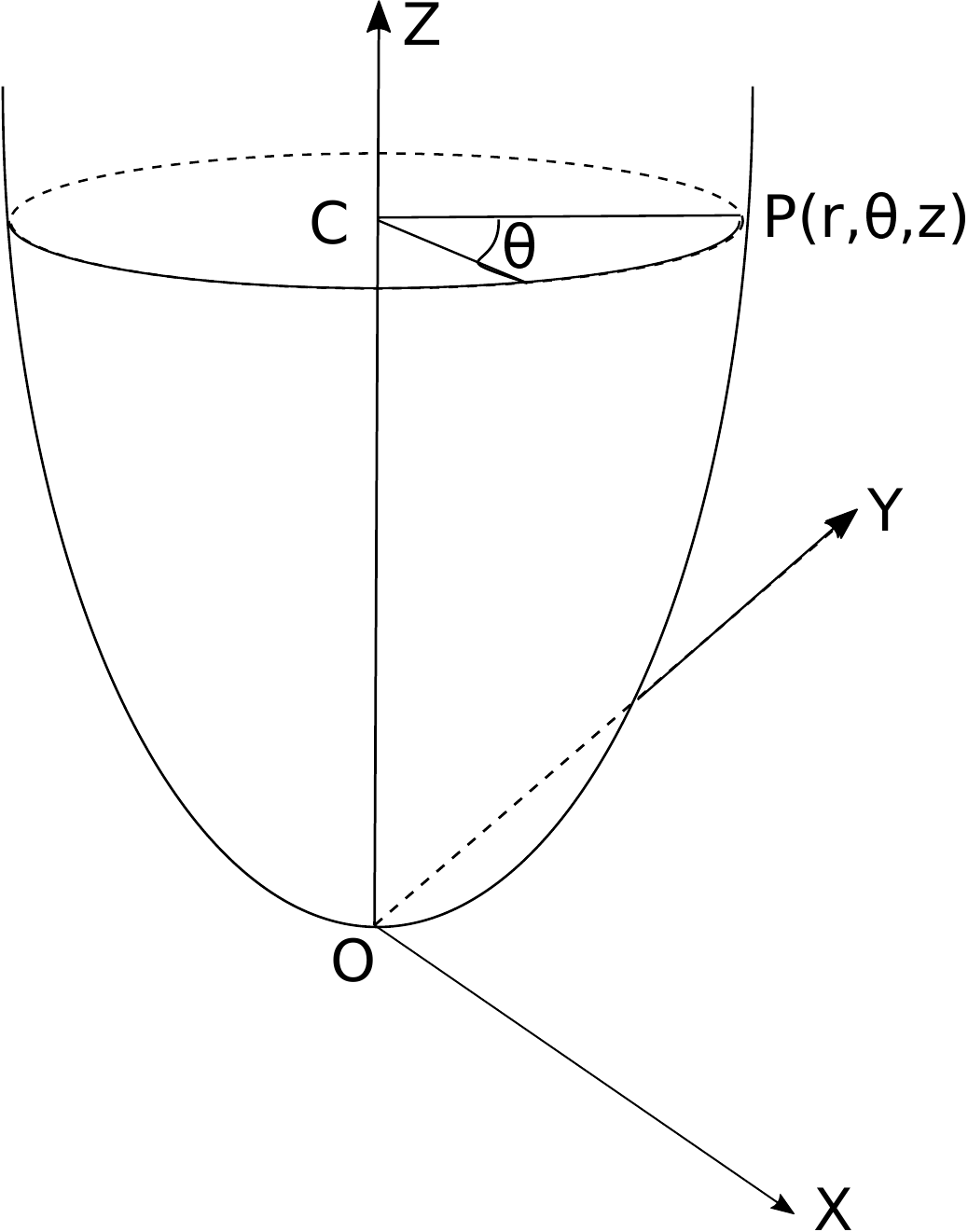}
	\begin{center}
		Fig. 3.13
	\end{center}
\end{wrapfigure}

Let the axis $oz$ be taken along the axis of revolution and let the equation of meridian section of the surface through the particle $P(r,\theta,z)$ at any time be
$$z=\phi(r),~r=\sqrt{x^2+y^2}=CP$$
Since the surface is smooth the reaction of the surface will be along the normal to the surface. Hence the work done by the reaction in any displacement of the particle on the surface is zero. So the equation of energy gives
\begin{equation}\label{eq3.16}
	\mathrm{d}\left(\frac{1}{2}mv^2\right)=X\mathrm{d}x+Y\mathrm{d}y+Z\mathrm{d}z
\end{equation}
where $m$ is the mass of the particle, $v$ is velocity at any time $t$, $X$, $Y$, $Z$ are the components of the external forces along $x$, $y$, $z$ axes. \\

As the reaction $R$ of the surface is acting along the normal to the surface, so $R$ lies on the meridian plane passing through the particle. Hence the moment of $R$ about the axis of revolution is zero. The components of the velocity of $P$ along $(r,\theta,z)$ direction are $\dot{r}$, $r\dot{\theta}$ and $\dot{z}$  respectively. As $\dot{r}$ and $\dot{z}$ lie on the meridian plane and $r\dot{\theta}$ is perpendicular to the meridian plane.  Hence the angular momentum of $P$ about $oz$ is $mr^2\dot{\theta}$. So by the principle of angular momentum
\begin{equation}\label{eq3.17}
	\frac{\mathrm{d}}{\mathrm{d}t}(mr^2 \dot{\theta})=-Xy+Yx=r(Y\cos\theta-X\sin\theta)
\end{equation}

 Thus equations (\ref{eq3.16}) and (\ref{eq3.17}) will determine the motion of the particle on the surface.\\
 
We now assume that the components of external forces are derived from a force function $U(r,z)$ where $U(r,z)$ is symmetrical about the axis of evolution.  Then we get from (\ref{eq3.16})
\begin{equation}\label{eq3.18}
	\frac{1}{2}mv^2=U+\frac{1}{2}mc.~~\left(\mbox{$c$, the constant of integration}\right)
\end{equation}

As the components of the external force along $r$, $\theta$, $z$ axes are $\dfrac{\partial U}{\partial r}$, $0$, $\dfrac{\partial U}{\partial z}$ respectively so the force lies on the meridian plane passing through the particle. Hence the moment of the external force about the axis of revolution is zero.  So from equation (\ref{eq3.17}) we get
$$\frac{\mathrm{d}}{\mathrm{d} t}\left(mr^2\dot{\theta}\right)=0$$
which an integration gives 
\begin{equation}\label{eq3.19}
	r^2\dot{\theta}=h, ~~\left(\mbox{a constant independent of $t$}\right)
\end{equation}

As the particle moves on the surface $z=\phi(r)$, so
$$\dot{z}=\frac{\mathrm{d}\phi}{\mathrm{d}r}\dot{r}=\phi'(r)\dot{r}.$$

Hence the velocity of the particle is given by
\begin{eqnarray}
	v^2&=&\dot{r}^2+r^2\dot{\theta}^2+z^2\nonumber\\&=&\left[1+\{\phi'(r)\}^2\right]\dot{r}^2+\frac{h^2}{r^2}\nonumber
\end{eqnarray}

So from (\ref{eq3.18}) i.e. $v^2=\dfrac{2U}{m}+c$
we obtain
$$\frac{\mathrm{d}r}{\mathrm{d}t}=\pm\sqrt{\frac{\frac{2U}{m}+c-\frac{h^2}{r^2}}{1+\{\phi'(r)\}^2}}$$
\hfill(the sign on the right hand side is to be chosen suitably)
\begin{equation}\label{eq3.20}
	\mbox{i.e. }\mathrm{d}t=\pm\sqrt{\frac{1+\{\phi'(r)\}^2}{\frac{2U}{m}+c-\frac{h^2}{r^2}}}~\mathrm{d}r
\end{equation}

Also from (\ref{eq3.19}) we get
\begin{equation}\label{eq3.21}
	\mathrm{d}\theta=\frac{h}{r^2}\mathrm{d}t=\pm\frac{h}{r^2}\sqrt{\frac{1+\{\phi'(r)\}^2}{\frac{2U}{m}+c-\frac{h^2}{r^2}}}~\mathrm{d}r
\end{equation}

Integrating (\ref{eq3.20}) and (\ref{eq3.21}) we get $t$ and $\theta$ as functions of $r$. Thus the solution of the problem of the motion of a particle on smooth surface of revolution under the action of external forces which can be derived from a force function symmetrical about the axis of revolution is reduced to two quadratures.\\

\subsection{Motion of a heavy particle on a smooth surface of revolution the axis of which is vertical}

We take the vertical axis of revolution as $z$ axes and use cylindrical co-ordinates $(r,\theta,z)$. If $v$ be the velocity of the particle at any time $t$ then
$$v^2=\dot{r}^2+r^2\dot{\theta}^2+\dot{z}^2$$

The energy equation gives $$v^2=c-2gz$$

$\left(\dfrac{\mathrm{d}}{\mathrm{d}t}\left(\dfrac{1}{2}mv^2\right)=-mg\mathrm{d}z,\mbox{ on integration, }\dfrac{1}{2}mv^2=-mgz+\dfrac{1}{2}mc\right)$

\begin{wrapfigure}[13]{r}{0.35\textwidth}
	\centering	\includegraphics[height=5 cm , width=4 cm ]{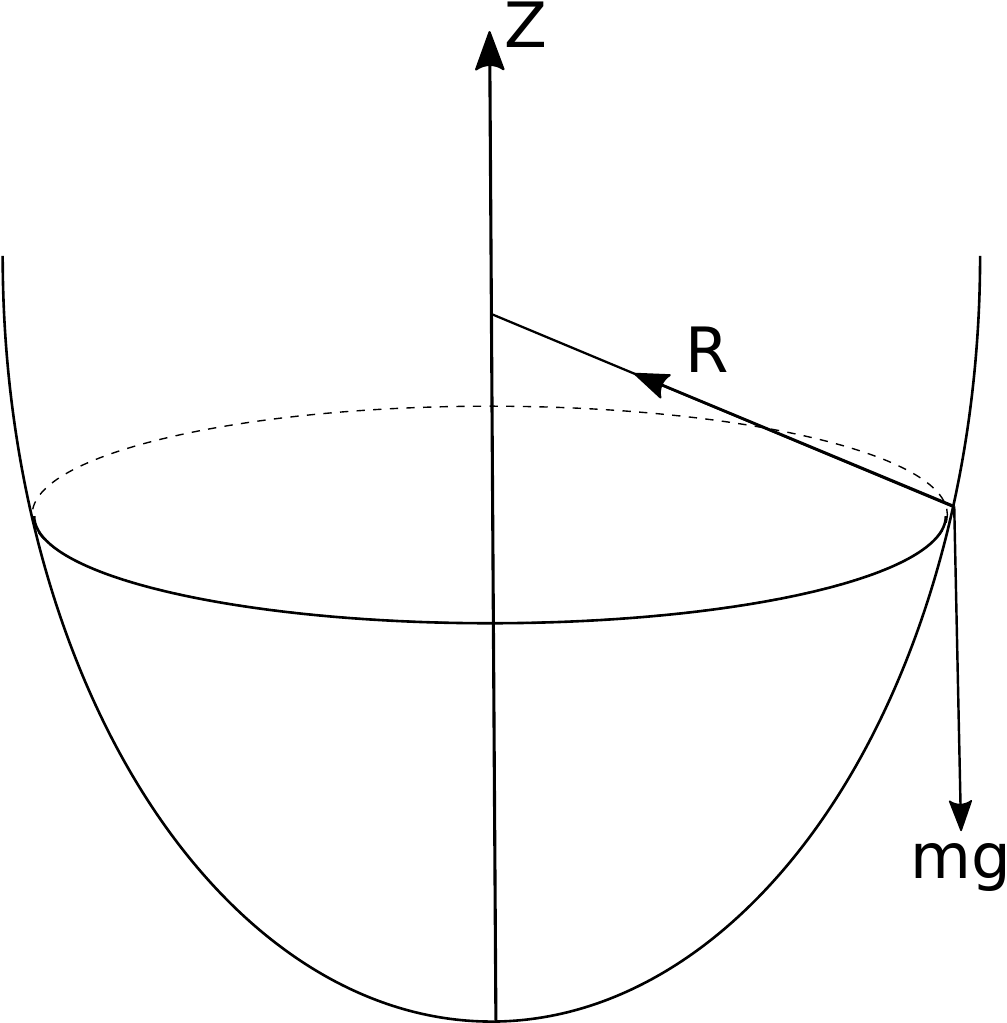}
	\begin{center}
		Fig. 3.14
	\end{center}
\end{wrapfigure}

Since the forces acting on the particle are the force of gravitation acting parallel to the axis (of revolution i.e., $z$ axis) and the normal reaction of the surface intersects the axis of revolution, hence the moment of momentum about the axis is a constant. Thus $r^2\dot{\theta}=h$, a constant. Let the equation of the meridian curve through the particle be $z=\phi(r)$, so that $\dot{z}=\phi'(r)\dot{r}$.\\

\begin{equation}
\therefore~v^2=\left(1+\{\phi'(r)\}^2\right)\dot{r}^2+\dfrac{h^2}{r^2}=c-2gz~.\nonumber
\end{equation}

$\left(1+\{\phi'(r)\}^2\right)\dot{r}^2=c-2gz-\dfrac{h^2}{r^2}$\\

Also $r^4\dot{\theta}^2=h^2$.\\

$\therefore~\dfrac{\left(1+\{\phi'(r)\}^2\right)}{r^4}\left(\dfrac{\mathrm{d}r}{\mathrm{d}\theta}\right)^2=\dfrac{c}{h^2}-\dfrac{2g}{h^2}z-\dfrac{1}{r^2}$\\

This is the differential equation of the path of the particle on the horizontal plane.\\

Let us now find the condition that the path of the particle on the surface is a circle of radius $r_0$ (say).\\

 Let $v_0$ be the velocity of the particle in its circular path at $P$. The acceleration of the particle is $\dfrac{v_0^2}{r_0}$ and it is maintained by its weight $mg$ and the normal reaction of the surface $R$. So we have 
  \begin{wrapfigure}[11]{r}{0.35\textwidth}
 	\centering	\includegraphics[height=4 cm , width=5 cm ]{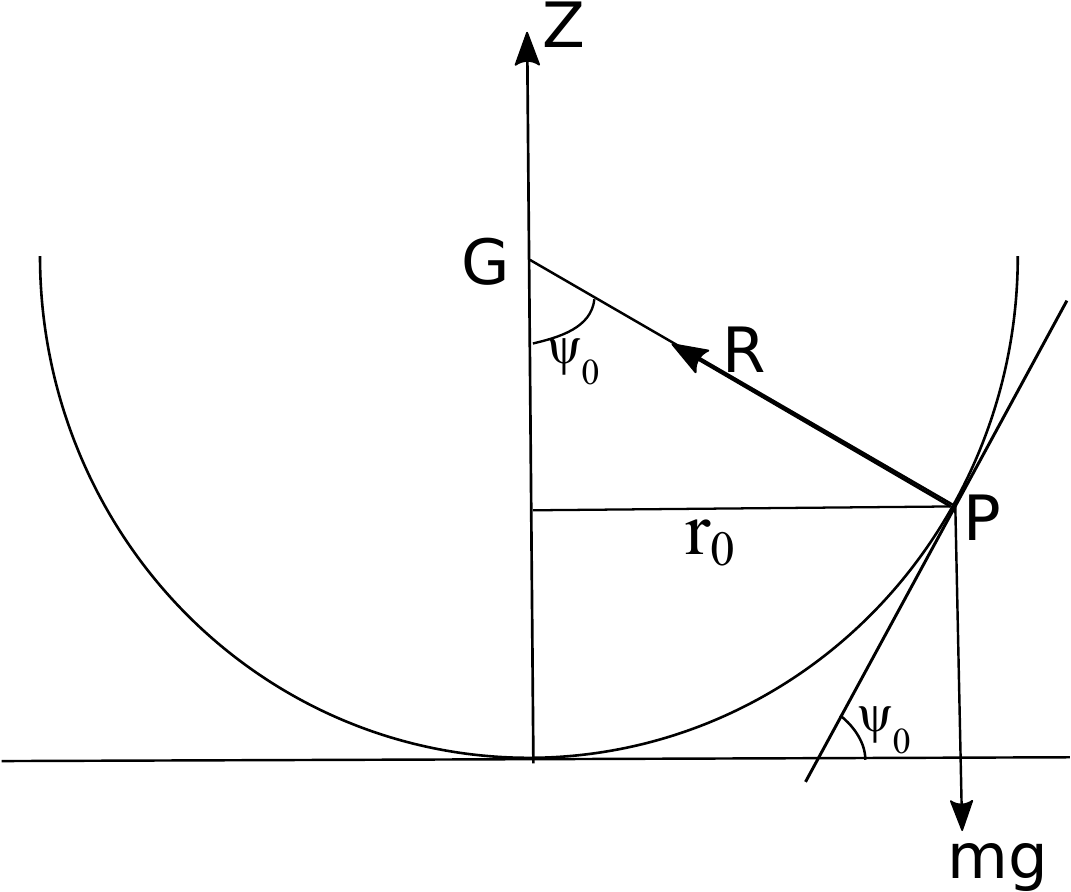}
 	\begin{center}
 		Fig. 3.15
 	\end{center}
 \end{wrapfigure}
 $$m\dfrac{v_0^2}{r_0}=R\sin\psi_0,~~mg=R\cos\psi_0~.$$
 
 Hence $\dfrac{v_0^2}{gr_0}=R\tan\psi_0$.\\
 
 But $\tan\psi_0=\dfrac{\mathrm{d}z}{\mathrm{d}r}{\bigg|}_{r=r_0}=\phi'(r_0)$.\\
 
 $\therefore~ \dfrac{v_0^2}{r_0}=\phi'(r_0)\mbox{ i.e., }v_0^2=gr_0\phi'(r_0)$.\\
 
 This is the condition that has to be satisfied to maintain a circular path.\\
 
 \subsection{Determine whether a heavy particle will rise or fall when it is projected horizontally on the surface of revolution}
 
 Let the vertical axis of revolution is taken as $z$ axis and $v_0$ be the initial velocity of projection from a point having $r$, $z$ co-ordinates $r_0$ and $z_0$ respectively. Suppose $r=f(z)$ be the equation of the meridian section of the surface then the energy equation gives
 $$v^2=c-2gz$$
 and the equation of angular momentum gives
 $$r^2\dot{\theta}=h$$
 
 Using initial conditions we have 
 $$v_0^2=c-2gz_0,~h=v_0r_0,~r_0=f(z_0)$$
 
 Thus we get $$v^2=v_0^2+2g(z_0-z)$$
 and $$r^2\dot{\theta}=v_0r_0$$
 
 But $v^2=\dot{r}^2+r^2\dot{\theta}^2+\dot{z}^2=\dot{z}^2\{1+[f'(z)]^2\}+\dfrac{v_0^2r_0^2}{r^2}$\\
 
 Thus we obtain
 \begin{eqnarray}
 	&&\dot{z}^2\{1+[f'(z)]^2\}+\dfrac{v_0^2r_0^2}{\{f(z)\}^2}=v_0^2+2g(z_0-z)\nonumber\\
 	\mbox{i.e., }&&\dot{z}^2\{1+[f'(z)]^2\}=v_0^2\left[1-\left\{\frac{f(z_0)}{f(z)}\right\}^2\right]+2g(z_0-z)=F(z)\mbox{ (say)}\nonumber
 \end{eqnarray}

Hence the particle will rise or fall if $\dot{z}\neq0$ and hence $F(z)>0$ with $F(z_0)=0$.\\

Using mean value theorem of differential calculus on $f(z)$ we obtain
\begin{eqnarray}
	&&\frac{F(z)-F(z_0)}{z-z_0}=F'\{z_0+(z-z_0)\theta\},~~0<\theta<1\nonumber\\
	\mbox{i.e., }&&F(z)=(z-z_0)F'\{z_0+(z-z_0)\theta\}>0\nonumber
\end{eqnarray}

This shows that $z-z_0$ and $F'\{z_0+(z-z_0)\theta\}$ will have same sign, however small $z-z_0$ may be. The particle will rise or fall according as $z-z_0$ is positive or negative i.e., according as $F'\{z_0+(z-z_0)\theta\}$ is positive or negative. Now making $z\to z_0$, the particle will rise or fall according as $F'(z_0)$ is positive or negative.\\

Now, $F'(z)=2\dfrac{\{f(z_0)\}^2}{\{f(z)\}^3}f'(z)v_0^2-2g$\\

So $F'(z_0)=2\left[\dfrac{f'(z_0)}{f(z_0)}v_0^2-g\right]$\\

Hence the particle will rise or fall according as
$$\dfrac{f'(z_0)}{f(z_0)}v_0^2~>\mbox{ or }<g$$
\vspace{4.5cm}

\subsection{Initial radius of curvature}

  \begin{wrapfigure}[11]{r}{0.35\textwidth}
	\centering	\includegraphics[height=4 cm , width=5 cm ]{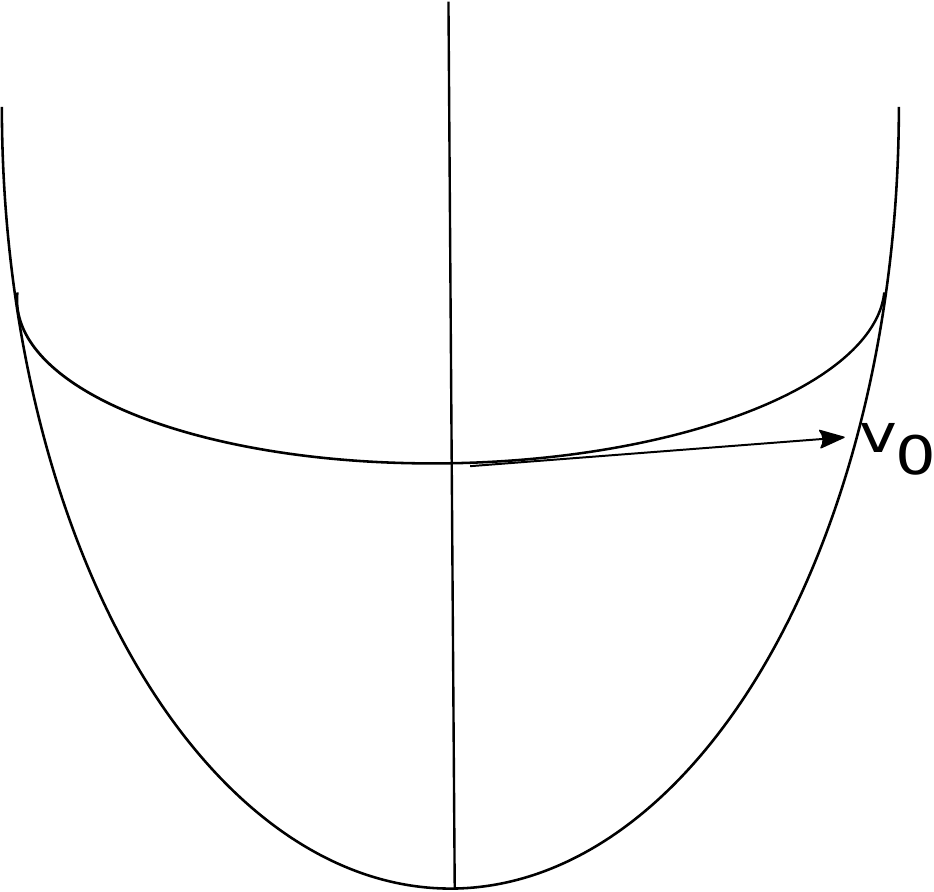}
	\begin{center}
		Fig. 3.16
	\end{center}
\end{wrapfigure}

Let a particle be projected horizontally with velocity $v_0$ from a point on the surface whose $r$ and $z$ co-ordinates are $r_0$, $z_0$ respectively. Suppose $\rho$ be the initial radius of curvature of the path and $\rho_0$ be the radius of curvature of the normal section of the surface through the tangent to the path at the pint under consideration. Let $\chi$ be the angle between the normal section at that point and the osculating plane. Let $v$ be the velocity of the particle at that point then we have the equation of motion 
$$m\frac{v^2}{\rho_0}\tan\chi=H$$
where $H$ is the component of the external forces along the tangent to the surface but perpendicular to the tangent to the path.\\

  \begin{wrapfigure}[11]{r}{0.35\textwidth}
	\centering	\includegraphics[height=4 cm , width=5 cm ]{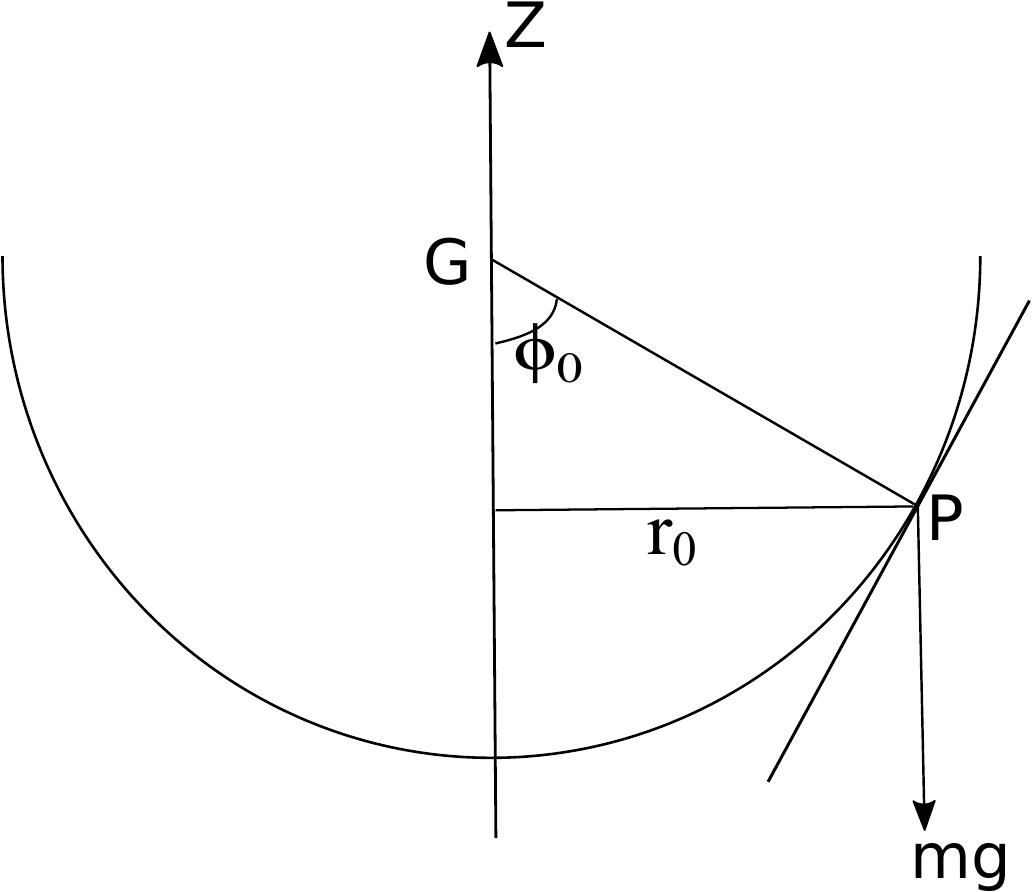}
	\begin{center}
		Fig. 3.17
	\end{center}
\end{wrapfigure}

In the figure $PG$ is the normal to the surface at $P$ and intersects the $z$ axis at $G$. Hence the radius of curvature of the normal section passing through the tangent to the curve at P will be $\rho_0=PG=\dfrac{r_0}{\sin\phi_0}$.\\

Hence $H=mg\sin\phi_0$, as $H$ is acting along the tangent to the meridian curve.\\

Thus we have
\begin{eqnarray}
&&	m\frac{v^2}{\rho_0}\sin\phi_0\tan\chi=mg\sin\phi_0\nonumber\\
\mbox{i.e., }&&\frac{v^2}{\rho_0}\tan\chi=g ~~(\because\sin\phi_0\neq0)\nonumber\\
\mbox{i.e., }&&\tan\chi=\frac{gr_0}{v^2}\nonumber\\
\therefore&&\rho=\rho_0\cos\chi=\frac{r_0v^2}{\sin\phi_0\sqrt{v^4+g^2r_0^2}}\nonumber
\end{eqnarray}

\subsection{Motion of a heavy particle on the surface of a smooth sphere}
  \begin{wrapfigure}[11]{r}{0.35\textwidth}
	\centering	\includegraphics[height=4 cm , width=4 cm ]{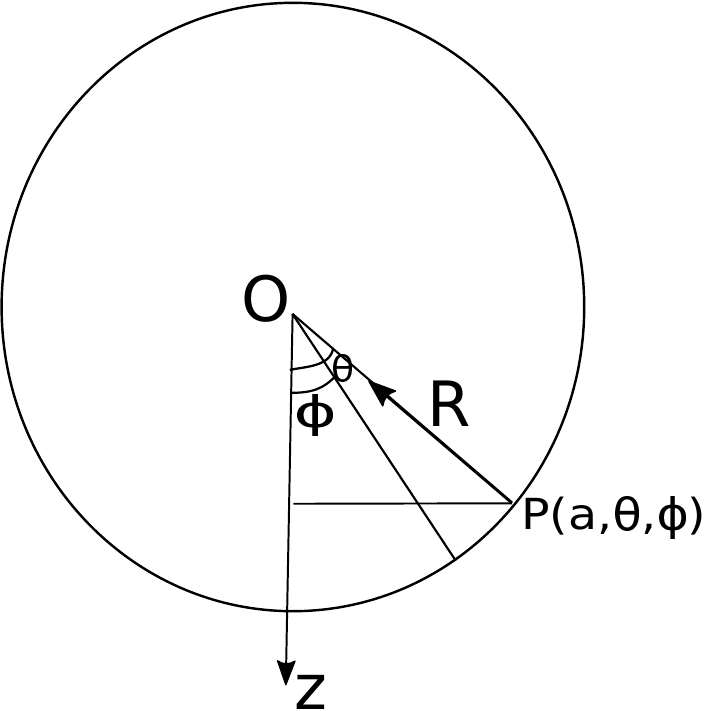}
	\begin{center}
		Fig. 3.18
	\end{center}
\end{wrapfigure}

We take the centre of the sphere $O$ as origin and $z$ axis vertically downwards. Let $a$ be the radius of the sphere and $P(a,\theta,\phi)$ be the position of the particle at any time $t$. The energy equation gives
\begin{eqnarray}
	&&\frac{1}{2}mv^2=mgz+\frac{1}{2}mc\nonumber\\
	\mbox{i.e., }&&v^2=c+2gz\nonumber\\
	\mbox{i.e., }&&v^2=c+2ga\cos\theta\label{eq3.22}
\end{eqnarray}

The conservation of angular momentum is written as
\begin{eqnarray}
	&&\frac{\mathrm{d}}{\mathrm{d}t}\left(a^2\dot{\phi}\sin^2\theta\right)=0\nonumber\\
	\mbox{i.e. }&&a^2\sin^2\theta\dot{\phi}=h\mbox{  ~~~~~(a constant)}\label{eq3.23}
\end{eqnarray}

The equation of motion along the normal  $PO$ is
\begin{equation}\label{eq3.24}
	m\frac{v^2}{\rho_0}=G+R
\end{equation}
where $G$ is the component of the external force along the inward drawn normal, $R$ is the reaction of the surface and $\rho_0$ is the radius of curvature of the normal section of the surface through a plane passing through the tangent to the path at $P$.\\

As any normal section is a great circle for a sphere so $\rho_0=a$ and $G=-mg\cos\theta$.\\

$\therefore~R=m\dfrac{v^2}{a}+mg\cos\theta=mg\left(\dfrac{c}{ag}+3\cos\theta\right)$.\\

Also $v^2=\dot{r}^2+r^2\dot{\theta}^2+r^2\sin^2\theta\dot{\phi}^2=a^2\dot{\theta}^2+a^2\sin^2\theta\dot{\phi}^2$.\\

Now using equations (\ref{eq3.22}) and (\ref{eq3.23}) we get
$$a^4\sin^2\theta\dot{\theta}^2=a^2\sin^2\theta(c+2ga\cos\theta)-h^2$$

Putting $z=a\cos\theta$, the above equation becomes
\begin{eqnarray}
	a^2\dot{z}^2&=&(a^2-z^2)(c+2gz)-h^2\nonumber\\
	&=&2g(a^2-z^2)(c_0+z)-h^2=F(z)\mbox{ ~~(say)}\nonumber
\end{eqnarray}	
where $c_0=\dfrac{c}{2g}$.\\

Note that, $F(+\infty)=-\mbox{ve}$, $F(a)=-\mbox{ve}$, $F(-a)=-\mbox{ve}$,  $F(-\infty)=+\mbox{ve}$. As for any actual position of the particle $z=z_0$ (say) on the surface of the sphere, we have $\dot{z}^2>0$ i.e. $F(z_0)>0$, so we have 
$$F(a)=-\mbox{ve}, F(z_0)=+\mbox{ve}, F(-a)=-\mbox{ve}, F(-\infty)=+\mbox{ve}$$

Thus there exists three real roots of $F(z)=0$ namely $z_1$, $z_2$ and $-z_3$ within the intervals $(z_0,a)$, $(-a,z_0)$ and $(-\infty,-a)$ respectively. Hence we write
\begin{eqnarray}
	\dot{z}^2=F(z)=2g(z_1-z)(z-z_2)(z_3+z)\nonumber\\\mbox{i.e. }z=\pm\int\sqrt{2g(z_1-z)(z-z_2)(z_3+z)}\mathrm{d}t\nonumber
\end{eqnarray} 

This gives $z$ i.e. $\theta$ in terms of $t$.\\

Note that $\dot{z}$ is real if $z_1>z>z_2$ and $\dot{z}=0$ at $z=z_1$ and $z_2$. Thus the motion of the particle is confined to a zone bounded by the circles $z=z_1$ and $z=z_2$.

Further, from (\ref{eq3.23}), $\dot{\phi}=\dfrac{h}{a^2\sin^z\theta}=\dfrac{h}{(a^2-z^2)}$,

which gives the angular velocity of the meridian plane through the particle about $OZ$. So knowing $\theta$ in terms of $t$, one can obtain $\phi$ as a function of time after integrating once.\\\\

\section{Problems}

{\bf 1. } A particle is projected horizontally with velocity $\sqrt{2ga}$ along the smooth surface of a sphere of radius $a$ at the at the level of the centre. Prove that the motion is confined between two horizontal planes at a distance $\dfrac{1}{2}(\sqrt{5}-1)a$ apart.\\

{\bf Solution: } We choose the centre of the sphere as origin and $z$-axis vertically downward. Let $P(a,\theta,\phi)$ be the position of the particle at time $t$ in spherical polar co-ordinates.\\

\begin{wrapfigure}[8]{r}{0.35\textwidth}
	\centering	\includegraphics[height=4 cm , width=4 cm ]{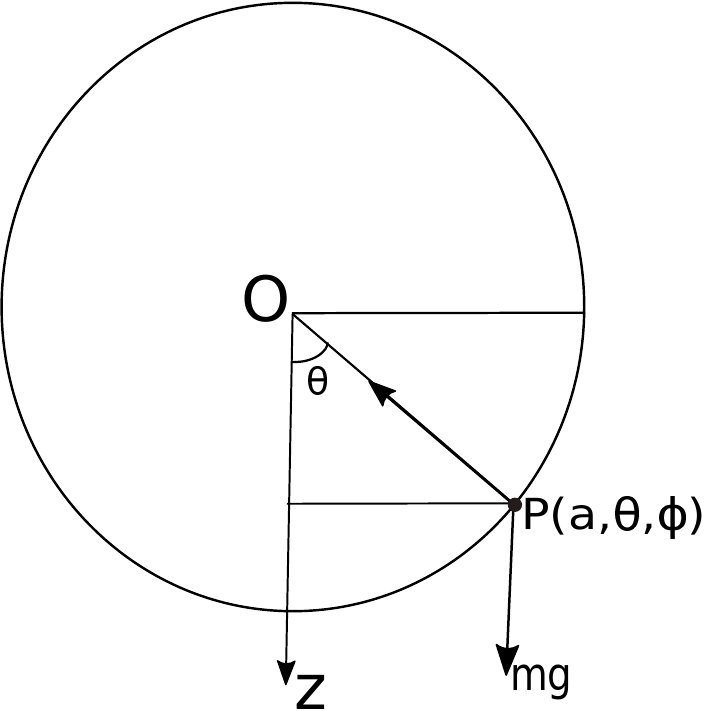}
	\begin{center}
		Fig. 3.19
	\end{center}
\end{wrapfigure}

Then the equation of energy gives
\begin{eqnarray}
	\frac{1}{2}mv^2=\frac{1}{2}mc+mgz\nonumber\\\mbox{i.e. }v^2=c+2gz=c+2ag\cos\theta\nonumber
\end{eqnarray}

Initially, $v=\sqrt{2ga}$, $\theta=\dfrac{\pi}{2}$ 

$\Rightarrow c=2ga$. \\

$\therefore v^2=2g(a+z)$ where $z=a\cos\theta$, gives the velocity of the particle at $P$.\\

Now, the equation of motion along the normal to the surface at $P$ is 
$$m\frac{v^2}{\rho_0}=R-mg\cos\theta$$
where $\rho_0$ is the radius of curvature of the normal section of the surface through a plane passing through the tangent to the path of $P$.\\

As any normal section is a great circle in case of a sphere so $\rho_0=a$. Hence we obtain
\begin{equation}
	R=m\left(\frac{v^2}{a}+\frac{gz}{a}\right)=\frac{mg}{a}(3z+2a)
\end{equation}

Thus $R$ vanishes when $z=-\dfrac{2}{3}a$.\\

Again, $v^2=\dot{r}^2+r^2\dot{\theta}^2+r^2\sin^2\theta\dot{\phi}^2=a^2\dot{\theta}^2+a^2\sin^2\theta\dot{\phi}^2$,

so the equation of angular momentum gives
\begin{eqnarray}
&&	a^2\sin^2\theta\dot{\phi}=h=a\sqrt{2ga}\nonumber\\
	\therefore&&a\sin\theta\dot{\phi}=\frac{\sqrt{2ga}}{\sin\theta}\nonumber\\
	\therefore&&v^2=a^2\dot{\theta}^2+\frac{2ga}{\sin^2\theta}=2ga+2gz\nonumber\\
	\therefore&&a^2\dot{\theta}^2=2ga+2gz-\frac{2ga}{1-\frac{z^2}{a^2}}\nonumber\\
		\mbox{i.e., }&&\dot{z}^2=\frac{2g}{a^2}\left[(a^2-z^2)(z+a)-a^3\right]=F(z)\mbox{ (say)}\nonumber
\end{eqnarray}

For maximum and minimum value of $z$, $\dot{z}=0$ and the values are given by
\begin{eqnarray}
	&&(a^2-z^2)(z+a)-a^3=0\nonumber\\\mbox{i.e., }&&z^3-a^2z+az^2=0\nonumber\\
	\implies&& z=0 \mbox{ or }z=\frac{a}{2}\left(-1\pm\sqrt{5}\right)\nonumber
\end{eqnarray}

As $z=-\dfrac{a}{2}\left(1+\sqrt{5}\right)<-a$, which is inadmissible so $\dot{z}=0$ at $z=0$ and $z=\dfrac{a}{2}\left(\sqrt{5}-1\right)$.\\

Thus the particle moves within the region bounded by the planes $z=0$, $z=\dfrac{a}{2}\left(\sqrt{5}-1\right)$. The distance between the plane is $\dfrac{a}{2}\left(\sqrt{5}-1\right)$.\\

\textbf{2.} A particle is projected horizontally under gravity with a velocity $V$ from a point on the inner surface of a smooth sphere at an angular distance $\alpha$ from the lowest point. Prove that the pressure on the surface when it is at angular distance $\theta$ from the lowest point is $$mg\left(3\cos\theta-2\cos\alpha+\frac{V^2}{ag}\right)$$
where $m$ is the mass of the particle and $a$ is the radius of the sphere. Prove that in the subsequent motion the particle will leave the surface if $3\sin\alpha<1$ and $2\dfrac{V^2}{ag}-7\cos\alpha$ lies between $\pm3\sqrt{1-9\sin^2\alpha}$.\\

\begin{wrapfigure}[11]{r}{0.35\textwidth}
	\centering	\includegraphics[height=5 cm , width=4 cm ]{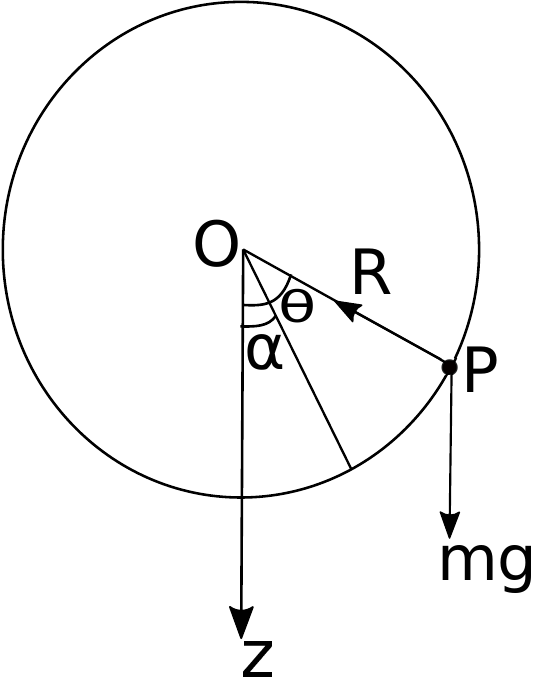}
	\begin{center}
		Fig. 3.20
	\end{center}
\end{wrapfigure}

\textbf{Solution: }The equation of energy gives
$$v^2=c+2gz=c+2ga\cos\theta$$

So $V^2=c+2ga\cos\alpha$.
\begin{equation}
	\therefore v^2=V^2+2ga(\cos\theta-\cos\alpha)
\end{equation}

The equation of angular momentum is
\begin{eqnarray}
	a^2\sin^2\theta\dot{\phi}=h=a\sin\alpha V\nonumber\\
	\mbox{i.e., }a\sin\theta\dot{\phi}=\frac{V\sin\alpha}{\sin\theta}
\end{eqnarray}

Now, $v^2=a^2\dot{\theta}^2+a^2\sin^2\theta\dot{\phi}^2=a^2\dot{\theta}^2+\dfrac{V^2\sin^2\alpha}{\sin^2\theta}$.\\

The equation of motion along the normal to the surface at $P$ is
\begin{eqnarray}
	m\frac{v^2}{a}&=&R-mg\cos\theta\label{eq3.2.3}\\
	\mbox{i.e., }~R&=&m\left(\frac{v^2}{a}+g\cos\theta\right)\nonumber\\&=&m\left\{\frac{V^2}{a}+2g(\cos\theta-\cos\alpha)+g\cos\theta\right\}\nonumber\\&=&mg\left(\frac{V^2}{ag}+3\cos\theta-2\cos\alpha\right)
\end{eqnarray}

From equation (\ref{eq3.2.3}) it is clear that if $\cos\theta$ is positive i.e., if $0<\theta\leq\dfrac{\pi}{2}$, i.e., the particle is within the lower hemisphere then $R>0$, so that it will not leave the surface of the sphere so long as it is within the region.\\

So, 
\begin{eqnarray}
	v^2&=&a^2\dot{\theta}^2+\frac{V^2\sin^2\alpha}{\sin^2\theta}=V^2+2ag(cos\theta-\cos\alpha)\nonumber\\
	\implies a^2\dot{\theta}^2&=&\frac{V^2}{\sin^2\theta}\left[\cos^2\alpha-\cos^2\theta\right]+2ag(\cos\theta-\cos\alpha)\nonumber\\
	&=&\frac{(\cos\alpha-\cos\theta)}{\sin^2\theta}\left[V^2(\cos\alpha+\cos\theta)-2ag\sin^2\theta\right]\nonumber\\
	&=&\frac{(\cos\alpha-\cos\theta)}{\sin^2\theta}f(cos\theta)\nonumber
\end{eqnarray}
where $f(\cos\theta)=2ga\cos^2\theta+V^2(\cos\alpha+\cos\theta)-2ag$.\\

For maximum and minimum value of $\theta$, $\dot{\theta}=0$ and it gives either $\theta=\alpha$ or $f(\cos\theta)=0$ where\\

$f(\cos0)=V^2(1+\cos\alpha)>0$.\\

$f(\cos\alpha)=2(V^2\cos\alpha-ag\sin^2\alpha)$.\\

$f(\cos\pi)=V^2(\cos\alpha-1)<0$.\\

Now, if $V^2\cos\alpha-ag\sin^2\alpha>0$, then the root of the equation $f(\cos\theta)=0$ will lie between $\cos\alpha$ and $-1$.   Hence in this case the particle projected horizontally at $\theta=\alpha$ will rise. However, if $V^2\cos\alpha-ag\sin^2\alpha<0$, the root of the equation $f(\cos\theta)=0$ will lie between $\cos\alpha$ and $1$. Hence the particle projected horizontally at $\theta=\alpha$ will fall. For $V^2\cos\alpha-ag\sin^2\alpha=0$, the particle will always at $\theta=\alpha$.\\

As $R=mg\left[\dfrac{V^2}{ag}+3\cos\theta-2\cos\alpha\right]$ , so the particle will leave the surface at a point where $R$ vanishes, provided that the particle reaches that point.\\

Now, 
\begin{eqnarray}
&&	f(\cos\theta)=0\nonumber\\\implies&&\cos^2\theta+2n^2(\cos\theta+\cos\alpha)-1, \mbox{ putting }\frac{v^2}{ag}=4n^2\nonumber\\
	\mbox{ i.e., }&&\cos\theta=-n^2\pm\sqrt{n^4+1-2n^2\cos\alpha}\nonumber
\end{eqnarray}

As the lower sign is inadmissible so $\dot{\theta}=0$ at $\theta_1=0$, where
$$\cos\theta=-n^2+\sqrt{n^4+1-2n^2\cos\alpha}$$ 

Now, if $\theta_1>\pi-\alpha$ then 
\begin{eqnarray} 
&&	\cos\theta_1<\cos(\pi-\alpha)=-\cos\alpha\nonumber\\
	\mbox{i.e., }&&-n^2+\sqrt{n^4+1-2n^2\cos\alpha}<-\cos\alpha\nonumber\\
	\mbox{i.e., }&&n^4+1-2n^2\cos\alpha<n^4+\cos^2\alpha-2n^2\cos\alpha\nonumber\\
	\mbox{i.e., }&&\cos^2\alpha>1,\mbox{ which is impossible}\nonumber
	\end{eqnarray}

Hence $\theta_1<\pi-\alpha$.\\

Let $R=0$ at $\theta=\theta_2$, given by
$$\cos\theta_2=\frac{2}{3}\cos\alpha-\frac{4}{3}n^2$$

The particle will leave the surface if $\theta_2<\theta_1$ i.e., $\cos\theta_2>\cos\theta_1$
\begin{eqnarray}
	\implies&&\frac{2}{3}\cos\alpha-\frac{4}{3}n^2>-n^2+\sqrt{n^4+1-2n^2\cos\alpha}\nonumber\\
	\mbox{i.e., }&&2\cos\alpha-n^2>3\sqrt{n^4+1-2n^2\cos\alpha}\nonumber\\
		\mbox{i.e., }&&4\cos^2\alpha+n^4-4n^2\cos\alpha>9\left(n^4+1-2n^2\cos\alpha\right)\nonumber\\
		\mbox{i.e. }&&8n^4-14n^2\cos\alpha-4\cos^2\alpha+9<0\nonumber\\
		\mbox{i.e. }&&\left(4n^2-\frac{7}{2}\cos\alpha\right)^2<\frac{81-81\sin^2\alpha-72}{4}=\frac{9}{4}\left(1-9\sin^2\alpha\right)\label{eq3.3.30}
\end{eqnarray}

The inequality will be true provided $1-9\sin^2\alpha>0$ i.e., $\sin\alpha<\dfrac{1}{3}$.\\

So, if $\sin\alpha<\dfrac{1}{3}$ then $\left|4n^2-\dfrac{7}{2}\cos\alpha\right|<\dfrac{3}{2}\sqrt{1-9\sin^2\alpha}$\\

i.e., $\left|2\dfrac{V^2}{ag}-7\cos\alpha\right|<3\sqrt{1-9\sin^2\alpha}$\\

If $\sin\alpha>\dfrac{1}{3}$, then inequality (\ref{eq3.3.30}) can not be satisfied by any real $v$. This means that the particle will not leave the surface if $\sin\alpha>\dfrac{1}{3}$.\\

{\bf3. } A heavy particle $P$ moves on the smooth inner surface of a fixed sphere of centre $O$ and the angle between OP and downward vertical is denoted by $\theta$. The particle is projected horizontally on the surface from a point at which $\theta=\alpha\left(<\dfrac{\pi}{2}\right)$. Prove that whatever be the velocity of projection $\theta$ will not exceed $(\pi-\alpha)$ in the subsequent motion and that if $\sin\alpha>\dfrac{1}{3}$ the particle will not leave the surface.\\

{\bf Solution: } Similar to the previous problem.\\

{\bf 4. } A particle is projected with a velocity $V$ along the inside of a smooth sphere of radius $a$ along a $\parallel$ whose latitude is $\lambda$. It is repealed from the point upon the sphere whose latitude is $90^o$ by a force per unit mass equal to $\mu$ times the distance from that point. Find the rate of increase of latitude and longitude of the particle when its latitude is $\theta$ and show that its path cuts the equator at 
$$\tan^{-1}\left\{\tan\lambda\left(1+\frac{2\mu a^2}{V^2}\mbox{cosec}\lambda\right)^\frac{1}{2}\right\}$$

{\bf Solution: } Let $O'P=r'$; so that force is $\mu r'$. If $\psi$ be the potential of the force then 
$$-\dfrac{\partial\psi}{\partial r'}=\mu r'\implies\psi=-\frac{1}{2}\mu r'^2+c$$

\begin{wrapfigure}[6]{r}{0.35\textwidth}
	\centering	\includegraphics[height=4 cm , width=4 cm ]{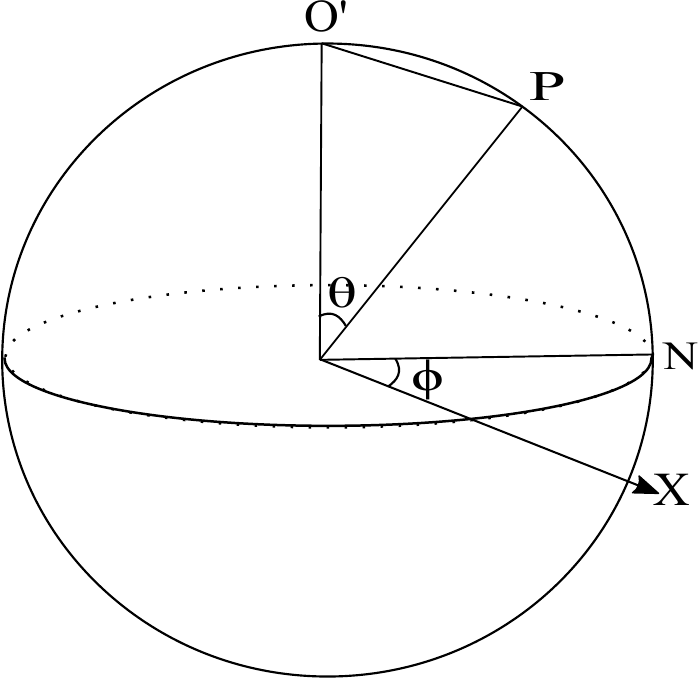}
	\begin{center}
		Fig. 3.21
	\end{center}
\end{wrapfigure}
 
 Let $\theta$ and $\phi$ be the co-latitude and longitude of the particle.\\
 
 From the energy equation we get $\dfrac{1}{2}v^2+\psi=$ constant, where $ v$  is the velocity of the particle at $P$.\\
 
 Thus we have, $\dfrac{1}{2}v^2-\dfrac{1}{2}\mu r'^2=$ constant.\\
 
 But $r'^2=a^2+a^2-2a^2\cos\theta=2a^2(1-\cos\theta)$.\\

Hence we get,
\begin{eqnarray}
	\dfrac{1}{2}v^2+\mu a^2\cos\theta&=&\mbox{constant=Initial value}\nonumber\\
	&=&\dfrac{1}{2}V^2+\mu a^2\sin\lambda\nonumber\\
	\therefore ~v^2~=~V^2&+&2\mu a^2(\sin\lambda-\cos\theta)\nonumber
\end{eqnarray}

As there is no force perpendicular to $ZOP$, so angular momentum is constant, i.e.,
\begin{eqnarray}
	a^2\sin^2\theta\dot{\phi}&=&\mbox{constant}=Va\cos\lambda\nonumber\\
	\mbox{i.e., }a\sin^2\theta\dot{\phi}&=&V\cos\lambda\nonumber
\end{eqnarray}

This gives the rate of increase of longitude when $\theta$ is known.\\

As,
\begin{eqnarray}
	v^2&=&a^2\dot{\theta}^2+a^2\sin^2\theta\dot{\phi}^2=a^2\dot{\theta}^2+\frac{V^2\cos^2\lambda}{\sin^2\theta}\nonumber\\
	&=&V^2+2\mu a^2(\sin\lambda-\cos\theta)\nonumber\\
	\therefore~a^2\dot{\theta}^2&=&V^2\left(1-\frac{\cos^2\lambda}{\sin^2\theta}\right)+2\mu a^2(\sin\lambda-\cos\theta)\nonumber
\end{eqnarray}

This gives the rate of increase of co-latitude of the particle and hence the latitude of the particle.\\

At the equator, $\theta=\dfrac{\pi}{2}$, so $a\dot{\phi}=V\cos\lambda$ and 
\begin{eqnarray}
	a^2\dot{\theta}^2&=&V^2\sin^2\lambda+2\mu a^2\sin\lambda\nonumber\\&=&V^2\sin^2\lambda\left\{1+\frac{2\mu a^2}{V^2}\mbox{cosec}\lambda\right\}\nonumber\\\therefore~\dfrac{\dot{\theta}^2}{\dot{\phi}^2}&=&\tan^2\lambda\left\{1+\frac{2\mu a^2}{V^2}\mbox{cosec}\lambda\right\}\nonumber
\end{eqnarray}

If the path cuts the equator at an angle $\alpha$, we have,
\begin{eqnarray}
	V\cos\alpha=a\dot{\phi},~V\sin\alpha=a\dot{\theta}~~(\because\theta=\frac{\pi}{2}\mbox{ at the equator})\nonumber\\
	\therefore~\dfrac{\dot{\theta}}{\dot{\phi}}=\tan\alpha=\tan\lambda\left\{1+\frac{2\mu a^2}{V^2}\mbox{cosec}\lambda\right\}^\frac{1}{2}\nonumber
\end{eqnarray}

{\bf 5. } A heavy particle is projected horizontally along the inner surface of a smooth sphere with a velocity due to a fall from the level of the centre to the point of projection. Show that the radius of curvature of its path when it is at an angular distance $\theta$ from the lowest point of the sphere is $\dfrac{a}{\left(1+\frac{1}{4}\sin^2\alpha\cos\alpha\sec^3\theta\right)^\frac{1}{2}}$, where $\alpha$ is the initial value of $\theta$ and $a$ is the radius of the sphere.\\

{\bf Solution: } Let $P(a,\theta,\phi)$ be the position of the particle at time $t$ and $v$ be the velocity at $P$.
$$\therefore~v^2=a^2\dot{\theta}^2+a^2\sin^2\theta\dot{\phi}^2$$

\begin{wrapfigure}[12]{r}{0.35\textwidth}
	\centering	\includegraphics[height=4 cm , width=4 cm ]{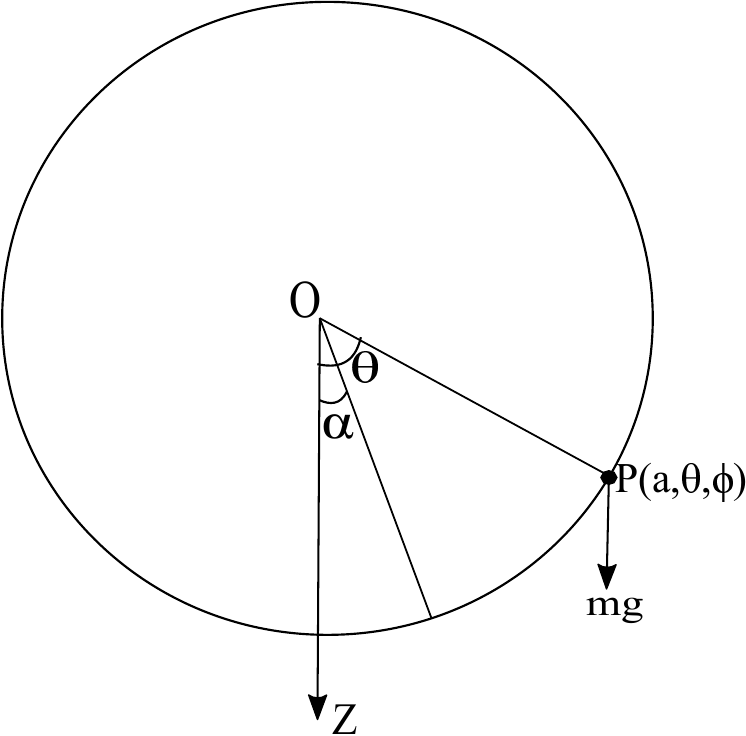}
	\begin{center}
		Fig. 3.22
	\end{center}
\end{wrapfigure}

As there are no external forces perpendicular to the $ZOP$ plane so
\begin{eqnarray}
	a^2\sin^2\theta\dot{\phi}=\mbox{constant}=\mbox{initial value}=Va\sin\alpha\nonumber\\
	\therefore~a\sin\theta\dot{\phi}=\mbox{horizontal velocity}=u=\frac{V\sin\alpha}{\sin\theta}\nonumber
\end{eqnarray}

The energy conservation equation gives 
$$v^2=2ag\cos\theta+c$$

Initially, $V^2=2ag\cos\alpha\implies c=0$
$$\therefore v^2=2ag\cos\theta$$

Also we have the equation of motion
$$m\frac{v^2}{\rho_0}\tan\chi=H,~~\rho_0=a\mbox{ in the present problem}$$

Let the tangent to the path at $P$ makes an angle $\psi$ with the meridian curve so
$$H=mg\sin\theta\sin\psi$$

Also $v\sin\psi=u=\sqrt{2ag\cos\alpha}\dfrac{\sin\alpha}{\sin\theta}$.
\begin{eqnarray}
	\therefore ~ m\frac{v^2}{a}\tan\chi&=&mg\sin\theta\frac{u}{v}\nonumber\\\therefore~\tan\chi&=&\frac{ag\sin\theta\sqrt{2ag\cos\alpha}}{2ag\cos\theta\sqrt{2ag\cos\theta}}=\frac{\sqrt{\sin^2\alpha\cos\alpha}}{\sqrt{4\cos^3\theta}}\nonumber\\\therefore~\frac{\sin\chi}{\sqrt{\sin^2\alpha\cos\alpha}}&=&\frac{\cos\chi}{\sqrt{4\cos^3\theta}}=\frac{1}{\sqrt{\sin^2\alpha\cos\alpha+4\cos^3\theta}}\nonumber\\\therefore~\cos\chi&=&\frac{1}{\sqrt{1+\frac{1}{4}\sin^2\alpha\cos\alpha\sec^3\theta}}\nonumber\\\therefore~\rho&=&\mbox{radius of curvature of the path at }P=\rho_0\cos\chi=a\cos\chi\nonumber\\&=&\frac{a}{\left(1+\frac{1}{4}\sin^2\alpha\cos\alpha\sec^3\theta\right)^\frac{1}{2}}\nonumber
\end{eqnarray}

{\bf 6. } A particle tied to a fixed point by an inextensible string is projected under gravity in any manner so that the string remains taut in the subsequent motion. Find the differential equation of the projection of the path on a horizontal plane and prove that the pedal equation of this curve is of the form
$$\frac{l^2}{p^2}=1+A\left(l^2-r^2\right)+B\left(l^2-r^2\right)^\frac{3}{2}$$ 

{\bf Solution: } Sine the string is taut throughout so the particle moves over a sphere. So the meridian section through the particle is a great circle:
\begin{equation}\label{eq3.6.1}
	z^2+r^2=l^2
\end{equation}

The differential equation of the projection of the path is
\begin{equation}
	\frac{\left(1+\{\phi'\}^2\right)}{r^4}\left(\frac{\mathrm{d}r}{\mathrm{d}\theta}\right)^2=\frac{c^2}{h^2}-\frac{2g}{h^2}z-\frac{1}{r^2}\nonumber
\end{equation}

Now, $\dot{\phi}=\dfrac{\mathrm{d}z}{\mathrm{d}r}=-\dfrac{r}{z}$ (using (\ref{eq3.6.1}))
\begin{eqnarray}
	\therefore ~\frac{l^2}{r^4z^2}\left(\frac{\mathrm{d}r}{\mathrm{d}\theta}\right)^2=\frac{c^2}{h^2}-\frac{2g}{h^2}z-\frac{1}{r^2}\nonumber\\\implies\frac{1}{r^2}\left(\frac{\mathrm{d}r}{\mathrm{d}\theta}\right)^2=\frac{c^2z^2r^2}{l^2h^2}-\frac{2gz^3r^2}{h^2l^2}-\frac{z^2}{l^2}\nonumber
\end{eqnarray}

As $p=r\sin\phi,~\tan\phi=r\dfrac{\mathrm{d}\theta}{\mathrm{d}r}$,
So,\begin{eqnarray}
	\frac{r^2}{p^2}&=&1+\frac{1}{r^2}\left(\frac{\mathrm{d}r}{\mathrm{d}\theta}\right)^2=\left(1-\frac{z^2}{l^2}\right)+\frac{c^2z^2r^2}{l^2h^2}-\frac{2gz^3r^2}{h^2l^2}\nonumber\\&=&\frac{r^2}{l^2}+\frac{c^2z^2r^2}{l^2h^2}-\frac{2gz^3r^2}{h^2l^2}\nonumber\\\implies\frac{l^2}{p^2}&=&1+A\left(l^2-r^2\right)+B\left(l^2-r^2\right)^\frac{3}{2}\nonumber
\end{eqnarray}

{\bf 7. } A heavy particle is projected horizontally with velocity $u$ along the inner surface of a smooth vertical circular cylinder of radius $a$. Show that the radius of curvature at any point of the path is $\dfrac{av^3}{u\sqrt{u^2v^2+a^2g^2}}$, $v$ being the velocity at the point.\\

{\bf Solution: } Let $\psi$ be the angle that the path of the particle makes with the vertical generator through $P$. Suppose $\rho_1$ and $\rho_2$ are the principal radius of curvature. Here $\rho_1=\infty$, $\rho_2=a$. Suppose $\rho_0$ be the radius of curvature of the normal section of the cylinder passing through that tangent to the curve at $P$. So we have,
$$\frac{1}{\rho_0}=\frac{\cos^2\psi}{\rho_1}+\frac{\sin^2\psi}{\rho_2}\implies\rho_0=a\mbox{cosec}^2\psi$$
\begin{wrapfigure}[12]{r}{0.35\textwidth}
	\centering	\includegraphics[height=4.5 cm , width=4 cm ]{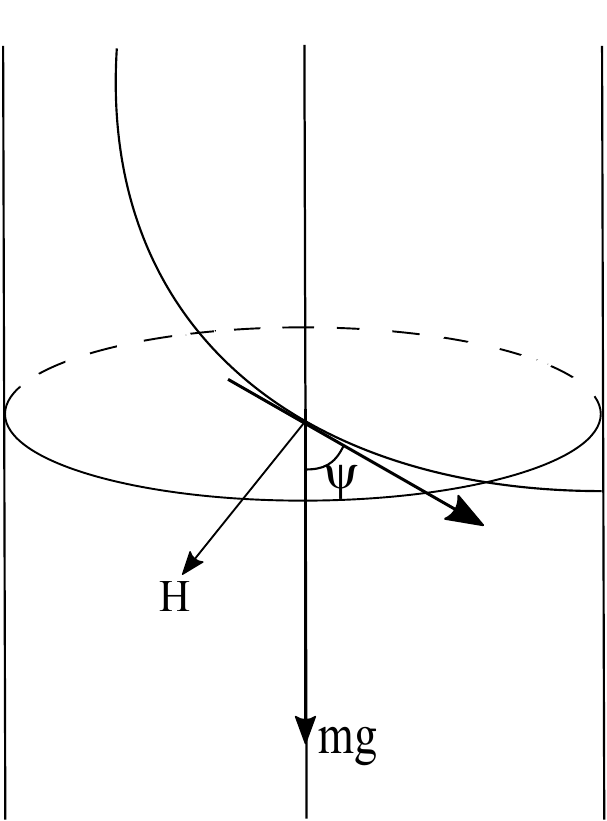}
	\begin{center}
		Fig. 3.23
	\end{center}
\end{wrapfigure}

Since the only force is the weight $mg$, so the horizontal component of velocity remains constant.
$$\therefore~v\sin\psi=u\implies\mbox{cosec}\psi=\frac{v}{u}$$

If $H$ be the component of the force along the tangent to the surface perpendicular to the tangent to the path then,
\begin{eqnarray}
	H&=&mg\sin\psi\nonumber\\\therefore~m\frac{v^2}{\rho_0}\tan\chi&=&mg\sin\psi\implies\tan\chi=\frac{ag}{uv}\nonumber\\\therefore~\frac{\sin\chi}{ag}&=&\frac{\cos\chi}{uv}=\frac{1}{\sqrt{a^2g^2+u^2v^2}}\nonumber\\\therefore~\rho&=&\mbox{radius of curvature of the path of the particle at }P\nonumber\\&=&\rho_0\cos\chi=a\mbox{cosec}^2\psi\cos\chi=\frac{av^3}{u\sqrt{u^2v^2+a^2g^2}}\nonumber
\end{eqnarray}

{\bf 8. } A heavy particle of mass $m$ moves on the smooth inner surface of a sphere of radius $a$ and its greatest and least depths below the centre are $\dfrac{a}{2}$ and $\dfrac{a}{4}$ respectively. Show that when the deep below the centre is $z$, the normal reaction of the sphere is $\dfrac{3mg}{a}\left(z+\frac{a}{2}\right)$. Also show that the time from maximum to minimum depth is $\sqrt{\dfrac{a}{g}}\int\limits_0^\frac{\pi}{2}\dfrac{\mathrm{d}\phi}{\sqrt{1-\frac{1}{8}\sin^2\phi}}$\\
\begin{wrapfigure}[12]{r}{0.35\textwidth}
	\centering	\includegraphics[height=5 cm , width=4 cm ]{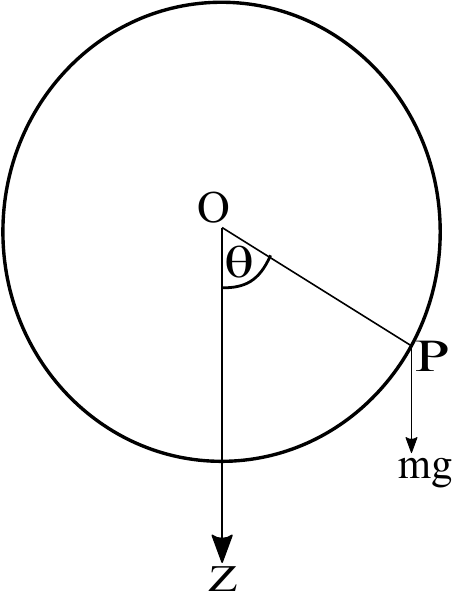}
	\begin{center}
		Fig. 3.24
	\end{center}
\end{wrapfigure}
{\bf Solution: } Let $v_1$ and $v_2$ be the horizontal velocities at the points $z=\dfrac{a}{2}$ and $z=\dfrac{a}{4}$. Then from the energy equation we have,
\begin{eqnarray}
	v^2&=&c+2gz\nonumber\\\therefore~v_1^2=c+2g\frac{a}{2}&,&v_2^2=c+2g\frac{a}{4}\nonumber\\\therefore~v_1^2-v_2^2&=&\frac{ag}{2}
\end{eqnarray} 

The equation of angular momentum gives
\begin{eqnarray}
	a^2\sin^2\theta\dot{\phi}=\mbox{constant}&=&v_1a\frac{\sqrt{3}}{2}=v_2a\frac{\sqrt{15}}{4}\nonumber\\\therefore~v_1=v_2\frac{\sqrt{5}}{2}\nonumber&&
\end{eqnarray} 

Thus, $v_2^2=2ag$, $v_1^2=\dfrac{5ag}{2}$ and $c=\dfrac{3}{2}ag$.
$$\therefore~h=	a^2\sin^2\theta\dot{\phi}=v_1a\frac{\sqrt{3}}{2}=\sqrt{\dfrac{5ag}{2}}a\frac{\sqrt{3}}{2}=\sqrt{\frac{15}{8}a^3g}$$ 

Now,
\begin{eqnarray}
	m\frac{v^2}{a}=R-mg\cos\theta&&\implies R=mg\left(\cos\theta+\frac{v^2}{ag}\right)\nonumber\\\therefore~R= mg\left(\frac{z}{a}+\frac{c}{ag}+\frac{2z}{a}\right)&=&mg\left(\frac{3z}{a}+\frac{3}{2}\right)=3\frac{mg}{a}\left(z+\frac{a}{2}\right)\nonumber
\end{eqnarray}

Also
\begin{eqnarray}
	v^2&=&a^2\dot{\theta}^2+a^2\sin^2\theta\dot{\phi}^2=2gz+c=2gz+\frac{3}{2}ag\nonumber\\\therefore&&a^2\dot{\theta}^2+a^2\sin^2\theta\cdot\frac{15}{8}\frac{a^3g}{a^4\sin^4\theta}=2gz+\frac{3}{2}ag\nonumber\\\therefore&&a^2\sin^2\theta\dot{\theta}^2+\frac{15}{8}ag=\left(1-\cos^2\theta\right)\left(2gz+\frac{3}{2}ag\right)\nonumber\\\therefore~\dot{z}^2&=&\frac{2g}{a^2}(a^2-z^2)\left(z+\frac{3a}{4}\right)-\frac{15}{8}ag\nonumber\\&=&\frac{2g}{a^2}\left[a^2z-z^3-\frac{3}{4}z^2a-\frac{3}{16}a^3\right]\nonumber
\end{eqnarray}

Since maximum and minimum depths are $\dfrac{a}{2}$ and $\dfrac{a}{4}$, so $\left(\dfrac{a}{2}-z\right)$ and $\left(z-\dfrac{a}{4}\right)$ are two factors. The remaining factor be $\left(z+\dfrac{3a}{2}\right)$. $$\dot{z}^2=\frac{2g}{a^2}\left[\left(\dfrac{a}{2}-z\right)\left(z-\dfrac{a}{4}\right)\left(z+\dfrac{3a}{2}\right)\right]$$

If $T$ the required time, then 
\begin{eqnarray}
	T&=&-\sqrt{\frac{a^2}{2g}}\int\limits_\frac{a}{2}^\frac{a}{4}\frac{\mathrm{d}z}{\sqrt{\left(\dfrac{a}{2}-z\right)\left(z-\dfrac{a}{4}\right)\left(z+\dfrac{3a}{2}\right)}}\nonumber\\
&&	\mbox{(Note that $\dot{z}<0$ in computing time from maximum to minimum depth)}\nonumber
\end{eqnarray}

Let us substitute
\begin{eqnarray}
z&=&\frac{a}{4}\cos^2\theta+\frac{a}{2}\sin^2\theta,~\mbox{i.e., }~\mathrm{d}z=\frac{a}{2}\sin\theta\cos\theta\mathrm{d}\theta\nonumber\\\therefore~T&=&\sqrt{\frac{a^2}{2g}}\int\limits_0^\frac{\pi}{2}\frac{\frac{a}{2}\sin\theta\cos\theta\mathrm{d}\theta}{\sqrt{\frac{a}{4}\cos^2\theta\cdot\frac{a}{4}\sin^2\theta\left(\frac{a}{4}\cos^2\theta+\frac{a}{2}\sin^2\theta+\frac{3a}{2}\right)}}\nonumber\\&=&\sqrt{\frac{a^2}{2g}}\int\limits_0^\frac{\pi}{2}\frac{\mathrm{d}\theta}{\sqrt{\frac{a}{16}\left(\cos^2\theta+2\sin^2\theta+6\right)}}\nonumber\\&=&\sqrt{\frac{a^2}{2g}}\int\limits_0^\frac{\pi}{2}\frac{\mathrm{d}\theta}{\sqrt{\frac{a}{16}\left(8-\cos^2\theta\right)}}=\sqrt{\dfrac{a}{g}}\int\limits_0^\frac{\pi}{2}\dfrac{\mathrm{d}\phi}{\sqrt{1-\frac{1}{8}\sin^2\phi}}\mbox{ (Substituting }\phi=\frac{\pi}{2}-\theta)\nonumber
\end{eqnarray}

{\bf 9. } A heavy particle moves on the surface of a paraboloid of revolution whose axis is vertical and vertex downwards being projected horizontally with velocity $\sqrt{2gh_2}$ at a height $h_1$ above the vertex. Show that the orbit touches alternatively the two horizontal circles at heights $h_1$, $h_2$ above the vertex and that the least value of the angle at which the orbit cuts the Meridian is $\tan^{-1}\dfrac{2\sqrt{h_1h_2}}{\left|h_1-h_2\right|}$\\

{\bf Solution: } In cylindrical co-ordinate system the equation of the Meridian section through the particle be $$r^2=2az$$ 

Also the equation of energy gives
\begin{eqnarray}
	v^2&=&c-2gz\nonumber\\\therefore~2gh_2&=&c-2gh_1\implies c=2h\left(h_1+h_2\right)\nonumber\\\therefore~v^2&=&2g\left(h_1+h_2-z\right)\nonumber
\end{eqnarray} 

Equation of angular momentum gives
\begin{eqnarray}
	r^2\dot{\theta}&=&h=\sqrt{2gh_2\cdot4ah_1}\nonumber\\\therefore v^2&=&\dot{r}^2+\dot{z}^2+r^2\dot{\theta}^2\nonumber\\&=&\frac{4a^2}{r^2}\dot{z}^2+\dot{z}^2+\frac{8agh_1h_2}{r^2}=2g\left(h_1+h_2-z\right)\nonumber\\\therefore\dot{z}^2\left(\frac{a}{z}+1\right)&=&2g\left(h_1+h_2-z\right)\frac{2gh_1h_2}{z}\nonumber\\\implies(a+z)\dot{z}^2&=&2g\left[\left(h_1+h_2\right)z-z^2-h_1h_2\right]=2g\left[\left(h_1-z\right)\left(z-h_2\right)\right]\nonumber
\end{eqnarray}
i.e., $\dot{z}$ vanishes  at $z=h_1$ and $z=h_2$.\\

Now, $r\dot{\theta}=\dfrac{8agh_1h_2}{4az}=\dfrac{2gh_1h_2}{z}=$ horizontal velocity.\\

If the path cuts the Meridian at an angle $\psi$ then,
$$r\dot{\theta}=v\sin\psi\implies\sin\psi=\sqrt{\frac{h_1h_2}{z\left(h_1+h_2-z\right)}}$$

So for least value of $\psi$, $z\left(h_1+h_2-z\right)$ should have a maximum. \\

This occurs when $z=\dfrac{h_1+h_2}{2}$, and then $\sin\psi=\dfrac{2\sqrt{h_1h_2}}{h_1+h_2}$.
$$\therefore~\cos\psi=\dfrac{\left|h_1-h_2\right|}{\left(h_1+h_2\right)} \mbox{~ i.e., ~}\tan\psi=\dfrac{2\sqrt{h_1h_2}}{\left|h_1-h_2\right|}$$

{\bf 10. } A particle is projected horizontally along the inner surface of a smooth cone whose access is vertical and vertex downwards. Find the pressure at any point in terms of the depth below the vertex and show that the particle leaves the cone at a depth $\left(\dfrac{V^2h^2}{g\tan^2\alpha}\right)^\frac{1}{3},$ where $h$ is the initial depth, $V$ is the initial velocity and $\alpha$ is the semi-vertical angle of the cone. Find also the radius of curvature at any point of the path.\\

{\bf Solution: } Using the cylindrical coordinates with $z$-axis vertically downward, let $P(r,\theta,z)$ be the position of the particle at any time $T$. If $v$ be velocity of the particle at $P$ then equation of energy gives $$v^2=c+2gz$$ 

\begin{wrapfigure}[13]{r}{0.35\textwidth}
	\centering	\includegraphics[height=5 cm , width=5 cm ]{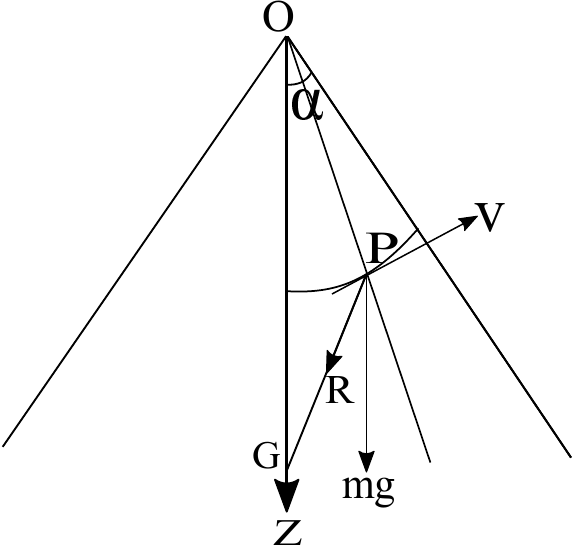}
	\begin{center}
		Fig. 3.25
	\end{center}
\end{wrapfigure}

From the initial condition
\begin{eqnarray}
	V^2=c+2gh\nonumber\\\therefore~v^2=V^2+2g(z-h)\nonumber
\end{eqnarray}

Equation of angular momentum gives
\begin{eqnarray}
	r^2\dot{\theta}&=&\mbox{constant}=Vh\tan\alpha\nonumber\\\therefore~r\dot{\theta}&=&\frac{Vh\tan\alpha}{r}=\frac{Vh\tan\alpha}{z\tan\alpha}=\frac{Vh}{z}=u \mbox{ (say)}\nonumber
\end{eqnarray}  

Resolving along the normal to the surface
$$m\frac{v^2}{\rho_0}=R+mg\sin\alpha$$ where $\dfrac{1}{\rho_0}=\dfrac{\cos^2\psi}{\rho_1}+\dfrac{\sin^2\psi}{\rho_2}$ with $\rho_1=\infty$, $\rho_2=PG$.\\

Here $\psi$ is the angle which the tangent to the path at $P$ makes with the generator of the cone through P.
\begin{eqnarray}
	\therefore~\frac{1}{\rho_0}=\frac{\sin^2\psi}{PG}&=&\frac{\sin^2\psi}{z\tan\alpha\sec\alpha}~~(\because PG\cos\alpha=r=z\tan\alpha)\nonumber\\\therefore~\frac{mv^2\sin^2\psi}{z\tan\alpha\sec\alpha}&=&R+mg\sin\alpha\nonumber
\end{eqnarray} 

Again, $v\sin\psi$= horizontal component of velocity= $u=V\dfrac{h}{z}$.
$$\therefore~R=\frac{mV^2h^2}{z^3\tan\alpha\sec\alpha}-mg\sin\alpha=\frac{mg}{\tan\alpha\sec\alpha}\left[\frac{V^2h^2}{z^3g}-\tan^2\alpha\right].$$ 

So R vanishes at a depth $z$ given by
\begin{equation}
	z=\left(\dfrac{V^2h^2}{g\tan^2\alpha}\right)^\frac{1}{3}
\end{equation}

Again $v^2=\dot{r}^2+r^2\dot{\theta}^2+\dot{z}^2=V^2+2g(z-h)$.
\begin{eqnarray}
	\implies \dot{z}^2\sec^2\alpha=-\frac{V^2h^2}{z^2}+V^2+2g(z-h)\nonumber\\=(z-h)\left[\frac{V^2}{z^2}(z+h)+2g\right]\nonumber
\end{eqnarray}

So for real value of $\dot{z}$ we should have $z>h$.\\

Thus the particle will leave the surface of the cone at a depth $z$, provided $z>h$ i.e., $\left(\dfrac{V^2h^2}{g\tan^2\alpha}\right)^\frac{1}{3}>h$ i.e., $\dfrac{V^2}{g\tan^2\alpha}>h$.\\

Now from the expression of $R$, we have if $R_0$ be the reaction of the surface at the initial position, then $R_0>0$, which implies $\dfrac{V^2}{gh}-\tan^2\alpha>0$ i.e., $\dfrac{V^2}{g\tan^2\alpha}>h$.\\

Hence the particle will leave the surface.\\

Also we have
\begin{eqnarray}
	m\frac{v^2}{\rho_0}\tan\chi&=&H=mg\cos\alpha\sin\psi\nonumber\\\mbox{i.e., }\frac{mv^2\sin^2\psi\cos\alpha}{z\tan\alpha}\tan\chi&=&mg\cos\alpha\sin\psi\nonumber\\\mbox{i.e., }\frac{u}{r}\tan\chi&=&\frac{g}{v}\nonumber\\\mbox{i.e., }\tan\chi&=&\frac{gr}{uv}\nonumber\\\therefore~\cos\chi&=&\frac{uv}{\sqrt{g^2r^2+u^2v^2}}\nonumber\\\therefore~\rho=\rho_0\cos\chi&=&\frac{r}{\frac{u^2}{v^2}\cos\alpha}\frac{uv}{\sqrt{g^2r^2+u^2v^2}}=\frac{r v^3\sec\alpha}{u\sqrt{g^2r^2+u^2v^2}}\nonumber
\end{eqnarray}

{\bf 11. } A heavy particle is projected with velocity $V$ from one end of the horizontal diameter of a smooth sphere of radius $a$ along the inner surface, the direction of projection making an angle $\beta$ with the equator. If the particle never leaves the surface, prove that $$3\sin^2\beta<2+\left(\frac{V^2}{3ag}\right)^2$$ 

{\bf Solution: } The equation of energy gives $v^2=c+2gz$. As $v=V$ at $z=0$, so $c=V^2$ and we have
\begin{equation}\label{eq3.11.1}
	v^2=V^2+2gz
\end{equation}

The conservation of angular momentum gives
\begin{equation}
	a^2\sin^2\theta\dot{\phi}=Va\cos\beta
\end{equation}

The equation of motion along the normal gives
$$m\frac{v^2}{a}=R-mg\cos\theta\mbox{ ~i.e., ~}R=mg\left[\frac{v^2}{ag}+\cos\theta\right]$$
 or using (\ref{eq3.11.1}) we have $$R=\frac{mg}{a}\left[\frac{V^2}{g}+3z\right]$$ 
 
 So $R=0$ when $z=-\dfrac{V^2}{3g}$.\\
 
 Now, \begin{eqnarray}
 	v^2=a^2\sin^2\theta\dot{\phi}^2+a^2\dot{\theta}^2=\frac{V^2\cos^2\beta}{\sin^2\theta}+a^2\dot{\theta}^2=V^2+2gz\nonumber\\\therefore~\dot{z}^2=\left(V^2+2gz\right)\frac{(a^2-z^2)}{a^2}-V^2\cos^2\beta,~~~~~~~~~~~z=a\cos\theta\nonumber
 \end{eqnarray}

So the particle will never leave the surface if
\begin{eqnarray}
	&&\dot{z}^2<0\mbox{ at }z=-\frac{V^2}{3g}\nonumber\\\mbox{i.e., }&&\left(V^2-\frac{2}{3}V^2\right)\left\{a^2-\left(\frac{V^2}{3g}\right)^2\right\}<a^2V^2\cos^2\beta\nonumber\\\mbox{i.e., }&&a^2-\left(\frac{V^2}{3g}\right)^2<3a^2\cos^2\beta=3a^2\left(1-\sin^2\beta\right)\nonumber\\\mbox{i.e., }&&1-\left(\frac{V^2}{3ag}\right)^2<3-3\sin^2\beta\nonumber\\\mbox{i.e., }&&3\sin^2\beta<2+\left(\frac{V^2}{3ag}\right)^2\nonumber
\end{eqnarray} 

{\bf 12. } A particle is projected horizontally under gravity with a velocity $V$ from a point on the inner surface of a smooth sphere at an angular distance $\alpha$ from the lowest point. Show that the $z$ co-ordinate of the highest point of the path of the particle on the surface is the smaller of the values of $z_1$ and $z_2$ given by the equation $z_1^2+\dfrac{V^2}{ag}(z_1+z_0)-a^2=0$ and $3z_2-2z_0+\dfrac{V^2}{g}=0$ where $z_0=a\cos\alpha$.\\

{\bf Solution: } The equation of energy gives
\begin{eqnarray}
	v^2&=&c+2gz\nonumber\\\therefore~V^2&=&c+2ag\cos\alpha\nonumber\\\therefore~v^2&=&V^2+2ag(\cos\theta-\cos\alpha)=V^2+2g(z-z_0)\nonumber
\end{eqnarray}

\begin{wrapfigure}[12]{r}{0.35\textwidth}
	\centering	\includegraphics[height=5 cm , width=4 cm ]{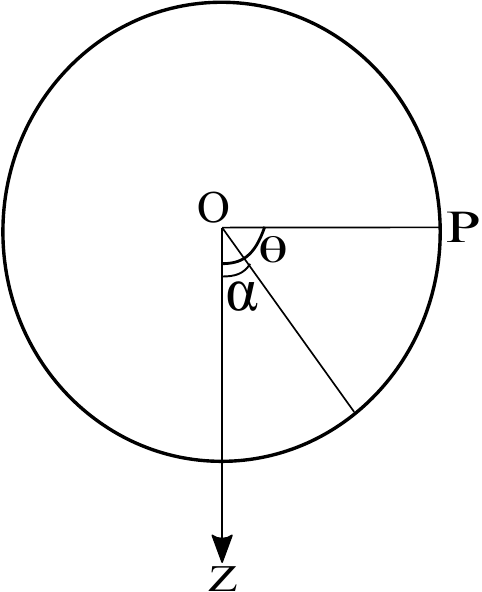}
	\begin{center}
		Fig. 3.26
	\end{center}
\end{wrapfigure} 

The conservation of angular momentum gives
$$a^2\sin^2\theta\dot{\phi}=Va\sin\alpha$$

The equation of motion along the normal gives
\begin{eqnarray}
	m\frac{v^2}{a}&=&R-mg\cos\theta\nonumber\\\therefore~R&=&m\left\{\frac{V^2+2g(z-z_0)}{a}+\frac{gz}{a}\right\}=\frac{mg}{a}\left[\frac{V^2}{g}+3z-2z_0\right]\nonumber
\end{eqnarray}

So $R=0$ at a height $z=z_2$ given by
\begin{equation}
	\frac{V^2}{g}+3z_2-2z_0=0
\end{equation} 

Again,
 \begin{eqnarray}
	v^2&=&a^2\sin^2\theta\dot{\phi}^2+a^2\dot{\theta}^2=\frac{V^2\sin^2\alpha}{\sin^2\theta}+a^2\dot{\theta}^2=V^2+2g(z-z_0)\nonumber\\&=&V^2\frac{(z_0^2-z^2)}{a^2}+2g(z-z_0)\frac{(a^2-z^2)}{a^2}\nonumber\\\therefore~\dot{z}^2&=&\frac{2g}{z}(z-z_0)\left[a^2-z^2-\frac{V^2}{2g}(z+z_0)\right]\nonumber
\end{eqnarray}

Now, $\dot{z}=0$ when $z=z_0$ initially and when
\begin{equation}
	a^2-z_1^2-\frac{V^2}{2g}(z_1+z_0)=0
\end{equation}
so the particle will rise to the smaller of $z_1$ and $z_2$.\\

{\bf 13. } A particle is projected horizontally along the inner surface of a smooth hemisphere whose axis is vertical and whose vertex is downwards, the point of projection being at an angular distance $\beta$ from the lowest point. Show that the initial velocity so that the particle may just ascend to the rim of the hemisphere is $\sqrt{2ag\sec\beta}$.\\

{\bf Solution: } Let $V$ be the velocity of projection. The energy equation gives 
$$v^2=c+2gz=V^2+2g(z-a\cos\beta)$$.

The angular momentum conservation equation gives $$a^2\sin^2\theta\dot{\phi}=Va\sin\beta$$

So, \begin{eqnarray}
	v^2=a^2\sin^2\theta\dot{\phi}^2+a^2\dot{\theta}^2&=&\frac{V^2\sin^2\beta}{\sin^2\theta}+a^2\dot{\theta}^2=V^2+2g(z-a\cos\beta)\nonumber\\\therefore~\dot{z}^2+V^2\sin^2\beta&=&\frac{(a^2-z^2)}{a^2}\left[V^2+2g(z-a\cos\beta)\right], ~~~~~z=a\cos\theta\nonumber
\end{eqnarray}

As $\dot{z}=0$ when $z=0$, so $V^2\sin^2\beta=V^2-2ga\cos\beta$. $$V^2=2ag\sec\beta$$.\vspace*{-.7cm}
\begin{eqnarray}
	\therefore~R&=&m\left(\frac{v^2}{a}+g\frac{z}{a}\right)=m\left[\frac{V^2}{a}+\frac{2g}{a}(z-a\cos\beta)+\frac{gz}{a}\right]\nonumber\\&=&m\left[2g(\sec\beta-\cos\beta)+\frac{3gz}{a}\right]\nonumber\\&=&m\left[2g\sec\beta\sin^2\beta+\frac{3gz}{a}\right]>0,\mbox{ in the hemisphere.}\nonumber
\end{eqnarray}

So the particle will not leave the surface.\\

{\bf 14. } A particle moves on the interior of a smooth sphere of radius $a$ under a force producing an acceleration $\mu\omega^n$ along the perpendicular $\omega$ drawn to the fixed diameter. It is projected with velocity $V$ along the great circle to which this diameter is perpendicular and is slightly disturbed from its path. Show that the new path will cut the old one $m$ times in a revolution, where $$m^2=4\left[1-\frac{\mu a^{n+1}}{V^2}\right]$$

{\bf Solution: } We take the centre of the sphere as origin and the fixed diameter along the $z$-axis. Let $P(r,\theta,z)$ be the position of the particle in cylindrical coordinates. Suppose $v$ be the velocity of the particle at $P$ and $\psi$ is the potential of the force. The energy condition gives
\begin{equation}\label{eq3.14.1}
	\frac{1}{2}v^2+\psi=\mbox{constant}
\end{equation}

\begin{wrapfigure}[12]{r}{0.35\textwidth}
	\centering	\includegraphics[height=5 cm , width=4 cm ]{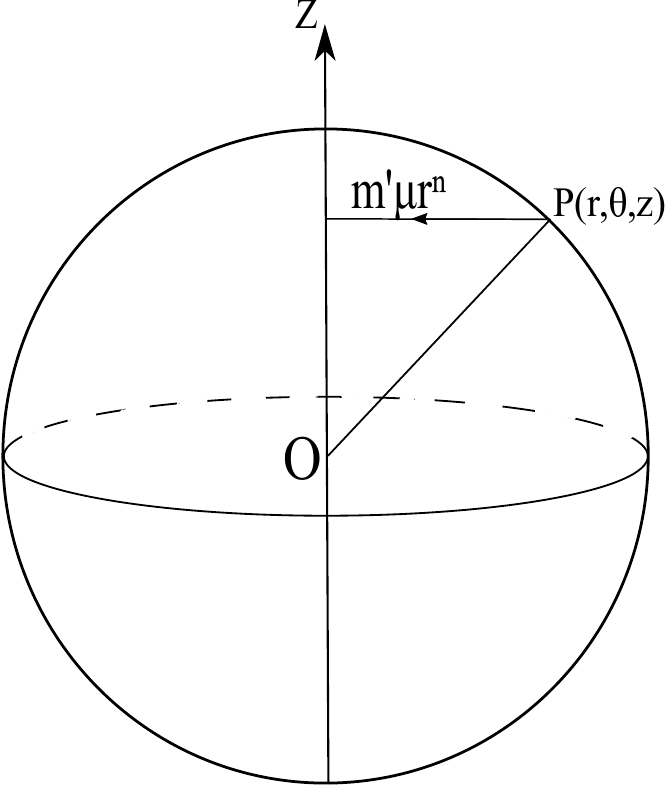}
	\begin{center}
		Fig. 3.27
	\end{center}
\end{wrapfigure} 

Here $-\dfrac{\partial \psi}{\partial r}=-\mu r^n$ ($\because$ the force is in the negative direction)\\

$\mbox{i.e., }\psi=\dfrac{\mu r^{n+1}}{n+1}+\mbox{constant}$.\\

So from (\ref{eq3.14.1}) $\dfrac{1}{2}v^2+\dfrac{\mu r^{n+1}}{n+1}=\mbox{constant}=\dfrac{1}{2}V^2+\dfrac{\mu a^{n+1}}{n+1}$ $$\mbox{i.e.,} v^2=V^2+\frac{2\mu}{n+1}\left(a^{n+1}-r^{n+1}\right)$$

The equation of the sphere in cylindrical coordinate is
$$x^2+y^2+z^2=a^2\mbox{ i.e., }r^2+z^2=a^2$$ with, $x=r\cos\theta$, $y=r\sin\theta$, $0\leq\theta\leq2\pi$.\\

Now differentiating with respect to $t$ we have
\begin{eqnarray}
	r\dot{r}+z\dot{z}&=&0\mbox{ ~i.e.,~ }\dot{r}^2=\frac{z^2\dot{z}^2}{r^2}\nonumber\\
	\therefore~v^2=\dot{r}^2+r^2\dot{\theta}^2+\dot{z}^2&=&\dot{z}^2\left(1+\frac{z^2}{r^2}\right)+r^2\dot{\theta}^2=\frac{a^2}{r^2}\dot{z}^2+r^2\dot{\theta}^2\nonumber
\end{eqnarray} 

The conservation of angular momentum gives $r^2\dot{\theta}=Va$
\begin{eqnarray}
	\therefore~v^2&=&\frac{a^2}{r^2}\dot{z}^2+\frac{V^2a^2}{r^2}=V^2+\frac{2\mu}{n+1}\left(a^{n+1}-r^{n+1}\right)\nonumber\\\therefore~\dot{z}^2&=&-\frac{V^2z^2}{a^2}+\frac{2\mu}{a^2(n+1)}\left[a^{n+1}(a^2-z^2)-{(a^2-z^2)}^\frac{n+3}{2}\right]=f(z) \mbox{ (say)}\nonumber\\
	\mbox{i.e., }\ddot{z}&=&\frac{1}{2}f'(z)\nonumber
\end{eqnarray}

 To consider the disturbed motion of the particle at $z=0$ we put $z=0+\epsilon$.
 \begin{eqnarray}
 	\therefore~\ddot{z}=\ddot{\epsilon}&=&\frac{1}{2}f'(\epsilon)=\frac{1}{2}\left[f'(0)+\epsilon f''(0)+\cdots\right]\nonumber\\&\simeq&\frac{1}{2}\left[f'(0)+\epsilon f''(0)\right]
 \end{eqnarray}

Here it is assumed that $\epsilon$ is so small that $\epsilon^2$ and other higher powers of $\epsilon$ are neglected.\\

Now
\begin{eqnarray}
	f'(z)&=&-\frac{2V^2z}{a^2}+\frac{2\mu}{a^2(n+1)}\left[-2za^{n+1}+(n+3)z{\left(a^2-z^2\right)}^\frac{n+1}{2}\right]\nonumber\\\therefore~f''(z)&=&-\frac{2V^2}{a^2}+\frac{2\mu}{a^2(n+1)}\left[-2a^{n+1}+(n+3){\left(a^2-z^2\right)}^\frac{n+1}{2}-(n+3)(n+1)z^2{(a^2-z^2)}^\frac{n-1}{2}\right]\nonumber
\end{eqnarray}

So, $f'(0)=0$, $f''(0)=-2\dfrac{V^2}{a^2}+\dfrac{2\mu a^{n+1}}{a^2}$ \\

Thus $\ddot{\epsilon}=-\dfrac{1}{a^2}\left(V^2-\mu a^{n+1}\right)\epsilon=-K^2\epsilon$\\
where for stability $K^2=\dfrac{V^2-\mu a^{n+1}}{a^2}>0$.\\

Hence if $V^2>\mu a^{n+1}$, the motion is a simple harmonic motion, having period of oscillation $T=\dfrac{2\pi}{K}$.\\

Now, $r^2\dot{\theta}=Va$\\

 i.e., $\dot{\theta}=\dfrac{Va}{r^2}=\dfrac{Va}{a^2-z^2}\simeq\dfrac{Va}{a^2}=\dfrac{V}{a}$ (when $z$ very small).\\

So on integration, $$\theta=\frac{V}{a}t+c$$

Assuming $\theta=0$ at $t=0$ $\implies c=0$ and we have $\theta=\dfrac{V}{a}t$.\\

Let $\theta_1$ and $\theta_2$ be the values of $\theta$ where the new path of the particle cuts the old part at two consecutive points. If $t_1$ and $t_2$ be the corresponding times then
$$\theta_1-\theta_2=\frac{V}{a}(t_1-t_2)$$

Since the new path cuts the old path $m$ times, so we have 
\begin{eqnarray}
	\theta_1-\theta_2&=&\dfrac{2\pi}{m} \mbox{ ~and~ } t_1-t_2=\frac{T}{2}\nonumber\\\therefore~\frac{2\pi}{m}&=&\frac{V}{a}\frac{T}{2}=\frac{V}{a}\frac{\pi}{K}\nonumber\\\therefore~m&=&\dfrac{2aK}{V}\implies m^2=\frac{4a^2}{V^2}K^2=4\left(1-\frac{\mu a^{n+1}}{V^2}\right)\nonumber
\end{eqnarray}

{\bf 15. } A particle is moving under gravity in a horizontal circle on the inner surface of a smooth sphere. A slant disturbance is given to the particle keeping the angular momentum unaltered. Discuss the stability of the motion\\

{\bf Solution: } We take the centre of the sphere as origin and $z$-axis vertically download. Let $v$ the velocity of the particle at $P(a,\theta,\phi)$ (in spherical polar coordinate). Then the equation of energy gives $$v^2=c+2gz=c+2ag\cos\theta$$

\begin{wrapfigure}[11]{r}{0.35\textwidth}
	\centering	\includegraphics[height=4.5 cm , width=4 cm ]{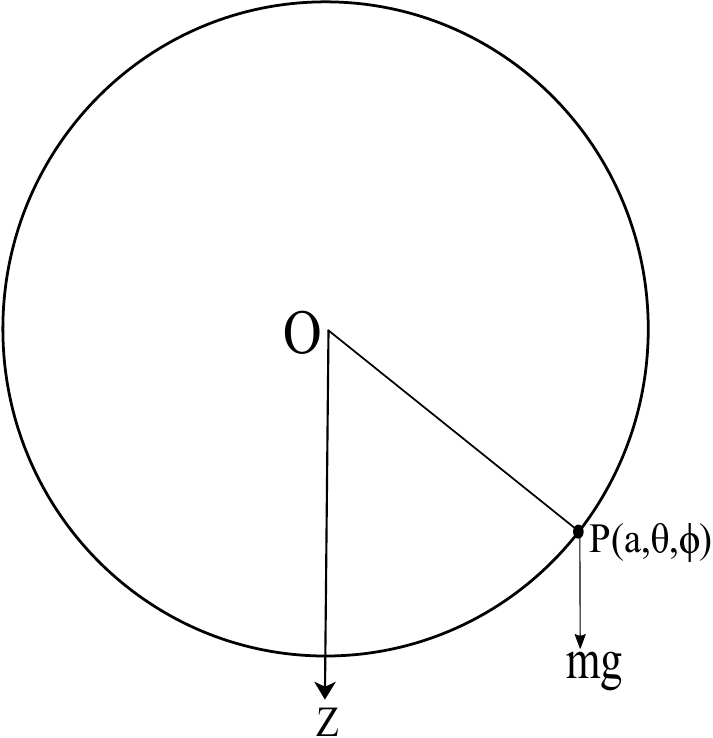}
	\begin{center}
		Fig. 3.28
	\end{center}
\end{wrapfigure}  

The equation of angular momentum gives
\begin{eqnarray}
	a^2\sin^2\theta\dot{\phi}&=&h\nonumber\\\therefore~v^2&=&a^2\dot{\theta}^2+a^2\sin^2\theta\dot{\phi}^2=a^2\dot{\theta}^2+\frac{h^2}{a^2\sin^2\theta}\nonumber\\\therefore~a^2\dot{\theta}^2&=&c-\frac{h^2}{a^2\sin^2\theta}+2ag\cos\theta\nonumber
\end{eqnarray}

Now differentiating both side with respect to $\theta$ gives
\begin{equation}\label{eq3.15.1}
	a^2\ddot{\theta}=\frac{h^2\cos\theta}{a^2\sin^3\theta}-ag\sin\theta
\end{equation}

If the particle describes a horizontal circle at $\theta=\alpha$ then $\dot{\theta}=0=\ddot{\theta}$ at $\theta=\alpha$. So from (\ref{eq3.15.1}),
\begin{equation}\label{eq3.15.2}
	0=\frac{h^2\cos\alpha}{a^2\sin^3\alpha}-ag\sin\alpha\mbox{~ i.e., ~}h^2\cos\alpha=a^3g\sin^4\alpha
\end{equation}

Let a slight disturbance is given to the particle so that its angular momentum remains unaltered. If $\theta$ differs form $\alpha$ by a small quantity $\xi$ where $\xi$ is small compared to $\alpha$, i.e., $\theta=\alpha+\xi$, then from (\ref{eq3.15.1})
\begin{eqnarray}
	a^2\ddot{\xi}&=&\frac{h^2\cos(\alpha+\xi)}{a^2\sin^3(\alpha+\xi)}-ag\sin(\alpha+\xi)\nonumber\\&=&\frac{h^2}{a^2}\frac{(\cos\alpha-\xi\sin\alpha)}{(\sin\alpha+\xi\cos\alpha)^3}-ag(\sin\alpha+\xi\cos\alpha)\nonumber\\&&\mbox{~~~~~~~ (since $\xi$ is small so $\cos\xi\approx1$ and $\sin\xi\approx\xi$)}\nonumber\\&=&\frac{h^2}{a^2\sin^3\alpha}(\cos\alpha-\xi\sin\alpha)(1+\xi\cot\alpha)^{-3}-ag(\sin\alpha+\xi\cos\alpha)\nonumber\\&=&\frac{a^3g\sin^4\alpha}{a^2\sin^3\alpha\cos\alpha}(\cos\alpha-\xi\sin\alpha)(1-3\xi\cot\alpha)-ag(\sin\alpha+\xi\cos\alpha)\nonumber\\&=&-\frac{ag}{\cos\alpha}\left(1+3\cos^2\alpha\right)\xi \mbox{~~~ (neglecting square and higher powers of $\xi$)}\nonumber
\end{eqnarray}

As from (\ref{eq3.15.2}), $\cos\alpha$ is positive, so
$$\ddot{\xi}=-\frac{g}{a\cos\alpha}\left(1+3\cos^2\alpha\right)\xi$$ 

Hence the disturbed motion of the particle is simple harmonic in nature and the path is a stable path. The period of oscillation is
$$T=\frac{2\pi}{\sqrt{\frac{g}{a\cos\alpha}\left(1+3\cos^2\alpha\right)}}=\frac{2\pi\sqrt{a\cos\alpha}}{\sqrt{g\left(1+3\cos^2\alpha\right)}}$$

Now, $a^2\sin^2\theta\dot{\phi}=h$ $\implies$ $\dot{\phi}=\dfrac{h}{a^2\sin^2\theta}\simeq\dfrac{h}{a^2\sin^2\alpha}$.
$$\therefore~\phi=\frac{h}{a^2\sin^2\alpha}t+c$$

Assuming $\phi=0$ at $t=0$, we get $c=0$. $$\therefore~\phi=\frac{h}{a^2\sin^2\alpha}t$$

If $\theta_1$, $\theta_2$ are the values of $\phi$ corresponding to two consecutive and minimum values of $\theta$ and $t_1$, $t_2$ are the values of $t$ at this two positions then 
\begin{eqnarray}
	\phi_1-\phi_2&=&\frac{h}{a^2\sin^2\alpha}\left(t_1-t_2\right)=\frac{h}{a^2\sin^2\alpha}\frac{T}{2}\nonumber\\&=&\frac{\pi}{\sqrt{1+3\cos^2\alpha}} \mbox{~~~ (Using (\ref{eq3.15.2}))}\nonumber
\end{eqnarray}

This may be called the apsidal angle of the path.\\

{\bf 15. } A particle tied at one end of a fine string of length $a$ whose other end is attached to a fixed point, is projected horizontally with velocity $V$ with the string inclined at an acute angle $\alpha$ with the downward vertical and is again moving horizontally when the string makes at acute angle $\beta$ with the upward vertical without the string going slack. Find $V$ and prove that (i) $\beta>\alpha$, (ii) $2\sin^2\alpha\geqslant\cos\beta(\cos\alpha-\cos\beta)$.\\

{\bf Solution: } From the equation of energy we have $$v^2=c+2gz=c+2ga\cos\theta$$ 

\begin{wrapfigure}[8]{r}{0.3\textwidth}
	\centering	\includegraphics[height=4.50 cm , width=4 cm ]{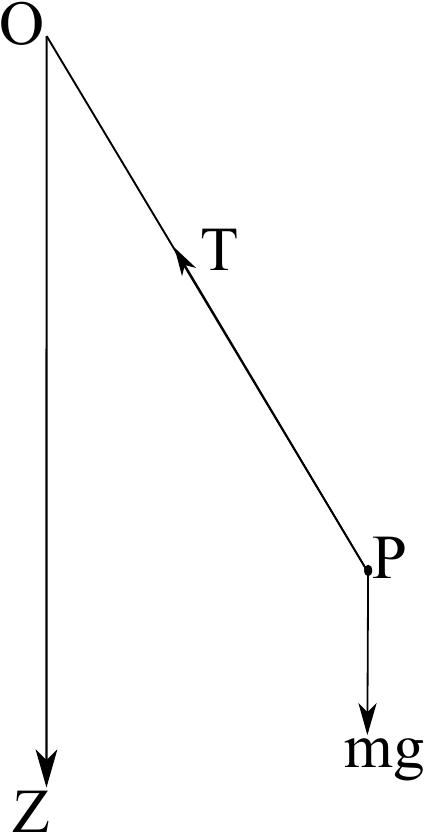}
	\begin{center}
		Fig. 3.29
	\end{center}
\end{wrapfigure}  

As $v=V$ when $z=a\cos\alpha$, so $$v^2=V^2+2ga(\cos\theta-\cos\alpha)$$

Equation of angular momentum gives
\begin{eqnarray}
	a^2\sin^2\theta\dot{\phi}&=&\mbox{constant}=Va\sin\alpha\nonumber\\\therefore~v^2&=&a^2\dot{\theta^2}+a^2\sin^2\theta\dot{\phi}^2\nonumber\\&=&a^2\dot{\theta}^2+\frac{V^2\sin^2\alpha}{\sin^2\theta}=V^2+2ga(\cos\theta-\cos\alpha)\nonumber\\\therefore~a^2\dot{\theta}^2&=&V^2\frac{\left(\sin^2\theta-\sin^2\alpha\right)}{\sin^2\theta}+2ga(\cos\theta-\cos\alpha)\nonumber\\&=&V^2\frac{\left(\cos^2\alpha-\cos^2\theta\right)}{\sin^2\theta}-2ag(\cos\alpha-\cos\theta)\nonumber\\&=&(\cos\alpha-\cos\theta)\left[V^2\frac{(\cos\alpha+\cos\theta)}{\sin^2\theta}-2ag\right]\nonumber
\end{eqnarray} 

As $\dot{\theta}=0$ at $\theta=\alpha$ and $\theta=\pi-\beta$, so we have
$$V^2\frac{(\cos\alpha-\cos\beta)}{\sin^2\beta}-2ag=0$$

So we must have $\cos\alpha>\cos\beta$ i.e., $\alpha<\beta$ and
$$V^2=\frac{2ag\sin^2\beta}{(\cos\alpha-\cos\beta)}$$

Resolving along $OP$,
\begin{eqnarray}
	&&m\frac{v^2}{a}=T-mg\cos\theta\nonumber\\\implies&& m\left[\frac{V^2+2ag(\cos\theta-\cos\alpha)}{a}\right]=T-mg\cos\theta\nonumber\\\implies&& T=\frac{m}{a}\left[=\frac{2ag\sin^2\beta}{(\cos\alpha-\cos\beta)}+2ag(\cos\theta-\cos\alpha)\right]+mg\cos\theta\nonumber
\end{eqnarray}

As $T$ is positive throughout the motion and $T$ is positive at $\theta=\pi-\beta$, so
\begin{eqnarray}
	&&\frac{2\sin^2\beta}{(\cos\alpha-\cos\beta)}-2(\cos\alpha+\cos\beta)-\cos\beta\geqslant0\nonumber\\\mbox{i.e., }&&2\sin^2\alpha\geqslant\cos\beta(\cos\alpha-\cos\beta)\nonumber
\end{eqnarray} 

{\bf 16. } The surface of a smooth funnel is given by the equation $$z=c^2\left(x^2+y^2\right)^{-\frac{1}{2}}$$
 the positive axis of $z$ being vertically downwards. A particle is projected horizontally, along the inner surface, at the level $z=c$ with such a velocity that it is again moving horizontally at the level $z=2c$. Prove that the principal normal, at the highest point of the path, makes an angle $\tan^{-1}\left(\dfrac{1}{5}\right)$ with the horizontal.\\
 
 {\bf Solution: } The equation of energy gives $$v^2=V^2+2g(z-c)$$ 
 
 Equation of angular momentum gives $$r^2\dot{\theta}=Vc$$
 
 Equation of Meridian plane is $zr=c^2$ $$\mbox{i.e., }z\dot{r}+r\dot{z}=0\mbox{ ~i.e.,~ }\dot{r}=-\frac{r\dot{z}}{z}=-\frac{c^2\dot{z}}{z^2}$$
 	
 	Now,\begin{eqnarray}
 		v^2&=&\dot{r}^2+r^2\dot{\theta}^2+\dot{z}^2\nonumber\\&=&\frac{c^4}z^4\dot{z}^2+\dot{z}^2+\frac{V^2c^2}{r^2}=V^2+2g(z-c)\nonumber\\\therefore~\dot{z}^2\left(1+\frac{c^4}{z^4}\right)&=&V^2\frac{(c^2-z^2)}{c^2}+2g(z-c)\nonumber
 	\end{eqnarray}
 	
 	Hence $\dot{z}=0$ gives $$V^2\frac{(c^2-z^2)}{c^2}+2g(z-c)=0$$
 	
 	As $\dot{z}=0$ when $z=c$ and $z=2c$,so we have $$2g=\frac{3cV^2}{c^2}\mbox{~ i.e., ~}V^2=\frac{2}{3}gc$$
 	
 	Now, $rz=c^2$, so $\dfrac{\mathrm{d}r}{\mathrm{d}z}=-\dfrac{c^2}{z^2}=-1=\cot\theta$ (say) at $z=c$.\\
 	
 	Also, $$\tan\chi=\frac{gr_0}{v_0}=\frac{gc}{V}=\frac{gc}{\frac{2}{3}gc}=\frac{3}{2}$$
 	
 	Hence the required angle $=\theta+\chi$
 	\begin{eqnarray}
 	\therefore&&\tan(\theta+\chi)=\frac{\frac{3}{2}-1}{1+\frac{3}{2}}=\frac{1}{5}\nonumber\\\therefore&&\tan A=\tan\left(\chi-(90^o-\theta)\right)=-\cot(\theta+\chi)=-5\nonumber	
 	\end{eqnarray}
 
 {\bf 17. } A particle is moving under gravity in a horizontal circle in the inner surface of a smooth paraboloid of revolution with its axis vertical. A slight disturbance is given to the particle keeping the angular momentum about the axis unaltered. Discuss the  stability of motion of the particle.\\
 
 {\bf Solution: } We take origin at the vertex of the paraboloid, $z$-axis along the axis of the paraboloid and vertically upward. Let $P(r,\theta,z)$ be the position of the particle at time $t$ in cylindrical co-ordinates. Then the velocity $v$ of the particle is given by
 \begin{equation}\label{eq3.17.1}
 	v^2=\dot{r}^2+r^2\dot{\theta}^2+\dot{z}^2
 \end{equation}
 
 \begin{wrapfigure}[14]{r}{0.35\textwidth}
 	\centering	\includegraphics[height=5 cm , width=7 cm ]{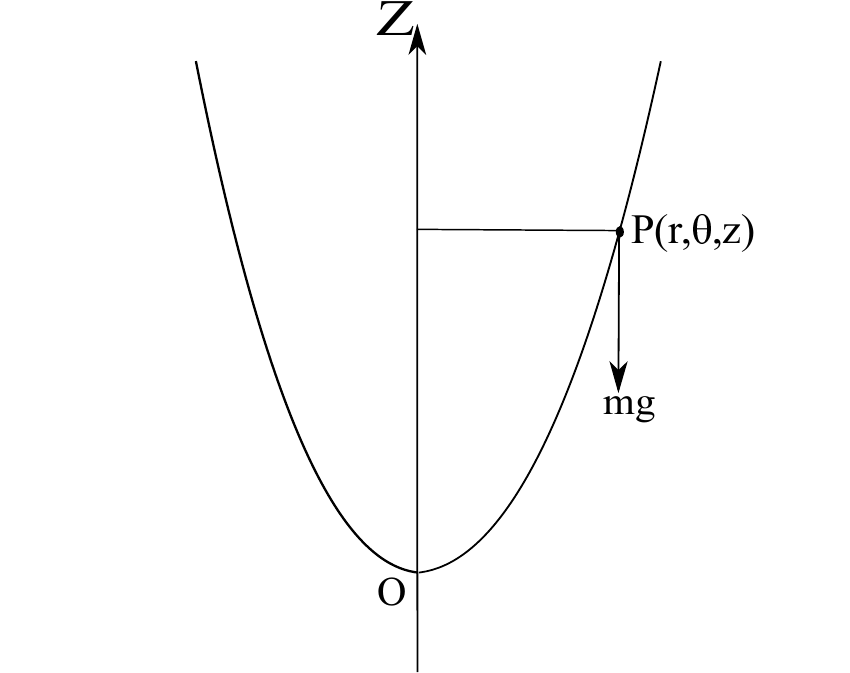}
 	\begin{center}
 		Fig. 3.30
 	\end{center}
 \end{wrapfigure}  

Suppose $z=f(r)$ be the equation of the meridian curve. From the equation of energy, we get
\begin{equation}\label{eq3.17.2}
	v^2=c-2gz
\end{equation}
and from the equation of angular momentum
\begin{equation}\label{eq3.17.3}
	r^2\dot{\theta}=h
\end{equation}

Using (\ref{eq3.17.2}) and (\ref{eq3.17.3}) in (\ref{eq3.17.1}) and also using the equation of the meridian plane we get
$$\dot{r}^2\left(1+\left\{f'(r)\right\}^2\right)+\frac{h^2}{r^2}=c-2gf(r)$$

Now, differentiating both side with respect to $r$, we have
\begin{eqnarray}
	2\ddot{r}\left(1+\left\{f'(r)\right\}^2\right)+2\dot{r}^2f'(r)f''(r)-2\frac{h^2}{r^3}=-2gf'(r)\nonumber\\\mbox{i.e., }\left(1+\left\{f'(r)\right\}^2\right)\ddot{r}+\dot{r}^2f'(r)f''(r)-\frac{h^2}{r^3}=-gf'(r)\label{eq3.17.4}
\end{eqnarray}

If the particle describes the circle $r=a$, then $\dot{r}=0=\ddot{r}$ at $r=a$, so from (\ref{eq3.17.4}) we have
\begin{equation}
	\frac{h^2}{a^3}=gf'(a)
\end{equation}

Let us now suppose that a slight disturbance is given to the particle such that the angular momentum about the axis remains unaltered. For this disturbed motion, $r=a+\xi$, where $\xi$ is small compared to $a$ so that we can neglect $\xi^2$ and higher power of $\xi$. We also assume that we can neglect $\dot{\xi}^2$, then from (\ref{eq3.17.4}) we get
\begin{eqnarray}
	&&\left(1+\left\{f'(a+\xi)\right\}^2\right)\ddot{\xi}+\dot{\xi}^2f'(a+\xi)f''(a+\xi)-\frac{h^2}{(a+\xi)^3}=-gf'(a+\xi)\nonumber\\\implies&&\ddot{\xi}\left(1+\{f'(a)\}^2+2\xi f'(a)f''(a)\right)-\frac{h^2}{a^3}\left(1-3\frac{\xi}{a}\right)=-g\left\{f'(a)+\xi f''(a)\right\}\nonumber\\&&\mbox{~~~~~~~~~~~~~~~ (neglecting square and higher powers of $\xi$ and its derivatives)}\nonumber\\\implies&&\ddot{\xi}\left(1+{\{f'(a)\}}^2\right)=-\frac{g}{a}\xi\left\{3f'(a)+af''(a)\right\}\nonumber\\\mbox{i.e., }&&\ddot{\xi}=-\frac{g}{a}\left\{\frac{3f'(a)+af''(a)}{1+{\{f'(a)\}}^2}\right\}\xi\nonumber
\end{eqnarray}

It follows that if, $3f'(a)+af''(a)>0$, the disturbed motion will be stable otherwise it is unstable.\\
	
	If the motion is stable, the period of complete oscillation is $\dfrac{2\pi}{K}$, where $K^2=\dfrac{3f'(a)+af''(a)}{1+{\{f'(a)\}}^2}$.\\
	
	Hrom (\ref{eq3.17.3}), $r^2\dot{\theta}=h$, i.e., $\dot{\theta}=\dfrac{h}{r^2}\simeq\dfrac{h}{a^2}$.\\
	
	Thus the change in azimuth form the maximum $r$ to the consecutive minimum is
	$$\frac{h}{a^2}\frac{T}{2}=\frac{\sqrt{ga^3f'(a)}}{a^2}\pi\sqrt{\frac{a}{g}\frac{{\left\{1+{\{f'(a)\}}^2\right\}}}{\{3f'(a)+af''(a)\}}}=\pi\sqrt{\frac{f'(a){\left\{1+{\{f'(a)\}}^2\right\}}}{\{3f'(a)+af''(a)\}}}$$
	
{\bf 18. } Discuss the motion of a heavy particle on the surface of a smooth right circular cone with axis vertical and vertex downward.\\

{\bf Solution: } The equation of motion along the line $OP$ is
$$\left(\ddot{r}-r\dot{\theta}^2\right)\sin\alpha+\ddot{z}\cos\alpha=-g\cos\alpha$$

Equation of angular momentum gives
$$r^2\dot{\theta}=h,~r=z\tan\alpha$$

So we have
\begin{eqnarray}
	\ddot{r}\left(sin\alpha+\frac{\cos^2\alpha}{\sin\alpha}\right)-\frac{h^2}{r^3}\sin\alpha=-g\cos\alpha\nonumber\\\mbox{i.e., }\ddot{r}-\frac{h^2}{r^3}\sin^2\alpha=-g\cos\alpha\sin\alpha\label{eq3.18.1}
\end{eqnarray}

 \begin{wrapfigure}[9]{r}{0.35\textwidth}
	\centering	\includegraphics[height=5 cm , width=5 cm ]{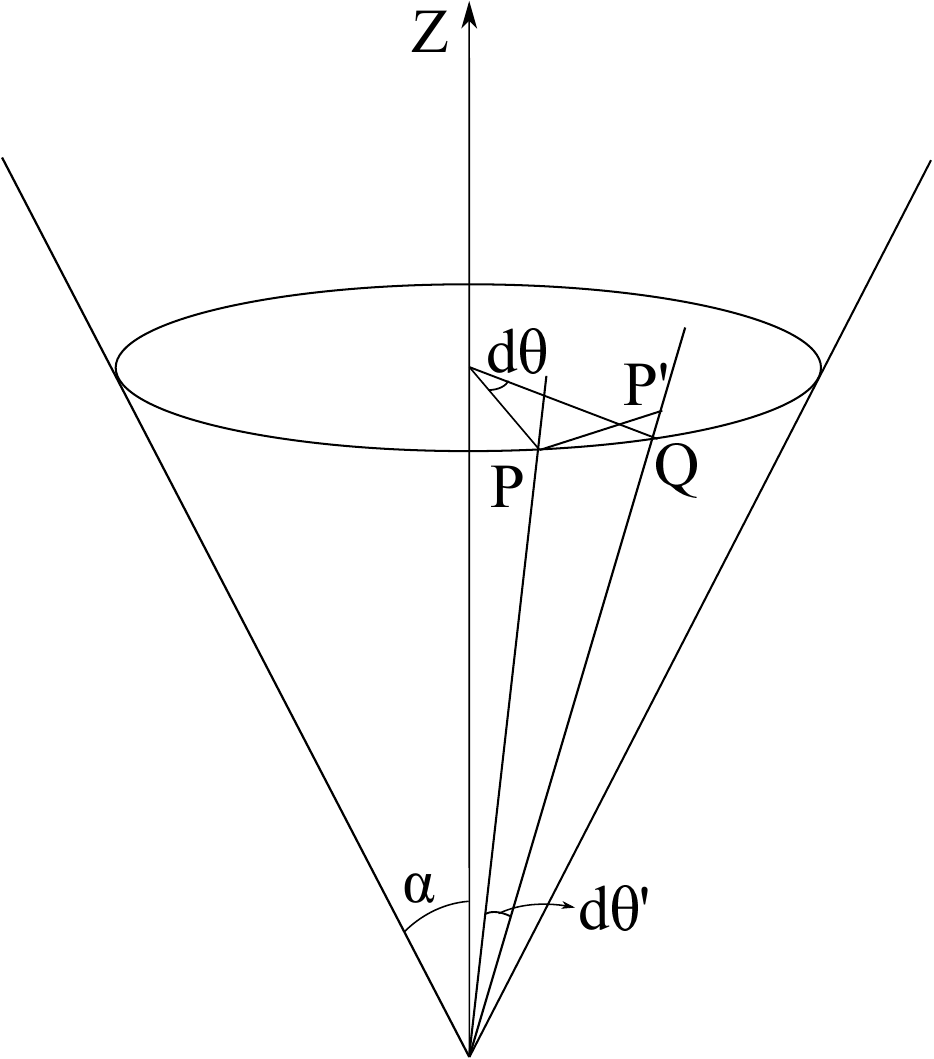}
	\begin{center}
		Fig. 3.31
	\end{center}
\end{wrapfigure}  

Choosing $u=\dfrac{1}{r}$ we have
\begin{eqnarray}
	&&\dot{r}=-\frac{1}{u^2}\frac{\mathrm{d}u}{\mathrm{d}t}=-\frac{1}{u^2}\frac{\mathrm{d}u}{\mathrm{d}\theta}\dot{\theta}=-h\frac{\mathrm{d}u}{\mathrm{d}\theta}\nonumber\\
	&&\ddot{r}=-h\frac{\mathrm{d}^2u}{\mathrm{d}\theta^2}hu^2=-h^2u^2\frac{\mathrm{d}^2u}{\mathrm{d}\theta^2}\nonumber\\
	\therefore~&&-h^2u^2\frac{\mathrm{d}^2u}{\mathrm{d}\theta^2}-h^2u^3\sin^2\alpha=-g\cos\alpha\sin\alpha\nonumber\\
	\mbox{i.e., }&&\frac{\mathrm{d}^2u}{\mathrm{d}\theta'^2}
\end{eqnarray}

This is the differential equation of the path of the particle. Also from (\ref{eq3.18.1}) if $\dot{r}=0=\ddot{r}$ when $r=a$ then
\begin{eqnarray}
	&&-\frac{h^2}{a^3}\sin^2\alpha=-g\cos\alpha\sin\alpha\nonumber\\\mbox{i.e., }&&\frac{h^2}{a^3}=g\cot\alpha\nonumber
\end{eqnarray}

For small disturbance $r=a+\xi$, we get from (\ref{eq3.18.1})
\begin{eqnarray}
	&&\ddot{\xi}-\frac{h^2\sin^2\alpha}{a^3}\left(1+\frac{\xi}{a}\right)^{-3}=-g\cos\alpha\sin\alpha\nonumber\\\mbox{i.e., }&&\ddot{\xi}-g\cos\alpha\sin\alpha\left(1-3\frac{\xi}{a}\right)=-g\cos\alpha\sin\alpha\nonumber\\&&~~~~~~~~~\mbox{(neglecting square and higher powers of $\xi$)}\nonumber\\\mbox{i.e., }&&\ddot{\xi}+3\frac{g}{a}\sin\alpha\cos\alpha\xi=0\nonumber
\end{eqnarray}

So the disturbed motion will be stable with the period of complete oscillation $\dfrac{2\pi}{K}$, where $K^2=3\dfrac{g}{a}\sin\alpha\cos\alpha$.\\

To discuss the vertical motion, let $OP=r'$ and $\angle POQ=\mathrm{d}\theta'$. So we have
\begin{eqnarray}
	&&r=r'\sin\alpha,~PQ=r\mathrm{d}\theta=r\mathrm{d}\theta'\nonumber\\\mbox{i.e., }&&\mathrm{d}\theta=\frac{r'}{r}\mathrm{d}\theta'=\frac{1}{\sin\alpha}\mathrm{d}\theta'\nonumber\\\therefore~&&\frac{\mathrm{d}u}{\mathrm{d}\theta}=\sin\alpha\frac{\mathrm{d}u}{\mathrm{d}\theta'}=\frac{\mathrm{d}u}{\mathrm{d}\theta'}(u\sin\alpha)=\frac{\mathrm{d}u'}{\mathrm{d}\theta'}~~~\left(u'=\frac{1}{r'}\right)\nonumber\\\therefore~&&\frac{\mathrm{d}^2u}{\mathrm{d}\theta^2}=\frac{\mathrm{d}}{\mathrm{d}\theta}\left(\frac{\mathrm{d}u'}{\mathrm{d}\theta'}\right)=\sin\alpha\frac{\mathrm{d}^2u}{\mathrm{d}{\theta'}^2}\nonumber
\end{eqnarray}

So the path of the particle becomes
\begin{eqnarray}
	&&\sin\alpha\frac{\mathrm{d}^2u'}{\mathrm{d}{\theta'}^2}+u'\sin\alpha=\frac{g\sin^3\alpha\cos\alpha}{h^2{u'}^2}\nonumber\\\mbox{i.e., }&&\frac{\mathrm{d}^2u'}{\mathrm{d}{\theta'}^2}+u'=\frac{g\sin^2\alpha\cos\alpha}{h^2{u'}^2}\nonumber
\end{eqnarray}

It follows that if the cone be developed into a plane the trace of the path on the surface will be the same as if a particle moves in a plane under the action of a constant central force.\\

Further, from the energy equation,
$$\dot{r}^2+r^2\dot{\theta}^2+\dot{z}^2=c-2gz$$
with $r=z\tan\alpha$, $r^2\dot{\theta}=h$.\\

So we get $$\dot{z}^2\sec^2\alpha+\frac{h^2}{z^2}\cot^2\alpha+2gz=c$$

This gives the vertical motion of the particle.\\

{\bf 19. } A particle is moving under gravity in a horizontal circle in the inner surface of a smooth cone with axis vertical and vertex downwards. A slight disturbance is given to the particle, keeping the angular momentum about the axis unaltered. Discuss the stability of motion of the particle.\\

{\bf Solution: } We take the vertex of the cone $O$ as origin and $z$-axis along the axis of the cone and vertically upward. Using spherical polar co-ordinate $P(r,\alpha,\phi)$ the velocity of it at time $t$ is given by $$v^2=\dot{r}^2+r^2\sin^2\alpha~\dot{\phi}^2$$

 \begin{wrapfigure}[9]{r}{0.35\textwidth}
	\centering	\includegraphics[height=5 cm , width=5 cm ]{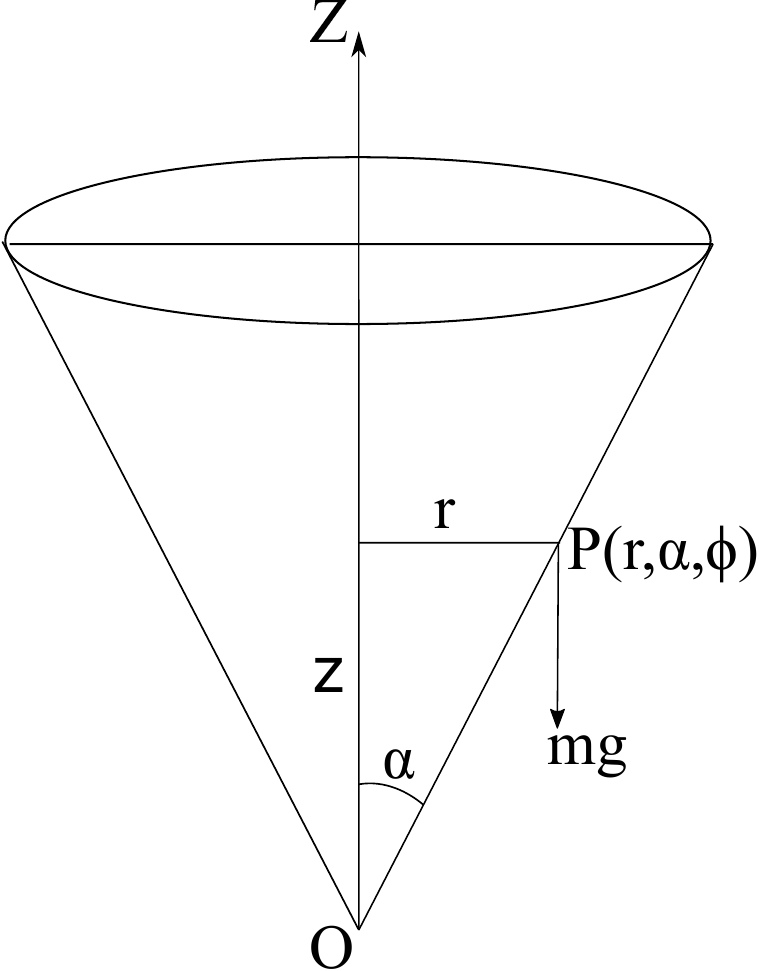}
	\begin{center}
		Fig. 3.32
	\end{center}
\end{wrapfigure}  

The energy momentum gives 
\begin{equation}\label{eq3.19.1}
	v^2=c-2gz=c-2gr\cos\alpha
\end{equation}

The conservation of angular momentum gives
\begin{equation}\label{eq3.19.2}
	r^2\sin^2\alpha~\dot{\phi}=h
\end{equation}

Using (\ref{eq3.19.1}) and (\ref{eq3.19.2}) in the expression for velocity we get
\begin{eqnarray}
	\dot{r}^2+r^2\sin^2\alpha~\dot{\phi}^2=c-2gr\cos\alpha\nonumber\\\mbox{i.e.,~ }\dot{r}^2+\frac{h^2}{r^2\sin^2\alpha}=c-2gr\cos\alpha\nonumber
\end{eqnarray}

Now, differentiating both side with respect to $r$
\begin{equation}\label{eq3.19.3}
	\ddot{r}-\frac{h^2}{r^3\sin^2\alpha}=-g\cos\alpha
\end{equation}

If the particle describes the circle $r=a$, i.e., $\dot{r}=0=\ddot{r}$ when $r=a$, then from (\ref{eq3.19.3})
\begin{equation}
	\frac{h^2}{a^3}=g\sin^2\alpha\cos\alpha
\end{equation}

Let us now suppose that a slight disturbance is given to the particle such that the angular momentum about the axis remains unaltered. For this disturbed motion $r=a+\xi$, where $\xi$ is small compared to $a$ so that we can neglect $\xi^2$ and higher powers of $\xi$. Hence from (\ref{eq3.19.3})
\begin{eqnarray}
	&&	\ddot{\xi}-\frac{h^2}{(a+\xi)^3\sin^2\alpha}=-g\cos\alpha\nonumber\\\mbox{i.e., }&&	\ddot{\xi}-\frac{h^2}{a^3\sin^2\alpha}\left(1-3\frac{\xi}{a}\right)=-g\cos\alpha\nonumber\\
	\mbox{i.e., }&&\ddot{\xi}+3\frac{g}{a}\cos\alpha~\xi=0
\end{eqnarray}

As $K^2=3\dfrac{g}{a}\cos\alpha>0$ so the disturbed motion is stable and the period of oscillation is $\dfrac{2\pi}{K}$.\\

From (\ref{eq3.19.2}), $\dot{\phi}=\dfrac{h}{r^2\sin^2\alpha}\simeq\dfrac{h}{a^2\sin^2\alpha}$.\\

Tf $\phi_1$, $\phi_2$ are the values of $\phi$ corresponding to two consecutive maximum and minimum value of $r$ and $t_1$, $t_2$ are the times at this two position then
\begin{eqnarray}
	\phi_1-\phi_2&=&\dfrac{h}{a^2\sin^2\alpha}(t_1-t_2)=\dfrac{h}{a^2\sin^2\alpha}\frac{T}{2}\nonumber\\&=&\frac{\pi}{\sqrt{\frac{3g\cos\alpha}{a}}}\frac{\left(a^3g\sin^2\alpha\cos\alpha\right)^\frac{1}{2}}{a^2\sin^2\alpha}=\frac{\pi}{\sqrt{3}\sin\alpha},\nonumber
\end{eqnarray} 
represents the apsidal angle..\\

{\bf 20. } A heavy particle moves on the external surface of a smooth circular cone, which is fixed with its axis vertical and vertex upwards. The velocity of the particle is that which it would acquire by sliding from the vertex. Show that the path of the particle, when developed into a plane, has a polar equation of the form $$r^3=a^3\sec^2\left(\frac{3\theta}{2}\right)$$
referred to the vertex as origin.\\

{\bf Solution: } Let $P(r,\alpha,\phi)$ be the spherical polar co-ordinates of the particle on the cone at time $t$. Let $O$, $P$, $Q$ on the cone becomes $O'$, $P'$, $Q'$ on the developed plane. Then assuming arc $PQ=\mathrm{d}s$ and arc $P'Q'=\mathrm{d}s'$ we have $\mathrm{d}s=\mathrm{d}s'$. \\

Now, $r=OP=O'P'=r'$ (say)\\

So,\begin{eqnarray}
	&&\mathrm{d}s^2=\mathrm{d}r^2+r^2\sin^2\alpha\mathrm{d}\phi^2=(\mathrm{d}r')^2+r'^2(\mathrm{d}\theta')^2\nonumber\\\mbox{i.e., }&&\sin\alpha\mathrm{d}\phi=\mathrm{d}\theta'
\end{eqnarray}

Resolving the velocity perpendicular to the $ZOP$ plane we obtain
\begin{equation}
	r^2\sin^2\alpha\dot{\phi}=h~~\mbox{ (constant)}
\end{equation}

\begin{wrapfigure}[9]{r}{0.35\textwidth}
	\centering	\includegraphics[height=6 cm , width=6 cm ]{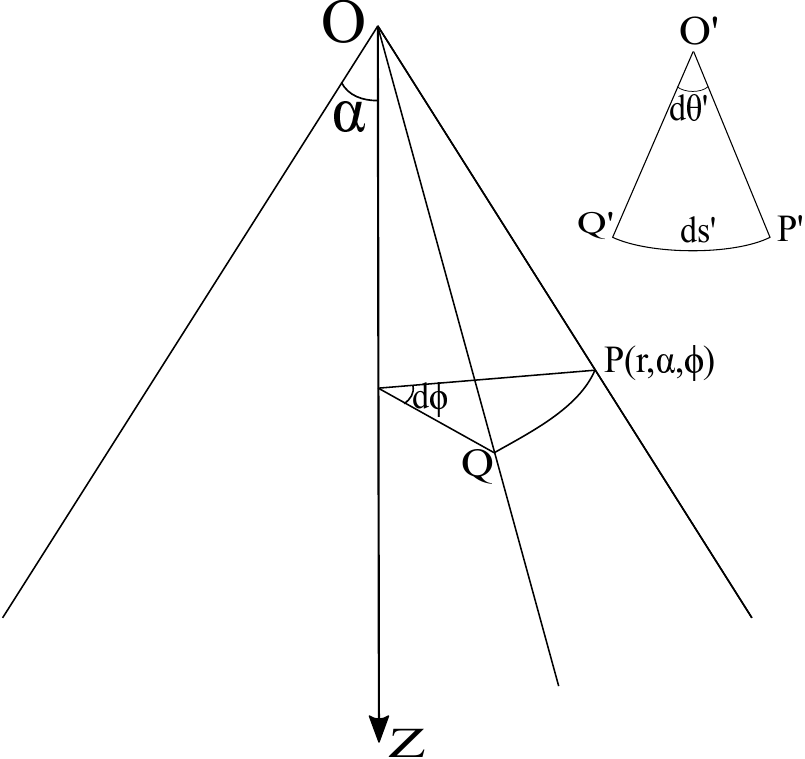}
	\begin{center}
		Fig. 3.33
	\end{center}
\end{wrapfigure}  

The energy equation conservation equation gives
\begin{equation}
	v^2=c+2gz=c+2gr\cos\alpha\nonumber
\end{equation}

But $v=0$ at $z=0$, $\implies c=0$.
\begin{eqnarray} 
\therefore&&v^2=2gr\cos\alpha\nonumber\\\implies&&\dot{r}^2+r^2\sin^2\alpha\dot{\phi}^2=2gr\cos\alpha\nonumber\\\mbox{i.e.,}&&\dot{r}^2=2gr\cos\alpha-\frac{h^2}{r^2\sin^2\alpha}\nonumber\\&&~~~=\frac{h^2r}{\sin^2\alpha}\left[\frac{2g\cos\alpha\sin^2\alpha}{h^2}-\frac{1}{r^3}\right]\nonumber\\&&~~~=\frac{h^2r}{\sin^2\alpha}\left(\frac{1}{a^3}-\frac{1}{r^3}\right), ~~~~~a^3=\frac{h^2}{2g\cos\alpha\sin^2\alpha}\nonumber
\end{eqnarray}

As, $r^4\sin^4\alpha\dot{\phi}^2=h^2$, so
\begin{eqnarray}
	&&\frac{1}{r^5\sin^2\alpha}\left(\frac{\mathrm{d}r}{\mathrm{d}\phi}\right)^2=\frac{1}{a^3}-\frac{1}{r^3}=\frac{1}{a^3}\left(1-\frac{a^3}{r^3}\right)\nonumber\\\implies&&\frac{\frac{a^\frac{3}{2}}{r^\frac{5}{2}}\mathrm{d}r}{\sqrt{1-\frac{a^3}{r^3}}}=\pm\sin\alpha\mathrm{d}\phi\nonumber
\end{eqnarray}

Hence on the developed plane
\begin{eqnarray}
	\frac{\frac{a^\frac{3}{2}}{{r'}^\frac{5}{2}}\mathrm{d}r'}{\sqrt{1-\frac{a^3}{{r'}^3}}}&=&\pm\mathrm{d}\theta'\nonumber\\\implies\cos^{-1}\left(\frac{a}{r'}\right)^\frac{3}{2}&=&\pm\frac{3}{2}\left(\theta'+k\right)\nonumber
\end{eqnarray}

Assuming $r'=a$, $\theta=0$ $\implies$ $k=0$
\begin{equation}
	\therefore\left(\frac{a}{r'}\right)^3=\cos^2\left(\frac{3\theta'}{2}\right)\nonumber
\end{equation}

{\bf 21. } A particle moves on a smooth right circular cone under the action of a force from the vertex, the law of repulsion being  $$\mu\left(\frac{a\cos^2\alpha}{r^3}-\frac{1}{2r^2}\right)$$ where $\alpha$ is the semi-vertical angle of the cone. Prove that if it be projected from an apse at a distance $a$ with velocity $\sqrt{\dfrac{\mu}{a}}\sin\alpha$, the path will be a parabola.\\

{\bf Solution: } Let $P(r,\alpha,\phi)$ be the spherical polar co-ordinates of the particle at time $t$ and $v$ be the velocity at $P$.\\

\begin{wrapfigure}[9]{r}{0.35\textwidth}
	\centering	\includegraphics[height=5 cm , width=5 cm ]{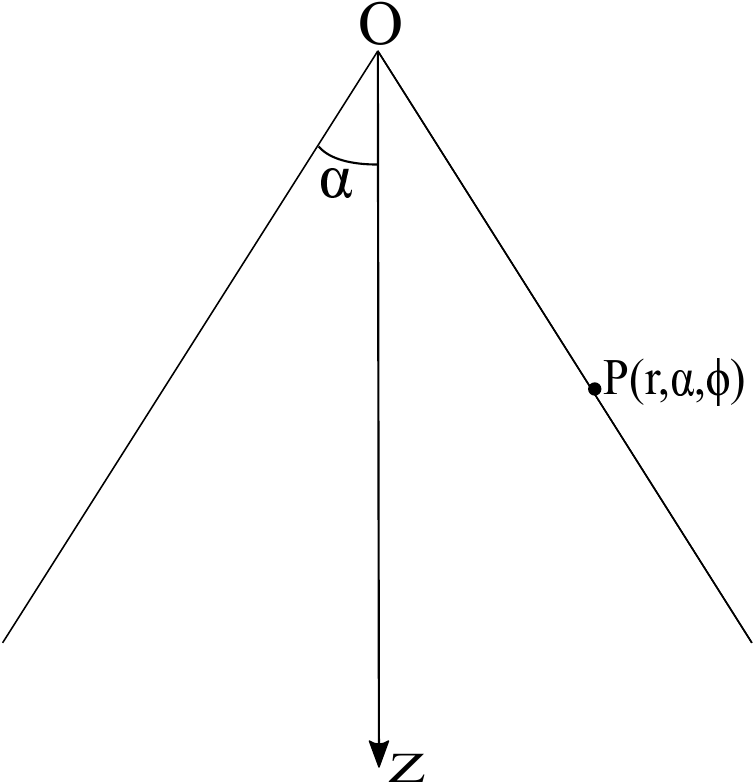}
	\begin{center}
		Fig. 3.34
	\end{center}
\end{wrapfigure} 

Now resolving the velocity perpendicular to $ZOP$-plane we get
\begin{eqnarray}
	&&r^2\sin^2\alpha\dot{\phi}=\mbox{constant}=\sqrt{\dfrac{\mu}{a}}\sin\alpha\cdot a\sin\alpha\nonumber\\\mbox{i.e., }&&r^2\dot{\phi}=\sqrt{\mu a}
\end{eqnarray}

If $\psi$ be the potential of the force, then
\begin{eqnarray}
	-\frac{\partial\psi}{\partial r}&=&\mu\left(\frac{a\cos^2\alpha}{r^3}-\frac{1}{2r^2}\right)\nonumber\\\mbox{i.e., }\psi&=&\frac{\mu}{2}\left(\frac{a\cos^2\alpha}{r^2}-\frac{1}{r}\right)+\mbox{ constant}\nonumber
\end{eqnarray} 

So the energy equation gives
\begin{eqnarray}
	&&\frac{1}{2}v^2+\psi=\mbox{constant}\nonumber\\\mbox{i.e.,}&&v^2+\mu\left(\frac{a\cos^2\alpha}{r^2}-\frac{1}{r}\right)-\mbox{constant}=\frac{\mu}{a}\sin^2\alpha+\left(\frac{\cos^2\alpha}{a}-\frac{1}{a}\right)=0\nonumber\\\mbox{i.e.,}&&\dot{r}^2+r^2\sin^2\alpha\dot{\phi}^2=\mu\left(\frac{1}{r}-\frac{a\cos^2\alpha}{r^2}\right)\nonumber\\\mbox{i.e.,}&&\dot{r}^2=\left(\frac{1}{r}-\frac{a\cos^2\alpha}{r^2}\right)-\frac{\mu a\sin^2\alpha}{r^2}=\mu\left(\frac{1}{r}-\frac{a}{r^2}\right)\nonumber
\end{eqnarray}

Also $r^4\dot{\phi}^2=\mu a$.
\begin{eqnarray}
	\therefore&&\frac{1}{r^4}\left(\frac{\mathrm{d}r}{\mathrm{d}\phi}\right)^2=\frac{1}{a}\left(\frac{1}{r}-\frac{a}{r^2}\right)=\frac{1}{ar^2}(r-a)\nonumber\\\mbox{i.e.,}&&\frac{a}{r^2(r-a)}\left(\frac{\mathrm{d}r}{\mathrm{d}\phi}\right)^2=1\nonumber\\\mbox{i.e.,}&&\frac{\frac{\sqrt{a}}{r^\frac{3}{2}}}{\sqrt{1-\frac{a}{r}}}\mathrm{d}r=\pm\mathrm{d}\phi\nonumber\\\mbox{i.e.,}&&2\cos^{-1}\sqrt{\frac{a}{r}}=\pm(\phi+k)\nonumber
\end{eqnarray}

Assuming $\phi=0$ at $r=a$ i.e., $k=0$.
\begin{eqnarray}\label{eq3.21.2}
	\therefore&&\frac{a}{r}=\cos^2\frac{\phi}{2}=\frac{1+\cos\phi}{2}\nonumber\\\mbox{i.e.,}&&2a=r+r\cos\phi 
\end{eqnarray} 

Now on the surface of the cone, the Cartesian co-ordinates $x$, $y$, $z$ are given by
$$x=r\sin\alpha\cos\phi,~y=r\sin\alpha\sin\phi,~z=r\cos\alpha$$

So equation (\ref{eq3.21.2}) can be written as
\begin{equation}\label{eq3.21.3}
	2a=\frac{z}{\cos\alpha}+\frac{x}{\sin\alpha}
\end{equation}
 which represents a plane. Hence the path of the particle on the cone is a plane curve. The direction cosine of the normal to the plane are proportional to $(1,0,\tan\alpha)$ and this normal is perpendicular to the generator
 \begin{equation}\label{eq3.21.4}
 	\frac{x}{\sin\alpha}=\frac{y}{0}=\frac{z}{-\cos\alpha}
 \end{equation} 

Hence the plane (\ref{eq3.21.3}) is parallel to the generator (\ref{eq3.21.4}). Since the path of the particle is the intersection of the plane (\ref{eq3.21.3}) with the cone, the path must be a parabolic path.\\

{\bf 22. } A particle moves along the smooth surface of a right circular cone under the action of a force parallel to the axis of the cone and proportional to the distance of the particle from the axis, its initial velocity being that which it would acquire in moving from the vertex. Prove that its path when the cone is developed into a plane is a rectangular hyperbola.\\

{\bf Solution: } If $\psi$ be the potential of the potential function then
\begin{eqnarray}
	&&-\frac{\partial\psi}{\partial z}=\lambda z\tan\alpha\nonumber\\\mbox{i.e.,}&&\psi=-\frac{\lambda z^2}{2}\tan\alpha\nonumber
\end{eqnarray}

\begin{wrapfigure}[12]{r}{0.35\textwidth}
	\centering	\includegraphics[height=5 cm , width=5 cm ]{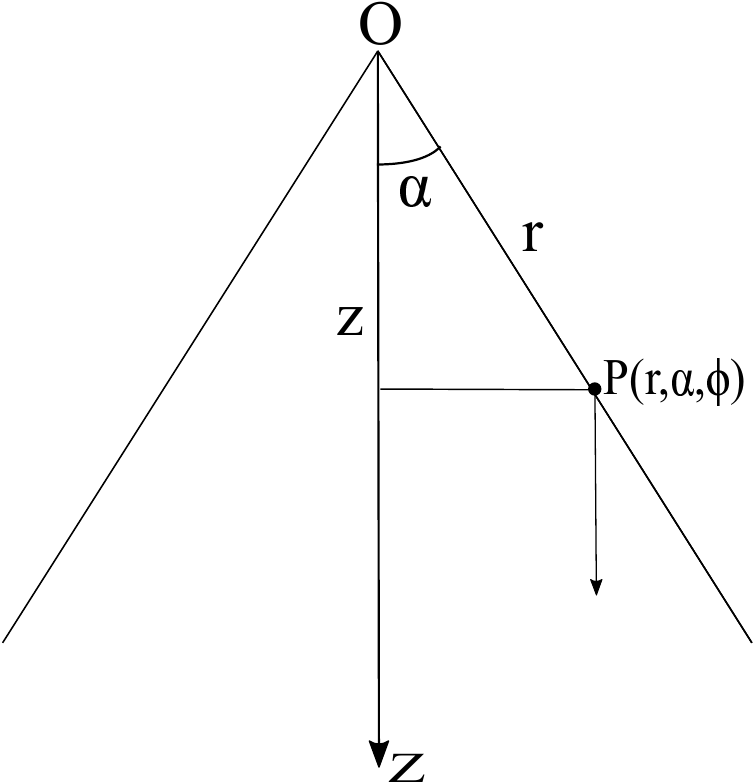}
	\begin{center}
		Fig. 3.35
	\end{center}
\end{wrapfigure} 

From the energy equation
\begin{eqnarray}
	&&\frac{1}{2}v^2+\psi=\mbox{constant}\nonumber\\\mbox{i.e.,}&&v^2-\lambda z^2\tan^2\alpha=c, \mbox{ a constant}\nonumber
\end{eqnarray} 

As $v=0$ at $z=0$ so $c=0$. \\

The equation of the angular momentum gives $r^2\sin^2\alpha\dot{\phi}=h$.
\begin{eqnarray}
	\therefore&&v^2=\dot{r}^2+r^2\sin^2\alpha\dot{\phi}^2=\lambda z^2\tan\alpha=\lambda r^2\sin\alpha\cos\alpha\nonumber\\\therefore&&\dot{r}^2+\frac{h^2}{r^2\sin^2\alpha}=\lambda r^2\sin\alpha\cos\alpha\nonumber\\\mbox{i.e.,}&&\dot{r}^2=\lambda r^2\sin\alpha\cos\alpha-\frac{h^2}{r^2\sin^2\alpha}\nonumber
\end{eqnarray} 

Also $r^4\sin^4\alpha\dot{\phi}^2=h^2$.
\begin{eqnarray}
	\therefore&&\frac{1}{r^6\sin^2\alpha}\left(\frac{\mathrm{d}r}{\mathrm{d}\phi}\right)^2=A-\frac{B}{r^4}\nonumber\\\mbox{i.e.,}&&\frac{\frac{1}{r^3}\mathrm{d}r}{\sqrt{A-\frac{B}{r^4}}}\mathrm{d}r=\sin\alpha\mathrm{d}\phi\nonumber
\end{eqnarray} 

So in developed plane
\begin{eqnarray}
	&&\frac{\frac{1}{{r'}^3}\mathrm{d}r'}{\sqrt{A-\frac{B}{{r'}^4}}}\mathrm{d}r'=\mathrm{d}\theta'\nonumber\\\mbox{i.e.,}&&\sin^{-1}\left(\frac{B}{\sqrt{A}{r'}^2}\right)=2k\theta'+c\nonumber\\\mbox{i.e.,}&&\frac{B}{\sqrt{A}{r'}^2}=\sin(2k\theta')\mbox{~~~~ (choosing $r'$, $\theta'$ in such a way that $c=0$)}\nonumber\\\mbox{i.e.,}&&x'y'=\mbox{constant, a rectangular hyperbola.}\nonumber
\end{eqnarray} 

 {\bf 23. } A heavy particle moves on a smooth right circular cone with axis vertical and vertex downwards, being projected with the velocity due to a fall from a height $z_0$ above the vertex to the point of projection. Prove that, if $z_1$, $z_2$ are the greatest and lowest heights attained then $$z_1^2+z_1z_2+z_2^2=z_0\left(z_1+z_2\right)$$
 
 {\bf Solution: } The energy equation gives
 $$v^2+2gz=c=2g(z_0-h)+2gh=2gz_0$$ 
 
 The equation of angular momentum can be written as
 $$r^2\sin^2\alpha\dot{\phi}=h$$
 
 Thus 
 \begin{eqnarray}
 	&&\dot{r}^2+r^2\sin^2\alpha\dot{\phi}^2=v^2=2g(z_0-z)\nonumber\\\mbox{i.e.,}&&\dot{z}^2\sec^2\alpha+\frac{h^2}{z^2\tan^2\alpha}=2g(z_0-z),~~~~~~~~~z=r\cos\alpha\nonumber
 \end{eqnarray}

Now, for maximum or minimum value, $\dot{z}=0$
\begin{eqnarray}\label{eq3.23.1}
	\mbox{i.e., }\frac{h^2}{z^2\tan^2\alpha}=2g(z_0-z)\nonumber\\\mbox{i.e., }\frac{h^2}{2g\tan^2\alpha}=z^2(z_0-z)
\end{eqnarray}

Let $z_1$, $z_2$ are two solutions of (\ref{eq3.23.1}). So we have 
\begin{eqnarray}
	&&\frac{h^2}{2g\tan^2\alpha}=z_1^2(z_0-z_1)=z_2^2(z_0-z_2)\nonumber\\\mbox{i.e.,}&&z_1^3-z_2^3=z_0\left(z_1^2-z_2^2\right)\nonumber\\\mbox{i.e.,}&&z_1^2+z_1z_2+z_2^2=z_0\left(z_1+z_2\right)\nonumber
\end{eqnarray}

As the vertex is downward and the particle moves on the surface so reaction can never be zero. \\
 
 {\bf 24. } Prove that if particles move on a right circular cone under no force, the projections of their paths on a plane perpendicular to the axis are similar curves of the type $r\sin n\theta=c$, whatever be their initial velocities.\\
 
 {\bf Solution: } In the cylindrical coordinates $(r,\theta,z)$ we have
 \begin{eqnarray}
 	&&v^2=c,~r^2\dot{\theta}=h, r=z\tan\alpha\nonumber\\\therefore&&\dot{r}^2+\dot{z}^2+r^2\dot{\theta}^2=c\nonumber\\\mbox{i.e.,}&&\dot{r}^2\mbox{cosec}^2\alpha+\frac{h^2}{r^2}=c\nonumber\\\mbox{i.e.,}&&\dot{r}^2=c\sin^2\alpha-\frac{h^2\sin^2\alpha}{r^2}=h^2\sin^2\alpha\left(\frac{1}{a^2}-\frac{1}{r^2}\right)\nonumber\\\mbox{i.e.,}&&\dot{r}=h\sin\alpha\left(\frac{1}{a^2}-\frac{1}{r^2}\right)^\frac{1}{2}\nonumber\\&&\dot{\theta}=\frac{h}{r^2}\nonumber\\\therefore&&\frac{\mathrm{d}r}{\mathrm{d}\theta}=r^2\sin\alpha\left(\frac{1}{a^2}-\frac{1}{r^2}\right)^\frac{1}{2}\nonumber\\\mbox{i.e.,}&&\frac{\frac{1}{r^2}\mathrm{d}r}{\sqrt{\frac{1}{a^2}-\frac{1}{r^2}}}=\sin\alpha\mathrm{d}\theta\nonumber\\\mbox{i.e.,}&&a=r\sin(n\theta),~~~n=\sin\alpha\nonumber\\&&~~~~~~~~~~\mbox{ and choosing $\theta$ in such a way that integration constant to be zero.}\nonumber
 \end{eqnarray} 

\section{Motion of a particle on a rough surface}

Let $\mu$ be the coefficient of friction and $R$ be the normal reaction of the surface, then equations of motion are
\begin{eqnarray}
	mv\frac{\mathrm{d}v}{\mathrm{d}s}&=&F-\mu R\label{eq3.61}\\m\frac{v^2}{\rho_0}&=&R+G\label{eq3.62}\\m\frac{v^2}{\rho_0}\tan\chi&=&H\label{eq3.63}
\end{eqnarray} 
where $F$, $G$, $H$ are the components of the external force acting on the particle along the path, along the normal to the surface, and along the tangent to the surface but perpendicular to the tangent to the path respectively. $\rho_0$ is the radius of curvature of the normal section of the surface through the tangent to the path and $v$ is the velocity of the particle at time $t$. Now eliminating $R$ between (\ref{eq3.61}) and (\ref{eq3.62}) we get
\begin{equation}
	mv\frac{\mathrm{d}v}{\mathrm{d}s}=F\left(\frac{v^2}{\rho_0}-G\right)\label{eq3.64}
\end{equation}

Thus equations (\ref{eq3.63}) and (\ref{eq3.64}) determine the motion of the particle on the surface. Further, knowing $v$ we can determine the normal reaction $R$ from equation (\ref{eq3.62}).\\

In the absence of external forces the equations of motion are
\begin{equation}
		mv\frac{\mathrm{d}v}{\mathrm{d}s}=-\mu R,~~m\frac{v^2}{\rho_0}=R,~~m\frac{v^2}{\rho_0}\tan\chi=0\nonumber
\end{equation}

So the last equation gives $\chi=0$ which shows that the path of the particle is a geodesic on the surface. The velocity of the particle is given by $$v\frac{\mathrm{d}v}{\mathrm{d}s}=-\mu\frac{v^2}{\rho_0}$$ 
and knowing $v$ we can determine $R$. \\

\subsection{Motion of a particle on a rough sphere under no external forces}

In the case of a sphere $\rho_0=a$, the radius of the sphere. The equations of motion are
\begin{equation}
	v\frac{\mathrm{d}v}{\mathrm{d}s}=-\frac{\mu}{m} R,~~\frac{v^2}{a}=\frac{R}{m},~~m\frac{v^2}{a}\tan\chi=0\nonumber
\end{equation}

So $\chi=0$ and the path is a geodesic i.e., a great circle.\\

 Also \begin{eqnarray}
 	&&v\frac{\mathrm{d}v}{\mathrm{d}s}=\frac{\mathrm{d}v}{\mathrm{d}t}=-\mu\frac{v^2}{a}\nonumber\\\mbox{i.e.,}&&\frac{1}{v}-\frac{1}{v_0}=\frac{\mu}{a}t\label{eq3.65}
 \end{eqnarray}
where we assume $v=v_0$ at $t=0$.
$$v=\frac{\mathrm{d}s}{\mathrm{d}t}=\frac{1}{\frac{1}{v_0}+\frac{\mu}{a}t}$$ 

So integrating once more
$$\frac{a}{\mu}\log\left(\frac{1}{v_0}+\frac{\mu}{a}t\right)=s+c$$

Choosing $s=0$ at $t=0$, we have
\begin{equation}\label{eq3.66}
	s=\frac{a}{\mu}\log\left(1+\frac{\mu v_0}{a}t\right)
\end{equation}

From equation (\ref{eq3.65}) we see that the velocity of the particle is decreasing with time and tends to zero as $t\to\infty$. Equation (\ref{eq3.66}) shows that $s$ is increasing with $t$ and tends to infinity as $t\to\infty$.\\

Thus the motion of the particle is along a great circle and the particle moves around the great circle indefinite number of times until its velocity vanishes.\\

\subsection{Motion of a particle on a right circular cylinder under no external forces}

We take $z$-axis along the axis of the cylinder and $P(r,\theta,z)$ be the position of the particle of the surface in cylindrical coordinates at time $t$. Since there is no external force acting on the particle the path of the particle on the surface will be a geodesic.\\

\begin{wrapfigure}[10]{r}{0.35\textwidth}
	\centering	\includegraphics[height=4 cm , width=4 cm ]{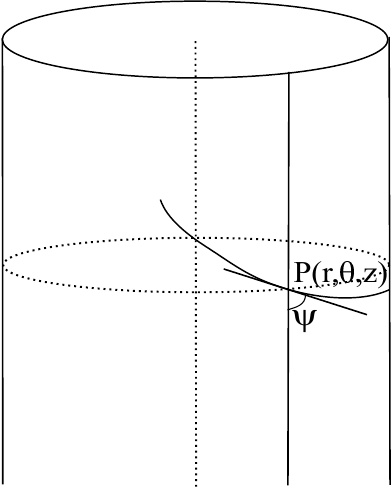}
	\begin{center}
		Fig. 3.36
	\end{center}
\end{wrapfigure} 

Before, discussing the motion of the particle we shall discuss an important result for a geodesic on the surface of the evolution.\\

Let $x^2+y^2=\phi(z)$ be the equation of the surface of revolution. The direction ratio of the normal to the surface at $(x,y,z)$ are $(2x,2y,-\phi'(z))$. The direction ratio of the principal normal to the path are $(x'',y'',z'')$ $\left(x''=\dfrac{\mathrm{d}^x}{\mathrm{d}s^2}\right)$. Thus for a geodesic the principal normal coincides with the normal to the surface. Therefore,
\begin{eqnarray} 
	&&\frac{x''}{2x}=\frac{y''}{2y}=\frac{z''}{-\phi'(z)}\nonumber\\\mbox{i.e.,}&&xy''-yx''=0\nonumber\\\mbox{i.e.,}&&xy'-yx'=\mbox{constant}\nonumber
\end{eqnarray}

\begin{wrapfigure}[9]{l}{0.35\textwidth}
	\centering	\includegraphics[height=3 cm , width=6 cm ]{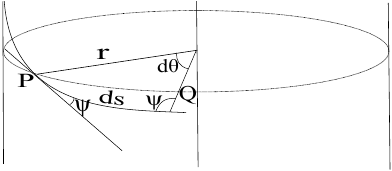}
	\begin{center}
		Fig. 3.37
	\end{center}
\end{wrapfigure} 

Using $x=r\cos\theta$, $y=r\sin\theta$, we get $$r^2\frac{\mathrm{d}\theta}{\mathrm{d}s}=\mbox{constant}$$

From the figure, arc $PQ=r\mathrm{d}\theta=\sin\psi\mathrm{d}s$ where, $\psi$ is the angle at which the path cuts the meridian
\begin{eqnarray}
	\therefore&&r\frac{\mathrm{d}\theta}{\mathrm{d}s}=\sin\psi\nonumber\\\mbox{i.e.,}&&r\sin\psi=\mbox{constant}
\end{eqnarray}

In case of cylinder $r=a=$constant, so $\sin\psi=$constant. This means that the path of the particle cuts all the generators at a constant angle, Hence the path is a helix.\\

So the equation of motion are
\begin{eqnarray}
	mv\frac{\mathrm{d}v}{\mathrm{d}s}&=&-\mu R\label{eq3.67}\\m\frac{v^2}{\rho_0}&=&R\label{eq3.68}\\\mbox{and~~}m\frac{v^2}{\rho_0}\tan\chi&=&0\label{eq3.69}
\end{eqnarray} 

Also $\dfrac{1}{\rho_0}=\dfrac{\cos^2\psi}{\infty}+\dfrac{\sin^2\psi}{a}=\dfrac{\sin^2\psi}{a}$.\\

From (\ref{eq3.67}) and (\ref{eq3.68}) \begin{eqnarray}
	v\frac{\mathrm{d}v}{\mathrm{d}s}&=&-\mu\frac{v^2}{\rho_0}=-\mu\frac{v^2}{a}\sin\psi\nonumber\\\mbox{i.e., }\frac{\mathrm{d}v}{\mathrm{d}t}&=&-\mu\frac{v^2}{a}\sin^2\psi\nonumber
\end{eqnarray}

On integration $\dfrac{1}{v}-\dfrac{1}{v_0}=\dfrac{\mu}{a}\sin^2\psi~t$

assuming $v=v_0$ at $t=0$, $$\therefore~v=\frac{\mathrm{d}s}{\mathrm{d}t}=\frac{1}{\frac{1}{v_0}+\frac{\mu}{a}\sin^2\psi~t}$$

So integrating once more (assuming $s=0$ at $t=0$) we have
$$s=\frac{a}{\mu\sin^2\psi}\log\left[1+\frac{\mu v_0}{a}\sin^2\psi~t\right]$$

\subsection{Discuss the motion of a heavy particle on a smooth surface of revolution having vertical axis of revolution}

We choose $O$ as the origin and vertical direction $O$ through as the $z$- axis. $P$ is the position of the particle at any time $t$. Then in cylindrical coordinates the components of the velocity and acceleration of the particle are:\\

\begin{wrapfigure}[12]{r}{0.35\textwidth}
	\centering	\includegraphics[height=4 cm , width=7 cm ]{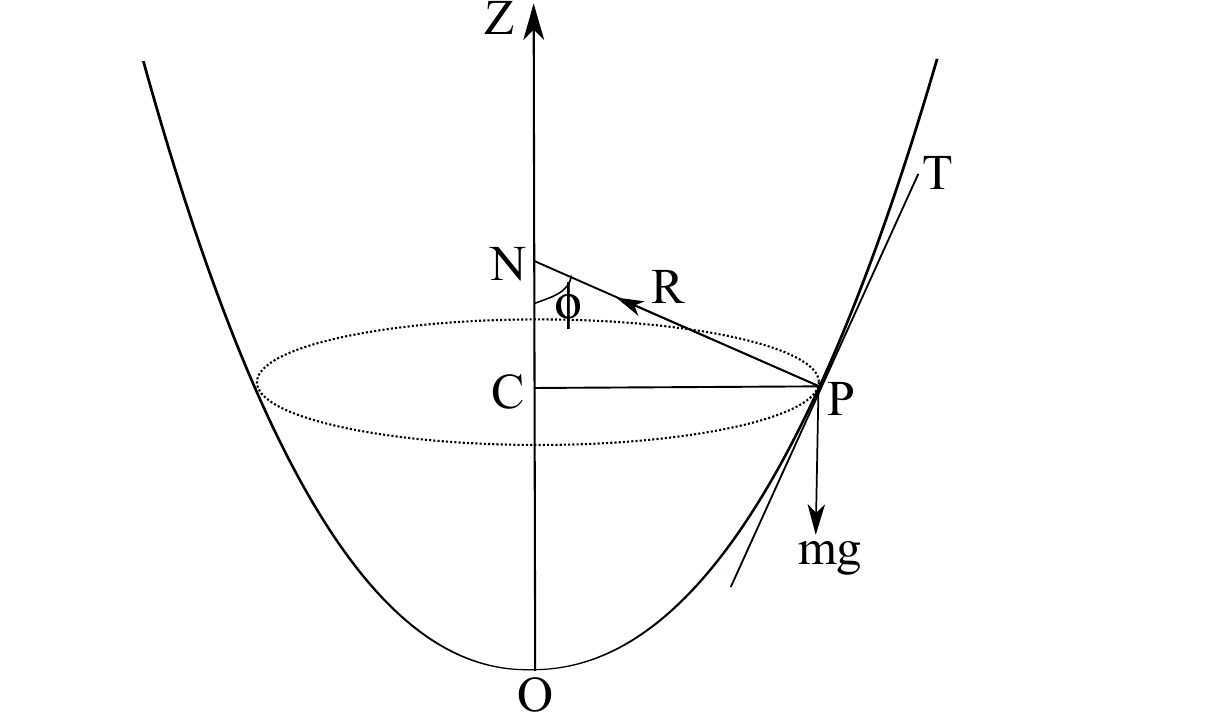}
	\begin{center}
		Fig. 3.38
	\end{center}
\end{wrapfigure} 

 velocity: $\dot{r}$, $r\dot{\theta}$, and $\dot{z}$ along radial, cross radial and along the $z$-asis. The component of the acceleration along these directions are $\ddot{r}-r\dot{\theta}^2$, $\dfrac{1}{r}\dfrac{\mathrm{d}}{\mathrm{d}t}\left(r^2\dot{\theta}\right)$ and $\ddot{z}$ respectively.\\
 
 As gravity is the only force acting on the particle so the equation of energy gives
 \begin{equation}\label{eq3.71}
 	\frac{1}{2}m\left(\dot{r}^2+r^2\dot{\theta}^2+\dot{z}^2\right)=c-mgz
 \end{equation}
 
If $\phi$ be the angle which the normal $PN$ at $P$ makes with the z-axis then equation of motion along the tangent $PT$ to the meridian curve gives
\begin{equation}\label{eq3.72}
	m\left\{\left(\ddot{r}-r\dot{\theta}^2\right)\cos\phi+\ddot{z}\sin\phi\right\}=-mg\sin\phi
\end{equation}

Also the equation of motion along the normal direction $PN$ gives
\begin{equation}\label{eq3.73}
	m\left[\ddot{z}\cos\phi-\left(\ddot{r}-r\dot{\theta}^2\right)\sin\phi\right]=R-mg\cos\theta
\end{equation}
 
 As gravity is acting on the meridian plane so there is no force along the cross redial direction (i.e., perpendicular to the meridian plane) and we have
 \begin{eqnarray}
 	&&\frac{1}{r}\frac{\mathrm{d}}{\mathrm{d}t}\left(r^2\dot{\theta}\right)=0\nonumber\\\mbox{i.e.,}&&r^2\dot{\theta}=h, \mbox{ a constant}\label{eq3.74}
 \end{eqnarray} 
which can be interpreted as the moment of momentum along the axis of revolution is conserved. Let $z=f(r)$ be the equation of the surface of revolution, then from the figure
\begin{equation}\label{eq3.75}
	\tan\phi=\frac{\mathrm{d}z}{\mathrm{d}r}=f'(r)
\end{equation}

Now using equation (\ref{eq3.74}) to eliminate $\theta$ and equation (\ref{eq3.75}) to eliminate $\phi$ from equation (\ref{eq3.72}) we have
\begin{equation}\label{eq3.76}
	\left(1+\{f'(r)\}^2\right)\frac{\mathrm{d}^2r}{\mathrm{d}t^2}+f'(r)f''(r)\left(\frac{\mathrm{d}r}{\mathrm{d}t}\right)^2-\frac{h^2}{r^3}=-gf'(r)
\end{equation}

Further writing
\begin{eqnarray}\label{eq3.77}
	\dot{r}&=&\frac{\mathrm{d}r}{\mathrm{d}\theta}\dot{\theta}=\frac{h}{r^2}\frac{\mathrm{d}r}{\mathrm{d}\theta}\nonumber\\\mbox{and }\ddot{r}&=&-\frac{2h^2}{r^5}\left(\frac{\mathrm{d}r}{\mathrm{d}\theta}\right)^2+\frac{h^2}{r^4}\frac{\mathrm{d}^2r}{\mathrm{d}\theta^2}
\end{eqnarray}

Equation (\ref{eq3.76}) reduces to
\begin{equation}\label{eq3.78}
	\frac{\left(1+\{f'(r)\}^2\right)}{r}\frac{\mathrm{d}^2r}{\mathrm{d}\theta^2}+\left\{f'(r)f''(r)-\frac{2}{r}\left[1+\{f'(r)\}^2\right]\right\}\frac{1}{r}\left(\frac{\mathrm{d}r}{\mathrm{d}\theta}\right)^2=1-\frac{gr^3}{h^2}f'(r)
\end{equation}

A first integral of the above second order equation gives
\begin{equation}\label{eq3.79}
	\left(1+\{f'(r)\}^2\right)\left(\frac{1}{r}\frac{\mathrm{d}r}{\mathrm{d}\theta}\right)^2+\left\{1+\frac{2g}{h^2}r^2f(r)\right\}=\lambda
\end{equation}
$\lambda$, being the constant of integration.\\

Note that equation (\ref{eq3.79}) can also be obtained from the energy equation (\ref{eq3.71}) by suitable elimination of the other variables involved.\\

Now, if the path of the particle is supposed to be a circle of radius $a$ then from equation (\ref{eq3.78}) we have
\begin{equation}
	ga^3f'(a)=h^2
\end{equation}

This condition for circular path can also be obtained from the following:\\

The non-zero acceleration is only along $PC$ of magnitude $\dfrac{h^2}{a^3}$ and it is maintained by the weight of the particle and the normal reaction. So equation of motion along $PC$ and $z$ axis gives
\begin{eqnarray}
	\frac{h^2}{a^3}&=&\frac{R}{m}\sin\phi\mbox{~~ and~~ }0~=~R\cos\phi-mg\nonumber\\
	\mbox{i.e.,}&&\tan\phi=\frac{h^2}{a^3g}\nonumber\\\mbox{i.e.,}&&h^2=a^3gf'(a)\nonumber
\end{eqnarray}

\section{Problems}

{\bf 25. } A particle is moving under gravity in contact with the inside of the smooth surface of revolution: $z=f(r)$, the positive $z$-direction being vertically upward. Originally the particle is describing a horizontal circle of radius $a$, its angular momentum about $z$-axis being $h$. It is then slightly disturbed so that the angular momentum in the subsequent motion is $h+\delta h$. Prove that the particle oscillates with period $\dfrac{2\pi}{n}$ where $n^2=\dfrac{g}{a}\left\{\dfrac{3f'(a)+af''(a)}{1+{\{f'(a)\}}^2}\right\}$ about the circle $r+\delta a$ where $$\delta a=\frac{2~\delta h~af'(a)}{h\{3f'(a)+af''(a)\}}$$

{\bf Solution: } From equation (\ref{eq3.76}) of the previous article
\begin{equation}\label{eq3.25.1}
	\left(1+\{f'(r)\}^2\right)\frac{\mathrm{d}^2r}{\mathrm{d}t^2}+f'(r)f''(r)\left(\frac{\mathrm{d}r}{\mathrm{d}t}\right)^2-\frac{h^2}{r^3}=-gf'(r)
\end{equation}

As the particle describe the circle: $r=a$ so we have
\begin{equation}
	\frac{h^2}{a^3}=gf'(a)
\end{equation}

Now for infinitesimal variation we have $r=a+\xi$, where $\xi$ is small and $h$ changes to $h+\delta h$ so we get from (\ref{eq3.25.1})
\begin{equation}
	\ddot{\xi}\left(1+\{f'(a+\xi)\}^2\right)-\frac{(h^2+2h\delta h)}{(a+\xi)^3}=-gf'(a+\xi)
\end{equation}
 where we have neglected $\dot{\xi}^2$ and $\delta h^2$.\\
 
 Thus we obtain
 \begin{eqnarray}
 	\ddot{\xi}&=&\frac{h^2+2h\delta h-g(a^3+3a^2\xi)\left[f'(a)+\xi f''(a)\right]}{(a^3+3a^2\xi)\left[1+\{f'(a)\}^2+2\xi f'(a)f''(a)\right]}\nonumber\\
 	&=&\frac{\cancel{a^3gf'(a)}+2h\delta h-\cancel{ga^3f'(a)}-3a^2g\xi f'(a)-ga^3\xi f''(a)}{a^3\left(1+\{f'(a)\}^2\right)+3a^2\xi\left(1+\{f'(a)\}^2\right)+2a^3\xi f'(a)f''(a)}\nonumber\\
 	&=&\frac{2h\delta h-a^2g\xi\left[3f'(a)+af''(a)\right]}{a^3\left(1+\{f'(a)\}^2\right)\left[1+\xi\left\{\frac{3}{a}+\frac{2f'(a)f''(a)}{\left(1+\{f'(a)\}^2\right)}\right\}\right]}\nonumber\\
	&=&\frac{2h\delta h-a^2g\xi\left[3f'(a)+af''(a)\right]}{a^3\left(1+\{f'(a)\}^2\right)}\left[1-\xi\left\{\frac{3}{a}+\frac{2f'(a)f''(a)}{\left(1+\{f'(a)\}^2\right)}\right\}\right]\nonumber\\
	&=&-\frac{g}{a}\frac{\left\{3f'(a)+af''(a)\right\}}{\left(1+\{f'(a)\}^2\right)}\left[\xi-\frac{2h\delta h}{a^2g\left\{3f'(a)+af''(a)\right\}}\right]\nonumber
 \end{eqnarray}

But it is given that
\begin{eqnarray}
	\delta a&=&\frac{2~\delta h~af'(a)}{h\{3f'(a)+af''(a)\}}=\frac{2h~\delta h~af'(a)}{h^2\{3f'(a)+af''(a)\}}\nonumber\\&=&\frac{2h~\delta h~af'(a)}{a^3gf'(a)\{3f'(a)+af''(a)\}}=\frac{2h~\delta h}{a^2g\{3f'(a)+af''(a)\}}\nonumber\\\therefore~\ddot{\xi}&=&-\frac{g}{a}\frac{\left\{3f'(a)+af''(a)\right\}}{\left(1+\{f'(a)\}^2\right)}\left(\xi-\delta a\right)\nonumber
\end{eqnarray} 

Putting $\eta=\xi-\delta a=r-a-\delta a$, we have
$$\ddot{\eta}=-\frac{g}{a}\frac{\left\{3f'(a)+af''(a)\right\}}{\left(1+\{f'(a)\}^2\right)}\eta$$
which represents a simple harmonic motion for which the time period is given by
$$T=\frac{2\pi}{n},~~~~~~n^2=\frac{g}{a}\left\{\frac{3f'(a)+af''(a)}{1+{\{f'(a)\}}^2}\right\}$$

{\bf 26. } A particle is moving under gravity be in contact with the inside of a rough circular cylinder with vertical generators. At time $t$ the particle is moving with velocity $V$ in a direction that makes an acute angle $\phi$ with the downward vertical. Establish the equation
$$\frac{1}{V}\frac{\mathrm{d}V}{\mathrm{d}\phi}=\frac{\mu V^2}{ag}\sin\phi-\cot\phi$$ 
where $a$ is the radius of the cylinder, $\mu$ is the coefficient of friction between the particle and the cylinder. If the velocity of the particle was $U$ when it was moving horizontally, prove that at time $t$ the horizontal component $u$ of its velocity is given by the equation
$$\frac{1}{U^2}-\frac{1}{u^2}=\frac{2\mu}{ag}\log\tan\left(\frac{\phi}{2}\right)$$

{\bf Solution: } We take $z$-axis vertical downward and along the axis of the cylinder. Let $P(a,\theta,z)$ be the cylindrical co-ordinates of the particle $P$ at time $t$. Then equation of motion are
\begin{eqnarray}
		mV\frac{\mathrm{d}V}{\mathrm{d}s}&=&-\mu R+mg\cos\phi\label{eq3.26.1}\\m\frac{V^2}{\rho_0}&=&R\label{eq3.26.2}\\m\frac{V^2}{\rho_0}\tan\chi&=&mg\sin\phi\label{eq3.26.3}
\end{eqnarray}
 where $R$ is the normal reaction of the surface. Also
 \begin{equation}\label{eq3.26.4}
 	\frac{1}{\rho_0}=\frac{\sin^2\phi}{a}
 \end{equation} 
(Note that the component of acceleration are $\ddot{r}-r\dot{\theta}^2=-a\dot{\theta}^2$, $\dfrac{1}{r}\dfrac{\mathrm{d}}{\mathrm{d}t}\left(r^2\dot{\theta}\right)=a\ddot{\theta}$ and $\ddot{z}$)

\begin{wrapfigure}[10]{r}{0.35\textwidth}
	\centering	\includegraphics[height=4 cm , width=5 cm ]{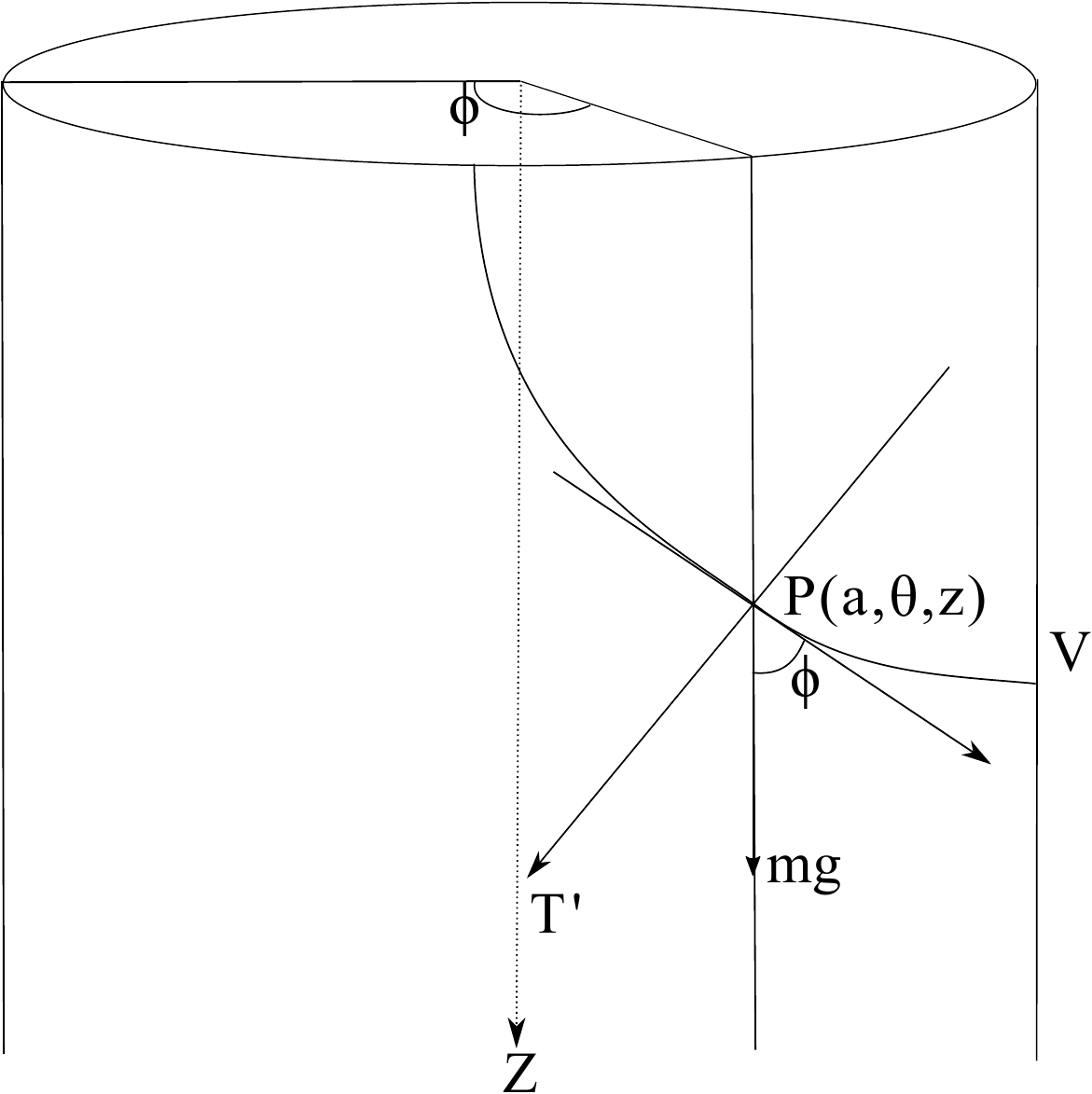}
	\begin{center}
		Fig. 3.39
	\end{center}
\end{wrapfigure}

Now\begin{eqnarray}
	\frac{V^2}{\rho_0}\tan\chi&=&\mbox{acceleration along }PT'\nonumber\\&=&\ddot{z}\sin\phi+a\ddot{\theta}\cos(\pi-\theta)\nonumber\\&=&\ddot{z}\sin\phi-a\ddot{\theta}\cos\theta\nonumber
\end{eqnarray}

But $\dot{z}=V\cos\phi$, $a\dot{\theta}=V\sin\phi$.
\begin{eqnarray}
	\therefore~	\frac{V^2}{\rho_0}\tan\chi&=&\sin\phi\frac{\mathrm{d}}{\mathrm{d}t}(V\cos\phi)-\cos\phi\frac{\mathrm{d}}{\mathrm{d}t}(V\sin\phi)\nonumber\\&=&-V\frac{\mathrm{d}\phi}{\mathrm{d}t}=g\sin\phi\mbox{~ (by equation (\ref{eq3.26.3}))}\nonumber\\
	\mbox{i.e., }\dot{\phi}&=&\dot{\phi}=-\frac{g}{V}\sin\phi
\end{eqnarray} 

From (\ref{eq3.26.1})
\begin{eqnarray}
	&&\frac{\mathrm{d}V}{\mathrm{d}t}=V\frac{\mathrm{d}V}{\mathrm{d}s}-\mu\frac{R}{m}+g\cos\phi\nonumber\\\implies&&\frac{\mathrm{d}V}{\mathrm{d}\phi}\dot{\phi}=-\mu\frac{V^2}{a}\sin^2\phi+g\cos\phi\mbox{~~(using (\ref{eq3.26.2}) and (\ref{eq3.26.4}))}\nonumber\\\implies&&-\frac{g}{V}\sin\phi\frac{\mathrm{d}V}{\mathrm{d}\phi}=-\mu\frac{V^2}{a}\sin^2\phi+g\cos\phi\nonumber\\\mbox{i.e.,}&&\frac{1}{V}\frac{\mathrm{d}V}{\mathrm{d}\phi}=\frac{\mu V^2}{ag}\sin\phi-\cot\phi\label{eq3.26.6}
\end{eqnarray}

\underline{\textbf{2nd Part : }} $u=V\sin\phi$\\

Now taking log and differentiating with respect to $\phi$ we get
\begin{eqnarray}
	\frac{1}{u}\frac{\mathrm{d}u}{\mathrm{d}\phi}&=&\frac{1}{V}\frac{\mathrm{d}V}{\mathrm{d}\phi}+\cot\phi=\frac{\mu V^2}{ag}\sin\phi \mbox{~ (by equation (\ref{eq3.26.6}))}\nonumber\\&=&\frac{\mu u^2}{ag}\mbox{ cosec}\phi\nonumber\\\mbox{i.e., }\frac{2}{u^3}\mathrm{d}u&=&\frac{2\mu}{ag}\mbox{ cosec}\phi~\mathrm{d}\phi\nonumber
\end{eqnarray} 

So integrating once we obtain
\begin{eqnarray}
	\int_U^u\frac{2}{u^3}\mathrm{d}u=\frac{2\mu}{ag}\int_{\frac{\pi}{2}}^{\phi}\mbox{cosec}\phi~\mathrm{d}\phi\nonumber\\\mbox{i.e.,~ }\frac{1}{U^2}-\frac{1}{u^2}=\frac{2\mu}{ag}\log\tan\left(\frac{\phi}{2}\right)\nonumber
\end{eqnarray}

{\bf 27. } A heavy particle moves on a rough vertical circular cylinder and is projected horizontally with a velocity $V$. Prove that at the point where the path cuts the generator at an angle $\phi$, the velocity $v$ is given by
$$\frac{ag}{v^2\sin^2\phi}=\frac{ag}{V^2}+2\mu\log(\cot\phi+\mbox{cosec }\phi)$$ 
and that the azimuthal angle $\theta$ and vertical decend $z$ are given by
$$ag\theta=-\int v^2~\mathrm{d}\phi\mbox{~ and~ }gz=-\int v^2\cot\phi~\mathrm{d}\phi$$ 
(the limits being $\dfrac{\pi}{2}$ to $\phi$)\\

{\bf Solution: } From the previous problem (see equation (\ref{eq3.26.6}))
\begin{equation}\label{eq3.27.1}
	\frac{1}{v}\frac{\mathrm{d}v}{\mathrm{d}\phi}=\frac{\mu v^2}{ag}\sin\phi-\cot\phi
\end{equation}

Now, \begin{eqnarray}
	\mathrm{d}\theta&=&\frac{\mathrm{d}\theta}{\mathrm{d}\phi}=\frac{\dot{\theta}}{\dot{\phi}}\mathrm{d}\phi\nonumber\\\mbox{i.e., }a\mathrm{d}\theta&=&a\frac{\dot{\theta}}{\dot{\phi}}\mathrm{d}\phi=\frac{v\sin\phi}{-\frac{g}{v}\sin\phi}\mathrm{d}\phi\nonumber\\\mbox{i.e., }ag\mathrm{d}\theta&=&-v^2\mathrm{d}\phi\nonumber
\end{eqnarray}

Integrating we get
\begin{eqnarray}
	ag\int_0^\theta\mathrm{d}\theta&=&-\int_\frac{\pi}{2}^\phi v^2~\mathrm{d}\phi\nonumber\\\mbox{i.e., ~~}ag\theta&=&-\int_\frac{\pi}{2}^\phi v^2~\mathrm{d}\phi\nonumber
\end{eqnarray}

Also, $\mathrm{d}z=\dfrac{\mathrm{d}z}{\mathrm{d}\phi}\mathrm{d}\phi=\dfrac{\dot{z}}{\dot{\phi}}\mathrm{d}\phi=\dfrac{V\cos\phi}{-\frac{g}{V}\sin\phi}\mathrm{d}\phi=-\dfrac{V^2}{g}\cot\phi$
\begin{eqnarray}
	\therefore&&g\int_0^z\mathrm{d}z=-\int_\frac{\pi}{2}V^2\cot\phi\mathrm{d}\phi\nonumber\\\mbox{i.e.,}&&gz=-\int_\frac{\pi}{2}V^2\cot\phi\mathrm{d}\phi\nonumber
\end{eqnarray}

Also from (\ref{eq3.27.1})
$$\frac{1}{v^3}\frac{\mathrm{d}v}{\mathrm{d}\phi}+\frac{1}{v^2}\cot\phi=\frac{\mu }{ag}\sin\phi$$

Put $\lambda=\dfrac{1}{v^2}$, $\dfrac{\mathrm{d}\lambda}{\mathrm{d}\phi}=-\dfrac{2}{v^3}\dfrac{\mathrm{d}v}{\mathrm{d}\phi}$
\begin{eqnarray}
	\therefore&&\frac{\mathrm{d}\lambda}{\mathrm{d}\phi}-2\lambda\cot\phi=-\frac{2\mu}{ag}\sin\phi\nonumber\\\implies&&\lambda e^{-\int2\cot\phi~\mathrm{d}\phi}=-\frac{2\mu}{ag}\int\sin\phi e^{-\int2\cot\phi~\mathrm{d}\phi}~\mathrm{d}\phi+c\nonumber\\\implies&&\lambda\mbox{ cosec}^2\phi=-\frac{2\mu}{ag}\int\mbox{ cosec}\phi~\mathrm{d}\phi+c\nonumber\\\implies&&\frac{1}{v^2}\mbox{ cosec}^2\phi=\frac{2\mu}{ag}\log(\mbox{cosec }\phi+\cot\phi)+c\nonumber
\end{eqnarray}

But $v=V$, $\phi=\dfrac{\pi}{2}$ $\implies$ $c=\dfrac{1}{v^2}$.
$$\therefore~\frac{ag}{v^2\sin^2\phi}=\frac{ag}{V^2}+2\mu\log(\cot\phi+\mbox{cosec }\phi)$$

\section{Discuss the motion of a particle on a rough right circular cone under no external force}

We take vertex of the cone as origin, $z$-axis along the axis of the cone and $(r,\theta,z)$ be the cylindrical coordinates of the position of the particle at $P$, on the surface of the cone at any time $t$.\\

\begin{wrapfigure}[14]{r}{0.35\textwidth}
	\centering	\includegraphics[height=5 cm , width=4.5 cm ]{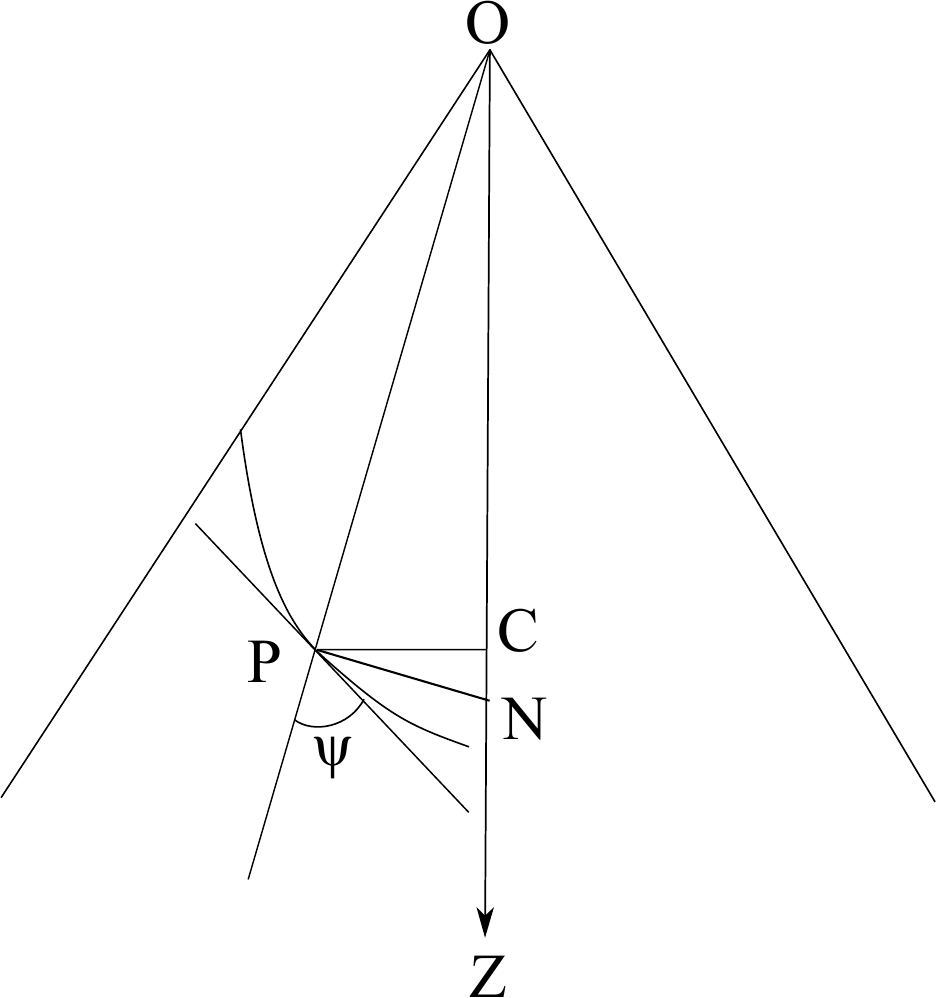}
	\begin{center}
		Fig. 3.40
	\end{center}
\end{wrapfigure}

Let the path of the particle at $P$ cuts the generator $OP$ of the cone at an angle $\psi$. As there are no external forces acting on the particle, so the path must be geodesic on the cone. Also we have
\begin{equation}
	r\sin\psi=\mbox{constant}=K(\mbox{say})\label{eq3.91}
\end{equation}

The equations of motion are
\begin{equation}
		mv\frac{\mathrm{d}v}{\mathrm{d}s}=-\mu\mbox{~ (along the path)} \label{eq3.92}
	\end{equation}and
\begin{equation} 
	m\frac{v^2}{\rho_0}=R\mbox{~ (along the normal to the surface)}\label{eq3.93}
\end{equation}
 where $R$ is the reaction of the surface, $\mu$ is the coefficient of friction, $v$ is the velocity of the particle at $P$ and $\rho_0$ is the radius of curvature of the normal section of the surface through tangent to the path at $P$.\\
 
Therefore,
\begin{equation}
	\frac{1}{\rho_0}=\frac{\cos^2\psi}{\infty}+\frac{\sin^2\psi}{PN}=\frac{\sin^2\psi}{r\sec\alpha}\label{eq3.94}
\end{equation}

Now eliminating $R$ between (\ref{eq3.92}) and (\ref{eq3.93}) and using $\rho_0$ from (\ref{eq3.94}) we get 
\begin{eqnarray}
	mv\frac{\mathrm{d}v}{\mathrm{d}s}&=&-\mu mv^2\frac{\sin^2\psi}{r\sec\alpha}\nonumber\\\mbox{i.e., }v\frac{\mathrm{d}v}{\mathrm{d}s}&=&-\mu v^2\cos\alpha\frac{K^2}{r^3}\label{eq3.95} ~~\mbox{using (\ref{eq3.92})}
\end{eqnarray}

\begin{wrapfigure}[10]{r}{0.35\textwidth}
	\centering	\includegraphics[height=4 cm , width=4.5 cm ]{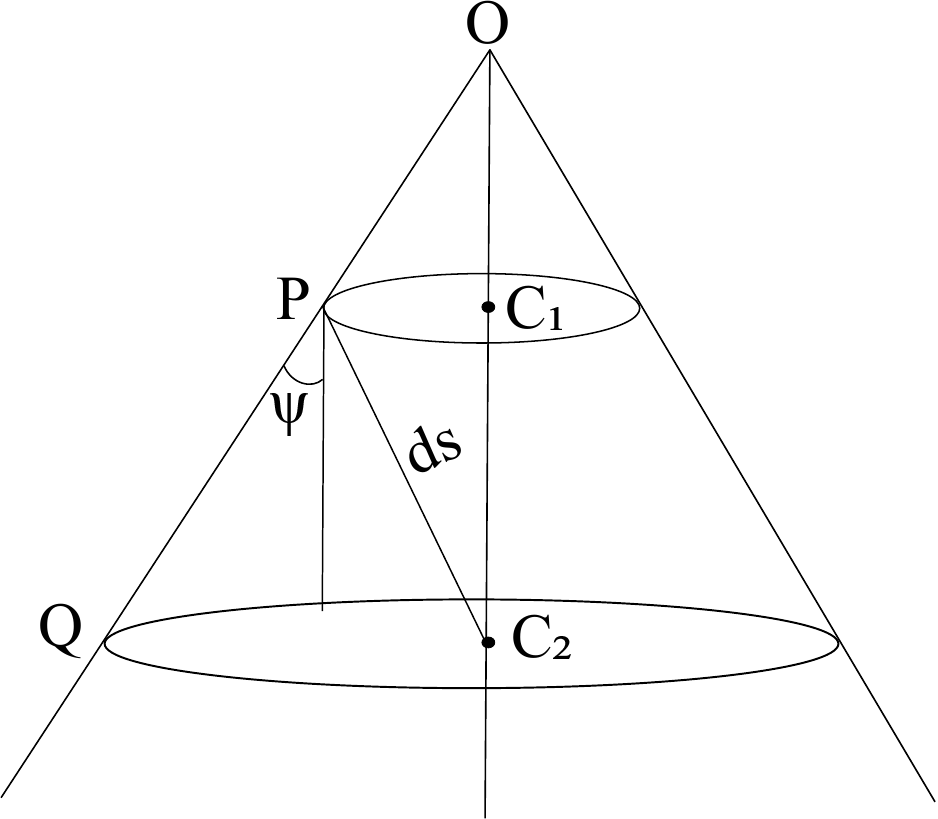}
	\begin{center}
		Fig. 3.41
	\end{center}
\end{wrapfigure}

Let $OP=r'$, $C_1P=r$, $PQ=\mathrm{d}r'$.\\

So $r=r'\sin\alpha$ ~i.e.,~ $\mathrm{d}r=\mathrm{d}r'\sin\alpha$.
\begin{eqnarray}
	\therefore&&\mathrm{d}s\cos\psi=\mathrm{d}r'=\mathrm{d}r\mbox{ cosec }\alpha\nonumber\\\therefore&&\mathrm{d}s=\frac{\mathrm{d}r\mbox{ cosec }\alpha}{\sqrt{1-\frac{K^2}{r^2}}}=\mbox{cosec }\alpha\frac{r\mathrm{d}r}{\sqrt{r^2-K^2}}\nonumber
\end{eqnarray}

on integration,
\begin{equation}\nonumber
	s=\sqrt{r^2-K^2}\mbox{cosec }\alpha+K'
\end{equation}

If we assume $s=0$, $r=K$ then $K'=0$
\begin{equation}
	\therefore~s^2\sin^2\alpha+K^2=r^2\nonumber
\end{equation}

Using this expression for $r$ in  (\ref{eq3.95}) we get
\begin{eqnarray}
	v\frac{\mathrm{d}v}{\mathrm{d}s}&=&-\mu v^2\cos\alpha\frac{K^2}{(s^2\sin^2\alpha+K^2)^\frac{3}{2}}\nonumber\\\mbox{i.e., }\frac{\mathrm{d}v}{v}&=&-\mu K^2 \cos\alpha\frac{\mathrm{d}s}{(s^2\sin^2\alpha+K^2)^\frac{3}{2}}\nonumber
\end{eqnarray} 

Now integrating
\begin{eqnarray}
	\int_{v_0}^v\frac{\mathrm{d}v}{v}&=&-\mu K^2 \cos\alpha\int_0^s\frac{\mathrm{d}s}{(s^2\sin^2\alpha+K^2)^\frac{3}{2}}\nonumber\\\mbox{i.e., }\ln\frac{v}{v_0}&=&\frac{\mu s}{\tan\alpha}\left(s^2+d^2\right)^{-\frac{1}{2}}\nonumber
\end{eqnarray}

{\bf 28. } A particle moves on a rough surface under no force but the reaction of the surface. The surface is one of revolution who is meridian curves are catenaries with the axis for the ditectrix and the particle is projected from the equator making an angle $\alpha$ with the meridian. Prove that the velocity of the particle when it crosses the meridian at an angle $\phi$ satisfies the equation
\begin{equation}
	\left(\frac{\mathrm{d}v}{\mathrm{d}\phi}\right)^2=\frac{2\mu^2v^2\cos^22\phi}{\cos2\phi-\cos2\alpha}, ~~~\mu\mbox{ being the coefficient of friction}\nonumber
\end{equation}

{\bf Solution: } The equation of the meridian curve be
\begin{equation}
	r=c\sec\psi
\end{equation}

\begin{wrapfigure}[10]{r}{0.35\textwidth}
	\centering	\includegraphics[height=4.5 cm , width=4.5 cm ]{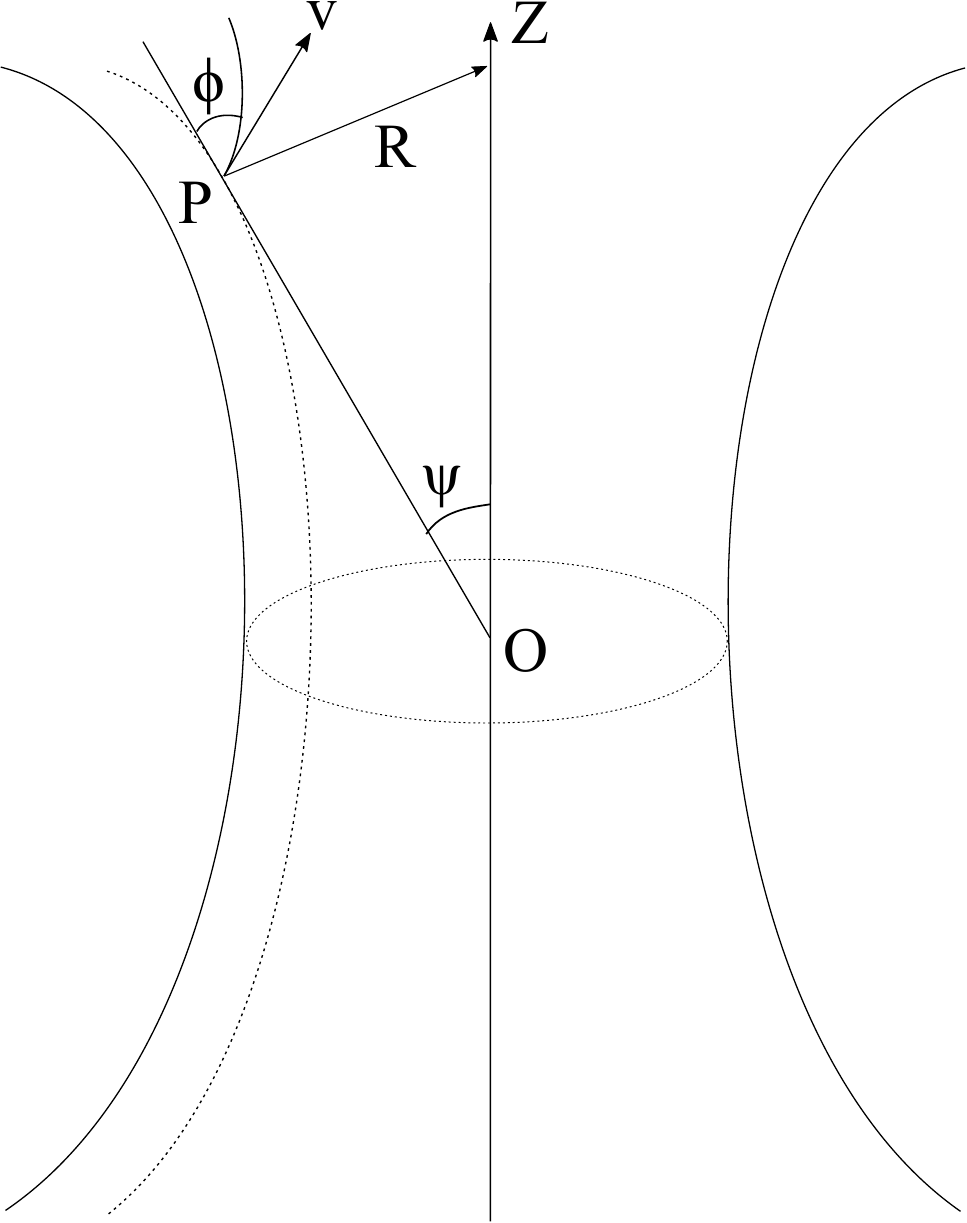}
	\begin{center}
		Fig. 3.42
	\end{center}
\end{wrapfigure}

Since the particle is under the action of no forces so the path will be a geodesic on the surface. If the path at $P$ cuts the meridian section at an angle $\phi$ then
\begin{eqnarray}
	&&r\sin\phi=\mbox{constant}=c\sin\alpha\nonumber\\\therefore&&\cos\psi=\frac{c}{r}=\frac{\sin\phi}{\sin\alpha}
\end{eqnarray}

The equation of motion along the path is
\begin{equation}
	mv\frac{\mathrm{d}v}{\mathrm{d}s}=-\mu R
\end{equation}
 and that along the axis ($z$-axis) is
 \begin{equation}
 	m\ddot{z}=-R\sin\psi-\mu R\cos\phi\cos\psi
 \end{equation}
 
Now, $\dot{z}=v\cos\phi\cos\psi$
 
so $\ddot{z}=\dot{v}\cos\phi\cos\psi+v\dfrac{\mathrm{d}}{\mathrm{d}t}(\cos\phi\cos\psi)$

Therefore 
\begin{eqnarray}
	m\dot{v}\cos\phi\cos\psi&=&m\ddot{z}-mv\dfrac{\mathrm{d}}{\mathrm{d}t}(\cos\phi\cos\psi)\nonumber\\&=&-R\left[\sin\psi+\mu\cos\phi\cos\psi\right]-mv\dfrac{\mathrm{d}}{\mathrm{d}t}(\cos\phi\cos\psi)\nonumber\\&=&\frac{m\dot{v}}{\mu}\left(\sin\psi+\mu\cos\phi\cos\psi\right)-mv\dfrac{\mathrm{d}}{\mathrm{d}t}\left(\cos\phi\frac{\sin\phi}{\sin\alpha}\right)\nonumber\\\mbox{i.e.,~ }\dot{v}\sin\psi&=&\mu v\dfrac{\mathrm{d}}{\mathrm{d}t}\left(\cos\phi\frac{\sin\phi}{\sin\alpha}\right)\nonumber\\\mbox{i.e.,~ }\frac{\mathrm{d}v}{\mathrm{d}\phi}\dot{\phi}\sin\psi&=&\frac{\mu v}{\sin\alpha}\cos2\phi\dot{\phi}\nonumber\\\therefore~\left(\frac{\mathrm{d}v}{\mathrm{d}\phi}\right)^2&=&\frac{\mu^2v^2\cos^22\phi}{\sin^2\alpha(1-\cos^2\psi)}=\frac{\mu^2v^2\cos^22\phi}{\sin^2\alpha-\sin^2\phi}=\frac{2\mu^2v^2\cos^22\phi}{\cos2\phi-\cos2\alpha}\nonumber
\end{eqnarray}

{\bf 29. } A particle moves on a helical wire whose axis is vertical. Prove that the velocity $v$ after describing an arc $s$ is given by the equation
$$v^2=ag\sec\alpha\sinh\phi~, ~~\frac{\mathrm{d}s}{\mathrm{d}\phi}=\frac{a}{2}\frac{\sec^2\alpha\cosh\phi}{(\tan\alpha-\mu\cosh\phi)}$$ where, $a$ is the radius of the cylinder on which the helix lies, $\alpha$ is the inclination of the helix to the horizon and $\mu$ is the coefficient of friction.\\

{\bf Solution: }  Resolving the forces acting on the particle along the tangent, principal normal and binormal to the path, we have the equations of motion
\begin{eqnarray}
	mv\frac{\mathrm{d}v}{\mathrm{d}s}&=&mg\sin\alpha-\mu R\nonumber\\m\frac{v^2}{\rho}=R_N&,&~~0=R_B-mg\cos\alpha\nonumber
\end{eqnarray} where $R$ is the reaction of the wire on the particle and $R_N$, $R_B$ the component of $R$ along the principal normal and the binormal respectively. As the path is helix so the principal normal and normal to the surface at $P$ coincide.\\

\begin{wrapfigure}[10]{r}{0.35\textwidth}\vspace{-.8cm}
	\centering	\includegraphics[height=4.7 cm , width=4.5 cm ]{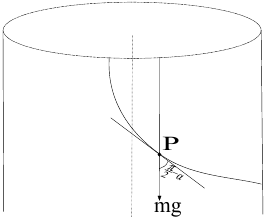}
	\begin{center}
		Fig. 3.43
	\end{center}
\end{wrapfigure}

 Now, $\rho=\rho_0\cos\psi=\rho_0$, the radius of curvature of the normal section of the cylinder through $P$ so we have
 \begin{eqnarray}
 	\frac{1}{\rho}&=&\frac{1}{\rho_0}=\frac{\cos^2\left(\frac{\pi}{2}-\alpha\right)}{\infty}+\frac{\sin^2\left(\frac{\pi}{2}-\alpha\right)}{a}=\frac{\cos^2\alpha}{a}\nonumber\\\therefore R_N&=&m\frac{v^2}{a}\cos\alpha, R_B=mg\cos\alpha\nonumber\\\therefore R^2&=&R_N^2+R_B^2=m^2g^2\cos^2\alpha\left(1+\frac{v^4\cos^2\alpha}{a^2g^2}\right)\nonumber
 \end{eqnarray}
 
 As $v^2=ag\sec\alpha\sinh\phi$, so $R=mg\cos\alpha\cosh\phi$.\\
 
 Hence $v\dfrac{\mathrm{d}v}{\mathrm{d}s}=\dfrac{ag}{2}\sec\alpha\cosh\phi\dfrac{\mathrm{d}\phi}{\mathrm{d}s}$.\\
 
 Thus from the equation of motion along the path we get
 \begin{eqnarray}
 	v\frac{\mathrm{d}v}{\mathrm{d}s}&=&g\sin\alpha-\mu g\cos\alpha\cosh\phi\nonumber\\\mbox{i.e., }\frac{\mathrm{d}s}{\mathrm{d}\phi}&=&\frac{a}{2}\frac{\sec\alpha\cosh\phi}{\sin\alpha-\mu\cos\alpha\cosh\phi}=\frac{a}{2}\frac{\sec^2\alpha\cosh\phi}{(\tan\alpha-\mu\cosh\phi)}\nonumber
 \end{eqnarray}
 
\section{Motion of a particle on a smooth revolving surface}

\begin{wrapfigure}[11]{r}{0.35\textwidth}
	\centering	\includegraphics[height=4.5 cm , width=4.5 cm ]{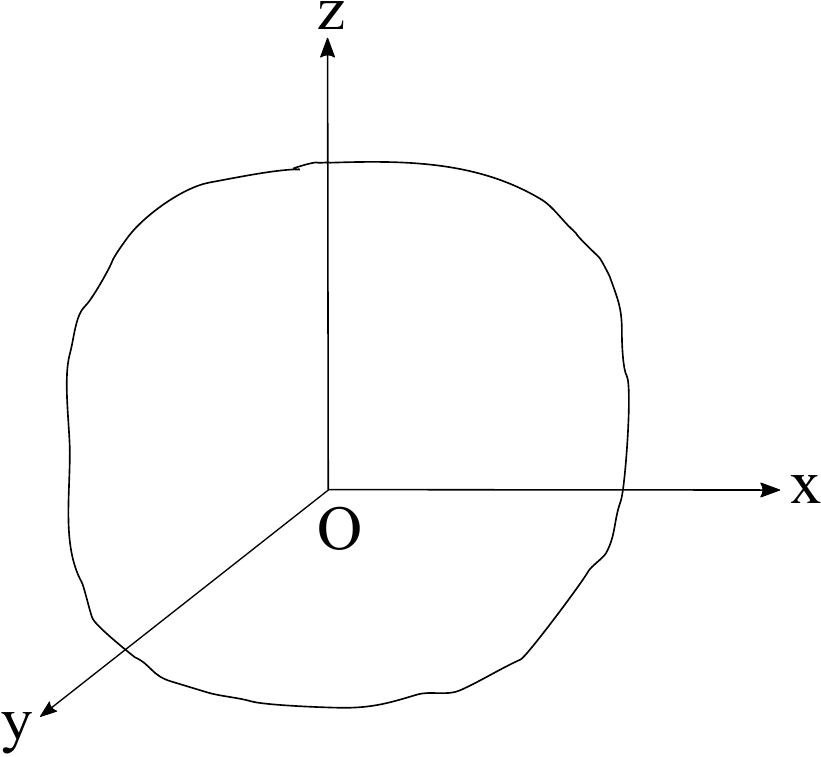}
	\begin{center}
		Fig. 3.44
	\end{center}
\end{wrapfigure}

Let $ox$, $oy$, $oz$ be the coordinate axes fixed relative to the surface and the surface is moving about an axis through $O$ with an angular velocity $\omega$. Let $\phi(x,y,z)=0$ be the equation of the surface and $\vec{r}=x\vec{i}+y\vec{j}+z\vec{k}$ be the position of the particle at time $t$, where $\left(\vec{i},\vec{j},\vec{k}\right)$ are the unit vectors along the co-ordinate axes. Let $m$ be the mass of the particle, then the equation of motion is $$m\frac{\mathrm{d}^2\vec{r}}{\mathrm{d}t^2}=\vec{F}+\vec{R}$$
where $\vec{F}$ is the external force and $\vec{R}$ is the reaction of the surface surface. Suppose $\vec{\omega}=\left(\omega_x,\omega_y,\omega_z\right)$ be the angular velocity of rotation then
$$\frac{\mathrm{d}\vec{r}}{\mathrm{d}t}=\frac{\partial\vec{r}}{\partial t}+\vec{\omega}\times\vec{r}$$

So,
\begin{eqnarray}
	\dfrac{\mathrm{d}^2\vec{r}}{\mathrm{d}t^2}&=&\frac{\mathrm{d}}{\mathrm{d}t}\left(\frac{\mathrm{d}\vec{r}}{\mathrm{d}t}\right)=\frac{\partial}{\partial t}\left(\frac{\mathrm{d}\vec{r}}{\mathrm{d}t}\right)+\vec{\omega}\times\frac{\mathrm{d}\vec{r}}{\mathrm{d}t}\nonumber\\&=&\frac{\partial}{\partial t}\left(\frac{\partial\vec{r}}{\partial t}+\vec{\omega}\times\vec{r}\right)+\vec{\omega}\times\left(\frac{\partial\vec{r}}{\partial t}+\vec{\omega}\times\vec{r}\right)\nonumber\\&=&\frac{\partial^2\vec{r}}{\partial t^2}+\dot{\vec{\omega}}\times\vec{r}+2\vec{\omega}\times\frac{\partial\vec{r}}{\partial t}+\vec{\omega}\times\left(\vec{\omega}\times\vec{r}\right)\nonumber\\&=&\left(\ddot{x},\ddot{y},\ddot{z}\right)+\begin{vmatrix}
		\vec{i}&\vec{j}&\vec{k}\\\dot{\omega}_x&\dot{\omega}_y&\dot{\omega}_z\\x&y&z
	\end{vmatrix}+2\begin{vmatrix}
	\vec{i}&\vec{j}&\vec{k}\\\omega_x&\omega_y&\omega_z\\\dot{x}&\dot{y}&\dot{z}
\end{vmatrix}+\begin{vmatrix}
\vec{i}&\vec{j}&\vec{k}\\\omega_x&\omega_y&\omega_z\\z\omega_y-y\omega_z&x\omega_z-z\omega_x&y\omega_x-x\omega_y
\end{vmatrix}\nonumber\\&=&\left(f_x,f_y,f_z\right)\nonumber
\end{eqnarray}

Hence the equations of motion are
$$mf_x=F_x+R_x~,~~mf_y=F_y+R_y~,~~mf_z=F_z+R_z$$

 \section{Motion of a particle on a smooth revolving plane}

Let $OI$ be the axis of rotation and $OZ$ be the normal to the plane at $O$ making an angle $\alpha$ with $OI$. Let the plane $IOZ$ cut the given plane along $OX$. We take $X$-axis along $OX$ and $Z$-axis along $OZ$, the axis $Y$ is taken perpendicular to $OX$ and $OZ$. Let the plane rotate about $OI$ with uniform angular velocity $\omega$ so that $\dot{\omega}=0$. Let $P(x,y,0)$ be the position of the particle at time $t$. So we write

\begin{wrapfigure}[14]{r}{0.35\textwidth}\vspace{-.5cm}
	\centering	\includegraphics[height=4.5 cm , width=4.5 cm ]{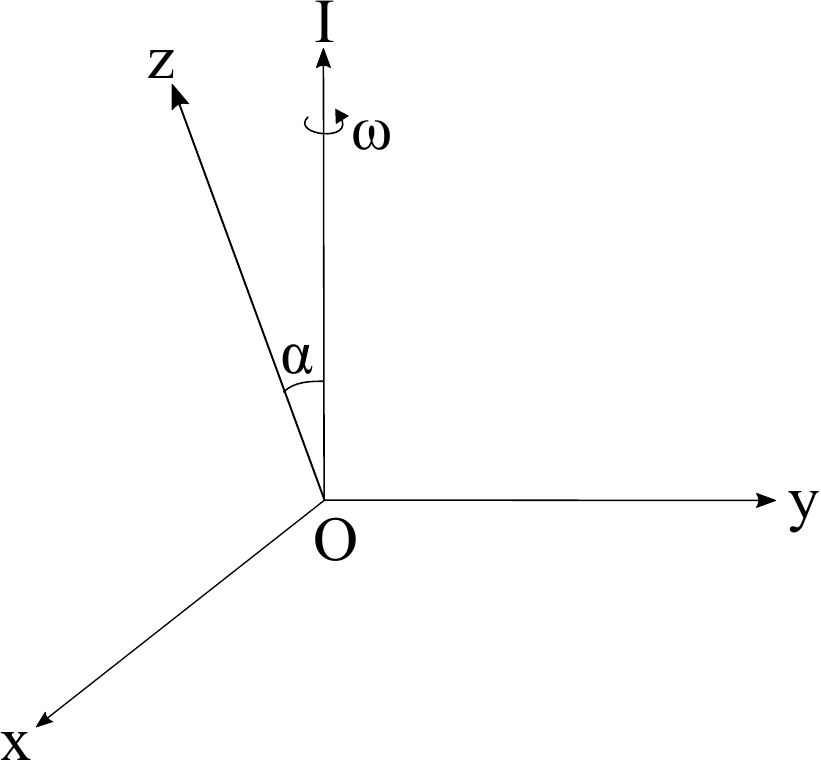}
	\begin{center}
		Fig. 3.45
	\end{center}
\end{wrapfigure}

$$\vec{r}=x\vec{i}+y\vec{j}+0\vec{k}$$
be the position of the particle and 
$$\vec{\omega}=\left(\omega_x,0,\omega_z\right)~,~~\omega_x=-\omega\sin\alpha~,~~\omega_z=\omega\cos\alpha$$
be the angular velocity of rotation.\\

Now, \begin{eqnarray}
	\dfrac{\mathrm{d}^2\vec{r}}{\mathrm{d}t^2}&=&\frac{\partial^2\vec{r}}{\partial t^2}+\dot{\vec{\omega}}\times\vec{r}+2\vec{\omega}\times\frac{\partial\vec{r}}{\partial t}+\vec{\omega}\times\left(\vec{\omega}\times\vec{r}\right)\nonumber\\&=&\left(\ddot{x},\ddot{y},0\right)+0+2\begin{vmatrix}
		\vec{i}&\vec{j}&\vec{k}\\\omega_x&\omega_y&\omega_z\\\dot{x}&\dot{y}&\dot{z}
	\end{vmatrix}+\begin{vmatrix}
		\vec{i}&\vec{j}&\vec{k}\\\omega_x&\omega_y&\omega_z\\-y\omega_z&x\omega_z&y\omega_x-x\omega_y
	\end{vmatrix}\nonumber\\&=&\left(f_x,f_y,f_z\right)\nonumber
\end{eqnarray}

So the equations of motion are $$mf_x=F_x+R_x~,~~mf_y=F_y+R_y~,~~mf_z=F_z+R_z$$
 
{\bf 30. } A particle slides on a smooth tube which is made to rotate with uniform angular velocity $\omega$ about a vertical axis. If the particle starts from relative rest from the point where the shortest distance between the axis and the tube meets the tube. Show that in time $t$ the particle has moved through a distance $\dfrac{2g}{\omega^2}\cot\alpha\mbox{ cosec }\alpha\sinh^2\left(\dfrac{1}{2}\omega\sin\alpha t\right)$ where $\alpha$ is the inclination of the tube to the vertical.\\

{\bf Solution: } Let the line of shortest distance between the axis of rotation and the tube be taken as axis of $y$. $z$-axis is taken parallel to the tube through $o$. $z$-axis is along $oz$ making an angle $\dfrac{\pi}{2}-\alpha$ with the axis of rotation. So the components of angular velocity are
$$\omega_x=-\omega\cos\alpha~,~~\omega_y=0~,~~\omega_z=\omega\sin\alpha$$ 

Let $OO'=a$, so the position of the particle at time $t$ is give by $(x,a,0)$. Then equation of motion about $x$-axis is
$$mf_x=mg\cos\alpha~,~~\mbox{i.e., ~}f_x=g\cos\alpha$$

Now
\begin{eqnarray}
	(f_x,f_y,f_z)=\frac{\mathrm{d}^2\vec{r}}{\mathrm{d}t^2}&=&\left(\ddot{x}\vec{i}+\ddot{y}\vec{j}+z\vec{k}\right)+2\begin{vmatrix}
		\vec{i}&\vec{j}&\vec{k}\\\omega_x&\omega_y&\omega_z\\\dot{x}&\dot{y}&\dot{z}
	\end{vmatrix}+\begin{vmatrix}
		\vec{i}&\vec{j}&\vec{k}\\\omega_x&\omega_y&\omega_z\\-y\omega_z&x\omega_z&y\omega_x-x\omega_y
	\end{vmatrix}\nonumber\\\therefore~f_x&=&\ddot{x}-\omega\sin\alpha(x\omega\sin\alpha)=\ddot{x}-\omega^2\sin^2\alpha x\nonumber\\\therefore&&\ddot{x}-\omega^2\sin^2\alpha x=g\cos\alpha\nonumber
\end{eqnarray}

\begin{wrapfigure}[10]{r}{0.35\textwidth}
	\centering	\includegraphics[height=4.5 cm , width=4.5 cm ]{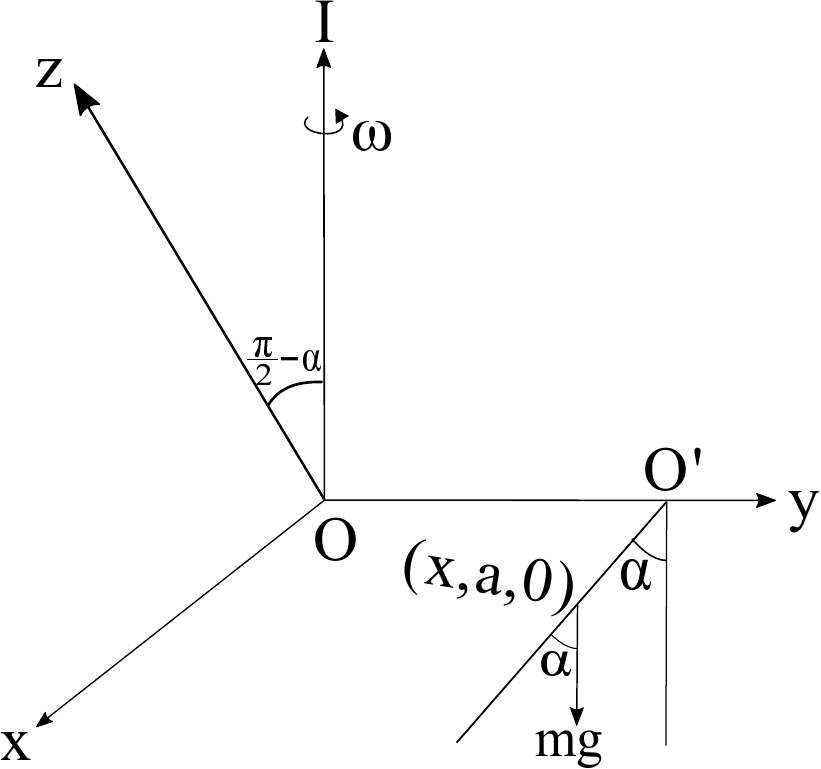}
	\begin{center}
		Fig. 3.46
	\end{center}
\end{wrapfigure}

The general solution is
\begin{eqnarray}
	x&=&A\cosh(\omega\sin\alpha t)+B\sinh(\omega\sin\alpha t)-\frac{g}{\omega^2}\mbox{ cosec }\alpha\cot\alpha\nonumber\\\therefore~\dot{x}&=&\omega\sin\alpha\left[A\sinh(\omega\sin\alpha t)+B\cosh(\omega\sin\alpha t)\right]\nonumber
\end{eqnarray}

Initial conditions: $x=0$, $\dot{x}=0$ at $t=0$.
\begin{eqnarray}
	\therefore0&=&A-\frac{g}{\omega^2}\mbox{ cosec }\alpha\cot\alpha\implies A=\frac{g}{\omega^2}\mbox{ cosec }\alpha\cot\alpha\nonumber\\0&=&B\nonumber\\
	\therefore~x&=&\frac{g}{\omega^2}\mbox{ cosec }\alpha\cot\alpha\left[\cosh(\omega\sin\alpha t)-1\right]\nonumber\\&=&\dfrac{2g}{\omega^2}\cot\alpha\mbox{ cosec }\alpha\sinh^2\left(\dfrac{1}{2}\omega\sin\alpha t\right)\nonumber
\end{eqnarray}

{\bf 31. } A particle is revolving on a smooth plane about a centre of force, with a force $\mu$ times the distance from it. When the body arrives at an apse the plane begins to revolve with an angular velocity $\dfrac{\sqrt{3\mu}}{2}$ about the apsidal line. Show that the subsequent orbit describe on the plane will be a portion of a parabola and that when the particle leaves the plane its velocity will be $\sqrt{3}$ times velocity at the vertex of the parabola.\\

{\bf Solution: } Let $O$ be the centre of force and $A$ be the apse. $OA=a$ be the apsidal distance. We choose the apse line $OA$ as the $x$-axis and $OY$ in the plane but perpendicular to $x$-axis is considered as $y$-axis. $oz$ is normal to the plane and is taken as $z$-axis. Let $P(x,y,0)$ be the position of the particle at any time $t$. Thus
\begin{eqnarray}
	\vec{F}&=&(-m\mu x,-m\mu y,0)\mbox{ be the external force,}\nonumber\\
		\vec{R}&=&(0,0,R)\mbox{ be the normal reation,}\nonumber\\
			\vec{r}&=&\overrightarrow{OP}=(x,y,0)\mbox{ be the position vector of $P$,}\nonumber\\
				\vec{\omega}&=&(\omega,0,0)\mbox{ be the angular velocity of rotation.}\nonumber
\end{eqnarray}

\begin{wrapfigure}[9]{r}{0.35\textwidth}
	\centering	\includegraphics[height=4.5 cm , width=4.5 cm ]{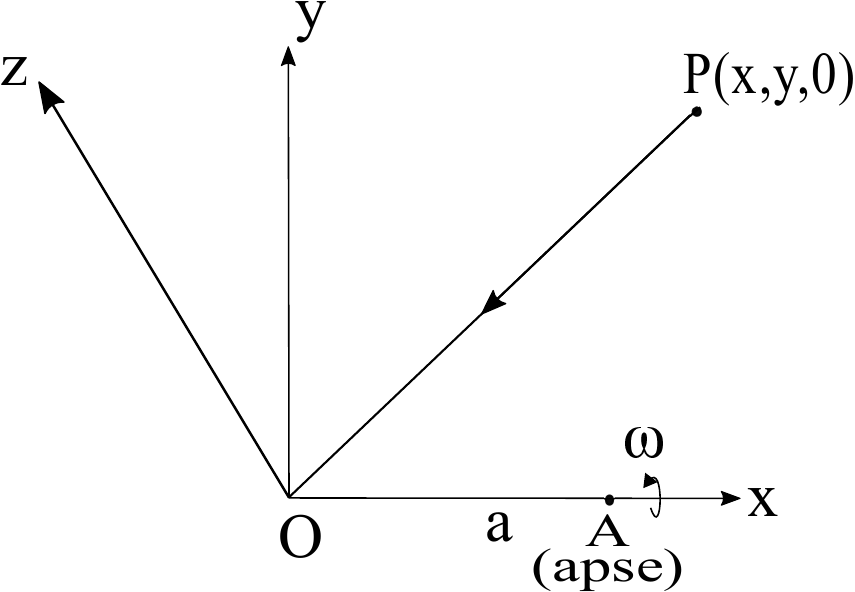}
	\begin{center}
		Fig. 3.47
	\end{center}
\end{wrapfigure}

Thus the equation of motion is
\begin{equation}
	m\frac{\mathrm{d}^2\vec{r}}{\mathrm{d}t^2}=\vec{F}+\vec{R}\nonumber
\end{equation}

Now,\begin{eqnarray}
	\frac{\mathrm{d}\vec{r}}{\mathrm{d}t}&=&\frac{\partial\vec{r}}{\partial t}+\vec{\omega}\times\vec{r}=(\dot{x},\dot{y},0)+\begin{vmatrix}
		\vec{i}&\vec{j}&\vec{k}\\\omega&0&0\\x&y&0
	\end{vmatrix}\nonumber\\&=&(\dot{x},\dot{y},\omega y)\nonumber\\\therefore~	\frac{\mathrm{d}^2\vec{r}}{\mathrm{d}t^2}&=&(\ddot{x},\ddot{y},\omega\dot{y})+\begin{vmatrix}
	\vec{i}&\vec{j}&\vec{k}\\\omega&0&0\\\dot{x}&\dot{y}&\omega y
\end{vmatrix}\nonumber\\&=&(\ddot{x},\ddot{y}-\omega^2y,2\omega\dot{y})\nonumber
\end{eqnarray}

Hence the component wise equations of motion are
\begin{eqnarray}
	m\ddot{x}&=&-m\mu x\label{eq3.31.1}\\m(\ddot{y}-\omega^2y)&=&-m\mu y\label{eq3.31.2}\\\mbox{and~~ ~}2m\omega\dot{y}&=&R\label{eq3.31.3}
\end{eqnarray}

Solution of (\ref{eq3.31.1}) gives: 
\begin{equation} x=A\cos(\sqrt{\mu}t)+B\sin(\sqrt{\mu}t)\label{eq3.31.4}
\end{equation}

Solution of (\ref{eq3.31.2}) gives: 
\begin{equation} y=A'\cos\left(\frac{\sqrt{\mu}}{2}t\right)+B'\sin\left(\frac{\sqrt{\mu}}{2}t\right)\label{eq3.31.5}
\end{equation}

The initial conditions are:
$$x=a,~y=0,~\dot{x}=0,~\dot{y}=V\mbox{ (say) at } t=0$$

This gives from (\ref{eq3.31.4}) and (\ref{eq3.31.5})
\begin{eqnarray}
	A=a,~B=0,~A'=0,~B'=\frac{2V}{\sqrt{\mu}}\nonumber\\\therefore~x=a\cos\sqrt{\mu}t~,y=\frac{2V}{\sqrt{\mu}}\sin\left(\frac{\sqrt{\mu}}{2}t\right)\nonumber
\end{eqnarray} 

So eliminating $t$ we obtain
$$x-a=-\dfrac{\mu ay^2}{2V^2}$$
which describes a parabola with $(a,0)$ as vertex. \\

Now, when when the particle leaves the surface then $R=0$, so from (\ref{eq3.31.4}) $\dot{y}=0$ i.e., $\cos\left(\dfrac{\sqrt{\mu}}{2}t\right)=0$ i.e., $t=\dfrac{\pi}{\sqrt{\mu}}$.\\

Thus when the particle leaves the surface, then $\dot{x}=0=\dot{y}$ and the  velocities are $(\dot{x},\dot{y},\omega y)=\left(0,0,\dfrac{\sqrt{3\mu}}{2}\dfrac{2V}{\sqrt{\mu}}\right)=\sqrt{3}V=\sqrt{3}$ times the velocity at the vertex.

\section{Motion of a particle in a plane wire or a fine tube rotating about a point in its plane}

Let $C$ be the plane curve and let the plane rotate about an axis perpendicular to the plane with angular velocity $\omega$. This axis of rotation is chosen as $z$-axis and axes $x$ and $y$ are taken on the plane fixed relative to the curve. Let at time $t$ the $x$-axis $ox$ makes an angle $\psi$ with a fixed line $ox'$ in the plane. By condition, $\dot{\psi}=\omega$ and suppose $\vec{r}=\vec{r}(s)$ be the equation of the curve with $\vec{r}=\overrightarrow{OP}$. Thus the equation of motion of the particle is
\begin{equation}
	m\frac{\mathrm{d}^2\vec{r}}{\mathrm{d}t^2}=\vec{F}+\vec{R}\label{eq3.105}
\end{equation}
 where $\vec{F}$ is the external force and $\vec{R}$ is the reaction of the wire.\\
 
 \begin{wrapfigure}[9]{r}{0.35\textwidth}
 	\centering	\includegraphics[height=4.5 cm , width=4.5 cm ]{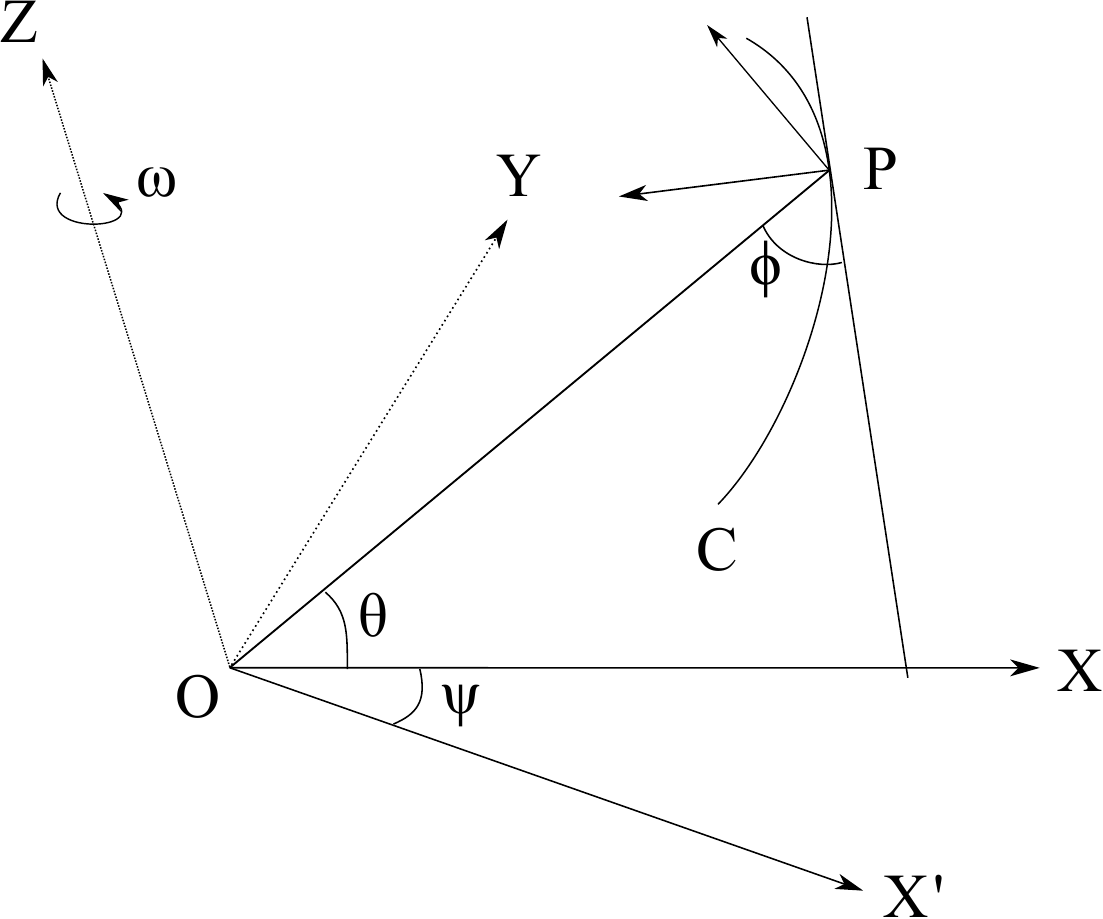}
 	\begin{center}
 		Fig. 3.48
 	\end{center}
 \end{wrapfigure}
 
 Now,\begin{eqnarray}
 	\frac{\mathrm{d}\vec{r}}{\mathrm{d}t}&=&\frac{\partial\vec{r}}{\partial t}+\vec{\omega}\times\vec{r}\nonumber\\
	\dfrac{\mathrm{d}^2\vec{r}}{\mathrm{d}t^2}&=&\frac{\partial^2\vec{r}}{\partial t^2}+\dot{\vec{\omega}}\times\vec{r}+2\vec{\omega}\times\frac{\partial\vec{r}}{\partial t}+\vec{\omega}\times\left(\vec{\omega}\times\vec{r}\right)\nonumber\\&=&\frac{\partial^2\vec{r}}{\partial t^2}+\dot{\omega}r\vec{e}+2\dot{s}\omega\vec{N}+(\vec{\omega}\cdot\vec{r})\vec{\omega}-\omega^2\vec{r}\nonumber \\ 	
	\label{eq3.106}
 \end{eqnarray} 
where $\vec{e}$ is the unit vector along the cross radial direction, $\vec{N}$ is the unit vector along the normal to the curve and $\vec{T}$ is the unit vector along the tangent to the curve. Using (\ref{eq3.106}) in (\ref{eq3.105}) we have
\begin{eqnarray}
	m\frac{\partial^2\vec{r}}{\partial t^2}&=&\vec{F}+\vec{R}-m\dot{\omega}r\vec{e}-2m\dot{s}\omega\vec{N}+m\omega^2\vec{r}\nonumber\\&&~~~~(\vec{\omega}\cdot\vec{r}=0 \mbox{ as $\vec{\omega}$ and $\vec{r}$ are orthogonal})\nonumber
\end{eqnarray} 

This gives the motion of the particle relative to the curve. Thus the motion of the particle in a revolving curve is the same as if the curve were at rest and the following forces are added to the acting forces namely (i) a radial force across $m\omega^2r$, (ii) a  cross-radial force $-m\dot{\omega}r$ and (iii) a normal force $-2m\dot{s}\omega$. Hence we write
\begin{eqnarray}
	m\frac{\partial^2\vec{r}}{\partial t^2}&=&mv\frac{\mathrm{d}v}{\mathrm{d}s}\vec{T}+m\frac{v^2}{\rho}\vec{N}\nonumber\\\mbox{and }\vec{F}&=&F_T\vec{T}+F_N\vec{N}\nonumber
\end{eqnarray}
 where $\rho$is the radius of curvature of the curve at $P$.\\
 
 Thus\begin{eqnarray}
 	mv\frac{\mathrm{d}v}{\mathrm{d}s}\vec{T}+m\frac{v^2}{\rho}\vec{N}&=&F_T\vec{T}+F_N\vec{N}+R\vec{N}-m\dot{\omega}r\sin\phi\vec{T}-m\dot{\omega}r\cos\phi\vec{N}\nonumber\\&&-2m\dot{s}\omega\vec{N}+m\omega^2r\cos\phi\vec{T}-m\omega^2r\sin\phi\vec{N}\nonumber
 \end{eqnarray} 
or in component along the tangential and normal we have
\begin{eqnarray}
		mv\frac{\mathrm{d}v}{\mathrm{d}s}&=&F_T-m\dot{\omega}r\sin\phi+m\omega^2r\cos\phi\nonumber\\	m\frac{v^2}{\rho}&=&F_N+R-m\dot{\omega}r\cos\phi-2m\dot{s}\omega-m\omega^2r\sin\phi\nonumber
\end{eqnarray}

For uniform angular velocity we have $\dot{\omega}=0$ and hence
\begin{equation}
	mv\frac{\mathrm{d}v}{\mathrm{d}s}=F_T+m\omega^2r\cos\phi=F_T+m\omega^2r\frac{\mathrm{d}r}{\mathrm{d}s}\nonumber
\end{equation} 

Further if  the external forces are conservative i.e., $\vec{F}=-\mbox{grad } V$ then $F_T=-\dfrac{\partial V}{\partial s}$. So the above tangential equation of motion becomes
\begin{equation}
		mv\frac{\mathrm{d}v}{\mathrm{d}s}=-\dfrac{\partial V}{\partial s}+m\omega^2r\frac{\mathrm{d}r}{\mathrm{d}s}\nonumber
\end{equation} 

on integration
\begin{eqnarray}
	&&\frac{1}{2}mv^2=-V+\frac{1}{2}m\omega^2r^2+c\nonumber\\\mbox{i.e., }&&\frac{1}{2}mv^2+V'=c~,~~~V'=V-\frac{1}{2}m\omega^2r^2\nonumber
\end{eqnarray}

Here $V'$ is known as the modified potential energy and the above equation is known as the modified equation of the energy.\\

{\bf 32. } A bead that moves on a circular wire and initially at rest at a point $A$. The wire is made to rotate uniformly in its own plane with angular velocity $\omega$ about the other end of the diameter through $A$. Show that the pressure between the bead and the wire vanishes at a time $\dfrac{1}{\omega}\log_e\left(\dfrac{3+\sqrt{5}}{2}\right)$ after the start.\\

{\bf Solution: } When the curve is reduced to rest the equation of motion along the tangent and normal to the curve at $P$ are
\begin{eqnarray}
	mv\frac{\mathrm{d}v}{\mathrm{d}s}&=&m\omega^2r\frac{\mathrm{d}r}{\mathrm{d}s}\label{eq3.32.1}\\m\frac{v^2}{a}&=&R-2m\omega v-m\omega^2r\cos\theta\label{eq3.32.2}
\end{eqnarray}

 \begin{wrapfigure}[9]{r}{0.35\textwidth}
	\centering	\includegraphics[height=4.5 cm , width=4.5 cm ]{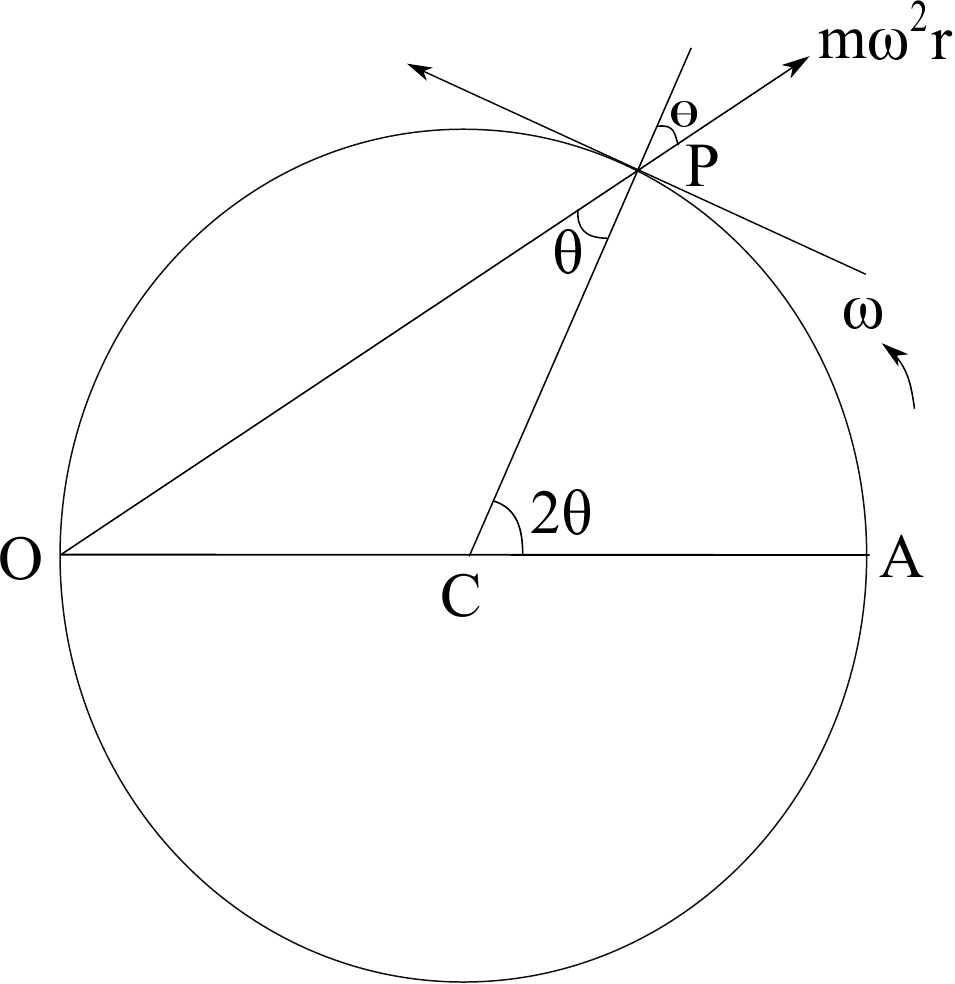}
	\begin{center}
		Fig. 3.49
	\end{center}
\end{wrapfigure}

Integrating (\ref{eq3.32.1}) gives $v^2=\omega^2 r^2+c$.\\

Initially, the particle was at rest at $A$, when the wire is made to rotate about $O$ with angular velocity $\omega$, the velocity of the bead  relative to the wire is $-2a\omega$. Thus $v=-2a\omega$, $r=2a$ $\implies$ $c=0$. \\

$\therefore v=\pm\omega r$, but $v=-2a\omega$ initially, so $v=-\omega r$ at any time $t$.\\

From equation (\ref{eq3.32.2}), the reaction vanishes when
\begin{eqnarray}
	\frac{v^2}{a}&=&-2\omega v-\omega^2r\cos\theta\nonumber\\\mbox{i.e., }\frac{\omega^2r^2}{a}&=&2\omega^2r-\omega^2r\frac{r}{2a} ~~~(\because r=2a\cos\theta, \mbox{ is the equation of the circle})\nonumber\\\mbox{i.e., }r&=&\frac{4a}{3}\nonumber
\end{eqnarray}

Further,\begin{eqnarray}
	v&=&-\omega r\nonumber\\\mbox{i.e., }\frac{\mathrm{d}s}{\mathrm{d}r}\cdot\frac{\mathrm{d}r}{\mathrm{d}t}&=&-\omega r\nonumber\\\mbox{i.e., }\frac{1}{-\sin\theta}\cdot\frac{\mathrm{d}r}{\mathrm{d}t}&=&-\omega r~~~\left(\frac{\mathrm{d}r}{\mathrm{d}s}=\cos\phi=\cos\left(\frac{\pi}{2}+\theta\right)=-\sin\theta\right.\nonumber\\&&~~~~~~ \left.\mbox{Note that as $\theta$ is negative so $\sin\theta$ is negative.}\right)\nonumber\\\therefore~\frac{\mathrm{d}r}{\mathrm{d}t}&=&\omega r\sqrt{\frac{4a^2-r^2}{4a^2}}\nonumber\\\mbox{i.e. }\frac{\omega}{2a}T&=&\int_{2a}^{\frac{4a}{3}}\frac{\mathrm{d}r}{r\sqrt{4a^2-r^2}}=\int_{r=2a}^{r=\frac{4a}{3}}\frac{-\frac{1}{u^2}\mathrm{d}u}{\frac{1}{u^2}\sqrt{4a^2u^2-1}}~~~~~\left(\mbox{Putting } r=\frac{1}{u}\right)\nonumber\\&=&-\frac{1}{2a}\int_{r=2a}^{r=\frac{4a}{3}}\frac{\mathrm{d}u}{\sqrt{u^2-\frac{1}{4a^2}}}\nonumber=-\frac{1}{2a}\left.\log\left[u+\sqrt{u^2-\frac{1}{4a^2}}\right]\right|_{r=2a}^{r=\frac{4a}{3}}\nonumber\\&=&-\frac{1}{2a}\left.\log\left[\frac{1}{r}+\sqrt{\frac{1}{r^2}-\frac{1}{4a^2}}\right]\right|_{r=2a}^{r=\frac{4a}{3}}\nonumber\\\therefore~T&=&\dfrac{1}{\omega}\log_e\left(\dfrac{3+\sqrt{5}}{2}\right)\nonumber
\end{eqnarray}

\section{Motion of a particle on a plane wire rotating about an axis in its plane}

 \begin{wrapfigure}[9]{r}{0.35\textwidth}
	\centering	\includegraphics[height=4.5 cm , width=4.5 cm ]{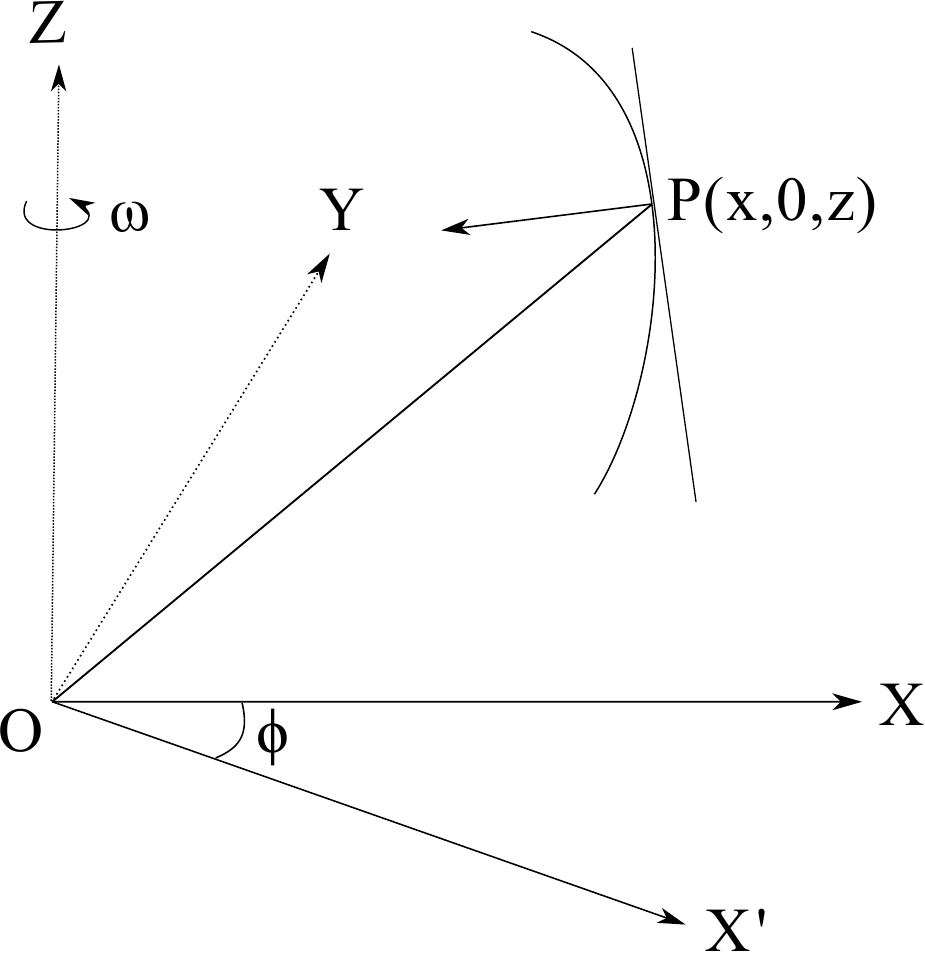}
	\begin{center}
		Fig. 3.50
	\end{center}
\end{wrapfigure}

Let the plane of the curve be the plane $zox$ and the pane is rotating with angular velocity $\omega$ about $oz$. let the equation of the curve be $z=f(x)$. Initially, let the plane of the curve coincides with a fixed plane $zox'$ in space. If at time $t$, $\phi$ be the angle between these two planes then $\omega=\dot{\phi}$. If $\vec{i}$, $\vec{j}$, $\vec{k}$ be the unit vectors along $ox$, $oy$ and $oz$ respectively and $P\equiv(x,0,z)$, then $\vec{r}=\overrightarrow{OP}=x\vec{i}+z\vec{k}$. The equation of motion of the particle is
\begin{equation}
	m\frac{\mathrm{d}^2\vec{r}}{\mathrm{d}t^2}=\vec{F}+\vec{R}\nonumber
\end{equation}
where $\vec{F}$ is the external force and $\vec{R}$ is the reaction on the wire.\\

Now,\begin{eqnarray}
		\frac{\mathrm{d}\vec{r}}{\mathrm{d}t}&=&\frac{\partial\vec{r}}{\partial t}+\vec{\omega}\times\vec{r}=(\dot{x},0,\dot{z})+\begin{vmatrix}
		\vec{i}&\vec{j}&\vec{k}\\0&0&\dot{\phi}\\x&0&z
	\end{vmatrix}\nonumber\\&=&(\dot{x},x\dot{\phi},\dot{z})\nonumber\\\therefore~	\frac{\mathrm{d}^2\vec{r}}{\mathrm{d}t^2}&=&(\ddot{x},x\ddot{\phi}+\dot{x}\dot{\phi},\ddot{z})+\begin{vmatrix}
		\vec{i}&\vec{j}&\vec{k}\\0&0&\dot{\phi}\\\dot{x}&x\dot{\phi}&\dot{z}
	\end{vmatrix}\nonumber\\&=&(\ddot{x}-x\dot{\phi}^2)\vec{i}+(2\dot{x}\dot{\phi}+x\ddot{\phi})\vec{j}+\ddot{z}\vec{k}\nonumber\\&=&\frac{\partial^2r}{\partial t^2}-x\omega^2\vec{i}+\frac{1}{x}\frac{\mathrm{d}}{\mathrm{d}t}\left(x^2\omega\right)\vec{j}\nonumber
\end{eqnarray}

Hence from the equation of motion
\begin{equation}
	m\frac{\partial^2r}{\partial t^2}=\vec{F}+\vec{R}+mx\omega^2\vec{i}-\frac{m}{x}\frac{\mathrm{d}}{\mathrm{d}t}\left(x^2\omega\right)\vec{j}\nonumber
\end{equation}

Thus the motion of the particle relative to the curve is the same as if the curve were at rest and the following forces are acting on the particle.\\

(i) a force $m\omega^2x$ is acting perpendicular to the axis of rotation and away from it.\\

(ii) a force $\dfrac{m}{x}\dfrac{\mathrm{d}}{\mathrm{d}t}\left(x^2\omega\right)$ perpendicular to the plane of the curve is acting in a sense opposite to the sense of rotation of the plane.\\

Now resolving along the tangent, principal normal and binormal to the curve at $P$ we get
\begin{eqnarray}
	mv\frac{\mathrm{d}v}{\mathrm{d}s}&=&F_T+m\omega^2x\frac{\mathrm{d}x}{\mathrm{d}s}\label{eq3.109}\\m\frac{v^2}{\rho}&=&F_N+R_N-m\omega^2x\frac{\mathrm{d}z}{\mathrm{d}s}\\0&=&F_B+R_B-\dfrac{m}{x}\dfrac{\mathrm{d}}{\mathrm{d}t}\left(x^2\omega\right)
\end{eqnarray}

Further, for conservative external forces, $F_T=-\dfrac{\partial V}{\partial s}$ and we have from (\ref{eq3.109})
\begin{equation}
	mv\frac{\mathrm{d}v}{\mathrm{d}s}=-\frac{\partial V}{\partial s}+m\omega^2x\frac{\mathrm{d}x}{\mathrm{d}s}\nonumber
\end{equation}
which on integration gives (assuming $\omega$ to be constant)
\begin{eqnarray}
	\frac{1}{2}mv^2+V-\frac{1}{2}m\omega^2x^2&=&\mbox{constant}\nonumber\\\mbox{i.e., }	\frac{1}{2}mv^2+V'&=&\mbox{constant}\longrightarrow\mbox{modified energy equation}\nonumber
\end{eqnarray}
with $V'=V-\dfrac{1}{2}m\omega^2x^2$ as the effective potential.\\

{\bf 33. } A smooth circular wire of radius $a$ rotates with uniform angular velocity $\omega$ around the tangent at the end of the diameter $AB$ of the wire. A small ring of unit mass is free to slide on the wire from $A$, $r$ being the distance from $A$. If $\omega^2=7\mu$ and if the ring is initially at $B$ with a velocity $4a\sqrt{2\mu}$ relative to the wire, show that after a time $t$ its angular distance from $O$ is $2\tan^{-1}\left(2\sqrt{2}\sinh(\sqrt{\mu}t)\right)$.\\

{\bf Solution: } When the curve is reduced  to rest the equation of motion along the tangent to the wire at $P$ is \begin{equation}
	v\frac{\mathrm{d}v}{\mathrm{d}s}=\mu r\frac{\mathrm{d}r}{\mathrm{d}s}+\omega^2x\frac{\mathrm{d}x}{\mathrm{d}s},\nonumber
\end{equation}
which on integration gives
\begin{equation}
	v^2=\mu r^2+\omega^2 x^2+c\nonumber
\end{equation}

\begin{wrapfigure}[9]{r}{0.35\textwidth}\vspace{-2cm}
	\hfill	\includegraphics[height=4.5 cm , width=4.5 cm ]{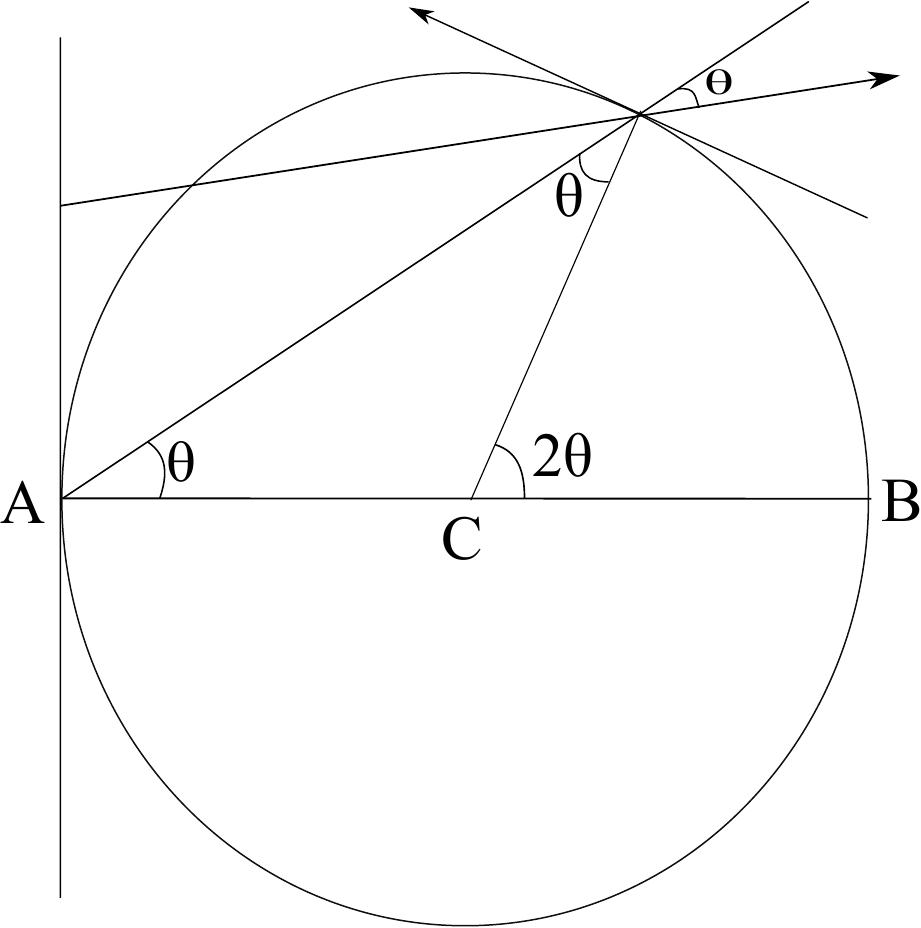}
	\begin{center}
		Fig. 3.51
	\end{center}
\end{wrapfigure}

Initially, $v=4a\sqrt{2\mu}$, $r=2a$, $x=2a$, so $c=0$.
\begin{eqnarray}
	\therefore~v^2&=&\mu r^2+\omega^2 x^2=\mu\cdot4a^2\cos^2\theta+7\mu\cdot4a^2\cos^4\theta\nonumber\\\therefore~\frac{\mathrm{d}s}{\mathrm{d}\theta}\frac{\mathrm{d}\theta}{\mathrm{d}t}&=&\sqrt{\mu}2a\cos\theta\sqrt{1+7\cos^2\theta} ~~~~(\because \theta\mbox{ increases with time})\nonumber\\\mbox{i.e., }\frac{\mathrm{d}\theta}{\mathrm{d}t}&=&\sqrt{\mu}\cos\theta\sqrt{1+7\cos^2\theta}~~\left(\because \frac{\mathrm{d}s}{\mathrm{d}(2\theta)}=a\mbox{ i.e., }\frac{\mathrm{d}s}{\mathrm{d}\theta}=2a\right)\nonumber
\end{eqnarray}

Now integrating
\begin{eqnarray}
	\mu t&=&\int_0^\theta\frac{\sec^2\theta\mathrm{d}\theta}{\sqrt{\tan^2\theta-8}}=\sinh^{-1}\left(\frac{\tan\theta}{2\sqrt{2}}\right)\nonumber\\\therefore~\tan\theta&=&2\sqrt{2}\sinh(\sqrt{\mu}t)\nonumber\\\theta&=&\tan^{-1}\left(2\sqrt{2}\sinh(\sqrt{\mu}t)\right)\nonumber
\end{eqnarray}

So the angular distance $=2\theta=2\tan^{-1}\left(2\sqrt{2}\sinh(\sqrt{\mu}t)\right)$\\

{\bf 34. } A  smooth tube in the form of a parabola of latus rectum $4a$, rotates with constant angular velocity $\omega$ about its axis which is vertical. A particle is at rest at a height $x_0$ above the vertex. Show that if the angular velocity is reduced to $\omega'$, the particle will oscillates from the level $x_0$ in one limb to the same level in the other limb in time 
$$\frac{2}{\sqrt{\omega^2-{(\omega')}^2}}\int_0^\frac{\pi}{2}\sqrt{1+\frac{x_0}{a}\sin^2\theta}\mathrm{d}\theta$$

{\bf Solution: } Let the equation of the parabola is $$y^2=4ax$$

When the curve is reduced to rest the equation of motion along the tangent at $P$ is
\begin{equation}
	v\frac{\mathrm{d}v}{\mathrm{d}s}=\omega^2y\frac{\mathrm{d}y}{\mathrm{d}s}-g\frac{\mathrm{d}x}{\mathrm{d}s}\nonumber
\end{equation}
i.e., on integration, $$v^2=\omega^2 y^2-2gx+c$$

\begin{wrapfigure}[10]{r}{0.35\textwidth}
	\hfill	\includegraphics[height=4.5 cm , width=4.5 cm ]{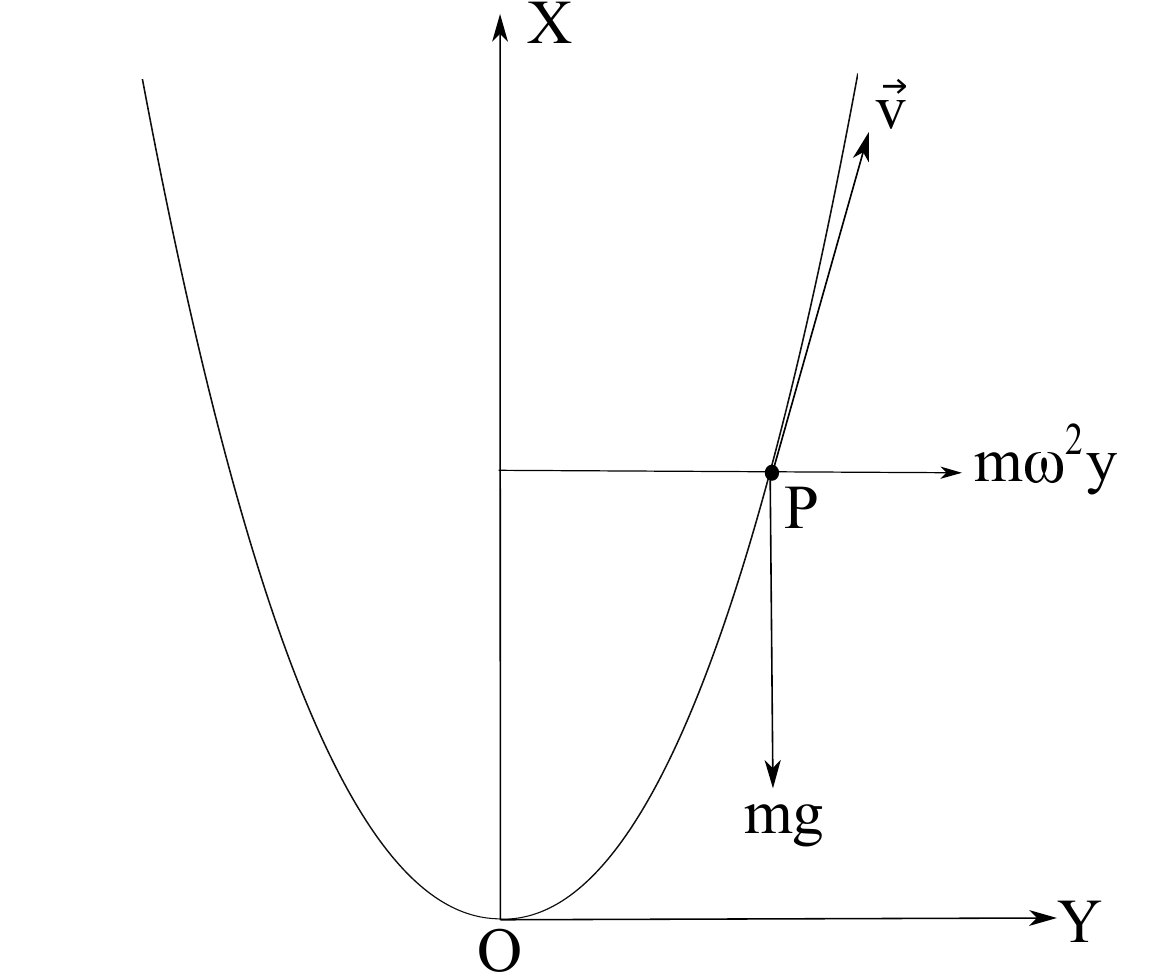}
	\begin{center}
		Fig. 3.52
	\end{center}
\end{wrapfigure}

But $v^2=\dot{x}^2+\dot{y}^2=\dot{x}^2\left\{1+\left(\dfrac{\mathrm{d}y}{\mathrm{d}x}\right)^2\right\}=\dot{x}^2\left\{1+\dfrac{a}{x}\right\}$.
\begin{equation}
	\therefore~\dot{x}^2\left(1+\frac{a}{x}\right)=4a\omega^2x-2gx+c\nonumber
\end{equation}

Now, differentiating both sides with respect to $x$, we have
\begin{equation}
	2\ddot{x}\left(1+\frac{a}{x}\right)-a\frac{\dot{x}^2}{x^2}=4a\omega^2-2g\nonumber
\end{equation}

For relative equilibrium at $x=x_0$, $\dot{x}=0=\ddot{x}$ at $x=x_0$.
$$\therefore~4a\omega^2=2g$$

When the angular velocity is $\omega'$, the equation of motion of the particle is
\begin{eqnarray}
	&&v\frac{\mathrm{d}v}{\mathrm{d}s}=\omega'^2y\frac{\mathrm{d}y}{\mathrm{d}s}-g\frac{\mathrm{d}x}{\mathrm{d}s}\nonumber\\\therefore&&v^2=\dot{x}^2+\dot{y}^2=\dot{x}^2\left(1+\frac{a}{x}\right)=\omega'^2 y^2-2gx+c'\nonumber\\\mbox{i.e.,}&&\dot{x}^2\left(1+\frac{a}{x}\right)-4ax\omega'^2+2gx=c'=0-4ax_0\omega'^2+2gx_0\nonumber\\\mbox{i.e.,}&&\dot{x}^2\left(1+\frac{a}{x}\right)=\left(2g-4a{\omega'}^2\right)(x_0-x)=4a\left(\omega^2-{\omega'}^2\right)(x_0-x)\nonumber
\end{eqnarray}

As $\omega>\omega'$ and $x>0$ so the motion of the particle will be confined to the region $x\leqslant x_0$ i.e., the particle will move from a level $x_0$ in one limb to the same level to the other. If $T$ be the corresponding time then $T$ can be evaluated as follows:
\begin{eqnarray}
	\dot{x}&=&-\sqrt{4a\left(\omega^2-{\omega'}^2\right)x\frac{(x_0-x)}{(a+x)}}~~~(\because x \mbox{ decreases with time})\nonumber\\
	\therefore-\int_{x_0}^0\sqrt{\frac{a+x}{x(x_0-x)}}\mathrm{d}x&=&\sqrt{4a\left(\omega^2-{\omega'}^2\right)}\int_0^\frac{T}{2}\mathrm{d}t\nonumber\\
	\mbox{i.e.,~ }\sqrt{a\left(\omega^2-{\omega'}^2\right)}T&=&\int_0^\frac{\pi}{2}\frac{\sqrt{a+x_0\sin^2\theta}}{x_0\sin\theta\cos\theta}2x_0\sin\theta\cos\theta\mathrm{d}\theta~~~(\mbox{Putting	 }x=x_0\sin^2\theta)\nonumber\\
	&=&2\int_0^\frac{\pi}{2}\sqrt{a+x_0\sin^2\theta}\mathrm{d}\theta\nonumber\\
	\therefore~T&=&\frac{2}{\sqrt{\left(\omega^2-{\omega'}^2\right)}}\int_0^\frac{\pi}{2}\sqrt{1+\frac{x_0}{a}\sin^2\theta}\mathrm{d}\theta\nonumber
\end{eqnarray}

{\bf 35. } A smooth wire in the form of the curve $y=b\cos\left(\frac{x}{a}\right)$, which lies between $x=\pm a\pi$, $a$ and $b$ being positive numbers, rotates with constant angular velocity $\omega$ about the axis of $y$ which is vertically downward. Prove that, if $a^2\omega^2<bg$, there are three positions of relative equilibrium for a bead which can slide on the wire and that two of them are unstable.\\

{\bf Solution: } The equation of motion is
\begin{equation}
	mv\frac{\mathrm{d}v}{\mathrm{d}s}=m\omega^2x\frac{\mathrm{d}x}{\mathrm{d}s}+mg\frac{\mathrm{d}y}{\mathrm{d}s}\nonumber
\end{equation}
so on integration 
\begin{eqnarray}
	&&v^2=\omega^2x^2+2gy+c\nonumber\\\mbox{i.e., }&&\dot{x}^2\left\{1+\left(\dfrac{\mathrm{d}y}{\mathrm{d}x}\right)^2\right\}=\omega^2x^2+2gb\cos\left(\frac{x}{a}\right)+c\nonumber\\\mbox{i.e., }&&\dot{x}^2\left\{1+\frac{b^2}{a^2}\sin^2\left(\frac{x}{a}\right)\right\}=\omega^2x^2+2gb\cos\left(\frac{x}{a}\right)+c\nonumber
\end{eqnarray}

\begin{wrapfigure}[12]{r}{0.35\textwidth}
	\hfill	\includegraphics[height=4.8 cm , width=4.5 cm ]{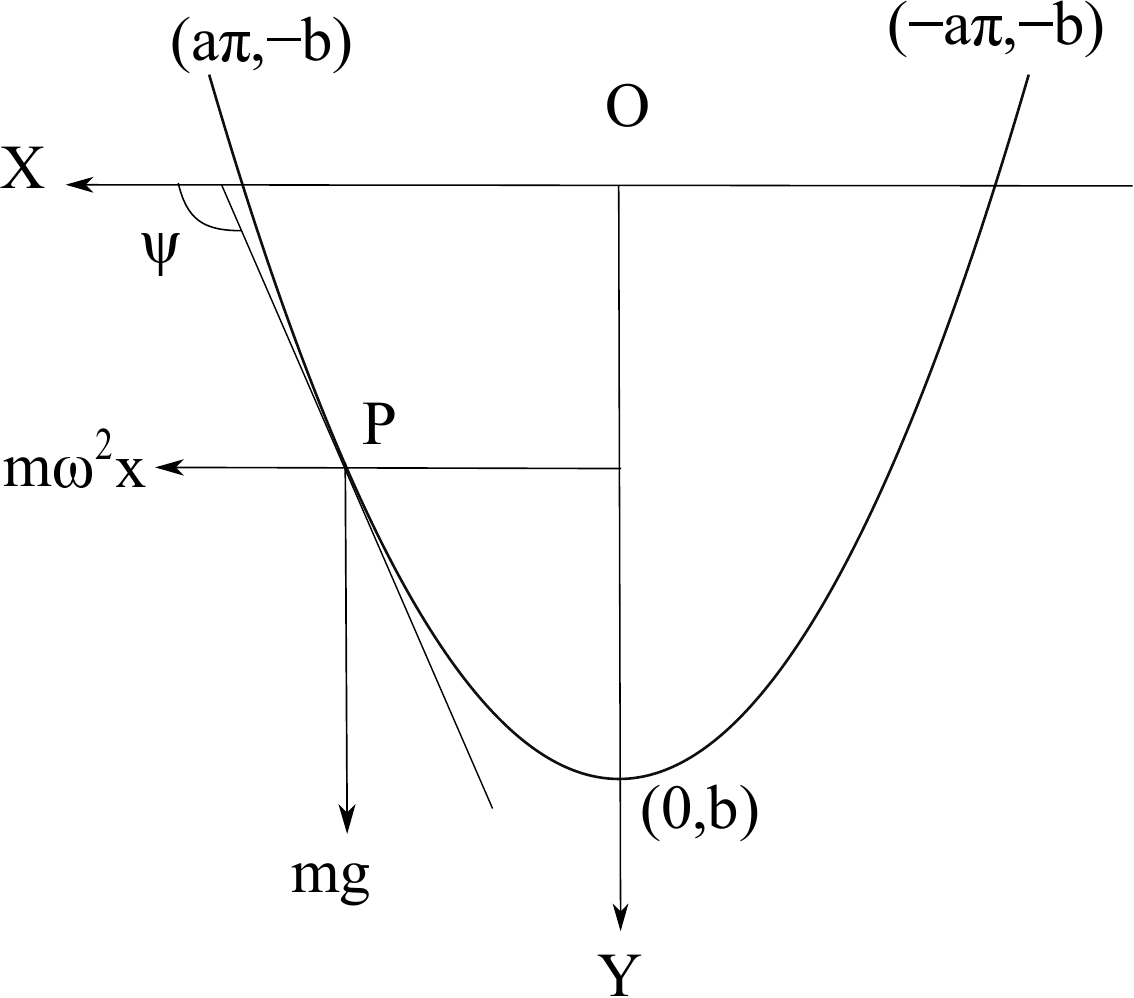}
	\begin{center}
		Fig. 3.53
	\end{center}
\end{wrapfigure}

Now differentiating with respect to $x$ we have
\begin{equation}
	\ddot{x}\left\{1+\frac{b^2}{a^2}\sin^2\left(\frac{x}{a}\right)\right\}+\dot{x}^2\frac{b^2}{a^3}\sin\left(\frac{x}{a}\right)\sin\left(\frac{x}{a}\right)=\omega^2x-g\frac{b}{a}\sin\left(\frac{x}{a}\right)\nonumber
\end{equation}

So for relative equilibrium, $\dot{x}=0=\ddot{x}$ i.e., the position of relative equilibrium is given by
\begin{equation}\label{eq3.35.1}
	\omega^2x=\frac{gb}{a}\sin\left(\frac{x}{a}\right)
\end{equation}

Clearly, $x=0$ is a solution of equation (\ref{eq3.35.1}). Also $x=\pm x_0$ will be a solution of equation (\ref{eq3.35.1})if the line $y=\omega^2x$ intersects the curve $y=\dfrac{gb}{a}\sin\left(\dfrac{x}{a}\right)$ at $x_0$ $(0<x_0<a\pi)$. For $0<x<a\pi$, the gradient of the curve $y=\dfrac{gb}{a}\sin\left(\dfrac{x}{a}\right)$ is $\dfrac{gb}{a^2}\cos\left(\dfrac{x}{a}\right)$, which is monotonically decreasing and the gradient of the straight line is $\omega^2$. The straight line will intersect the curve at $\pm x_)$ if the gradient of the curve at $x=0$ is greater than the gradient of the straight line i.e., $\dfrac{gb}{a^2}>\omega^2$.\\

Thus if $a^2\omega^2<gb$ there are three positions of relative equilibrium. Also we have at $\pm x_0$ the gradient of the curve is less than that of the straight line i.e., $\dfrac{gb}{a^2}\cos\left(\dfrac{x_0}{a}\right)<\omega^2$ i.e, $a^2\omega^2-gb\cos\left(\dfrac{x_0}{a}\right)>0$.\\

Stability Criteria:\\

Case-I: $x=0$ is the equilibrium position. Let $x=0+\epsilon$ where $\epsilon$ is very small, be a displaced position. So $\epsilon^2$, $\dot{\epsilon}^2$, $\cdots$ are neglected. Now,
\begin{eqnarray}
	&&\ddot{x}\left\{1+\frac{b^2}{a^2}\sin^2\left(\frac{x}{a}\right)\right\}+\dot{x}^2\frac{b^2}{a^3}\sin\left(\frac{x}{a}\right)\sin\left(\frac{x}{a}\right)=\omega^2x-g\frac{b}{a}\sin\left(\frac{x}{a}\right)\nonumber\\\mbox{i,e., }&&\ddot{\epsilon}\left\{1+\frac{b^2}{a^2}\sin^2\left(\frac{\epsilon}{a}\right)\right\}=\omega^2\epsilon-\frac{gb}{a}\sin\left(\frac{\epsilon}{a}\right)\nonumber\\\mbox{i.e., }&&\ddot{\epsilon}\left(1+\frac{b^2}{a^2}\cdot\frac{\epsilon^2}{a^2}\right)=-\left(\frac{gb}{a^2}-\omega^2\right)\epsilon\nonumber\\\mbox{i.e., }&&\ddot{\epsilon}=-\left(\frac{gb}{a^2}-\omega^2\right)\epsilon\nonumber
\end{eqnarray}

As $\dfrac{gb}{a^2}-\omega^2>0$ so the motion is a simple harmonic motion. Hence the equilibrium position is stable.\\

Case-II: $x=\pm x_0$ are the equilibrium position.\\

Let $x=\pm x_0+\epsilon$.
\begin{eqnarray}
	\therefore&&\ddot{\epsilon}\left\{1+\frac{b^2}{a^2}\sin^2\left(\frac{\pm x_0+\epsilon}{a}\right)\right\}=\omega^2(\pm x_0+\epsilon)-\frac{gb}{a}\sin\left(\frac{\pm x_0+\epsilon}{a}\right)\nonumber\\\mbox{i.e.,}&&\ddot{\epsilon}\left[1+\frac{b^2}{a^2}\left\{\pm\sin\frac{x_0}{a}+\frac{\epsilon}{a}\cos\frac{x_0}{a}\right\}^2\right]=\cancel{\pm\omega^2 x_0}+\omega^2\epsilon-\frac{gb}{a}\left[\cancel{\pm\sin\frac{x_0}{a}}+\frac{\epsilon}{a}\cos\frac{x_0}{a}\right]\nonumber\\&&~~~~~~~~~~~~~~~~~~~~~~~~~~~~~~~~~~~~~~~~~~~~~~~~~~~\left(\mbox{As }\omega^2x_0=\frac{gb}{a}\sin\frac{x_0}{a}\right)\nonumber\\\mbox{i.e,}&&\ddot{\epsilon}\left[1+\frac{\omega^4x_0^2}{g^2}\pm2\frac{\omega^2x_0b\epsilon}{a^2g}\cos\frac{x_0}{a}\right]=\left(\omega^2-\frac{gb}{a^2}\cos\frac{x_0}{a}\right)\epsilon\nonumber\\\therefore&&\ddot{\epsilon}=\frac{\left(\omega^2-\frac{gb}{a^2}\cos\frac{x_0}{a}\right)}{\left(1+\frac{\omega^4x_0^2}{g^2}\right)}\epsilon\left[1\pm\frac{2\frac{\omega^2x_0b\epsilon}{a^2g}\cos\frac{x_0}{a}}{1\pm\frac{\omega^4x_0^2}{g^2}}\right]^{-1}\nonumber\\&&~~~=\frac{\left(\omega^2-\frac{gb}{a^2}\cos\frac{x_0}{a}\right)}{\left(1+\frac{\omega^4x_0^2}{g^2}\right)}\epsilon ~~~~~~~~~~~\left(a^2\omega^2>gb\cos\left(\dfrac{x_0}{a}\right)\right)\nonumber
\end{eqnarray}

Hence the motion is not simple harmonic motion. So $x=\pm x_0$ are unstable position of equilibrium.\\

\section{Motion relative to Earth}

Let $SN$ be the axis drawn from South to North, $O$ be any point on or near the surface of the Earth. $B$ is the foot of the perpendicular from $O$ on $SN$. $\vec{I}$ is the unit vector along $\overrightarrow{BO}$. The triad of unit vectors $\vec{i}$, $\vec{j}$, $\vec{k}$ fixed relative to earth are defined as:\\

\begin{wrapfigure}[12]{r}{0.35\textwidth}
	\hfill	\includegraphics[height=5 cm , width=4.9 cm ]{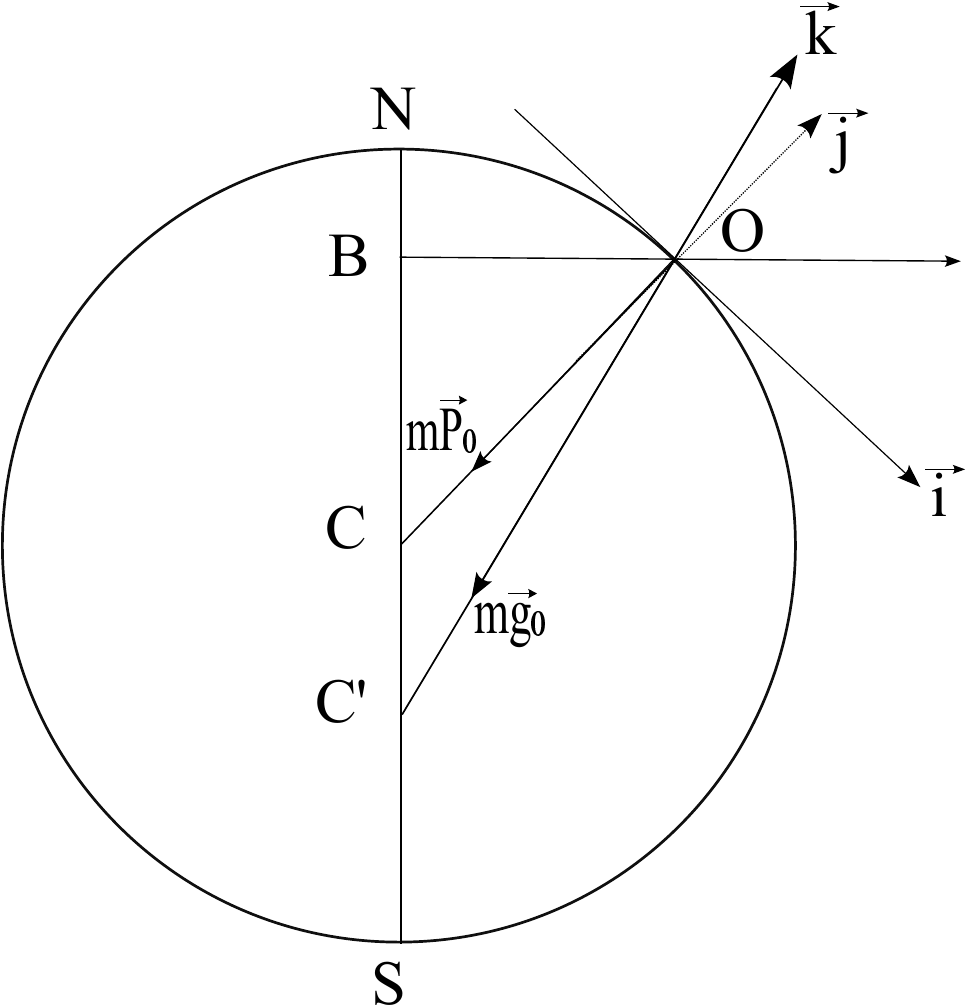}
	\begin{center}
		Fig. 3.54
	\end{center}
\end{wrapfigure}

$\vec{i}$ is horizontal and points South\\

$\vec{j}$ is horizontal and points East\\

$\vec{k}$ is vertical and points upwards.\\

Let $\overrightarrow{BO}=\vec{r_0}$ and $\vec{r}$ be the position vector of a particle relative to $O$. Then the position vector of the particle relative to $B$ is $$\vec{r}_B=\vec{r}_0+\vec{r}$$

The absolute acceleration of the particle is
$$\ddot{\vec{r}}_B=\ddot{\vec{r}}_0+\ddot{\vec{r}}=\omega^2\vec{r}_0+\ddot{\vec{r}}$$

If $m$ be the mass of the particle then as the force of gravity is proportional to $m$, let it be $m\vec{P}$ and let $\vec{F}$ be the resultant of all other forces. Then the equation of motion of the particle is
\begin{eqnarray}
	m\ddot{\vec{r}}_B&=&m\vec{P}+\vec{F}\nonumber\\\mbox{i.e.,~ }m\ddot{\vec{r}}&=&m\vec{P}+\vec{F}+m\omega^2\vec{r}_0=\vec{F}+m\vec{g}\label{eq3.113}
\end{eqnarray} where 
\begin{equation}
	m\vec{g}=m\vec{P}+m\omega^2\vec{r}_0
\end{equation}
is termed as the apparent force of gravity.

Thus in considering the motion of a particle relative to $O$, we can assume the point $O$ at rest provided we replace the force of gravity by the apparent force of gravity.\\

Suppose we consider a plumb line in the equilibrium with its bob at $O$. Then $\vec{F}$ will be not the tension in the string along the unit vector $\vec{k}$. Since at $O$, $\vec{r}=0$, so from (\ref{eq3.113}) $\vec{F}+m\vec{g}_0=\vec{0}$, where $m\vec{g}_0=m\vec{P}_0+m\omega^2\vec{r}_0$ and $m\vec{P}_0$ is the force of gravity at $O$. Since $m\vec{g}_0=-\vec{F}$, $m\vec{g}_0$ is acting vertically downward at $O$ and the vertically downward direction at $O$ will be along $OC'$. The angles $OC'N$ and $OCN$ are called geographical and geocentric co-latitude of the point $O$ respectively.

\subsection{Acceleration of a particle}

Here $C$ is not the centre. Let $\lambda$ be the geographical latitude of the place at $O$, a point on near the surface of the earth. Let $\omega$ be the angular velocity of the Earth about its axis. We choose the axes as follows:\\

\begin{wrapfigure}[12]{r}{0.35\textwidth}\vspace{-.3cm}
	\hfill	\includegraphics[height=4.5 cm , width=4.5 cm ]{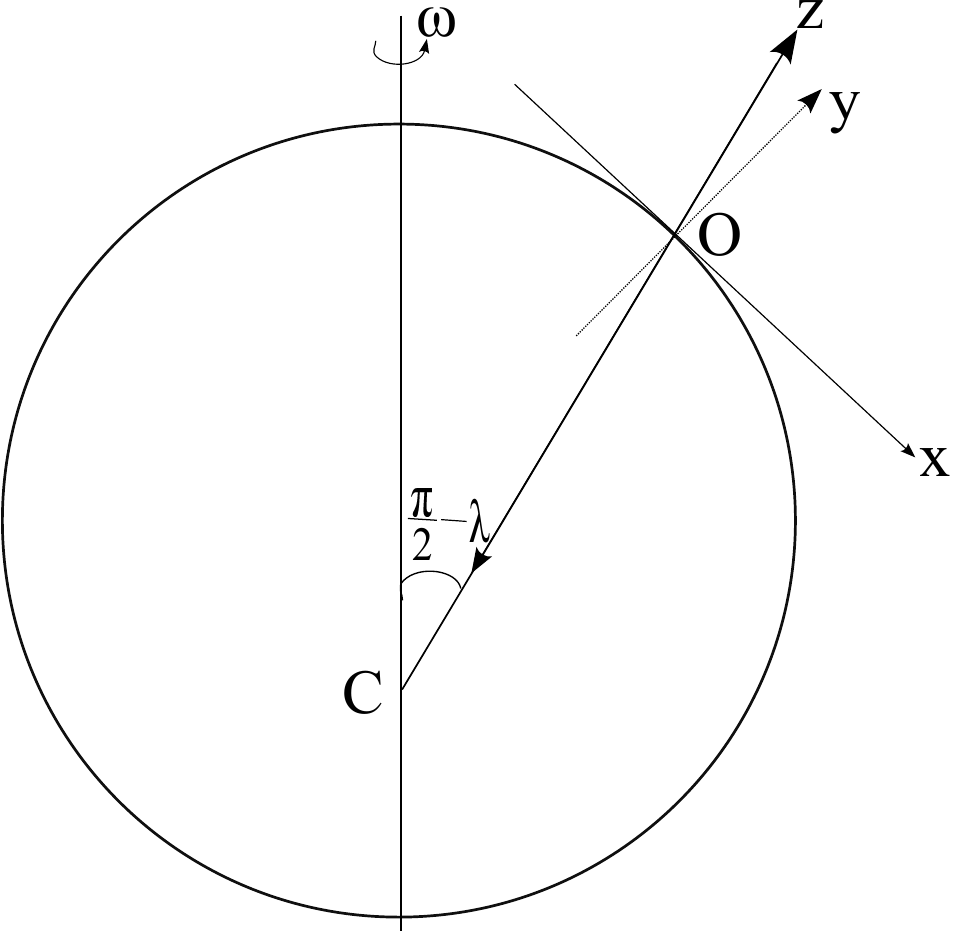}
	\begin{center}
		Fig. 3.55
	\end{center}
\end{wrapfigure}

$z$-axis is vertical and points upwards\\

$x$-axis is horizontal and points south\\

$y$-axis is horizontal and points east.\\

Then components of $\omega$ about $x$, $y$, $z$ axes are 
$$\omega_x=\omega\cos(\pi-\lambda)=-\omega\cos\lambda~,~~\omega_y=0~,~~\omega_z=\omega\sin\lambda.$$

Now referred to $ox$, $oy$, $oz$ as axes let $(x,y,z)$ be the co-ordinates of the particle at time $t$. Suppose $(u_x,u_y,u_z)$ and $(f_x,f_y,f_z)$ are the components of velocity and acceleration of the particle. Then $$(u,v,w)=(\dot{x},\dot{y},\dot{z})+\begin{vmatrix}
	\vec{i}&\vec{j}&\vec{k}\\-\omega\cos\lambda&0&\omega\sin\lambda\\x&y&z
\end{vmatrix}$$
 where $\vec{i}$, $\vec{j}$, $\vec{k}$ are the unit vectors along the three axes $ox$, $oy$, $oz$ respectively.\begin{eqnarray}
 	\therefore~u_x=\dot{x}-\omega y\sin\lambda~,~~u_y=\dot{y}+\omega x\sin\lambda+\omega z\cos\lambda~,~~u_z=\dot{z}-\omega y\cos\lambda\nonumber
 \end{eqnarray}

Similarly\begin{eqnarray}
	&&(f_x,f_y,f_z)=(\dot{u_x},\dot{u_y},\dot{u_z})+\begin{vmatrix}
		\vec{i}&\vec{j}&\vec{k}\\-\omega\cos\lambda&0&\omega\sin\lambda\\u&v&w
	\end{vmatrix}\nonumber\\\mbox{i.e.,}&&f_x=\ddot{x}-2\omega\dot{y}\sin\lambda-\omega^2x\sin^2\lambda-\omega^2z\sin\lambda\cos\lambda\nonumber\\&&f_y=\ddot{y}+2\omega \dot{x}\sin\lambda+2\omega\dot{z}\cos\lambda-\omega^2y\nonumber\\&&f_z=\ddot{z}-2\omega\dot{y}\cos\lambda-\omega^2x\sin\lambda\cos\lambda-\omega^2z\cos^2\lambda\nonumber
\end{eqnarray}

\section{Motion of a freely falling body}

Suppose a particle of mass $m$ falls freely from a height $h$ above $O$. Let $m\vec{g}$ be the constant apparent force of gravity. Then the equation of motion of the particle are \begin{eqnarray}
	&&mf_x=0~,~~mf_y=0~,~~mf_z=-mg\nonumber\\\mbox{i.e., }&&\ddot{x}-2\omega\dot{y}\sin\lambda-\omega^2x\sin^2\lambda-\omega^2z\sin\lambda\cos\lambda=0\nonumber\\&&\ddot{y}+2\omega \dot{x}\sin\lambda+2\omega\dot{z}\cos\lambda-\omega^2y=0\nonumber\\&&\ddot{z}-2\omega\dot{y}\cos\lambda-\omega^2x\sin\lambda\cos\lambda-\omega^2z\cos^2\lambda=-g\label{eq3.115}
\end{eqnarray}

We get first approximation of the solution by noting that $\dot{x}$, $\dot{y}$ are small compared to $\dot{z}$ and neglecting $\omega^2$ so that the set of equations (\ref{eq3.115}) become
\begin{equation}\label{eq3.116}
	\ddot{x}=0~,~~\ddot{y}=-2\omega\dot{z}\cos\lambda~,~~\ddot{z}=-g
\end{equation}

The initial conditions at $t=0$ are $x=0=y$, $z=h$, $\dot{x}=0=\dot{y}=\dot{z}$, so the solution of equation (\ref{eq3.116}) gives
\begin{equation}\label{eq3.117}
	x=0~,~~y=\frac{1}{3}g\omega t^3\cos\lambda~,~~z=h-\frac{1}{2}gt^2
\end{equation}

Thus in the first approximation the time of falling from the height $h$ is $t=\sqrt{\dfrac{2h}{g}}$ and then $y=\dfrac{1}{3}g\omega\cos\lambda\sqrt{\dfrac{8h^3}{g^3}}$.\\

This shows that the particle will have a easterly deviation as it reaches the surface of the earth.\\

Now to find the second approximation of the solution of equation (\ref{eq3.115}) we substitute the solution (\ref{eq3.117}) in (\ref{eq3.115}) and neglect $\omega^3$. Hence we obtain
\begin{eqnarray}
	&&\ddot{x}-2\omega^2 gt^2\cos\lambda\sin\lambda-\omega^2\sin\lambda\cos\lambda\left(h-\frac{1}{2}gt^2\right)=0\nonumber\\\mbox{i.e.}&&\ddot{x}-\frac{3}{2}\omega^2 gt^2\cos\lambda\sin\lambda-h\omega^2\sin\lambda\cos\lambda=0,\nonumber\\&&\ddot{y}-2\omega gt\cos\lambda=0,\nonumber\\\mbox{and}&&\ddot{z}-2\omega^2gt^2\cos^2\lambda-\omega^2\cos^2\lambda\left(h-\frac{1}{2}gt^2\right)=-g\nonumber\\\mbox{i.e.,}&&\ddot{z}-\frac{3}{2}\omega^2gt^2\cos^2\lambda-h\omega^2\cos^2\lambda=-g\nonumber
\end{eqnarray}

So on integration using the initial conditions
\begin{eqnarray}
	x&=&\frac{1}{8}\omega^2 gt^4\cos\lambda\sin\lambda+\frac{1}{2}h\omega^2t^2\sin\lambda\cos\lambda\nonumber\\y&=&\frac{1}{3}g\omega t^3\cos\lambda\nonumber\\z&=&h-\frac{1}{2}gt^2+\frac{1}{8}\omega^2gt^4\cos^2\lambda+\frac{1}{2}h\omega^2t^2\cos^2\lambda\nonumber
\end{eqnarray}

When the particle reaches the ground then $z=0$ and the last equation gives the corresponding time. If $t_0$be the time of flight in the 1st approximation then $t_0^2=\dfrac{2h}{g}$. To get the second approximation of $t$, let $t^2=t_0^2+\epsilon$, where $\epsilon$ is small compared to $t_0^2$ and we neglect $\epsilon^2$. Thus we get
\begin{eqnarray}
	0&=&h-\frac{1}{2}g(t_0^2+\epsilon)+\frac{1}{8}\omega^2g(t_0^2+\epsilon)^2\cos^2\lambda+\frac{1}{2}h\omega^2(t_0^2+\epsilon)\cos^2\lambda\nonumber\\\mbox{i.e.,~ }0&=&-\frac{1}{2}g\epsilon+\frac{1}{8}\omega^2g\left(\frac{4h^2}{g^2}+\frac{2h}{g}\epsilon\right)\cos^2\lambda+\frac{1}{2}h\omega^2\left(\frac{2h}{g}+\epsilon\right)\cos^2\lambda\nonumber\\\mbox{i.e.,~ }\epsilon&=&\frac{3\omega^2h^2\cos^2\lambda}{g^2}\nonumber
\end{eqnarray}

Hence the time of flight $T^2=\dfrac{2h}{g}+\dfrac{3\omega^2h^2\cos^2\lambda}{g^2}$.\\

Then $x$ and $y$ are given by
\begin{eqnarray}
	y&=&\frac{1}{3}g\omega\cos\lambda\sqrt{\frac{8h^3}{g^3}}\left\{1+\frac{3}{2}\frac{3\omega^2h\cos^2\lambda}{g}\right\}^\frac{3}{2}\nonumber\\&=&\frac{1}{3}g\omega\cos\lambda\sqrt{\frac{8h^3}{g^3}}\nonumber\\x&=&\frac{1}{8}\omega^2gt^4\cos\lambda\sin\lambda+\frac{1}{2}h\omega^2t^2\sin\lambda\cos\lambda\nonumber\\&=&\frac{1}{8}\omega^2g\cos\lambda\sin\lambda\left\{\dfrac{2h}{g}+\dfrac{3\omega^2h^2\cos^2\lambda}{g^2}\right\}^2+\frac{1}{2}h\omega^2\sin\lambda\cos\lambda\left\{\dfrac{2h}{g}+\dfrac{3\omega^2h^2\cos^2\lambda}{g^2}\right\}\nonumber\\&=&\frac{1}{2g}h^2\omega^2\cos\lambda\sin\lambda+\frac{1}{g}h^2\omega^2\sin\lambda\cos\lambda\nonumber\\&=&\frac{3}{2g}h^2\omega^2\sin\lambda\cos\lambda\nonumber
\end{eqnarray}

\section{Motion of a particle projected vertically upwards}

Let a particle be projected vertically upward from the point $O$ on or near the surface of the earth, with initial velocity $\omega_0$ along the $z$-axis. The equations of motion, when $\omega^2$ is neglected with the natural assumption that $\dot{x}$, $\dot{y}$ are small compare to $\dot{z}$ are
$$\ddot{x}=0~,~~\ddot{y}+2\omega\dot{z}\cos\lambda=0~,~~\ddot{z}=-g$$

The initial conditions are $\dot{x}=0=\dot{y}$, $\dot{z}=\omega_0$, $x=0=y=z$ at $t=0$. So the solution of the above equations of motion are
$$x=0~,~~z=\omega_0t-\frac{1}{2}gt^2~,~~y=-g\omega \cos\lambda\left(\omega_0t^2-\frac{1}{3}gt^3\right)$$

So when the particle again reaches the level of projection then the time of flight is $T=\dfrac{2\omega_0}{g}$ and then $$y=-\frac{4\omega_0^2}{g^2}\omega\cos\lambda\left(\omega_0-\frac{2}{3}\omega_0\right)=-\frac{4}{3}\frac{\omega_0^3}{g^2}\omega\cos\lambda$$

Hence after reaching the level of projection the particle is deviated towards west by the amount $\dfrac{4}{3}\frac{\omega_0^3}{g^2}\omega\cos\lambda$. Further, from the solution we note that if $t>\dfrac{3\omega_0}{g}$ then $y>0$. Therefore, if the particle allows to fall below the level of projection and the particle does not reach the surface of earth before time $\dfrac{3\omega_0}{g}$, then there will be a deviation towards east.\\

\subsection{Motion of a projectile}\label{projectile}

Let the projectile be projected from $O$ with a velocity whose components along the axes are $u_0$, $v_0$, $w_0$. The equations of motion of the projectile (when $\omega^2$ is neglected) are
\begin{eqnarray}
	&&\ddot{x}-2\omega\dot{y}\sin\lambda=0\nonumber\\&&\ddot{y}+2\omega \dot{x}\sin\lambda+2\omega\dot{z}\cos\lambda=0\nonumber\\&&\ddot{z}-2\omega\dot{y}\cos\lambda=-g\nonumber
	\end{eqnarray}

\begin{wrapfigure}[10]{r}{0.3\textwidth}
	\hfill	\includegraphics[height=4.5 cm , width=4.5 cm ]{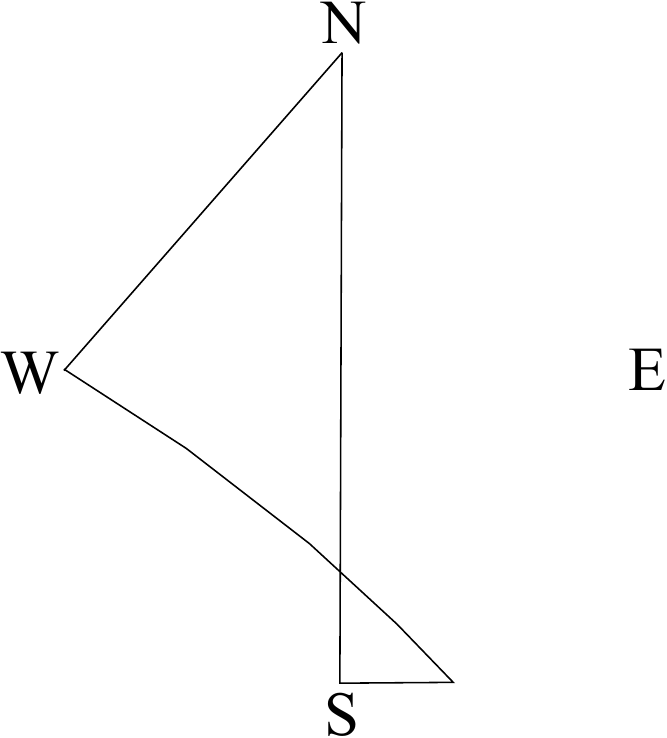}
	\begin{center}
		Fig. 3.56
	\end{center}
\end{wrapfigure}

The initial conditions are $x=y=0=z$, $\dot{x}=u_0$, $\dot{y}=v_0$, $\dot{z}=w_0$ at $t=0$.\\

Thus neglecting $\omega^2$ and higher powers and using the above initial conditions the solution to the above equations of motion for the projectile are
\begin{eqnarray}
	x&=&u_0t+(\omega v_0\sin\lambda)t^2\nonumber\\y&=&v_0t+\frac{1}{3}g\omega t^3\cos\lambda-\omega t^2(u_0\sin\lambda+w_0\cos\lambda)\nonumber\\z&=&w_0t-\frac{1}{2}gt^2+\omega v_0 t^2\nonumber
\end{eqnarray}

So when the particle reaches the level of projection again then $z=0$ at $t=T$ and is given by
\begin{eqnarray}
	T&=&\frac{w_0}{\frac{g}{2}-\omega v_0\cos\lambda}=\frac{2w_0}{g}\left(1-\frac{2v_0\omega}{g}\cos\lambda\right)^{-1}\nonumber\\&\simeq&\frac{2w_0}{g}\left(1+\frac{2v_0\omega}{g}\cos\lambda\right)~,~~~\mbox{ neglecting $\omega^2$ and higher powers}\nonumber
\end{eqnarray}

If there is no rotation then the corresponding time of flight is $T_0=\dfrac{2w_0}{g}$. This shows that the time of flight is increased due to the rotation by an amount $T_0\times\dfrac{2v_0\omega}{g}\cos\lambda$.\\

\begin{wrapfigure}[10]{r}{0.3\textwidth}
	\hfill	\includegraphics[height=4.5 cm , width=4.5 cm ]{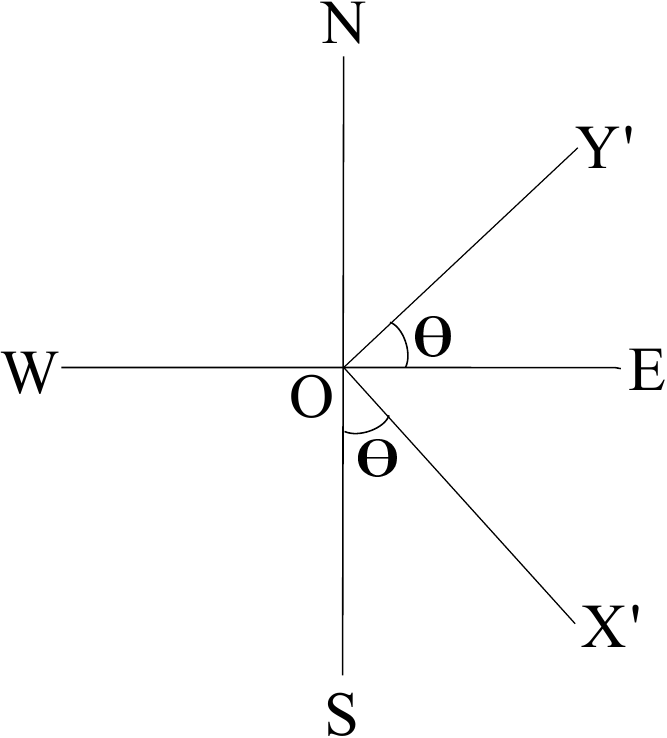}
	\begin{center}
		Fig. 3.57
	\end{center}
\end{wrapfigure}

Suppose the vertical plane of projection of the projectile be $\theta$ east of south. Let $OX'$ be the horizontal line along this direction. $OY'$ is perpendicular to $OX'$ and horizontal, i.e., $OY'$ is horizontal and $\theta$ north of east. Then referred to $OX'$, $OY'$ and $OZ$ as axes let the co-ordinate of the projectile at time $t$ be $(x',y',z')$ where
\begin{eqnarray}
	x'&=&x\cos\theta+y\sin\theta\nonumber\\y'&=&-x\sin\theta+y\cos\theta\nonumber
\end{eqnarray} 

If $v$ be the horizontal component of initial velocity of projection then $u_0=v\cos\theta$, $v_0=v\sin\theta$
$$\mbox{i.e., ~}U=u_0\cos\theta+v_0\sin\theta \mbox{~ and ~}u_0\sin\theta=v_0\cos\theta$$

Thus\begin{eqnarray}
	x'&=&x\cos\theta+y\sin\theta\nonumber\\&=&(u_0t+\omega v_0\sin\lambda t^2)\cos\theta+\left[v_0t+\frac{1}{3}g\omega t^3\cos\lambda-\omega t^2(u_0\sin\lambda+w_0\cos\lambda)\right]\sin\theta\nonumber\\&=&Ut+\left(\frac{1}{3}g\omega\cos\lambda\sin\theta\right)t^3-(\omega w_0\cos\lambda\sin\theta)t^2\nonumber
\end{eqnarray}

Similarly\begin{eqnarray}
	y'&=&-x\sin\theta+y\cos\theta\nonumber\\&=&\omega t^2\left[\frac{1}{3}gt\cos\lambda\cos\theta-(U\sin\lambda+w_0\cos\lambda\cos\theta)\right]\nonumber
\end{eqnarray}

Thus the particle will have a deviation of amount $$\omega t^2(U\sin\lambda+w_0\cos\lambda\cos\theta)-\frac{1}{3}\omega gt^3\cos\lambda\cos\theta$$
to the right of vertical plane of projection when observed from $O$. So the amount of deviation when the particle reaches the ground i.e., when $t=\dfrac{2w_0}{g}\left(1+\dfrac{2v_0\omega}{g}\cos\lambda\right)=T$ will be
\begin{eqnarray}
	&=&\omega(U\sin\lambda+w_0\cos\lambda\cos\theta)\frac{4w_0^2}{g^2}-\frac{1}{3}\omega g\cos\lambda\cos\theta\frac{8w_0^3}{g^3}\nonumber\mbox{~~(neglecting $\omega^2$ and higher powers)}\\&=&\frac{4\omega w_0^2}{g^2}\left(U\sin\lambda+\frac{1}{3}w_0\cos\lambda\cos\theta\right)\nonumber
\end{eqnarray}

Similarly,
\begin{eqnarray}
	x'{\bigg|}_{t=T}=\frac{2w_0}{g}\left[U(1+\dfrac{2v_0\omega}{g}\cos\lambda)-\frac{2w_0^2}{3g}\omega \cos\lambda\sin\theta\right]\nonumber\mbox{~~(neglecting $\omega^2$ and higher powers)}
\end{eqnarray}

So the horizontal range \begin{eqnarray}
	R&=&\sqrt{(x')^2+(y')^2}=x'\sqrt{1+\left(\frac{y'}{x'}\right)^2}\nonumber\\	&\simeq&x'{\bigg|}_{t=T}\mbox{~~ (neglecting $\omega^2$ and higher powers)}\nonumber\\\therefore~R&=&\frac{2w_0U}{g}+\frac{4\omega w_0\cos\lambda}{g^2}\left[v_0U-\frac{1}{3}w_0^2\sin\theta\right]\nonumber
\end{eqnarray}

If there is no rotation then the range $R_0=\dfrac{2w_0U}{g}$,

 so increase in length = $\dfrac{4\omega w_0\cos\lambda}{g^2}\left[v_0U-\dfrac{1}{3}w_0^2\sin\theta\right]$.\\

\subsection{Horizontal pressure on railway lines}

Let $M$ be the mass of a train ($OA$) moving on a straight railway track in a direction $\theta$ east of south as shown in the figure. Let $R$ and $S$ denote the horizontal pressure of the rails (measured in the sense of $\theta$ north of east) and the vertical reaction.\\

\begin{wrapfigure}[12]{r}{0.3\textwidth}
	\hfill	\includegraphics[height=4.5 cm , width=4.5 cm ]{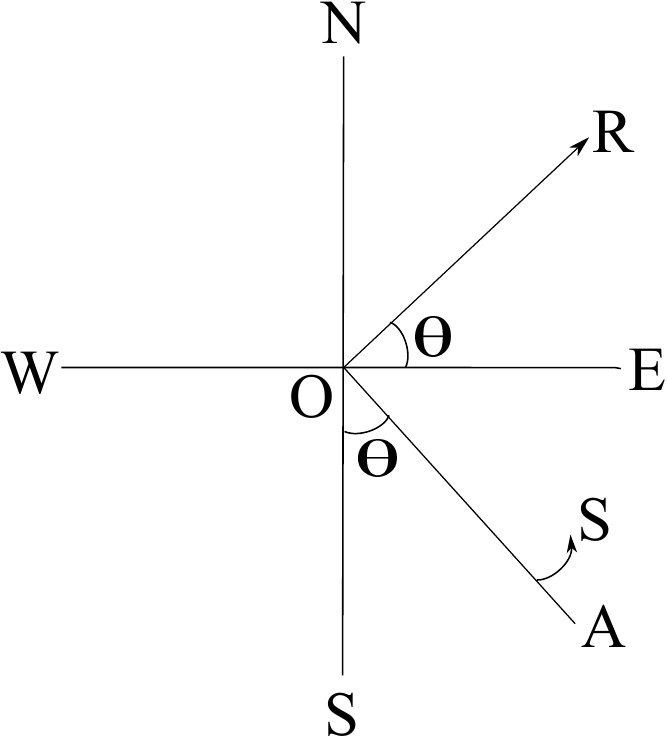}
	\begin{center}
		Fig. 3.58
	\end{center}
\end{wrapfigure}

Then the equations of motion are
\begin{eqnarray}
	&&M(\ddot{x}-2\omega\dot{y}\sin\lambda-\omega^2x\sin^2\lambda)=-R\sin\theta\nonumber\\&&M(\ddot{y}+2\omega \dot{x}\sin\lambda-\omega^2y)=R\cos\theta\nonumber\\\mbox{and}&&M(-2\omega\dot{y}\cos\lambda-\omega^2x\sin\lambda\cos\lambda)=S-mg\nonumber
\end{eqnarray}

If $V$ be the uniform velocity of the train relative to the rail then we have
$$\ddot{x}=0=\ddot{y} \mbox{ ~and~ } \dot{x}=V\cos\theta~,~~\dot{y}=V\sin\theta$$

So at $x=0$, $y=0$ we have
\begin{eqnarray}
	R\sin\theta&=&2M\omega V\sin\theta\sin\lambda\nonumber\\R\cos\theta&=&2M\omega V\cos\theta\sin\lambda\nonumber\\\mbox{i.e.,~ }R&=&2M\omega v\sin\lambda\nonumber
\end{eqnarray}
gives pressure on the rails to the right in the direction of motion.\\

 Also $S=Mg-2M\omega V\sin\theta\cos\lambda$ gives the vertical reaction of the rail.\\

{\bf 36. } A particle is projected in latitude $\lambda$ with velocity $V$ at an elevation $\gamma$ in a direction $\theta$ east of south, Prove that due to earth's rotation with angular velocity $\omega$ the time of flight is increased by $2VT\left(\dfrac{\omega}{g}\right)\cos\lambda\cos\gamma\sin\theta$, and that the particle falls at a distance $\dfrac{4\omega\sin^2\gamma}{g^2}V^3\left(\cos\gamma\sin\lambda+\dfrac{1}{3}\sin\gamma\cos\lambda\cos\theta\right)$ to the right of the vertical plane of projection and that the range is increased by $4\omega\sin\gamma\cos\lambda\cos\theta\left(\cos^2\gamma-\dfrac{1}{3}\sin^2\gamma\right)\dfrac{V^3}{g^2}$ (approximately).\\

{\bf Solution: } In \ref{projectile}, we shall have to choose $U=V\cos\gamma$, $w_0=V\sin\gamma$, $u_0=U\cos\theta=V\cos\theta\cos\gamma$, $v_0=U\sin\theta=V\cos\gamma\sin\theta$.\\

When Earth's rotation is considered then time of flight $$T'=\frac{2w_0}{g}+\frac{4w_0v_0\omega}{g^2}\cos\lambda$$

$T=\dfrac{2w_0}{g}$ = time of flight when there is no rotation of earth.\\

Thus increase in time of flight = $\dfrac{4w_0v_0\omega}{g^2}\cos\lambda=\dfrac{2TV\omega}{g}\cos\gamma\cos\lambda\sin\theta$.\\

The amount of deviation to the right of the vertical plane of projection
\begin{eqnarray}
	&=&\frac{4\omega w_0^2}{g^2}\left(U\sin\lambda+\frac{1}{3}w_0\cos\lambda\cos\theta\right)\nonumber\\&=&\frac{4\omega V^2\sin^2\gamma}{g^2}\left(V\cos\gamma\sin\lambda+\frac{1}{3}V\sin\gamma\cos\lambda\cos\theta\right)\nonumber\\&=&\frac{4\omega V^3\sin^2\gamma}{g^2}\left(\cos\gamma\sin\lambda+\frac{1}{3}\sin\gamma\cos\lambda\cos\theta\right)\nonumber
\end{eqnarray}

Increase in the range
\begin{eqnarray}
	&=&\frac{4\omega w_0\cos\lambda}{g^2}\left[v_0U-\frac{1}{3}w_0^2\sin\theta\right]\nonumber\\&=&\frac{4\omega V\sin\gamma\cos\lambda}{g^2}\left[V^2\cos^2\gamma\sin\theta-\frac{1}{3}V^2\sin^2\gamma\sin\theta\right]\nonumber\\&=&\frac{4\omega V^3}{g^2}\sin\gamma\cos\lambda\sin\theta\left(\cos^2\gamma-\frac{1}{3}\sin^2\gamma\right)\nonumber
\end{eqnarray}

{\bf 37. } A shot is fired from a point on the earth's surface, the angle between the meridian and the plane of projection being $\theta$, measured from north to east. Show that, when the shot again reaches the earth's surface, its deviation from the plane of projection, due to the rotation of the earth, is to be the right or to the left according as $$3\tan\lambda-\tan\gamma\cos\theta$$
is positive or negative, where $\lambda$ is the latitude and $\gamma$ is the angle of projection. The square of the angular velocity of the earth is neglected, and the earth's surface within range is taken to be plane.\\

{\bf Solution: } In the previous problem, $\theta\rightarrow\pi-\theta$

So deviation = $\dfrac{4\omega w_0^2}{g^2}\left(U\sin\lambda+\dfrac{1}{3}w_0\cos\lambda\cos\theta\right)$.

Now $U=V\cos\gamma$, $w_0=V\sin\gamma$,
\begin{eqnarray}
	\therefore~\mbox{deviation}&=&\frac{4\omega V^2\sin^2\gamma}{g^2}\left(V\cos\gamma\sin\lambda+\frac{1}{3}V\sin\gamma\cos\lambda\cos(\pi-\theta)\right)\nonumber\\&=&\frac{4\omega V^3\sin^2\gamma}{g^2}\left(\cos\gamma\sin\lambda-\frac{1}{3}\sin\gamma\cos\lambda\cos\theta\right)\nonumber
\end{eqnarray}

Hence deviation will be to the right if
\begin{eqnarray}
	&&\cos\gamma\sin\lambda-\frac{1}{3}\sin\gamma\cos\lambda\cos\theta>0\nonumber\\\mbox{i.e.,}&&3\tan\lambda-\tan\gamma\cos\theta>0\nonumber
\end{eqnarray}
otherwise the deviation is to the left.\\

\section{Foucault's Pendulum}

\begin{wrapfigure}[12]{r}{0.3\textwidth}
	\hfill	\includegraphics[height=4.2 cm , width=4.5 cm ]{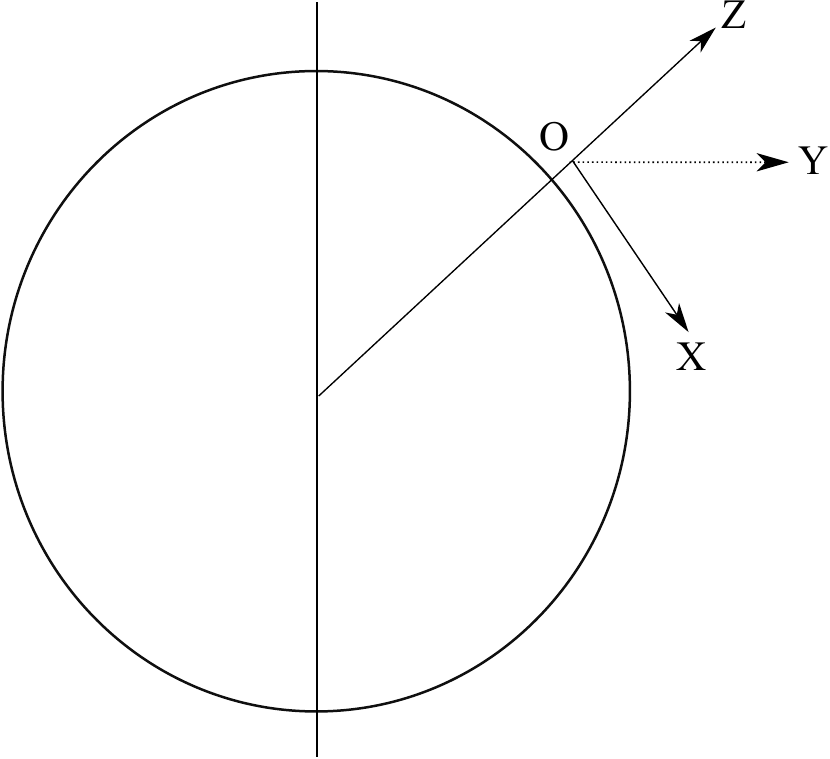}
	\begin{center}
		Fig. 3.59
	\end{center}
\end{wrapfigure}

Let a pendulum be set up at earth's north pole. If started properly it may vibrate as a simple pendulum in a vertical plane. Since the earth under the pendulum turns around the axis with angular velocity $\omega$, the vertical plane of vibration appears to an observer at the surface of the earth to turn around the vertical with angular velocity $-\omega$. Foucault was the first to point out that a pendulum could be used to demonstrate earth's rotation. It is not necessary that the pendulum should be placed at the pole but the rotation of the vertical plane of oscillation of the pendulum may be observed from any place except on the equator.\\

\begin{wrapfigure}[12]{l}{0.3\textwidth}
\centering	\includegraphics[height=4.2 cm , width=4.5 cm ]{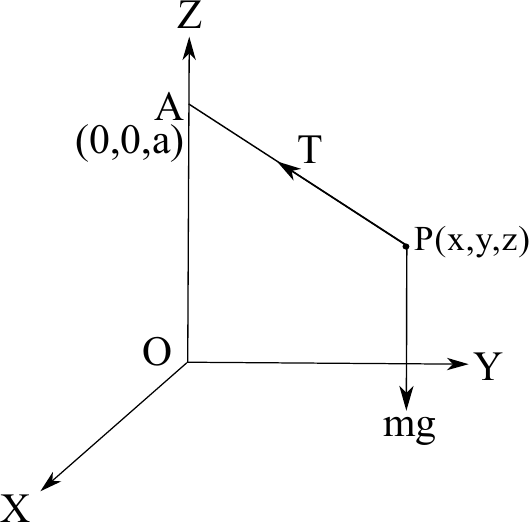}
	\begin{center}
		Fig. 3.60
	\end{center}
\end{wrapfigure}

Let the origin be taken at $O$. a point on or near the surface of earth. The axes $X$, $Y$ and $Z$ are taken along the south, east and vertically upwards respectively. Let $\lambda$ be the geographical latitude of the place $O$. Let the pendulum consists of a particle of mass $m$ attached by light string of length $a$ from a point $(0,0,a)$ so that in equilibrium position the particle is at $O$. We shall discuss small oscillation of the pendulum about this position considering rotation of earth. Let $AP=a$, so the direction cosine of $PA$ are $\left(-\dfrac{x}{a},-\dfrac{y}{a},\dfrac{a-z}{a}\right)$. The focus acting on the particle are (i) apparent force of gravity acting vertically downward, (ii) tension $T$ in the string. Now components of $T$ along the axes are $$-T\dfrac{x}{a}~,~-T\dfrac{y}{a}~,~T\dfrac{a-z}{a}$$

We shall now consider two separate approximations namely (i) smallness of $\omega$ and (ii) smallness of oscillation. Due to first approximation we shall neglect $\omega^2$ and higher powers while second approximation tells us that $x$, $y$ and their derivatives are small so that $z$ and its derivatives are small quantities of second order and hence will be neglected.\\

The equation of motion of the particle are
\begin{eqnarray}
	&&m(\ddot{x}-2\omega\dot{y}\sin\lambda)=-T\frac{x}{a}\nonumber\\&&m(\ddot{y}+2\omega \dot{x}\sin\lambda)=-T\frac{y}{a}\nonumber\\\mbox{and}&&m(-2\omega\dot{y}\cos\lambda)=-mg+T\frac{(a-z)}{a}\simeq T-mg\nonumber\mbox{ ~~(neglecting $z$)}
\end{eqnarray}

So $T=mg-2m\omega\dot{y}\cos\lambda$\\

Using this expression for $T$ into the equations of motion along $X$ and $Y$ directions give
\begin{eqnarray}
	&&\ddot{x}-2\omega\dot{y}\sin\lambda=-g\frac{x}{a}\label{eq3.118}\\&&\ddot{y}+2\omega \dot{x}\sin\lambda=-g\frac{y}{a}\label{eq3.119}
\end{eqnarray}

Now, (\ref{eq3.118}) + $i\times$ (\ref{eq3.119})  gives $$\ddot{\xi}+2i\omega\sin\lambda\dot{\xi}+p^2\xi=0$$
with $\xi=x+iy$, $p^2=\dfrac{g}{a}$.\\

The solution takes the form (neglecting $\omega^2$)
$$\xi=Ae^{-i\omega t\sin\lambda+ipt}+Be^{-i\omega t\sin\lambda-ipt}$$
where the complex constants $A$ and $B$ depend on the initial conditions. Also the above solution can be written as $$(x+iy)e^{i\omega t\sin\lambda}=Ae^{ipt}+Be^{-ipt}$$

\begin{wrapfigure}[12]{r}{0.3\textwidth}
	\hfill	\includegraphics[height=5.2 cm , width=4 cm ]{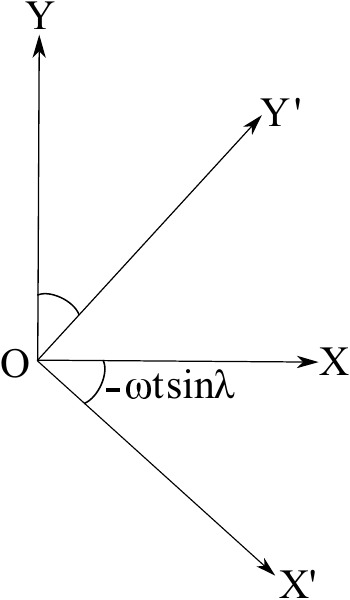}
	\begin{center}
		Fig. 3.61
	\end{center}
\end{wrapfigure}

Let us now choose $OX'$, $OY'$ in the plane $XOY$ such that the axis $OX'$ makes an angle $-\omega t\sin\lambda$ with $OX$ at time $t$. Thus the set of axes $(OX,OY')$  is rotating about $OZ$ with angular velocity $-\omega\sin\lambda$. If $(x',y',z')$ be the position of the particle at time $t$ referred to $OX'$, $OY'$, $OZ$ axes then 
\begin{eqnarray}
	x&=&x'\cos(\omega t\sin\lambda)+y'\sin(\omega t\sin\lambda)\nonumber\\y&=&-x'\sin(\omega t\sin\lambda)+y'\cos(\omega t\sin\lambda)\nonumber
\end{eqnarray}

Thus \begin{eqnarray}
	(x+iy)&=&(x'+iy')\cos(\omega t\sin\lambda)-i(x'+iy')\sin(\omega t\sin\lambda)\nonumber\\&=&(x'+iy')e^{-i\omega t\sin\lambda}\nonumber\\\mbox{i.e., }(x'+iy')&=&(x+iy)e^{i\omega t\sin\lambda}=Ae^{ipt}+Be^{-ipt}\nonumber
\end{eqnarray}

Suppose the particle was initially oscillating in the plane $ZOX$, i.e., the initial position of the plane $ZOX'$. We also assume that the particle was initially pulled to a distance $x_0$ and then let go, so that initially $x'=x_0$, $y'=0$, $\dot{x}'=0$, $\dot{y}'=0$ at $t=0$.\\

Hence $$(x'+iy')=(A_1+iA_2)e^{ipt}+(B_1+iB_2)e^{-ipt}$$
 where $A_i$'s and $B_i$'s $(i=1,2)$ are real constants.\\
 
  Now using the initial conditions
 \begin{eqnarray}
 	x_0=(A_1+B_1)+i(A_2+B_2)\nonumber\\0=(A_1-B_1)+i(A_2-B_2)\nonumber\\\mbox{i.e.,~ }A_1=\frac{x_0}{2}=B_1~,~~A_2=0=B_2\nonumber
 \end{eqnarray} 

So, \begin{eqnarray}
	(x'+iy')&=&\frac{x_0}{2}\left(e^{ipt}+e^{-ipt}\right)=x_0\cos pt\nonumber\\\mbox{i.e.,}&&x'=x_0\cos pt~, ~~y'=0\nonumber
\end{eqnarray}

This shows that the motion of the particle in the particle plane $ZOX'$ is simple harmonic and the vertical plane is rotating about the vertical at $O$ with angular velocity $\omega\sin\lambda$ from east to west through south.\\

Further, as the component of the angular velocity of Earth about the vertical through $O$ is $\omega\sin\lambda$ from west to east through south so that the vertical plane of oscillation of the pendulum turns round the vertical through $O$ with an angular velocity equal and opposite to that of Earth. Equivalently, the vertical plane of oscillation is fixed in space and earth is rotating wih an angular velocity $\omega\sin\lambda$ about the vertical through $O$.\\

However, in general, $$(x'+iy')=(A_1+iA_2)e^{ipt}+(B_1+iB_2)e^{-ipt}$$

Now, separating real and imaginary parts we can show that the projection of the path of the particle on the $X'Y'$ plane is an ellipse and this ellipse is rotating about the vertical with angular velocity $\omega\sin\lambda$.\\

\section {Nonlinear oscillations} 

Many important physical problems have evolution equation of the following general form:
\begin{equation}\label{eq3.120}
	\ddot{x}+\mu f(x,\dot{x})+\omega^2x=0
\end{equation}
 where $\mu$ is a small parameter. This type of equation is known as quasi-linear differential equation. There is no general solution of such differential equation -- only approximate solutions can be obtained for such type of differential equation. In the context of nonlinear oscillation we shall discuss the solution of (\ref{eq3.120}) using the method of Krylov and Bogoliubov.\\
 
Note that equation (\ref{eq3.120}) reduces to simple harmonic oscillation: $\ddot{x}+\omega^2x=0$ when $\mu=0$. So for $\mu=0$ the solution can be taken as
\begin{equation}\label{eq3.121}
	x=a\sin(\omega t+\phi)
\end{equation}
 where $a$ and $\phi$ are arbitrary constants.\\
 
 We now assume that the solution of equation (\ref{eq3.120}) is of the form (\ref{eq3.121}) with $a=a(t)$ and $\phi=\phi(t)$ i.e.,
 \begin{equation}\label{eq3.122}
 	x(t)=a(t)\sin\{\omega t+\phi(t)\}
 \end{equation}
 as well as the first order derivatives has the form
 \begin{equation}\label{eq3.123}
 	\dot{x}(t)=a\omega\cos(\omega t+\phi)
 \end{equation}

Now differentiating (\ref{eq3.122}) with respect to $t$ gives
\begin{equation}
	\dot{x}(t)=\dot{a}(t)\sin\{\omega t+\phi(t)\}+a(t)\omega\cos\{\omega t+\phi(t)\}+a(t)\dot{\phi}(t)\cos\{\omega t+\phi(t)\}\nonumber
\end{equation}

Now comparing with equation (\ref{eq3.123}) we get
\begin{equation}\label{eq3.124}
	\dot{a}\sin\{\omega t+\phi\}+a\dot{\phi}\cos\{\omega t+\phi\}=0
\end{equation}

Also differentiating (\ref{eq3.123}) gives
\begin{equation}
		\ddot{x}=\dot{a}\omega\cos\{\omega t+\phi\}-a\omega^2\sin\{\omega t+\phi\}-a(t)\omega\dot{\phi}\sin\{\omega t+\phi\}\nonumber
\end{equation}

Substituting $\theta=\omega t+\phi$ we obtain
\begin{eqnarray}\label{eq3.125}
	x=a\sin\theta~,~~\dot{x}=a\omega\cos\theta~,~~\ddot{x}=\dot{a}\omega\cos\theta-a\omega^2\sin\theta-a\omega\dot{\phi}\sin\theta
\end{eqnarray}

Using (\ref{eq3.125}) in (\ref{eq3.120}) gives
\begin{equation}\label{eq3.126}
	\dot{a}\omega\cos\theta-a\omega\dot{\phi}\sin\theta=-\mu f(a\sin\theta,a\omega\cos\theta)
\end{equation}

Now solving for $\dot{a}$ and $\dot{\phi}$ from equations (\ref{eq3.124}) and (\ref{eq3.126}) we have
\begin{eqnarray}
	\dot{a}&=&-\frac{\mu}{\omega} f(a\sin\theta,a\omega\cos\theta)\cos\theta\label{eq3.127}\\\dot{\phi}&=&\frac{\mu}{a\omega} f(a\sin\theta,a\omega\cos\theta)\sin\theta\label{eq3.128}
\end{eqnarray}
which are nothing but the time variation of amplitude (i.e., $a$) and phase (i.e., $\phi$) with time. Note that till now we have not made any approximation so equations (\ref{eq3.127}) and \ref{eq3.128} are exact in nature. As $\mu$ is small so $\dot{a}$ and $\dot{\phi}$ are small i.e., $a$ and $\phi$ are slowly varying function of $t$. Also in time $\dfrac{2\pi}{\omega}$, $\theta=\omega t+\phi$ will increase by an amount $2\pi$ approximately without appreciable change in $a$ and $\phi$.\\ 

According to Krylov and Bogoliubov, the right hand side of equations (\ref{eq3.127}) and (\ref{eq3.128}) can be replaced by their average values over the range of $\theta$ from $)$ to $2\pi$ as their first approximation, considering the amplitude $a$ as constant. Hence equations (\ref{eq3.127}) and (\ref{eq3.128}) become (in the first approximation)
\begin{eqnarray}
	\dot{a}&=&-\frac{\mu}{2\pi\omega}\int_0^{2\pi} f(a\sin\theta,a\omega\cos\theta)\cos\theta\mathrm{d}\theta\label{eq3.129}\\\dot{\phi}&=&\frac{\mu}{2\pi a\omega}\int_0^{2\pi} f(a\sin\theta,a\omega\cos\theta)\sin\theta\mathrm{d}\theta\label{eq3.130}
\end{eqnarray}

Thus equations (\ref{eq3.129}) and (\ref{eq3.130}) represent the general manner in which the amplitude $a$ and phase $\phi$ vary for various choices of $f(x,\dot{x})$.\\

As a particular case, if $f(x,\dot{x})=x^n$ then form (\ref{eq3.129}) it follows that $\dot{a}=0$ and $\dot{\phi}\neq0$ from equation (\ref{eq3.130}). Hence amplitude is constant But phase  varies. But reverse is the situation if $f(x,\dot{x})=\dot{x}^n$ i.e., $\phi$ is constant while the amplitude varies.\\

{\bf Example 1:} $$\ddot{x}+\mu x^3+\omega^2x=0\mbox{ , ~~$\mu$ is small}$$

We shall determine approximate solution of this equation using the method of Krylov and Bogoliubov. The approximate solution is given by
$$x=a(t)\sin\theta(t)~,~~\theta(t)=\omega t+\phi(t)$$ 
where the amplitude $a$ and phase $\phi$ change with time as
\begin{eqnarray}
	\dfrac{\mathrm{d}a}{\mathrm{d}t}&=&-\frac{\mu}{2\pi\omega}\int_0^{2\pi} f(a\sin\theta,a\omega\cos\theta)\cos\theta\mathrm{d}\theta\nonumber\\\dfrac{\mathrm{d}\phi}{\mathrm{d}t}&=&\frac{\mu}{2\pi a\omega}\int_0^{2\pi} f(a\sin\theta,a\omega\cos\theta)\sin\theta\mathrm{d}\theta\nonumber
\end{eqnarray}

In the present problem, $f(x,\dot{x})=x^3=a^3\sin^3\theta$, hence
\begin{eqnarray}
	\dfrac{\mathrm{d}a}{\mathrm{d}t}&=&-\frac{\mu a^3}{2\pi\omega}\int_0^{2\pi} \sin^3\theta\cos\theta\mathrm{d}\theta=0\nonumber\\\dfrac{\mathrm{d}\phi}{\mathrm{d}t}&=&\frac{\mu a^2}{2\pi \omega}\int_0^{2\pi} \sin^4\theta\mathrm{d}\theta=\frac{\mu a^2}{2\pi \omega}\cdot4\cdot\frac{3}{4}\cdot\frac{1}{2}\cdot\frac{\pi}{2}\nonumber\\&=&\frac{3\mu a^2}{8\omega}\nonumber\\\mbox{i.e., ~}\phi&=&\frac{3\mu a^2}{8\omega}t+\phi_0\mbox{~~(assuming $a$ remains constant and $\phi=\phi_0$ at $t=0$)}\nonumber\\\therefore~x&=&a\sin(\omega t+\phi)=a\sin\left\{\omega t+\frac{3\mu a^2}{8\omega}t+\phi_0\right\}\nonumber\\&=&a\sin\left\{\left(\omega +\frac{3\mu a^2}{8\omega}\right)t+\phi_0\right\}\nonumber
\end{eqnarray}

So the period of oscillation is $$\frac{2\pi}{\omega\left(1+\frac{3\mu a^2}{8\omega^2}\right)}=\frac{2\pi}{\omega}\left(1-\frac{3\mu a^2}{8\omega^2}\right)$$ 

Thus if $\mu>0$, the period decreases with the increase of amplitude $a$.\\ 

{\bf Example 2:} $$\ddot{x}+\mu (1-x^2)\dot{x}+x=0$$

Here $\ddot{x}+x=0$ has solution $$x=a\sin\theta~,~~\theta=t+\phi$$  

So the time variation of $a$ and $\phi$ are given by
\begin{eqnarray}
	\dfrac{\mathrm{d}a}{\mathrm{d}t}&=&-\frac{\mu}{2\pi}\int_0^{2\pi} (a^2\sin^2\theta-1)a\cos^2\theta\mathrm{d}\theta\nonumber\\&=&-\frac{\mu a^3}{2\pi}\int_0^{2\pi} \sin^2\theta\cos^2\theta\mathrm{d}\theta+\frac{\mu a}{2\pi}\int_0^{2\pi} \cos^2\theta\mathrm{d}\theta\nonumber\\&=&-\frac{\mu a^3}{2\pi}\cdot4\frac{\Gamma\left(\frac{3}{2}\right)~\Gamma\left(\frac{1}{2}\right)}{2\Gamma(3)}+\frac{\mu a}{2\pi}\frac{2\pi}{2}=\frac{\mu a}{2}\left(1-\frac{a^2}{4}\right)\nonumber\\\dfrac{\mathrm{d}\phi}{\mathrm{d}t}&=&\frac{\mu }{2\pi a}\int_0^{2\pi} (a^2\sin^2\theta-1)a\sin\theta\cos\theta\mathrm{d}\theta\nonumber\\&=&\frac{\mu a^2}{2\pi}\int_0^{2\pi} \sin^3\theta\cos\theta\mathrm{d}\theta-\frac{\mu}{2\pi}\int_0^{2\pi} \sin\theta\cos\theta\mathrm{d}\theta=0\nonumber\\\therefore~\phi=\phi_0~, \mbox{ a constant.}\nonumber\\\mbox{Also }\dfrac{\mathrm{d}a}{\mathrm{d}t}&=&\frac{\mu a}{2}\left(1-\frac{a^2}{4}\right)\nonumber\\\implies\int\frac{\mathrm{d}a}{a\left(1-\frac{a^2}{4}\right)}&=&\int\frac{\mu}{2}\mathrm{d}t\nonumber\\\mbox{i.e., ~}\frac{2a}{\sqrt{4-a^2}}&=&e^{\left(\frac{\mu}{2}t+c\right)}\nonumber\\\mbox{i.e.,~ } a^2&=&\frac{4e^{\left(\frac{\mu}{2}t+c\right)}}{4+e^{\left(\frac{\mu}{2}t+c\right)}}\nonumber\rightarrow\mbox{ this gives the explicit time dependence of }a.
\end{eqnarray}

Note that $a\rightarrow 2$ as $t\rightarrow\infty$. Also $\dfrac{\mathrm{d}a}{\mathrm{d}t}=0$ for $a=2$. Hence the oscillatory motion is steady i.e., a steady state is reached after infinite time. Thus the solution is $$x(t)=a(t)\sin(t+\phi_0)$$

{\bf Example 3:} $$\ddot{x}+\epsilon \dot{x}|\dot{x}|+x=0$$

For $\epsilon=0$ the solution is $$x=a\sin\theta~,~~\dot{x}=a\cos\theta~,~~\theta=t+\phi$$ 

Here $$f(x,\dot{x})=\dot{x}|\dot{x}|=\left\{\begin{array}{cl}
	\dot{x}^2&,~\dot{x}>0\\-\dot{x}^2&,~\dot{x}<0
\end{array}\right.$$

\begin{eqnarray}
	\dfrac{\mathrm{d}a}{\mathrm{d}t}&=&-\frac{\mu}{2\pi}\int_0^{2\pi}f(x,\dot{x})\cos\theta\mathrm{d}\theta\nonumber\\&=&-\frac{\epsilon}{2\pi}\left[\int_0^{\frac{\pi}{2}}a^2\cos^2\theta\cos\theta\mathrm{d}\theta-\int_{\frac{\pi}{2}}^{\frac{3\pi}{2}}a^2\cos^2\theta\cos\theta\mathrm{d}\theta+\int_{\frac{3\pi}{2}}^{2\pi}a^2\cos^2\theta\cos\theta\mathrm{d}\theta\right]\nonumber\\&=&-\frac{\epsilon a^2}{2\pi}\left[\frac{2}{3}+\frac{4}{3}+\frac{2}{3}\right]=-\frac{4\epsilon a^2}{3\pi}\nonumber\\\mbox{i.e., ~}\frac{1}{a}&=&\frac{4\epsilon}{3\pi}\frac{1}{t}+\frac{1}{a_0}\nonumber
\end{eqnarray} 

As before $\dfrac{\mathrm{d}\phi}{\mathrm{d}t}=0$. Hence $$x=a\sin(t+\phi_0)~,~~a=\left(\frac{4\epsilon}{3\pi}\frac{1}{t}+\frac{1}{a_0}\right)^{-1}.$$

{\bf Example 4:} $$\ddot{x}-2k\dot{x}+c\dot{x}^3+\omega^2x=0~,~~k,c \mbox{ are small positive constants}$$

So solution will be of the form $$x=a(t)\sin\theta~,~~\dot{x}=a\omega\cos\theta~,~~\theta=\omega t+\phi(t)$$
\begin{eqnarray}
	\dfrac{\mathrm{d}a}{\mathrm{d}t}&=&-\frac{c}{2\pi\omega}\int_0^{2\pi}\left(a^3\omega^3\cos^3\theta-\frac{2k}{c}a\omega\cos\theta\right)\cos\theta\mathrm{d}\theta\nonumber\\&=&-\frac{ca^3\omega^2}{2\pi}\frac{3\pi}{4}+\frac{ka}{\pi}\pi=-\frac{3a^3c\omega^2}{8}+ka\nonumber\\\dfrac{\mathrm{d}\phi}{\mathrm{d}t}=0\nonumber\\\therefore~\int\frac{\mathrm{d}a}{a\left(k-\frac{3}{8}a^2c\omega^2\right)}&=&\int\mathrm{d}t\nonumber\\\mbox{i.e., ~}\frac{k}{a^2}-\frac{3}{8}c\omega^2&=&a_0e^{-2t}\nonumber\\a&=&\frac{1}{\sqrt{k}}\left(a_0e^{-2t}+\frac{3}{8}c\omega^2\right)^{-\frac{1}{2}}\nonumber
\end{eqnarray}

Hence $a\rightarrow \dfrac{2\sqrt{2}}{\sqrt{3kc}}\dfrac{1}{\omega}$ as $t\rightarrow\infty$.\\

Hence there will be a steady motion after infinite time.\\

\section{Problem of three bodies: (three body problem)} 

We consider the motion of three bodies having masses $m_1$, $m_2$ and $m_3$. Suppose they are in motion under their mutual gravitational attraction. Let $(x_1,y_1,z_1)$, $(x_2,y_2,z_2)$ and $(x_3,y_3,z_3)$ be the coordinates of the bodies (i.e., coordinates of the centre of gravity of the bodies) at any time. The distances between the masses i.e., $(m_1,m_2)$, $(m_1,m_3)$ and $(m_2,m_3)$ are chosen as $r_{12}$, $r_{13}$ and $r_{23}$ respectively. So we have
$$r_{ij}^2=(x_i-x_j)^2+(y_i-y_j)^2+(z_i-z_j)^2~, ~~i,j=1,2,3$$

The potential energy of the masses $m_2$ and $m_3$ at the position of $m_1$ is 
$$V_1=G\frac{m_2}{r_{12}}+G\frac{m_3}{r_{13}}$$
 where $G$ is the gravitational constant. So the equations of motion of mass $m_1$ are 
 $$\frac{\mathrm{d}^2x_1}{\mathrm{d}t^2}=\frac{\partial V_1}{\partial x_1}~,~~\frac{\mathrm{d}^2y_1}{\mathrm{d}t^2}=\frac{\partial V_1}{\partial y_1}~,~~\frac{\mathrm{d}^2z_1}{\mathrm{d}t^2}=\frac{\partial V_1}{\partial z_1}$$
 
Note that if we define  $V=G\left(\dfrac{m_1m_2}{r_{12}}+\dfrac{m_2m_3}{r_{23}}+\dfrac{m_3m_1}{r_{31}}\right)$
  then in the above equations of motion we can replace $V_1$ by $V$ and similarly we can write down the equations of motion for the masses $m_2$ and $m_3$. In particular, the equations of motion of these three masses can be written in compact form as 
  $$\frac{\mathrm{d}^2x_i}{\mathrm{d}t^2}=\frac{\partial V}{\partial x_i}~,~~\frac{\mathrm{d}^2y_i}{\mathrm{d}t^2}=\frac{\partial V_1}{\partial y_i}~,~~\frac{\mathrm{d}^2z_i}{\mathrm{d}t^2}=\frac{\partial V_1}{\partial z_i}~,~~i=1,2,3$$
 
Since there are no external forces acting on the system except their mutual gravitational attraction, so the centre of mass (c.m) of the system moves uniformly and we write 
$$\sum\limits_{i=1}^3m_i\frac{\mathrm{d}x_i}{\mathrm{d}t}=a_x~,~~\sum\limits_{i=1}^3m_i\frac{\mathrm{d}y_i}{\mathrm{d}t}=a_y~,~~\sum\limits_{i=1}^3m_i\frac{\mathrm{d}z_i}{\mathrm{d}t}=a_z$$
 where $(a_x,a_y,a_z)$ are constants.\\
 
Now integrating the above equations of motion of the centre of mass we obtain 
$$\sum\limits_{i=1}^3m_ix_i=a_xt+b_x~,~~\sum\limits_{i=1}^3m_iy_i=a_yt+b_y~,~~\sum\limits_{i=1}^3m_iz_i=a_zt+b_z$$
 
 These are termed as integrals of the centre of mass of the system. Since the moments of the forces acting on the particles about the co-ordinate axes vanish therefore the angular momentum of the system about the coordinate axes remains constant. Thus we have 
 $$\sum\limits_{i=1}^3m_i\left(y_i\frac{\partial z_i}{\partial t}-z_i\frac{\partial y_i}{\partial t}\right)=c,$$ 
 and similar two equations. These are the integrals of the angular momentum of the system of bodies. The kinetic energy of the system is $$T=\frac{1}{2}\sum\limits_{i=1}^3m_i\left\{\left(\frac{\mathrm{d}x_i}{\mathrm{d}t}\right)^2+\left(\frac{\mathrm{d}y_i}{\mathrm{d}t}\right)^2+\left(\frac{\mathrm{d}z_i}{\mathrm{d}t}\right)^2\right\}$$
 
There do not exist any other integrals of the general three body problem.\\
 
 In terms of Polar coordinates the position of the centre of gravity of the bodies can be written as $(r_1,\theta_1)$, $(r_2,\theta_2)$ and $(r_3,\theta_3)$. Then the kinetic energy is given by $$T=\frac{1}{2}m_1\left(\dot{r}_1^2+r_1^2\dot{\theta}_1^2\right)+\frac{1}{2}m_2\left(\dot{r}_2^2+r_2^2\left(\dot{\theta}_1^2+\dot{\theta}_2^2\right)\right)+\frac{1}{2}m_3\left(\dot{r}_3^2+r_3^2\left(\dot{\theta}_1^2+\dot{\theta}_3^2\right)\right)$$ and $$V=\frac{m_2m_3}{\sqrt{r_2^2+r_3^2-2r_2r_3\cos(\theta_2-\theta_3)}}+\frac{m_3m_1}{\sqrt{r_3^2+r_1^2-2r_3r_1\cos\theta_3}}+\frac{m_1m_2}{\sqrt{r_1^2+r_2^2-2r_1r_2\cos\theta_2}}$$
 
 Note that here $\theta_1$ is a cyclic co-ordinate i.e., $$\frac{\partial L}{\partial\theta_1}=0$$
 
Hence the three bodies move in a plane like a single particle with 2 degrees of freedom.


\chapter{Lagrangian and Hamiltonian formulation}

\section{Constraint system}
 
 In nature, there is no absolute concept of free motion. Every dynamical system is restricted by either a geometrical or a physical condition. Such a condition (or a restriction) is called a constraint and the force which gives rise this constraint is called constraint force. In fact, any force acting on constraint surface is called a constraint force.
 
 Consider a system of particles $P_k~(k=1,2,3,...,n.)$ referred to contain frame of reference. $\overrightarrow{r_k}$ is the radius vector drawn to the k-th particle and $\dot{\overrightarrow{r_k}}=\overrightarrow{v_k}$, is the velocity of the k-th particle (`$.$' indicates differentiation with respect to time).
 
 Let there be some constraints imposed on the particles as to their position and velocity. A system moving under no constraint is called a free system. Analytically, constraints are of the form 
 \begin{equation}
 	f(\overrightarrow{r_k},\dot{\overrightarrow{r_k}},f)=0\label{1}
 \end{equation}   
 (Note that this scalar equation has $6n+1$ variables $\overrightarrow{r_k}=(x_k,y_k,z_k)$)
 
 If the constraint equation is independent of the velocity then it is known as finite or geometric constraint. So $f(\overrightarrow{r_k},t)=0$ is an example of these constraints. Usually, constraints of the type (\ref{1}) are known as differential or kinematic constraints. We shall consider those differential constraints which are linear in velocities i.e.,
 \begin{equation}
 	l_1\dot{\overrightarrow{r_1}}+l_2\dot{\overrightarrow{r_2}}+...+l_n\dot{\overrightarrow{r_n}}+D=0\label{2}
 \end{equation}
 Here $l_1,l_2,...,l_3,D$ are scalar functions of $t$ and $\dot{\overrightarrow{r}}$. 
 
 It is to be noted that a finite constraint imposes restrictions as to the position of the system and time while with a differential constraint the system may occupy any arbitrary position in space and in time $'t'$ but in this position the velocities can not be arbitrary and there will be restriction on the velocities.
 
 A finite constraint like $f(\overrightarrow{r_k},t)=0$ implies a differential constraint which is obtained by termwise differentiation as 
 \begin{equation}
 	\Sigma_{k=1}^{n}\frac{\partial f}{\partial r_k}\dot{\overrightarrow{r_k}} +\frac{\partial f}{\partial t}=0\label{3}
 \end{equation}  
 But this differential constraint is equivalent to the finite constraint $f(\overrightarrow{r_k},t)=c$ (which is the original finite constraint if $c=0$ only). The differential constraints (\ref{3}) is called integrable.
 
 The constraints can be classified as follows :
 
 A) : (Scleronomic or Rheonomic)
 
 B) :(Holonomic or Non-holonomic)
 
 C): (Conservative or Dissipative)
 
 D): (Bilateral or Unilateral)
 
 A constraint which does not depend explicitly on time is called a Scleronomic constraint, otherwise it is called a Rheonomic constraint. So for a scleronomic constraint `$f$',~~$\frac{\partial f}{\partial t}=0$. 
 
 So equation (\ref{3}) becomes $~\Sigma_{k=1}^{n}\frac{\partial f}{\partial r_k}\dot{\overrightarrow{r_k}}=0~$, which is linear and homogeneous in velocities. Now comparing this with the differential constraint (\ref{2}) we see that a differential constraint will be stationary if $D=0$ and the vector $\overrightarrow{l}=(l_1,l_2,...,l_3)$ is not an explicit function of `$t$'. 
 
 \vspace{.5cm}
 
 $\bullet${\textbf{Example}} : If we consider the motion of a simple pendulum of constant length then the position of the bob at any time can be characterized by the constraint
 \begin{equation}
 	x^2+y^2+z^2=l^2\label{4}
 \end{equation}
 (l is the constant length of the pendulum)
 
 This constraint is called a scleronomic constraint. But if the length of the pendulum changes with time then the above constraint Still holds with $l=l(t)$. So such a constraint is called a rheonomic constraint.
 
 If a constraint relation does not contain any term depending on velocity then it is called a holonomic constraint, otherwise it s non-holonomic. The constraint in equation (\ref{4}) is an example of  holonomic constraint.
 
 If a dynamical system preserves the total mechanical energy of the system then we say that the dynamical system is in conservative form (here the constraint forces do not work). The non-conservation of the total energy leads a dynamical system to be dissipative (in this case the constraint forces do work).
 
 In principle, a simple pendulum motion is conservative while if its bob slides on a rough circular track, then the system is dissipative (due to friction there will be a loss of energy).
 
 If on the constraint surface, it is possible for the dynamical system to have forward and backward motion then the constraint is called a bilateral constraint. Here the constraints are expressed mathematically in the form of the equations of the type $f(\overrightarrow{r_k},\dot{\overrightarrow{r_k}},t)=0$. On the otherhand, if at some point on the constraint surface the motion is uni-directional then the constrain is called a unilaleral constraint . Usually, the unilateral constraints are expressed in the form of inequalities of the form : $f(r_k,\dot{r_k},t)\geq0$. In the motion of a simple pendulum the constraint surface is the surface of a sphere and the motion is possible in all directions. So the constraint is bilateral. The motion of a gas confined to a spherical container of radius $R$, the constraint is $|r|\leq R$ and hence it is unilateral.
 
 A particle is constrained to move over a surface. Suppose the surface is given by $f(x,y,z)=0$. Then it is a finite stationary constraint.
 
 If the surface is undergoing deformation or is moving , then the equation of the surface will be of the form $f(x,y,z,t)=0$. This is a non-stationary but finite constraint.
 
 The equation of constraint of two particle connected by a rod of constant length is $|\overrightarrow{r_1}-\overrightarrow{r_2}|=l,$   i.e. $(x_1-x_2)^2+(y_1-y_2)^2+(z_1-z_2)^2=l^2$. This is a holonomic, scleronomic system.
 
 But if the length of the rod be variable the equation of constraint will be
 \begin{equation}
 	(x_1-x_2)^2+(y_1-y_2)^2+(z_1-z_2)^2=[f(t)]^2=l^2\nonumber
 \end{equation} 
 This is a holonomic but rheonomic constraint.
 
 Two particles in a plane are connected by a rod of constant length $'l'$ and are constraint to move in a manner such that the velocity of the middle point of the rod is in the direction of the length of the rod. The constraint equations are $z_1=0=z_2~,~(x_1-x_2)^2+(y_1-y_2)^2=l^2$ and $\dfrac{\dot{x_1}+\dot{x_2}}{x_1-x_2}=\dfrac{\dot{y_1}-\dot{y_2}}{y_1-y_2}$.
 
 This is a non-holonomic system because the last equation is a non-integrable differential constraint.
 
 Suppose two particles are connected by a thread and is expressed as $l^2-|r_1-r_2|^2\geq0$ (unilateral constraint). The condition of equality sign is the condition of taut. So a system with unilateral constraint may be regarded in such a way that a part of the constraint refers to the condition of tout i.e., bilateral constraint and the other part is such that there is no such constraint. Thus a unilateral constraint is either a bilateral constraint or is eliminated altogether. Hence one can use only the bilateral constraint.
 
 \vspace{.5cm}
 
 \section{{{Generalised Coordinates} :}}
 
 \vspace{.25cm}

 The least number of independent variables (compatible with the constraints) which can characterise the position and configuration of a dynamical system are called the degrees of freedom of the dynamical system. Also these independent variables (coordinate) are termed as generalised coordinates.
 
 It may be noted that the cartesian co-ordinates are not necessarily be the generalised coordinates but they may be functions of them. In fact, the particular set of independent variables which is suitable for describing a dynamical system is called the generalised coordinate system.
 
 For example, if we consider the motion of a particle revolving around a fixed attracting  centre (central force problem) then the polar coordinates namely $(r,\theta,\phi)$ are generalised coordinates and the form of the cartesian coordinates are $x=r\cos\phi\sin\theta~, ~y=r\sin\phi\sin\theta~,~z=r\cos\theta$ .
 
 \vspace{.5cm}
 
 \section{{{Lagrange's equation of motion}} :}
 
 \vspace{.25cm}
 Suppose a system is defined at the instant $'t'$ by $k$ generalised coordinates $q_1,q_2,...,q_k$. These coordinates $q$ define position of each member of the system at the instant and hence if $\overrightarrow{r}$ be the position vector of a material point of mass $m$
 of the system at time $t$ relative to some base point, then 
 \begin{equation}
 	\overrightarrow{r}=\overrightarrow{r}(q_1,q_2,...,q_k;t)=\overrightarrow{r}(q_k;t)=\overrightarrow{r}(\overrightarrow{q},t)~~j=1,2,...k.\nonumber
 \end{equation}
 We assume that these function $\overrightarrow{r}(\overrightarrow{q},t)$ together with their partial derivatives are continuous in a certain region of $q$ and $t$. We shall now consider the following results :
 
 \vspace{.5cm}
 
 $\bullet$ \textbf{{Lemma I}} :

 \begin{equation}
 	\frac{\partial\dot{\overrightarrow{r}}}{\partial\dot{\overrightarrow{q_j}}}= \frac{\partial\overrightarrow{r}}{\partial q_j}\label{5}
 \end{equation}
 \textbf{Proof} :
 
 As  $\overrightarrow{r}=\overrightarrow{r}(\overrightarrow{q};t)$ so $\dot{\overrightarrow{r}}=\frac{d\overrightarrow{r}}{dt}=\frac{\partial\overrightarrow{r}}{\partial t}+ \Sigma_{j=1}^{k}\frac{\partial\overrightarrow{r}}{\partial q_j}\dot{q_j}$;
 
 Now differentiating partially with respect to $\dot{q_j}$ we have $\frac{\partial\dot{\overrightarrow{r}}}{\partial\dot{\overrightarrow{q_j}}}= \frac{\partial\overrightarrow{r}}{\partial q_j}$ .
 
 \vspace{.5cm}
 
 $\bullet$ \textbf{{Lemma II}} : Show that 
 \begin{equation}
 	\frac{d}{dt}\left(\frac{\partial\overrightarrow{r}}{\partial q_j}\right) =\frac{\partial}{\partial q_j}\left(\frac{d\overrightarrow{r}}{dt}\right)\label{6}
 \end{equation}
 \textbf{Proof} : As $\overrightarrow{r}$ is a function of $q_j$ and $t$ so the partial derivative $\frac{\partial\overrightarrow{r}}{\partial q_j}$ is also function of $q_j$'s and $t$. Hence 
 \begin{align}
 	\frac{d}{dt}\left(\frac{\partial\overrightarrow{r}}{\partial q_j}\right) &=	\frac{\partial}{\partial t}\left(\frac{\partial\overrightarrow{r}}{\partial q_j}\right) +\Sigma_{p=1}^{k}\frac{\partial^2\overrightarrow{r}}{\partial q_j\partial q_p}\dot{q_p}\nonumber\\
 	&=\frac{\partial}{\partial q_j}\left(\frac{\partial\overrightarrow{r}}{\partial t}\right) + \frac{\partial}{\partial q_j}\left(\Sigma\frac{\partial\overrightarrow{r}}{\partial q_p}\dot{q_p}\right)\nonumber\\
 	&=\frac{\partial}{\partial q_j}\left[\frac{\partial\overrightarrow{r}}{\partial t}+ \Sigma_p\frac{\partial\overrightarrow{r}}{\partial q_p}\dot{q_p} \right]\nonumber\\
 	&=\frac{\partial}{\partial q_j}\left(\frac{d\overrightarrow{r}}{dt}\right).\nonumber
 \end{align}
 Let $\overrightarrow{F_i}$ be the external force at time t on the particle $m_i$ of the system having position vector $\overrightarrow{r_i}(\overrightarrow{q},t)$. Then according to D'Alembert's principle, the system of forces $(\overrightarrow{F_i}-m_i\ddot{\overrightarrow{r_i}})$  acting at different points of the system, are in equilibrium. Hence, for an arbitrary virtual displacement of the system consistent with the constraint, the total work done will be zero i.e., 
 \begin{equation}
 	(\overrightarrow{F_i}-m_i\ddot{\overrightarrow{r_i}})\delta\overrightarrow{r_i}=0 \label{7}
 \end{equation}
 (note that the virtual work done by the reactions of the constraints due to any arbitrary virtual displacement consistent with the constraint is zero). Let $\delta w$ be the virtual work done by the forces $\overrightarrow{F_i}$ on the virtual displacement $\delta\overrightarrow{r_i}$, so we have 
 \begin{align}
 	\delta w=\sum_{i=1}\overrightarrow{F_i}.\delta\overrightarrow{r_i}&= \sum_{i=1}\overrightarrow{F_i}.\sum_{j}\frac{\partial\overrightarrow{r_i }}{\partial q_j}\delta q_j\nonumber\\
 	&=\sum_{j}\delta q_j\sum_{i}\frac{\partial\overrightarrow{r_i }}{\partial q_j}\dot{\overrightarrow{F_i}}\nonumber\\
 	&=\sum_{j}\delta q_j.Q_j\label{8}
 \end{align} 
 where $Q_j$ is called the generalised force associated with the generalised coordinate $q_j$ with expression 
 \begin{equation}
 	Q_j=\sum_{i}\overrightarrow{F_i}.\frac{\partial\overrightarrow{r_i}}{\partial q_j}= \frac{\partial w}{\partial q_j}\label{9}
 \end{equation} 
 Now,
 \begin{align}
 	\sum_{i}m_i\ddot{\overrightarrow{r_i}}\delta\overrightarrow{r_i} &=\sum_{i}m_i\ddot{\overrightarrow{r_i}}\Sigma_{j}\frac{\partial\overrightarrow{r_i }}{\partial q_j}\delta q_j\nonumber\\
 	&=\sum_{j}\delta q_j\left[\frac{d}{dt}\left(\sum_{i}m_i\dot{\overrightarrow{r_i}} \frac{\partial\overrightarrow{r_i}}{\partial q_j}\right)-\sum_{i}m_i \dot{\overrightarrow{r_i}}\frac{d}{dt}\left(\frac{\partial \overrightarrow{r_i}}{\partial q_j}\right)\right]\nonumber\\
 	&=\sum_{j}\delta q_j\left[\frac{d}{dt}\left(\sum_{i}m_i\dot{\overrightarrow{r_i}} \frac{\partial\dot{\overrightarrow{r_i}}}{\partial\dot{q_j}}\right)-\sum_{i}m_i \dot{\overrightarrow{r_i}}\frac{\partial\dot{\overrightarrow{r_i}}}{\partial q_j}\right]	\label{10}
 \end{align}
 (using lemma I in the 1st term and lemma II in the 2nd term)
 
 Let $T$ is the kinetic energy of the system i.e., 
 \begin{align}
 	T&=\frac{1}{2}\sum_{i}m_i\dot{\overrightarrow{r_i}}^2\nonumber\\
 	\frac{\partial T}{\partial\dot{q_j}}&=\sum_{i}m_i\dot{\overrightarrow{r_i}}\frac{\partial \dot{\overrightarrow{r_i}}}{\partial\dot{q_j}}\nonumber
 \end{align}
 and
 \begin{equation}
 	\frac{\partial T}{\partial {q_j}}=\sum_{i}m_i\dot{\overrightarrow{r_i}}\frac{\partial \dot{\overrightarrow{r_i}}}{\partial {q_j}}\nonumber
 \end{equation}
 Using these partial derivatives in (\ref{10}) we have
 \begin{equation}
 	\sum_{i}m_i\ddot{\overrightarrow{r_i}}\delta\overrightarrow{r_i}=\sum_{j}\delta q_j \left[\frac{d}{dt}\left(\frac{\partial T}{\partial\dot{q_j}}\right)-\frac{\partial T}{\partial q_j} \right]\label{11}
 \end{equation}
 Now substituting (\ref{8}) and (\ref{11}) in (\ref{7}) we have
 \begin{equation}
 	\sum_{j=1}^{K}\delta q_j \left[Q_j-\frac{d}{dt}\left(\frac{\partial T}{\partial\dot{q_j}}\right)+\frac{\partial T}{\partial q_j} \right]=0\label{12}
 \end{equation}
 
 \vspace{.25cm}
 
 \textbf{\underline{Case I}} : {Unconnected holonomic system} :
 
 In this case all the generalised coordinates $q_1,q_2,...,q_k$ are all independent and consequently, they will allow arbitrary independent variations.Hence the coefficients of $\delta q_j$ in (\ref{12}) must vanish separately i.e.,
 \begin{equation}
 	\frac{d}{dt}\left(\frac{\partial T}{\partial\dot{q_j}}\right)-\frac{\partial T}{\partial q_j}= Q_j~,~~j=1,2,...,k\label{13}
 \end{equation}
 These are the $'k'$ second order equations and are known as Lagrange's equations for a holonomic unconnected system. Integration of these equations determine q's as a function of $'t'$. Further, if the external field is a potential field then we can write 
 \begin{align}
 	\delta w=-\delta V(q,t)=-\sum_{j}\frac{\partial V}{\partial q_j}\delta q_j\nonumber\\
 	\therefore\frac{\partial w}{\partial q_j}=-\frac{\partial V}{\partial q_j}=Q_j\nonumber
 \end{align}
 Hence the above Lagrange's equation becomes
 \begin{align}
 	\frac{d}{dt}\left(\frac{\partial T}{\partial\dot{q_j}}\right)-\frac{\partial T}{\partial q_j}=-\frac{\partial V}{\partial q_j}\nonumber\\
 	i.e.,\frac{d}{dt}\left(\frac{\partial T}{\partial\dot{q_j}}\right)-\frac{\partial (T-V)}{\partial q_j}=0\nonumber\\
 	i.e.,\frac{d}{dt}\left(\frac{\partial L}{\partial\dot{q_j}}\right)-\frac{\partial L}{\partial q_j}=0\nonumber
 \end{align}
 where $L=T-V$ is called the Lagrangian function ( or Kinetic potential).
 
 \vspace{.25cm}
 
 \textbf{\underline{Case II}} :  {Connected holonomic system} :
 
 In this case, the number of degrees of freedoms being less than the no. of generalised coordinates i.e., all the coordinates $q_1,q_2,...,q_k$ are not all independent. As the system is holonomic and connected the relations between the coordinates are of the form 
 \begin{equation}
 	f_i(q,t)=0,~i=1,2,...,r<k.\nonumber
 \end{equation}
 Thus $r$ coordinates of the system are not all independent. Now without any loss of generality, we assume that first r-coordinates of the k-coordinates are not independent. For virtual variations of the coordinates at time $t$, we can write 
 \begin{equation}
 	\sum_{j=1}^{k}\frac{\partial f_i}{\partial q_j}\delta q_j=0~,~i=1,2,...,r\nonumber
 \end{equation}
 From this it follows that $r$ of the variations $\delta q_1,\delta q_2,...,\delta q_k$ depend on the remaining variations which are assumed to be independent. We now multiply these equations by unknown arbitrary multipliers $\lambda_i$ and sum over $i$ to get
 \begin{eqnarray}
 	\sum_{i=1}^{r}\lambda_i\left(\sum_{j=1}^{k}\frac{\partial f_i}{\partial q_j}\delta q_j \right)=0\nonumber\\
 	i.e.,\sum_{j=1}^{k}\left(\sum_{i=1}^{r}\lambda_i\frac{\partial f_i}{\partial q_j}\right) \delta q_j=0\nonumber
 \end{eqnarray}
 Adding these to equation (\ref{12}) we get
 \begin{equation}
 	\sum_{j=1}^{k}\left[Q_j+\sum_{i=1}^{r}\lambda_i\frac{\partial f_i}{\partial q_j}-\frac{d}{dt}\left(\frac{\partial T}{\partial\dot{q_j}}\right)+ \frac{\partial T}{\partial q_j} \right]\delta q_j=0\label{14}
 \end{equation}
 It is possible to choose $\lambda_i$'s such that the coefficients of first r-dependent variables $\delta q_1,\delta q_2,...,\delta q_r$ vanish separately. Hence with this choice one gets 
 \begin{equation}
 	\frac{d}{dt}\left(\frac{\partial T}{\partial\dot{q_j}}\right)- \frac{\partial T}{\partial q_j} =Q_j+\sum_{i=1}^{r}\lambda_i\frac{\partial f_i}{\partial q_j}\label{15} 
 \end{equation}
 Thus equation (\ref{14}) reduces to
 \begin{equation}
 	\sum_{j=r+1}^{k}\left[Q_j+\sum_{i=1}^{r}\lambda_i\frac{\partial f_i}{\partial q_j}-\frac{d}{dt}\left(\frac{\partial T}{\partial\dot{q_j}}\right)+ \frac{\partial T}{\partial q_j} \right]\delta q_j=0\label{16}
 \end{equation}
 In the above equation (\ref{16}) the variations $\delta q_j~(j=r+1,...,k)$ are all independent and consequently the coefficients must vanish separately to give 
 \begin{equation}
 	\frac{d}{dt}\left(\frac{\partial T}{\partial\dot{q_j}}\right)- \frac{\partial T}{\partial q_j} =Q_j+\sum_{i=1}^{r}\lambda_i\frac{\partial f_i}{\partial q_j}~,~j=r+1,...,k.\label{17}
 \end{equation}
 Thus combining these two cases, one gets
 \begin{equation}
 	\frac{d}{dt}\left(\frac{\partial T}{\partial\dot{q_j}}\right)- \frac{\partial T}{\partial q_j} =Q_j+\sum_{i=1}^{r}\lambda_i\frac{\partial f_i}{\partial q_j}\label{18}
 \end{equation}
 with $i=1,2,...,r<k$ and $j=1,2,...,k.$
 
 These are the system of  $k$ equations with $k+r$ variables (namely $'k'$ no. of $'q'$s and $'r'$ no of $\lambda$'s). These system of equations (\ref{18}) together with the constraint equations are the Lagrange's equation of connected holonomic system defined by $k$ generalised coordinates with $r$ conditions (constrains). For detail about the Lagrangian formulation see Appendix III.
 
 \vspace{.5cm}
 
 \section{Lagrange's equations in non-holonomic system.}
 
 \vspace{.25cm}
 
 The derivation of Lagrange's equation for holonomic system required that the generalised coordinates be independent. For a non-holonomic system however there will be more generalised coordinates than the no of degrees of freedom. Therefore, $\delta q$'s are no longer independent if one assume a virtual displacement consistent with the constraint.
 
 Let $q_1,q_2,...,q_k$ be the $k$ generalised coordinates of the system subject to $r$ non-integrable differential constraints.
 \begin{equation}
 	\sum_{j=1}^{k}a_{ij}\dot{q_j}+a_{i0}=0~,~i=1,2,...,r<k.\label{19}
 \end{equation} 
 where $a'$s are functions of $q$ and $t$ alone. For a virtual variation at time $t$ one gets
 \begin{equation}
 	\sum a_{ij}\delta q_j=0~,~i=1,2,...,r.\label{20}
 \end{equation}
 ($t$ is not allowed to vary)
 
 Now, multiplying equation (\ref{20}) by arbitrary undetermine multiplier $\lambda_i$ and summing over $i$ results
 \begin{eqnarray}
 	\sum_{i}\lambda_{i}\sum_{j} a_{ij}\delta q_j=0\nonumber\\
 	i.e.,\sum_{j}\left(\sum_{i}\lambda_{i}a_{ij}\right)\delta q_j=0\label{21}
 \end{eqnarray}
 Adding equation (\ref{21}) to equation (\ref{6}) one gets
 \begin{equation}
 	\sum_{j}\left[Q_j+\sum_{i}\lambda_ia_{ij} -\frac{d}{dt}\left(\frac{\partial T}{\partial\dot{q_j}}\right)+ \frac{\partial T}{\partial q_j} \right]\delta q_j=0\label{22}
 \end{equation}
 As before $\lambda_{i}$'s are chosen in such a manner that the coefficients of  $\delta q_1,\delta q_2,,...,\delta q_r$ in (\ref{22}) vanish separately. then the remaining variation $\delta q_{r+1},...,\delta q_k$ in (\ref{22}) are perfectly arbitrary and independent and consequently their coefficient must also vanish separately. As a result one gets the following system of equations;
 \begin{equation}
 	\frac{d}{dt}\left(\frac{\partial T}{\partial\dot{q_j}}\right)- \frac{\partial T}{\partial q_j}= Q_j+\sum_{i}\lambda_ia_{ij},
 	~~i=1,2,...,r;~~~i=1,2,...,k.\label{23}
 \end{equation} 
 These are the $k$ Lagrange's equations of motion of non-holonomic system whose constraints are defined by equation (\ref{19}).
 
 \vspace{.5cm}
 
 $\bullet$ \textbf{{Examples}} :
 
 \vspace{.25cm}
 
 \textbf{1}. \underline{Planetary motion} :
 For a planet moving in an elliptic orbit around the sun the K.E.
 \begin{equation}
 	T=\frac{1}{2}m(\dot{r}^2+r^2\dot{\theta}^2)\nonumber
 \end{equation}
 and $\overrightarrow{F}=-\dfrac{m\mu}{r^3}\overrightarrow{r}$ i.e., $V=-\dfrac{m\mu}{r}$
 
 $\therefore L=T-V=\dfrac{1}{2}m(\dot{r}^2+r^2\dot{\theta}^2)+\dfrac{m\mu}{r}$.
 
 Here $m$ is the mass of the planet and that of sun is $\mu$. The two polar coordinates $(r,\theta)$ are the generalised coordinates here. Then Lagrange's equations of motion are
 \begin{equation}
 	\frac{d}{dt}\left(\frac{\partial L}{\partial\dot{r}}\right)-\frac{\partial L}{\partial r}=0\mbox{~~and~~}\frac{d}{dt}\left(\frac{\partial L}{\partial\dot{\theta}}\right)-\frac{\partial L}{\partial \theta}=0\nonumber
 \end{equation}
 \begin{equation}
 	i.e.,\ddot{r}-r\dot{\theta}^2=-\frac{\mu}{r^2}\mbox{~~and~~}\frac{d}{dt}(r^2\theta)=0,\nonumber
 \end{equation}
 which are the well known radial and cross radial equations of motion.
 
 \vspace{.25cm}
 
 \textbf{2}. The Lagrangian $L$ for the motion of a particle of unit mass is given by 
 \begin{equation}
 	L=\frac{1}{2}(\dot{x}^2+\dot{y}^2+\dot{z}^2)-V+\dot{x}A+\dot{y}B+\dot{z}C\nonumber
 \end{equation}
 where $V,A,B,C$ are functions of $x,y,z$. Show that the equations of motion are 
 \begin{equation}
 	\ddot{x}=-\frac{\partial V}{\partial x}+\dot{y}\left(\frac{\partial B}{\partial x}- \frac{\partial A}{\partial y}\right)-\dot{z}\left(\frac{\partial A}{\partial z}-\frac{\partial C}{\partial x} \right)\nonumber
 \end{equation}
 and similar two equations.
 
 \vspace{.25cm}
 
 \textbf{3}. The Lagrangian of a dynamical system is given by 
 \begin{equation}
 	L=m(\dot{q_4}^2-\dot{q_1}^2-\dot{q_2}^2-\dot{q_3}^2)^{\frac{1}{2}}+e\sum_{k=1}^{4}A_k\dot{q_k}\nonumber
 \end{equation}
 where $q_i$'s $(i=1,2,3,4)$ are the generalised coordinates, $A_i$'s are functions of $q_i$'s only and $e,m$ are constant. Show that the Lagrange's equations of motion are
 \begin{equation}
 	m\frac{d}{dt}(\lambda\dot{q_r})=e\sum_{k=1}^{4}\left(\frac{\partial A_r}{\partial q_k} -\frac{\partial A_k}{\partial q_r} \right)\dot{q_r}~,~r=1,2,3\nonumber
 \end{equation}
 and
 \begin{equation}
 	m\frac{d}{dt}(\lambda\dot{q_4})=e\sum_{k=1}^{4}\left(\frac{\partial A_k}{\partial q_4} -\frac{\partial A_4}{\partial q_k} \right)\dot{q_k}\nonumber
 \end{equation}
 where $\lambda^{-2}=\dot{q_4}^2-\dot{q_1}^2-\dot{q_2}^2-\dot{q_3}^2$.
 
 Also evaluate $p_i=\frac{\partial L}{\partial\dot{q_i}},~~(i=1,2,3,4)$ and show that 
 \begin{equation}
 	p_4=eA_4+\sqrt{\sum_{k=1}^{4}(p_k-eA_k)^2+m^2}.\nonumber
 \end{equation}
 
 \vspace{.5cm}
 
 \section{Expression for K.E. of a dynamical system :}
 
 \vspace{.25cm}
 
 The K.E. of a moving system is by definition
 \begin{equation}
 	T=\frac{1}{2}\Sigma m\dot{\overrightarrow{r}}^2~,~~\mbox{where}~\overrightarrow{r}=\overrightarrow{r}(q,t) \nonumber
 \end{equation}
 Here $q$ stands for $k$ generalised coordinates $q_1,q_2,...,q_k,$ so 
 \begin{equation}
 	\dot{\overrightarrow{r}}=\frac{d\overrightarrow{r}}{dt}=\frac{\partial\overrightarrow{r}}{\partial t}+\sum_{j=1}^{k}\frac{\partial\overrightarrow{r}}{\partial q_j}\dot{q_j}\nonumber
 \end{equation}
 Thus
 \begin{eqnarray}
 	T&=&\frac{1}{2}\sum m\left(\frac{\partial\overrightarrow{r}}{\partial t}+\sum_{j=1}^{k}\frac{\partial\overrightarrow{r}}{\partial q_j}\dot{q_j}\right)^2\nonumber\\
 	&=&\frac{1}{2}\sum m\left[\left(\frac{\partial\overrightarrow{r}}{\partial t}\right)^2+ 2\frac{\partial\overrightarrow{r}}{\partial t}\sum_{j=1}^{k}\frac{\partial\overrightarrow{r}}{\partial q_j}\dot{q_j}+ \left(\sum_{j=1}^{k}\frac{\partial\overrightarrow{r}}{\partial q_j}\dot{q_j}\right)^2 \right]\nonumber\\
 	&=&\frac{1}{2}\sum m\left(\frac{\partial\overrightarrow{r}}{\partial t}\right)^2 + \sum_{j=1}^{k}\left(\sum m\frac{\partial\overrightarrow{r}}{\partial t}\frac{\partial\overrightarrow{r}}{\partial q_j}\right)\dot{q_j} +\frac{1}{2}\sum_{i}\sum_{j}\left(\sum m\frac{\partial\overrightarrow{r}}{\partial q_i}\frac{\partial\overrightarrow{r}}{\partial q_j}\right)\dot{q_i}\dot{q_j}\nonumber
 \end{eqnarray}
 Here the terms $\sum m\frac{\partial\overrightarrow{r}}{\partial t}\frac{\partial\overrightarrow{r}}{\partial q_j}$ and $\sum m\frac{\partial\overrightarrow{r}}{\partial q_i}\frac{\partial\overrightarrow{r}}{\partial q_j}$ are summation over the system (i.e., over $m$ and $r$) and not on $i$ and $j$. So they can be symbolically written as $b_j$ and $a_{ij}$ respectively, and we write 
 \begin{equation}
 	T=\frac{1}{2}\sum_{i,j}a_{ij}\dot{q_i}\dot{q_j}+\sum_{i}b_i\dot{q_i}+\frac{1}{2}C \label{24}
 \end{equation}
 where $a_{ij}=\sum m\frac{\partial\overrightarrow{r}}{\partial q_i}\frac{\partial\overrightarrow{r}}{\partial q_j}=a_{ji}~,~b_i=\sum m\frac{\partial\overrightarrow{r}}{\partial t}\frac{\partial\overrightarrow{r}}{\partial q_j}~,~C=\sum m\left(\frac{\partial\overrightarrow{r}}{\partial t}\right)^2.$
 
 In general, $a,b,c$ are functions of $\overrightarrow{r}$ i.e., $q$ and $t$. But in particular, if the system is scleronomic in nature then $b$ and $c$ will be zero and we have 
 \begin{equation}
 	T=\frac{1}{2}\sum_{i}\sum_{j}a_{ij}\dot{q_i}\dot{q_j}
 \end{equation}
 
 \vspace{.5cm}
 
 $\bullet$ \textbf{{Some results}} :
 
 \vspace{.25cm}
 
 \textbf{I.} If the K.E. is homogeneous and quadratic of the components of velocities then 
 \begin{equation}
 	\sum\dot{q}\frac{\partial T}{\partial\dot{q}}-L=T+V\nonumber
 \end{equation}
 \textbf{Proof :} By Euler's theorem on homogeneous function $\sum\dot{q}\frac{\partial T}{\partial\dot{q}}=2T.$
 
 Hence, L.H.S.$=2T-(T-V)=T+V$.
 
 \vspace{.25cm}
 
 \textbf{II.} Show that $\left(\sum\dot{q}\frac{\partial T}{\partial\dot{q}}-L\right)$ is a constant of motion.
 
 \textbf{Proof} : We know that if $A$ is a constant of motion, then $\frac{dA}{dt}=0$.
 
 Here, 
 \begin{eqnarray}
 	\frac{d}{dt}\left[\sum\dot{q}\frac{\partial T}{\partial\dot{q}}-L \right]&=&\frac{d}{dt}\left[ \sum\dot{q}\frac{\partial (T-V)}{\partial\dot{q}}-L\right]~(V\mbox{is independent of }\dot{q})\nonumber\\
 	&=&\frac{d}{dt}\left[\sum\dot{q}\frac{\partial L}{\partial\dot{q}}-L \right]\nonumber\\
 	&=&\cancel{\sum\ddot{q}\frac{\partial L}{\partial\dot{q}}}+\sum\dot{q}\frac{d}{dt}\left( \frac{\partial L}{\partial\dot{q}}\right)-\cancel{\sum\frac{\partial L}{\partial\dot{q}}\ddot{q}} - \sum\frac{\partial L}{\partial q}\dot{q}\nonumber\\
 	&=&\sum\dot{q} \frac{\partial L}{\partial{q}}- \sum\dot{q}\frac{\partial L}{\partial q}=0,~\mbox{by Lagrange's equation of motion.}\nonumber
 \end{eqnarray}
 Thus in a scleronomic system the total energy of the system remains constant during the motion.
 
 \vspace{.5cm}
 
 $\bullet$\textbf{{Problem}} : Determine the motion of a dynamical system whose Lagrangian $L$ is given by 
 \begin{equation}
 	L=\frac{ma^2}{7^2}(16\dot{q_1}^2+20\dot{q_1}\dot{q_2}+25\dot{q_2}^2) - \frac{mga}{2} \left(\frac{1}{3}{q_1}^2+\frac{5}{6}{q_2}^2\right),\nonumber
 \end{equation}
 given that $\dot{q_1}=0=\dot{q_2}$ and ${q_1}=\beta={q_2}$ at $t=0$.
 
 \vspace{.5cm}
 
 \section{Equation of energy :}
 
 \vspace{.25cm}
 
 We have
 \begin{equation}
 	T=\frac{1}{2}\sum_{i}\sum_{j}a_{ij}\dot{q_i}\dot{q_j}=T(q,\dot{q})\nonumber
 \end{equation}
 We assume that the geometrical equations do not depend on $t$ explicitly. So $T$ is a homogeneous quadratic function of velocity and in the above $a_{ij}=a_{ji}$, is a function of $q$'s only. Now,
 \begin{equation}
 	\frac{dT}{dt}=\sum_{j}\frac{\partial T}{\partial q_j}\dot{q_j}+ \sum_{j}\frac{\partial T}{\partial \dot{q_j}}\ddot{q_j}\nonumber
 \end{equation} 
 The Lagrange's equations of motion are 
 \begin{equation}
 	\frac{d}{dt}\left(\frac{\partial T}{\partial\dot{q_j}}\right)-\frac{\partial T}{\partial q_j}= Q_j~,~~ j=1,2,...,k.\nonumber
 \end{equation}
 Now, multiplying these equations by $\dot{q_j}$ and summing over $j$ we have
 \begin{eqnarray}
 	\sum\left[\dot{q_j}\frac{d}{dt}\left(\frac{\partial T}{\partial\dot{q_j}}\right)-\dot{q_j}\frac{\partial T}{\partial q_j}\right]&=&\sum Q_j\dot{q_j}\nonumber\\
 	i.e.,\sum\left[\frac{d}{dt}\left(\dot{q_j}\frac{\partial T}{\partial\dot{q_j}}\right)-\ddot{q_j} \frac{\partial T}{\partial\dot{q_j}}-\dot{q_j}\frac{\partial T}{\partial q_j} \right]&=&\sum Q_j\dot{q_j}\nonumber\\
 	i.e.,2\frac{dT}{dt}-\frac{dT}{dt}&=&\sum Q_j\dot{q_j}~\mbox{(By Euler's theorem)} \nonumber\\
 	i.e.,\frac{dT}{dt}&=&\sum Q_j\dot{q_j}\label{26}
 \end{eqnarray}
 Thus the rate of change of K.E. of a holonomic scleronomic dynamical system with frictionless bilateral constraints is equal to the rate at which work has been done by the external forces. If the external forces are derived form a force function $U,$ independent of $t$ then 
 \begin{equation}
 	U=U(q)~\mbox{and}~Q_i=\frac{\partial U}{\partial q_i}\nonumber
 \end{equation}
 \begin{equation}
 	\mbox{and~}\sum Q_i\dot{q_i}=\sum_{i}\frac{\partial U}{\partial q_i}.\frac{dq_i}{dt}= \frac{dU}{dt}\nonumber
 \end{equation}
 Thus the above energy equation takes the form $\frac{dT}{dt}=\frac{dU}{dt}$ i.e., $T-U=$Constant i.e.,$T+V=$Constant, where V=-U is the potential energy.
 
 \vspace{.5cm}
 
 \section{Gerneralised momentum and cyclic co-ordinate :}
 
 \vspace{.25cm}
 
 Let $L=L(q,\dot{q},t)$ be the Lagrangian of a mechanical system of $n$ d.f. defined at any instant $'t'$ by $n$ generalised coordinates $q_1,q_2,...,q_n$. The generalised or conjugate momentum $p_i$ associated with the generalised coordinate $q_i$ is defined as 
 \begin{equation}
 	p_i=\frac{\partial L}{\partial\dot{q_i}}\label{27}
 \end{equation}
 In terms of the generalised momentum Lagrange's equations of motion becomes 
 \begin{equation}
 	\dot{p_i}=\frac{\partial L}{\partial{q_i}}\label{28}
 \end{equation}
 A coordinate $q_i$ is said to be cyclic or ignorable when $L$ does not involve this coordinate explicitly. The velocity corresponding to cyclic coordinate may remain in $L$. If $q_i$ be cyclic then $\frac{\partial L}{\partial q_i}=0$ and hence $\dot{p_i}=0$ i.e., $p_i=$constant. Thus generalised momentum associated with cyclic coordinate is a conserved quantity.
 
 \vspace{.5cm}
 
 \section{Lagrangian system with cyclic or ignorable coordinates :}
 
 \vspace{.25cm}
 
 Let us consider a holonomic dynamical system with $k$ d.f. and $q_1,q_2,...,q_k$ be the generalised coordinates of the system. If the external forces are derived from a force function $U$, then the Lagrange's equations of motion are
 \begin{equation}
 	\frac{d}{dt}\left(\frac{\partial L}{\partial\dot{q_i}}\right)-\frac{\partial L}{\partial q_i}=0,~~ i=1,2,...,k\label{29}
 \end{equation}
 Suppose $\mu(<k)$ of these coordinates say $q_1,q_2,...,q_\mu$ are ignorable then we have $\mu$ first integrals of Lagrange's equations in the form 
 \begin{equation}
 	\frac{\partial L}{\partial\dot{q_i}}=\mbox{constant}=\beta_i,~~i=1,2,...,\mu\label{30}
 \end{equation}
 As $L=T-V$ and $T$ is a quadratic function of generalised velocities $\dot{q_1},\dot{q_2},...,\dot{q_k},$ so equation (\ref{30}) are linear in $\dot{q_1},\dot{q_2},...,\dot{q_\mu}$. Therefore from these $\mu-$equations (\ref{30}), we can express $\dot{q_1},\dot{q_2},...,\dot{q_\mu}$ in terms of $\dot{q_{\mu+1}},\dot{q_{\mu+2}},...,\dot{q_k}, \\{q_{\mu+1}},...,{q_k},\beta_1,...,\beta_k$ and $t$ i.e.,
 
 $\dot{q_i}=\dot{q_i}(\dot{q}_{\mu+1},\dot{q}_{\mu+2},...,\dot{q_k},\dot{q}_{\mu+1},...,\dot{q_k},\beta_1,...,\beta_{\mu},t),~~i=1,2,...,\mu.$
 Let us now introduce a function $R$ defined as 
 \begin{equation}
 	R=L-\sum_{i=1}^{\mu}\dot{q_i}\frac{\partial L}{\partial\dot{q_i}}=L-\sum_{i=1}^{\mu} \beta_i\dot{q_i}\nonumber
 \end{equation}
 i.e.,
 \begin{eqnarray}
 	R({q}_{\mu+1},{q}_{\mu+2},...,{q_k},\dot{q}_{\mu+1},...,\dot{q_k},\beta_1,...,\beta_{\mu},t)~~~~~~~~~~~~~~~~~~~~~~~~~~~~~~~~~~~~~~~~\nonumber\\	=L({q}_{\mu+1},{q}_{\mu+2},...,{q_k},\dot{q}_{1},...,\dot{q_k},\beta_1,...,\beta_{\mu},t)-\sum_{i=1}^{\mu}\beta_i\dot{q_i}\label{31}
 \end{eqnarray}
 Consider a virtual variation of $R$ at any instant $t$
 \begin{equation}
 	\delta R=\delta\left(L-\sum_{i=1}^{\mu}\dot{q}_i\beta_i \right)\nonumber
 \end{equation}
 \begin{eqnarray}
 	i.e.,&~&\sum_{i=\mu+1}^{k}\frac{\partial R}{\partial q_i}\delta q_i + \sum_{i=\mu+1}^{k}\frac{\partial R}{\partial\dot{q}_i}\delta\dot{q}_i + \sum_{i=1}^{\mu}\frac{\partial R}{\partial\beta_i}\delta\beta_i\nonumber\\
 	&~&=\sum_{i=\mu+1}^{k}\frac{\partial L}{\partial q_i}\delta q_i +\cancel{ \sum_{i=1}^{\mu}\frac{\partial L}{\partial\dot{q}_i}\delta\dot{q}_i} + \sum_{i=\mu+1}^{k}\frac{\partial L}{\partial\dot{q}_i}\delta\dot{q}_i -\sum_{i=1}^{\mu} \dot{q}_i\delta\beta-\cancel{\sum_{i=1}^{\mu}\beta_i\delta\dot{q}_i}\nonumber\\
 	i.e.,&~&\sum_{i=\mu+1}^{k}\left(\frac{\partial R}{\partial q_i}-\frac{\partial L}{\partial q_i} \right)\delta q_i + \sum_{i=\mu+1}^{k}\left(\frac{\partial R}{\partial\dot{q}_i}-\frac{\partial L}{\partial\dot{q}_i} \right)\delta\dot{q}_i + \sum_{i=1}^{\mu}\left(\frac{\partial R}{\partial\beta_i}+\dot{q}_i\right)\delta\beta_i=0\nonumber
 \end{eqnarray}
 This relation is true for arbitrary variations, therefore,
 \begin{eqnarray}
 	\mbox{therefore,~}\frac{\partial R}{\partial q_i}&=&\frac{\partial L}{\partial q_i},~~i=\mu+1,...,k\nonumber\\
 	\frac{\partial R}{\partial\dot{q}_i}&=&\frac{\partial L}{\partial\dot{q}_i},~~i=\mu+1,...,k\nonumber\\
 	\mbox{and~~}\dot{q}_i&=&-\frac{\partial R}{\partial\beta_i},~~i=1,2,...,\mu\label{32}
 \end{eqnarray}
 Putting these relations in Lagrange's equations of motion i.e.,
 \begin{equation}
 	\frac{d}{dt}\left(\frac{\partial L}{\partial\dot{q_i}}\right)-\frac{\partial L}{\partial q_i}=0,~~ i=\mu+1,...,k\nonumber
 \end{equation}
 we obtain
 \begin{equation}
 	\frac{d}{dt}\left(\frac{\partial R}{\partial\dot{q_i}}\right)-\frac{\partial R}{\partial q_i}=0,~~ i=\mu+1,...,k\label{33}
 \end{equation}
 Thus we have reduced the original Lagrange's equations with $'k'$ degrees of freedom to another set of Lagrange's equations with $k-\mu$ degrees of freedom. These $k-\mu$ equations will determine $q_{\mu+1},...,q_k$ as function of $t$. Subsequently, we can determine $q_1,q_2,...,q_{\mu}$ from the equations 
 \begin{equation}
 	\frac{\partial R}{\partial\beta_i}=-\dot{q}_i~\mbox{~i.e.,~}~q_i=-\int\frac{\partial R}{\partial\beta_i}dt~,~i=1,2,...,\mu\nonumber
 \end{equation} 
 Here the function $R$ defined in equation (\ref{31}) is called the modified Lagrangian function or Routhian.
 
 \vspace{.5cm}
 
 $\bullet$ \textbf{{Problem :}}
 
 \vspace{.25cm}
 
 Discuss the planetary motion in terms of Routhian.
 
 For planetary motion,
 \begin{equation}
 	T=\frac{1}{2}(\dot{r}^2+r^2\dot{\theta}^2)~,~V=-\frac{\mu}{r}\mbox{~~so~~}L=\frac{1}{2}(\dot{r}^2+r^2\dot{\theta}^2)+\frac{\mu}{r}\nonumber
 \end{equation}
 Here $\theta$ is a cyclic coordinate and hence $\frac{\partial L}{\partial\dot{\theta}} =r^2\dot{\theta}=\beta$, a constant.
 
 Thus the Routhian function is given by 
 \begin{equation}
 	R=L-\dot{\theta}\beta=\frac{1}{2}\dot{r}^2-\frac{1}{2}r^2\dot{\theta}^2+\frac{\mu}{r}=\frac{1}{2}\dot{r}^2-\frac{1}{2}\frac{\beta^2}{r^2}+\frac{\mu}{r}\nonumber
 \end{equation}
 By Lagrange's equation of motion for $r$
 \begin{eqnarray}
 	\frac{d}{dt}\left(\frac{\partial R}{\partial\dot{r}}\right)-\frac{\partial R}{\partial r}&=&0\nonumber\\
 	\mbox{i.e.,~~}\ddot{r}-\left(\frac{B^2}{r^3}-\frac{\mu}{r^2}\right)&=&0\nonumber
 \end{eqnarray}
 Integrating once
 \begin{eqnarray}
 	\dot{r}^2&=&-\frac{\beta^2}{r^2}+\frac{2\mu}{r}+h\nonumber\\
 	\therefore\frac{dr}{d\theta}=\frac{\dot{r}}{\dot{\theta}}&=&\pm\frac{r^2}{\beta}\sqrt{h-\frac{\beta^2}{r^2}+\frac{2\mu}{r}}\nonumber\\
 	\mbox{i.e.,~}\pm d\theta=\frac{\left(\frac{\beta}{r^2}\right)dr}{\sqrt{h-\frac{\beta^2}{r^2}+\frac{2\mu}{r}}}&=&\frac{-d\left(\frac{\beta}{r}\right)}{\sqrt{\left(h+\frac{\mu^2}{\beta^2}\right)-\left\{ \frac{\beta}{r}-\frac{\mu}{\beta}\right\}^2}}\nonumber\\
 	&=&\frac{-d\left(\frac{\beta}{r}-\frac{\mu}{\beta}\right)}{\sqrt{\left(h+\frac{\mu^2}{\beta^2}\right)-\left(\frac{\beta}{r}-\frac{\mu}{\beta}\right)^2}}\nonumber\\
 	\mbox{i.e.,~~}\pm(\theta+\alpha)&=&\arccos\left[\frac{\left(\frac{\beta}{r}-\frac{\mu}{\beta}\right)}{\sqrt{\left(h+\frac{\mu^2}{\beta^2}\right)}}\right]\nonumber\\
 	\mbox{i.e.,~~}\frac{\beta}{r}&=&\frac{\mu}{\beta}+\sqrt{h+\frac{\mu^2}{\beta^2}}\cos(\theta+\alpha)\nonumber\\
 	\mbox{i.e.,~~}\frac{\left(\frac{\beta^2}{\mu}\right)}{r}&=&1+\sqrt{1+\frac{\beta^2h}{\mu^2}} \cos(\theta+\alpha)\nonumber
 \end{eqnarray}
 $\rightarrow$The elliptic path of the planet.
 
 \vspace{.25cm}
 
 \textbf{Problem: 1}
 
 The Lagrangian of a dynamical system is 
 \begin{equation}
 	L=\frac{q_1^2}{aq_2+b}+\frac{1}{2}\dot{q}_2^2+2q_2^3+cq_2\nonumber
 \end{equation}
 where $a,b,c$ are given constants. Find an integral giving $q_2$ as a function of $t$.
 
 \vspace{.25cm}
 
 \textbf{Problem: 2}
 
 In a dynamical system with $2$ d.f., the K.E. is given by 
 \begin{equation}
 	L=\frac{\dot{q}_1^2}{2(aq_2+b)}+\frac{1}{2}q_2^2\dot{q}_2^2~,~V=c+dq_2\nonumber
 \end{equation}	
 where $a,b,c$ and $d$ are given constants. Show that the value of $q_2$ in terms of time is given by the equation of the form 
 \begin{equation}
 	(q_2-k)(q_2+2k)^2=h(t-t_0)^2\nonumber
 \end{equation}
 with $h,k,t_0$ as constants.
 
 \vspace{.25cm}
 
 \textbf{Problem: 3}
 
 The energies of a dynamical system with $2$ d.f. are given by
 \begin{equation}
 	L=\frac{\dot{q}_1^2}{2(aq_2+b)}+\frac{1}{2}q_2^2\dot{q}_2^2~,~V=c+dq_2^2\nonumber
 \end{equation} 
 Find $q_1$ and $q_2$ by the method of ignorations of coordinates.
 
 \vspace{.5cm}
 
\section{Lioville's theorem for a dynamical system :}
 
 \vspace{.25cm}
 
 $\bullet$ \textbf{Statement } : If for a dynamical system $T$ and $V$ are of the form 
 \begin{eqnarray}
 	2T&=&\{u_1(q_1)+...+u_n(q_n)\}\{v_1(q_1)\dot{q}_1^2+...+v_n(q_n) \dot{q}_n^2\}\nonumber\\
 	&=&u\sum_{r=1}^{n}v_r(q_r)\dot{q}_r^2~,~~u=\sum_{r=1}^{n}u_r(q_r)\label{34}	
 \end{eqnarray}
 \begin{equation}
 	V=-\phi(\mbox{force function})=\frac{w_1(q_1)+...+w_n(q_n)}{u_1(q_1)+...+u_n(q_n)}=\frac{1}{u}\sum_{r=1}^{n}w_r(q_r)\nonumber\\
 \end{equation}
 where $q_1,q_2,...,q_n$ are independent parameters defining the position of system then the solution of the problem can be obtained by a quadrature.
 
 \textbf{{Proof} :}
 
 Let us make a change of variables from$q_1,q_2,...,q_n$ to $Q_1,Q_2,...,Q_n$ such that 
 \begin{equation}
 	\dot{Q}_r^2=v_r(q_r)\dot{q}_r^2\label{35}
 \end{equation}
 \begin{eqnarray}
 	\mbox{i.e.,~}Q_r&=&\int\sqrt{v_r(q_r)}dq_r~,~r=1,2,...,n.\nonumber\\
 	\mbox{So~}u=\sum u_r(q_r)&=&\sum U_r(Q_r)=U\nonumber\\
 	V=\frac{1}{u}\sum w_r(q_r)&=&\frac{1}{U}\sum W_r(Q_r)\nonumber\\
 	\mbox{Hence,~}T&=&\frac{1}{2}U\sum_{r=1}^{n}\dot{Q}_r^2\nonumber
 \end{eqnarray}
 The Lagrange's equation of motion 
 \begin{equation}
 	\frac{d}{dt}\left(\frac{\partial T}{\partial\dot{Q_r}}\right)-\frac{\partial T}{\partial Q_r}=-\frac{\partial V}{\partial Q_r}~,~r=1,2,...,r\nonumber
 \end{equation}
 now becomes 
 \begin{eqnarray}
 	\frac{d}{dt}(U\dot{Q}_r)-\frac{1}{2}\frac{\partial U}{\partial Q_r}\sum\dot{Q}_r^2&=&-\frac{\partial V}{\partial Q_r}\nonumber\\
 	\mbox{i.e.,}2U\dot{Q}_r\frac{d}{dt}(U\dot{Q}_r)-U\dot{Q}_r\frac{\partial U}{\partial Q_r}\sum\dot{Q}_r^2&=&-2U\dot{Q}_r\frac{\partial V}{\partial Q_r} \nonumber\\
 	\mbox{i.e.,}\frac{d}{dt}(U^2\dot{Q}_r^2)-\dot{Q_r}\left[U\frac{\partial{U}}{\partial{Q_r}}\sum\dot{Q_r}^2-2U\frac{\partial{V}}{\partial{Q_r}}\right]&=&0\nonumber
 \end{eqnarray}
 Form the energy equation : $T+V=h$, a constant.
 
 We have $\dfrac{1}{2}U\sum\dot{Q_r}=h-V$
 
 So using this result in the above equation we obtained
 \begin{eqnarray}
 	\frac{d}{dt}(U^2\dot{Q}_r^2)-\dot{Q_r}\left[\frac{\partial{U}}{\partial{Q_r}}(2h-2V)-2U\frac{\partial{V}}{\partial{Q_r}}\right]&=&0\nonumber\\
 	\mbox{i.e.,}\frac{d}{dt}(U^2\dot{Q}_r^2)-2\dot{Q_r}\left[h\frac{\partial{U}}{\partial{Q_r}}-\frac{\partial}{\partial{Q_r}}\sum{W_r(Q_r)}\right]&=&0\nonumber\\
 	\mbox{i.e.,}\frac{d}{dt}(U^2\dot{Q}_r^2)-2\dot{Q_r}\left[h\frac{d{U_r}}{d{Q_r}}-\frac{d{W_r}}{d{Q_r}}\right]&=&0\nonumber\\
 	\mbox{i.e.,}\frac{d}{dt}\left(U^2\dot{Q}_r^2-2hU_r+2W_r\right)&=&0\nonumber\\
 	\mbox{i.e.,}U^2\dot{Q}_r^2-2hU_r+2W_r&=&\mbox{Constant}\nonumber\\
 	\mbox{i.e.,}\frac{1}{2}u^2v_r(q_r)\dot{q_r}^2&=&hu_r-w_r+\gamma_r~\mbox{(say)}\nonumber\\
 	&=&\chi_r(q_r)\nonumber\\
 	&~&~~~~~~\mbox{(in the old coordinates $q_r$)}\nonumber\\
 	\mbox{i.e.,}\frac{\sqrt{2}}{u}dt&=&\sqrt{\frac{v_r(q_r)}{\chi_r(q_r)}}dq_r\label{36}
 \end{eqnarray}
 Thus we have 
 \begin{eqnarray}
 	\int\sqrt{\frac{v_1(q_1)}{\chi_1(q_1)}}dq_1&=&\int\sqrt{\frac{v_2(q_2)}{\chi_2(q_2)}}dq_2+\beta_2\nonumber\\
 	&=&...\nonumber\\
 	&=&...\nonumber\\
 	&=&\int\sqrt{\frac{v_r(q_r)}{\chi_r(q_r)}}dq_r+\beta_r\nonumber\\
 	&=&...\nonumber\\
 	&=&...\nonumber\\
 	&=&\int\sqrt{\frac{v_n(q_n)}{\chi_n(q_n)}}dq_n+\beta_n\label{37}
 \end{eqnarray}
 Now, multiplying equation (\ref{36}) by $u_r(q_r)$ and summing over `$r$' from $1$ to $n$ we get
 \begin{equation}
 	\frac{2}{u}\sum{u_r}(q_r)dt=\sum_{r=1}^{n}u_r(q_r)\sqrt{\frac{v_r(q_r)}{\chi_r(q_r)}}dq_r\nonumber
 \end{equation}
 Integrating once we have
 \begin{equation}
 	\sqrt{2}t+c=\int\sum{u_r(q_r)}\sqrt{\frac{v_r(q_r)}{\chi_r(q_r)}}dq_r\label{38}
 \end{equation} 
 Equation (\ref{37}) gives the coordinates in terms of any one of them and then equation (\ref{38}) determines all the coordinates as a function of $t$. Thus equation (\ref{37}) and (\ref{38}) give the solution of the problem subject to the condition that the solution contain $2n$ arbitrary constant. Again from the equation 
 \begin{equation}
 	\frac{1}{2}u^2v_r(q_r)\dot{q_r}^2=hu_r-w_r+\gamma_r\nonumber
 \end{equation} 
 we have on addition, 
 \begin{eqnarray}
 	u\frac{u}{2}\sum{v_r}\dot{q_r}^2&=&h\sum{u_r}-\sum{w_r}+\sum{\gamma_r}\nonumber\\
 	\mbox{i.e.,~}u.T&=&h.u-u.V+\sum{\gamma_r}\nonumber\\
 	\mbox{i.e.,~}u(T+V)&=&u.h+\sum{\gamma_r}\nonumber\\
 	\mbox{i.e.,~}\sum{\gamma_r}&=&0.\nonumber
 \end{eqnarray} 
 Hence we have $2n-1$ arbitrary constants in the solutions.
 
 \vspace{.5cm}
 
 $\bullet$ \textbf{{Problem: 1} }
 
 \vspace{.25cm}
 
 Show that the dynamical system for which 
 \begin{equation}
 	2T=q_1q_2(\dot{q_1}^2+\dot{q_2}^2),~V=\frac{1}{q_1}+\frac{1}{q_2}\nonumber
 \end{equation}
 can be expressed as one of Liouville's type.
 
 \vspace{.25cm}
 
 \textbf{{Solution} :}
 \begin{eqnarray}
 	\mbox{~~~~~~~~~~~~~~~~~~~~Let,~} q_1=r_1+r_2,~q_2&=&r_1-r_2\nonumber\\
 	\therefore{2T}&=&(r_1^2-r_2^2)\left[2\dot{r_1}^2+2\dot{r_2}\right]\nonumber\\
 	\mbox{i.e.,}~T&=&(r_1^2-r_2^2)\left(\dot{r_1}^2+\dot{r_2}\right)\nonumber\\
 	V&=&\frac{2r_1}{\dot{r_1}^2-\dot{r_2}}\nonumber
 \end{eqnarray}
 So the dynamical problem is of Liouville's type.
 
 \vspace{.5cm}
 
 $\bullet$ \textbf{{Problem: 2} }
 
 \vspace{.25cm}
 
 The K.E. and P.E. of a dynamical system is given by 
 \begin{equation}
 	T=\frac{1}{2}({q_1}^2+{q_2}^2)(\dot{q_1}^2+\dot{q_2}^2),~V=\frac{1}{{q_1}^2+{q_2}^2}\nonumber
 \end{equation}
 Show by Liouville's theorem that the relation between $q_1$ and $q_2$ is $a^2q_1^2+b^2q_2^2+2abq_1q_2\cos\gamma=\sin^2\gamma$ where $a,~b,~\gamma$ are constants.
 
 \vspace{.25cm}
 
 \textbf{{Solution} :}
 
 \vspace{.25cm}
 
 Here, $u_1=q_1^2,~u_2=q_2^2;~v_1=1,~v_2=1; w_1=1,w_2=0~$ so the dynamical system is a Liouville's type.
 \begin{eqnarray}
 	\mbox{Here,~}\chi_1&=&hu_1-w_1+\gamma=hq_1^2-1+\gamma\nonumber\\
 	\chi_2&=&hu_2-w_2-\gamma=hq_2^2-\gamma\nonumber
 \end{eqnarray}
 
 Hence, the first integral can be written as 
 
 \begin{eqnarray}
 	\int\sqrt{\frac{v_1}{\chi_1}}dq_1&=&\int\sqrt{\frac{v_2}{\chi_2}}dq_2+\beta\nonumber\\
 	\mbox{i.e.,}\int\frac{dq_1}{\sqrt{hq_1^2+\gamma-1}}&=&\int\frac{dq_2}{\sqrt{hq_2^2-\gamma}}+\beta\label{39A}\\
 	\mbox{i.e.,}\int\frac{dq_1}{\sqrt{q_1^2+\frac{\gamma-1}{h}}}&=&\int\frac{dq_2}{\sqrt{q_2^2-\frac{\gamma}{h}}}+\beta\sqrt{h}\nonumber\\
 	\mbox{i.e.,}\ln\left|{q_1+\sqrt{q_1^2+\frac{\gamma-1}{h}}}\right|&=&\ln\left|{q_2+\sqrt{q_2^2-\frac{\gamma}{h}}}\right|+\beta\sqrt{h}\nonumber\\
 	\mbox{i.e.,}\frac{q_1+\sqrt{q_1^2+\frac{\gamma-1}{h}}}{q_2+\sqrt{q_2^2-\frac{\gamma}{h}}}&=&e^{\beta\sqrt{h}}=c\mbox{~(say)}\nonumber\\
 	\mbox{i.e.,}q_1-cq_2&=&\left(c\sqrt{q_2^2-\frac{\gamma}{h}}-\sqrt{q_1^2+\frac{\gamma-1}{h}}\right)\nonumber
 \end{eqnarray}
 
 \begin{eqnarray}
 	\mbox{i.e.,}q_1^2+c^2q_2^2-2cq_1q_2&=&c^2\left(q_2^2-\frac{\gamma}{h}\right)+q_1^2+\frac{\gamma-1}{h}-2c\sqrt{\left(q_2^2-\frac{\gamma}{h}\right)\left(q_1^2+\frac{\gamma-1}{h}\right)}\nonumber\\
 	\mbox{i.e.,}\frac{c^2\gamma}{h}-\frac{\gamma-1}{h}-2cq_1q_2&=&-2c\sqrt{\left(q_2^2-\frac{\gamma}{h}\right)\left(q_1^2+\frac{\gamma-1}{h}\right)}\nonumber\\
 	\mbox{i.e.,}k-2cq_1q_2&=&-2c\sqrt{q_1^2q_2^2-\frac{\gamma{q_1}^2}{h}+\frac{\gamma-1}{h}q_2^2-\frac{\gamma(\gamma-1)}{h^2}},~~~~k=\frac{c^2\gamma-\gamma+1}{h}\nonumber
 \end{eqnarray}
 squaring both sides
 \begin{eqnarray}
 	\mbox{i.e.,~}k^2-2ckq_1q_2+4c^2q_1^2q_2^2&=&4c^2q_1^2q_2^2-4c^2\frac{\gamma{q_1}^2}{h}+4c^2\frac{\gamma-1}{h}q_2^2-4c^2\frac{\gamma(\gamma-1)}{h^2}\nonumber\\
 	\mbox{i.e.,~}4c^2\frac{\gamma{q_1}^2}{h}+4c^2\frac{1-\gamma}{h}q_2^2-4ckq_1q_2&=&4c^2\frac{\gamma(1-\gamma)}{h^2}-k^2\nonumber\\
 	\mbox{i.e.,}a^2q_1^2+b^2q_2^2+2abq_1q_2\cos\gamma&=&\sin^2\gamma\nonumber
 \end{eqnarray}
 where, $a^2=\dfrac{4c^2\gamma}{h},~b^2=\dfrac{4c^2(1-\gamma)}{h},~4ck=4c\left[\dfrac{c^2\gamma}{h}-\dfrac{\gamma-1}{h}\right]=\dfrac{4c}{h}(c^2\gamma-\gamma+1)$
 
 i.e.,$4ck=2\sqrt{\dfrac{c^2\gamma}{h}}2\sqrt{\dfrac{c^2(1-\gamma)}{h}}.\dfrac{c^2\gamma-\gamma+1}{c\sqrt{\gamma(1-\gamma)}}=2ab\cos\gamma$
 
 $\therefore\cos\gamma=\dfrac{c^2\gamma-\gamma+1}{2c\sqrt{\gamma(1-\gamma)}}~\&~\sin^2\gamma=1-\dfrac{(c^2\gamma-\gamma+1)^2}{4c^2\gamma(1-\gamma)}$
 
 If $h<0$ then from (\ref{39A})
 \begin{eqnarray}
 	&~&-\int\frac{dq_1}{\sqrt{(\frac{1-\gamma}{-h})-q_1^2}}=-\int\frac{dq_2}{\sqrt{(\frac{\gamma}{-h})-q_2^2}}+\beta\sqrt{-h}\nonumber\\
 	&~&\mbox{i.e,}~~\cos^{-1}\frac{q_1}{\sqrt{\left(\frac{1-\gamma}{-h}\right)}}=\cos^{-1}\frac{q_2}{\sqrt{\left(\frac{\gamma}{-h}\right)}}+\beta\sqrt{-h}\nonumber\\
 	&~&\mbox{i.e,}~~\cos^{-1}\left[\frac{q_1q_2}{\sqrt{\frac{(1-\gamma)\gamma}{h^2}}}-\sqrt{\left(1-\frac{q_1^2h}{1-\gamma}\right)\left(1-\frac{q_2^2h}{\gamma}\right)}\right]=\pi-\gamma\nonumber
 \end{eqnarray}

 ~~~~~~~~~~~~~~~~~~~~~~~~~~~~~~~~~~~~~~~~~~~~~~~~~~~~~~~~~~~~~~~~~~~~~~~~~~~~~~~~~~~~~~~~~~~~~~~~~~~~~~~~~~~~~~~~~~~~~~~~~~~~~~~~~~~~~~~~~~~~~~~~~~~~~~~~~~~~~~~~~~~~~~~~~~~~~~~~~~~~~~~~~~~~~~~~~~~~~~~~~~~~~~~~~~~~~~~~~~~
 \vspace{.5cm}
 
 $\bullet$ \textbf{{Problem: 3} }
 
 \vspace{.25cm}
 
 The K.E. and P.E. of a particle moving in a plane are given by 
 \begin{equation}
 	2T=\dot{x}^2+\dot{y}^2,~V(r)=-\frac{\mu}{r}-\frac{\mu'}{r'}\nonumber
 \end{equation}
 where $r,r'$ are the distances of the particle from the points $(c,0)$ and $(-c,0)$ respectively. Prove by using Liouville's theorem that the problem can be solved completely.
 \begin{figure}[h!]
 	\centering
 	\includegraphics[width=0.6\textwidth]{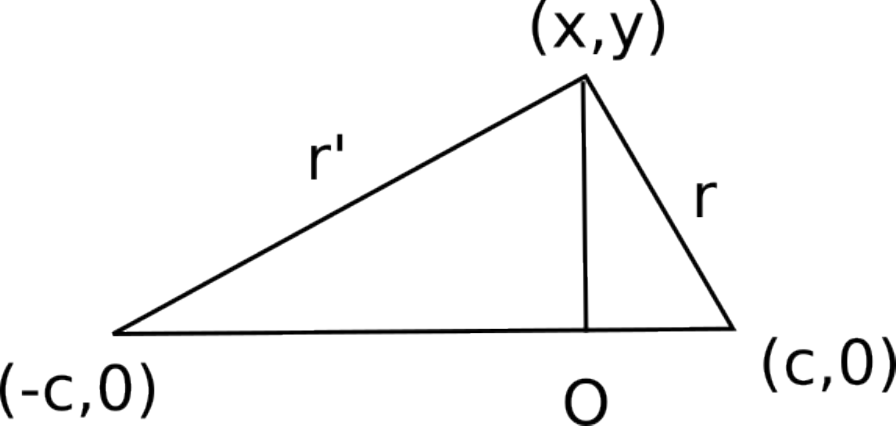}\\
 	\label{fig1}
 \end{figure}
 
 \textbf{{Solution} :}
 
 \vspace{.25cm}
 
 Here $r^2=(x-c)^2+y^2,~{r'}^2=(x-c)^2+y^2$. 
 
 Let, $r=q_1+r_2,~r'=q_1-q_2$. 
 
 Now,
 \begin{eqnarray}
 	V&=&\frac{-\mu(q_1-q_2)-\mu'(q_1+q_2)}{(q_1^2-q_2^2)}\nonumber\\
 	&=&-\frac{-\left[(\mu+\mu')q_1+(\mu'-\mu)q_2\right]}{(q_1^2-q_2^2)}\nonumber
 \end{eqnarray}
 \begin{eqnarray}
 	r^2-r'^2&=&-4cx~~~
 	\mbox{i.e.,~}(q_1-q_2)^2-(q_1+q_2)^2=4cx\nonumber\\
 	\mbox{i.e.,~}x&=&-\frac{q_1q_2}{c}~~~
 	\implies\dot{x}=-\frac{1}{c}(\dot{q_1}q_2+q_1\dot{q_2})\nonumber
 \end{eqnarray}
 Similarly,
 \begin{eqnarray}
 	r^2+r'^2&=&2(x^2+c^2)+2y^2\nonumber\\
 	\mbox{i.e.,~}2(q_1^2+q_2^2)&=&2\left(\frac{q_1^2q_2^2}{c^2}+c^2\right)+2y^2\nonumber\\
 	\therefore{y}&=&\frac{1}{c}\left[(c^2-q_1^2)^{\frac{1}{2}}(q_2^2-c^2)^{\frac{1}{2}}\right]\nonumber
 \end{eqnarray}
 \begin{eqnarray}
 	\therefore\dot{y}&=&\frac{1}{c}\left[-\frac{q_1\dot{q_1}}{\sqrt{c^2-q_1^2}}\sqrt{q_2^2-c^2}+\frac{q_2\dot{q_2}}{\sqrt{q_2^2-c^2}}\sqrt{c^2-q_1^2}\right]\nonumber
 \end{eqnarray}
 \begin{eqnarray}
 	\therefore\dot{x}^2+\dot{y}^2&=&\frac{1}{c^2}(\dot{q_1}q_2+q_1\dot{q_2})^2+\frac{1}{c^2}\left[\frac{q_2^2\dot{q_2}^2(c^2-q_1^2)}{q_2^2-c^2}+\frac{1}{c^2}\frac{q_1^2\dot{q_1}^2(q_2^2-c^2)}{c^2-q_1^2}-\frac{1}{c^2}2q_1q_2\dot{q_1}\dot{q_2}\right]\nonumber\\
 	&=&\frac{1}{c^2}\left[\dot{q_1}^2q_2^2+\dot{q_2}^2{q_1}^2+\frac{q_1^2\dot{q_1}^2(q_2^2-c^2)}{(c^2-q_1^2)}+\frac{q_2^2\dot{q_2}^2(c^2-q_1^2)}{(q_2^2-c^2)}\right]\nonumber\\
 	&=&\frac{1}{c^2}\left[\dot{q_1}^2\left\{\frac{(c^2-q_1^2)q_2^2+q_1^2(q_2^2-c^2)}{(c^2-q_1^2)}\right\}+\dot{q_2}^2\left\{\frac{q_1^2(q_2^2-c^2)+q_2^2(c^2-q_1^2)}{(q_2^2-c^2)}\right\}\right]\nonumber\\
 	&=&\frac{1}{c^2}\left[\frac{\dot{q_1}^2c^2(q_1^2-q_2^2)}{(c^2-q_1^2)}+\frac{\dot{q_2}^2c^2(q_1^2-q_2^2)}{(q_2^2-c^2)}\right]\nonumber\\
 	&=&(q_1^2-q_2^2)\left[\frac{\dot{q_1}^2}{c^2-q_1^2}+\frac{\dot{q_2}^2}{q_2^2-c^2}\right]\nonumber\\
 	\therefore{2T}=\dot{x}^2+\dot{y}^2&=&(q_1^2-q_2^2)\left[\frac{\dot{q_1}^2}{c^2-q_1^2}+\frac{\dot{q_2}^2}{q_2^2-c^2}\right]\nonumber
 \end{eqnarray}
 Hence the problem is of Lioville's type and hence can solved.
 
 \vspace{.5cm}
 
 $\bullet$ \textbf{{Problem: 4} }
 
 \vspace{.25cm}
 
 The K.E. of a particle whose rectangular co-ordinates $(x,y)$ is $\frac{1}{2}(\dot{x}^2+\dot{y}^2)$ and its P.E. is $\frac{A}{x^2}+\frac{A'}{y^2}+\frac{B}{r}+\frac{B'}{r'}+c(x^2+y^2),$  where $A,~B,~A',~B'$ and $C$ are constants $r,~r'$ are the distances of the particle form the points whose co-ordinates are $(\pm{c},0),~c$ being a constant. Show that when the quantities $\frac{1}{2}(r\pm{r'})$ are taken as new variables the system is of Lioville's type.
 
 \vspace{.25cm}
 
 \textbf{{Solution} :}
 
 \vspace{.25cm}
 
 Let $r=q_1+q_2,~r'=q_1-q_2$ so as in previous problem 
 \begin{eqnarray}
 	{2T}&=&(q_1^2-q_2^2)\left[\frac{\dot{q_1}^2}{c^2-q_1^2}+\frac{\dot{q_2}^2}{q_2^2-c^2}\right]\nonumber\\
 	V&=&\frac{A}{x^2}+\frac{A'}{y^2}+\frac{B}{r}+\frac{B'}{r'}+c(x^2+y^2)\nonumber\\
 	&=&\frac{Ac^2}{q_1^2q_2^2}+\frac{A'c^2}{(c^2-q_1^2)(q_2^2-c^2)}+\frac{(B+B')q_1+(B'-B)q_2}{q_1^2-q_2^2}+C(q_1^2+q_2^2-c^2)\nonumber\\
 	&=&\frac{Ac^2(q_1^2-q_2^2)}{q_1^2q_2^2(q_1^2-q_2^2)}+\frac{A'c^2(q_1^2-q_2^2)}{(c^2-q_1^2)(q_2^2-c^2)(q_1^2-q_2^2)}\nonumber\\
 	&~&~~~~~~~~~~~+\frac{(B+B')q_1+(B'-B)q_2}{(q_1^2-q_2^2)}+\frac{c(q_1^2-q_2^2)(q_1^2+q_2^2-c^2)}{(q_1^2-q_2^2)}\nonumber\\
 	&=&\frac{Ac^2\left(\frac{1}{q_2^2}-\frac{1}{q_1^2}\right)}{(q_1^2-q_2^2)}+\frac{A'c^2\left(-\frac{1}{c^2-q_1^2}-\frac{1}{q_2^2-c^2}\right)}{(q_1^2-q_2^2)}+\frac{(B+B')q_1+(B'-B)q_2}{q_1^2-q_2^2}\nonumber\\
 	&~&~~~~~~~~~~~~~~~~~~~~~~~~~~~~~~~~~~~~~~~~~~~~~~~~~~~~~~~~+\frac{c(q_1^4-c^2q_1^2)}{(q_1^2-q_2^2)}+\frac{c(q_2^4-c^2q_2^2)}{(q_1^2-q_2^2)}\nonumber
 \end{eqnarray}
 i.e., $V$ is of the form $\frac{w_1(q_1)+w_2(q_2)}{v_1(q_1)+v_2(q_2)}$ where 
 \begin{equation}
 	w_1(q_1)=-\frac{Ac^2}{q_1^2}+\frac{A'c^2}{q_1^2-c^2}+(B+B')q_1+c(q_1^4-c^2q_1^2)\nonumber
 \end{equation}
 and
 \begin{equation}
 	w_2(q_2)=\frac{Ac^2}{q_2^2}+\frac{A'c^2}{c^2-q_2^2}+(B-B')q_2+c(q_2^4-c^2q_2^2)\nonumber
 \end{equation}
 Hence the problem is Lioville's type.
 
 \vspace{.5cm}
 
\section{Hamiltonian of a Mechanical system: } 
 
 \vspace{.25cm}

 Suppose a dynamical system with $n$ degrees of freedom is described by the Lagrangian $L(q,\dot{q},t)$. The generalised momentum $p$ associated with the generalised coordinate $q$ is given by 
 \begin{equation}
 	p=\frac{\partial{L}}{\partial\dot{q}}\label{39}
 \end{equation}
 We introduce a function $H$ such that
 \begin{equation}
 	H=\sum_{i=1}^{n}p_i\dot{q}_i-L\label{40}
 \end{equation}
 As $L=T-V$, contains terms of 2nd degree in $\dot{q}$'s, so $p$'s are linear function of  $\dot{q}$'s. Therefore, solving the equation (\ref{39}) for $\dot{q}$'s we can express them as function of ${q}$'s, $p$'s and $t$. Thus $H$ is a function of $q$'s, $p$'s and $t$ i.e., $H=H(q,p,t)$. The function so constructed is called the Hamiltonian of a dynamical system.
 
 \vspace{.5cm}
 
 \section{Hamilton's canonical equations of motion:}
 
 \vspace{.25cm}
 
 Consider a dynamical system described at any instant by the Hamiltonian $H$ so that 
 \begin{equation}
 	H=\sum_{i=1}^{n}p_i\dot{q}_i-L\nonumber
 \end{equation} 
 where $p_i=\frac{\partial{L}}{\partial\dot{q}_i}$.
 
 Let us consider a virtual variation of $H$ at time instant $t$ i.e., we have 
 \begin{eqnarray}
 &~&\delta{H}=\delta\left(\sum_{i=1}^{n}p_i\dot{q}_i\right)-\delta{L}\nonumber\\
 	&~&\mbox{i.e.,}\sum_{i=1}^{n}\frac{\partial{H}}{\partial\dot{q}_i}\delta{q_i}+\sum_{i=1}^{n}\frac{\partial{H}}{\partial{p}_i}\delta{p_i}+\frac{\partial{H}}{\partial{t}}\delta{t}=\cancel{\sum_{i=1}^{n}p_i\delta\dot{q_i}}+\sum_{i=1}^{n}\dot{q}_i\delta{p_i}-\sum_{i=1}^{n}\frac{\partial{L}}{\partial{q}_i}\delta{q_i}-\cancel{\sum_{i=1}^{n}\frac{\partial{L}}{\partial\dot{q}_i}\delta{\dot{q}_i}}-\frac{\partial{L}}{\partial{t}}\delta{t}\nonumber
 \end{eqnarray}
 As $p_i=\frac{\partial{L}}{\partial\dot{q}_i}$ and from Lagrange's equation of motion i.e., $\frac{d}{dt}\left(\frac{\partial{L}}{\partial{\dot{q}_i}}\right)=\frac{\partial{L}}{\partial{q}_i}$ we have $\dot{p}_i=\frac{\partial{L}}{\partial{q}_i}$ so the equation simplifies to 
 \begin{eqnarray}
 	\sum_{i=1}^{n}\frac{\partial{H}}{\partial{q}_i}\delta{q_i}+\sum_{i=1}^{n}\frac{\partial{H}}{\partial{p}_i}\delta{p_i}+\frac{\partial{H}}{\partial{t}}\delta{t}=\sum_{i=1}^{n}\dot{q}_i\delta{p_i}-\sum_{i=1}^{n}\dot{p}_i\delta{q_i}-\frac{\partial{L}}{\partial{t}}\delta{t}\nonumber\\
 	\mbox{i.e.,}\sum_{i=1}^{n}\left(\frac{\partial{H}}{\partial{q_i}}+\dot{p}_i\right)\delta{q}_i+\sum_{i=1}^{n}\left(\frac{\partial{H}}{\partial{p_i}}-\dot{q}_i\right)\delta{p}_i+\left(\frac{\partial{H}}{\partial{t}}+\frac{\partial{L}}{\partial{t}}\right)\delta{t}=0\nonumber
 \end{eqnarray}
 As $q$ and $p$'s are independent so we have 
 \begin{equation}
 	\frac{\partial{H}}{\partial{q}_i}=-\dot{p}_i,~\frac{\partial{H}}{\partial{p}_i}=\dot{q}_i,~~i=1,2,...,n.\nonumber
 \end{equation}
 and $\dfrac{\partial{H}}{\partial{t}}=-\dfrac{\partial{L}}{\partial{t}}$.
 
 The above set of $2n$ equations of 1st order in $2n$ unknown $p_i$'s and $q_i$'s of a system having $n$ degrees of freedom are called the Hamilton's canonical equations of motion.
 
 \vspace{.5cm}
 
 \section{Physical significance of $H$ in case of a Sceleronomic system:}
 
 \vspace{.25cm}
 
 As in a Sceleronomic system $T$ is a homogeneous quadratic in $\dot{q}$'s and $V$ is independent of $\dot{q}$'s so $L=T-V$ is also quadratic in $\dot{q}$'s.
 
 Thus
 \begin{eqnarray}
 	H&=&\sum{p}_i\dot{q}_i-L=\sum\dot{q}_i\frac{\partial{L}}{\partial{\dot{q}_i}}-L=\sum\dot{q}_i\frac{\partial{T}}{\partial{\dot{q}_i}}-L=2T-(T-V)=T+V\nonumber\\
 	&=&\mbox{Total energy of the system.}\nonumber
 \end{eqnarray}
 
 \textbf{\underline{Note-I} :}
 
 As $H=\sum{p}_i\dot{q}_i-L$, so if a coordinate $q_k$ is absent in Lagrangian it also absent in $H$ i.e., cyclic coordinates do not occur in $H$.
 
 \textbf{\underline{Note-II} :}
 
 In a Sceleronomic system
 \begin{eqnarray}
 	\frac{dH}{dt}&=&\sum\left(\frac{\partial{H}}{\partial{q}_i}.\frac{dq_i}{dt}+\frac{\partial{H}}{\partial{p}_i}.\frac{dp_i}{dt}\right)=\sum_{i}\left(\dot{q}_i\frac{\partial{H}}{\partial{q}_i}+\dot{p}_i\frac{\partial{H}}{\partial{p}_i}\right)\nonumber\\
 	&=&\sum_{i}\left(\frac{\partial{H}}{\partial{p}_i}.\frac{\partial{H}}{\partial{q}_i}-\frac{\partial{H}}{\partial{q}_i}.\frac{\partial{H}}{\partial{p}_i}\right)=0\nonumber
 \end{eqnarray}
 So $H$ is constant.
 
 As in a Sceleronomic system $H=T+V$, hence K.E + P.E. is constant for such system. However, if $H$ has explicit time dependence then 
 \begin{equation}
 	\frac{dH}{dt}=\frac{\partial{H}}{\partial{t}}.\nonumber
 \end{equation}

 \section{Problems}

 $\bullet$ \textbf{Problem:} If the Hamiltonian of a dynamical system is given by $H=p_{1}q_{1}-p_{2}q_{2}-aq_{1}^{2}+bq_{2}^{2}$ where $a$, $b$ are constants, then show that $\dfrac{p_{2}-bq_{2}}{q_{1}}$ is constant and also find the position of the system.
 
 \vspace{.5cm}
 
 $\bullet$ \textbf{Problem:} If $H=qp^{2}-qp+bp$, be the Hamiltonian of a system then find $q$ and $p$ as function of $t$.
 
 \vspace{.5cm}
 
 $\bullet$ \textbf{Problem:} If all the coordinates of a system are cyclic then prove that the coordinate may be found by integration.
 
 \vspace{.25cm}
 
 \textbf{Solution:} If all the coordinates are ignorable then $H=H(p)$ so $\dot{p_{i}} =-\dfrac{\partial H}{\partial q_{i}}=0$ i.e $p_{i}$'s are constant, and $\dot{q_{i}}=\dfrac{\partial H}{\partial p_{i}}=f_{i}(t)$ ($\because p_{i}$'s are constants). Therefore $q_{i}=\int f_{i}(t)dt + \alpha_{i}$, $i=1,2,...,n$. Further, if the system is scelronomic, then $H=H(p_{i})$ and $\dot{q_{i}}=\dfrac{\partial H}{\partial p_{i}}=K_{i}$ (say) i.e $q_{i}=K_{i}t+\alpha_{i}, i=1,2,...,n$. So generalized coordinates are linear functions of $t$.
 
 \vspace{.5cm}
 
 $\bullet$ \textbf{Problem:} If the K.E. $T(q,\dot{q})$ of a system is a homogeneous quadratic in the velocities $\dot{q}$'s and $T'(q,p)$ is what $T$ becomes when expressed in terms of variables $q$ and $p$, then prove that
 
 (i) $\dot{q_{i}}=\dfrac{\partial T'}{\partial p_{i}}$, (ii) $\dfrac{\partial T}{\partial q_{i}}+\dfrac{\partial T'}{\partial q_{i}}=0$

 (iii) $T'$ is a homogeneous quadratic in p

 (iv) $T+T'=\sum p_{i}\dot{q_{i}}$
 
 \vspace{.25cm}
 
 \textbf{Solution:} (i) $\dot{q_{i}}=\dfrac{\partial H}{\partial p_{i}}=\dfrac{\partial (T'+V)}{\partial p_{i}}=\dfrac{\partial T'}{\partial p_{i}}$ ($\because V$ is a function of $q$'s only)
 
 (ii) $\dfrac{\partial H}{\partial q_{i}}=-\dot{p_{i}}=-\dfrac{\partial L}{\partial q_{i}}$. This implies $\dfrac{\partial (T'+V)}{\partial q_{i}}=-\dfrac{\partial(T-V)}{\partial q_{i}}$ i.e $\dfrac{\partial T'}{\partial q_{i}}+\dfrac{\partial T}{\partial q_{i}}=0$.
 
 (iii) $T$ is a homogeneous quadratic function of $\dot{q_{i}}$'s i.e $2T=\sum\dot{q_{i}}\dfrac{\partial T}{\partial \dot{q_{i}}}=\sum \dot{q_{i}}\dfrac{\partial L}{\partial \dot{q_{i}}}=\sum_{i} p_{i}\dot{q_{i}}$.
 Therefore $2T'=\sum p_{i}\dfrac{\partial T'}{\partial p_{i}}$ (using the result of (i)) which shows that $T'$ is a homogeneous quadratic in $p_{i}$'s.
 
 (iv) $H=\sum p_{i}\dot{q_{i}}-L$, i.e $T'+V=\sum p_{i}\dot{q_{i}}-(T-V)$, i.e $T+T'=\sum p_{i}\dot{q_{i}}$.
 
 \vspace{.5cm}
 
 $\bullet$ \textbf{Problem:} If the K.E of a scleronomic system is $T(q,\dot{q_{i}})=\dfrac{1}{2}\sum a_{ij}\dot{q_{i}}\dot{q_{j}}$ and if $T'(q,p)$ is what $T$ becomes when expressed in terms of $q$ and $p$ then show that $2T'+\dfrac{D}{\Delta}=0$ where $\Delta=det(a_{ij})$, 
 \[
 D =
 \begin{vmatrix}
 	a_{11} & a_{12}  & \dots & a_{1n} & p_{1} \\ 
 	a_{21} & a_{22}  & \dots & a_{2n} & p_{2} \\
 	\hdotsfor{5} \\
 	a_{n1} &  a_{n2} & \dots & a_{nn} & p_{n}\\
 	p_{1} & p_{2} &  \dots & p_{n} & 0
 \end{vmatrix}
 \]
 
 \textbf{Proof:} By definition
 $p_{i}=\dfrac{\partial T}{\partial \dot{q_{i}}}=\sum_{i} a_{ij}\dot{q_{j}}$, i.e $\sum_{j=1}^{n} a_{ij}\dot{q_{j}}-p_{i}=0$ or in explicit form
 \begin{eqnarray}
 	a_{11}\dot{q_{1}}+a_{12}\dot{q_{2}}+...+a_{1n}\dot{q_{n}}-p_{1}=0\nonumber\\ 
 	a_{21}\dot{q_{1}}+a_{22}\dot{q_{2}}+...+a_{2n}\dot{q_{n}}-p_{2}=0\nonumber\\ 
 	......	...................................................\nonumber\\
 	a_{n1}\dot{q_{1}}+a_{n2}\dot{q_{2}}+...+a_{nn} \dot{q_{n}}-p_{n}=0\nonumber
 \end{eqnarray}
 Also, $2T'=\sum_{i}p_{i}\dot{q_{i}}$ i.e
 \begin{equation}
 	p_{1}\dot{q_{1}}+p_{2}\dot{q_{2}}+...+p_{n}\dot{q_{n}}-2T'=0.\nonumber\end{equation} Now eliminating $\dot{q_{1}}. $$\dot{q_{2}}$,...,$\dot{q_{n}}$ we have
 \[
 \begin{vmatrix}
 	a_{11} & a_{12}  & \dots & a_{1n} & p_{1} \\ 
 	a_{21} & a_{22}  & \dots & a_{2n} & p_{2} \\
 	\hdotsfor{5} \\
 	a_{n1} &  a_{n2} & \dots & a_{nn} & p_{n}\\
 	p_{1} & p_{2} &  \dots & p_{n} & 2T'
 \end{vmatrix}=0
 \] i.e
 \[
 \begin{vmatrix}
 	a_{11} & a_{12}  & \dots & a_{1n} & p_{1}+0 \\ 
 	a_{21} & a_{22}  & \dots & a_{2n} & p_{2}+0 \\
 	\hdotsfor{5} \\
 	a_{n1} &  a_{n2} & \dots & a_{nn} & p_{n}+0\\
 	p_{1} & p_{2} &  \dots & p_{n} & 0+2T'
 \end{vmatrix}=0\] i.e
 \[D+
 \begin{vmatrix}
 	a_{11} & a_{12}  & \dots & a_{1n} & 0 \\ 
 	a_{21} & a_{22}  & \dots & a_{2n} & 0 \\
 	\hdotsfor{5} \\
 	a_{n1} &  a_{n2} & \dots & a_{nn} & 0\\
 	p_{1} & p_{2} &  \dots & p_{n} & 2T'
 \end{vmatrix}=0\] i.e
 $~~~~~~~~~~~~~~~~~~~~~~~~~~~~~~~~~~~~~~~~~D+2T' \Delta=0$ or, $2T'+\dfrac{D}{\Delta}=0$.
 
 \vspace{.25cm}
 
 $\bullet$ \textbf{Problem:} Solution of Planetary motion using Hamilton's canonical equations:\\
 The K.E of a planet in polar coordinates is 
 \begin{eqnarray}
 	T=\dfrac{m}{2}(\dot{r}^{2}+r^{2}\dot{\theta}^{2})-\dfrac{\mu}{r},~~ V=-\dfrac{\mu}{r}.\nonumber
 \end{eqnarray}
 $~~~~~~~~~~~~~~~~~~~~~~ \therefore H=T+V=\dfrac{m}{2}(\dot{r}^{2}+r^{2}\dot{\theta}^{2})-\dfrac{\mu}{r}$.\\
 Let $p_{r}$, $p_{\theta}$ are the generalized momenta corresponding to $r$ and $\theta$ variables then 
 
 $p_{r}=\dfrac{\partial T}{\partial \dot{r}}=m\dot{r}$,~~ $p_{\theta}=\dfrac{\partial T}{\partial \dot{\theta}}=mr^{2}\dot{\theta}$.
 
 $\therefore \dot{r}= \dfrac{p_{r}}{m},~~ \dot{\theta}=\dfrac{p_{\theta}}{mr^{2}}$.
 
 $H=\dfrac{m}{2}\left(\dfrac{p_{r}^{2}}{m^{2}}+r^{2} \dfrac{p_{\theta}^{2}}{m^{2}r^{4}}\right)-\dfrac{\mu}{r}=\dfrac{1}{2m}\left(p_{r}^{2}+\dfrac{p_{\theta}^{2}}{r^{2}}\right)-\dfrac{\mu}{r}$. 
 
 The Hamilton's canonical equations are
 \begin{eqnarray}
 	\dot{r}=\dfrac{\partial H}{\partial p_{r}}=\dfrac{p_{r}}{m},\nonumber\\
 	\dot{\theta}=\dfrac{\partial H}{\partial p_{\theta}}=\dfrac{p_{\theta}}{mr^{2}}.\nonumber
 \end{eqnarray} 
 
 Also, $\dfrac{\partial H}{\partial r}=\dfrac{-p_{\theta}^{2}}{mr^{3}}+\dfrac{\mu}{r^{2}}=-\dot{p_{r}}$
 
 and $\dfrac{\partial H}{\partial \theta}=0$ i.e $\dot{p_{\theta}}=0$ or $p_{\theta}=\mbox{Constant}=h~(\mbox{say}),$. 
 
 (note that $\theta$ is a cyclic coordinate).
 
 Thus, $r^{2}\dot{\theta}=\dfrac{h}{m}=c$(say) i.e, $\dot{\theta}=\dfrac{c}{r^{2}}$
 
 and $\dot{p_{r}}=\dfrac{h^{2}}{mr^{3}}-\dfrac{\mu}{r^{2}}$. 
 
 i.e, 
 \begin{equation}
 	\dfrac{d^{2}r}{dt^{2}}=\dfrac{h^{2}}{m^{2}r^{3}}-\dfrac{\mu}{mr^{2}}=\dfrac{c^{2}}{r^{3}}-\dfrac{l}{r^{2}}\nonumber
 \end{equation}
 i.e, 
 \begin{equation}
 	m(\ddot{r}-r\dot{\theta}^{2})=-\dfrac{\mu}{r^{2}},\nonumber
 \end{equation} 
 
 which is the radial equation of motion.


\chapter{Dynamics from Hamilton's principle}

\section{Hamilton's principle}
Let $q_{1}, q_{2},..., q_{n}$ be the independent coordinates of a holonomic dynamical system with $n$ degrees of freedom. Let us suppose that we have a point P in an ($n+1$)-dimensional space whose coordinates are ($q_{1}, q_{2},..., q_{n}, t$). This point is called a representative point of the configuration of the system. Let $P_{0}$ be the representative point of the configuration of the system at time $t=t_{0}$ and $P_{1}$ be the representative point at time $t=t_{1}$. Then as the system moves the representative point describes a curve $C$ joining the points $P_{0}$ and $P_{1}$. This curve is called the trajectory of the system. Let us suppose that we have another curve $C'$ in this space, passing through the points $P_{0}$ and $P_{1}$ and lying infinitely near to the curve $C$.
\begin{figure}
	\centering
	\includegraphics[width=0.4\textwidth]{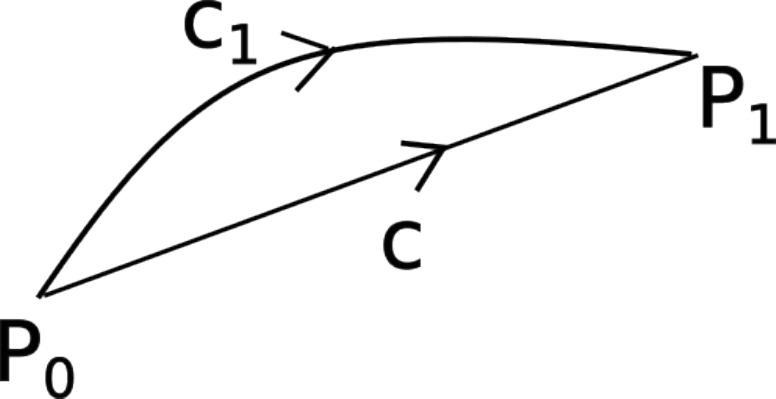}\\
	\label{madhu1}
\end{figure}

All along the curve $C$, the equations of motion of the system are satisfied and we call it a dynamically possible curve. The curve $C'$ is geometrically possible but dynamically impossible i.e the values of $q_{1}, q_{2}, ..., q_{n}$ and $\dot{q_{1}}, \dot{q_{2}}, ..., \dot{q_{n}}$ as obtained from the curve $C'$ do not satisfy the equation of motion.

Let
\begin{equation}
S=\int_{t_{0}}^{t_{1}}\left(T+U\right)~dt=\int_{t_{0}}^{t_{1}}\left(T-V\right)~dt=\int_{t_{0}}^{t_{1}}~L~dt,\nonumber
\end{equation} where $T$ is the K.E of the system, $U$ is the force  function from which the external forces are derived and $L$ is the Lagrangian of the system. The function $S$ is called the Hamilton's principle function for the motion of the system. When we consider a case of $(n+1)$ dimensions the integral $S$ in an actual motion of the system is taken along the curve $C$ and in any other geometrically possible motion it is taken along the curve $C'$.

Hamilton's principle states that the principle function $S$ is stationary when the integral is taken along the actual trajectory of the system as compare to all other geometrically possible trajectories differ infinitesimally from the actual trajectory $C$ but having the same terminal points as the actual trajectory.

Of all possible motion of a mechanical system co-terminus in space and time, the actual motion is that for which
\begin{equation}
\delta\int_{t_{0}}^{t_{1}}~L~dt=0.\nonumber
\end{equation}
\section{Verification of Hamilton's principle from D'Alembert's principle}
According to D'Alembert's principle among all motions of the system consistent with the constraints the actual motion is that which satisfies at every moment the equation
\begin{equation}
\sum (m{\ddot{\overrightarrow{r}}}-\overrightarrow{F}). \delta \overrightarrow{r} =0.\nonumber
\end{equation} 
Now, $\sum \overrightarrow{F}.\delta \overrightarrow{r}=\delta U$, $U$ is the work function,
\begin{eqnarray}
 \mbox{and}~~~\sum m{\ddot{\overrightarrow{r}}} \delta \overrightarrow{r}&=&\dfrac{d}{dt}~(\sum m{\dot{\overrightarrow{r}}} \delta \overrightarrow{r})-\sum m\dot{\overrightarrow{r}}.\delta \dot{\overrightarrow{r}}\nonumber\\
 &=&\dfrac{d}{dt} (\sum m{\dot{\overrightarrow{r}}}.\delta \overrightarrow{r})-\delta T.\nonumber
 \end{eqnarray}
\begin{eqnarray}
 \therefore~~\sum \left(m{\ddot{\overrightarrow{r}}}-\overrightarrow{F}\right).\delta \overrightarrow{r}&=&\dfrac{d}{dt} ~\sum ( m{\dot{\overrightarrow{r}}} \delta \overrightarrow{r})-\delta T-\delta U\nonumber\\
 &=&\dfrac{d}{dt} (\sum m{\dot{\overrightarrow{r}}} \delta \overrightarrow{r})-\delta L.\nonumber
 \end{eqnarray}
\begin{eqnarray}
\mbox{ Hence}~~ \sum (m{\ddot{\overrightarrow{r}}}-\overrightarrow{F})\delta \overrightarrow{r}=0\nonumber\\
 \implies \dfrac{d}{dt}(\sum m \dot{\overrightarrow{r}}.\delta \overrightarrow{r})-\delta L=0.\nonumber
 \end{eqnarray} Now integrating both sides w.r.t $t$ from $t_{0}$ to $t_{1}$ we have
\begin{equation}
\int_{t_{0}}^{t_{1}}~\delta L~ dt=\int_{t_{0}}^{t_{1}}\dfrac{d}{dt}\left(\sum m~{\dot{\overrightarrow{r}}}\delta \overrightarrow{r}\right)dt=\sum m \dot{\overrightarrow{r}}\delta \overrightarrow{r}|^{t_{1}}_{t_{0}}=0\nonumber
\end{equation} as $\delta \overrightarrow{r}=0$ when $t=t_{0}, t_{1}.$

 $\therefore \delta\int_{t_{0}}^{t_{1}}Ldt=0$ i.e. $\delta S=0$, which is Hamilton's principle. \\
\section{Hamilton's equations from Hamilton's principle}
Let $(q_{1}, q_{2},...,q_{n})$ and $(q_{1}+\delta q_{1}, q_{2}+\delta q_{2},...,q_{n}+\delta q_{n})$ be the generalized coordinates of the system corresponding to the paths $C$ and $C_{1}$ at the same time $t$ with $\delta q_{i}=\epsilon \eta_{i}(t)$. By construction, $\delta q_{1},...,\delta q_{n}$ are quite arbitrary except that $\delta q_{_{i}}=0$ at $t=t_{0}$ and $t=t_{1}$, but there is no restrictions on $\delta\dot{ q_{i}}$'s. As $p_{i}=\dfrac{\partial L}{\partial \dot{ q_{i}}}$ is linear in $\dot{ q_{i}}$'s so $\delta p_{i}$'s are all arbitrary. Now
\begin{equation}
\delta S=\int_{t_{0}}^{t_{1}} \delta L~dt=\int_{t_{0}}^{t_{1}}\delta \left[\sum p_{i}~\dot{ q_{i}}-H\right]~dt\nonumber
\end{equation}
As $\dot{ q_{i}}=\dfrac{d}{dt}(\delta q_{i})$ (see appendix I) so we have
\begin{equation}
\delta \left(\sum p_{i}~\dot{ q_{i}}\right)=\dfrac{d}{dt}(\sum p_{i}~\delta q_{i})+\sum \dot{ q_{i}} \delta p_{i}-\sum_{i} \dot{p_{i}}~\delta q_{i},\nonumber
\end{equation} and $\delta H=\sum_{i} \dfrac{\partial H}{\partial p_{i}}~\delta p_{i}+\sum_{i}\dfrac{\partial H}{\partial q_{i}}~\delta q_{i}$. Also we have $\delta \dot{ q_{i}}=\dfrac{d}{dt}(\delta q_{i})$. Hence,
\begin{eqnarray}\nonumber
\delta S&=&\int_{t_{0}}^{t_{1}}\dfrac{d}{dt}(\sum p_{i} \delta q_{i})+\sum \dot{ q_{i}}\delta p_{i}-\sum \dot{p_{i}}\delta q_{i}-\sum \dfrac{\partial H}{\partial p_{i}}\delta p_i-\sum \dfrac{\partial H}{\partial q_{i}}\delta q_{i}\\&=&\int_{t_{0}}^{t_{1}}\sum_{i}\left[(q_{i}-\dfrac{\partial H}{\partial p_{i}})\delta p_{i}-(\dot{p_{i}}+\dfrac{\partial H}{\partial q_{i}})\delta q_{i}\right]~dt+\sum p_{i}\delta q_{i}|^{t_{1}}_{t_{0}}\nonumber\\&=&\int_{t_{0}}^{t_{1}}\sum_{i}\left[(\dot{ q_{i}}-\dfrac{\partial H}{\partial p_{i}})\delta p_{i}-(\dot{p_{i}}+\dfrac{\partial H}{\partial q_{i}})\delta q_{i}\right]dt \label{eq2..1}
\end{eqnarray} ~~~~~~~~~~~~~~~~~~~~~~~~~~~~~~~~~~~~~~~~~~~~~~~~~~~($\because \delta q_{i}=0$ at $t=t_{0}$ and $t=t_{1}$).

 If we assume Hamilton's equation of motion then R.H.S will be zero and hence $\delta S=0$. This verifies Hamilton's principle on the assumption of Hamilton's canonical equations.

\vspace{0.0cm}

On the other hand, if $\delta S=0$ for all arbitrary variation $\delta q_{i}$'s and $\delta p_{i}$'s then equation (\ref{eq2..1}) shows that the integral on the R.H.S vanishes for all these arbitrary variations. Thus we get $$\dot{ q_{i}}=\dfrac{\partial H}{\partial p_{i}}, \dot{p_{i}}=-\dfrac{\partial H}{\partial q_{i}}.$$ These are Hamilton's canonical equations. Thus Hamilton's canonical equations can be obtained from Hamilton's principle.\\
\section{Lagrange's equation from Hamilton's principle}
Let $q_{1}, q_{2},...,q_{n}$ be the independent coordinates of the holonomic dynamical system with $n$ degrees of freedom, moving under the action of external forces derived from a force function $U$ and let $L=T+U$ be the Lagrangian function of the system. Then the Hamilton's principle function is\\
$~~~~~~~~~~~~~~~~~~~~~~~~~~~~~~~~~~~~~~~~~~~~~~~~~~~~~S=\int_{t_{0}}^{t_{1}}L dt$,~~~~~~~~ $L=L(q,\dot{q},t)$.\\
Thus $\delta S=\int_{t_{0}}^{t_{1}}\delta L dt=\int_{t_{0}}^{t_{1}}\sum_{i}\left(\dfrac{\partial L}{\partial q_{i}}\delta q_{i}+\dfrac{\partial L}{\partial \dot{ q_{i}}}\delta\dot{q_i}\right)dt$.

 But $\delta \dot{ q_{i}}=\dfrac{d}{dt}(\delta q_{i})$, so
\begin{eqnarray}
\int_{t_{0}}^{t_{1}} \dfrac{\partial L}{\partial \dot{ q_{i}}}\delta \dot{ q_{i}}dt&=&\int_{t_{0}}^{t_{1}}\dfrac{\partial L}{\partial \dot{ q_{i}}}\dfrac{d}{dt}(\delta q_{i})dt\nonumber\\
&=&\int_{t_{0}}^{t_{1}}\dfrac{d}{dt}\left(\dfrac{\partial L}{\partial \dot{ q_{i}}}\delta q_{i}\right)dt-\int_{t_{0}}^{t_{1}}\delta q_{i}\dfrac{d}{dt}\left(\dfrac{\partial L}{\partial \dot{ q_{i}}}\right)dt\nonumber\\
&=&
\dfrac{\partial L}{\partial \dot{ q_{i}}} \delta q_{i}|^{t_{1}}_{t_{0}}-\int_{t_{0}}^{t_{1}}\dfrac{d}{dt}\left(\dfrac{\partial L}{\partial \dot{ q_{i}}}\right)\delta q_{i}dt\nonumber\\
&=&
-\int_{t_{0}}^{t_{1}}\dfrac{d}{dt}\left(\dfrac{\partial L}{\partial \dot{ q_{i}}}\right)\delta q_{i}dt,\nonumber
\end{eqnarray} $~~~~~~~~~~~~~~~~~~~~~~~~~~~~~~~~~~~~~~~~~~~~~~~~~~~~~~~~~~~~\because \delta q_{i}=0$ at $t=t_{0},t_{1}$.\\
$~~~~~~~~~~~~~~~~~~~~~~~~~~~~~~~~~~~~\therefore \delta S= \int_{t_{0}}^{t_{1}}\left[\sum_{i}\dfrac{\partial L}{\partial q_{i}}-\dfrac{d}{dt}\left(\dfrac{\partial L}{\partial \dot{ q_{i}}}\right)\right]\delta q_{i}dt.$\\
For $\delta S=0$, when $S$ is taken along the actual trajectory, it is necessary and sufficient that the integral on the R.H.S vanishes. Since $\delta q_{i}$'s are all arbitrary the necessary and sufficient condition for vanishing of $\delta S$ is that
\begin{equation}
\dfrac{\partial L}{\partial q_{i}}-\dfrac{d}{dt}\left(\dfrac{\partial L}{\partial \dot{ q_{i}}}\right)=0,~~~~~i=1,2,...,n. \nonumber
\end{equation} These are the Lagrange's equations of motion.\\
On the other hand, if we assume Lagrange's equations of motion then $\delta S=0$ when taken actual trajectory of the system as compared to any infinitely near trajectory so that Hamilton's principle is proved.\\
\section{Principle of least action:}
The action of a mechanical system is defined at the instant $t$ by the integral
\begin{equation}
A=\int_{t_{0}}^{t}\sum p\dot{q}dt,\nonumber
\end{equation} where $'p'$ is the generalized momentum corresponding to generalized coordinate $q$.\\
The Principle of least action states that the motion of a scelronomic conservative system between two prescribed configuration occurs in such a way that $\Delta A=0$, for the actual path as compared to the neighbouring paths provided the total energy of the system is the same constant value in the neighbouring motions as in the actual motion. Now, $A=\int_{t_{0}}^{t_{1}}\sum p~\dot{q}~dt$.\\
\begin{eqnarray}
\therefore\Delta A&=&\Delta\int_{t_{0}}^{t}\left(\sum p~\dot{q}\right)dt=\int_{t_{0}}^{t}\left[\delta (\sum p \dot{q})+\dfrac{d}{dt}(\sum p \dot{q})\Delta t\right]dt\nonumber\\
&=&\int_{t_{0}}^{t_{1}}\left[\sum p \delta \dot{q}+\sum \dot{q} \delta p\right]dt+\sum p \dot{q}\Delta t|^{t}_{t_{0}}.\nonumber
\end{eqnarray}
 We have $\Delta q=\delta q+\dot{q}\Delta t$ i.e. $\dot{q}\Delta t=\Delta q-\delta q.$
\begin{eqnarray}
\therefore\Delta  A&=&\sum p (\Delta q-\delta q)|^{t}_{t_{0}}+\int_{t_{0}}^{t}\left[\dfrac{d}{dt}\sum p\delta q-\sum \dot{p}\delta q\right]dt+\int_{t_{0}}^{t}\sum \dot{q}~\delta p ~dt\nonumber\\
&=&
\sum p(\Delta q-\delta q)|^{t}_{t_{0}}+(\sum p ~\delta q)|^{t}_{t_{0}}+\int_{t_{0}}^{t}\sum(\dot{q}~\delta p-\dot{p}~\delta q)dt\nonumber\\
&=&\sum p\Delta q|^{t}_{t_{0}}+\int_{t_{0}}^{t}\sum\left(\frac{\partial H}{\partial p}\delta p+\frac{\partial H}{\partial q}\delta q\right)dt\nonumber\\
&=&\sum p\Delta q|^{t}_{t_{0}}+\int_{t_{0}}^{t}\delta H~dt.\nonumber
\end{eqnarray}
As $\Delta q(t_{0})=\Delta q(t)=0$ and $H$ is the total energy which is conserved. So one gets $\Delta A=0$.\\
\section{Poisson Bracket}
Let $f(q,p,t)$ and $g(q,p,t)$ be two functions of the variables $q_{1},...,q_{n}$ and $p_{1},...,p_{n}$ and $t$. Then the expression
$$\sum_{i=1}^{n}\left(\dfrac{\partial f}{\partial q_{i}}~\dfrac{\partial g}{\partial p_{i}}-\dfrac{\partial f}{\partial p_{i}}~\dfrac{\partial g}{\partial q_{i}}\right)$$ is called the Poisson's bracket of the two quantities $f$ and $g$ with respect to the variables $q$ and $p$ and is denoted by $\{f,g\}_{(q,p)}$. 

\vspace{.25cm}

$\bullet$ \textbf{{Some important properties of Poisson bracket:}}\\
\textbf{~I.} $\{f,g\}=-\{g,f\}$\\
\textbf{II.} $\{f,c\}=0, c$ is any constant\\
\textbf{III.} $\{\sum_{i=1}^{n}a_{i}f_{i},g\}=\sum_{i=1}^{n}a_{i}\{f_{i},g\}$\\
\textbf{IV.} $\{f_{1}f_{2},g\}=f_{1}\{f_{2},g\}+f_{2}\{f_{1},g\}$\\
\textbf{V.} $\dfrac{\partial}{\partial t}\{f,g\}=\{\dfrac{\partial f}{\partial t},g\}+\{f,\dfrac{\partial g}{\partial t}\}$\\
$~~~~\dfrac{d}{dt}\{f,g\}=\{\dfrac{df}{dt},g\}+\{f,\dfrac{dg}{dt}\}$\\
\textbf{VI.} $\{f,q_{j}\}=-\dfrac{\partial f}{\partial p_{j}}, ~\{f,p_{j}\}=\dfrac{\partial f}{\partial q_{j}}$ \\
So $\{q_{i},q_{j}\}=0=\{p_{i},p_{j}\}, \{p_{i},q_{j}\}=-\delta_{ij}=\{q_{i},p_{j}\}$\\
Now if we write $H$ for $f$ then we get\\
$\dot{ q_{j}}=\dfrac{\partial H}{\partial p_{j}}=\{q_{j},H\}, \dot{p_{j}}=-\dfrac{\partial H}{\partial q_{j}}=\{p_{j},H\}$. These are the most symmetrical form of the canonical equations of motion of a mechanical system.\\
\textbf{VII. {Jacobi's identity:}} For any three functions $f,g$ and $h$ which are functions of $q,p,t$ the following identity holds:\\
$~~~~~~~~~~~~~~~~~~~~~~~~~~~~~~~~~~~~~~~~~~~~\{f,\{g,h\}\}+\{g,\{h,f\}\}+\{h,\{f,g\}\}=0$\\
Proof: By definition,
\begin{equation}
\{F,H\}=\sum_{i=1}^{n}\left(\dfrac{\partial F}{\partial q_{i}}~\dfrac{\partial H}{\partial p_{i}}-\dfrac{\partial F}{\partial p_{i}}~\dfrac{\partial H}{\partial q_{i}}\right)= D_{F}(H),\label{eq2..2}
\end{equation} where $D_{F}=\sum_{i=1}^{n}\left(\dfrac{\partial F}{\partial q_{i}}\dfrac{\partial}{\partial p_{i}}-\dfrac{\partial F}{\partial p_{i}}~\dfrac{\partial}{\partial q_{i}}\right)=\sum_{i=1}^{2n}F_{i}\dfrac{\partial}{\partial \xi_{i}}$\\
 with $\xi_{i}=p_{i},~\xi_{n+i}=q_{i},~F_{i}=\dfrac{\partial F}{\partial q_{i}},~F_{n+i}=-\dfrac{\partial F}{\partial p_{i}}$.\\
Now, $\{f,\{g,h\}\}+\{g,\{h,f\}\}=\{f,\{g,h\}\}-\{g,\{f,h\}\}=\{f,\lambda\}-\{g,\mu\}$, where $\lambda=\{g,h\}=D_{g}(h)$ and $\mu=\{f,h\}=D_{f}(h)$.

\begin{eqnarray}\therefore&& \{f,\{g,h\}\}+\{g,\{h,f\}\}+\{h,\{f,g\}\}\nonumber
	\\&=&D_{f}(\lambda)-D_{g}(\mu)=\sum_{j=1}^{2n}f_{j}\dfrac{\partial}{\partial \xi_{j}}(\lambda)-\sum_{i=1}^{2n}g_{i}\dfrac{\partial}{\partial \xi_{i}}(\mu)\nonumber\\
	&=&\sum_{j=1}^{2n}f_{j}\dfrac{\partial}{\partial \xi_{j}}\left(\sum_{i=1}^{2n}g_{i}~\dfrac{\partial h}{\partial \xi_{i}}\right)-\sum_{i=1}^{2n}g_{i}\dfrac{\partial}{\partial \xi_{i}}\left(\sum_{j=1}^{2n}f_{j}\dfrac{\partial h}{\partial \xi_{j}}\right)\nonumber\\
&=&\sum_{i,j=1}^{2n}\sum f_{j} \dfrac{\partial}{\partial \xi_{j}}\left(g_{i}\dfrac{\partial h}{\partial \xi_{i}}\right)-\sum_{i,j=1}^{2n}g_{i}\dfrac{\partial}{\partial \xi_{i}}\left(f_{j}\dfrac{\partial h}{\partial \xi_{j}}\right)\nonumber\\
&=&\sum_{i,j=1}^{2n}\left[f_{j}\left(\dfrac{\partial g}{\partial \xi_{j}}\dfrac{\partial h}{\partial \xi_{i}}+g_{i}\dfrac{\partial^{2}h}{\partial \xi_{i}\partial \xi_{j}}\right)-g_{i}\left(\dfrac{\partial f_{j}}{\partial \xi_{i}}\dfrac{\partial h}{\partial \xi_{j}}+f_{j}\dfrac{\partial^{2}h}{\partial \xi_{i}\partial \xi_{j}}\right)\right]\nonumber\\
&=&\sum_{i,j=1}^{2n}\left[f_{j}\dfrac{\partial g_{i}}{\partial \xi_{j}}\dfrac{\partial h}{\partial \xi_{i}}-g_{j}\dfrac{\partial f_{i}}{\partial \xi_{j}}\dfrac{\partial h}{\partial \xi_{i}}\right]\nonumber\\
&=&\sum_{i,j=1}^{2n}\left[f_{j}\dfrac{\partial g_{i}}{\partial \xi_{j}}\dfrac{\partial h}{\partial \xi_{i}}-g_{i}\dfrac{\partial f_{j}}{\partial \xi_{i}}\dfrac{\partial h}{\partial \xi_{j}}\right]\nonumber\\ 
&&\mbox{(interchanging $i$ and $j$ in the 2nd term. This is possible since $i,j$ are dummy indices)}\nonumber\\
&=&\sum_{i=1}^{2n}\dfrac{\partial h}{\partial \xi_{i}}\sum_{j=1}^{2n}\left(f_{j}\dfrac{\partial g}{\partial \xi_{j}}-g_{j}\dfrac{\partial f_{i}}{\partial \xi_{i}}\right)=\sum_{i=1}^{2n}C_{i}\dfrac{\partial h}{\partial \xi_{j}}\nonumber
\end{eqnarray} where,
$C_{i}=\sum_{j=1}^{2n}\left(f_{j}\dfrac{\partial g}{\partial \xi_{j}}-g_{j}\dfrac{\partial f_{i}}{\partial \xi_{j}}\right)$.\\
Therefore,
\begin{eqnarray}
\{f,\{g,h\}\}+\{g,\{h,f\}\}&=&\sum_{i=1}^{n}C_{i}\dfrac{\partial h}{\partial \xi_{i}}+\sum_{i=n+1}^{2n}C_{i}\dfrac{\partial h}{\partial \xi_{i}}\nonumber\\&=&\sum_{i=1}^{n}\left(A_{i}\dfrac{\partial h}{\partial p_{i}}+B_{i}\dfrac{\partial h}{\partial q_{i}}\right)\label{eq2..*}
\end{eqnarray} where $A$ and $B$ are functions of $f$ and $g$ and their derivatives but is independent explicitly of $h$. So when $h$ is varied $A$ and $B$ remain unchanged. Now putting $h=p_{j}$ in the above equation, the R.H.S becomes $A_{j}$ and L.H.S is
\begin{equation}
\{f,\{g,p_{j}\}\}+\{g,\{p_{j},f\}\}=\{f,\dfrac{\partial g}{\partial q_{j}}\}-\{g,\dfrac{\partial f}{\partial q_{j}}\}=\{f,\dfrac{\partial g}{\partial q_{j}},g\}=\dfrac{\partial}{\partial q_{j}}\{f,g\}\label{eq2..**}
\end{equation}
Similarly putting $h=q_{j}$, the R.H.S of (\ref{eq2..*}) becomes $B_{j}$ and the L.H.S is
\begin{eqnarray}
\{f,\{g,q_{j}\}\}+\{g,\{q_{j},f\}\}&=&\left\{f,-\dfrac{\partial g}{\partial p_{j}}\right\}+\left\{g,\dfrac{\partial f}{\partial p_{j}}\right\}\nonumber\\&=&-\left\{f,\dfrac{\partial g}{\partial p_{j}}\right\}-\left\{\dfrac{\partial f}{\partial p_{j}},g\right\}\nonumber\\&=&-\dfrac{\partial}{\partial p_{j}}\{f,g\}\label{eq2..***}
\end{eqnarray}
Putting the values of $A_{i}$'s and $B_{i}$'s from (\ref{eq2..**}) and (\ref{eq2..***}) in (\ref{eq2..*}) we have,\\
$\{f,\{g,h\}\}+\{g,\{h,f\}\}=\sum_{i=1}^{n}\left[\dfrac{\partial}{\partial q_{i}}\{f,g\}.\dfrac{\partial h}{\partial p_{i}}-\dfrac{\partial}{\partial p_{i}}\{f,g\}\dfrac{\partial h}{\partial q_{i}}\right]=-\{h,\{f,g\}\}$\\
$~~~~~~~~~~~~~~~~~~~~~~~~~~\therefore \{g,\{g,h\}\}+\{g,\{h,f\}\}+\{h,\{f,g\}\}=0$.
\section{Constants of motion}
Let $f(q,p,t)$ be any function of the dynamical variables $q$ and $p$ of a mechanical system defined by the Hamiltonian $H(q,p,t)$. Let us calculate the total derivative of $f$ with respect to $t$ i.e.\\
\begin{eqnarray}
\dfrac{df}{dt}&=&\dfrac{\partial f}{\partial t}+\sum_{i=1}^{n}\left(\dfrac{\partial f}{\partial q_{i}}\dot{ q_{i}}+\dfrac{\partial f}{\partial p_{i}}\dot{p_{i}}\right)\nonumber\\
	&=&\dfrac{\partial f}{\partial t}+\sum_{i=1}^{n}\left(\dfrac{\partial f}{\partial q_{i}}\dfrac{\partial H}{\partial p_{i}}-\dfrac{\partial f}{\partial p_{i}}\dfrac{\partial H}{\partial q_{i}}\right)\nonumber\\
	&=&\dfrac{\partial f}{\partial t}+\{f,H\}\nonumber.
\end{eqnarray} Those functions of the canonical variables which remain constant during the motion are called constants of motion. Thus the condition that $f(q,p,t)$ will be a constant of motion is that
\begin{equation}
\dfrac{df}{dt}=0 ~i.e~\dfrac{\partial f}{\partial t}+\{f,H\}=0.\nonumber
\end{equation} However, if $f$ is explicitly independent of $t$ then the condition simplifies to $\{f,H\}=0$.

\vspace{.25cm}

$\bullet$ \textbf{Note I:} If $f(q,p,t)$ and $g(q,p,t)$ be two constants of motion of a mechanical system then their Poisson bracket is also a constant of motion.
\begin{eqnarray}
\dfrac{d}{dt}\{f,g\}&=&\dfrac{\partial}{\partial t}\{f,g\}+\{\{f,g\},H\}\nonumber\\
&=&\left\{\dfrac{\partial f}{\partial t},g\right\}+\left\{f,\dfrac{\partial g}{\partial t}\right\}+\{f,\{g,H\}\}+\{g,\{H,f\}\}\nonumber\\
&&~~~~~~~~~~~~~~~~~~~~~~~~~~~~~~~~~~~~~~~~~~~~~~~\mbox{(by Jacobi's identity)}\nonumber\\
&=&\left\{\dfrac{\partial f}{\partial t}+\{f,H\},g\right\}+\left\{f,\dfrac{\partial g}{\partial t}+\{g,H\}\right\}\nonumber\\&=&\left\{\dfrac{df}{dt},g\right\}+\left\{f,\dfrac{dg}{dt}\right\}=\{0,g\}+\{f,0\}=0.\nonumber
\end{eqnarray} Thus given any two constants of motion we can generate all constants of motion by considering Poisson's bracket.

\vspace{.25cm}

$\bullet$ \textbf{Note II:} \textbf{Lagrange's bracket:} For any two quantities $f(q,p,t)$ and $g(q,p,t)$ the sum
\begin{equation}
\sum_{i=1}^{n}\left(\dfrac{\partial q_{i}}{\partial f}\dfrac{\partial p_{i}}{\partial g}-\dfrac{\partial p_{i}}{\partial f}\dfrac{\partial g_{i}}{\partial g}\right)\nonumber
\end{equation} is called the Lagrange's bracket of the two quantities $f$ and $g$ with respect to $q$ and $p$ and is denoted by $[f,g]_{q,p}$.\\
$\bullet$ \textbf{{Some properties of Lagrange's bracket:}}\\
{\textbf{I.}} $[f,g]=-[g,f]$\\
\textbf{II.} $[f,q_j]=-\dfrac{\partial p_j}{\partial f}$\\
\textbf{III.} $[f,p_j]=\dfrac{\partial q_j}{\partial f}$\\
\textbf{IV.} $[q_i,q_j]=0=[p_i,p_j]$\\
\textbf{V.} $[q_i,p_j]=\delta_{ij}$.

\vspace{.25cm}

$\bullet$ \textbf{{Connection  between two types of brackets:}}\\
Let $u_{i}(q,p,t) (i=1,2,,,,2n)$ be $2n$ independent functions of the variables $q_{1},q_{2},...,q_{n}$ and $p_{1},p_{2},...,p_{n}$ so that they can be solved to express $q_{1},q_{2},...,q_{n}$ as functions of $u,t$ i.e. $q_{i}=q_{i}(u,t)$ and similarly $p_{i}=p_{i}(u,t)$. Then we shall show that
\begin{equation}
\sum_{i=1}^{2n} \{u_{i},u_{j}\}[u_{i},u_{s}]=\delta_{js}\nonumber
\end{equation}
\textbf{Proof:} We have $\{u_i,u_j\}=\sum_{r=1}^{n}\left(\dfrac{\partial q_{r}}{\partial u_{i}}\dfrac{\partial p_{r}}{\partial u_{j}}-\dfrac{\partial q_{r}}{\partial u_{j}}\dfrac{\partial p_{r}}{\partial u_{i}}\right)$ and $[u_i,u_s]=\sum_{i=1}^{n}\left(\dfrac{\partial u_{i}}{\partial q_{k}}\dfrac{\partial u_{s}}{\partial p_{k}}-\dfrac{\partial u_i}{\partial p_k}\dfrac{\partial u_s}{\partial q_k}\right)$\\
Therefore L.H.S =
\begin{eqnarray}
&&\sum_{i=1}^{2n}\sum_{r=1}^{n}\left(\dfrac{\partial q_{r}}{\partial u_{i}}\dfrac{\partial p_{r}}{\partial u_{j}}-\dfrac{\partial q_{r}}{\partial u_{j}}\dfrac{\partial p_{r}}{\partial u_{i}}\right)\sum_{i=1}^{n}\left(\dfrac{\partial u_{i}}{\partial q_{k}}\dfrac{\partial u_{s}}{\partial p_{k}}-\dfrac{\partial u_i}{\partial p_k}\dfrac{\partial u_s}{\partial q_k}\right)\nonumber\\
&=&\sum_{r=1}^{n}\sum_{k=1}^{n}\dfrac{\partial p_{r}}{\partial u_{j}}\dfrac{\partial u_{s}}{\partial p_{k}}\sum_{i=1}^{2n}\dfrac{\partial p_{r}}{\partial u_{i}}\dfrac{\partial u_{i}}{\partial q_{k}}-\sum_{r=1}^{n}\sum_{k=1}^{n}\dfrac{\partial q_{r}}{\partial u_{j}}\dfrac{\partial u_{s}}{\partial p_{k}}\sum_{i=1}^{2n}\dfrac{\partial p_{r}}{\partial u_{i}}\dfrac{\partial u_i}{\partial p_k}\nonumber\\ &&~~~~~~~~~~~~~~~~~~-\sum_{r}\sum_{k}\dfrac{\partial p_{r}}{\partial u_{j}}\dfrac{\partial u_{s}}{\partial q_{k}}\sum_{i=1}^{2n}\dfrac{\partial q_{r}}{\partial u_{i}}\dfrac{\partial u_i}{\partial p_k}+\sum_{r}\sum_{k}\dfrac{\partial q_{r}}{\partial u_{j}}\dfrac{\partial u_{i}}{\partial p_{k}}
\sum_{i=1}^{2n}\dfrac{\partial p_{r}}{\partial u_{i}}\dfrac{\partial u_{i}}{\partial p_{k}}\nonumber\\
&=&\sum_{r=1}^{n}\dfrac{\partial p_{r}}{\partial u_{j}}\dfrac{\partial u_{s}}{\partial p_{r}}-\sum_{r=1}^{n}\dfrac{\partial q_{r}}{\partial u_{j}}\dfrac{\partial u_s}{\partial p_r}.0-0+\sum_{r}\dfrac{\partial q_{r}}{\partial u_{j}}\dfrac{\partial u_s}{\partial q_r}\nonumber\\
&=&\sum_{r=1}^{n}\left(\dfrac{\partial p_{r}}{\partial u_{j}}\dfrac{\partial u_{s}}{\partial p_{r}}+\dfrac{\partial q_{r}}{\partial u_{j}}\dfrac{\partial u_s}{\partial q_r}\right)=\delta_{js}.\nonumber
\end{eqnarray}
\section{Legendre's (dual) transformation:}
Let $Q(q)$ be a given function of the variables $q_{1},q_{2},...,q_{n}$ and let the set of new variables $p_{1},p_{2},...,p_{n}$ be defined by the transformation $p_{i}=\dfrac{\partial Q}{\partial q_{i}}$. Then a new function $P(p)$ can be constructed so as to define the transformation $q_{i}=\dfrac{\partial P}{\partial p_{i}}$. Further, if $Q$ contains the independent parameter $t$ then $P$ does contain $t$ and $\dfrac{\partial Q}{\partial t}=-\dfrac{\partial P}{\partial t}$. (The variable $q$, which actively participate in the transformation is called active variable while $t$ is called a passive variable.)\\
\textbf{Proof:} Let $Q(q,t)$ be a given function of the independent variables $q_{1},q_{2},...,q_{n}$ and independent parameter $t$. We define a set of new independent variables $p_{1},p_{2},...,p_{n}$ by 
\begin{equation}
p_{i}=\dfrac{\partial Q}{\partial q_{i}}\label{eq2..3}
\end{equation}
Since $p_{i}$'s are all independent, the equation (\ref{eq2..3}) can be inverted to give\\
\begin{equation}
q_{i}=q_{i}(p_{i},t)\label{eq2..4}
\end{equation}
provided the Jacobian of the transformation is not equal to zero. We now construct a function $P$ by\\
\begin{equation}
P=\sum p.q- Q(q,t)\label{eq2..5}.
\end{equation} We eliminate from the R.H.S of (\ref{eq2..5}) the variables $q$'s by the corresponding $p$'s with the help of (\ref{eq2..4}) to express $P$ as a function of $p$'s and $t$. Now consider a variation of (\ref{eq2..5}) due to an infinitesimal variation of the variables involved in that equation i.e.\\
\begin{eqnarray}
&&\delta P=\delta \left(\sum p.q-Q\right)\nonumber\\
&&\mbox{ or,}~~
\sum \dfrac{\partial P}{\partial p}\delta p+\dfrac{\partial P}{\partial t}\delta t=\cancel{\sum p.\delta q}+\sum q.\delta p-\cancel{\sum\frac{\partial Q}{\partial q}}\delta q-\dfrac{\partial Q}{\partial t}\delta t. ~~~(\because p=\dfrac{\partial Q}{\partial q}).\nonumber\\
&& \mbox{Consequently},~~
q=\dfrac{\partial P}{\partial p},~~\dfrac{\partial Q}{\partial t}=-\dfrac{\partial P}{\partial t}.\nonumber
\end{eqnarray}
Legendre's transformation changes a given function $Q$ of a given set of variables $q$ into a new function $P$ of a new set of variables $p$.
\section{Transformation of canonical variables}
Let $q_{1},q_{2},...,q_{n}$ be $n$ independent variables and $t$ be an independent character. We define new variables $Q_{i}$ by means of the transformation\\
\begin{equation}
Q_{i}=Q_{i}(q,t)\label{eq2..6}
\end{equation}
and assume that the new variable $Q_{i}$'s are all independent so that the inverse transformation $q_{i}=q_{i}(Q,t)$ exists. We can associate a definite geometrical picture with the transformation (\ref{eq2..6}). We can think of an $n$- dimensional space in which the variables $q$'s are the axial variables of a set of $n$ rectangular axes. We denote this space by $S_{q}$. Similarly we can think of an $n$-dimensional space $S_{Q}$. To a point of $S_{q}$ corresponds a point $S_{Q}$ and for this reason the transformation (\ref{eq2..6}) is called a point transformation.\\
$~~~~~~~~~~~~~~~~~~~~~~~~~~~~~~~~~~~~~~~~~~~~~~~~$Let us now try to extend the transformation (\ref{eq2..6}) by allowing the variables $(q,p)$ take part in equations of transformation (\ref{eq2..6}). Let us consider the transformation of the type: 
\begin{equation}
Q=Q(q,p,t)~~~~~\mbox{and}~~~~ P=P(q,p,t)\label{eq2..7}
\end{equation}
and assume that the new variables $Q$'s and $P$'s are all independent so that the inverse transformation $q=q(Q,P,t),~~p=p(Q,P,t)$ exists. The canonical forms of the equation of motion will not always be preserved under a general transformation of the type (2). We restrict ourselves to those transformation of the type (\ref{eq2..7}) which produce new canonical variables $Q$ and $P$ so that the transformation changes the canonical equations to
\begin{eqnarray}
\dot{q}=\dfrac{\partial H}{\partial p},~~\dot{p}=-\dfrac{\partial H}{\partial q}\label{eq2..8} \\to\nonumber\\
\dot{Q}=\dfrac{\partial K}{\partial P},~~\dot{P}=-\dfrac{\partial K}{\partial Q}\label{eq2..9}
\end{eqnarray}
where $K(Q,P,t)$ is new Hamiltonian of the system. Transformation (\ref{eq2..7}) is then called a canonical or contact transformation i.e. the equations of canonical transformation. The variables $q,p$ satisfy Hamilton's equation of motion (\ref{eq2..8}) and hence satisfy Hamilton's principle i.e.
\begin{eqnarray}
&&\delta \int_{t_{0}}^{t_{1}}\left(\sum p~\dot{q}-H\right)dt=0\nonumber\\ \mbox{or,}~~&&\delta\int_{t_{0}}^{t_{1}}\left(\sum pdq-Hdt\right)=0\label{eq2..10}
\end{eqnarray}
If $Q$, $P$ are also canonical variables and satisfy equation (\ref{eq2..9}) then the Hamilton's principle takes the form
\begin{equation}
\delta \int_{t_{0}}^{t_{1}}\left(\sum PdQ-Kdt\right)=0\label{eq2..11}
\end{equation}
 where $K$ is the Hamiltonian associated with $Q,p$ at moment $t$. The equations (\ref{eq2..10}) and (\ref{eq2..11}) must simultaneously be valid for one implies the other. Hence
 \begin{equation}
\delta \int_{t_{0}}^{t_{1}}\left[\left(\sum pdq-Hdt\right)-C\left(\sum PdQ-Kdt\right)\right]=0\label{eq2..12}
\end{equation}
 where $C$ is a constant. This means that
 \begin{equation}
\sum pdq-Hdt=C(\sum PdQ-Kdt)+dF\label{eq2..13}
\end{equation}
 where $F$ is some arbitrary function of the variables at time $t$.
 
 \vspace{.25cm}
 
(\textbf{Note:} The simultaneous validity of (\ref{eq2..10}) and (\ref{eq2..11}) does not mean that the integrands in the two integrals are equal but they can differ at most by a total derivative of arbitrary function $F$. The integrand between these two end points of such a difference term is then $\int_{t_{0}}^{t_{1}}\dfrac{\partial F}{\partial t}dt=F(t_{1})-F(t_{0})$ and the variation of these integral is automatically zero for any function $F$ since the variations vanish at the end points.)

\vspace{.25cm}

The function $F$ is called the generating function and $C$ is the valency of the canonical transformation. If $C=1$, the canonical transformation is called univalent. Once $F$ is given, the transformation (\ref{eq2..7}) is completely specified.\\
$~~~~~~~~~~~~~~~~~~~~~~~~~~~~~~~$ In order to effect the transformation between the two sets of canonical variables $F$ must be a function of both the old and the new variables besides the time $t$. The generating function   may thus be a function of $4n$ variables in all. But only $2n$ of these are independent because the two sets of co-ordinates are connected by $2n$ transformation equations (\ref{eq2..7}). The generating function can therefore be written as a function of independent variables in one of the four forms:\\
$F_{1}(q,Q,t),~~F_{2}(q,P,t),~~F_{3}(p,Q,t),~~F_{4}(p,P,t)$\\
The selection of the particular form depends on the specific problem under consideration. If in (\ref{eq2..13}) we put $C=1$ then we have $p=\dfrac{\partial F}{\partial q},~~P=-\dfrac{\partial F}{\partial Q},~~K-H=\dfrac{\partial F}{\partial t}$. So if $F$ is independent of $t$ then $K=H$ i.e. the form of $H$ is conserved. Again given $F$ we can find $p$ and $P$. Thus in the case of contact transformation when $(q,p)\rightarrow(Q,P)$ and $K=H$, we have
$$ \sum pdq-\sum PdQ=dF$$. i.e to say in a contact transformation $\sum pdq-\sum PdQ$ is a perfect differential.

\section{Formula for canonical transformation}

\vspace{.5cm}

\textbf{I.} Let $F=F_{1}(q,Q,t)$ then from (\ref{eq2..13}) putting $C=1$, we have 
\begin{equation}
\sum pdq-\sum PdQ+(K-H)dt=dF_{1}(q,Q,t)=\sum \dfrac{\partial F_{1}}{\partial q}dq+\sum \dfrac{\partial F_{1}}{\partial Q}dQ+\dfrac{\partial F_{1}}{\partial t}dt\label{eq2..14}
\end{equation}
Now equating coefficients of like differentials we have
\begin{eqnarray}
p_{i}&=&\dfrac{\partial F_{1}(q,Q,t)}{\partial q_{i}}\label{eq2..15}\\
P_{i}&=&-\dfrac{\partial F_{1}(q,Q,t)}{\partial Q_{i}}\label{eq2..16}\\
\mbox{and}~~K&=&H+\dfrac{\partial F_{1}}{\partial t}\label{eq2..17}
\end{eqnarray} The $n$ equations (\ref{eq2..15}) can be solved for $n$ $Q$'s. As 
\begin{equation}
	Q_{i}=Q_{i}(q,p,t)\label{eq2..18}
\end{equation} so inserting these values of $Q$ in (\ref{eq2..16}) we have
\begin{equation} 
	P_{i}=P_{i}(q,p,t)\label{eq2..19}
	\end{equation} (\ref{eq2..18}) and (\ref{eq2..19}) are the equations of canonical transformation and $(\ref{eq2..17})$ connects $K$ and $H$.\\
\textbf{II.} Let the generating function be $F=F_{2}(q,p,t)$. From (\ref{eq2..14})\\
\begin{eqnarray}
\sum pdq-\sum PdQ+(K-H)dt=dF_{1}(q,Q,t)\nonumber\\
or,~~\sum p dq+\sum QdP+(K-H)dt=d(F_{1}+\sum PQ)\label{eq2..20}
\end{eqnarray} Now, we express the quantity $F_{1}(q,Q,t)+\sum PQ$ in terms of the variables $q,P,t$ by the relation (\ref{eq2..16}) and write $F_{2}$ for the transformed quantity i.e.
\begin{equation}
F_{2}(q,P,t)=F_{1}(q,Q,t)+\sum PQ\label{eq2..21}
\end{equation} where $P_{i}=-\dfrac{\partial F_{1}}{\partial Q_{i}}$. In terms of $F_{2}$ equation (\ref{eq2..20}) becomes 
\begin{equation}
\sum pdq+\sum QdP+(K-H)dt=dF_{2}(q,P,t)=\sum \dfrac{\partial F_{2}}{\partial q}dq+\sum \dfrac{\partial F_{2}}{\partial P}dP+\dfrac{\partial F_{2}}{\partial t}dt,\nonumber
\end{equation} so the transformation equations become
\begin{eqnarray}
p_{i}&=&\dfrac{\partial F_{2}}{\partial q_{i}}\label{eq2..22}\\
Q_{i}&=&\dfrac{\partial F_{2}}{\partial P_{i}}\label{eq2..23}\\
and,&&\nonumber\\
K&=&H+\dfrac{\partial F_{2}}{\partial t}\label{eq2..24}
\end{eqnarray} (\ref{eq2..22}) can be solved for P to give
\begin{equation}
 P_{i}=P_{i}(q,p,t)\label{eq2..25}
 \end{equation} and then from (\ref{eq2..23}) 
\begin{equation}
Q_{i}=Q_{i}(q,p,t)\label{eq2..26}
\end{equation}
Equations (\ref{eq2..25}) and (\ref{eq2..26}) are the equations of canonical transformation for the generating function of the type $F_{2}$.\\
\textbf{III.} Let the generating function $F=F_{3}(p,Q,t)$\\
We write eq (\ref{eq2..14}) in this case as 
\begin{equation}
-\sum qdp-\sum PdQ+\sum (K-H)dt=d(F_{1}-\sum pdq)\label{eq2..27}
\end{equation} Now express $F_{1}-\sum pq$ in terms of the variables $p,Q,t
$ by (\ref{eq2..15}) and write $F_{3}(p,Q,t)$ as
\begin{equation}
	 F_{3}(p,Q,t)=F_{1}(q,Q,t)-\sum pq\label{eq2..28}
	 \end{equation} where $p_{i}=\dfrac{\partial F_{1}(q,Q,t)}{\partial q_{i}}$. In terms of $F_{3}$ equation (\ref{eq2..27}) now becomes,
\begin{eqnarray}
\sum qdp-\sum PdQ+(K-H)dt=dF_{3}=\sum \dfrac{\partial F_{3}}{\partial p}dp+\sum \dfrac{\partial F_{3}}{\partial Q}dQ+\dfrac{\partial F_{3}}{\partial t}dt\nonumber
\end{eqnarray}
$\therefore$
\begin{eqnarray}
q_{i}&=&-\dfrac{\partial F_{3}}{\partial p_{i}}\label{eq2..29}\\
P_{i}&=&-\dfrac{\partial F_{3}}{\partial Q_{i}}\label{eq2..30}\\
K&=&H+\dfrac{\partial F_{3}}{\partial t}\label{eq2..31}
\end{eqnarray} Equations (\ref{eq2..29}) and (\ref{eq2..30}) constitute the transformation equations for the generating function $F_{3}(p,Q,t)$ and (\ref{eq2..31}) is the relation between $K$ and $H$.\\
\textbf{IV.} $F=F_{4}(p,P,t)$, the we have from (\ref{eq2..14})
\begin{eqnarray}
\sum pdq-\sum PdQ+(K-H)dt=dF_{1}\nonumber\\
or~~~~-\sum qdp+\sum QdP+(K-H)dt=d\left[F_{1}(q,Q,t)+\sum PQ-\sum pq\right]\nonumber
\end{eqnarray} By equation, (\ref{eq2..15}) we can express $Q$ in terms of $p$ and $q$ and then by (\ref{eq2..16}) we can express $Q$ in terms of $P$ and $p$. So by (\ref{eq2..15}) and (\ref{eq2..16}) we can write
\begin{equation}
F_{1}(q,Q,t)+\sum PQ-\sum pq= F_{4}(p,P,t)\nonumber
\end{equation}  Thus we write
\begin{equation}
-\sum qdp+\sum QdP+(K-H)dt=dF_{4}(p,P,t)=\sum \dfrac{\partial F_{4}}{\partial p}dp+\sum \dfrac{\partial F_{4}}{\partial P}dP+\dfrac{\partial F_{4}}{\partial t}dt\nonumber
\end{equation} Hence the transformation equations are
\begin{eqnarray}
q_{i}&=&-\dfrac{\partial F_{4}}{\partial p_{i}}\label{eq2..32}\\
Q_{i}&=&\dfrac{\partial F_{4}}{\partial P_{i}}\label{eq2..33}\\
K&=&H+\dfrac{\partial F_{4}}{\partial t}\label{eq2..34}
\end{eqnarray}
Equations (\ref{eq2..32}) and (\ref{eq2..33}) constitute the canonical transformation and (\ref{eq2..34}) is the relation between $K$ and $H$.

\vspace{.5cm}

$\bullet$ \textbf{Problem I:} Prove that the transformation $Q=p$, $P=-q$ is a contact transformation.

\vspace{.25cm}

\textbf{Solution:} Here $pdq-PdQ=pdq+qdp=d(p,q)$. So it is a contact transformation.

\vspace{.5cm}

$\bullet$ \textbf{Problem II:} $P=\dfrac{1}{2}(p^{2}+q^{2}),~~Q=tan^{-1}\left(\frac{q}{p}\right)$.  So $pdq-PdQ=pdq-\dfrac{1}{2}(p^{2}+q^{2}).\dfrac{1}{1+\frac{q^{2}}{p^{2}}}d(\frac{q}{p})=pdq-\frac{1}{2}\dfrac{p^{2}(pdq-qdp)}{p^{2}}=\frac{1}{2}(pdq+qdp)=d(\frac{1}{2}pq)$.

\vspace{.5cm}

$\bullet$ \textbf{Problem III:} $P=logsinp,~~Q=q~tanp$.

\vspace{.5cm}

$\bullet$ \textbf{Problem IV:} $Q=log(\frac{1}{2}cosp),~~P=q~cot p$.

\vspace{.5cm}

$\bullet$ \textbf{Problem V:} $Q=log(1+\sqrt{q} cosp), ~~P=2(1+\sqrt{q}cos p)\sqrt{q}sin p$.

\vspace{.5cm}

$\bullet$ \textbf{Problem VI:} Show that the transformation $Q=\sqrt{q}cos p,~P=\sqrt{q} sin p$ represents a canonical transformation with valency not equal to unity. Solve for the new Hamiltonian of the system for which $T=\frac{1}{2}m\dot{q}^{2},~~V=\frac{1}{2}\mu {q}^{2}$.

\vspace{.25cm}

\textbf{Solution:} $PdQ=\sqrt{q} sin p\left[\frac{1}{2}q^{-\frac{1}{2}} ~cosp~ dq-sin p~ \sqrt{q}~ dp\right]=\dfrac{1}{2}sin p ~cos p ~dq-q~ sin^{2}p~dp$\\
$\therefore Cpdq-PdQ=(Cp-\frac{1}{2} ~sinp~cosp)dq+q~sin^{2}p~dp=udq+vdp$ (say)\\
$\dfrac{\partial u}{\partial p}=C-\frac{1}{2}(cos^{2}p-sin^{2}p)=C-\frac{1}{2}(1-2sin^{2}p)$\\
$\dfrac{\partial u}{\partial q}=sin^{2}p$. Hence for exactness: $\dfrac{\partial u}{\partial p}=\dfrac{\partial v}{\partial q}\implies C-\frac{1}{2}+sin^{2}p=sin^{2}p\implies C=\frac{1}{2}$\\
$H=T+V=\frac{1}{2}m\dot{q}^{2}+\frac{1}{2}\mu {q}^{2}$.\\
$p=\dfrac{\partial L}{\partial \dot{q}}=\dfrac{\partial T}{\partial \dot{q}}=m\dot{q}$. $\therefore H=\dfrac{1}{2}m\dfrac{p^{2}}{m^{2}}+\dfrac{1}{2}\mu q^{2}=\dfrac{p^{2}}{2m}+\dfrac{1}{2}\mu q^{2}$.
Now, $Q=\sqrt{q}cos p,~~ P=\sqrt{q}sin p. \therefore Q^{2}+P^{2}=q,~~\dfrac{P}{Q}=tan p$. $\therefore K=CH=\dfrac{1}{2}\left[\dfrac{1}{2m}\left(tan^{-1}(\frac{P}{Q})\right)^{2}+\dfrac{1}{2}\mu(Q^{2}+P^{2})^{2}\right]$ is the new Hamiltonian.

\vspace{.5cm}

$\bullet$ \textbf{Problem VII:} The K.E and P.E  of one dimensional harmonic oscillator is $T=\dfrac{1}{2}\dot{q}^{2},~~V=\dfrac{1}{2}\mu^{2}q^{2}.$. Solve the problem using canonical transformation $F_{1}=\dfrac{1}{2}\mu q^{2} cot Q$.

\vspace{.25cm}

\textbf{Solution:} $p=\dfrac{\partial T}{\partial \dot{q}}=\dot{q}$. $\therefore$ Hamiltonian $H=T+V=\dfrac{1}{2}p^{2}+\dfrac{1}{2}\mu q^{2}=\dfrac{1}{2}(\mu q^{2}+p^{2})$\\
The canonical equations for $F_{1}$ are\\
$p_{i}=\dfrac{\partial F_{1}}{\partial q_{i}},~~P_{i}=-\dfrac{\partial F_{1}}{\partial Q_{i}}$.\\
$\therefore p=\mu q cotQ,~~P=\dfrac{1}{2}\mu q^{2}cosec^{2}Q$\\
$\therefore q^{2}=\dfrac{2P}{\mu cosec^{2}Q},~~p^{2}=\mu^{2}\dfrac{2cot^{2}Q}{\mu cosec^{2}Q}=2\mu P cos^{2}Q$\\
$\therefore K=H=\dfrac{1}{2}\left[\mu.2.P.sin^{2}Q+2.\mu.P.cos^{2}Q\right]=\mu P.$\\
$\therefore-\dot{P}=\dfrac{\partial K}{\partial Q}=0,~~\dot{Q}=\dfrac{\partial K}{\partial P}=\mu$\\
i.e. $P=const=\dfrac{E}{\mu}~(say),~~Q=\mu t+\epsilon$.\\
$\therefore q=\dfrac{\sqrt{2E}}{\mu}sin(\mu t+\epsilon),~~p=\sqrt{2E}cos(\mu t+\epsilon)$.
\section{Infinitesimal contact transformation}
Consider the canonical transformation for which the generating function $F_{2}(q,P)=\sum q_{i}P_{i}$, then the transformation equations are
\begin{equation}
p_{i}=\dfrac{\partial F_{2}}{\partial q_{i}}=P_{i},~~Q_{i}=\dfrac{\partial F_{2}}{\partial P_{i}}=q_{i},~~K=H.\nonumber
\end{equation} As the new and old co-ordinates being the same so $F_{2}$ generates the identity transformation.\\
If the new co-ordinates differ from the old ones only by infinitesimals it is called infinitesimal contact transformation. Here only 1st order terms are retained in these infinitesimal. This is represented by $Q_{i}=q_{i}+\delta q_{i},~~P_{i}=p_{i}+\delta p_{i}$, where $\delta q_{i}$, $\delta p_{i}$ are the infinitesimal changes in the co-ordinates. It is clear that in this case the generating function will differ only by an infinitesimal amount from the above identity transformation. Thus we can write,
\begin{equation}
F_{2}=\sum q_{i} P_{i}+\epsilon G(q,P)\nonumber
\end{equation} where $\epsilon$ is some infinitesimal parameter of the transformation. The transformation equations are
\begin{eqnarray}
p_{i}=\dfrac{\partial F}{\partial q_{i}}=P_{i}+\epsilon\dfrac{\partial G}{\partial q_{i}}\nonumber\\
\therefore\delta p_{i}=P_{i}-p_{i}=-\epsilon\dfrac{\partial G}{\partial q_{i}}\label{eq2..35}
\end{eqnarray} Again, $Q_{i}=\dfrac{\partial F}{\partial P_{i}}=q_{i}+\epsilon\dfrac{\partial G}{\partial P_{i}}$. The second term on the R.H.S contains the infinitesimal parameter and since from (\ref{eq2..35}) $P$ differs from $p$ only an infinitesimal so we can replace $P_{i}$ by $p_{i}$ in the derivative $\dfrac{\partial G}{\partial P_{i}}$ to have the result correct up to 1st order. Thus
\begin{equation}
\delta q_{i}=Q_{i}-q_{i}=\epsilon\left(\dfrac{\partial G}{\partial p_{i}}\right)\label{eq2..36}
\end{equation} Now we consider an infinitesimal contact transformation in which $\epsilon$ is the small time interval $dt$ and $G$ is $H(q,p)$, then we have
\begin{eqnarray}
\delta q_{i}=dt.\dfrac{\partial H}{\partial p_{i}}=\dot{ q_{i}}dt=dq_{i}\nonumber\\
\delta p_{i}=-dt\dfrac{\partial H}{\partial q_{i}}=\dot {p_{i}} dt=dp_{i}\nonumber
\end{eqnarray} Thus the transformation changes the co-ordinates and momenta at the time $t$ to the values they have at the time $t+\delta t$. Thus the motion of the system in a time interval $\delta t$ can be described by an infinitesimal contact transformation generated by the Hamiltonian.\\
~~~~~~~~~~~~~~~Extending this result we can say that the system motion in a finite interval from $t_{0}$ to $t$ is represented by a succession of infinitesimal contact transformation. Thus the values of $p$ and $q$ at any time $t$ can be attained from their initial values by a canonical transformation, which is a constant function of time. Thus the Hamiltonian is the generator of motion of the system in time translation.\\
~~~~~~~~~~~~~Now consider the change in some function $u=u(q,p)$ as a result of this transformation. If the infinitesimal canonical transformation is generated by the Hamiltonian, the result for substituting the new variable for the old is to change $u$ from its value at $t$ to the value it has at a time $dt$ later. The change in a function $'u'$ as a result of an infinitesimal canonical transformation is
\begin{eqnarray}
\delta u&=& u(q_{i}+\delta q_{i}, p_{i}+\delta p_{i})-u(q_{i},p_{i})\nonumber\\
&=&\sum_{i}(\delta q_{i}\dfrac{\partial u}{\partial q_{i}}+\delta p_{i}\dfrac{\partial u}{\partial p_{i}})~~~~~\mbox{(upto~~ 1st ~~~order)}\nonumber\\
&=&\epsilon\sum_{i}\left(\dfrac{\partial u}{\partial q_{i}}\dfrac{\partial G}{\partial p_{i}}-\dfrac{\partial u}{\partial p_{i}}\dfrac{\partial G}{\partial q_{i}}\right)~~~~~\mbox{(by (\ref{eq2..35}) ~and~ (\ref{eq2..36}))}\nonumber\\
&=&\epsilon[u,G]\nonumber
\end{eqnarray} If $u$ is replaced by the Hamiltonian then $\delta H=\epsilon [H,G]$. Now, if $G$ be a constant of motion its Poisson bracket with the Hamiltonian is zero. i.e $[H,G]=0$, which implies $\delta H=0$. Thus the constants of motion are the generating functions of the infinitesimal canonical transformation which leave the Hamiltonian invariant.
\section{Invariance of Poisson bracket under Canonical transformation}
The Poisson's bracket of two arbitrary functions $F$ and $G$ with respect to the variables $q$, $p$ is defined as
\begin{equation}
\{F,G\}_{q,p}=\sum_{i}\left(\dfrac{\partial F}{\partial q_{i}}\dfrac{\partial G}{\partial p_{i}}-\dfrac{\partial F}{\partial p_{i}}\dfrac{\partial G}{\partial q_{i}}\right)\label{eq2..37}
\end{equation} Now considering $q_{i}$ and $p_{i}$ as functions of the transformed set of variables $Q_{k}$ and $P_{k}$, the equation (\ref{eq2..37}) can be written as
\begin{eqnarray}
\{F,G\}_{q,p}&=&\sum_{i,k}\left[\dfrac{\partial F}{\partial q_{i}}\left(\dfrac{\partial G}{\partial Q_{k}}\dfrac{\partial Q_{k}}{\partial p_{i}}+\dfrac{\partial G}{\partial P_{k}}\dfrac{\partial P_{k}}{\partial p_{i}}\right)-\dfrac{\partial F}{\partial p_{i}}\left(\dfrac{\partial G}{\partial Q_{k}}\dfrac{\partial Q_{k}}{\partial q_{i}}+\dfrac{\partial G}{\partial P_{k}}\dfrac{\partial P_{k}}{\partial q_{i}}\right)\right]\nonumber\\
&=&\sum_{k}\left[\dfrac{\partial G}{\partial Q_{k}}\{\sum_{i}\left(\dfrac{\partial F}{\partial q_{i}}\dfrac{\partial Q_{k}}{\partial p_{i}}-\dfrac{\partial F}{\partial p_{i}}\dfrac{\partial Q_{k}}{\partial q_{i}}\right)\}+\dfrac{\partial G}{\partial P_{k}}\{\sum_{i}\left(\dfrac{\partial F}{\partial q_{i}}\dfrac{\partial P_{k}}{\partial p_{i}}-\dfrac{\partial F}{\partial p_{i}}\dfrac{\partial P_{k}}{\partial q_{i}}\right)\}\right]\nonumber\\
&=&\sum_{k}\left[\dfrac{\partial G}{\partial Q_{k}}\{F,Q_{k}\}_{q,p}+\dfrac{\partial G}{\partial P_{k}}\{F,P_{k}\}_{q,p}\right]\label{eq2..38}
\end{eqnarray}
We shall use this formula (\ref{eq2..38}) to evaluate $\{F,Q_{k}\}_{q,p}$ and $\{F,P_{k}\}_{q,p}$. Now,
\begin{eqnarray}
\{Q_{k},F\}_{q,p}=\sum_{j}\left\{\dfrac{\partial F}{\partial Q_{j}}\{Q_{k},Q_{j}\}_{q,p}+\dfrac{\partial F}{\partial P_{j}}\{Q_{k},P_{j}\}_{q,p}\right\}\nonumber\\=
\sum_{j}\left(\dfrac{\partial F}{\partial Q_{j}}.0+\dfrac{\partial F}{\partial P_{j}}.\delta_{kj}\right)=\dfrac{\partial F}{\partial P_{k}}\nonumber
\end{eqnarray} $\therefore \{F,Q_{k}\}_{q,p}=-\dfrac{\partial F}{\partial P_{k}}$. \\
Similarly, we find $\{P_{k},F\}_{q,p}=-\dfrac{\partial F}{\partial Q_{k}}$ i.e. $\{F,P_{k}\}_{q,p}=\dfrac{\partial F}{\partial Q_{k}}$. \\
Hence from (\ref{eq2..38})
\begin{equation}
\{F,G\}_{q,p}=\sum_{i}\left(-\dfrac{\partial G}{\partial Q_{i}}\dfrac{\partial F}{\partial P_{i}}+\dfrac{\partial G}{\partial P_{i}}\dfrac{\partial F}{\partial Q_{i}}\right)=\{F,G\}_{(Q,P)}\nonumber
\end{equation} Therefore Poisson bracket remains invariant under canonical transformation.

\vspace{.5cm}

$\bullet$ Show that the Jacobian of any canonical transformation is unity i.e. $J=\dfrac{\partial (Q_{1},Q_{2},...,Q_{n},P_{1},P_{2},...,P_{n})}{\partial (q_{1},q_{2},...,q_{n},p_{1},p_{2},...,p_{n})}=1$

\vspace{.25cm}

\textbf{Proof:} We shall use $F_{2}(q,P,t)$ to give the canonical transformation from $(q,p)\rightarrow(Q,P)$, where $Q_{i}=Q_{i}(q,p,t)$ and $P_{i}=P_{i}(q,p,t)$. The transformation equations are $p_{i}=\dfrac{\partial F_{2}}{\partial q_{i}},~~Q_{i}=\dfrac{\partial F_{2}}{\partial P_{i}}$.\\
Now we can write,
\begin{eqnarray}
J&=&\dfrac{\partial (Q_{1},Q_{2},...,Q_{n},P_{1},P_{2},...,P_{n})}{\partial (q_{1},q_{2},...,q_{n},p_{1},p_{2},...,p_{n})}\nonumber\\
&=&\dfrac{\partial (Q_{1},Q_{2},...,Q_{n},P_{1},P_{2},...,P_{n})}{\partial (q_{1},q_{2},...,q_{n},P_{1},P_{2},...P_{n})}\times \dfrac{\partial (q_{1},q_{2},...,q_{n},P_{1},P_{2},...,P_{n})}{\partial (q_{1},q_{2},...,q_{n},p_{1},p_{2},...,p_{n})}\nonumber\\&=&\dfrac{\partial (Q_{1},Q_{2},...,Q_{n})}{\partial (q_{1},q_{2},...,q_{n})}|_{P=const.}\times \dfrac{\partial (P_{1},P_{2},...,P_{n})}{\partial (p_{1},p_{2},...,p_{n})}|_{q=const.}\label{eq2..39}
\end{eqnarray}
By definition, $
\dfrac{\partial (Q_{1},Q_{2},...,Q_{n})}{\partial (q_{1},q_{2},...,q_{n})}=$ \[
\begin{vmatrix}
\frac{\partial Q_{1}}{\partial q_{1}} & \frac{\partial Q_{2}}{\partial q_{1}}  & \dots & \frac{\partial Q_{n}}{\partial q_{1}} \\ 
\frac{\partial Q_{1}}{\partial q_{2}} & \frac{\partial Q_{2}}{\partial q_{2}}  & \dots & \frac{\partial Q_{n}}{\partial q_{2}} \\
\hdotsfor{4} \\
\frac{\partial Q_{1}}{\partial q_{n}} & \frac{\partial Q_{2}}{\partial q_{n}}  & \dots & \frac{\partial Q_{n}}{\partial q_{n}} 
\end{vmatrix}\]
\[=
\begin{vmatrix}
\frac{\partial^{2}F_{2}}{\partial q_{1} \partial P_{1}} & \frac{\partial^{2} F_{2}}{\partial q_{1} \partial P_{2}}  & \dots & \frac{\partial^{2} F_{2}}{\partial q_{1} \partial P_{n}} \\ 
\frac{\partial^{2}F_{2}}{\partial q_{2} \partial P_{1}} & \frac{\partial^{2} F_{2}}{\partial q_{2}\partial P_{1}}  & \dots & \frac{\partial_{2} F_{2}}{\partial q_{2}\partial P_{n}} \\
\hdotsfor{4} \\
\frac{\partial^{2}F_{2}}{\partial q_{n} \partial P_{1}} & \frac{\partial^{2} F_{2}}{\partial q_{n} \partial P_{2}}  & \dots & \frac{\partial^{2}F_{2}}{\partial q_{n} \partial P_{n}} 
\end{vmatrix}\]
\[=
\begin{vmatrix}
\frac{\partial}{\partial P_{1}} (\frac{\partial F_{2}}{\partial q_{1}}) & \frac{\partial}{\partial P_{2}}(\frac{\partial F_{2}}{\partial q_{1}})  & \dots & \frac{\partial}{\partial P_{n}}(\frac{\partial F_{2}}{\partial q_{1}}) \\ 
\frac{\partial }{\partial P_{1}}(\frac{\partial F_{2}}{\partial q_{2}}) & \frac{\partial}{\partial P_{2}}(\frac{\partial F_{2}}{\partial q_{2}})  & \dots & \frac{\partial}{\partial P_{n}} (\frac{\partial F_{2}}{\partial q_{2}})\\
\hdotsfor{4} \\
\frac{\partial}{\partial P_{1}} (\frac{\partial F_{2}}{\partial q_{n}})& \frac{\partial}{\partial P_{2}}(\frac{\partial F_{2}}{\partial q_{n}})  & \dots & \frac{\partial}{\partial P_{n}}(\frac{\partial F_{2}}{\partial q_{n}}) 
\end{vmatrix}\]
\[=
\begin{vmatrix}
\frac{\partial p_{1}}{\partial P_{1}} & \frac{\partial p_{1}}{\partial P_{2}}  & \dots & \frac{\partial p_{1}}{\partial P_{n}} \\ 
\frac{\partial p_{2}}{\partial P_{1}} & \frac{\partial p_{2}}{\partial P_{2}}  & \dots & \frac{\partial p_{2}}{\partial P_{n}} \\
\hdotsfor{4} \\
\frac{\partial p_{n}}{\partial P_{1}} & \frac{\partial p_{n}}{\partial P_{2}}  & \dots & \frac{\partial p_{n}}{\partial P_{n}} 
\end{vmatrix}\]\\
$$=\frac{\partial(p_1,p_2,...,p_n)}{\partial(P_1,P_2,...,P_n)}$$
Hence from (\ref{eq2..39}) $J=1$. (proved)
\section{Phase space:}
For each degree of freedom of a dynamical system there are two quantities namely $(q_{k},p_{k})$ that assume independent role. Imagine a space of $2m$ dimension in which $p_{1},p_{2},...,p_{m}$ and $q_{1},q_{2},...,q_{m}$ are the co-ordinates associated with a point called representative point. As $q_{k},p_{k}$ change in physical space with time the representative point moves in this $2m$ dimensional phase space.
\section{Liouviell's Theorem}
\textbf{Statement:} The density of an element in the phase space corresponding to a system of particles remains constant during the motion i.e. the total increment in $\rho$ as the state of the system varies is zero.

\vspace{.25cm}

\textbf{Proof:} Suppose that a large no. of identical one dimensional system are present each having a representative point in the $pq$-plane. At a given point in the $pq$-plane and within an area $dp.dq$ there will be many representative points at any instant. Let the density of such points be $\rho(q,p,t)$. By the density we mean the number of points divided by the elementary area as the latter approaches to zero. The no. of representative points moving into $dpdq$ on the left edge is $\rho ~\dot{q}~dp$ per unit time. Thus the  no. moving out of $dpdq$ through its right edge is $\{\rho~\dot{q}+\dfrac{\partial}{\partial q}(\rho~\dot{q})dq\}dp$. Hence the net increase in $\rho$ in the element is 
\begin{equation}
-\left(\{\rho~\dot{q}+\dfrac{\partial}{\partial q}(\rho~\dot{q})dq\}dp\right)+\rho\dot{q}=-\dfrac{\partial}{\partial q}(\rho \dot{q})~dq.dp\label{eq2..40}
\end{equation} In a similar way the net gain due to flow in the vertical direction is
\begin{equation}
-\dfrac{\partial}{\partial p}(\rho \dot{p})~dq.dp\label{eq2..41}
\end{equation} But the sum of these (\ref{eq2..40}) and (\ref{eq2..41}) equals $\left(\dfrac{\partial \rho}{\partial t}\right)dpdq$.
\begin{eqnarray}
\therefore \dfrac{\partial \rho}{\partial t}+\dfrac{\partial}{\partial q}(\rho \dot{q})+\dfrac{\partial}{\partial \rho}(\rho \dot{p})=0\nonumber\\
or, ~~\dfrac{\partial \rho}{\partial t}+\{\dot{q}\dfrac{\partial \rho}{\partial q}+\dot{p}\dfrac{\partial \rho}{\partial p}\}+\rho\{\dfrac{\partial \dot{q}}{\partial q}+\dfrac{\partial \dot{p}}{\partial p}\}=0\nonumber
\end{eqnarray} If $H$ be the Hamiltonian of the system then $\dot{p}=-\frac{\partial H}{\partial q},~\dot{q}=\frac{\partial H}{\partial p}$. So the 3rd term in the R.H.S within curly bracket vanishes identically. Hence we have,
\begin{eqnarray}
\dfrac{\partial \rho}{\partial t}+\{\dot{q}\dfrac{\partial \rho}{\partial q}+\dot{p}\dfrac{\partial \rho}{\partial p}\}=0\nonumber\\
or,~\dfrac{\partial \rho}{\partial t}dt+\dfrac{\partial \rho}{\partial p}\dot{p}dt+\dfrac{\partial \rho}{\partial q}\dot{q}dt=0\nonumber\\
i.e. ~~\dfrac{\partial \rho}{\partial t}=0\nonumber
\end{eqnarray} , Hence the theorem.\\
The result deduced above for a system with 1 degree of freedom may be generalized for a system with $m$ degrees of freedom. If the representative point in phase space be defined by the co-ordinates $q_{1},q_{2},...,q_{m},p_{1},p_{2},...,p_{m}$ the Liouviell's theorem becomes
\begin{equation}
\dfrac{\partial \rho}{\partial t}+\sum_{k=1}^{m}\left(\dot{q_{k}}\dfrac{\partial \rho}{\partial q_{k}}+\dot{p_{k}}\dfrac{\partial \rho}{\partial p_{k}}\right)=0\nonumber
\end{equation} If $H$ be the Hamiltonian of the system then we have
\begin{eqnarray}
\dfrac{\partial \rho}{\partial t}+\sum_{k=1}^{m}\left(\dfrac{\partial H}{\partial p_k}\dfrac{\partial \rho}{\partial q_{k}}-\dfrac{\partial H}{\partial q_{k}}\dfrac{\partial \rho}{\partial p_{k}}\right)=0\nonumber\\
or,~~	\dfrac{\partial \rho}{\partial t}+\{\rho,H\}=0~~ i.e. ~~\dfrac{d\rho}{dt}=0\nonumber
\end{eqnarray} Thus $\rho$ is a constant of motion. The density of the systems in the neighbourhood of some given system in phase space remains constant in time.

\begin{figure}
	\centering
	\includegraphics[width=0.4\textwidth]{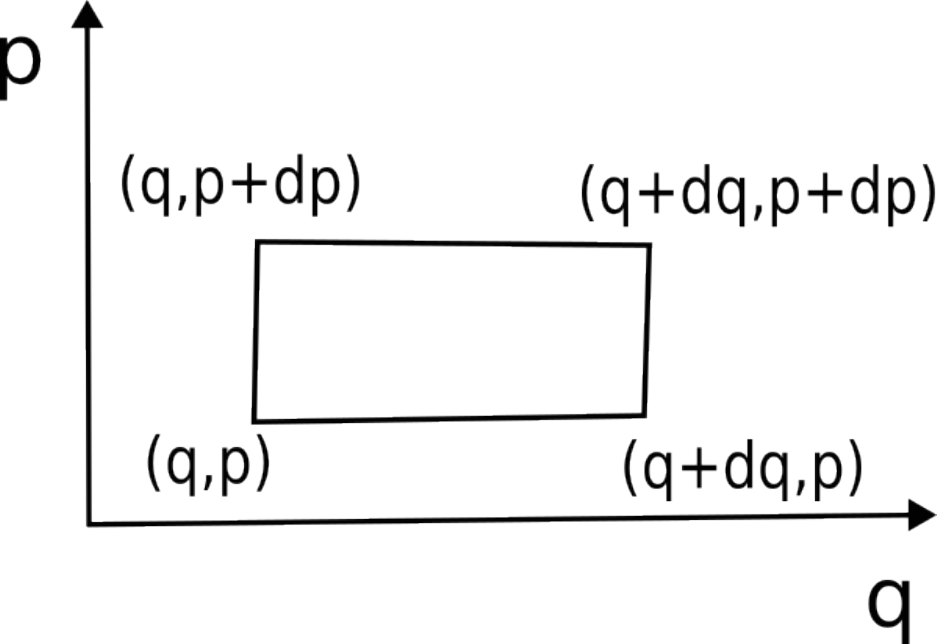}\\
	\label{liovell_2_13}
\end{figure}

\section{Integral Invariance of Poincare}
The following integrals
\begin{eqnarray}
J_{1}=\int_{S_{2}}\sum_{i}dq_{i}dp_{i}\nonumber\\
J_{2}=\int_{S_{4}}\sum_{i,k}dq_{i}dp_{i}dq_{k}dp_{k}\nonumber\\
---------------\nonumber\\
---------------\nonumber\\
J_{n}=\int_{S_{2n}}dq_{1}dq_{2}...dq_{n}dp_{1}dp_{2}...dp_{n}\nonumber
\end{eqnarray} are invariant under canonical transformation, where $S_{2}$ is a two dimensional surface in phase space, $S_{4}$ is a four dimensional surface in phase space,.... $S_{2n}$ is a $2n$ dimensional surface in phase space, enclosing completely an arbitrary region of space.\\
Proof: Due to complexity we shall show only the invariance of $J_{1}$.\\
Here $S_{2}$ is any two dimensional surface of phase space. Since the point in $2D$ can be represented uniquely by two independent variables $u$ and $v$ (say). We can take on the surface\\
$q_{i}=q_{i}(u,v),~p_{i}=p_{i}(u,v)$.\\
Now,
\begin{eqnarray}
\sum dq_{i}dp_{i}=\sum_{i}\dfrac{\partial (q_{i},p_{i})}{\partial (u,v)}dudv\nonumber\\
=\sum_{i}\left(\dfrac{\partial q_{i}}{\partial u}\dfrac{\partial p_{i}}{\partial v}-\dfrac{\partial q_{i}}{\partial v}\dfrac{\partial p_{i}}{\partial u}\right)dudv\nonumber\\
=[u,v]_{q,p}dudv\nonumber\\
\therefore J_{1}=\int \int_{S_{2}}\sum dq_{i} dp_{i}=\int \int_{S_{2}} [u,v]_{q,p}dudv.\nonumber
\end{eqnarray} To show the invariance of $J_{1}$ under canonical transformation we shall have to show that
\begin{equation}
\int \int_{S_{2}}\sum_{i}\dfrac{\partial (q_{i},p_{i})}{\partial (u,v)}dudv=\int \int_{S_{2}} \sum_{k} \dfrac{\partial (Q_{k},P_{k})}{\partial (u,v)}dudv=\int \int_{S_{2}}[u,v]_{q,p}dudv\nonumber
\end{equation} where $Q_{i}=Q_{i}(q,p,t),~~P_{i}=P_{i}(q,p,t)$ are also canonical variables. As the region of integration is arbitrary so to prove the invariance we shall have to show
\begin{equation}
\sum_{i}\dfrac{\partial (q_{i},p_{i})}{\partial (u,v)} =\sum_{k} \dfrac{\partial (Q_{k},P_{k})}{\partial (u,v)}\label{eq2..42}
\end{equation}
Now for canonical transformation from $(q,p)$ to $(Q,P)$ we take the help of $F_{2}=F_{2}(q,P,t)$. The transformation equations are
\begin{equation}
p_{i}=\dfrac{\partial F_{2}}{\partial q_{i}},~~Q_{i}=\dfrac{\partial F_{2}}{\partial P_{i}}\label{eq2..43}
\end{equation} Thus the L.H.S of (\ref{eq2..42}) is
\[\sum_{i}
\begin{vmatrix}
\dfrac{\partial q_{i}}{\partial u} & \dfrac{\partial p_{i}}{\partial u} \\ 
\dfrac{\partial q_{i}}{\partial v} & \dfrac{\partial p_{i}}{\partial v} 
\end{vmatrix}
\] Using (\ref{eq2..43}) we can write,
\begin{eqnarray}
\dfrac{\partial p_{i}}{\partial u}=\dfrac{\partial}{\partial u}\left(\dfrac{\partial F_{2}}{\partial q_{i}}\right)=\sum_{k} \dfrac{\partial^{2}F_{2}}{\partial P_{k}\partial q_{i}}.\dfrac{\partial P_{k}}{\partial u}+\sum_{k}\dfrac{\partial^{2}F_{2}}{\partial q_{k}\partial q_{i}}.\dfrac{\partial q_{k}}{\partial u}\nonumber\\
and,~~	\dfrac{\partial p_{i}}{\partial v}=\dfrac{\partial}{\partial v}\left(\dfrac{\partial F_{2}}{\partial q_{i}}\right)=\sum_{k} \dfrac{\partial^{2}F_{2}}{\partial P_{k}\partial q_{i}}.\dfrac{\partial P_{k}}{\partial v}+\sum_{k}\dfrac{\partial^{2}F_{2}}{\partial q_{k}\partial q_{i}}.\dfrac{\partial q_{k}}{\partial v}\nonumber
\end{eqnarray} So the L.H.S of (\ref{eq2..42}) becomes
\[
\sum_{i}
\begin{vmatrix}
\dfrac{\partial q_{i}}{\partial u} & \sum_{k} \dfrac{\partial^{2}F_{2}}{\partial P_{k}\partial q_{i}}.\dfrac{\partial P_{k}}{\partial u}+\sum_{k}\dfrac{\partial^{2}F_{2}}{\partial q_{k}\partial q_{i}}.\dfrac{\partial q_{k}}{\partial u} \\ 
\dfrac{\partial q_{i}}{\partial v} & \sum_{k} \dfrac{\partial^{2}F_{2}}{\partial P_{k}\partial q_{i}}.\dfrac{\partial P_{k}}{\partial v}+\sum_{k}\dfrac{\partial^{2}F_{2}}{\partial q_{k}\partial q_{i}}.\dfrac{\partial q_{k}}{\partial v}
\end{vmatrix}
\]
\[=\sum_{i} \sum_{k} \dfrac{\partial^{2}F_{2}}{\partial P_{k}\partial q_{i}}
\begin{vmatrix}
\dfrac{\partial q_{i}}{\partial u} & \dfrac{\partial P_{k}}{\partial u} \\ 
\dfrac{\partial q_{i}}{\partial v} & \dfrac{\partial P_{k}}{\partial v} 
\end{vmatrix}+\sum_{i}\sum_{k} \dfrac{\partial^{2}F_{2}}{\partial q_{k}\partial q_{i}}
\begin{vmatrix}
\dfrac{\partial q_{i}}{\partial u} & \dfrac{\partial q_{k}}{\partial u} \\ 
\dfrac{\partial q_{i}}{\partial v} & \dfrac{\partial q_{k}}{\partial v} 
\end{vmatrix}
\] (The second term on the R.H.S is zero by interchanging $i$ and $k$. We replace this zero term by a similar zero term as follows:)
\[=
\sum_{i} \sum_{k} \dfrac{\partial^{2}F_{2}}{\partial P_{k}\partial q_{i}}
\begin{vmatrix}
\dfrac{\partial q_{i}}{\partial u} & \dfrac{\partial P_{k}}{\partial u} \\ 
\dfrac{\partial q_{i}}{\partial v} & \dfrac{\partial P_{k}}{\partial v} 
\end{vmatrix}+\sum_{i}\sum_{k} \dfrac{\partial^{2}F_{2}}{\partial P_{k}\partial P_{i}}
\begin{vmatrix}
\dfrac{\partial P_{i}}{\partial u} & \dfrac{\partial P_{k}}{\partial u} \\ 
\dfrac{\partial P_{i}}{\partial v} & \dfrac{\partial P_{k}}{\partial v} 
\end{vmatrix}
\]
\[=\sum_{k}
\begin{vmatrix}
\sum_{i}  \dfrac{\partial^{2}F_{2}}{\partial P_{k}\partial q_{i}}	\dfrac{\partial q_{i}}{\partial u}+\sum \dfrac{\partial^{2}F_{2}}{\partial P_{k}\partial P_{i}}\dfrac{\partial P_{i}}{u} &&&\dfrac{\partial P_{k}}{\partial u} \\ 
\sum_{i}  \dfrac{\partial^{2}F_{2}}{\partial P_{k}\partial q_{i}}	\dfrac{\partial q_{i}}{\partial v}+ \sum \dfrac{\partial^{2}F_{2}}{\partial P_{k}\partial P_{i}}\dfrac{\partial P_{i}}{v} &&& \dfrac{\partial P_{k}}{\partial v} 
\end{vmatrix}
\]
\[=
\sum_{k}
\begin{vmatrix}
\dfrac{\partial}{\partial u}\left(\dfrac{\partial F_{2}}{\partial P_{k}}\right) & \dfrac{\partial P_{k}}{\partial u} \\ 
\dfrac{\partial}{\partial v}\left( \dfrac{\partial F_{2}}{\partial P_{k}}\right)& \dfrac{\partial P_{k}}{\partial v}
\end{vmatrix}
\]
\[=
\sum_{k}
\begin{vmatrix}
\dfrac{\partial Q_{k}}{\partial u} & \dfrac{\partial P_{k}}{\partial u} \\ 
\dfrac{\partial Q_{k}}{\partial v} & \dfrac{\partial P_{k}}{\partial v}
\end{vmatrix}=\sum_{k} \dfrac{\partial (Q_{k}, P_{k})}{\partial (u,v)}=[u,v]_{(Q,P)}
\]$~~~~~~~~~~~~~~~~~~~~~~~~~~~~~~~~~~~~~~$ Hence the proof.


\chapter{Action principle $\&$ consequences}

\section{Hamilton's Principle Function}
Let, $$S=\int_{t_0}^{t_1}(T-V)dt.$$
Suppose a given problem is solved. `$S$' may be regarded as a function of $t_0$, $t_1$, $a_1$, $a_2$,...,$a_n$ and $\dot{a_1}$, $\dot{a_2}$,...,$\dot{a_n}$, where $a$'s and $\dot{a}$'s are the initial values of $q$'s and $\dot{q}$'s at $t=t_0$. The co-ordinates $q_1$, $q_2$,..., $q_n$ at time $t$ may be regarded as a function of the same quantities so that we may consider that we have eliminated $\dot{a_1}$, $\dot{a_2}$,...,$\dot{a_n}$ and regard `$S$' as a function of $t_0$, $t_1$, $a_1$, $a_2$,...,$a_n$, $q_1$, $q_2$,..., $q_n$. `$S$' thus expressed is called Hamilton's principle function.

\section{Hamilton-Jacobi partial differential equation}
Let us consider $$S=\int_{t_0}^{t}(T-V)dt=\int_{t_0}^{t}Ldt$$
The value of the Lagrangian function $L$ on the varied path is $L+\delta L$.

\begin{figure}[h!]
	\centering
	\includegraphics[scale=0.4]{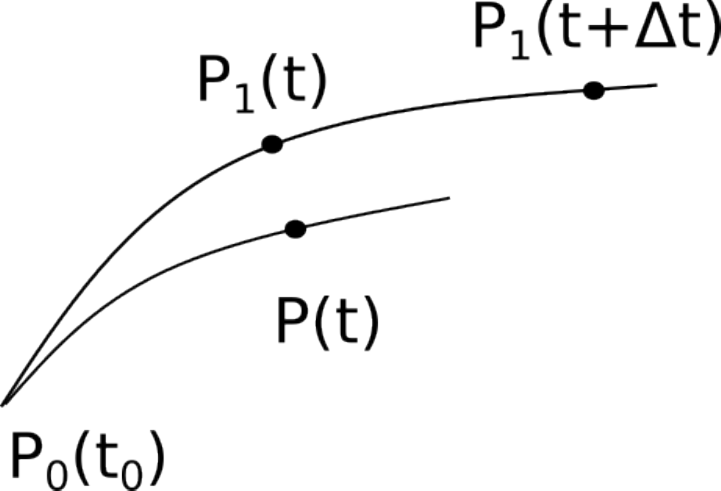}
\end{figure}

Now, 
\begin{eqnarray}
\Delta S=\Delta\int_{t_0}^{t}Ldt&=&\int_{t_0}^{t+\Delta t}(L+\delta L)dt-\int_{t_0}^{t}Ldt\nonumber\\
&=&\int_{t_0}^{t}(L+\delta L)dt+\int_{t}^{t+\Delta t}(L+\delta L)dt-\int_{t_0}^{t}Ldt\nonumber\\
&=&\int_{t}^{t+\Delta t}(L+\delta L)dt+\int_{t_0}^{t}\delta Ldt\nonumber\\
&=&L.\Delta t+\int_{t_0}^{t}\sum\left(\frac{\partial L}{\partial q_i}\delta q_i+\frac{\partial L}{\partial\dot{q_i}}\delta\dot{q_i}\right)dt\nonumber\\
&~&~~~~~~~~~~~~~~~~~~~~~~~~~~~~~~~~~~~~~~~~~\mbox{(by mean value theorem)}\nonumber\\
&=&L.\Delta t+\int_{t_0}^{t}\sum\left\{\frac{\partial L}{\partial q_i}-\frac{d}{dt}\left(\frac{\partial L}{\partial\dot{q_i}}\right)\right\}\delta q_idt+\sum\frac{\partial L}{\partial\dot{q_i}}\delta q_i\bigg|^t_{t_0}\nonumber\\
&=&L.\Delta t+\sum\left(\frac{\partial L}{\partial\dot{q_i}}\delta q_i\right)\bigg|^t\nonumber\\
&~&~~~~~(\mbox{using Lagrange's equation of motion and}~~\delta q_i=0~~\mbox{at}~~t=t_0)\nonumber\\
&=&L.\Delta t+\sum\frac{\partial L}{\partial\dot{q_i}}\left(\Delta q_i-\dot{q_i}\Delta t\right)\nonumber\\
&~&~~~~~~~~~~~~~~~~~~~~~~~~~~~~~~~~~~~~~~~~~(\because \Delta q=\delta q+\dot{q}\Delta t)\nonumber\\
&=&\left(L-\sum\frac{\partial L}{\partial\dot{q_i}}\dot{q_i}\right)\Delta t+\sum\frac{\partial L}{\partial\dot{q_i}}\Delta q_i\nonumber
\end{eqnarray}

\begin{equation}
\therefore \Delta S=-H\Delta t+\sum\frac{\partial L}{\partial\dot{q_i}}\Delta q_i\nonumber
\end{equation}

\begin{equation}
\therefore \frac{\partial S}{\partial t}=-H,~~~~~~\frac{\partial S}{\partial q_i}=\frac{\partial L}{\partial\dot{q_i}}=p_i,~~~~~~i=1,2,...n.\nonumber
\end{equation}

\begin{eqnarray}
\frac{\partial S}{\partial t}&=&-H(t,q_1,q_2,...,q_n,p_1,p_2,...,p_n)\nonumber\\
&=&-H\left(t,q_1,q_2,...,q_n,\frac{\partial S}{\partial q_1},\frac{\partial S}{\partial q_2},...,\frac{\partial S}{\partial q_n}\right)\nonumber
\end{eqnarray}
or, $$\frac{\partial S}{\partial t}+H\left(t,q_1,q_2,...,q_n,\frac{\partial S}{\partial q_1},\frac{\partial S}{\partial q_2},...,\frac{\partial S}{\partial q_n}\right)=0$$
This is called the Hamilton-Jacobi partial differential equation.

\vspace{.5cm}

$\bullet$ \textbf{Problem I}: A particle oscillates in a straight line about a centre of force which varies as the distance. Show that the Hamilton's principle function is $$S=\dfrac{\sqrt{\mu}}{2}\left[\dfrac{\left(x_0^2+x^2\right)\cos\left\{\sqrt{\mu}(t-t_0)\right\}-2xx_0}{\sin\left\{\sqrt{\mu}(t-t_0)\right\}}\right]$$.

\vspace{.25cm}

\textbf{Solution:} The equation of motion is $\ddot{x}=-n^2x$.

$\therefore x=A\cos nt+B\sin nt$.

Suppose at $t=t_0$, $x=x_0$, $\dot{x}=\dot{x_0}$.

$\therefore x_0=A\cos nt_0+B\sin nt_0$.

and $\dfrac{\dot{x_0}}{n}=-A\sin nt_0+B\cos nt_0$.

$\therefore A=\left(x_0\cos nt_0-\dfrac{\dot{x_0}}{n}\sin nt_0\right)$, $B=\left(x_0\sin nt_0+\dfrac{\dot{x_0}}{n}\cos nt_0\right)$.

\begin{equation}
\therefore x=x_0\cos n\tau+\dfrac{\dot{x_0}}{n}\sin n\tau,~~~~~~~~~~\tau=t-t_0.\label{eq3..1}
\end{equation}
If $V$ be the potential, then $~~-\dfrac{\partial V}{\partial x}=-n^2x$.

i.e, $V=\dfrac{n^2x^2}{2}$ and $T=\dfrac{1}{2}\dot{x}^2$.

\begin{eqnarray}
V=\frac{1}{2}n^2\left[x_0^2\cos^2n\tau+\frac{\dot{x_0}^2}{n^2}\sin^2n\tau+\frac{2x_0\dot{x_0}}{n}\sin n\tau\cos n\tau\right]\nonumber.\\
T=\frac{1}{2}n^2\left[x_0^2\sin^2n\tau+\frac{\dot{x_0}^2}{n^2}\cos^2n\tau-\frac{2x_0\dot{x_0}}{n}\sin n\tau\cos n\tau\right]\nonumber.
\end{eqnarray} 
\begin{eqnarray}
\therefore S&=&\int_{0}^{t}(T-V)dt=\int_{0}^{\tau}(T-V)d\tau\nonumber\\
&=&\frac{n^2}{2}\int_{0}^{\tau}\left[\frac{\dot{x_0}^2}{n^2}\cos 2n\tau-x_0^2\cos 2n\tau-\frac{2x_0\dot{x_0}}{n}\sin 2n\tau\right]d\tau\nonumber\\
&=&\frac{n^2}{2}\left[\left(\frac{\dot{x_0}^2}{n^2}-x_0^2\right)\left\{\frac{\sin 2n\tau}{2n}\right\}_0^\tau+\frac{2x_0\dot{x_0}}{n}\left\{\frac{\cos 2n\tau}{2n}\right\}_0^\tau\right]\nonumber\\
&=&\frac{n}{4}\left(\frac{\dot{x_0}^2}{n^2}-x_0^2\right)\sin 2n\tau+\frac{x_0\dot{x_0}}{2}(\cos 2n\tau-1)\nonumber
\end{eqnarray}

From (\ref{eq3..1}), $$\frac{\dot{x_0}}{n}=\frac{x-x_0\cos n\tau}{\sin n\tau}$$

\begin{eqnarray}
\therefore S=\frac{n}{4}\left[\frac{x^2+x_0^2\cos^2n\tau-2xx_0\cos n\tau-x_0^2\sin^2n\tau}{\sin^2n\tau}\right]\sin 2n\tau\nonumber\\
~~~~~~~~~~~~~~~~~~~~~~+\frac{x_0n}{2}\frac{(x-x_0\cos n\tau)(\cos 2n\tau-1)}{\sin n\tau}\nonumber\\
=\frac{1}{2}\sqrt{\mu}\left[\frac{(x^2+x_0^2)\cos(\sqrt{\mu}\tau)-2xx_0}{\sin(\sqrt{\mu}\tau)}\right],~~~~~~~~~~~~~~~~~~~~~~\nonumber\\
~~~~~~~~~~~~~~~~~~~~~~~~~~~~~~~~~~~~~~~~~~~~~\tau=t-t_0, ~~n=\sqrt{\mu}\nonumber.
\end{eqnarray}

\vspace{.5cm}

$\bullet$ \textbf{Problem II}: Prove that in case of a particle of unit mass moving in a plane $xy$ under a central acceleration $n^2r$, the value of the Hamilton's principal function is $S=\dfrac{n}{2\sin n\tau}\left\{\left(x_0^2+y_0^2+x^2+y^2\right)\cos n\tau-2(xx_0+yy_0)\right\}$ and $\tau=t-t_0$.

\vspace{.5cm}

\section{Virial Theorem}
Let us consider the motion of a system subject to a force $\overrightarrow{F_i}$ in the position $\overrightarrow{r_i}$.Then the kinetic energy of the system is
$$T=\frac{1}{2}\sum m\dot{\overrightarrow{r}}^2.$$
\begin{eqnarray}
\therefore 2T&=&\sum m\dot{\overrightarrow{r}}^2=\sum m\dot{\overrightarrow{r}}\dot{\overrightarrow{r}}=\frac{d}{dt}\left(\sum m\dot{\overrightarrow{r}}.\overrightarrow{r}\right)-\sum\overrightarrow{r}\frac{d}{dt}\left(m\dot{\overrightarrow{r}}\right)\nonumber\\
&=&\frac{d}{dt}\left(\sum m\dot{\overrightarrow{r}}.\overrightarrow{r}\right)-\sum\overrightarrow{r}.\overrightarrow{F},~~~~\overrightarrow{F}=\frac{d}{dt}\left(m\dot{\overrightarrow{r}}\right)=m\ddot{\overrightarrow{r}}.\nonumber
\end{eqnarray} 
or, 
\begin{equation}
2T+\sum\overrightarrow{r}.\overrightarrow{F}=\frac{d}{dt}\left(\sum m\dot{\overrightarrow{r}}.\overrightarrow{r}\right).\label{eq3..2}
\end{equation}
$\big(\sum\overrightarrow{r}.\overrightarrow{F}$ is called the virial of the system$\big).$

Let us now average the equation (\ref{eq3..2}) over an arbitrary interval of time $(0,\tau)$, then 
\begin{eqnarray}
\frac{1}{\tau}\int_{0}^{\tau}\left(2T+\sum\overrightarrow{r}.\overrightarrow{F}\right)dt=\frac{1}{\tau}\left[\sum m\dot{\overrightarrow{r}}.\overrightarrow{r}\right]_0^\tau~~~~~~~~~~~~~~~~~~~~~~~~~~~~~~~~\nonumber\\
~~~~~~~~~~~~~~~~~~~~~~~~~~~~~=\frac{1}{\tau}\left[\left(\sum m\dot{\overrightarrow{r}}.\overrightarrow{r}\right)\bigg|_\tau-\left(\sum m\dot{\overrightarrow{r}}.\overrightarrow{r}\right)\bigg|_{\tau=0}\right]\label{eq3..3}
\end{eqnarray} 
Let us now assume that the system executed a finite motion in a finite region of space, then the numerator of the right hand side of equation (\ref{eq3..3}) is finite for all values of $\tau$, since it is bounded for all values of $\tau$. Consequently, taking the limit as $\tau\rightarrow\infty$ we get from (\ref{eq3..3})
\begin{eqnarray}
\lim_{\tau\rightarrow\infty}\frac{1}{\tau}\int_{0}^{\tau}\left(2T+\sum\overrightarrow{r}.\overrightarrow{F}\right)dt=0\nonumber\\
\mbox{i.e,}~~2\overline{T}+\overline{\left(\sum\overrightarrow{r}.\overrightarrow{F}\right)}=0\label{eq3..4}
\end{eqnarray}
where $\overline{T}=\lim_{\tau\rightarrow\infty}\frac{1}{\tau}\int_{0}^{\tau}Tdt$ and $\overline{\left(\sum\overrightarrow{r}.\overrightarrow{F}\right)}=\lim_{\tau\rightarrow\infty}\frac{1}{\tau}\int_{0}^{\tau}\left(\overrightarrow{r}.\overrightarrow{F}\right)dt$.

Equation (\ref{eq3..4}) states that for finite motion of a system in a finite region of space twice the time average value of the kinetic energy is equal to $(-1)\times$ time average value of the virial of the system. This is known as \underline{virial theorem}.

If the field be conservative, then $\overrightarrow{F}=-\mbox{grad}V=-\dfrac{\partial V}{\partial\overrightarrow{r}}$; so equation (\ref{eq3..4}) becomes 
$$2\overline{T}=\sum\overrightarrow{r}.\frac{\partial V}{\partial\overrightarrow{r}}$$
In particular, if $V$ is homogeneous in the co-ordinates of degree $n$ then 
\begin{eqnarray}
\sum\overrightarrow{r}.\frac{\partial V}{\partial\overrightarrow{r}}=nV\nonumber\\
\therefore 2\overline{T}=n\overline{V}\label{eq3..5}
\end{eqnarray}

We obtain from the equation of energy

$$\overline{T}+\overline{V}=\mbox{constant}=E~~(\mbox{say})$$
\begin{eqnarray}
\therefore\overline{V}&=&E-\overline{T}\nonumber\\
\mbox{or},~~~2\overline{T}&=&n\left(E-\overline{T}\right)\nonumber\\
\mbox{or},~~~\overline{T}&=&\frac{n}{n+2}E~~~\mbox{and}~~~\overline{V}=\frac{2}{n+2}E\nonumber
\end{eqnarray}

\section{Vibration of String and Membrence:}
$\star$ \textbf{Transverse Vibration of a Stretched String:}

We take $x-$ axis along $AB$ the equilibrium position of the string and the end points at $x=0$ and $x=l$. We consider the plane of vibration as $xy-$ plane. The transverse displacement is $y$. We assume it to be so small that the tangent to the string at any point in the displaced position makes a small inclination with the $x-$ axis. Now we apply Hamilton's principle to deduce the equation of motion. Let $\rho$ be the density of the material of the string. The kinetic energy is 
$$T=\frac{1}{2}\int\rho\dot{y}^2dx$$
The integration being taken along the length of the string. The potential energy of the string is given by 
$$V=\int T_1(ds-dx),$$
taken over the length of the string, $T_1$ being the constant tension. Now 
$$ds^2=dx^2+dy^2~~\mbox{i.e,}~~\frac{ds}{dx}=\sqrt{1+\left(\frac{dy}{dx}\right)^2}\approx 1+\frac{1}{2}\left(\frac{dy}{dx}\right)^2$$
so that the potential energy is 
$$V\approx\frac{1}{2}\int T_1\left(\frac{dy}{dx}\right)^2dx$$
Now by Hamilton's principle, we have 
\begin{eqnarray}
\int_{t_1}^{t_2}\delta(T-V)dt=0~~~~~~~~~~~~~~~~~~~~~~~~\nonumber\\
\mbox{or,}~~\int_{t_1}^{t_2}dt\int\left[\rho\dot{y}\frac{d}{dt}\delta y-T_1\frac{dy}{dx}\frac{d}{dx}\delta y\right]dx=0\nonumber\\
\mbox{or,}~~\rho\dot{y}\delta ydx\bigg|_{t_1}^{t_2}-\int_{t_1}^{t_2}dt\int\left[\rho\ddot{y}\delta y+T_1\frac{dy}{dx}\delta y\right]dx=0\nonumber
\end{eqnarray}
~~~~~~~~~~~~~~~~~~~~~~~~~~~~~~~~~~~~~~~~~~~~~(Integrating by parts with respect to $t$ alone)

As $\delta y=0$ at $t=t_1$ and $t=t_2$, so again integrating by parts with respect to `$x$' we have
$$\int T_1\frac{dy}{dx}\delta ydt\bigg|_0^l+\int_{t_1}^{t_2}dt\int\left[\rho\ddot{y}-T_1\frac{d^2y}{dx^2}\right]\delta ydx=0$$
As $\delta y=0$ at $x=0$ and $x=l$ so
$$\int_{t_1}^{t_2}dt\int\left[\rho\ddot{y}-T_1\frac{\partial^2y}{\partial x^2}\right]\delta ydx=0$$
In order to satisfy this we have in local form
\begin{eqnarray}
\rho\ddot{y}-T_1\frac{\partial^2y}{\partial x^2}=0~~~~~~~~~~~~~~~\nonumber\\
\mbox{or,}~~\frac{\partial^2y}{\partial t^2}=c^2\frac{\partial^2y}{\partial x^2},~~~c^2=\frac{T_1}{\rho}\nonumber
\end{eqnarray}
This is the differential equation of vibrating string and is known as wave equation.

To find a solution of the differential equation we put, $\xi_1=x-ct$, $\xi_2=x+ct$.
\begin{eqnarray}
\therefore\frac{\partial}{\partial x}=\frac{\partial}{\partial\xi_1}+\frac{\partial}{\partial\xi_2},~~~\frac{\partial}{\partial t}=-c\frac{\partial}{\partial\xi_1}+c\frac{\partial}{\partial\xi_2}\nonumber\\
\therefore\left(-c\frac{\partial}{\partial\xi_1}+c\frac{\partial}{\partial\xi_2}\right)^2y=c^2\left(\frac{\partial}{\partial\xi_1}+\frac{\partial}{\partial\xi_2}\right)^2y\nonumber\\
\mbox{i.e,}~~\frac{\partial^2y}{\partial\xi_1\partial\xi_2}=0~~~~~~~~~~~~~~~~~~~~~~~~~~~~~~~~\nonumber\\
\mbox{or,}~~\frac{\partial y}{\partial\xi_2}=F(\xi_2)\nonumber~~~~~~~~~~~~~~~~~~~~~~~~~~~~~~~\\
\mbox{or,}~~y=f_1(\xi_1)+f_2(\xi_2)=f_1(x-ct)+f_2(x+ct)\nonumber
\end{eqnarray}

\begin{figure}[h!]
	\centering
	\includegraphics[ height=0.6 \textheight, width=0.6\textheight]{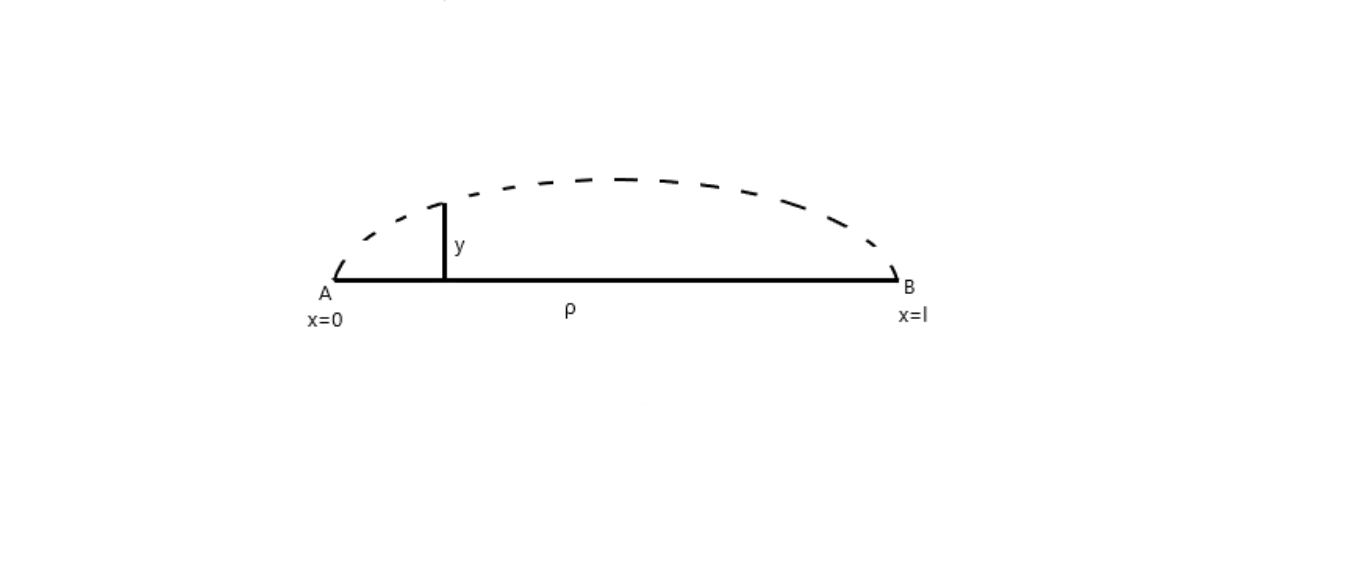}
\end{figure}

Let us assume, $y=f(x)\cos(nt+\epsilon)$ as a solution of the equation. Then from the wave equation
\begin{eqnarray}
f''(x)+\frac{n^2}{c^2}f(x)=0~~~~~~~~~~~~~~~~~~~~~\nonumber\\
\mbox{i.e,}~~f(x)=A\cos\left(\frac{nx}{c}\right)+B\sin\left(\frac{nx}{c}\right)\nonumber
\end{eqnarray}
Hence the solution is 
$$y=\left[A\cos\left(\frac{nx}{c}\right)+B\sin\left(\frac{nx}{c}\right)\right]\cos(nt+\epsilon)$$
As $y=0$ when $x=0$ and $x=l$ so we have 
$$A=0~~~\mbox{and}~~~0=B\sin\left(\frac{nl}{c}\right)$$
(Note that $B=0$ gives the trivial solution $y=0$)

\begin{eqnarray}
\therefore\sin\left(\frac{nl}{c}\right)=0=\sin s\pi~~~~~~(s~\mbox{is an integer including zero})\nonumber\\
\therefore n=\frac{s\pi c}{l}~~~~~~~~~~~~~~~~~~~~~~~~~~~~~~~~~~~~~~~~~~~~~~~~~~~~~~~~~~~~\nonumber\\
\therefore y=B_s\sin\left(\frac{s\pi x}{l}\right)\cos\left(\frac{s\pi c}{l}t+\epsilon\right)~~~~~~~~~~~~~~~~~~~~~~~~~~~~\nonumber
\end{eqnarray}
Hence the general solution is 
\begin{equation}
y=\sum_{s=1}^{\infty}B_s\sin\left(\frac{s\pi x}{l}\right)\cos\left(\frac{s\pi c}{l}t+\epsilon_s\right)\nonumber
\end{equation}
The period of the $s$th mode is $\dfrac{2\pi}{n}=\dfrac{2l}{sc}=\dfrac{2l}{s}\sqrt{\dfrac{\rho}{T_1}}$~~~~~~$(s=1,~2,...)$.

The frequency of the $s$-th mode is $\dfrac{s}{2l}\sqrt{\dfrac{T_1}{\rho}}$.

The frequency for greatest mode (i.e, $s=1$) is $\dfrac{1}{2l}\sqrt{\dfrac{T}{\rho}}$.

To determine the constants $B_s$ and $\epsilon_s$ we write the general solution as 
\begin{equation}
y=\sum_{s=1}^{\infty}\sin\left(\frac{s\pi x}{l}\right)\left[C_s\cos\left(\frac{s\pi ct}{l}\right)+D_s\sin\left(\frac{s\pi ct}{l}\right)\right]\nonumber
\end{equation}
where $C_s$ and $D_s$ are arbitrary constants, which will be determined from initial conditions. The vibration given by
\begin{equation}
y=\sin\left(\frac{s\pi x}{l}\right)\left[C_s\cos\left(\frac{s\pi ct}{l}\right)+D_s\sin\left(\frac{s\pi ct}{l}\right)\right]\nonumber
\end{equation}
is called the normal mode of vibration of the string. The general vibration of the string is obtained by superposition of normal modes. For a fixed `$s$', $y$ vanishes, when 
$$x=\frac{l}{s},~~\frac{2l}{s},~~\frac{3l}{s},...,\frac{(s-1)l}{s}$$
besides the points $x=0$ and $x=l$.

These points are called the nodes. For $s=2$, $x=\dfrac{l}{2}$ is the only node i.e, the middle point of the string is the only node. Now we shall determine the constants $C_s$ and $D_s$ as follows:

Let initially i.e, at $t=0$, $y=y_0(x)$ and $\dfrac{\partial y}{\partial t}=\dot{y_0}(x)$.

So when $t=0$, we have 
$$y_0(x)=\sum_{s=1}^{\infty}C_s\sin\left(\frac{s\pi x}{l}\right)$$
By Fourier's method, multiplying both side by $\sin\left(\dfrac{s\pi x}{l}\right)$ and integrating with respect to $x$ between the limits $x=0$ and $x=l$, we have
\begin{eqnarray}
\int_{0}^{l}y_0(x)\sin\frac{s\pi x}{l}dx=\int_{0}^{l}\sin^2\frac{s\pi x}{l}dx=\frac{C_s}{2}l\nonumber\\
\therefore C_s=\frac{2}{l}\int_{0}^{l}y_0(x)\sin\left(\frac{s\pi x}{l}\right)dx~~~~~~~~~~~~~~~~\nonumber
\end{eqnarray}
Similarly differentiating the expression for $y$ with respect to `$t$' and putting $t=0$, we have 
$$\dot{y_0}(x)=\sum_{s=1}^{\infty}\left(\frac{s\pi c}{l}\right)D_s\sin\left(\frac{s\pi x}{l}\right)$$
Now proceeding as before, we have
$$D_s=\frac{2}{s\pi c}\int_{0}^{l}\dot{y_0}(x)\sin\left(\frac{s\pi x}{l}\right)dx$$

$\star$ \textbf{Plucked String:}

\begin{figure}[h!]
	\centering
	\includegraphics[scale=0.5]{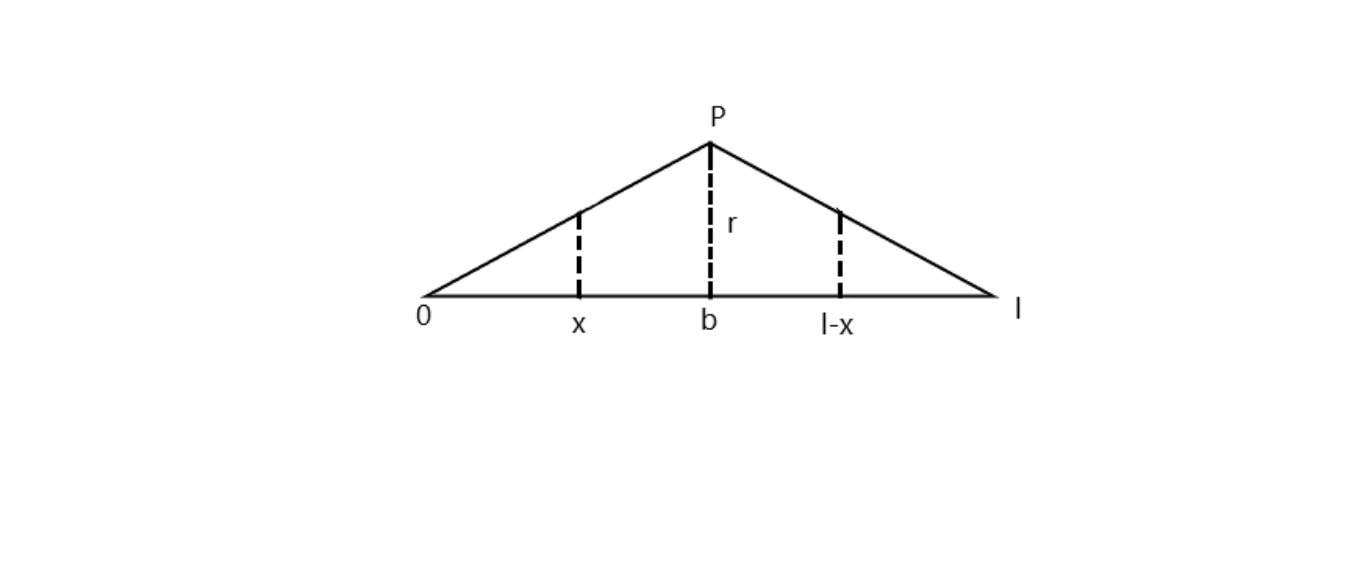}
\end{figure}

Let $x$-axis is along the string and the end points are at $x=0$, $x=l$. The equation of vibration is 
$$\frac{\partial^2y}{\partial t^2}=c^2\frac{\partial^2y}{\partial x^2},~~~\mbox{where}~~c^2=\frac{T_1}{\rho}$$
The motion starts from rest by displacing the point $x=b$ of the string through a distance `$r$' transversly and then letting it go. When $t=0$, $y=f(x)$ and $\dot{y}=0$, $f(x)$ defines the form of the string initially. Therefore $y=0$ when $x=0$ and $x=l$ and initially,
\begin{eqnarray}
f(x)&=&\frac{rx}{b},~~~0\leq x\leq b,~~~t=0\nonumber\\
&=&\frac{r(l-x)}{(l-b)},~~~b\leq x\leq l,~~~t=0\nonumber
\end{eqnarray}
As before, the solution of the wave equation is of the form 
$$y=\sum_{s=1}^{\infty}\sin\left(\frac{s\pi x}{l}\right)\left[C_s\cos\left(\frac{s\pi ct}{l}\right)+D_s\sin\left(\frac{s\pi ct}{l}\right)\right]$$
Differentiating term-by-term, we have
\begin{eqnarray}
0=\dot{y}\big|_{t=0}=\sum_{s=1}^{\infty}D_s\sin\left(\frac{s\pi x}{l}\right)\left(\frac{s\pi c}{l}\right)\nonumber\\
\therefore D_s=0~~~~~~~~~~~~~~~~~~~~~~~~~~~~~\nonumber
\end{eqnarray}
Also at $t=0$, $y=f(x)$ gives
$$f(x)=\sum_{s=1}^{\infty}C_s\sin\left(\frac{s\pi x}{l}\right)$$
To determine $C_s$ we multiply both side by $\sin\left(\dfrac{s\pi x}{l}\right)$ and integrating between $x=0$ to $x=l$ we have 
\begin{eqnarray}
C_s&=&\frac{2}{l}\int_{0}^{l}f(x)\sin\left(\frac{s\pi x}{l}\right)dx\nonumber\\
&=&\frac{2}{l}\left[\int_{0}^{b}\frac{rx}{b}\sin\left(\frac{s\pi x}{l}\right)dx+\int_{b}^{l}\frac{r(l-x)}{l-b}\sin\left(\frac{s\pi x}{l}\right)dx\right]\nonumber\\
&=&\frac{2rl^2}{b(l-b)s^2\pi^2}\sin\frac{s\pi b}{l}\nonumber
\end{eqnarray}
Hence the general solution is 
\begin{eqnarray}
y&=&\sum_{s=1}^{\infty}C_s\sin\frac{s\pi x}{l}\cos\frac{s\pi c}{l}t\nonumber\\
&=&\frac{2rl^2}{b(l-b)\pi^2}\sum_{s=1}^{\infty}\frac{1}{s^2}\sin\left(\frac{s\pi b}{l}\right)\sin\left(\frac{s\pi x}{l}\right)\cos\left(\frac{s\pi c}{l}t\right)\nonumber
\end{eqnarray}

\vspace{1cm}

$\bullet$ \textbf{Problem:} If a slightly elastic string is stressed between two fixed points and the motion is started by drawing aside through a distance `$b$', a point on the string distance $\dfrac{1}{5}$ of the length `$l$' of the string from one end, the displacement at any instant will be given by the equation 
$$y=\frac{25b}{2\pi^2}\sum_{n=1}^{\infty}\left[\frac{1}{n^2}\sin\left(\frac{n\pi}{5}\right)\sin\left(\frac{n\pi x}{l}\right)\cos\left(\frac{n\pi ct}{l}\right)\right]$$

$\star$ \textbf{Forced Vibration of a String:}

Suppose $x$-axis is along the length of the string with $x=0$ and $x=l$ are the two end points. Let the displacement of the point $x=b$ at any time is represented by $y=r\cos(pt+\epsilon)$. The equation of motion of the string is 
$$\frac{\partial^2y}{\partial t^2}=c^2\frac{\partial^2y}{\partial x^2}$$
We assume the solution of the form 
$$y=f(x)\cos(pt+\epsilon),$$
then from the above differential equation $f(x)$ has the solution of the form 
$$f(x)=A\cos\frac{p}{c}x+B\sin\frac{p}{c}x$$
Hence we write the solution in the form 
\begin{eqnarray}
y&=&\left(A\cos\frac{px}{c}+B\sin\frac{px}{c}\right)\cos(pt+\epsilon),~~~0\leq x\leq b\nonumber\\
&=&\left(C\cos\frac{px}{c}+D\sin\frac{px}{c}\right)\cos(pt+\epsilon),~~~b\leq x\leq l\nonumber
\end{eqnarray}
To find the constants we have $y=0$, at $x=0$ and $x=l$ and $y=r\cos(pt+\epsilon)$ at $x=b$, therefore,
\begin{eqnarray}
0=A,~~~~~0=C\cos\left(\frac{pl}{c}\right)+D\sin\left(\frac{pl}{c}\right)\nonumber\\
r\cos(pt+\epsilon)=B\sin\frac{pb}{c}\cos(pt+\epsilon)~~~~~~~\nonumber\\
\mbox{and}~~~r\cos(pt+\epsilon)=\left(C\cos\frac{pb}{c}+D\sin\frac{pb}{c}\right)\cos(pt+\epsilon)\nonumber\\
\therefore B=r~cosec\left(\frac{pb}{c}\right)~~~~~~~~~~~~~~~~~~~~~~~\nonumber
\end{eqnarray}
\begin{eqnarray}
\mbox{Also}~~~r&=&C\cos\frac{pb}{c}+D\sin\frac{pb}{c}=\frac{c}{\sin\left(\frac{pl}{c}\right)}\left[\cos\frac{pb}{c}\sin\frac{pl}{c}-\cos\frac{pl}{c}\sin\frac{pb}{c}\right]\nonumber\\
&=&\frac{c}{\sin\frac{pl}{c}}\sin\left[\frac{p}{c}(l-b)\right]\nonumber
\end{eqnarray}
\begin{eqnarray}
\therefore C=\frac{r\sin\left(\frac{pl}{c}\right)}{\sin\left[\frac{p}{c}(l-b)\right]},~~~~~D=\frac{-r\cos\left(\frac{pl}{c}\right)}{\sin\left[\frac{p}{c}(l-b)\right]}\nonumber
\end{eqnarray}
Thus
\begin{eqnarray}
y&=&\frac{r\sin\left(\frac{px}{c}\right)}{\sin\left(\frac{pb}{c}\right)}\cos(pt+\epsilon),~~~0\leq x\leq b\nonumber\\
&=&\frac{r\sin\left[\frac{p}{c}(l-x)\right]}{\sin\left[\frac{p}{c}(l-b)\right]}\cos(pt+\epsilon),~~~b\leq x\leq l\nonumber
\end{eqnarray}
If there be a force $F\cos(pt+\epsilon)$ at $x=b$, then 
\begin{equation}
F\cos(pt+\epsilon)=T_1\left(\frac{dy}{dx}\right)_{x=b^-}-T_1\left(\frac{dy}{dx}\right)_{x=b^+}\nonumber
\end{equation}
\begin{eqnarray}
\mbox{But,}~~~\frac{dy}{dx}&=&\frac{r\frac{p}{c}\cos\left(\frac{px}{c}\right)}{\sin\left(\frac{pb}{c}\right)}\cos(pt+\epsilon),~~~\mbox{for}~~0\leq x\leq b\nonumber\\
\mbox{and}~~~\frac{dy}{dx}&=&-\frac{r\frac{p}{c}\cos\left[\frac{p}{c}(l-x)\right]}{\sin\left[\frac{p}{c}(l-b)\right]}\cos(pt+\epsilon),~~~\mbox{for}~~b\leq x\leq l\nonumber
\end{eqnarray}
and so
\begin{eqnarray}
F&=&T_1\left(r\frac{p}{c}\right)\left[\cot\left(\frac{pb}{c}\right)-\cot\left(\frac{p(l-b)}{c}\right)\right]\nonumber\\
&=&T_1\frac{rp}{c}\frac{\sin\left(\frac{pl}{c}\right)}{\sin\left(\frac{pb}{c}\right)\sin\left[\frac{p(l-b)}{c}\right]}\nonumber
\end{eqnarray}
So $r$ can be obtained from above and $y$ can be uniquely determined.

\vspace{1cm}

$\bullet$ \textbf{Problem I:} A string of length $\left(l+l'\right)$ is stretch with tension between two fixed points. The length `$l$' has mass $m$ per unit length, the length `$l'$' has mass $m'$ per unit length. Prove that the possible period $T$ of transverse vibration are given by the equation
$$\frac{\tan\frac{2\pi l}{T}\sqrt{\frac{m}{P}}}{\tan\frac{2\pi l'}{T}\sqrt{\frac{m'}{P}}}+\sqrt{\frac{m}{m'}}=0$$

\vspace{.5cm}

\textbf{Solution:} 

\begin{figure}[h!]
	\centering
	\includegraphics[scale=0.5]{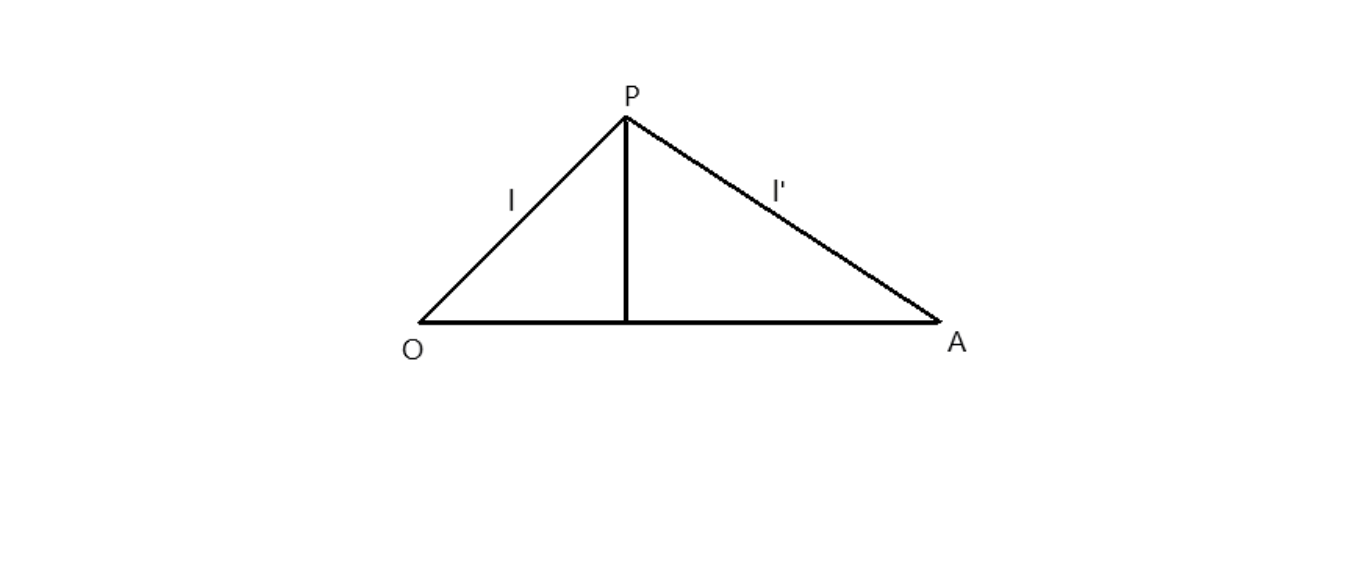}
\end{figure}

Let
\begin{eqnarray}
y_1&=&\left(A\cos\frac{px}{c_1}+B\sin\frac{px}{c_1}\right)\cos(pt+\epsilon),~~~\mbox{for}~0\leq x\leq l\nonumber\\
y_2&=&\left\{C\cos\frac{p\left(l+l'-x\right)}{c_2}+D\sin\frac{p\left(l+l'-x\right)}{c_2}\right\}\cos(pt+\epsilon),~~~\mbox{for}~l\leq x\leq l+l'\nonumber
\end{eqnarray}
\begin{eqnarray}
\mbox{Now,}~~~y_1&=&0,~~\mbox{at}~x=0\implies A=0~~~~~~~~~~~~~~~~~~~~~~~~~~\nonumber\\
y_2&=&0,~~\mbox{at}~x=l+l'\implies C=0\nonumber
\end{eqnarray}
\begin{eqnarray}
y_1=P~~\mbox{at}~x=l,~~~y_2=P~~\mbox{at}~x=l.~~~~~~~~~~\nonumber\\
\mbox{i.e,}~~y_1=y_2~~\mbox{at}~x=l~~\mbox{and}~~\frac{dy_1}{dx}=\frac{dy_2}{dx}~~\mbox{at}~x=l\nonumber
\end{eqnarray}
\begin{equation}
\therefore B\sin\frac{pl}{c_1}=D\sin\frac{pl'}{c_2}\label{eq3..6}
\end{equation}
\begin{eqnarray}
\mbox{Also}~~~\frac{dy_1}{dx}&=&B\frac{p}{c_1}\cos\frac{pl}{c_1}~~\mbox{at}~x=l\nonumber\\
\mbox{and}~~~\frac{dy_2}{dx}&=&-D\frac{p}{c_2}\cos\frac{pl'}{c_2}~~\mbox{at}~x=l\nonumber
\end{eqnarray}
\begin{equation}
\therefore \frac{B}{c_1}\cos\frac{pl}{c_1}=-\frac{D}{c_2}\cos\frac{pl'}{c_2}\label{eq3..7}
\end{equation}
\begin{equation}
(\ref{eq3..6})\div(\ref{eq3..7})\implies c_1\tan\frac{pl}{c_1}=-c_2\tan\frac{pl'}{c_2}\label{eq3..8}
\end{equation}
$$\mbox{But,}~~~t=\frac{2\pi}{p},~~c_1^2=\frac{P}{m},~~c_2^2=\frac{P}{m'}$$
Hence from (\ref{eq3..8}),
$$\frac{\tan\frac{2\pi l}{T}\sqrt{\frac{m}{P}}}{\tan\frac{2\pi l'}{T}\sqrt{\frac{m'}{P}}}+\sqrt{\frac{m}{m'}}=0$$

\vspace{1cm}

$\bullet$ \textbf{Problem II:} Three strings $AB$, $BC$, $CD$ are stretched in a straight line $AD$ and freely joined at $B$ and $C$ and vibrate transversely. Show that the frequency equation is given by 
$$\frac{\tan n_1l_1}{n_1}+\frac{\tan n_2l_2}{n_2}+\frac{\tan n_3l_3}{n_3}=n_2^2\frac{\tan n_1l_1}{n_1}.\frac{\tan n_2l_2}{n_2}.\frac{\tan n_3l_3}{n_3}$$
where, $n_r=\dfrac{p}{c_r}$, $c_r=\sqrt{\dfrac{T_1}{\rho_r}}$, $T_1$ is the tension of the string, $\rho_r$ is the line density and $l_1$, $l_2$ and $l_3$ are the lengths of $AB$, $BC$ and $CD$ respectively.

\vspace{.5cm}

\textbf{Solution:}

\begin{figure}[h!]
	\centering
	\includegraphics[scale=0.7]{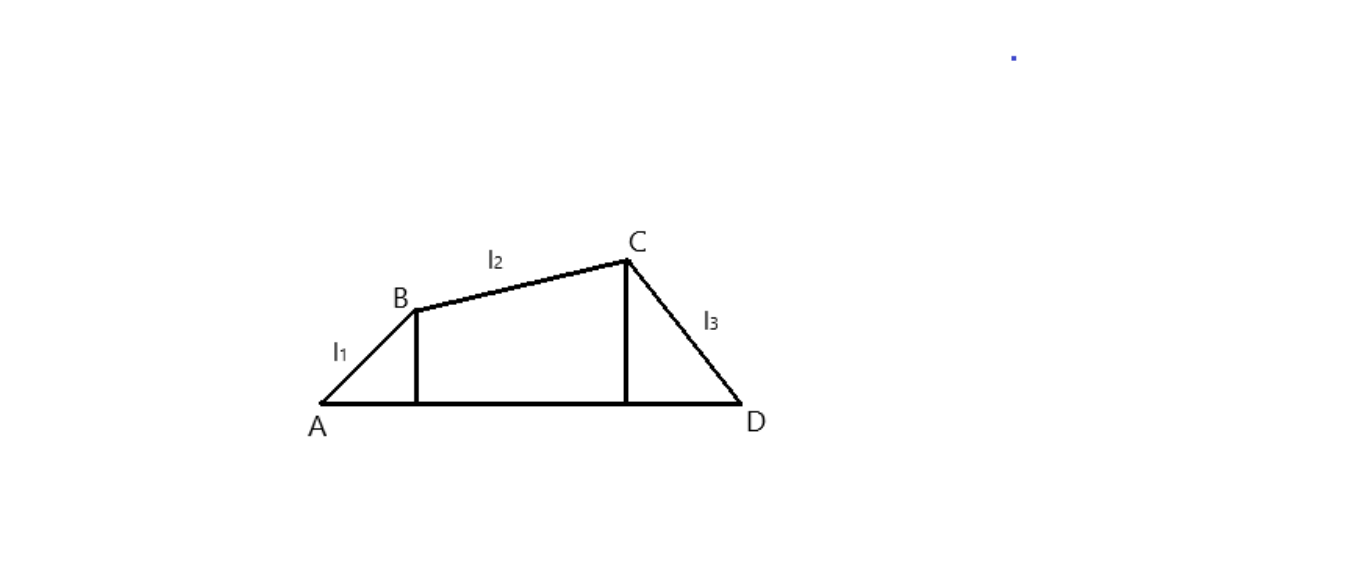}
\end{figure}

Let the solution be 
\begin{eqnarray}
y_1&=&\left(A\cos\frac{px}{c_1}+B\sin\frac{px}{c_1}\right)\cos(pt+\epsilon)~~~\mbox{for}~AB\nonumber\\
y_2&=&\left(C\cos\left\{\frac{p(l_1+l_2-x)}{c_2}\right\}+D\sin\left\{\frac{p(l_1+l_2-x)}{c_2}\right\}\right)\cos(pt+\epsilon)~~~\mbox{for}~BC\nonumber\\
\mbox{and}~~~y_3&=&\left[E\cos\left\{\frac{p(l_1+l_2+l_3-x)}{c_3}\right\}+F\sin\left\{\frac{p(l_1+l_2+l_3-x)}{c_3}\right\}\right]\cos(pt+\epsilon)~~~\mbox{for}~CD\nonumber
\end{eqnarray}
The boundary conditions are
\begin{eqnarray}
(i)~~y_1=0,~~x=0\implies A=0~~~~~~~~~~~~~\nonumber\\
(ii)~~y_3=0,~~x=l_1+l_2+l_3\implies E=0\nonumber\\
(iii)~~y_1=y_2~~\mbox{when}~x=l_1~~~~~~~~~~~~~~~~~~~\nonumber
\end{eqnarray}
\begin{equation}
\therefore B\sin\frac{pl_1}{c_1}=\left[C\cos\frac{pl_2}{c_2}+D\sin\frac{pl_2}{c_2}\right]\label{eq3..9}
\end{equation}
\begin{eqnarray}
(iv)~~y_2=y_3,~~\mbox{when}~x=l_1+l_2~~~~~~~~~~~~~\nonumber\\
\therefore C=F\sin\frac{pl_3}{c_3}~~~~~~~~~~~~~~~~~~~~~~~\label{eq3..10}
\end{eqnarray}
\begin{eqnarray}
(v)~~\frac{dy_1}{dx}=\frac{dy_2}{dx}~~\mbox{at}~x=l_1~~~~~~~~~~~~~~~~~~~~~~~~~\nonumber\\
\therefore \frac{Bp}{c_1}\cos\frac{pl_1}{c_1}=\frac{p}{c_2}\left[C\sin\frac{pl_2}{c_2}-D\cos\frac{pl_2}{c_2}\right]\label{eq3..11}
\end{eqnarray}
\begin{eqnarray}
(vi)~~\frac{dy_2}{dx}=\frac{dy_3}{dx},~\mbox{at}~x=l_1+l_2~~~~~~~~~~~~~~~~~~~~~\nonumber\\
\therefore -\frac{P}{c_2}D=-\frac{FP}{c_3}\cos\frac{pl_3}{c_3}\implies D=\frac{Fc_2}{c_3}\cos\frac{pl_3}{c_3}\label{eq3..12}
\end{eqnarray}
(\ref{eq3..9}) $\div$ (\ref{eq3..11}) using (\ref{eq3..10}) and (\ref{eq3..12}),
\begin{eqnarray}
c_1\tan\frac{pl_1}{c_1}&=&c_2\frac{\cos\frac{pl_2}{c_2}\sin\frac{pl_3}{c_3}+\frac{c_2}{c_3}\cos\frac{pl_3}{c_3}\sin\frac{pl_2}{c_2}}{\sin\frac{pl_2}{c_2}\sin\frac{pl_3}{c_3}-\frac{c_2}{c_3}\cos\frac{pl_2}{c_2}\cos\frac{pl_3}{c_3}}\nonumber\\
&=&c_2\frac{\tan\frac{pl_3}{c_3}+\frac{c_2}{c_3}\tan\frac{pl_2}{c_2}}{-\frac{c_2}{c_3}+\tan\frac{pl_2}{c_2}\tan\frac{pl_3}{c_3}}\nonumber
\end{eqnarray}
\begin{eqnarray}
\mbox{or,}~~-\frac{c_1c_2}{c_3}\tan\frac{pl_1}{c_1}+c_1\tan\frac{pl_1}{c_1}\tan\frac{pl_2}{c_2}\tan\frac{pl_3}{c_3}=c_2\tan\frac{pl_3}{c_3}+\frac{c_2^2}{c_3}\tan\frac{pl_2}{c_2}\nonumber\\
\mbox{or,}~~c_2\tan\frac{pl_3}{c_3}+\frac{c_2^2}{c_3}\tan\frac{pl_2}{c_2}+\frac{c_1c_2}{c_3}\tan\frac{pl_1}{c_1}=c_1\tan\frac{pl_1}{c_1}\tan\frac{pl_2}{c_2}\tan\frac{pl_3}{c_3}\nonumber\\
\mbox{or,}~~c_1\tan\frac{pl_1}{c_1}+c_2\tan\frac{pl_2}{c_2}+c_3\tan\frac{pl_3}{c_3}=\frac{c_1c_3}{c_2}\tan\frac{pl_1}{c_1}\tan\frac{pl_2}{c_2}\tan\frac{pl_3}{c_3}\nonumber\\
\mbox{or,}~~\frac{\tan n_1l_1}{n_1}+\frac{\tan n_2l_2}{n_2}+\frac{\tan n_3l_3}{n_3}=n_2^2\frac{\tan n_1l_1}{n_1}.\frac{\tan n_2l_2}{n_2}.\frac{\tan n_3l_3}{n_3}\nonumber
\end{eqnarray}

\vspace{1cm}

$\bullet$ \textbf{Problem III:} A uniform string whose length is $2l$ and mass $2lm$ is stretched at tension $T$ between two fixed points. The middle point of the string being displaced a small distance $b$ $\perp$ to the string and then released. Show that the subsequent motion of the string referred to the string is given by the equation
$$y=\frac{8b}{\pi^2}\sum_{r=0}^{\infty}\frac{1}{(2r+1)^2}\cos\left[\frac{(2r+1)}{2l}\pi x\right]\cos\left[\frac{(2r+1)}{2l}\pi ct\right]$$
where $c$ is given by the equation $mc^2=T$.

\vspace{.5cm}

\textbf{Solution:}

\begin{figure}[h!]
	\centering
	\includegraphics[scale=0.5]{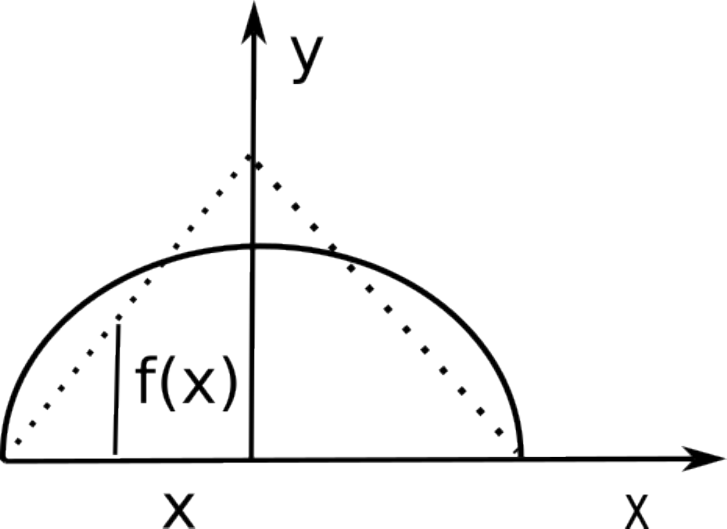}
\end{figure}

Let the solution 
$$y=\left(A\cos\frac{nx}{c}+B\sin\frac{nx}{c}\right)\cos(nt+\epsilon)$$
Since the string is symmetrical about $y$-axis i.e, $y$ is same for $+x$, $-x$.

$\therefore B=0$.

$\therefore$ $y=0$ when $x=\pm l$~~$\implies$~~$\cos\dfrac{nl}{c}=0=\cos(2r+1)\frac{\pi}{2}$

$\therefore\dfrac{nl}{c}=(2r+1)\dfrac{\pi}{2}$, $r$ is any integer even zero.
\begin{eqnarray}
\therefore y&=&A\cos\frac{nx}{c}\cos(nt+\epsilon)\nonumber\\
&=&\sum_{r=0}^{\infty}\cos\frac{(2r+1)\pi x}{2l}\left[C_r\cos\left\{\frac{(2r+1)}{2l}\pi ct\right\}+D_r\sin\left\{\frac{(2r+1)}{2l}\pi ct\right\}\right]\nonumber
\end{eqnarray}
Now at $t=0$, $\dfrac{dy}{dt}=0$ $\implies$ $D_r=0$.

Also at $t=0$, 
\begin{eqnarray}
y=f(x)&=&\frac{b}{l}(l-x),~~~0\leq x\leq l\nonumber\\
&=&\frac{b}{l}(l+x),~~~-l\leq x\leq 0\nonumber
\end{eqnarray}
\begin{eqnarray}
\therefore y&=&\sum_{r=0}^{\infty}C_r\cos\left[\frac{(2r+1)}{2l}\pi x\right]\cos\left[\frac{(2r+1)}{2l}\pi ct\right]\label{eq3..13}\\
\therefore f(x)&=&\sum_{r=0}^{\infty}C_r\cos\left[\frac{(2r+1)}{2l}\pi x\right]\nonumber
\end{eqnarray}
Now multiplying both sides by $\cos\frac{(2r+1)}{2l}\pi x$ and integrating over $x$ from $-l$ to $+l$ we have
\begin{eqnarray}
C_r&=&\frac{1}{l}\int_{-l}^{l}f(x).\cos\left[\frac{(2r+1)}{2l}\pi x\right]dx\nonumber\\
&=&\frac{b}{l}\left\{\int_{0}^{l}(l-x)\cos\left[\frac{(2r+1)}{2l}\pi x\right]dx+\int_{-l}^{0}(l+x)\cos\left[\frac{(2r+1)}{2l}\pi x\right]dx\right\}\nonumber\\
&=&\frac{2b}{l}\int_{0}^{l}(l-x)\cos\left[\frac{(2r+1)}{2l}\pi x\right]dx\nonumber\\
&=&\frac{2b}{l}\left\{\left[(l-x)\frac{2l}{(2r+1)\pi}\sin\frac{(2r+1)\pi x}{2l}\right]_0^l+\int_{0}^{l}\frac{2l}{(2r+1)\pi}\sin\left[\frac{(2r+1)\pi x}{2l}\right]dx\right\}\nonumber\\
&=&\frac{2b}{l}\frac{2l}{(2r+1)\pi}\frac{2l}{(2r+1)\pi}\left[-\cos\frac{(2r+1)\pi x}{2l}\right]_0^l\nonumber\\
&=&\frac{8b}{(2r+1)^2\pi^2}\nonumber
\end{eqnarray}
\begin{equation}
\therefore C_r=\frac{8b}{(2r+1)^2\pi^2}\nonumber
\end{equation}
\begin{equation}
\therefore y=\frac{8b}{\pi^2}\sum_{r=0}^{\infty}\frac{1}{(2r+1)^2}\cos\frac{(2r+1)\pi x}{2l}\cos\frac{(2r+1)\pi ct}{2l}\nonumber
\end{equation}
When $\rho$ is not constant then mass of the element $\delta x$ is $\rho\delta x$. For transverse vibration the acceleration of $\rho\delta x$ along $OX$ is zero and $\ddot{y}$ along $OY$. So the equation of motion will be 
\begin{equation}
\rho\delta x.0=-T\cos\psi+\left[T\cos\psi+\frac{\partial}{\partial y}(T\cos\psi).\delta y\right]\nonumber
\end{equation}
\begin{equation}
\therefore T\cos\psi=\mbox{constant}~~~\mbox{i.e,}~~T=\mbox{constant upto 1st order.}\nonumber
\end{equation}
\begin{eqnarray}
\rho\delta x\frac{\partial^2 y}{\partial t^2}=-T\sin\psi+\left[T\sin\psi+\frac{\partial}{\partial x}(T\sin\psi)\delta x\right]=T\frac{\partial}{\partial x}(\tan\psi)\delta x\nonumber\\
~~~~~~~~~~~~~~~~~~~~~~~~~(\because~~\mbox{for small}~\psi,~~\tan\psi=\sin\psi)\nonumber
\end{eqnarray}
\begin{equation}
\therefore\rho\frac{\partial^2y}{\partial t^2}=T\frac{\partial}{\partial x}\left(\frac{\partial y}{\partial x}\right)\nonumber
\end{equation}
\begin{equation}
\mbox{Hence,}~~~\frac{\partial^2y}{\partial t^2}=\frac{T}{\rho}\frac{\partial^2y}{\partial x^2}=c^2\frac{\partial^2y}{\partial x^2}\nonumber
\end{equation}
Note that here $c^2=\dfrac{T}{\rho}$ is not a constant.

\vspace{1cm}

$\bullet$ \textbf{Problem I:} If the density of a stressed string be $\dfrac{m}{x^2}$ where $x$ is the distance measured from a point in the line of propagation, the ends of the string being $x=l_1$, $x=l_2$. Show that the frequency equation is $$\dfrac{4p^2}{c^2}=1+\left\{\dfrac{2n\pi}{\log\left(\frac{l_2}{l_1}\right)}\right\}^2$$
where $c^2=\dfrac{T}{m}$, $T$ is the tension of the string and the vibration being transverse.

\vspace{.5cm}

\textbf{Solution:} The differential equation of vibrating string is
$$\frac{\partial^2y}{\partial t^2}=\frac{T^2}{\rho}\frac{\partial^2y}{\partial x^2}=c^2x^2\frac{\partial^2y}{\partial x^2}$$
Let us write the solution as
$$y=f(x)\cos(pt+\epsilon)$$
then the differential equation for $f(x)$ is
$$x^2\frac{d^2f}{dx^2}+\frac{p^2}{c^2}=0$$
putting $x=e^u$, the differential equation becomes
$$\frac{d^2f}{du^2}-\frac{df}{du}+\frac{p^2}{c^2}=0,$$
which has the solution 
$$f(x)=c_1e^{\frac{u}{2}}\cos\left(\sqrt{\mu}u+\epsilon\right)=c_1\sqrt{x}\cos\left(\sqrt{\mu}\log x+\epsilon\right),~~~~~\mu=\frac{p^2}{c^2}-\frac{1}{4}$$
\begin{eqnarray}
\mbox{As}~~y=0~~\mbox{at}~~x=l_1,~~\mbox{so}~~c_1\sqrt{l_1}\cos\left(\sqrt{\mu}\log l_1+\epsilon\right)=0\nonumber\\
\mbox{Also}~~y=0~~\mbox{at}~~x=l_2,~~\mbox{so}~~c_1\sqrt{l_2}\cos\left(\sqrt{\mu}\log l_2+\epsilon\right)=0\nonumber
\end{eqnarray}
\begin{eqnarray}
\mbox{Hence}~~\sqrt{\mu}\log l_1+\epsilon=(2s+1)\frac{\pi}{2}\nonumber\\
\sqrt{\mu}\log l_2+\epsilon=(2s'+1)\frac{\pi}{2}\nonumber
\end{eqnarray}
\begin{eqnarray}
\therefore \sqrt{\mu}\log\frac{l_1}{l_2}=2(s'-s)\frac{\pi}{2}=n\pi~~~~~~~~~~~~~~~~~~~~~~~\nonumber\\
\mbox{or},~~\left(\frac{p^2}{c^2}-\frac{1}{4}\right)^{\frac{1}{2}}=n\pi\log\frac{l_2}{l_1}\implies\frac{4p^2}{c^2}=1+\left\{\frac{2n\pi}{\log\frac{l_1}{l_2}}\right\}^2\nonumber\\
~~~~~~~~~~~~~~~~~~~~~~~~~~~~~~~~~~~~~~~~~~~~~~~~~~~~~~~~~~~~~~~~~~~~~~[\mbox{Proved}]\nonumber
\end{eqnarray}

\vspace{1cm}

$\bullet$ \textbf{Problem II:} If a string of length `$l$' and tension $T_0$ is stretched between two points and is not uniform but of line density $\dfrac{\rho_0}{(1+kx)^2}$ where $x$ is the distance from one end then show that the transverse vibration is of period $\dfrac{2\pi}{n}$ where $\sqrt{4n^2-k^2c^2}\log(1+kl)=2ick\pi~~~(4n^2>k^2c^2),~~~c^2=\dfrac{T_0}{\rho_0}$ and `$i$' is a +ve integer. Examine the case $i=0$.

\vspace{.5cm}

\textbf{Solution:} The differential equation of the vibrating string 
$$\frac{\partial^2y}{\partial t^2}=\frac{T^2}{\rho}\frac{\partial^2y}{\partial x^2}=c^2(1+kx)^2\frac{\partial^2y}{\partial x^2}=c^2k^2z^2\frac{\partial^2y}{\partial z^2},~~~1+kx=z$$
$$\mbox{or},~~~z^2\frac{\partial^2y}{\partial z^2}-\frac{1}{c^2k^2}\frac{\partial^2y}{\partial t^2}=0$$
Let the solution be $y=f(z)\cos(pt+\epsilon)$
$$\therefore z^2\frac{d^2f}{dz^2}+\frac{1}{c^2k^2}p^2=0$$
The solution is $f(z)=c_1\sqrt{z}\cos(\sqrt{\mu}\log z+\epsilon),~~~\mu=\dfrac{p^2}{c^2k^2}-\dfrac{1}{4}$
$$\therefore y=c_1\sqrt{1+kx}\cos\left\{\sqrt{\mu}\log(1+kx)+\epsilon\right\}$$
$$\mbox{As}~~y=0,~~x=0\implies\epsilon=(2s+1)\frac{\pi}{2}$$
$$\mbox{Also},~~y=0,~~x=l\implies\sqrt{\mu}\log(1+kl)+\epsilon=(2s'+1)\frac{\pi}{2}$$
$$\therefore\sqrt{\mu}\log(1+kl)=i\pi,~~~i~~\mbox{is a +ve integer.}$$
$$\therefore\sqrt{4p^2-k^2c^2}\log(1+kl)=2ick\pi$$
When $i=0$, then $4p^2=k^2c^2$ i.e, $p=\dfrac{kc}{2}$.

\vspace{1cm}

$\bullet$ \textbf{Problem III:} A transverse force $Y\sin pt$ acts at the point of junction of two strings of different mass per unit length which are joined and stretched between two points  at a distance `$l$' apart. The length of the strings being `$b$' and `$l-b$'. Prove that if $c_1$ and $c_2$ be the velocity of transverse wave in the two strings the displacement of the point of junction of the string at time `$t$' is $\dfrac{Y\sin pt}{\left[\dfrac{pT}{c_1}\cot\dfrac{pb}{c_1}+\dfrac{pT}{c_2}\cot\dfrac{p(l-b)}{c_2}\right]}$ where $T$ is the tension of the string. 

\vspace{.5cm}

\textbf{Solution:}

\begin{figure}[h!]
	\centering
	\includegraphics[scale=0.7]{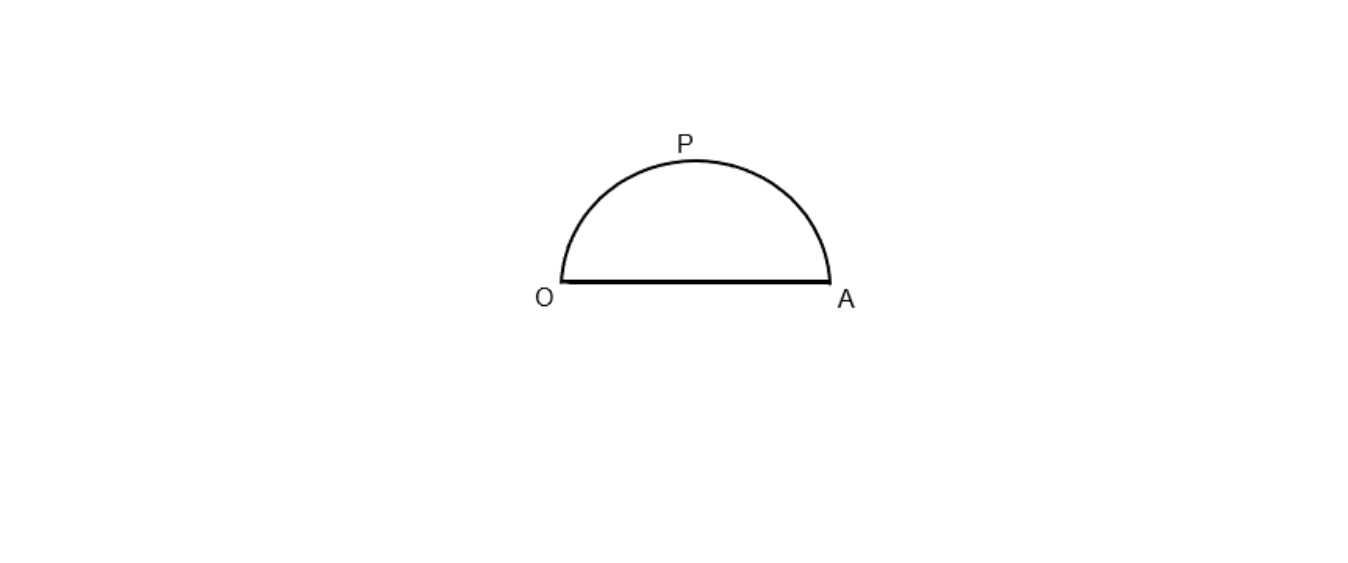}
\end{figure}

For the part $OP$ the solution is 
$$y_1=\left(A\cos\dfrac{px}{c_1}+B\sin\dfrac{px}{c_2}\right)\sin pt,~~~0\leq x\leq b$$
For the part $PA$ of the string the solution is 
$$y_2=\left[C\cos\frac{p(l-x)}{c_2}+D\sin\frac{p(l-x)}{c_2}\right]\sin pt,~~~b\leq x\leq l.$$
The boundary conditions: $y_1=0$ at $x=0$ gives $A=0$.

$y_2=0$ at $x=l$ gives $c=0.$
$$\therefore y_1=B\sin\frac{px}{c_1}\sin pt~~~\mbox{and}~~~y_2=D\sin\left\{\frac{p(l-x)}{c_2}\right\}\sin pt$$
The condition: $y_1=y_2$ at $x=b$ $\implies$
\begin{equation}
B\sin\frac{pb}{c_1}=D\sin\frac{p(l-b)}{c_2}\label{eq3..14}
\end{equation}
Now, 
$$Y\sin pt=T\left(\frac{dy_1}{dx}\right)_{x=b}-T\left(\frac{dy_2}{dx}\right)_{x=b}$$
\begin{eqnarray}
\therefore Y&=&pT\left[\frac{B}{c_1}\cos\frac{pb}{c_1}+\frac{D}{c_2}\cos\frac{p(l-b)}{c_2}\right]\nonumber\\
\mbox{or},~~~\frac{Y}{pT}&=&D\sin\frac{p(l-b)}{c_2}\left[\frac{1}{c_1}\cot\frac{pb}{c_1}+\frac{1}{c_2}\cot\left\{\frac{p(l-b)}{c_2}\right\}\right]\nonumber\\
\therefore D&=&\frac{Y}{\sin\frac{p(l-b)}{c_2}}\frac{1}{\left[\frac{pT}{c_1}\cot\frac{pb}{c_1}+\frac{pT}{c_2}\cot\left\{\frac{p(l-b)}{c_2}\right\}\right]}\nonumber\\
\therefore y_2&=&D\sin\frac{p}{c_2}(l-x)\sin pt\nonumber
\end{eqnarray}
$$\mbox{Hence},~~~\left(y_2\right)_b=\frac{Y\sin pt}{\left[\frac{pT}{c_1}\cot\frac{pb}{c_1}+\frac{pT}{c_2}\cot\left\{\frac{p(l-b)}{c_2}\right\}\right]}$$

\vspace{1cm}

$\bullet$ \textbf{Problem IV:} The ends of a stretched uniform string of length `$l$' are attached to two small rings without mass which can slide on two parallel rods at right angle to the string. The middle point of the string is acted on by the transverse force $F\sin pt$. Prove that the force vibration at a distance $\xi$ from either end is given by 
$$y=-\frac{cF}{2pT}\frac{\cos\frac{p\xi}{c}\sin pt}{\sin\left(\frac{pl}{2c}\right)}$$

\vspace{.5cm}

\textbf{Solution:}

\begin{figure}[h!]
	\centering
	\includegraphics[scale=0.7]{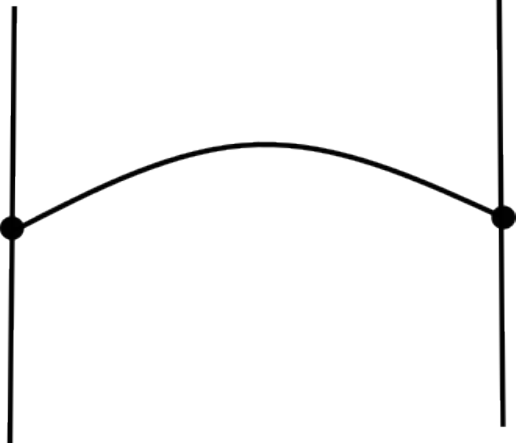}
\end{figure}

Let
\begin{eqnarray}
y_1&=&\left[A\cos\frac{px}{c}+B\sin\frac{px}{c}\right]\sin pt,~~~0\leq x\leq\frac{l}{2}\nonumber\\
y_2&=&\left[C\cos\frac{p(l-x)}{c}+D\sin\frac{p(l-x)}{c}\right]\sin pt,~~~\frac{l}{2}\leq x\leq l\nonumber
\end{eqnarray}
Since the rings are massless, we can imagine the string upto the rings so that $\dfrac{dy_1}{dx}=0$ at $x=0$, $\dfrac{dy_2}{dx}=0$ at $x=l$.
$$\therefore B=D=0$$.
$$y_1=y_2~~\mbox{at}~~x=\frac{l}{2}\implies A=C$$.
\begin{eqnarray}
\mbox{Also},~~~F\sin pt&=&\left[T\left(\frac{dy_1}{dx}\right)_{x=\frac{l}{2}}-T\left(\frac{dy_2}{dx}\right)_{x=\frac{l}{2}}\right]\nonumber\\
&=&-2T\frac{p}{c}A\sin\left(\frac{pl}{2c}\right)\sin pt\nonumber
\end{eqnarray}
$$\therefore A=-\frac{Fc}{2Tp}\frac{1}{\sin\left(\frac{pl}{2c}\right)}$$
$$\therefore y_1=-\frac{Fc}{2pT}\frac{\cos\left(\frac{px}{c}\right)\sin pt}{\sin\left(\frac{pl}{2c}\right)}$$
$$\therefore\mbox{at}~~x=\xi,~~\mbox{displacement}=-\frac{Fc}{2pT}\frac{\cos\left(\frac{p\xi}{c}\right)\sin pt}{\sin\frac{pl}{2c}}$$

\vspace{1cm}

$\bullet$ \textbf{Problem V:} If a stretched string held at its middle point be drawn aside at a point of quadric section and released from rest. Prove that in the ensuing vibration the energy in the harmonic of order `$r$' is proportional to $r^{-2}\sin^2\left(\dfrac{r\pi}{4}\right)\sin^4\left(\dfrac{r\pi}{8}\right)$.

\vspace{.5cm}

\textbf{Solution:}

\begin{figure}[h!]
	\centering
	\includegraphics[scale=0.4]{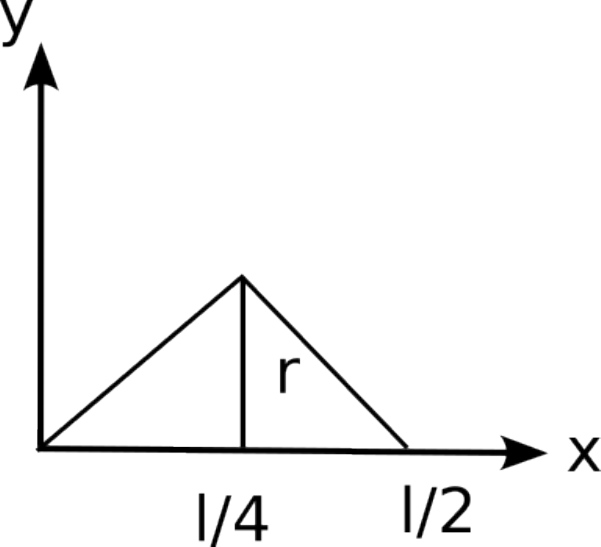}
\end{figure}

Let the solution be 
$$y=\left(A\cos\frac{px}{c}+B\sin\frac{px}{c}\right)\cos(pt+\epsilon)$$
\begin{eqnarray}
\mbox{As},~~~y&=&0,~~x=0\implies A=0\nonumber\\
y&=&0,~~x=l\implies\sin\frac{pl}{c}=0~~\mbox{i.e,}~~p=\frac{S\pi c}{l},~~~~~S~\mbox{is an integer including zero.}\nonumber
\end{eqnarray}
\begin{eqnarray}
\therefore y&=&B\sin\frac{S\pi x}{l}\cos\left(\frac{S\pi ct}{l}+\epsilon\right)\nonumber\\
&=&\sin\left(\frac{S\pi x}{l}\right)\left[C_s\cos\left(\frac{S\pi ct}{l}\right)+D_s\sin\left(\frac{S\pi ct}{l}\right)\right]\nonumber
\end{eqnarray}
$$\dot{y}=0~~\mbox{at}~~t=0~~\mbox{gives}~~D_s=0$$
\begin{equation}
\therefore y=C_s\sin\left(\frac{S\pi x}{l}\right)\cos\left(\frac{S\pi ct}{l}\right)\label{eq3..15}
\end{equation}
Initially at $t=0$, $y=y_0$, given by
\begin{eqnarray}
y_0&=&\frac{4\gamma x}{l},~~~0\leq x\leq\frac{l}{4}\nonumber\\
&=&\frac{\gamma\left(\frac{l}{2}-x\right)}{\frac{l}{4}},~~~\frac{l}{4}\leq x\leq\frac{l}{2}\nonumber\\
&=&0,~~~\frac{l}{2}\leq x\leq l\label{eq3..16}
\end{eqnarray}
From (\ref{eq3..15}), 
$$y_0=C_s\sin\left(\frac{S\pi x}{l}\right)$$
$$\mbox{or},~~~C_s\int\sin^2\left(\frac{S\pi x}{l}\right)dx=\int y_0\sin\left(\frac{S\pi x}{l}\right)dx$$
\begin{eqnarray}
C_s\frac{l}{2}&=&\int_{0}^{\frac{l}{4}}\frac{4\gamma x}{l}\sin\left(\frac{S\pi x}{l}\right)+\int_{\frac{l}{4}}^{\frac{l}{2}}\frac{4\gamma}{l}\left(\frac{l}{2}-x\right)\sin\left(\frac{S\pi x}{l}\right)dx\nonumber\\
&=&\frac{4\gamma}{l}\bigg\{-\left(\frac{l}{S\pi}\right)\left[x\cos\left(\frac{S\pi x}{l}\right)\right]_0^{\frac{l}{4}}+\int_{0}^{\frac{l}{4}}\left(\frac{l}{S\pi}\right)\cos\left(\frac{S\pi x}{l}\right)dx\nonumber\\
&~&+\left[-\left(\frac{l}{2}-x\right)\frac{l}{S\pi}\cos\left(\frac{S\pi x}{l}\right)\right]_{\frac{l}{4}}^{\frac{l}{2}}-\int_{\frac{l}{4}}^{\frac{l}{2}}\frac{l}{S\pi}\cos\left(\frac{S\pi x}{l}\right)dx\bigg\}\nonumber\\
&=&\frac{4\gamma}{l}\bigg\{\cancel{-\frac{l}{S\pi}\frac{l}{4}\cos\frac{S\pi}{4}}+\frac{l^2}{S^2\pi^2}\sin\left(\frac{S\pi}{4}\right)+\cancel{\frac{l}{S\pi}\frac{l}{4}\cos\frac{S\pi}{4}}\nonumber\\
&~&~~~~~~~~~~~~~~~~~~~~~~~~~~~~~~~~~~-\frac{l^2}{S^2\pi^2}\sin\frac{S\pi}{2}+\frac{l^2}{S^2\pi^2}\sin\frac{S\pi}{4}\bigg\}\nonumber\\
&=&\frac{4\gamma}{l}\left(\frac{l^2}{S^2\pi^2}\right)\left[2\sin\frac{S\pi}{4}-\sin\frac{S\pi}{2}\right]\nonumber\\
&=&\frac{4\gamma l}{S^2\pi^2}\left[2\sin\frac{S\pi}{4}-\sin\frac{S\pi}{2}\right]\nonumber
\end{eqnarray}
\begin{eqnarray}
\therefore	C_s&=&\frac{8\gamma}{S^2\pi^2}.2\sin\frac{S\pi}{4}.2\sin^2\frac{S\pi}{8}\nonumber\\
&=&\frac{32\gamma}{S^2\pi^2}\sin\left(\frac{S\pi}{4}\right)\sin^2\left(\frac{S\pi}{8}\right)\nonumber
\end{eqnarray}
The kinetic energy is $=\frac{1}{2}\rho\int_{0}^{l}\left(\frac{\partial y}{\partial t}\right)^2dx$, potential energy is $=\frac{1}{2}\rho c^2\int_{0}^{l}\left(\frac{\partial y}{\partial x}\right)^2dx$.
\begin{eqnarray}
\mbox{As},~~~y&=&C_s\sin\left(\frac{S\pi x}{l}\right)\sin\left(\frac{S\pi ct}{l}\right)\nonumber\\
\therefore\frac{\partial y}{\partial x}&=&C_s.\frac{S\pi}{l}.\cos\left(\frac{S\pi x}{l}\right)\sin\left(\frac{S\pi ct}{l}\right)=\lambda_1\cos\left(\frac{S\pi x}{l}\right)\sin\left(\frac{S\pi ct}{l}\right)\nonumber\\
\frac{\partial y}{\partial t}&=&C_s.\frac{S\pi c}{l}.\sin\left(\frac{S\pi x}{l}\right).\cos\left(\frac{S\pi ct}{l}\right)=\lambda_2\sin\left(\frac{S\pi x}{l}\right)\cos\left(\frac{S\pi ct}{l}\right)\nonumber\\
&~&~~~~~~~~~~~~~~~~~~~~~~~~~~~~~~~~~~~~~~~~~~~~~~~~~~~~~~~~~~~~~~~~~~~~~~~~~(\lambda_2=c\lambda_1)\nonumber
\end{eqnarray}
\begin{eqnarray}
\therefore\mbox{Kinetic energy (K.E.)}&=&\frac{1}{2}\rho\lambda_2^2\int_{0}^{l}\sin^2\left(\frac{S\pi x}{l}\right)\cos^2\left(\frac{S\pi ct}{l}\right)dx\nonumber\\
&=&\frac{1}{2}\rho\lambda_2^2\cos^2\left(\frac{S\pi ct}{l}\right).\frac{1}{2}\int_{0}^{l}\left[1+\cos\left(\frac{2S\pi x}{l}\right)\right]dx\nonumber\\
&=&\frac{1}{4}\rho\lambda_2^2\cos^2\left(\frac{S\pi ct}{l}\right).l\nonumber\\
&=&\frac{l}{4}\rho\lambda_2^2\cos^2\left(\frac{S\pi ct}{l}\right)\nonumber
\end{eqnarray}
\begin{eqnarray}
\mbox{Potential energy (P.E.)}&=&\frac{1}{4}\rho c^2\lambda_1^2\sin^2\left(\frac{S\pi ct}{l}\right)\int_{0}^{l}\left(1-\cos\left(\frac{S\pi x}{l}\right)\right)dx\nonumber\\
&=&\frac{l}{4}\rho c^2\lambda_1^2\sin^2\left(\frac{S\pi ct}{l}\right)\nonumber\\
&=&\frac{l}{4}\rho\lambda_2^2\sin^2\left(\frac{S\pi ct}{l}\right)\nonumber
\end{eqnarray}
$\therefore\mbox{K.E.+P.E.}=\dfrac{l}{4}\rho\lambda_2^2$.

\vspace{1cm}

$\bullet$ \textbf{Problem VI:} A uniformly stretched string of which the extremities are fixed starts from rest in the form $y=A\sin\left(\dfrac{m\pi x}{l}\right)$, where $m$ is an integer, $l$ the distance between two fixed ends. Prove that if the resistance of air be taken into account and be assumed to be $2K$ times the momentum per unit length then the displacement after any time $t$ is given by
$$y=Ae^{-Kt}\left(\cos m't+\frac{K}{m'}\sin m't\right)\sin\left(\frac{m\pi x}{l}\right)$$
where $m'^2=\dfrac{m^2\pi^2c^2}{l^2}-K^2$, with $c$ the velocity of the wave of transverse vibration of the string.

\vspace{.5cm}

\textbf{Solution:} In the resisting medium the equation of motion is 
$$\frac{\partial^2y}{\partial t^2}=c^2\frac{\partial^2y}{\partial x^2}-2K\frac{\partial y}{\partial t}$$
\begin{equation}
\mbox{We assume},~~~y=f(t).\sin\left(\frac{m\pi x}{l}\right)\nonumber
\end{equation}
So the differential equation becomes 
$$f''(t)\sin\left(\frac{m\pi x}{l}\right)=-c^2.\left(\frac{m^2\pi^2}{l^2}\right)\sin\left(\frac{m\pi x}{l}\right).f(t)-2Kf'(t)\sin\left(\frac{m\pi x}{l}\right)$$
$$\mbox{or},~~~f''(t)+2Kf'(t)+\frac{c^2m^2\pi^2}{l^2}f(t)=0$$
If $z=f(t)$ then 
$$\frac{d^2z}{dt^2}+2K\frac{dz}{dt}+\frac{c^2m^2\pi^2}{l^2}z=0$$
the auxiliary equation is
$$\alpha^2+2K\alpha+\frac{c^2m^2\pi^2}{l^2}=0$$
$$\mbox{or},~~~\alpha=\frac{-2K\pm\sqrt{4K^2-4\frac{c^2m^2\pi^2}{l^2}}}{2}=-K\pm im'$$
\begin{eqnarray}
\therefore z&=&f(t)=e^{-Kt}\left[c_1\cos m't+c_2\sin m't\right]\nonumber\\
\therefore y&=&e^{-Kt}\left[c_1\cos m't+c_2\sin m't\right]\sin\left(\frac{m\pi x}{l}\right)\nonumber
\end{eqnarray}
$$\mbox{But at}~~t=0,~~y=A\sin\frac{m\pi x}{l}\implies c_1=A$$
$\dot{y}=0$ at $t=0$
\begin{eqnarray}
\therefore\frac{dy}{dt}&=&-Ke^{-Kt}\left[c_1\cos m't+c_2\sin m't\right]\sin\left(\frac{m\pi x}{l}\right)\nonumber\\
&~&~~~~~~~~+e^{-Kt}\left[-c_1m'\sin m't+c_2m'\cos m't\right]\sin\left(\frac{m\pi x}{l}\right)\nonumber
\end{eqnarray}
$$\mbox{So at}~~t=0,~~\dot{y}=0\implies-Kc_1+c_2m'=0~~\mbox{i.e,}~~c_2=\frac{AK}{m'}$$
$$\therefore y=e^{-Kt}A\left(\cos m't+\frac{K}{m'}\sin m't\right)\sin\left(\frac{m\pi x}{l}\right)$$

\vspace{1cm}

$\bullet$ \textbf{Problem VII:} Two uniform strings are attached together and stretched in a straight line between two fixed points with tension $T$ and carry a particle of mass $M$ attached at the point of junction. The line densities are $\rho$ and $\rho'$ and lengths are $l$ and $l'$. Show that $T=c^2\rho=c'^2\rho'$. The period $\dfrac{2\pi}{n}$ of transverse vibration are given by 
$$M.n=c\rho\cot\left(\frac{nl}{c}\right)+c'\rho'\cot\left(\frac{nl'}{c'}\right)$$

\vspace{.5cm}

\textbf{Solution:} 

\begin{figure}[h!]
	\centering
	\includegraphics[scale=0.6]{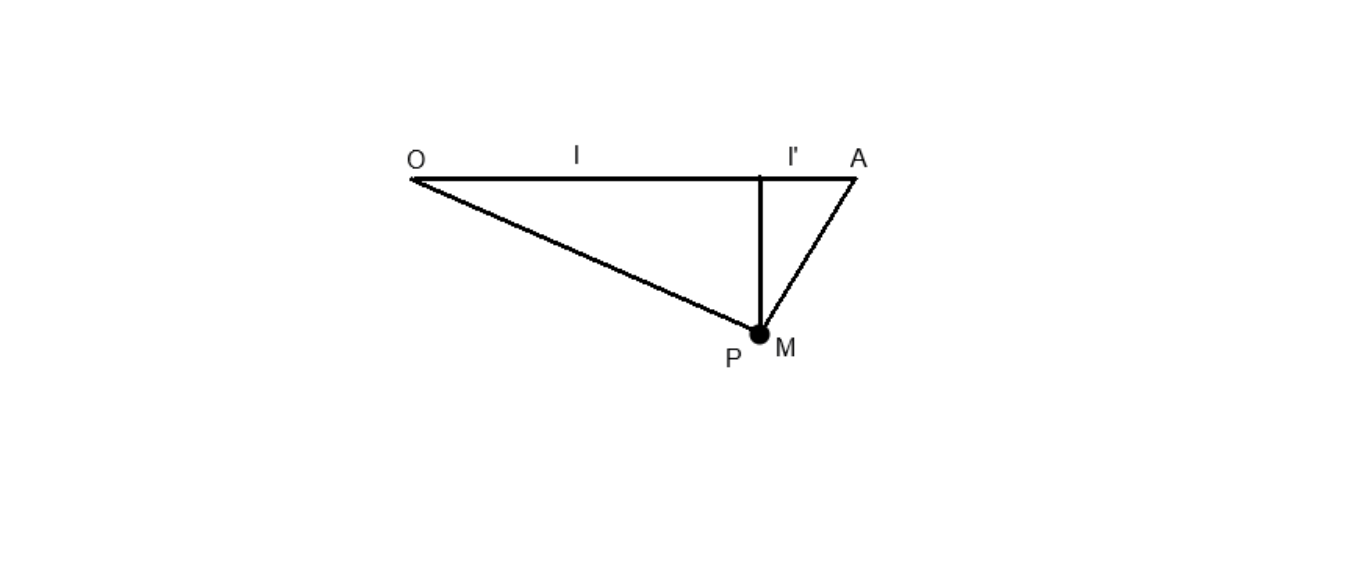}
\end{figure}

Let
\begin{eqnarray}
y_1&=&\left(A\cos\frac{px}{c_1}+B\sin\frac{px}{c_1}\right)\sin pt,~~~~0<x<l\nonumber\\
y_2&=&\left[C\cos\frac{p(l+l'-x)}{c_2}+D\sin\frac{p(l+l'-x)}{c_2}\right]\sin pt,~~~~l<x<l+l'\nonumber
\end{eqnarray}
be the solutions at the two parts of the strings.
\begin{eqnarray}
\mbox{Now},~~~y_1&=&0,~~x=l\implies A=0\nonumber\\
y_2&=&0,~~x=l+l'\implies C=0\nonumber\\
y_1&=&y_2~~\mbox{at}~~x=l\implies B\sin\frac{pl}{c_1}=D\sin\frac{pl'}{c_2}\nonumber
\end{eqnarray}
Now equation of motion of $M$ is 
$$M\left(\frac{d^2y}{dt^2}\right)_{x=l}=T\left[\left(\frac{dy_2}{dx}\right)_{x=l}-\left(\frac{dy_1}{dx}\right)_{x=l}\right]$$
$$\mbox{or},~~Mp\sin\frac{pl}{c_1}p^2\sin pt=T\left[\frac{Dp}{c_2}\cos\frac{pl'}{c_2}+\frac{Bp}{c_1}\cos\frac{pl}{c_1}\right]\sin pt$$
\begin{eqnarray}
\mbox{or},~~M.p&=&T\left[\frac{1}{c_1}\cot\frac{pl}{c_1}+\frac{1}{c_2}\cot\frac{pl'}{c_2}\right]\nonumber\\
&=&c_1\rho_1\cot\frac{pl}{c_1}+c_2\rho_2\cot\frac{pl'}{c_2}\nonumber
\end{eqnarray}

\section{Longitudinal Vibration of a Stretched Elastic String:}
We take the $x$-axis along the equilibrium position of the stretched string. Let the end points be then given by $x=0$ and $x=l$ so that `$l$' is the stretched equilibrium length. Let $\rho$ be the line density in the stretched equilibrium position. Suppose $l_0$ be the natural length of the string and $\rho_0$ is the line density when unstretched. Since the mass of the string remains unaltered we have $l_0\rho_0=l\rho$.

\begin{figure}[h!]
	\centering
	\includegraphics[scale=0.7]{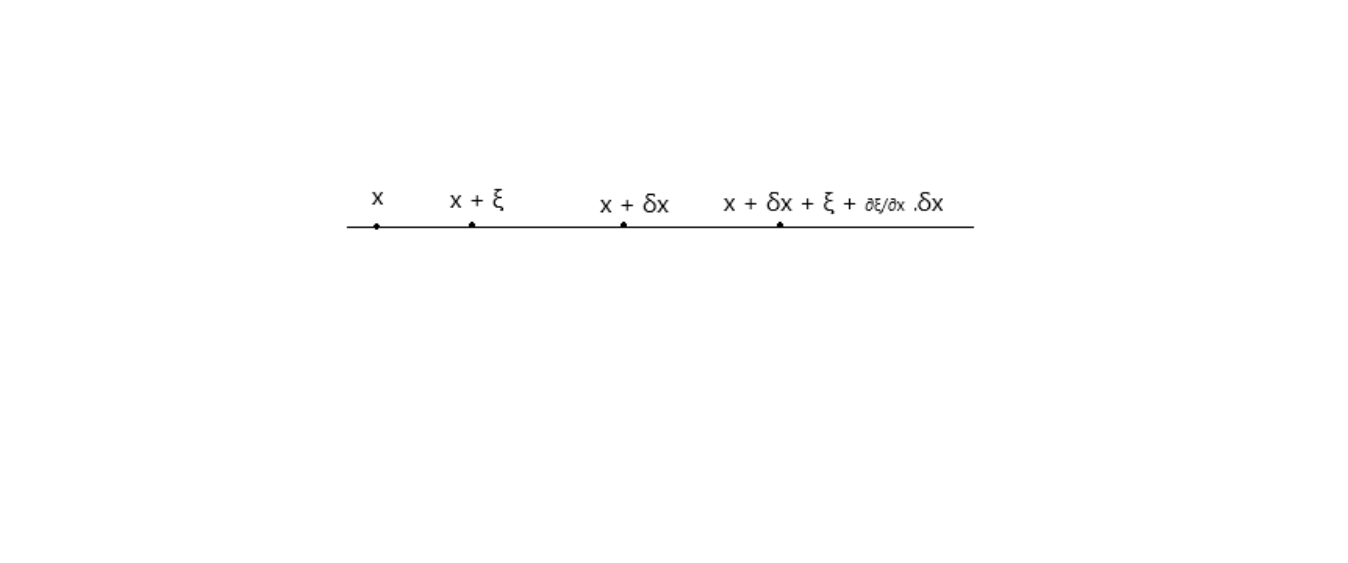}
\end{figure}

Let us consider an element of length $\delta x$ of the string in the stretched equilibrium position, the end points being given by $x$ and $x+\delta x$. Now, in course of vibration let the end point $x$ be shifted to the point $x+\xi$, ($\xi$ is very small) and the point $x+\delta x$ will be shifted to the position $x+\delta x+\xi+\dfrac{\partial \xi}{\partial x}\delta x$, so that the length of the element becomes $\delta x+\dfrac{\partial\xi}{\partial x}\delta x$. If $\delta x_0$ be the unstretched length of the element, whose length is $\delta x$ in the stretched equilibrium position then the tension of the string at any time is given by 
$$T=E\frac{\delta x+\frac{\partial\xi}{\partial x}\delta x-\delta x_0}{\delta x_0},$$
where $E$ is the Young's modulus of the material of the string.
$$\mbox{Now},~~T=E\frac{\delta x-\delta x_0}{\delta x_0}+E\frac{\partial\xi}{\partial x}\frac{\delta x}{\delta x_0}=T_0+E\frac{\partial\xi}{\partial x}\frac{\delta x}{\delta x_0}$$
where $T_0$ is the tension of the string in the stretched equilibrium position and has the same value for every point of the string.

As $\dfrac{\delta x}{\delta x_0}=\dfrac{l}{l_0}$, so putting $E\dfrac{l}{l_0}=E'$, we get $T=T_0+E'\dfrac{\partial\xi}{\partial x}$, $T_0$ is constant.

Let us write down the equation of motion of the element $\delta x$ of the string
$$\rho\delta x\ddot{\xi}=-T+T+\frac{\partial T}{\partial x}\delta x=\frac{\partial}{\partial x}\left[T_0+E'\frac{\partial\xi}{\partial x}\right]\delta x$$
$$\mbox{or},~~~\frac{\partial^2\xi}{\partial t^2}=\frac{E'}{\rho}\frac{\partial^2\xi}{\partial x^2}=a^2\frac{\partial^2\xi}{\partial x^2},~~~\mbox{with}~~a^2=\frac{E'}{\rho}.$$
Here `$a$' is the velocity of propagation of the disturbance refer to the stretched equilibrium position. Also we have 
$$a^2=\frac{E'}{\rho}=E\frac{l}{l_0}.\frac{1}{\frac{\rho_0l_0}{l}}=\left(\frac{E}{\rho_0}\right)\left(\frac{l}{l_0}\right)^2$$
The boundary conditions are $\xi=0$ when $x=0$ and $\xi=0$ when $x=l$.

Let $\xi=f(x)\cos(pt+\epsilon)$ be the solution, then 
$$f(x)=A\cos\frac{px}{a}+B\sin\frac{px}{a},~~~A~\mbox{and}~B~\mbox{are constants}.$$
$$\therefore\xi=\left(A\cos\frac{px}{a}+B\sin\frac{px}{a}\right)\cos(pt+\epsilon)$$
$$\xi=0,~~x=0\implies A=0.$$
$$\xi=0,~~x=l\implies\sin\frac{pl}{a}=0~~\mbox{i.e,}~~\frac{pl}{a}=s\pi\implies p=\frac{s\pi a}{l}$$
$$\therefore p=\frac{s\pi}{l}\sqrt{\frac{E}{\rho_0}}\frac{l}{l_0}=\frac{s\pi}{l_0}\sqrt{\frac{E}{\rho_0}}$$

\section{Transverse Vibration of a Stretched Membrence}
We take the plane of the membrence in the equilibrium position as the plane of $xy$. Assuming that the membrence is subjected to stretching. Let the tension be $T_1$ and $z$ be the normal displacement of the point of the membrence which occupy the position $(x,y)$ when in equilibrium. Let us construct a prism on the element of area $dxdy$ on the $xy$ plane and this prism intersects the surface of the displaced membrence in a figure which is approximately rectangular and lengths of the sides of this rectangle are 
$$\sqrt{1+\left(\frac{\partial z}{\partial x}\right)^2}dx,~~~\sqrt{1+\left(\frac{\partial z}{\partial y}\right)^2}dy$$ 
\begin{eqnarray}
\mbox{i.e,}~~\left[1+\frac{1}{2}\left(\frac{\partial z}{\partial x}\right)^2\right]dx~&~&\mbox{and}~~~\left[1+\frac{1}{2}\left(\frac{\partial z}{\partial y}\right)^2\right]dy\nonumber\\
&~&\left(\mbox{assuming }~\frac{\partial z}{\partial x}~\mbox{and}~\frac{\partial z}{\partial y}~\mbox{to be very small}\right)\nonumber
\end{eqnarray}
Therefore, the area of the portion is
$$\left[1+\frac{1}{2}\left\{\left(\frac{\partial z}{\partial x}\right)^2+\left(\frac{\partial z}{\partial y}\right)^2\right\}\right]dxdy,$$ 
neglecting the quantities of higher order than the 2nd. Therefore, in the displaced position the increment of the area of the portion of the membrence which occupy the position $dxdy$ in the equilibrium position is  $$\frac{1}{2}\left\{\left(\frac{\partial z}{\partial x}\right)^2+\left(\frac{\partial z}{\partial y}\right)^2\right\}dxdy.$$
Therefore, potential energy of the membrence which is the work done by the tension $T_1$ is 
$$V=\frac{1}{2}T_1\int\int\left\{\left(\frac{\partial z}{\partial x}\right)^2+\left(\frac{\partial z}{\partial y}\right)^2\right\}dxdy$$ 
The integral is taken over the area of the membrence in the equilibrium position. The kinetic energy of the membrence at any time is 
$$T=\frac{1}{2}\int\int\rho\left(\frac{\partial z}{\partial t}\right)^2dxdy,~~~(\rho~\mbox{is the uniform density of the membrence.})$$
from Hamilton's principle we have
$$\int_{t_1}^{t_2}dt.\delta(T-V)=0$$
\begin{eqnarray}
\mbox{or,}~~\int_{t_1}^{t_2}dt\int\int\left[\rho\frac{\partial z}{\partial t}\frac{\partial}{\partial t}\delta z-T_1\frac{\partial z}{\partial x}.\frac{\partial}{\partial x}(\delta z)-T_1\frac{\partial z}{\partial y}.\frac{\partial}{\partial y}(\delta y)\right]dxdy=0\nonumber\\
\mbox{or,}~~\int\int\left[\rho\frac{\partial z}{\partial t}\delta zdxdy\right]_{t_1}^{t_2}+\int_{t_1}^{t_2}dt\int\int\bigg[-\rho\frac{\partial^2 z}{\partial t^2}\delta z-T_1\frac{\partial}{\partial x}\left(\frac{\partial z}{\partial x}.\delta z\right)\nonumber\\
-T_1\frac{\partial}{\partial y}\left(\frac{\partial z}{\partial y}\delta z\right)+T_1\left(\frac{\partial^2z}{\partial x^2}+\frac{\partial^2z}{\partial y^2}\right)\delta z\bigg]dxdy=0\nonumber
\end{eqnarray}
the first term on the left hand side $=0$ since $\delta z=0$ when $t=t_1$ and $t=t_2$.
$$\mbox{Now,}~~\int\int\left[\frac{\partial}{\partial x}\left(\frac{\partial z}{\partial x}\delta z\right)+\frac{\partial}{\partial y}\left(\frac{\partial z}{\partial y}\delta z\right)\right]dxdy,$$
can be converted into surface integral integral taken over the boundary of the membrence as 
$$-\int\left(l\frac{\partial z}{\partial x}+m\frac{\partial z}{\partial y}\right)\delta zds$$
Here $(l,m,0)$ being the direction cosine of the inward drawn normal. As $\delta z=0$ on the boundary of the membrence (which is fixed) so the above surface integral vanishes. Thus we have 
$$\int_{t_1}^{t_2}dt\left[-\rho\frac{\partial^2z}{\partial t^2}+T_1\left(\frac{\partial^2z}{\partial x^2}+\frac{\partial^2 z}{\partial y^2}\right)\right]dxdy=0$$
Hence the equation of vibration of the membrence is
$$c^2\left(\frac{\partial^2z}{\partial x^2}+\frac{\partial^2z}{\partial y^2}\right)=\frac{\partial^2z}{\partial t^2},~~~~c^2=\frac{T_1}{\rho}$$

\section{Vibration of a rectangular membrence}

\begin{figure}[h!]
	\centering
	\includegraphics[scale=0.4]{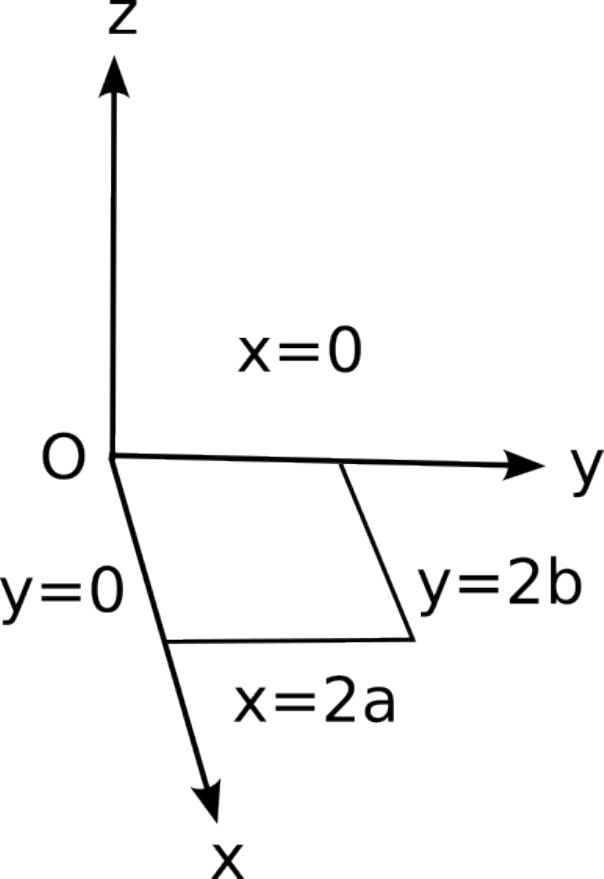}
\end{figure}

We consider the equation of the boundary as 
$$x=0,~~x=2a,~~y=0,~~y=2b.$$
We take the equilibrium plane of the membrence as $xoy$ plane. So the equation of transverse vibration of the membrence is given by 
$$\frac{\partial^2z}{\partial t^2}=c^2\left(\frac{\partial^2z}{\partial x^2}+\frac{\partial^2z}{\partial y^2}\right),~~~c^2=\frac{T_1}{\rho}$$ 
Boundary conditions are $z=0$ when $x=0$, $x=2a$ and $y=0$ and $y=2b$.

Let us assume separable form of the solution of the wave equation as
$$z=F_1(x).F_2(y).F_3(t)$$
and we have
$$\frac{1}{F_3}\frac{d^2F_3}{dt^2}=c^2\left[\frac{1}{F_1}\frac{d^2F_1}{dx^2}+\frac{1}{F_2}\frac{d^2F_2}{dy^2}\right]$$
Due to vibrating nature of the membrence let 
$$F_3(t)=\cos(pt+\epsilon),$$
so that the above equation becomes
$$-p^2=c^2\left[\frac{1}{F_1}\frac{d^2F_1}{dx^2}+\frac{1}{F_2}\frac{d^2F_2}{dy^2}\right]$$
$$\mbox{or,}~~\frac{1}{F_1}\frac{d^2F_1}{dx^2}=-\frac{p^2}{c^2}-\frac{1}{F_2}\frac{d^2F_2}{dy^2}=\mbox{constant}=-m^2~(\mbox{say})$$
Considering these two ordinary differential equation of the form 
$$\frac{d^2F_i}{dx_i^2}+m^2F_i=0,$$\vspace{-0.6 cm}\\
where $i=1,~2$ and $x_1=x, ~x_2=y$.
Then we have
\begin{eqnarray}
F_1&=&A\cos mx+B\sin mx\nonumber\\
F_2&=&A'\cos ny+B'\sin ny\nonumber
\end{eqnarray} 
Thus the complete solution is 
$$z=(A\cos mx+B\sin mx)(A'\cos ny+B'\sin ny)\cos(pt+\epsilon)$$
The boundary conditions give: 
$$(i)~~z=0~\mbox{at}~x=0\implies A=0~~~~~~~~~~~~~~~~~~~~~~~~~~~~~~~~~~~~~~~~~~~~~~~~~~~~~~~~~~$$
\begin{eqnarray}
(ii)~~z=0~\mbox{at}~x=2a\implies&~&B\sin 2am=0\nonumber\\
\implies&~&\sin 2am=0~~(\because B\neq 0~\mbox{for nontrivial solution})\nonumber\\
&~&\therefore 2am=s\pi~~\mbox{i.e,}~~m=\frac{s\pi}{2a},~~s~\mbox{is an integer}\nonumber
\end{eqnarray}
\begin{eqnarray}
(iii)~~z=0~\mbox{at}~y=0\implies A'=0~~~~~~~~~~~~~~~~~~~~~~~~~~~~~~~~~~~~~~~~~~~~~~~~~~~~~~~\nonumber
\end{eqnarray}
\begin{eqnarray}
(iv)~~z=0~\mbox{at}~y=2b\implies&~& 2bn=s'\pi~~~(s'~\mbox{is an integer})~~~~~~~~~~~~~~~~~~~~~~~~\nonumber\\
&~&\mbox{i.e,}~~n=\frac{s'\pi}{2b}\nonumber
\end{eqnarray}
Therefore the complete solution with the given boundary condition is
$$z=c'\sin\frac{s\pi x}{2a}\sin\frac{s'\pi y}{2b}\cos(pt+\epsilon)$$
Substituting this value of $z$ in the equation of vibration we have 
\begin{eqnarray}
p^2=\frac{c^2\pi^2}{4}\left(\frac{s^2}{a^2}+\frac{s'^2}{b^2}\right),~~~\mbox{with}~s,~s'~\mbox{as integers}.\nonumber
\end{eqnarray}
This gives the frequency of vibration and $\frac{2\pi}{p}$ gives the period of vibration.

\section{Circular Membrence}
The plane of the membrance is chosen as $xy$-plane. The circular boundary is characterised by $x^2+y^2=a^2$. Now the equation of transverse vibration of the membrence is
$$\frac{\partial^2z}{\partial t^2}=c^2\left[\frac{\partial^2z}{\partial x^2}+\frac{\partial^2z}{\partial y^2}\right].$$  
Let $(r,\theta)$ be the polar co-ordinates of a point in the plane of the membrence with the centre of the circle as pole. Then the boundary is characterised by $r=a$. In polar co-ordinate the equation of transverse vibration of the membrence becomes 
$$\frac{\partial^2z}{\partial t^2}=c^2\left[\frac{\partial^2z}{\partial r^2}+\frac{1}{r}\frac{\partial z}{\partial r}+\frac{1}{r^2}\frac{\partial^2z}{\partial\theta^2}\right]$$
where $x=r\cos\theta$, $y=r\sin\theta$ and $c^2=\dfrac{T_1}{\rho}$. Here $T_1$ is the uniform tension and $\rho$ is the uniform surface density. 

The boundary condition is $z=0$ when $r=a$. Now assuming separable form of the solution as
$$z=F(r).f(\theta)\cos(pt+\epsilon)$$
We have from the equation of vibration 
$$-r^2\left[\frac{p^2}{c^2}+\frac{1}{F}\left\{\frac{d^2F}{dr^2}+\frac{1}{r}\frac{dF}{dr}\right\}\right]=\frac{1}{f}\frac{d^2f}{d\theta^2}=\mbox{Constant}=-m^2~(\mbox{say})$$
$$\therefore \frac{d^2f}{d\theta^2}+m^2f=0~~~\mbox{i.e,}~~f=A\cos(m\theta+\epsilon')$$
$$\mbox{and}~~~\frac{d^2F}{dr^2}+\frac{1}{r}\frac{dF}{dr}+\left(\frac{p^2}{c^2}-\frac{m^2}{r^2}\right)F=0,~~~~~~~~$$
which is Bessel equation of order $m$. The integral of this equation regular at the origin is 
$$F(r)=c''J_m\left(\frac{pr}{c}\right)$$
$$\therefore z=c'J_m\left(\frac{pr}{c}\right)\cos(m\theta+\epsilon')\cos(pt+\epsilon)$$
Since the boundary is fixed at $r=a$ so $z=0$ at $r=a$. This gives $J_m\left(\dfrac{pa}{c}\right)=0$, which gives $p$ and hence the frequency of vibration.

\vspace{1cm}

$\bullet$ \textbf{Problem:} A circular membrence of radius $a$ is fixed at the edge and is subjected to uniform tension $T_1$. Prove that in a normal displacement $z$, symmetrical about the centre, the potential energy is $V=T_1\int_{0}^{a}\left(\dfrac{\partial z}{\partial r}\right)^2\pi rdr$, assuming $z=\zeta\left(1-\frac{r^2}{a^2}\right)$. Obtain the approximate period of vibration.

\vspace{.5cm}

\textbf{Solution:} The expression for potential energy is
$$V=\frac{1}{2}T_1\int\int\left[\left(\frac{\partial z}{\partial x}\right)^2+\left(\frac{\partial z}{\partial y}\right)^2\right]dxdy$$
$$\mbox{Now,}~~\frac{\partial z}{\partial x}=\frac{\partial z}{\partial r}.\frac{\partial r}{\partial x}+\frac{\partial z}{\partial\theta}.\frac{\partial\theta}{\partial x}=\frac{\partial z}{\partial r}.\frac{\partial r}{\partial x}~~\left(\because\frac{\partial z}{\partial\theta}=0~\mbox{as}~z~\mbox{is symmetrical about the centre}\right)$$
$$\mbox{As}~~r^2=x^2+y^2~~~~~~~~~~~~~~~~~~~~~~~~~~~~~~~~~~~~~$$
$$\therefore r\frac{\partial r}{\partial x}=x~~~~\therefore\frac{\partial r}{\partial x}=\frac{x}{r}~~~~~~~~~~~~~~~~~~~$$
$$\mbox{Similarly,}~~~\frac{\partial r}{\partial y}=\frac{y}{r}~~~~~~~~~~~~~~~~~~~~~~~~~~~~~~~~~~~~~~~~~~~~~~~$$
$$\therefore\left(\frac{\partial z}{\partial x}\right)^2+\left(\frac{\partial z}{\partial y}\right)^2=\frac{(x^2+y^2)}{r^2}\left(\frac{\partial z}{\partial r}\right)^2=\left(\frac{\partial z}{\partial r}\right)^2$$
\begin{eqnarray}
\therefore V&=&\frac{1}{2}T_1\int_{\theta=0}^{2\pi}\int_{r=0}^{a}\left(\frac{\partial z}{\partial r}\right)^2dr.rd\theta\nonumber\\
&=&T_1\int_{0}^{a}\left(\frac{\partial z}{\partial r}\right)^2\pi rdr\nonumber\\
&=&T_1\int_{0}^{a}\frac{4\zeta^2r^2}{a^4}\pi rdr\nonumber\\
&=&\frac{4\pi\zeta^2}{a^4}T_1\int_{0}^{a}r^3dr\nonumber\\
&=&\pi T_1\zeta^2\nonumber
\end{eqnarray}
\begin{eqnarray}
\mbox{Kinetic Energy}~(T)&=&\frac{1}{2}\rho\int_{0}^{2\pi}\int_{0}^{a}\left(\frac{\partial z}{\partial t}\right)^2rd\theta dr\nonumber\\
&=&\rho\pi\int_{0}^{a}\left(\frac{\partial z}{\partial t}\right)^2rdr\nonumber\\
&=&\rho\pi\int_{0}^{a}\dot{\zeta}^2(t)\left(1-\frac{r^2}{a^2}\right)^2rdr\nonumber\\
&=&\rho\pi\dot{\zeta}^2(t)\int_{0}^{a}\left(1-\frac{2r^2}{a^2}+\frac{r^4}{a^4}\right)rdr\nonumber\\
&=&\rho\pi\dot{\zeta}^2(t)\left[\frac{a^2}{2}-\frac{a^2}{2}+\frac{a^2}{6}\right]\nonumber\\
&=&\frac{\pi}{6}\rho\dot{\zeta}^2(t)a^2\nonumber
\end{eqnarray}
From Lagrange's equation of motion
$$\frac{d}{dt}\left(\frac{\partial T}{\partial\dot{\zeta}}\right)-\frac{\partial T}{\partial\zeta}=-\frac{\partial V}{\partial\zeta}$$
$$\mbox{or,}~~\rho\pi a^2\frac{\ddot{\zeta}}{3}-0=-2\pi T_1\zeta\implies\ddot{\zeta}=-\frac{6T_1\zeta}{\rho a^2},~~\mbox{a simple harmonic motion (S.H.M.)}$$
$$\therefore\mbox{Time period}=\frac{2\pi}{\sqrt{\frac{6\pi}{\rho a^2}}}=\frac{2\pi\sqrt{\rho a^2}}{\sqrt{6\pi}}$$

\vspace{1cm}

$\bullet$ \textbf{Problem:} If $z=\zeta\left(1-\dfrac{r^2}{a^2}\right)\left(1+\dfrac{\beta r^2}{a^2}\right)$, then show that the period will be $2\pi\sqrt{\dfrac{\rho a^2}{20T_1}\dfrac{(\beta^2+5\beta+10)}{(\beta^2+2\beta+3)}}$.

\vspace{.5cm}

\textbf{Solution:} 
$$z=\zeta\left(1-\frac{r^2}{a^2}\right)\left(1+\frac{\beta r^2}{a^2}\right)$$
$$\frac{\partial z}{\partial r}=\zeta\left[-\frac{2r}{a^2}\left(1+\frac{\beta r^2}{a^2}\right)+\left(1-\frac{r^2}{a^2}\right)\frac{2\beta r}{a^2}\right]$$
$$\frac{\partial z}{\partial t}=\dot{\zeta}\left(1-\frac{r^2}{a^2}\right)\left(1+\frac{\beta r^2}{a^2}\right)~~~~~~~~~~~~~~~~~~$$
\begin{eqnarray}
\therefore\mbox{K.E.}=T&=&\frac{1}{2}\rho\int_{0}^{2\pi}\int_{0}^{a}\dot{\zeta}^2\left(1-\frac{r^2}{a^2}\right)^2\left(1+\frac{\beta r^2}{a^2}\right)^2rdrd\theta\nonumber\\
&=&\pi\rho\dot{\zeta}^2\int_{0}^{a}\left(1-\frac{2r^2}{a^2}+\frac{r^4}{a^4}\right)\left(1+\frac{\beta^2r^4}{a^4}+\frac{2\beta r^2}{a^2}\right)rdr\nonumber\\
&=&\frac{\pi\rho\dot{\zeta}^2a^2}{60}\left(\beta^2+5\beta+10\right)\nonumber
\end{eqnarray}
\begin{eqnarray}
V&=&T_1\pi\zeta^2\int_{0}^{a}\left[\frac{2r}{a^2}\left(\beta-\frac{\beta r^2}{a^2}-1-\frac{\beta r^2}{a^2}\right)\right]^2rdr\nonumber\\
&=&\frac{4T_1\pi\zeta^2}{12}\left(\beta^2+2\beta+3\right)\nonumber
\end{eqnarray}
Now, from Lagrange's equation of motion
$$\frac{d}{dt}\left(\frac{\partial T}{\partial\dot{\zeta}}\right)-\frac{\partial T}{\partial\zeta}=-\frac{\partial V}{\partial\zeta}$$
$$\mbox{or,}~~\frac{2\pi\rho\ddot{\zeta}a^2}{60}\left(\beta^2+5\beta+10\right)=-\frac{8T_1\pi\rho}{12}\left(\beta^2+2\beta+3\right)$$
$$\therefore\ddot{\zeta}=-\frac{20\pi}{\rho a^2}\frac{\left(\beta^2+2\beta+3\right)}{\left(\beta^2+5\beta+10\right)}\zeta\implies T=2\pi\sqrt{\frac{\rho a^2}{20T_1}\frac{(\beta^2+5\beta+10)}{(\beta^2+2\beta+3)}}$$


\chapter{Small oscillation}

\section{Oscillation}

Let us consider a holonomic dynamical system with $n$ degrees of freedom in which the coordinates are independent of time. Let $q_1,q_2, ...q_n$ be the be the generalised coordinates of the system. Let the external forces be conservative and derived from a potential function $V$.  The position of equilibrium of the system is given by

\begin{equation}\label{eqn1}
\frac{\partial V}{\partial q_1} =0=\frac{\partial V}{\partial q_2} ... =	\frac{\partial V}{\partial q_n}~.
\end{equation}

Solving these equations we get the position of the equilibrium of the system. In the position of the equilibrium of the system, we can without any loss of generality consider $q_1,q_2, ...q_n$ to be zero. The equation (\ref{eqn1}) shows that  V is stationary in the position of the equilibrium. If further V is  minimum, then the equilibrium is stable. This means that if we give a small disturbance to the system, the coordinates $q_1,q_2, ...q_n$ and the velocity $\dot{q}_1,\dot{q}_2,..\dot{q}_n$  remain small. Since the  connection of the system is a quadratic expression in $\dot{q}_1,\dot{q}_2,..\dot{q}_n$  and  therefore,

\begin{equation}
T=\frac{1}{2}\sum_{i}\sum_{j} a_{i,j}\dot{q_i}\dot{q_j}\label{n2}
\end{equation}
where $a_{ij}=a_{ji}$,~~$i,j=1,2,...,n$.

Since the subsequent motion,  $q_1,q_2, ...q_n$ and as well as $\dot{q}_1,\dot{q}_2,..\dot{q}_n$ remain small, if we want to retain the quantities upto 2nd order in the expression of $T$, we can neglect the 1st order quantities in the coeff $a_{i,j}~ i.e.,$ we can put $q_1=q_2=...=q_n=0$ in the expression for $a_{ij}$. So we consider $a_{ij}$ as constants. Since $V$ is indeterminate to the extent of an additive constant we can choose the constant in such a manner that $V$ vanishes in the position of equilibrium. Expanding $V$ in Taylor series about the point $q_1=q_2=...=q_n=0$ and taking account of the relation (1) , we can put $V$ in the form 
\begin{eqnarray}
V=\frac{1}{2}\sum_{i}\sum_{j} c_{i,j} q_i q_j,\label{n3}
\end{eqnarray} 
 in which $c_{ij}=c_{ji},~ i,j=1,2,3,...n.$

In this expression for $V$ we have neglected the 3rd and higher order terms.

Now the Lagrange's equation of motion of the system are
\begin{equation}\label{Leq}
\frac{d}{dt}\left(\frac{\partial T}{\partial{\dot{q}_r}}\right)-\frac{\partial T}{\partial{q}_r} +\frac{\partial V}{\partial{{q}_r}}=0~ ,~~~r=1,2,3...n.\nonumber
\end{equation}

Substituting the value of $T$ and $V$ given in equations (\ref{n2}) and (\ref{n3}) respectively, we get
\begin{equation}
a_{r1}\ddot{q_1}+a_{r2}\ddot{q_2}+...+a_{rn}\ddot{q_n}+c_{r1}q_1 +c_{r2}q_2 +...+c_{rn}q_n=0 ~, ~~r=1,2,...n.\label{n4}
\end{equation}

This is a set of linear homogeneous equations in $q_r$'s. To solve it we choose $q_r=A_re^{\lambda t}$ as a trial solution ($A_r$ and $\lambda$ are constants). Therefore substituting in (\ref{n4}) one gets,

\begin{equation}
(a_{r1}\lambda^2+c_{r1})A_1 +(a_{r2}\lambda^2+c_{r2})A_2+ ... +(a_{rn}\lambda^2+c_{rn})A_n=0,~ r=1,2,...n.
\end{equation}

or in explicit form  

\begin{eqnarray}
(a_{11}\lambda^2+c_{11})A_1 +(a_{12}\lambda^2+c_{12})A_2+ ... +(a_{1n}\lambda^2+c_{1n})A_n&=&0,\nonumber\\ \nonumber
(a_{21}\lambda^2+c_{21})A_1 +(a_{22}\lambda^2+c_{22})A_2+ ... +(a_{2n}\lambda^2+c_{2n})A_n&=&0, \\ \nonumber
-----------------------  \\ \nonumber
-----------------------  \\ \nonumber
(a_{n1}\lambda^2+c_{n1})A_1 +(a_{n2}\lambda^2+c_{n2})A_2+ ... +(a_{nn}\lambda^2+c_{nn})A_n&=&0.
\end{eqnarray}	

Now eliminating $A_1$, $A_2$, ... $A_n$, we have

\begin{equation}
\Delta(\lambda)\equiv
\begin{bmatrix} 
(a_{11}\lambda^2+c_{11}) & (a_{12}\lambda^2+c_{12})&\cdots & (a_{1n}\lambda^2+c_{1n}) \\
(a_{21}\lambda^2+c_{21}) & (a_{22}\lambda^2+c_{22})&\cdots & (a_{2n}\lambda^2+c_{2n}) \\
\vdots & & \ddots & \vdots \\ 
(a_{n1}\lambda^2+c_{n1}) & (a_{n2}\lambda^2+c_{n2})&\cdots & (a_{nn}\lambda^2+c_{nn})\\
\end{bmatrix}
=0\nonumber
\end{equation}
This equation is known as frequency equation and $\Delta(\lambda)$ is called a Lagrange's determinant or harmonic determinant. It can be proved that the roots of the Lagrange's determinantal equation in $\lambda^2$ are all real. Further, if $V$ is essentially $+ve$ the roots of the Lagrange's determinantal equation in $\lambda^2$ are $-ve$, $i.e.,$ we write $\lambda^2=-\sigma^2$ where $\sigma$ is real and $+ve$.

If $\alpha_1$, $\alpha_2$, ... $\alpha_n$, be the minors of any one  row of $\Delta(\lambda)$, then we have
\begin{equation}
\frac{A_1}{\alpha_1}=\frac{A_2}{\alpha_2}=...=\frac{A_n}{\alpha_n}=H ~(say)\nonumber
\end{equation}
Since $\lambda$ appears as $\lambda^2$ in the determinant $\Delta(\lambda)$, therefore $\alpha_1$, $\alpha_2$, ... $\alpha_n$ have the same values when we put $\lambda=\pm i\sigma$. Then corresponding to a root $\lambda^2=- \sigma^2$, we have the solution
\begin{eqnarray}
q_r&=&\alpha_r(H e^{i\sigma t} + K e^{-i\sigma t}) \nonumber\\ \nonumber
&=& c \alpha_r cos(\sigma t +\sigma)~,~r=1,2,...,n. ( \because \alpha_r ~\text{have the same values for} \lambda=\pm \sigma)
\end{eqnarray}
The constants $c$ and $\epsilon$ are same for all values of $r$. This solution is said to constitute a normal mode of oscillation or a fundamental mode of oscillation of the system corresponding to $n$ values of $\lambda^2$ satisfying Lagrange's determinantal equation. For the $n$ normal modes of oscillation
\begin{equation}
q_r=c\alpha_r cos(\sigma t+\epsilon)+c^1\alpha_r cos(\sigma t+\epsilon^1)~,\nonumber
\end{equation}
to calculate  $\alpha_r$, one substitutes $\lambda^2=-\sigma^2$.

The constants $c,c^1 ...$ and $\epsilon, \epsilon^1 ...$etc. are to be determined from initial conditions. Since $c,c^1$ are small, therefore in the subsequent motion $q_1,q_2...q_n$ and $\dot{q_1},\dot{q_2}...\dot{q_n}$ remains small.

\vspace{.5cm}

\underline{ \textbf{Note:}} We can from the consideration of mechanics prove that the roots of the determinantal equation in $\lambda^2$ are real. On the assumption that the position of the equilibrium is characterized $q_1=0=q_2=...=q_n$ and this is a stable equilibrium if $V$ is minimum there. If possible, let $\lambda=\pm (\tau+i \delta)$ corresponding to a root of the determinantal equation. Since the co-efficient of this equation are all real, therefore $\lambda=\pm (\tau-i \delta)$ could also give the root of the equation. As a particular solution of the equation of motion we can then take
\begin{equation}
q_r=A_r(e^{\tau t}+e^{-\tau t})cos(\delta t)+B_r(e^{\tau t}-e^{-\tau t})sin(\delta t)~, r=1,2,...n.\nonumber
\end{equation}
The motion subsequent to the disturbance begins with small values of $q_r$'s and $\dot{q_r}$'s. But this co-ordinates and velocities increase indefinitely with time. This contradicts the assumption that the position of the equilibrium is stable. Thus the necessary condition for stability of equilibrium is that $\tau=0$, so that $\lambda^2=-\delta^2$. Thus the roots of the determinantal equation are not only real but $-ve$ also.

\vspace{.5cm}

$\bullet$ \textbf{Problem:} One point of a uniform circular hoop of mass $M$ and radius $a$ is fixed and hoop is free to move in a vertical plane through the fixed point. A bead of mass $m$ slides on the hoop and there is no friction. Prove that the periods of small oscillation about the position of the stable equilibrium are $2\pi \sqrt{\dfrac{2a}{g}}$ and $2\pi \sqrt{\dfrac{M}{m+M}\frac{a}{g}}$.

\vspace{.25cm}

\textbf{Solution:} The co-ordinates of the system are $\theta$ and $\phi$, the angles which $OC$ and $CP$ make with the downward vertical.

\begin{figure}[h!]
	\centering
	\includegraphics[scale=0.33]{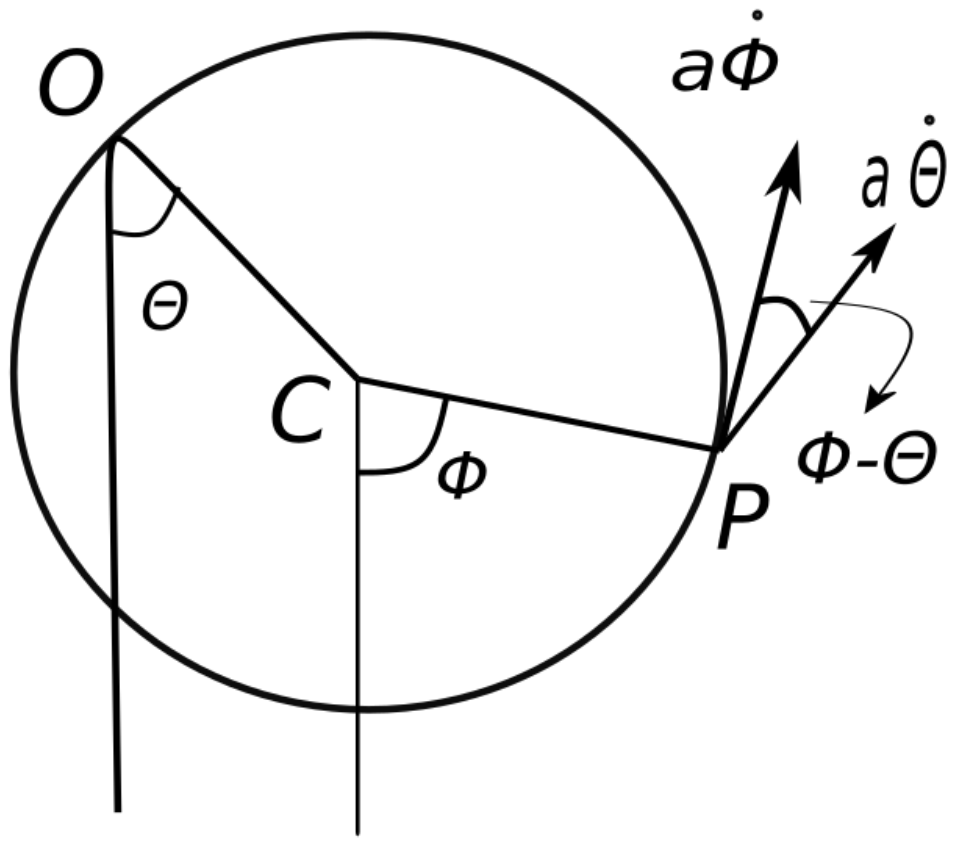}
\end{figure}

\begin{eqnarray}
\therefore ~T&=&\frac{1}{2}M(2 a^2)\dot{\theta}^2 + \frac{1}{2}m[a^2 \dot{\theta}^2+a^2 \dot{\phi}^2+2a^2 \dot{\theta}\dot{\phi}cos(\theta-\phi)]  \nonumber \\ 
V&=&-M g a cos\theta - mg(a cos\theta +a cos\phi)+V_0\nonumber
\end{eqnarray}
( $V_0,$ a constant, the initial P.E.)

Since $\theta$ and $\phi$ are very small, so $\theta \approx 1-\theta^2/2$, $cos\phi \approx 1-\phi^2/2$, $cos(\theta-\phi)\approx 1.$

\begin{eqnarray}
T&=&\frac{1}{2}M(2 a^2)\dot{\theta}^2 + \frac{1}{2}m[a^2 \dot{\theta}^2+a^2 \dot{\phi}^2+2a^2 \dot{\theta}\dot{\phi}] \nonumber \\
V&=&-(M+m)ga(1-\theta^2/2)-mga(1-\phi^2/2)+V_0\nonumber
\end{eqnarray}
The Lagrange's equations are 
\begin{eqnarray}
\frac{d}{dt}\left( \frac{\partial T}{\partial \dot{\theta}}\right) - \frac{\partial T}{\partial \theta} +\frac{\partial V}{\partial \theta}=0 \nonumber \\
\frac{d}{dt}\left( \frac{\partial T}{\partial \dot{\phi}}\right) - \frac{\partial T}{\partial \phi} +\frac{\partial V}{\partial \phi}=0~.\nonumber
\end{eqnarray}
$\therefore$

\begin{eqnarray}
\frac{d}{dt}[Ma^22\dot{\theta}+\frac{1}{2}m(2a^2\dot{\theta}+2a^2\dot{\phi})]+(m+M)ga\theta&=&0 \nonumber \\
or, ~ 2Ma^2 \ddot{\theta}+ma^2(\ddot{\theta}+\ddot{\phi}) +(m+M)ag\theta&=&0 \nonumber \\
or,~ \ddot{\theta}(2M+m)a +ma\ddot{\phi}+(m+M)g\theta&=&0 \nonumber \\
\text{Similarly,}~~~~~~~~~~~~~~~~~~~~~~~~~~~~\nonumber \\ ~~ (m\ddot{\theta} +M\ddot{\phi})a +mg\phi&=&0\nonumber
\end{eqnarray}

If $\theta =A_1e^{\lambda t},~\phi=A_2e^{\lambda t}$. Then the above two equations become
\begin{eqnarray}
\left[a(2M+m)\lambda^2 +(M+m)g\right]A_1 +A_2(ma \lambda^2)&=&0	\nonumber \\
(am\lambda^2)A_1+(am\lambda^2+mg)A_2&=&0\nonumber
\end{eqnarray}
Now eliminating $A_1$ and $A_2$,

\begin{eqnarray}
\begin{bmatrix} 
a(2M+m)\lambda^2 +(m+_M)g & ma\lambda^2  \\
ma\lambda^2 & ma\lambda^2+mg \\
\end{bmatrix} &=&0 \nonumber \\
a^2(2Mm +m^2)\lambda^4 + (Mm +m^2)g^2 +~~ \nonumber \\  ~ag\lambda^2(Mm+m^2+2Mm+m^2) -m^2 a^2 \lambda^4 &=&0 \nonumber \\
2Ma^2\lambda^4 +ag(3M+2m)\lambda^2 +g^2(m+M)&=&0 \nonumber \\
\therefore \lambda^2 =\frac{-ag(3M+2m)\pm\sqrt{a^2g^2(3M+2m)^2-8a^2g^2M(m+M)}}{4Ma^2}&& \nonumber \\
=\frac{-ag(3M+2m)\pm ag(M+2m)}{4Ma^2}=-\frac{g}{2a},~-\frac{g(M+m)}{aM}~~~~&&\nonumber
\end{eqnarray}
$\because \lambda=-\sigma^2 \Rightarrow \sigma=\sqrt{\dfrac{g}{2a}}, \text{or}~ \sqrt{\dfrac{g(M+m)}{aM}}$ $\therefore \theta=A cos\left (\sqrt{\dfrac{g}{2a}}t +\epsilon_1 \right) $,$\phi=Bcos\left (\sqrt{\dfrac{g(M+m)}{aM}}t+\epsilon_2 \right ). $
The time periods are $2\pi\sqrt{\dfrac{2a}{g}},~~2\pi \sqrt{\dfrac{aM}{g(M+m)}}.$

\vspace{.5cm}

$\bullet$ \textbf{Problem:} A smooth circular wire of mass $8m$ and radius $a$ swings in a vertical plane being suspended by an inextensible string of length a attached to one point of it. A particle of mass $m$ can slide on the wire. Prove that, the period of small oscillation about the stable equilibrium are $2\pi \sqrt{\dfrac{8a}{3g}},2\pi \sqrt{\dfrac{a}{3g}},\text{and}~ 2\pi \sqrt{\dfrac{8a}{9g}}$. Also show that the angle between the radius to the bead and the diameter through the pt of suspension is a normal coordinate. 

\vspace{.25cm}

\textbf{Solution:} The K.E of the circular wire $E_k=\dfrac{1}{2}Mv^2+\dfrac{1}{2}M K^2\dot{\theta}^2$, where $v$ is the velocity of the centre, $K$ is the radius of gyration of the wire about a line through $C\perp$ to the plane of motion.
$\therefore v^2=a^2\dot{\psi}^2 +a^2\dot{\theta}^2+2a^2\dot{\theta}\dot{\psi} cos(\theta-\psi)$.
Since $\theta$ and $\psi$ are very small, so $cos(\theta-\psi)=1$.
\begin{eqnarray}
v^2&=&a^2(\dot{\theta}+\dot{\psi})^2, k^2=a^2 \nonumber \\
\therefore, T&=&\frac{1}{2}M a^2 (\dot{\theta}+\dot{\psi})^2 +\frac{1}{2}M a^2 \dot{\theta}^2 +\frac{1}{2}m a^2 (\dot{\theta}+\dot{\phi}+\dot{\psi})^2 \nonumber \\
V&=&-9mga(2-\frac{\theta^2}{2}-\frac{\psi^2}{2})-mga (1-\phi^2/2)\nonumber
\end{eqnarray}

\begin{figure}[h!]
	\centering
	\includegraphics[scale=0.33]{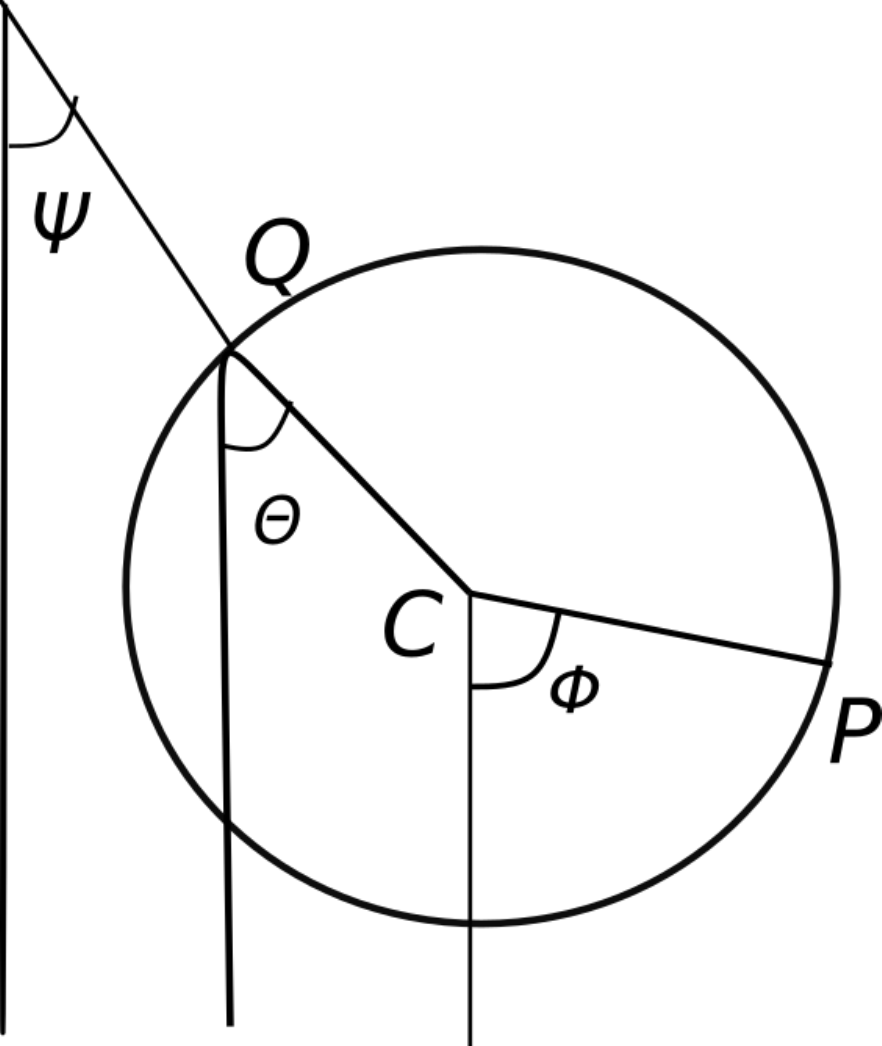}
\end{figure}

So the L's equations are
\begin{eqnarray}
\frac{d}{dt}\left(  \frac {\partial T}{\partial  \dot{\theta}} \right)- \frac {\partial T}{\partial \theta}+\frac {\partial V}{\partial \theta}=0 & \Rightarrow &  17 a \ddot{\theta}+a \ddot{\phi} +9 a\ddot{\psi}+9 g\theta=0 \label{n6}  \\
\frac{d}{dt}\left(  \frac {\partial T}{\partial  \dot{\phi}} \right)- \frac {\partial T}{\partial \phi}+\frac {\partial V}{\partial \phi}=0 & \Rightarrow&  a \ddot{\theta}+a \ddot{\phi} + a\ddot{\psi}+ g\phi=0 \label{n7} \\
\frac{d}{dt}\left(  \frac {\partial T}{\partial  \dot{\psi}} \right)- \frac {\partial T}{\partial \psi}+\frac {\partial V}{\partial \psi}=0 & \Rightarrow&  9 a \ddot{\theta}+a \ddot{\phi} + 9 a\ddot{\psi}+ 9g\psi=0 \label{n8}
\end{eqnarray}
Substitute,  $\theta=A_1 e^{\lambda t},\phi=A_2e^{\lambda t}, \psi=A_3e^{\lambda t}$. So the above equations (\ref{n6})-(\ref{n8}) reduce to 
\begin{eqnarray}\label{4eqn17}
(17 a \lambda^2 +9g)A_1 + (a\lambda^2)A_2+ (9a\lambda^2)A_3=0\nonumber \\
(a\lambda^2)A_1+(a\lambda^2+g)A_2+(a\lambda^2)A_3=0 \nonumber \\
(9 a\lambda^2)A_1+(a\lambda^2)A_2+(9a\lambda^2+9g)A_3=0~.
\end{eqnarray}
Eliminating, $A_1,A_2,A_3$ we have
\begin{eqnarray}
\begin{bmatrix} 
(17 a \lambda^2 +9g) & a\lambda^2&9a\lambda^2\\
a\lambda^2& a\lambda^2+g & a\lambda^2\\
9a\lambda^2 & a\lambda^2)&9a\lambda^2+9g\\
\end{bmatrix}
&=&0 \nonumber \\
i.e., (8a\lambda^2+3g) (8a\lambda^2+9g) (a\lambda^2+3g)& = &0\nonumber
\end{eqnarray}
$\therefore  \lambda^2=-\dfrac{3g}{8a},-\dfrac{9g}{8a},-\dfrac{3g}{a}$

So the solutions are\\
$\theta=\alpha cos\left(\sqrt{\dfrac{3g}{8a}}t+\epsilon \right),\phi=\beta cos\left(\sqrt{\dfrac{9g}{8a}}t+\epsilon \right), \psi=\gamma cos\left(\sqrt{\dfrac{3g}{a}}t+\epsilon \right)$
Hence, periods of small oscillations are\\
$2\pi \sqrt{\dfrac{8a}{3g}},~2\pi \sqrt{\dfrac{8a}{9g}},~2\pi \sqrt{\dfrac{a}{3g}}$.\\
From equation (\ref{4eqn17}) putting $\lambda^2=-\dfrac{3g}{8a}$, we have 

\begin{equation}
-	\frac{3g}{8} A_1 +\frac{5g}{8}A_2-\frac{3g}{8}A_3=0\nonumber
\end{equation}

$i.e.,$ $3A_1-5A_2+3A_3=0\Rightarrow A_1:A_2:A_3=5:6:5$

putting $\lambda^2=-\dfrac{9g}{8a}$ in  equation (\ref{4eqn17})\\

$	-\dfrac{9g}{8}A_1-\dfrac{g}{8}A_2-\dfrac{9g}{8}A_3=0  $ \\
$9A_1+A_2+9A3=0\Rightarrow A_1:A_2:A_3=1:-18:1$

For $\lambda^2=-\dfrac{3g}{a}$ from  equation (\ref{4eqn17})\\
$-3gA_1-2gA_2-3gA_3=0$

$i.e,~~3A_1+2A_2+3A_3=0\Rightarrow A_1:A_2:A_3=1:-3:1$

$\therefore~\psi=5c_1 cos\left(\sqrt{\dfrac{3g}{a}}t+\epsilon \right)+c_2 cos\left(\sqrt{\dfrac{3g}{8a}}t+\epsilon \right)+c_3 cos\left(\sqrt{\dfrac{9g}{8a}}t+\epsilon \right)$ 	 \\
$~ ~~\theta=6c_1 cos\left(\sqrt{\dfrac{3g}{a}}t+\epsilon \right)-18c_2 cos\left(\sqrt{\dfrac{3g}{8a}}t+\epsilon \right)-3c_3 cos\left(\sqrt{\dfrac{9g}{8a}}t+\epsilon \right)$ 	 	\\
$~~~ \phi=5c_1 cos\left(\sqrt{\dfrac{3g}{a}}t+\epsilon \right)+c_2 cos\left(\sqrt{\dfrac{3g}{8a}}t+\epsilon \right)+c_3 cos\left(\sqrt{\dfrac{9g}{8a}}t+\epsilon \right)$ 	\\

\vspace{.5cm}

\section{Normal Co-ordinates and Small Oscillation}

\vspace{.25cm}

If the K.E. and P.E. of a vibrating system are in the form \\
\begin{eqnarray}
2T&=&a_{11}\dot{q_1}^2+a_{22}\dot{q_2}^2+...+a_{nn}\dot{q_n}^2+2a_{12}\dot{q_1}\dot{q_2}+...\nonumber\\ \nonumber
V&=&V_0+\frac{1}{2}(b_{11}q_1^2+b_{22}q_2^2+...+b_{nn}q_n^2+2b_{12}q_1q_2+...)
\end{eqnarray}

then it is always possible to find a linear transformation of the co-ordinates such that the K.E. and P.E. when expressed in terms of new co-ordinates have the form

\begin{eqnarray}
2T&=&\dot{Q_1}^2+\dot{Q_2}^2+...+\dot{Q_n}^2\nonumber \\ \nonumber
V&=&V_0+\frac{1}{2}(\delta_{1}^2Q_1^2+\delta_{2}^2Q_2^2+...+\delta_{n}^2Q_n^2)
\end{eqnarray}
where $\delta_1,\delta_2,...\delta_n$ are constants. These new cordnitaes are called normal co-ordinate or principle co-orrodinates.

When we have two homogeneous quadratic functions of any no. of variables all of which are $+ve$ for all values of the variables , then by a linear transformation,we may eliminate the product terms from both the expressions and the same time it is possible to make the coefficients of the square terms to be unity in any of the expressions.

Let the linear transformation of the coordinates be 
\begin{eqnarray}
q_1=b_{11}Q_1+b_{12}Q_2+....+b_{1n}Q_n\nonumber \\
q_2=b_{21}Q_1+b_{22}Q_1+....+b_{2n}Q_n \nonumber \\
--------------\nonumber \\
q_n=b_{n1}Q_1+b_{n2}Q_1+....+b_{nn}Q_n \label{n10}
\end{eqnarray}
where $b's$ are constants. So, obviously, the velocity will transform by the same set of relations. Hence by the above result it is possible to have a real transformation so that K.E. and P.E are transform to

\begin{eqnarray}
2T&=&\dot{Q_1}^2+\dot{Q_2}^2+...+\dot{Q_n}^2\nonumber \\ \nonumber
2V&=&\lambda_{1}^2Q_1^2+\lambda_{2}^2Q_2^2+...+\lambda_{n}^2Q_n^2
\end{eqnarray}
$\lambda's$ are constant.

Thus the L's eq. of motion i.e.,

\begin{eqnarray}
\frac{d}{dt}\left( \frac{\partial T}{\partial \dot{Q_r}}\right) - \frac{\partial T}{\partial Q_r} =-\frac{\partial V}{\partial Q_r},~~~r =1,2,....n \nonumber 
\end{eqnarray}

gives ,
\\
$\ddot{Q_k}+\lambda {Q_k}=0,~~ k=1,2,...n$,\\
Assuming , $Q_k=A_ke^{\lambda t}$ i.e.,  $\ddot{Q}_k=\lambda^2Q_k$, Lagranges determinantal eq. becomes

\begin{eqnarray}
\begin{bmatrix} 
( \lambda^2 +\lambda_1) & 0&\dots&0\\
0& ( \lambda^2 +\lambda_2)& \dots&0\\
0 & 0&\dots&( \lambda^2 +\lambda_n)\\
\end{bmatrix}
=0\nonumber
\end{eqnarray}
It is shown in the theory of linear transformation of the variable that the roots of the determinantal equation are not affected by the transformation i.e., the periods are same for all systems of coordinates.

\vspace{.5cm}

\section{Effect of a constrain on the small oscillation of the system about equilibrium configuration }

\vspace{.25cm}

Let us consider the small oscillation of a holonomic dynamical system with n d.f. and let $q_1,q_2,...q_n$ be the normal coordinates of the system. Let us suppose that a constrain is introduced in the system and the constrain does no work. Then the equation of constrain is $f(q_1,q_2,...q_n)=0$. Since this relation is to be satisfied in the position of the equilibrium in which the coordinates vanish, it can be explained by the  Taylors theorem:

\begin{equation}
f_0(q_1,q_2,...q_n)+\left[ \left( \frac{\partial f}{\partial q_1}\right)_0  q_1 + \left( \frac{\partial f}{\partial q_2}\right)_0  q_2+... \right]+...=0\nonumber
\end{equation}

Neglecting the higher powers of $q_r$, we have
\begin{eqnarray}\label{4eqn26}
\left( \frac{\partial f}{\partial q_1}\right)_0  q_1 + \left( \frac{\partial f}{\partial q_2}\right)_0  q_2+...+ \left( \frac{\partial f}{\partial q_n}\right)_0  q_n=0 \nonumber \\
i.e., ~ A_1q_1+A_2q_2+...+A_nq_n=0
\end{eqnarray} 
Where $A_r=\left( \dfrac{\partial f}{\partial q_r}\right)_0,~~r=1,2,...n. $ \\
The K.E. and P.E. of the system during this small oscillation can be written in the form

$T=\dfrac{1}{2}\sum \dot{q}_r^2,~~~V=V_0+\dfrac{1}{2}\sum \delta_r^2 q_r^2$\\
subjected to the above constraint (\ref{4eqn26}).\\
The Lagrange's equation for a connected holonomic system is given by
\begin{equation}
\frac{d}{dt}\left( \frac{\partial T}{\partial \dot{q_r}}\right) - \frac{\partial T}{\partial q_r} =-\frac{\partial V}{\partial q_r}+\lambda \frac{\partial F}{\partial q_r},~~~r =1,2,....n  ~,\nonumber
\end{equation}
where $\lambda$ is a function of time and \\
$F(q_1,q_2,...q_n)=A_1q_1+A_2q_2+...+A_nq_n$, So, we have the equation of motion \\

\begin{equation}\label{4eqn28}
\ddot{q_r}=-\delta_r^2+\lambda A_r
\end{equation} 
Let us assume the solution 
\begin{eqnarray}
q_r=a_re^{i\delta t}\label{n13}\\
 \mbox{and}~~ ~ \lambda =\alpha e^{i\delta t}\label{n14}
\end{eqnarray}
where $\alpha$ is a constant. Substituting (\ref{n13}) and (\ref{n14}) in (\ref{4eqn28}),  we get

\begin{eqnarray}
(-\delta^2+\delta_r^2)q_r=\alpha A_r e^{i\delta t} \nonumber \\
or, ~(-\delta^2+\delta_r^2)a_r e^{i\delta t}=\alpha A_r e^{i\delta t} \nonumber \\
\therefore~ a_r=\frac{\alpha A_r}{(\delta^2-\delta_r^2)} ~~, r=1,2,..,n.
\end{eqnarray}

Therefore, from (\ref{4eqn26}) and (\ref{n13})
\begin{eqnarray}
A_1a_1e^{i\delta t}+A_2a_2e^{i\delta t}+...	+A_na_ne^{i\delta t}=0 \nonumber \\
i.e.,~ A_1a_1+A_2a_2+...+A_na_n=0 (\because e^{i\delta t} \neq 0) \nonumber \\
or, ~ \frac{A_1^2}{(\delta^2-\delta_1^2)}+\frac{A_2^2}{(\delta^2-\delta_2^2)}+...+\frac{A_n^2}{(\delta^2-\delta_n^2)}=0\nonumber \\
X(\delta^2) \equiv  A_1^2(\delta^2-\delta_2^2)...(\delta^2-\delta_n^2)+A_2^2(\delta^2-\delta_1^2)(\delta^2-\delta_3^2)...(\delta^2-\delta_n^2)
\nonumber \\
+ A_n^2(\delta^2-\delta_1^2)(\delta^2-\delta_2^2)...(\delta^2-\delta_{n-1}^2)\nonumber
\end{eqnarray}

This is the frequency equation of the constrain oscillation and the freq. eq. is of the degree $(n-1)$.

Let, $\delta_1>\delta_2>\delta_3>....>\delta_n$, then $ X(\delta_1^2)>0, X(\delta_2^2)<0, X(\delta_3^2)>0...   $ and so on. So, there is atleast one root of $ X(\delta^2)=0$ between $(\delta_1^2, \delta_2^2), (\delta_2^2, \delta_3^2), ...(\delta_{(n-1)}^2,\delta_n^2).$  Thus the root of the freq. eq. of the constrain motion are separated by the roots of the equation of the original motion and conversely. Consequently, the periods of the oscillation of the constrained motion are separated by the periods of the oscillation of the original motion and conversely.

In small oscillation about  a position of stable equilibrium let $q_1,q_2,...q_n$ be the normal coordinates of the holonomic dynamical system. Then we can write the K.E. and P.E. as
\begin{eqnarray}
T&=&\frac{1}{2}\left[\dot{q_1}^2+...+\dot{q_n}^2\right] \nonumber \\
V&=&V_0+\frac{1}{2}\left[\sigma_1^2q_1^2+...+\sigma_n^2q_n^2\right] \nonumber
\end{eqnarray}
Let us introduce $n-1$ constrains in the system which do not work and consistent with the position equilibrium. The eqs. of the constrains can therefore be represented by $(n-1)$ homogeneous linear eqs. in $q_1,q_2,...q_n$.

\begin{eqnarray}
A_{11}q_1+A_{12}q_2+...+A_{1n}q_n=0\nonumber \\
A_{21}q_1+A_{22}q_2+...+A_{2n}q_n=0 \nonumber \\
-------------\nonumber \\
------------- \nonumber \\
A_{(n-1)1}q_1+A_{(n-1)2}q_2+...+A_{(n-1)n}q_n=0\nonumber
\end{eqnarray}
Solving these we get 
$\dfrac{q_1}{B_1}=\dfrac{q_2}{B_2}=...=\dfrac{q_n}{B_n}=q~(say)$ i.e., $q_1=B_1q,q_2=B_2q,...q_n=B_nq.$
Here $B_1,...B_n$ are functions of $A_{11},A_{12},...,A_{(n-1)n}$ and are minors of the elements of a row of a determinant from the above coefficients. The constrain system have therefore only 1 d.f.  and it can be expressed by the single coordinate $q$. Thus in the constrain system

$T=\dfrac{1}{2}(\sum B_r^2)\dot{q}^2~,V=V_0+\dfrac{q^2}{2}(\sum B_r^2\sigma_r^2)$. 
Then the equation of motion of constrain system is 
\begin{eqnarray}
\frac{d}{dt}\left( \frac{\partial T}{\partial \dot{q}}\right) - \frac{\partial T}{\partial q} =-\frac{\partial V}{\partial q}\nonumber \\
\mbox{or,}~~\left(\sum B_r^2\right)\ddot{q}+q\sum B_r^2\sigma_r^2=0\nonumber \\
\mbox{or,}~~\ddot{q}+\frac{\sum B_r^2\sigma_r^2}{\sum B_r^2}q=0\nonumber
\end{eqnarray}

Now , if $\dfrac{2\pi}{\sigma}$ be the period of the  constrain motion then we get from the  above expression $\sigma^2=\dfrac{\sum B_r^2\sigma_r^2}{\sum B_r^2}$, which shows that $\sigma^2$ is the wt.a.m. of $\sigma_r^2$ with wt. $B_r^2, ~~ r=1,2,...,n$ and hence it lies between the greatest and the least of the nos $\sigma_1^2, \sigma_2^2,...\sigma_n^2$. Thus the period of the constrain motion $\dfrac{2\pi}{\sigma}$ lies between the greatest and the least of the periods of the oscillation $\dfrac{2\pi}{\sigma_r}~~(r=1,2,...,n)$ of the original motion.

If the K.E. and P.E. of a vibrating system are given by $T=\dfrac{1}{2}\sum\sum a_{rs}\dot{q_r}\dot{q_s}, V=\dfrac{1}{2}\sum\sum b_{rs} q_r q_s $, where $a_{rs}=a_{sr}$ and $b_{rs}=b_{sr}$, then it is always possible  to obtain a linear transformation of $q's$ to the new system of $Q's$ such that  

$T=\dfrac{1}{2}\sum \dot{Q_k}^2, V=\dfrac{1}{2}\sum \sigma_k^2 Q_k^2  $ \\
For the normal mode of oscillation of the system \\
\begin{equation}\label{4eqn35}
q_r=\sum_s \alpha_{rs}c_s cos(\sigma_s t+\epsilon_s)~, 
\end{equation}

where $\alpha_{rs}$ are the minors of any of the rows of the harmonic determinant. Let

\begin{eqnarray}\label{4eqn36}
q_r'= c_{s}cos(\sigma_s t+\epsilon_s) \nonumber \\
\therefore q_r=\sum_k \alpha_{rk}q_{k}' ~~, ~~ q_s=\sum_{l=1} ^n\alpha_{sl}q_{l}' 
\end{eqnarray}

substituting the values of $q_r$ and $q_s$ in the expression of K.E. in the form

\begin{eqnarray}
2T=\sum_{	
	r=1}^{n}\sum_{s=1}^{n} a_{rs}\sum_{k=1}^{n} \alpha_{rk}\dot{q_k}'\sum_{l=1}^{n} \alpha_{sl}\dot{q_l}' \nonumber \\
=\sum_{k=1}^{n}\sum_{l=1}^{n}\left[\sum_{	r=1}^{n}\sum_{s=1}^{n} a_{rs} \alpha_{rk} \alpha_{sl}\right]\dot{q_k}'\dot{q_l}' \nonumber
\end{eqnarray}
Introducing the notation
\begin{equation}\label{4eqn37}
2T(\alpha_k,\alpha_l)=\sum_{	r=1}^{n}\sum_{s=1}^{n}a_{rs} \alpha_{rk} \alpha_{sl}
\end{equation}

\begin{equation}\label{4eqn38}
2V(\alpha_k,\alpha_l)=\sum_{	r=1}^{n}\sum_{s=1}^{n}b_{rs} \alpha_{rk} \alpha_{sl}
\end{equation}

We can  write 
\begin{equation}\label{4eqn39}
2T=\sum_k\sum_l 2T(\alpha_k,\alpha_l)\dot{q_k}'\dot{q_l}' 
\end{equation}
\begin{equation}\label{4eqn40}
2V=\sum_k\sum_l 2V(\alpha_k,\alpha_l)\dot{q_k}'\dot{q_l}' 
\end{equation}

Now we shall prove  
\begin{eqnarray} \label{4eqn41}
T(\alpha_k,\alpha_l)=0 ~~\text{when}  ~~k\neq l  
\end{eqnarray}

\begin{eqnarray} \label{4eqn42}
\text{and}~~  \sigma_k^2 T(\alpha_k,\alpha_l)=V(\alpha_k,\alpha_l)
\end{eqnarray}
From  equation (\ref{4eqn39}) and  equation (\ref{4eqn41})

\begin{eqnarray}\label{4eqn43}
2T=\sum_k 2T(\alpha_k,\alpha_l)\dot{q_k}'^2 =\sum \dot{Q}^2_k \nonumber \\
\text{where},~~\dot{Q}^2_k= 2T(\alpha_k,\alpha_l)\dot{q_k}'^2 ~~i.e.,~~ \dot{Q}_k= \sqrt{2T(\alpha_k,\alpha_l)}\dot{q_k}'
\end{eqnarray}

From equation (\ref{4eqn40}) and equation  (\ref{4eqn42}), when $l=k$,

\begin{eqnarray}
2V=\sum_k \sigma_k^2 2T(\alpha_k,\alpha_k) q_k^2=\sum_k \sigma_k^2Q_k^2 \nonumber \\
\text{where},~~Q_k=\sqrt{2T(\alpha_k,\alpha_k)},~~q_k'=c_k\sqrt{2T(\alpha_k,\alpha_k)} cos(\sigma_k t+ \epsilon _k)
\end{eqnarray}

\textbf{{Proof of (\ref{4eqn41}) and (\ref{4eqn42}):}} \\

\vspace{.25cm}

From $L's$ equation of motion \\
$\sum_s \left(a_{rs}\frac{d^2}{dt^2}+b_{rs}\right)q_s=0~,~,r=1,2,...,n.$\\
put $q_s=\sum_{	k}\alpha_{sk}q'_{k}$, then we get \\

$\sum_s \left[   - a_{rs}\sigma^2_k +b_{rs}   \right]\alpha_{sk}q'_{k}=0,$   where   $q'_k=c_k cos(\sigma_k t+ \epsilon _k) \neq 0$

\begin{equation}\label{4eqn45}
\therefore\sum_s \left[   - a_{rs}\sigma^2_k +b_{rs}   \right]\alpha_{sk}=0
\end{equation}

Now multiplying  (\ref{4eqn45}) by $\alpha_{rl}$ for the $l^{th}$ mode and adding for $r$ from $r=1~to ~n$ and we get

\begin{eqnarray}
\sum_r  \alpha_{rl} \sum_s\left[   - a_{rs}\sigma^2_k +b_{rs}   \right]\alpha_{sk}=0,\nonumber \\
or, ~  -\left[\sum_r \sum_s   a_{rs}  \alpha_{rl} \alpha_{sk}\right]\sigma_k^2 +\sum_r\sum_s b_{rs}\alpha_{rl} \alpha_{sk}=0 \nonumber \\
\mbox{or,}~~-2T(\alpha_k,\alpha_l) \sigma_k^2+V(\alpha_l,\alpha_k)=0 ~\text{(by (\ref{4eqn37}) and(\ref{4eqn38}) )}
\end{eqnarray}

From symmetry,
$-2T(\alpha_l,\alpha_k) \sigma_k^2+V(\alpha_k,\alpha_l)=0$\\
Interchanging $k$ and $l$ we get similarly, \\ 
$-2T(\alpha_k,\alpha_l) \sigma_l^2+2V(\alpha_k,\alpha_l)=0$

 Subtracting these two, $(\sigma_k^2-\sigma_l^2)2T(\alpha_k,\alpha_l) =0$, But $(\sigma_k^2\neq \sigma_l^2)$,
$\therefore 2T(\alpha_k,\alpha_l) =0,~for~k\neq l$\\
If $k=l$, $-2T(\alpha_k,\alpha_k) \sigma_k^2+2V(\alpha_k,\alpha_k)=0$\\
$\implies V(\alpha_k,\alpha_k)=T(\alpha_k,\alpha_k) \sigma_k^2$\\

\vspace{.5cm}

$\bullet$ {\textbf{Problem}} A light string OAB is tied to a fixed pt. at $O$ and carries a mass $2m$ at A and  a mass $m$ at B. The lengths OA and AB are $ l/2$  and $ 3l/4$   respectively. The string is free to move in a vertical plane and the system oscillates about the position of equilibrium. The inclination of OA and AB to the vertical are denoted by $\theta $ and $\phi$ respectively. Find the normal coordinates. The system is held with a string straight and inclined  a small angle $\alpha$ to the vertical and is let go from rest from this position at the instant $t=0,$ show that at any subsequent time $\theta =\dfrac{\alpha}{3}(2 cos ~nt +cos ~2nt), \phi=\dfrac{\alpha}{3}(4 cos ~nt - cos ~2nt)$ where $n=\sqrt{\dfrac{g}{l}}$ .\\

\vspace{.25cm}

{\textbf{Solution}} 

$T=\dfrac{1}{2}.2m \left(\dfrac{1}{2}l\dot{\theta}\right)^2+\dfrac{1}{2}m\left[ \dfrac{1}{2}l\dot{\theta}+\dfrac{3}{4}l\dot{\phi}\right]^2 $ \\
$V=V_0-3 mg \dfrac{l}{2}(1-\theta^2/2)-mg \dfrac{3l}{4}(1-\phi^2/2)$\\

\begin{figure}[h!]
	\centering
	\includegraphics[scale=0.33]{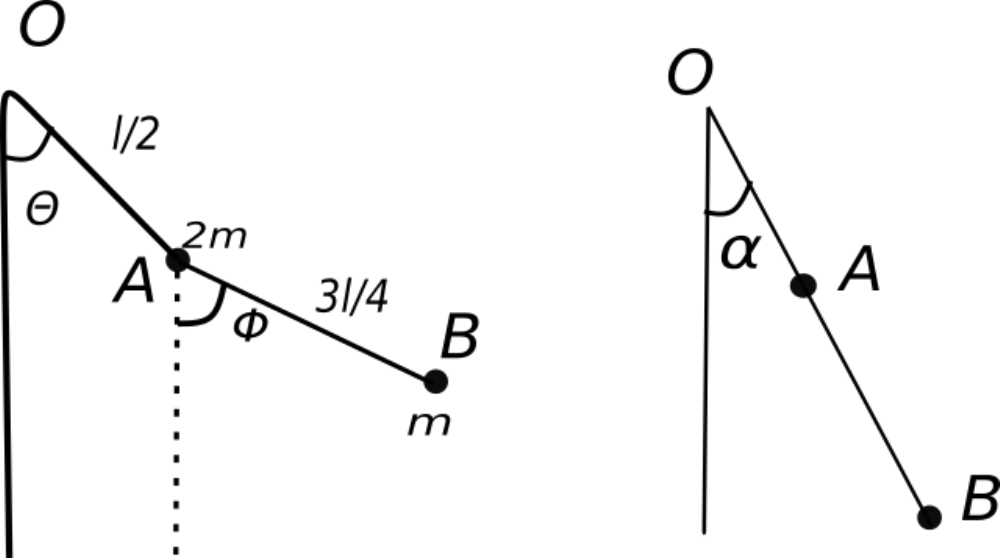}
\end{figure}

So The $L's$ equations are

\begin{eqnarray}
\frac{d}{dt}\left( \frac{\partial T}{\partial \dot{\theta}}\right) - \frac{\partial T}{\partial \theta} +\frac{\partial V}{\partial \theta} =0\nonumber \\
\frac{d}{dt}\left( \frac{\partial T}{\partial \dot{\phi}}\right) - \frac{\partial T}{\partial \phi} +\frac{\partial V}{\partial \phi} =0 \nonumber \\
\therefore \frac{d}{dt}\left[  2m   (\frac{1}{2}l)^2 \dot{\theta} +m(\frac{1}{2}l \dot{\theta}+\frac{3}{4}l \dot{\phi}) \frac{l}{2}\right]+\frac{3}{2}mg l \theta=0 \nonumber \\
or,~\frac{1}{2}l \ddot{\theta}+\frac{1}{4}l \ddot{\theta} +\frac{3}{8}l \ddot{\phi}+\frac{3}{2}g \theta=0\nonumber \\
or,~	\frac{3}{4}l \ddot{\theta} +\frac{3}{8}l \ddot{\phi}+\frac{3}{2}g \theta=0 \nonumber \\
\Rightarrow	\frac{1}{2}l \ddot{\theta} +\frac{1}{4}l \ddot{\phi}+g \theta=0
\end{eqnarray}
\begin{eqnarray}
\frac{d}{dt}\left[    m\frac{3}{4}l\left(\frac{1}{2}l\dot{\theta}+\frac{3}{4}l\dot{\phi}  \right)  \right] +mg \frac{3}{4}l  \phi=0 \nonumber \\
0r,~\frac{l\ddot{\theta}}{2} +\frac{3}{4} l\ddot{\phi} +g\phi=0
\end{eqnarray}

Let, $\theta=A_1e^{\lambda t}$, $\phi=A_2e^{\lambda t}$, as a solutions,
\begin{eqnarray}
\frac{1}{2}l\lambda^2A_1+ \frac{1}{4}l\lambda^2A_2 +gA_1=0\nonumber \\
or,~\left(\frac{1}{2}l\lambda^2 +g\right)A_1 +\frac{1}{4}l\lambda^2A_2=0  \label{4eq53} \\
\text{similarly},~~  \frac{1}{2} l \lambda^2 A_1 + \left( \frac{3}{4}l\lambda^2 +g \right)A_2=0
\end{eqnarray}
Eliminating $A_1 $ and $ A_2$
\begin{eqnarray}
\begin{bmatrix} 
\left(\frac{1}{2}l\lambda^2 +g\right) & \frac{1}{4}l\lambda^2 \\
\frac{1}{2} l \lambda^2  &  \left( \frac{3}{4}l\lambda^2 +g \right) \\
\end{bmatrix} =0 \nonumber \\
or,~ \frac{3}{8}l^2 \lambda^4 +\frac{3}{4}gl \lambda^2  +\frac{1}{2}lg \lambda^2 +g^2 - \frac{1}{8}l^2\lambda^4=0 \nonumber \\
or,~ \frac{1}{4}l^2\lambda^4+\frac{5}{4}gl \lambda^2+g^2=0 \nonumber \\
or,~\lambda^4+5 n^2\lambda^2+4 n^4=0 \nonumber \\
\Rightarrow  \lambda^2=-4n^2,-n^2  \nonumber \\
i.e., ~\sigma^2=4n^2,n^2 
\end{eqnarray}

So, the periods are $\dfrac{2\pi}{2n}, \dfrac{2\pi}{n}.$

 Equation (\ref{4eq53}) on simplification gives \\
$(2 \lambda^2+4n^2) A+\lambda^2 B=0$ \\
Putting,  $\lambda^2=-n^2$, $2A=B$  $\Rightarrow \frac{A}{1}=\frac{B}{2}$ \\
Putting,  $\lambda^2=-4n^2$, $A+B=0$  $\Rightarrow \frac{A}{1}=\frac{B}{-1}$ \\

$\therefore~ \theta=ccos(nt+\epsilon)+c'cos(2nt+\epsilon ')$\\
$~~~~~~~\phi=2ccos(nt+\epsilon)-c'cos(2nt+\epsilon ')$\\
Initially, when $t=0,$ $\theta=\phi=\alpha$  and  $\dot{\theta}=\dot{\phi}=0$

$\therefore~ \alpha=ccos\epsilon +c'cos\epsilon '$\\
$ \alpha=2c ~cos\epsilon - c'cos\epsilon '$ \\
$\Rightarrow 2 \alpha=3 c ~cos \epsilon$ \\
$\dot{\theta}=0 ~at ~t=0 \Rightarrow 0=nc~ sin \epsilon +2nc'~sin\epsilon'$ \\
$\dot{\phi}=0~\mbox{at}~t=0\implies 0=2nc\sin\epsilon-2nc'\sin\epsilon'$\\
$\therefore  ~sin \epsilon =0=sin \epsilon' ~i.e. ~\epsilon=\epsilon'=0$ \\
$\therefore~ c=\dfrac{2\alpha}{3}, ~\therefore~ c'=\alpha-\dfrac{2\alpha}{3}=\dfrac{\alpha}{3}$ \\
$\theta=\dfrac{2}{3}\alpha cos nt + \dfrac{1}{3}\alpha cos 2nt=\dfrac{1}{3}\alpha \left(  2 cos nt +cos 2nt \right) $\\
$\phi=\dfrac{1}{3}\alpha \left(   cos nt - cos 2nt \right)$\\
If we take $\xi $ and $\eta$ the normal co-ordinates then the eqs. of motion will be of the form \\
$\ddot{\xi}=-n^2 \xi, \ddot{\eta}=-4n^2\eta$  \\
The periods of these two oscillations will correspond to $\lambda^2=-n^2$ and ${\lambda}^2=-4n^2$. Even when expressed in terms of the normal co-ordinates. Thus the expressions for the normal co-ordinates will contain terms of the type $cos (nt+\epsilon)$  and $cos (2nt+\epsilon')$ constraint with the given system. Adding we get $\theta+\phi=3c~ cos(nt+\epsilon)=\xi$ and $2\theta-\phi=3c'cos(2nt+\epsilon')=\eta$

\vspace{.5cm}

$\bullet$ {\textbf{Problem:}} 
A circular  arc of  radius $a$ is fixed in a vertical plane and a uniform circular disc of mass $M$ and radius $a/4$ is placed inside so as to roll on the arc. When the disc is in the position of equilibrium, a particle of mass $M/3$ is fixed to it in the vertical diameter through the center at a distance $a/6$ from the centre. Show that the time of  small oscillation about the position of equilibrium is  $\dfrac{\pi}{6} \sqrt{\dfrac{83 a}{g}}$.

\vspace{.25cm}

{\textbf{Solution:}}  For rolling $\dfrac{a}{4}(\theta+\psi)=a\psi . ~i.e. ~ \psi=\theta/3$\\
\begin{eqnarray}
V&=&V_0 -Mg\frac{3}{4} a cos \psi - \frac{Mg}{3}\left(   \frac{3}{4}a cos \psi +\frac{a}{6} cos \theta \right) \nonumber \\
&=&V_0-Mga cos \psi -\frac{Mga }{18} cos \theta \nonumber \\
&=&V_0-Mga (1-\frac{\psi^2}{2})-\frac{Mga }{18}(1-\frac{\theta^2}{2}) \nonumber  \\
&=&V_0-Mga (1-\frac{\psi^2}{2})-\frac{Mga }{18}(1-\frac{9 \psi^2}{2}) \nonumber \\
&=&V_0-\frac{19 Mga }{18} +\frac{3Mga }{4} \psi^2 \nonumber
\end{eqnarray}

\vspace{0.7 cm}

\begin{figure}[h!]
	\centering
	\includegraphics[scale=0.33]{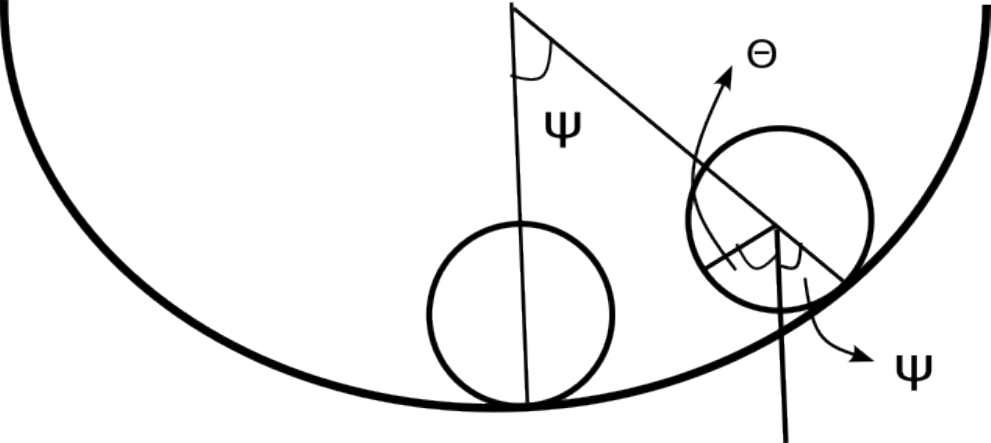}
\end{figure}

\vspace{0.5 cm}

K.E. of the plate =$\dfrac{1}{2 }M(\dfrac{3a}{4})^2\dot{\psi}^2 +\dfrac{1}{2}M\dfrac{a^2}{3^2}\dot{\theta}^2\\
~~~~~~~~~~~~~~~~~~~~=\dfrac{1}{2}M\dfrac{27}{32}a^2 \dot{\psi}^2$\\
Velocity of the particle $=\dfrac{3}{4}a \dot{\psi}$ along $||$ to the tangent to the arc upward $+\dfrac{1}{6}a\dot{\theta}$  along $||$ to the tangent to the plate downward.\\
$=\dfrac{3}{4}a \dot{\psi}-\dfrac{1}{6}a3\dot{\psi}=\dfrac{a}{4}\dot{\psi}$ (Neglecting the small angle between the two tangents.)\\
$\therefore $ K.E. of the particle $=\dfrac{1}{2}\dfrac{M}{3}\dfrac{a^2 \dot{\psi}^2}{16}$\\
So, the total K.E. $=\dfrac{1}{2}Ma^2\dot{\phi}\left(\dfrac{27}{32}+\dfrac{1}{48}\right)=\dfrac{1}{2}Ma^2\dot{\phi}\left(\dfrac{83}{96}\right)$\\
Thus the L's eq. of motion \\
$	\dfrac{d}{dt}\left( \dfrac{\partial T}{\partial \dot{\psi}}\right) - \dfrac{\partial T}{\partial \psi} +\dfrac{\partial V}{\partial \psi} =0 \Rightarrow Ma^2\dfrac{83}{96}\ddot{\psi}+\dfrac{3}{2}Mga\psi=0$ \\
$\therefore $ Time period $=\dfrac{2 \pi}{\sqrt{144 g/a}}=\dfrac{2\pi }{12}\sqrt{\dfrac{83a}{g}}=\dfrac{\pi}{6}\sqrt{\dfrac{83a}{g}}$

\vspace{.5cm}

$\bullet$ {\textbf{Problem:}}  Two heavy uniform rods AB and AC, each of mass $m$ and length $2a$ are hinged at A and placed symmetrically over a smooth cylinder of radius c whose axis is horizontal. If they are slightly and symmetrically displaced from the position of the equilibrium, show that the time of small oscillation is $2\pi \sqrt{\frac{a sin \alpha}{3g}\frac{(1+3sin^2\alpha)}{(1+2sin^2\alpha)}}$, where $a cos^3\alpha=c sin \alpha$.

{\textbf{Solution:}} AB,AC are two rods hinged at A and they are placed symmetrically over a circular cylinder whose axis is horizontal. Let $\theta$ be the angle which the rods make with the vertical through $O$. We take the centre of the cylinder as origin and x-axis horizontal in the plane of the rods and y-axis vertically downwards.

Let G be the C.G of the rod, so that  $AG=a$. If $(x,y)$ be the co-ordinates of G, then $x=a sin\theta,y=a cos\theta -c ~cosec\theta$. As the potential energy of the system is \\
$V=-2mgy+A$  (A is a constant)
so for the equilibrium position $\frac{dV}{d \theta}=0 \Rightarrow ~\frac{dy}{d \theta}=0  \Rightarrow ~-a sin \theta +c cosec \theta cot \theta=0, \therefore c cos \theta=a sin^3 \theta$.\\

\begin{figure}[h!]
	\centering
	\includegraphics[scale=0.33]{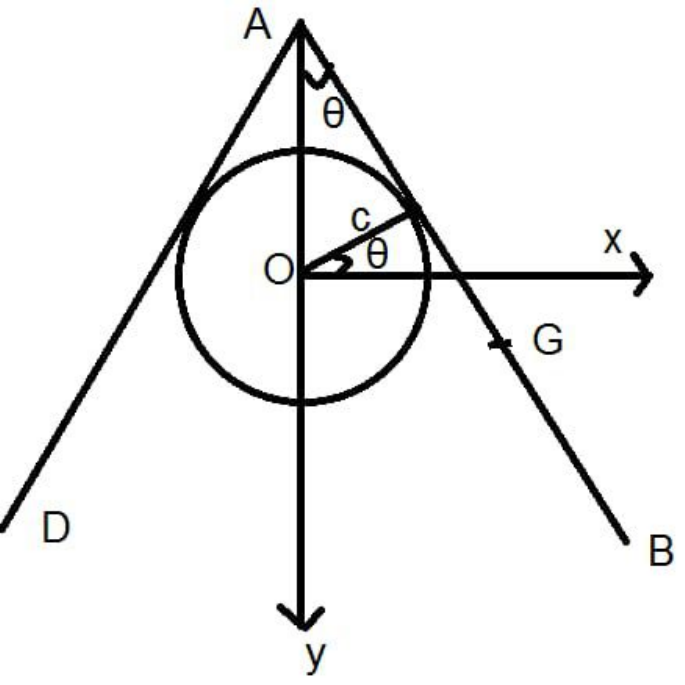}
\end{figure}

If $\theta=\alpha$ be the position of the equilibrium, then 
\begin{equation}
c~cos \alpha= a sin^3 \alpha\label{n33}
\end{equation}
Let us put, $\theta=\alpha+\xi$, where $\xi$ is very small. Now expanding $y$ near about the equilibrium position 

\begin{eqnarray}
y&=&y|_{\theta=\alpha}+\frac{\partial y}{\partial \theta}_{ | \theta=\alpha}\xi+{\frac{\partial  ^2 y}{\partial \theta ^2}}_{ | \theta=\alpha} \frac{\xi^2}{2!}+...\nonumber \\
\text{As} ~\frac{d^2 y}{d \theta ^2}&=&-a cos \theta -c cosec \theta~ cot^2 \theta -c cosec^3 \theta \nonumber \\
\therefore   \frac{d^2 y}{d \theta ^2}_{ | \theta=\alpha}&=&-a cos \alpha-c \frac{cos^2 \alpha}{sin^3 \alpha}- \frac{c}{sin^3 \alpha} \nonumber \\
&=&\left[ -a cos \alpha   sin^3 \alpha -c(1+cos^2\alpha) \right]/sin^3 \alpha\nonumber \\
&=&-\frac{a}{cos\alpha}(1+2cos^2\alpha) \nonumber \\
\therefore V&=& 2mg \frac{a}{cos\alpha}(1+2cos^2\alpha) \frac{\xi^2}{2} +A~\text{(Neglecting higher power of $\xi$)} \nonumber
\end{eqnarray}

As , $x=a sin \theta \Rightarrow \dot{x}=a cos \theta \dot{\theta}=acos(\alpha+\xi)\dot{\xi}=acos\alpha \dot{\xi}$ (neglecting 2nd order small quantities) and $\dot{y}=\frac{d y}{d \theta}\dot{\theta}=0$.

$\therefore$ K.E. of the system $T=\frac{1}{2}2m \left[   a^2 cos^2\alpha \dot{\xi}^2 +\frac{a^2}{3}\dot{\xi}^2  \right]$\\

The L's equation is  

\begin{eqnarray}
\frac{d}{dt}\left( \frac{\partial T}{\partial \dot{\xi}}\right) - \frac{\partial T}{\partial \xi} +\frac{\partial V}{\partial \xi} =0 \nonumber \\
\Rightarrow 2a^2 m(cos^2\alpha \ddot{\xi}+\frac{1}{3}\ddot{\xi}) +2mg a \frac{(1+2cos^2 \alpha)}{cos \alpha} \xi=0 \nonumber \\
\Rightarrow \ddot{\xi} =-\frac{g(1+2 cos^2 \alpha)}{a cos \alpha (\frac{1}{3}+cos^2 \alpha)}\xi \nonumber \\
T=2\pi \sqrt{\frac{a cos \alpha}{3g }\frac{(1+3 cos^2\alpha)}{(1+2cos^2 \alpha)}}\nonumber
\end{eqnarray}
$\bullet$ {\textbf{Problem IV:}} (small oscillation)
A rhombus formed of four equal rods freely joint is placed over a fixed smooth sphere in a vertical plane so that only the upper pair is in  contact with the sphere. Show that the time of symmetrical oscillation in the vertical plane is in contact with the sphere. Show that the time of symmetrical oscillation in the vertical plane is 
$2\pi\sqrt{\frac{2a cos \alpha}{3g (1+2cos^2\alpha)}}$ where $2 a$ is the length of each rod and $\alpha$ is the angle it makes with the vertical in the position of the equilibrium.

$\bullet$ {\textbf{Solution:}}  We choose the center of the sphere $O$ as origin and $x$ and $y$ axes as shown in the figure. The depth of C.G.  G below O is
\begin{eqnarray}
y=2a cos \theta -c cosec \theta \nonumber \\
\therefore \frac{dy}{d\theta}=-2a sin \theta +c cosec \theta cot \theta \nonumber \\
\therefore \frac{dy}{d\theta}=0~ at ~\theta=\alpha  ~\Rightarrow \frac{2a}{cos \alpha}=\frac{c}{sin^3 \alpha} \label{n34} \\
\therefore \theta=\alpha+\xi  ~~\text{(where $\xi$ is very small)}\nonumber \\
\text{K.E.}=\frac{1}{2}4m \frac{4a^2}{3}\dot{\xi}^2 \nonumber 
\end{eqnarray}

\begin{figure}[h!]
	\centering
	\includegraphics[scale=0.4]{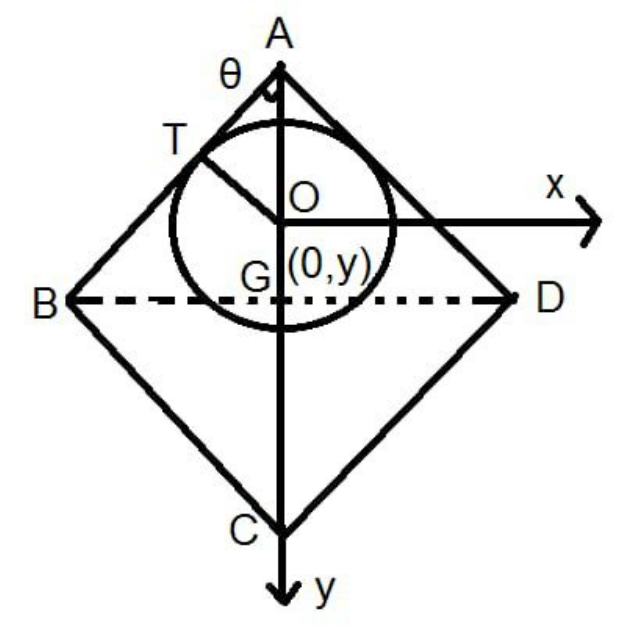}
\end{figure}

\begin{eqnarray}
V&=&-4mg y \nonumber \\
&=& -4mg \left[    y|_{\theta=\alpha}  +{\frac{dy}{d \theta}}|_{\theta=\alpha} \xi  +{\frac{d^2y}{d \theta^2}}|_{\theta=\alpha} \frac{\xi ^2}{2}+...\right]\label{n35}  \\
&=&-4mg \left[   (2a cos \alpha -\frac{c}{sin \alpha} ) +0+(-2a cos \theta-c cosec \theta cot^2 \theta -c cosec^3 \theta)_{|\theta=\alpha} \frac{\xi ^2}{2}  \right]  \nonumber \\
&=& -4mg \left[   (2a cos \alpha -\frac{c}{sin \alpha} ) - \frac{2a}{cos \alpha}(1+2 cos^2 \alpha)\frac{\xi ^2}{2} \right] \nonumber
\end{eqnarray}

So the Lagranges equation becomes \\

$\frac{d}{dt}\left( \frac{\partial T}{\partial \dot{\xi}}\right) - \frac{\partial T}{\partial \xi} +\frac{\partial V}{\partial \xi} =0$ \\
$\mbox{or}, ~ 4 \frac{4m}{3}a^2 \ddot{\xi}+4mg \frac{2a}{cos \alpha}(1+2cos^2\alpha)\xi=0$ \\
$\therefore T=2\pi\sqrt{\frac{2a cos \alpha}{3g (1+2cos^2\alpha)}}$

\vspace{.5cm}

$\bullet$ {\textbf{Problem V:}}  Two uniform rods of same mass and of same length $2a$ are freely jointed at a common extremity and rest upon two smooth pegs which are in the same horizontal plane so that  each rod is inclined at same angle $\alpha$ to the vertical. Show that the time of small oscillation when the join moves in a vertical  st. line through the centre of the line joining the pegs is $2\pi\sqrt{\frac{a}{9g}(\frac{1+3 cos^2 \alpha}{cos \alpha})}$.

\vspace{.25cm}

{\textbf{Solution:}}  Let $2c $ be the horizontal distance between the pegs. Take the mid-pt. of the line joining the pegs as the fixed origin and x-axis is along horizontal direction and y-axis is vertically downwards.

\begin{eqnarray}
\therefore y&=&a cos \theta -c cot \theta   \nonumber  \\
\frac{d y}{d \theta}&=&-a sin \theta +c cosec^2\theta \nonumber \\
\frac{d^2 y}{d \theta^2}&=&-a cos \theta -2c cosec^2\theta cot \theta\nonumber
\end{eqnarray}

At equilibrium position $\theta=\alpha,~  \frac{d y}{d \theta}=0$\\
$\therefore ~ a sin^3\alpha=c$\\

\begin{figure}[h!]
	\centering
	\includegraphics[scale=0.4]{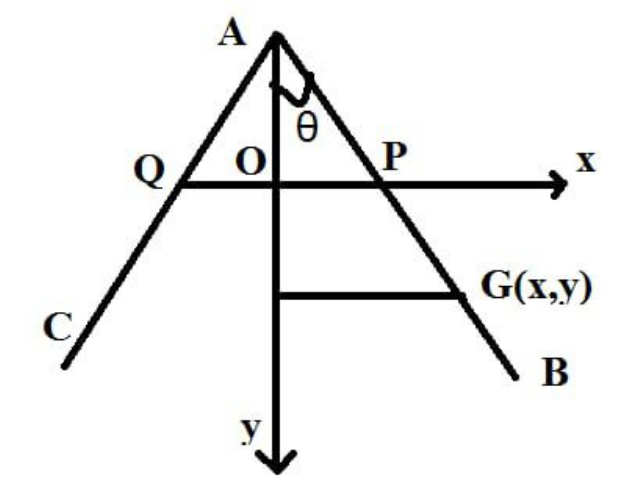}
\end{figure}

Let us put $\theta=\alpha+\xi$.

\begin{eqnarray}
\therefore~ y&=&\left[    y|_{\theta=\alpha}  +\frac{dy}{d \theta}_{|\theta=\alpha} \xi  +\frac{d^2y}{d \theta^2}_{|\theta=\alpha} \frac{\xi ^2}{2} +... \right] \nonumber \\
&=&A+\left[-a cos \alpha -2 c~cosec^2 cot \alpha \right]\frac{\xi ^2}{2} \nonumber \\
&=&A-\frac{3}{2}\xi^2 a~cos \alpha\nonumber
\end{eqnarray}

$\therefore V=V_0 +3 mg a ~cos \alpha ~\xi^2$,\\
 $x=a sin \theta,~~ \dfrac{dx}{dt}=a cos \theta \dot{\theta}=a cos \alpha ~\dot{\xi}$ \\
$\therefore T=\frac{1}{2}2 m \left[a^2 cos^2 \alpha \dot{\xi}^2  + \frac{1}{3} a^2  \dot{\xi}^2 \right]$  \\
$=\frac{ma^2}{3}(1+3 cos^2 \alpha)\dot{\xi}^2$

So, by L's eqs. we have 

\begin{eqnarray}
\frac{d}{dt}\left( \frac{\partial T}{\partial \dot{\xi}}\right) - \frac{\partial T}{\partial \xi} +\frac{\partial V}{\partial \xi} =0  \nonumber \\
or, ~~\frac{2ma^2}{3}(1+3 cos^2 \alpha)\ddot{\xi}+3mg ~2a~cos \alpha ~\xi=0  \nonumber \\
\therefore\ddot{\xi}=-\frac{9g cos \alpha}{a(1+3 cos^2 \alpha)}\xi    \nonumber \\
\therefore~ ~T=2\pi \sqrt{\frac{a(1+3 cos^2\alpha)}{9g ~cos \alpha}} \nonumber
\end{eqnarray}

$\bullet$ {\textbf{Problem VI:}} 
A uniform beam rests with one end on a smooth horizontal plane and the other end is supported by a string of length $l$ which is attached to a fixed pt. show that the time of small oscillation in the vertical plane is $2\pi\sqrt{\frac{2l}{g}}$.

{\textbf{Solution:}}
From the figure
\begin{eqnarray}
x&=&l sin \theta +a cos(\phi_0+\phi) \label{n36}\\
y&=&a sin(\phi_0+\phi)\label{n37} \\
h&=&CO=l cos \theta +2a sin(\phi_0+\phi) \label{n38}
\end{eqnarray}

\begin{figure}[h!]
	\centering
	\includegraphics[scale=0.4]{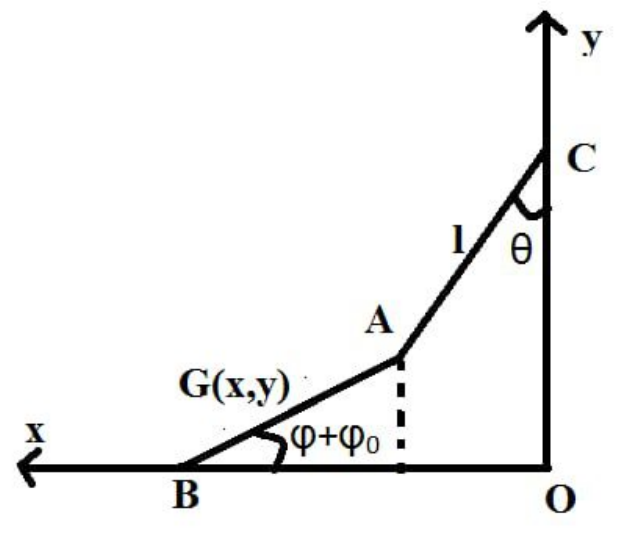}
\end{figure}

From equation (\ref{n38}), $O=dh=-lsin \theta d\theta + 2 a cos(\phi+\phi_0) d\phi$ \\
$\therefore 2 a ~cos\phi_0 d\phi=l \theta~ d\theta$ (for small $\theta $ and $\phi$)\\
which shows that $d\phi$ is a small quantity of 2nd order. 
\begin{eqnarray}
\dot{x}&=&l cos \theta \dot{\theta} - a sin(\phi_0+\phi) \dot{\phi} \approx l \dot{\theta} ~(\text{up to 1st order} )\nonumber \\
\dot{y}&=&a cos(\phi+\phi_0)  \dot{\phi} \approx 0~ (\text{up to 1st order})\nonumber \\
\therefore \text{P.E.} ~~ V&=&mgy=\frac{mg}{2}(h-l cos \theta) =\frac{mg}{2}\left(h-l(1-\frac{\theta^2}{2})\right) \nonumber \\
&=& V_0 +\frac{mgl}{4} \theta^2  \nonumber \\
\text{K.E.} ~~(T)&=&\frac{1}{2}m(\dot{x}^2+\dot{y}^2) + \frac{1}{2}m \frac{a^2}{3}\dot{\phi}^2 \nonumber \\
&=&\frac{1}{2}ml^2 \dot{\theta}^2 ~~(\text{up to 2nd  order})\nonumber
\end{eqnarray}

So by L's equation\\

$\frac{d}{dt}\left( \frac{\partial T}{\partial \dot{\theta}}\right) - \frac{\partial T}{\partial \theta} +\frac{\partial V}{\partial \theta} =0$\\
$\Rightarrow ml^2 \ddot{\theta} +\frac{mgl}{2}\theta=0~\Rightarrow T=2\pi\sqrt{\frac{2l}{g}}$

$\bullet$ {\textbf{Problem VII:}} 
Two equal uniform rods AB and BC each of length $l$ smoothly joined together at B are suspended from A and oscillates in a vertical plane through A. Show that the periods of normal oscillation are  $\frac{2\pi}{n},n^2=(3  \pm \frac{6}{\sqrt{7}})g/l$.

{\textbf{Solution:}}

\begin{figure}[h!]
	\centering
	\includegraphics[scale=0.4]{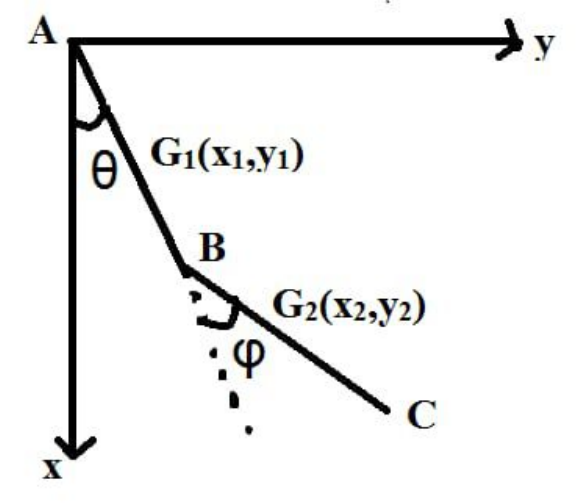}
\end{figure}

\begin{eqnarray}
x_1 &=& \frac{l}{2}cos \theta =  \frac{l}{2}(1-\theta^2/2)\nonumber \\
y_1 &=&  \frac{l}{2} sin\theta= \frac{l}{2} \theta   \nonumber \\
x_2 &=& l cos \theta + \frac{l}{2} cos \phi  \nonumber \\
&=&l(1-\theta^2/2) +\frac{l}{2}(1-\phi^2/2) \nonumber \\
y_2 &=& l sin \theta + \frac{l}{2} sin \phi=l\theta+\frac{l\phi}{2}  \nonumber \\
T&=&\frac{1}{2}m\left[\dot{x_1}^2+\dot{y_1}^2+\dot{x_2}^2+\dot{y_2}^2\right] +\frac{1}{2}m \left(\frac{l^2}{12} \dot{\theta}^2+\frac{l^2}{12} \dot{\phi}^2\right) \nonumber \\
&=&  \frac{m}{2}\left[\frac{l^2}{4} sin^2 \theta \dot{\theta}^2+\frac{l^2}{4} cos^2 \theta \dot{\theta}^2 +(lsin \theta \dot{\theta }+\frac{1}{2} sin \phi \dot{\phi})^2+\left(l\cos\theta\dot{\theta}+\frac{l}{2}\cos\phi\dot{\phi}\right)^2\right]+\frac{ml^2}{24}\left(\dot{\theta}^2+\dot{\phi}^2\right) \nonumber \\
&=& \frac{m}{2}\left[\frac{l^2}{4}  \dot{\theta}^2+{l^2} \dot{\theta}^2 +\frac{l^2}{4} \dot{\phi}^2+ l^2 \dot{\theta} \dot{\phi}
\right] + \frac{m l^2}{24}( \dot{\theta}^2 +  \dot{\phi}^2)\nonumber \\ 
&=& \frac{ml^2}{2}\left(\frac{16 \dot{\theta}^2 +4 \dot{\phi}^2 +12 \dot{\theta} \dot{\phi}}{12}\right)\nonumber \\
&=&\frac{1}{6}ml^2 (4\dot{\theta}^2+\dot{\phi}^2 + 3 \dot{\theta} \dot{\phi})\nonumber
\end{eqnarray} 

\begin{eqnarray}
V &=&-mg \frac{l}{2}\left(1-\frac{\dot{\theta}^2}{2}\right)-mg \left[ l (1-\frac{ \theta^2}{2}) +\frac{l}{2}\left(1-\frac{\phi^2}{2}\right)   \right] \nonumber \\
&=& V_0 + mgl \left( \frac{3 \theta^2}{4} +\frac{\phi^2}{4}  \right)\nonumber
\end{eqnarray}
So, the L's equations are 
\begin{eqnarray}\nonumber
\frac{d}{dt}\left( \frac{\partial T}{\partial \dot{\theta}}\right) - \frac{\partial T}{\partial \theta} +\frac{\partial V}{\partial \theta} =0  \end{eqnarray}

\begin{eqnarray}\label{4eqn65}
\frac{d}{dt}\left( \frac{\partial T}{\partial \dot{\phi}}\right) - \frac{\partial T}{\partial \phi} +\frac{\partial V}{\partial \phi} =0  
\end{eqnarray}

From Eqn. (\ref{4eqn65}), 
\begin{eqnarray}
\frac{1}{6}m l^2(8\ddot{\theta}+3 \ddot{\phi}) +\frac{3}{2} mgl \theta=0 \nonumber \\
\Rightarrow~~ 8\ddot{\theta}+3 \ddot{\phi} +\frac{9 g}{l} \theta =0
\end{eqnarray}

\begin{eqnarray}
\frac{1}{6}m l^2(2 \ddot{\phi} + 3\ddot{\theta}) +\frac{1}{2} mgl \phi =0 \nonumber \\
\Rightarrow~~ 2 \ddot{\phi} + 3\ddot{\theta}+ +\frac{3 g}{l} \phi =0 
\end{eqnarray}
Let $\theta=\theta_0e^{i\lambda t}$, $\phi=\phi_0 e^{i\lambda t}$

$\therefore (-8 \lambda^2 +\frac{9 g}{l} )\theta_0 - 3 \phi_0\lambda^2=0$ \\
$ \Rightarrow -3 \lambda^2 \theta_0 +(-2 \lambda^2 + \frac{3 g}{l})\phi_0=0$\\
Eliminating, $\theta_0$ and $\phi_0$, we have

\begin{eqnarray}
\begin{bmatrix} 
-8 \lambda^2 +\frac{9 g}{l} & - 3 \lambda^2 \\
- 3 \lambda^2  & - 2 \lambda^2 +\frac{3 g}{l}  \\
\end{bmatrix} &=&0 \nonumber \\
or, ~ 7\lambda^4 -42 \frac{42 g}{l}\lambda^2 +27 \frac{g^2}{l^2}&=&0 \nonumber \\
or, \lambda^2 = \frac{42 g/l \pm \sqrt{42^2 g^2/l^2 -4.7.27 g^2/l^2}}{2.7} \nonumber \\
=\frac{42 \frac{g}{l} \pm \sqrt{1008}\frac{g}{l}}{14}\nonumber
\end{eqnarray}
Hence the time periods $ =(3\pm \frac{6}{\sqrt{7}})g/l$.

\vspace{.5cm}

\section{Eulerian angles $\theta,~\phi$ and $\psi$: }

Consider a material system composed of $N$ particles. Let the material system rotates about a point O. The point O is fixed with respect to the material system as well as in the space. Suppose $ox,~oy,~oy$ be a rectangular coordinate system through O. These axes are fixed with respect to the material system. Let $\overrightarrow{i},~\overrightarrow{j},~\overrightarrow{k}$ be the unit vectors along the coordinate axes. Then the position of the rotating axes as well as the components of angular velocity with respect to the rotating axes can be expressed by the three angles $\theta,~\phi$ and $\psi$ and their derivatives. These angles are known as Eulerian angles. To define these angles we proceed as follows :

Since the end point of any unit vector having initial point O lies on the surface of a sphere of radius unity with centre at O. We restrict our discussion on the surface of this unit sphere (having center at O). Let $\overrightarrow{I},~\overrightarrow{J},~\overrightarrow{K}$ be the initial position of $\overrightarrow{i},~\overrightarrow{j}$ and $\overrightarrow{k}$ respectively. Let $\theta$ be the angle between $\overrightarrow{K}$ and $\overrightarrow{k}$. Let the grate circle through the terminal points of $\overrightarrow{K}$ and $\overrightarrow{k}$ meets the grate circle through the terminal points of $\overrightarrow{I}$ and $\overrightarrow{J}$ at the points A and B as shown in the figure. We denote the unit vector $\overrightarrow{OA}$ by $\overrightarrow{I}'$. Let $\phi$ be the angle between $\overrightarrow{I}$ and $\overrightarrow{I}'$. Let us consider the rotation of $\overrightarrow{I}$ and $\overrightarrow{J}$ through an angle $\phi$ in an anticlockwise sense in the plane of the grate circle through the terminal points of $\overrightarrow{I}$ and $\overrightarrow{J}$. Therefore, $\overrightarrow{I}'$ is the new position of $\overrightarrow{I}$. Therefore, $\overrightarrow{I}'$ is the new position of $\overrightarrow{I}$. Let $\overrightarrow{J}'$ be the new position of $\overrightarrow{J}$ and consequently, we get
\begin{equation}
	(\overrightarrow{I},~\overrightarrow{J},~\overrightarrow{K})\xrightarrow{\mbox{due to rotation}~\phi\overrightarrow{K}}(\overrightarrow{I}',~\overrightarrow{J}',~\overrightarrow{K})\nonumber
\end{equation} 
Next consider the rotation of $\overrightarrow{I}'$ and $\overrightarrow{K}$ through on angle $\theta$ in an anticlockwise sense in the plane of the grate circle through the terminal points of $\overrightarrow{I}'$ and $\overrightarrow{K}$. Therefore $\overrightarrow{I}''$ is the new position of $\overrightarrow{I}'$ and $\overrightarrow{k}$ be the new position of $\overrightarrow{K}$  and consequently we get  
\begin{equation}
	(\overrightarrow{I}',~\overrightarrow{J}',~\overrightarrow{K})\xrightarrow{\mbox{due to rotation}~\theta\overrightarrow{J}'}(\overrightarrow{I}'',~\overrightarrow{J}',~\overrightarrow{k})\nonumber
\end{equation} 
Finally, consider the rotation of $\overrightarrow{I}''$ and $\overrightarrow{J}'$ through on angle $\psi$ in an anticlockwise sense in the plane of the grate circle through the terminal points of $\overrightarrow{I}''$ and $\overrightarrow{J}'$. Therefore $\overrightarrow{i}$ is the new position of $\overrightarrow{I}'',~\overrightarrow{j}$ is the new position of $\overrightarrow{J}'$ and consequently we get
\begin{equation}
	(\overrightarrow{I}'',~\overrightarrow{J}',~\overrightarrow{k})\xrightarrow{\mbox{due to rotation}~\psi\overrightarrow{k}}(\overrightarrow{i},~\overrightarrow{j},~\overrightarrow{k})\nonumber
\end{equation} 
therefore, if $\overrightarrow{n}$ denotes the total rotation of $\overrightarrow{i},~\overrightarrow{j},~\overrightarrow{k}$ from the initial position $(\overrightarrow{I},~\overrightarrow{J},~\overrightarrow{K})$ then 
\begin{equation}
	\overrightarrow{n}=\phi\overrightarrow{K}+\theta\overrightarrow{J}'+\psi\overrightarrow{K}\nonumber
\end{equation}
So for infinitesimal rotation $\delta\overrightarrow{n}$ during the time interval $(t,t+\delta{t})$ we have 
\begin{equation}
	\delta\overrightarrow{n}=\delta\phi\overrightarrow{K}+\delta\theta\overrightarrow{J}'+\delta\psi\overrightarrow{K}\nonumber
\end{equation}
If $\overrightarrow{\omega}$ be the angular velocity of the material system about O then 
\begin{equation}
	\overrightarrow{\omega}=\lim_{\delta{t}\rightarrow{0}}\frac{\delta\overrightarrow{n}}{\delta{t}}=\dot{\phi}\overrightarrow{K}+\dot{\theta}\overrightarrow{J}'+\dot{\psi}\overrightarrow{K}\nonumber
\end{equation}
For the rotation : $\phi\overrightarrow{K}$ we have 
\begin{eqnarray}
	\overrightarrow{I}'=\overrightarrow{I}\cos\phi+\overrightarrow{J}\sin\phi,~\overrightarrow{J}'&=&\overrightarrow{I}\cos\left(\frac{\pi}{2}+\phi\right)+\overrightarrow{J}\sin\left(\frac{\pi}{2}+\phi\right)\nonumber\\
	&=&-\overrightarrow{I}\sin\phi+\overrightarrow{J}\cos\phi\nonumber
\end{eqnarray}
For the rotation : $\theta\overrightarrow{J}'$ we have 
\begin{eqnarray}
	\overrightarrow{k}=\overrightarrow{K}\cos\theta+\overrightarrow{I}'\sin\theta,~\overrightarrow{I}''&=&\overrightarrow{K}\cos\left(\frac{\pi}{2}+\theta\right)+\overrightarrow{I}'\sin\left(\frac{\pi}{2}+\theta\right)\nonumber\\
	&=&-\overrightarrow{K}\sin\theta+\overrightarrow{I}'\cos\theta\nonumber
\end{eqnarray}
For the rotation : $\psi\overrightarrow{k}$ we have 
\begin{eqnarray}
	\overrightarrow{i}=\overrightarrow{I}''\cos\psi+\overrightarrow{J}'\sin\psi,~\overrightarrow{j}&=&\overrightarrow{I}''\cos\left(\frac{\pi}{2}+\psi\right)+\overrightarrow{J}'\sin\left(\frac{\pi}{2}+\psi\right)\nonumber\\
	&=&-\overrightarrow{I}''\sin\psi+\overrightarrow{J}'\cos\psi\nonumber
\end{eqnarray}
Also we can write the above three rotations in the matrix form as 
\begin{eqnarray}
	\begin{pmatrix}
		\overrightarrow{I}'\\ \overrightarrow{J}'
	\end{pmatrix}
	&=&
	\begin{pmatrix}
		\cos\phi & \sin\phi\\ -\sin\phi & \cos\phi
	\end{pmatrix}
	\begin{pmatrix}
		\overrightarrow{I}\\ \overrightarrow{J}
	\end{pmatrix}\nonumber\\
	\begin{pmatrix}
		\overrightarrow{k}\\ \overrightarrow{I}''
	\end{pmatrix}
	&=&
	\begin{pmatrix}
		\cos\theta & \sin\theta\\ -\sin\theta & \cos\theta
	\end{pmatrix}
	\begin{pmatrix}
		\overrightarrow{K}\\ \overrightarrow{I}'
	\end{pmatrix}\nonumber\\
	\begin{pmatrix}
		\overrightarrow{i}\\ \overrightarrow{j}
	\end{pmatrix}
	&=&
	\begin{pmatrix}
		\cos\psi & \sin\psi\\ -\sin\psi & \cos\psi
	\end{pmatrix}
	\begin{pmatrix}
		\overrightarrow{I}''\\ \overrightarrow{J}'
	\end{pmatrix}\nonumber
\end{eqnarray}
As $\begin{pmatrix}
	\cos\alpha & \sin\alpha\\ -\sin\alpha & \cos\alpha
\end{pmatrix}\begin{pmatrix}
	\cos\alpha & -\sin\alpha\\ \sin\alpha & \cos\alpha
\end{pmatrix}=\begin{pmatrix}
	\cos\alpha & -\sin\alpha\\ \sin\alpha & \cos\alpha
\end{pmatrix}\begin{pmatrix}
	\cos\alpha & \sin\alpha\\ -\sin\alpha & \cos\alpha
\end{pmatrix}=\begin{pmatrix}
	1 & 0\\ 0 & 1
\end{pmatrix}$

i.e., $\begin{pmatrix}
	\cos\alpha & -\sin\alpha\\ \sin\alpha & \cos\alpha
\end{pmatrix}$ is the inverse of  $\begin{pmatrix}
	\cos\alpha & \sin\alpha\\ -\sin\alpha & \cos\alpha
\end{pmatrix}$. So we can write 
\begin{eqnarray}
	\begin{pmatrix}
		\cos\phi & -\sin\phi\\ \sin\phi & \cos\phi
	\end{pmatrix}
	\begin{pmatrix}
		\overrightarrow{I}'\\ \overrightarrow{J}'
	\end{pmatrix}
	=
	\begin{pmatrix}
		\overrightarrow{I}\\ \overrightarrow{J}
	\end{pmatrix}
	\implies
	\begin{matrix}
		\overrightarrow{I}=\overrightarrow{I}'\cos\phi-\overrightarrow{J}'\sin\phi\nonumber\\\overrightarrow{J}=\overrightarrow{I}'\sin\phi+\overrightarrow{J}'\cos\phi
	\end{matrix}\nonumber
\end{eqnarray}
\begin{eqnarray}
	\begin{pmatrix}
		\cos\theta & -\sin\theta\\ \sin\theta & \cos\theta
	\end{pmatrix}
	\begin{pmatrix}
		\overrightarrow{k}\\ \overrightarrow{I}''
	\end{pmatrix}
	=
	\begin{pmatrix}
		\overrightarrow{K}\\ \overrightarrow{I}'
	\end{pmatrix}
	\implies
	\begin{matrix}
		\overrightarrow{K}=\overrightarrow{k}\cos\theta-\overrightarrow{I}''\sin\theta\nonumber\\\overrightarrow{I}'=\overrightarrow{k}\sin\theta+\overrightarrow{I}''\cos\theta
	\end{matrix}\nonumber
\end{eqnarray}
\begin{eqnarray}
	\begin{pmatrix}
		\cos\psi & -\sin\psi\\ \sin\psi & \cos\psi
	\end{pmatrix}
	\begin{pmatrix}
		\overrightarrow{i}\\ \overrightarrow{j}
	\end{pmatrix}
	=
	\begin{pmatrix}
		\overrightarrow{I}''\\ \overrightarrow{J}'
	\end{pmatrix}
	\implies
	\begin{matrix}
		\overrightarrow{I}''=\overrightarrow{i}\cos\psi-\overrightarrow{j}\sin\psi\nonumber\\\overrightarrow{J}'=\overrightarrow{i}\sin\psi+\overrightarrow{j}\cos\psi
	\end{matrix}\nonumber
\end{eqnarray}
\begin{eqnarray}
	\therefore\overrightarrow{\omega}&=&\dot{\phi}\overrightarrow{K}+\dot{\theta}\overrightarrow{J}'+\dot{\psi}\overrightarrow{k}\nonumber\\
	&=&\dot{\phi}\left\{\overrightarrow{k}\cos\theta-(\overrightarrow{i}\cos\psi-\overrightarrow{j}\sin\psi)\sin\theta\right\}+\dot{\theta}\left\{\sin\psi\overrightarrow{i}+\cos\psi\overrightarrow{j}\right\}+\dot{\psi}\overrightarrow{k}\nonumber\\
	&=&\overrightarrow{i}\left\{\dot{\theta}\sin\psi-\dot{\phi}\cos\psi\sin\theta\right\}+\overrightarrow{j}\left\{\dot{\theta}\cos\psi+\dot{\phi}\sin\psi\sin\theta\right\}+\overrightarrow{k}\left\{\dot{\psi}+\dot{\phi}\cos\theta\right\}\nonumber
\end{eqnarray}
$\therefore\omega_1=\dot{\theta}\sin\psi-\dot{\phi}\cos\psi\sin\theta,~\omega_2=\dot{\theta}\cos\psi+\dot{\phi}\sin\psi\sin\theta,~\mbox{and}~\omega_3=\dot{\psi}+\dot{\phi}\cos\theta$

\vspace{.5cm}

\section{Motion of a symmetrical Top: }

A symmetrical top is a rigid body which is a solid of revolution about its axis of symmetry. Let the top is rotating about a point O on its axis of symmetry $oz$. We take $oz$ as the $z$-axis fixed with respect top. Choose $ox,~oy$ such that $ox,~oy$ and $oz$ forms a rectangular cartesian coordinate system at O such that they are fixed with respect to the top and assume them as the principle axes of inertia of the rigid body at O. Let, $A,~B,~C$ be the moment of inertia of the top with respect to $ox,~oy,~oz$ respectively. As the rigid body (Top) is symmetry about $oz$ so we have $A=B$. Let $\overrightarrow{i},~\overrightarrow{j}$ and $\overrightarrow{k}$ be the unit vectors along the three axes. Then the position of the rotating axes $ox,~oy,~oz$ at any time $t$ can be expressed with respect to the three Eulerian angles $\theta,~\phi$ and $\psi$. Also the components of angular velocity with respect to the rotating axes can be expressed with respect to the Eulerian angles and their derivatives $~\omega_1=\dot{\theta}\sin\psi-\dot{\phi}\cos\psi\sin\theta,~\omega_2=\dot{\theta}\cos\psi+\dot{\phi}\sin\psi\sin\theta,~\omega_3=\dot{\psi}+\dot{\phi}\cos\theta$.

As the rigid body is rotating about the point O with an an angular velocity $\omega$, its kinetic energy $T$ is given by 
\begin{eqnarray}
	T&=&\frac{1}{2}\left[A\omega_1^2+B\omega_2^2+C\omega_3^2\right]\nonumber\\
	&=&\frac{1}{2}\left[A\left(\dot{\theta}\sin\psi-\dot{\phi}\cos\psi\sin\theta\right)^2+A\left(\dot{\theta}\cos\psi+\dot{\phi}\sin\psi\sin\theta\right)^2+C\left(\dot{\psi}+\dot{\phi}\cos\theta\right)^2\right]\nonumber\\
	&=&\frac{1}{2}A\left(\dot{\theta}^2+\dot{\phi}^2\sin^2\theta\right)+\frac{1}{2}C\left(\dot{\psi}+\dot{\phi}\cos\theta\right)^2\nonumber
\end{eqnarray}
Let, $G$ be the centroid of the top and $OG=h$, then the potential energy of the top referred to the horizontal through O is given by $V=Mgh\cos\theta$, $M$-the mass of the top.

So the Lagrangian of the present system is 
\begin{equation}
	L=T-V=\frac{1}{2}A\left(\dot{\theta}^2+\dot{\phi}^2\sin^2\theta\right)+\frac{1}{2}C\left(\dot{\psi}+\dot{\phi}\cos\theta\right)^2-Mgh\cos\theta\nonumber
\end{equation}
Now, taking $\theta,~\phi$ and $\psi$ as the generalised coordinates, the Lagrange equation of motion can be written as 
\begin{equation}
	\frac{d}{dt}\left(\frac{\partial{L}}{\partial\dot{\theta}}\right)-\frac{\partial{L}}{\partial{\theta}}=0,~\frac{d}{dt}\left(\frac{\partial{L}}{\partial\dot{\phi}}\right)-\frac{\partial{L}}{\partial{\phi}}=0,~\frac{d}{dt}\left(\frac{\partial{L}}{\partial\dot{\phi}}\right)-\frac{\partial{L}}{\partial{\phi}}=0\nonumber
\end{equation}
\begin{eqnarray}
	\implies~A\ddot{\theta}-A\dot{\phi}^2\sin\theta\cos\theta+{C}\dot{\phi}\left(\dot{\psi}+\dot{\phi}\cos\theta\right)\sin\theta-Mgh\sin\theta&=&0\label{g1}
\end{eqnarray}
\begin{eqnarray}
	\mbox{and,~}A\frac{d}{dt}\left[\dot{\phi}\sin^2\theta\right]+C\frac{d}{dt}\left[\left(\dot{\psi}+\dot{\phi}\cos\theta\right)\cos\theta\right]&=&0\nonumber\\
	\implies~\frac{d}{dt}\left[A\dot{\phi}\sin^2\theta+C\left(\dot{\psi}+\dot{\phi}\cos\theta\right)\cos\theta\right]&=&0\label{g2}\\
	\mbox{and,~}\frac{d}{dt}\left[C\left(\dot{\psi}+\dot{\phi}\cos\theta\right)\right]&=&0\nonumber\\
	\mbox{i.e.,~}\left(\dot{\psi}+\dot{\phi}\cos\theta\right)=\mbox{Constant}&=&n(\mbox{say})\label{g3}
\end{eqnarray}
from (\ref{g2}) and (\ref{g3}) 
\begin{equation}
	A\dot{\phi}\sin^2\theta+Cn\cos\theta=\mbox{Constant}=D(\mbox{say})\label{g4}
\end{equation}
Using (\ref{g3}) and (\ref{g4}) in (\ref{g1}) we get
\begin{eqnarray}
	A\ddot{\theta}-A\dot{\phi}^2\sin\theta\cos\theta+{C}\dot{\phi}n\sin\theta-Mgh\sin\theta&=&0 \nonumber \\
	\label{g5} 
\end{eqnarray}
\begin{eqnarray}
	&~&A\ddot{\theta}-A\left[\frac{D-Cn\cos\theta}{A\sin^2\theta}\right]^2\sin\theta\cos\theta+{C}n\left[\frac{D-Cn\cos\theta}{A\sin^2\theta}\right]\sin\theta-Mgh\sin\theta=0\nonumber\\
	&~&\implies~A\ddot{\theta}+f(\theta)=0 \nonumber \\
	\label{g6}
\end{eqnarray}
where
\begin{eqnarray}
	f(\theta)&=&-\frac{1}{A}\frac{(D-Cn\cos\theta)^2}{\sin^3\theta}\cos\theta+\frac{Cn}{A}\frac{(D-Cn\cos\theta)}{\sin\theta}-Mgh\sin\theta\nonumber\\
	\mbox{i.e.,}A\sin^3\theta{f(\theta)}&=&-\cos\theta(D-Cn\cos\theta)^2+Cn\sin^2\theta(D-Cn\cos\theta)-AMgh\sin^4\theta\nonumber\\
	\label{g7}
\end{eqnarray}
Now, integrating (\ref{g6}) with respect to $\theta$ we have 
\begin{equation}
	\frac{1}{2}A\dot{\theta}^2+\Psi(\theta)=0\label{g8}
\end{equation}
Analyzing the real roots of $\Psi(\theta)=0$ and using the properties of Elliptic function one can easily find $\phi$ and $\psi$. Now we shall discuss the stability of the top when it executes steady motion at $\theta=\alpha=$constant and let $\dot{\phi}=\Omega$, a constant when $\theta=\alpha$. Form (\ref{g5}) 
\begin{equation}
	f(\alpha)=-\sin\alpha\left[A\Omega^2\cos\alpha-Cn\Omega+Mgh\right]=0\label{g9}
\end{equation}
Assuming $\theta\neq0$ (i.e.,$\alpha\neq0$) we have,
\begin{equation}
	A\Omega^2\cos\alpha-Cn\Omega+Mgh=0\label{g10}
\end{equation}
It has a pair of real and distinct roots $\Omega_1,~\Omega_2$ of $\Omega$ provided $C^2n^2>4AMgh\cos\alpha$.

Suppose that the above inequality holds good. As $f(\alpha)=0$ so if we put $\theta=\alpha+\epsilon$ where $\epsilon$ is small then $f(\theta)\approx\epsilon{f'(\alpha)}$ to the first order and hence (\ref{g6}) becomes
\begin{equation}
	A\ddot{\epsilon}+\epsilon{f'(\alpha)}=0\label{g11}
\end{equation}
Now, if we can show $f'(\alpha)>0$ then this equation will reduce to the form $\ddot{\epsilon}+\omega^2\epsilon=0$ where $\omega^2=\frac{f'(\alpha)}{A}$ showing stability about $\theta=\alpha$ for small perturbations. Now, differentiation (\ref{g7}) with respect to the $\theta$ and putting $\theta=\alpha$ and using ${f(\alpha)}=0$ we get 
\begin{eqnarray}
	A{f'(\alpha)}\sin^3\alpha&=&\sin\alpha(D-Cn\cos\alpha)^2-2Cn\cos\alpha\sin\alpha(D-Cn\cos\alpha)\nonumber \\ 
	&+&2Cn\cos\alpha\sin\alpha(D-Cn\cos\alpha)
	+C^2n^2\sin^3\alpha-4AMgh\sin^3\alpha\cos\alpha  \nonumber\\
	&=&\sin\alpha(D-Cn\cos\alpha)^2+C^2n^2\sin^3\alpha-4AMgh\sin^3\alpha\cos\alpha\nonumber
\end{eqnarray}
putting $\theta=\alpha$ in equation (\ref{g4}) we get
\begin{equation}
	D-Cn\cos\alpha=A\Omega\sin^2\alpha\nonumber
\end{equation}
So equation (\ref{g5}) $\implies~Cn=A\Omega\cos\alpha+\frac{Mgh}{\Omega}$. Hence,
\begin{eqnarray}
	A{f'(\alpha)}&=&A^2\Omega^2\sin^2\alpha+\left(A\Omega\cos\alpha+\frac{Mgh}{\Omega}\right)^2-4AMgh\cos\alpha\nonumber\\
	&=&A^2\Omega^2-2AMgh\cos\alpha+\left(\frac{Mgh}{\Omega}\right)^2\nonumber\\
	&=&\left(A\Omega-\frac{Mgh}{\Omega}\right)^2+2\frac{Mgh}{\Omega}(1-\cos\alpha)>0~\mbox{for}~\alpha\neq0\nonumber
\end{eqnarray} 
Hence for $\alpha\neq0$ the motion is S.H.M. type showing that the position $\theta=\alpha$ is one of stable configuration. 

$(\mbox{Alternatively,}$
\begin{eqnarray}
	f'(\alpha)&=&\frac{A^2\Omega^2\sin^5\alpha+C^2n^2\sin^3\alpha-4AMgh\sin^3\alpha\cos\alpha}{\sin^3\alpha}\nonumber\\
	&=&A^2\Omega^2\sin^2\alpha+\left(\frac{C^2n^2}{A}-4Mgh\cos\alpha\right)\nonumber\\
	&>&0\nonumber
\end{eqnarray}
Since $C^2n^2>4AMgh\cos\alpha$ for real roots of $\Omega$ in equation (\ref{g10}).$)$

\section{Principle of Linear Momentum  : Conservation of Linear Momentum: } 

If a physical system is under the action of a system of externally applied forces and subject to bilateral constraints only, the rate of change of linear momentum of the system is equal to the vector sum of all the externally applied forces acting on the system.

Suppose we have a physical system imposed of $N$ particles. Let the system is under the action of a system of externally applied forces and subject to bilateral constraints only. Let $P_i$ be the position of the $i$-th particle having mass $m_i$ and position vector $\overrightarrow{r_i}$ with respect to a fixed point $O~i.e.~\overrightarrow{OP}=\overrightarrow{r_i}$. Suppose $\overrightarrow{F_i}$ be the resultant applied forces acting on the $i$-th particle. According to D'Alembert principle

\begin{equation}
\sum_{i=1}^{N}\left\{\overrightarrow{F_i}+(-m_i\ddot{\overrightarrow{r_i}})\right\}=0~~(\mbox{basis principle of statics})\nonumber
\end{equation}
\begin{eqnarray}
\mbox{i.e.,}~\sum{m_i\ddot{\overrightarrow{r_i}}}&=&\sum_{i}\overrightarrow{F_i}\nonumber\\
\mbox{i.e.,}~\frac{d}{dt}\left(\sum{m_i\dot{\overrightarrow{r_i}}}\right)&=&\sum_{i}\overrightarrow{F_i}\nonumber
\end{eqnarray}
Thus the rate of change of linear momentum of a material system in a given direction is equal to the vector sum of all the external forces applied to the system in that direction.

In particular, if $\sum\overrightarrow{F_i}=0$ then $\frac{d}{dt}\left(\sum{m_i\dot{\overrightarrow{r_i}}}\right)=0$ 

i.e., $\sum{m_i\dot{\overrightarrow{r_i}}}=$ Constant.

So the total linear momentum of a material system is conserved. This is known as conservation of linear momentum.

\section{Principle of Angular Momentum  : Conservation of Angular Momentum: }	

If a physical system is under the action of a system of externally applied forces and subjected to bilateral constrains only, then the rate of change of angular momentum (i.e. moment of momentum) of the system in a given direction is equal to the vector sum of the moments of all the externally applied forces acting on the system in that direction.

Form the basic principle of statics if a system of external forces are in equilibrium then the sum of the moments of all forces about any point in space is equal to zero.

Thus taking moment about O
\begin{eqnarray}
\sum\overrightarrow{r_i}\times\left\{\overrightarrow{F_i}+(-m_i\ddot{\overrightarrow{r_i}})\right\}&=&0\nonumber\\
\mbox{i.e.,~}~\sum\overrightarrow{r_i}\times{m_i\ddot{\overrightarrow{r_i}}}&=&\sum_{i}\overrightarrow{r_i}\times{\overrightarrow{F_i}}\nonumber\\
\mbox{i.e.,~}~\frac{d}{dt}\left(\overrightarrow{r_i}\times{m_i\dot{\overrightarrow{r_i}}}\right)&=&\sum\overrightarrow{r_i}\times{\overrightarrow{F_i}}\nonumber\\
\mbox{i.e.,~}~\frac{d\overrightarrow{H}}{dt}&=&\sum\overrightarrow{r_i}\times{\overrightarrow{F_i}}=\overrightarrow{L}\nonumber
\end{eqnarray}
where $\overrightarrow{H}=\overrightarrow{r_i}\times{m_i\dot{\overrightarrow{r_i}}}$ is the angular momentum of the material system in a given direction and $\overrightarrow{L}$ is the vector sum of the moments of all externally applied forces acting on the system in that direction.

Now, if $\overrightarrow{L}=0$ then $\overrightarrow{L}=$Constant i.e., total angular momentum in that direction of the system is conserved. This is known as conservation of angular momentum.

\section{Euler's Dynamical Equation:}

Suppose a physical system rotate about O which is fixed with respect to the system as well as the space. Consider a rectangular coordinate system $ox,~oy,~oz$ through O. These axes are fixed with respect to the system. Let $\overrightarrow{i},~\overrightarrow{j}$ and $\overrightarrow{k}$ be the unit vector along these coordinate axes. Then $\overrightarrow{r}=x\overrightarrow{i}+y\overrightarrow{j}+z\overrightarrow{z}$ with $(x,y,z)$ be the coordinate of the point $P$ with respect to the coordinate system $ox,~oy,~oz$ which is fixed in the system . Now,
\begin{eqnarray}
\dot{\overrightarrow{r}}=\frac{d\overrightarrow{r}}{dt}&=&\frac{\partial{\overrightarrow{r}}}{\partial{t}}+\overrightarrow{\omega}\times{\overrightarrow{r}}\nonumber\\
&=&\frac{\partial}{\partial{t}}(x\overrightarrow{i}+y\overrightarrow{j}+z\overrightarrow{z})+\overrightarrow{\omega}\times{\overrightarrow{r}}\nonumber\\
&=&\dot{x}\overrightarrow{i}+\dot{y}\overrightarrow{j}+\dot{z}\overrightarrow{z}+\overrightarrow{\omega}\times{\overrightarrow{r}}\nonumber\\
&=&\overrightarrow{\omega}\times{\overrightarrow{r}}\nonumber
\end{eqnarray}

($\because{\dot{x}=\dot{y}=\dot{z}=0}$ as the axes are fixed in the rigid body)

Suppose $\overrightarrow{\omega}=\omega_1\overrightarrow{i}+\omega_2\overrightarrow{j}+\omega_3\overrightarrow{k}$ be the angular velocity of rotation through O. Thus
\begin{eqnarray}
\overrightarrow{H}=\sum\overrightarrow{r}\times({m_i\dot{\overrightarrow{r}}})&=&\sum{m}\overrightarrow{r}\times(\overrightarrow{\omega}\times{\overrightarrow{r}})\nonumber\\
&=&\sum m\left[(\overrightarrow{r}.\overrightarrow{r})\overrightarrow{\omega}-(\overrightarrow{r}.\overrightarrow{\omega})\overrightarrow{r}\right]\nonumber\\
&=&\sum{m}\Big{[}(x^2+y^2+z^2)(\omega_1\overrightarrow{i}+\omega_2\overrightarrow{j}+\omega_3\overrightarrow{k})\nonumber \\
&& -(x\omega_1+y\omega_2+z\omega_3)(x\overrightarrow{i}+y\overrightarrow{j}+z\overrightarrow{k})\Big{]}\nonumber
\end{eqnarray}

\begin{equation}
H=\sum\Big{[}\omega_1\{(y^2+z^2)\overrightarrow{i}-xy\overrightarrow{j}-xz\overrightarrow{k}\}+\omega_2\{(x^2+z^2)\overrightarrow{j}-xy\overrightarrow{i}-yz\overrightarrow{k}\} \nonumber \\
 +\omega_3\{(y^2+x^2)\overrightarrow{k}-xz\overrightarrow{i}-yz\overrightarrow{j}\}\Big{]}\nonumber
\end{equation}
As $\omega_1,~\omega_2,~\omega_3$ are same for all the particles of the material system so
\begin{eqnarray}
\overrightarrow{H}&=&\omega_1\left\{\left[\sum{m}(y^2+z^2)\right]\overrightarrow{i}-\left[\sum{mxy}\right]\overrightarrow{j}-\left[\sum{mxz}\right]\overrightarrow{k}\right\}\nonumber\\
&+&\omega_2\left\{\left[\sum{m}(x^2+z^2)\right]\overrightarrow{j}-\left[\sum{mxy}\right]\overrightarrow{i}-\left[\sum{myz}\right]\overrightarrow{k}\right\}\nonumber\\
&+&\omega_1\left\{\left[\sum{m}(y^2+x^2)\right]\overrightarrow{k}-\left[\sum{mxz}\right]\overrightarrow{i}-\left[\sum{mzy}\right]\overrightarrow{j}\right\}\nonumber
\end{eqnarray}
Now, if $A,~B,~C$ are the moments of inertia and $D,~E,~F$ are the product of inertia of the physical system with respect to the coordinate system $ox,~oy,~oz$ through O, then $A=\sum{m}(y^2+z^2),~B=\sum{m}(x^2+z^2),~C=\sum{m}(x^2+y^2),~D=\sum{mzy},~E=\sum{mzx},~F=\sum{mxy}$. Thus $\overrightarrow{H}=\omega_1\left(A\overrightarrow{i}-F\overrightarrow{j}-E\overrightarrow{k}\right)+\omega_2\left(B\overrightarrow{j}-F\overrightarrow{i}-D\overrightarrow{k}\right)+\omega_3\left(C\overrightarrow{k}-E\overrightarrow{i}-D\overrightarrow{j}\right)$.

Further, if the coordinate axes are the principal axes of inertia then $D=E=F=0$ and we get $\overrightarrow{H}=\omega_1A\overrightarrow{i}+\omega_2B\overrightarrow{j}+\omega_3C\overrightarrow{k}$

$\therefore\frac{dH}{dt}=\frac{\partial{\overrightarrow{H}}}{\partial{t}}+\overrightarrow{\omega}\times\overrightarrow{H}=A\dot{\omega}_1\overrightarrow{i}+B\dot{\omega}_2\overrightarrow{j}+C\dot{\omega}_3\overrightarrow{k}+\begin{vmatrix}
\overrightarrow{i} & \overrightarrow{j} & \overrightarrow{k}\\
\omega_1 & \omega_2 & \omega_3\\
H_1 & H_2 & H_3\\
\end{vmatrix}$

Also if $\overrightarrow{L}=L_1\overrightarrow{i}+L_2\overrightarrow{j}+L_3\overrightarrow{k}$, then
\begin{eqnarray}
L_1=A\dot{\omega}_1-(B-C)\omega_2\omega_3\nonumber\\
L_2=B\dot{\omega}_2-(C-A)\omega_3\omega_1\nonumber\\
L_3=C\dot{\omega}_3-(A-B)\omega_1\omega_2\nonumber
\end{eqnarray}
These are known as Euler's dynamical equations. Kinetic energy of a physical system with respect to a rotating coordinate system which is fixed with respect to the system can be derived as follows:
\begin{eqnarray}
T=\frac{1}{2}\sum{m}\left|\frac{d\overrightarrow{r}}{dt}\right|^2=\frac{1}{2}\sum{m}\frac{d\overrightarrow{r}}{dt}.\frac{d\overrightarrow{r}}{dt}\nonumber\\
\dot{\overrightarrow{r}}=\frac{d\overrightarrow{r}}{dt}=\frac{\partial{\overrightarrow{r}}}{\partial{t}}+\overrightarrow{\omega}\times{\overrightarrow{r}}=\overrightarrow{\omega}\times{\overrightarrow{r}}\nonumber
\end{eqnarray}
(as the axes are fixed in the system)
\begin{eqnarray}
\therefore{T}&=&\frac{1}{2}\sum{m}(\overrightarrow{\omega}\times\overrightarrow{r}).\frac{d\overrightarrow{r}}{dt}\nonumber\\
&=&\frac{1}{2}\sum{m}\overrightarrow{\omega}.\left(\overrightarrow{r}\times\frac{d\overrightarrow{r}}{dt}\right)\nonumber\\
&=&\frac{1}{2}\sum{m}\overrightarrow{\omega}.\left\{\overrightarrow{r}\times(\overrightarrow{\omega}\times\overrightarrow{r})\right\}\nonumber\\
&=&\frac{1}{2}\sum{m}\overrightarrow{\omega}.\left[(\overrightarrow{r}.\overrightarrow{r})\overrightarrow{\omega}-(\overrightarrow{r}.\overrightarrow{\omega})\overrightarrow{r}\right]\nonumber\\
&=&\frac{1}{2}\sum{m}\left[(\overrightarrow{r}.\overrightarrow{r})(\overrightarrow{\omega}.\overrightarrow{\omega})-(\overrightarrow{r}.\overrightarrow{\omega})(\overrightarrow{\omega}.\overrightarrow{r})\right]\nonumber\\
&=&\frac{1}{2}\sum{m}\left[(x^2+y^2+z^2)(\omega_1^2+\omega_2^2+\omega_3^2)-(x\omega_1+y\omega_2+z\omega_3)^2\right]\nonumber\\
&=&\frac{1}{2}\sum{m}\left[\omega_1^2(y^2+z^2)+\omega_2^2(x^2+z^2)+\omega_3^2(y^2+x^2)-2\omega_1\omega_2xy-2\omega_2\omega_3yz-2\omega_3\omega_1xz\right]\nonumber
\end{eqnarray}
As $\omega_1,~\omega_2$ and $\omega_3$ are same for all points so 
\begin{eqnarray}
{T}&=&\frac{1}{2}\Big{[}\omega_1^2\sum{m}(y^2+z^2)+\omega_2^2\sum{m}(x^2+z^2)+\omega_3^2\sum{m}(y^2+x^2)-2\omega_1\omega_2\sum{m}xy \nonumber \\
&-& 2\omega_2\omega_3\sum{m}zy\Big{]}
-\omega_1\omega_3\sum{m}xz\nonumber\\
&=&\frac{1}{2}\left[A\omega_1^2+B\omega_2^2+C\omega_3^2-2\omega_1\omega_2F-2\omega_2\omega_3D-2\omega_1\omega_3E\right]\nonumber
\end{eqnarray}
Further, if the axes are principal axes then
\begin{equation}
{T}=\frac{1}{2}\left[A\omega_1^2+B\omega_2^2+C\omega_3^2\right]\nonumber
\end{equation}


\chapter{Appendix}

\section{Appendix-I: A simple note on vector calculus}

~~~\textbf{Result I:} If $\vec{a}(t)$ is a vector of constant magnitude i.e. $|\vec{a}(t)|=$ constant, then $\dfrac{\mathrm{d}\vec{a}}{\mathrm{d}t}$ is orthogonal to $\vec{a}$.

\textbf{Proof:} As $|\vec{a}(t)|=$ constant,

So, $\vec{a}\cdot\vec{a}=|\vec{a}|^2=$ constant

$\therefore ~2\vec{a}\cdot\dfrac{\mathrm{d}\vec{a}}{\mathrm{d}t}=0$

Hence for non-zero vector $\vec{a}$,  $\dfrac{\mathrm{d}\vec{a}}{\mathrm{d}t}$ is orthogonal to $\vec{a}$.\\

\textbf{Remark:} A natural question arises for a unit vector :

If $\vec{a}$ is an unit vector then $\dfrac{\mathrm{d}\vec{a}}{\mathrm{d}t}$ is orthogonal to $\vec{a}$. But does $\dfrac{\mathrm{d}\vec{a}}{\mathrm{d}t}$ is also an unit vector?\\

\textbf{Result II:} If $\vec{a}$ is an unit vector then $\dfrac{\mathrm{d}\vec{a}}{\mathrm{d}\theta}$ is an unit vector.

\begin{wrapfigure}[8]{r}{0.48\textwidth}
	\includegraphics[height=4.5 cm , width=7 cm ]{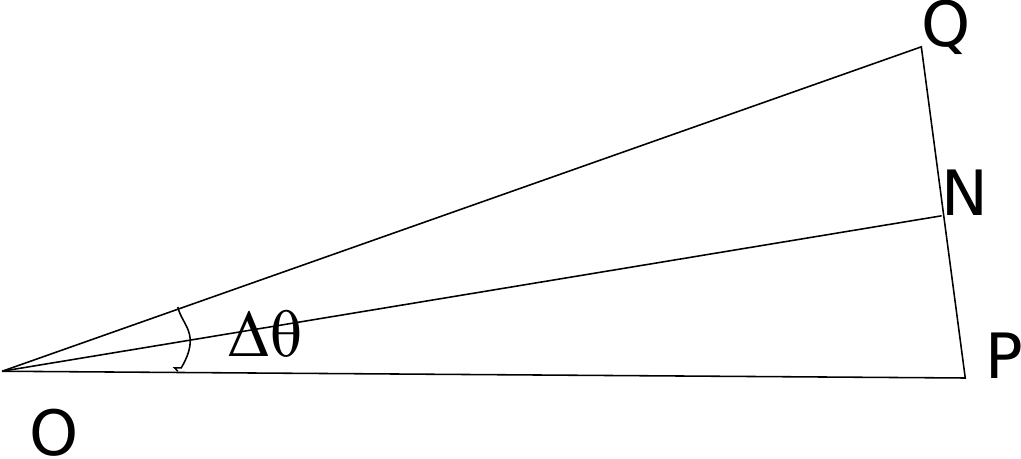}
	\begin{center}
		Fig. 1.1
	\end{center}\vspace{-\intextsep}
\end{wrapfigure}

\textbf{Proof:}  Let $\overrightarrow{OP}=\vec{a}$ be the position of the unit vector in time $t$ and $\overrightarrow{OQ}$ be the position of the unit vector in time $t+\Delta t$. As $\vec{a}$ is a unit vector, so, $OP=OQ=1$.

Let $\angle POQ=\Delta\theta$.  Draw $ON\perp PQ$.

Let $\overrightarrow{OQ}=\vec{a}+\overrightarrow{\Delta a}$,  so that $\overrightarrow{PQ}=\overrightarrow{\Delta a}$ and $\overrightarrow{NQ}=\dfrac{1}{2}\overrightarrow{\Delta a}$.

Now, in $\triangle ONQ$, ~$\dfrac{NQ}{OQ}=\sin\left(\dfrac{1}{2}\Delta\theta\right)$

$\therefore~\sin\left(\dfrac{1}{2}\Delta\theta\right)=NQ=|\overrightarrow{NQ}|=\dfrac{1}{2}|\overrightarrow{\Delta a}|$

$\therefore~\dfrac{\sin\left(\dfrac{1}{2}\Delta\theta\right)}{\dfrac{1}{2}\Delta\theta}=\dfrac{\dfrac{1}{2}|\overrightarrow{\Delta a}|}{\dfrac{1}{2}\Delta\theta}$

Now proceed to the limit as $\Delta\theta\rightarrow0$, we have $\big|\dfrac{\mathrm{d}\vec{a}}{\mathrm{d}\theta}=1\big|$.

Hence $\dfrac{\mathrm{d}\vec{a}}{\mathrm{d}\theta}$ is perpendicular to $\vec{a}$ and is of unit magnitude.

\section{Appendix-II: Some Results on Calculus of variation}

$\star$ \textbf{Calculus of Variation}

${\bullet}\textbf{{Fundamental lemma } :}$

If $f(x)$ is continuous in $[a,b]$ and  if $\int_{a}^{b}f(x)\phi(x)dx=0$ for all values of $\phi(x)$ which are continuous in $[a,b]$ and which vanishes at $x=a$ and $x=b$ then $f(x)\equiv0$.

\textbf{Proof } :  If possible let $f(x)$ be non-zero at a point in $[a,b]$. Without loss of generality we can take $f(x)$ to be positive at this point. Since $f(x)$ is continuous at the point and is positive there so we can find an interval $(x_0,~x_1)\subset[a,b]$ in which $f(x)$ is positive. Let $a<x_0<x_1<b$ and let us define $\phi(x)$ as follows :
\begin{eqnarray}
	\phi(x)&=&0,~a\le{x}\le{x_0}\nonumber\\
	&=&(x-x_0)^p(x_1-x)^p,~{x_0}\le{x}\le{x_1}\nonumber\\
	&=&0,~\le{x_1}\le{x}\le{b}\nonumber
\end{eqnarray}
Therefore, $\phi(x)$ is a function satisfying all the conditions of the theorem, where $p$ is a positive integer. Therefore by the condition of the theorem 
\begin{equation}
	\int_{x_0}^{x_1}f(x)(x-x_0)^p(x_1-x)^pdx=0\nonumber
\end{equation}
But this can not be true because the integrand is essentially positive in $[x_0,~x_1]$ except at the end points and the integrand is continuous. Hence our assumption that $f(x)\neq0$ is not correct and hence the lemma is true.

$\bullet${\textbf{Variation of the derivative}} :

Let us consider a curve $y=y(x)$ passing through the points $(a,b)$ and $(c,d)$. Let $y_1=y_1(x)$ be any arbitrary curve in the neighbourhood of the curve $y(x)$ and passing through the points $(a,b)$ and $(c,d)$. Let us denote the difference between the function $y_1(x)$ and $y(x)$ for the same value of $x$ as $\delta{y}$ i.e., $\delta{y}=y_1(x)-y(x)=\delta{y(x)}$, a function of $x$. This function $\delta{y}$ is called the variation of the function $y(x)$ for a given $x$.

Again '$dy$' refer to the differential change in the ordinate $y$ as we pass from a point on a curve $y=y(x)$ to a neighbouring point on the same curve, while $y_1(x)=y(x)+\delta{y(x)}$. In fact, $\delta{y(x)}$ is the infinitesimal virtual displacement suffered by the point $(x,y)$ on $y(x)$ for given $x$. Thus the points $\{x,~y(x)\}$ on the curve $y=y(x)$ shifts to the point $\{x,~y_1(x)\}$, on the neighbourhood curve, subject to the variation $\delta{y(x)}$. Similarly, variation $\delta{y'(x)}$ can define as 
\begin{equation}
	\delta{y'(x)}=y'_1(x)-y'(x)=\frac{d}{dx}(y_1-y)=\frac{d}{dx}(\delta{y})\nonumber
\end{equation}
Thus $\delta$ of $y'(x)=\frac{d}{dx}$ of $\delta{y}$ i.e., the variation of the derivative is the derivative of the variation.

$\bullet$ {\textbf{Variation of the integrals}} :

Let $I=\int_{t_0}^{t_1}f(q_1,...,q_n;\dot{q_1},...,\dot{q_n};t)dt$. Suppose by keeping $t$ fixed we may vary $q$'s such that they becomes $q_1+\delta{q_1},q_2+\delta{q_2},...,q_n+\delta{q_n}$, where $\delta{q_i=\epsilon\eta_i(t)},~i=1,2,...,n$. Here $\eta_i$'s  are continuous function of '$t$' possessing continuous derivative of 1st order and $\epsilon$ is an infinitesimal quantity. Further, it is assumed that the function $f$ together with its partial derivatives of 1st two orders are continuous. Now define 
\begin{equation}
	I_1=\int_{t_0}^{t_1}f(q_1+\epsilon{\eta_1},q_2+\epsilon{\eta_2},...,q_n+\epsilon{\eta_n};\dot{q_1}+\epsilon\dot{\eta},...,\dot{q_n}+\epsilon\dot{\eta_n};t)dt\nonumber
\end{equation}
so 
\begin{equation}
	I_1-I=\int_{t_0}^{t_1}\left[\sum_{i=1}^{n}\left\{\frac{\partial{f}}{\partial{q_i}}\epsilon\eta_i+\frac{\partial{f}}{\partial{\dot{q_i}}}\epsilon\dot{\eta_i}\right\}+R\right]dt\nonumber
\end{equation}
{(by Taylor's expansion)}

where $R$ is an infinitesimal in comparison to 1st term. Thus we call 
\begin{equation}
	I_1-I=\int_{t_0}^{t_1}\sum_{i=1}^{n}\left\{\frac{\partial{f}}{\partial{q_i}}\epsilon\eta_i+\frac{\partial{f}}{\partial{\dot{q_i}}}\epsilon\dot{\eta_i}\right\}dt\nonumber
\end{equation}
as the 1st variation of $I$ or simply the variation $I$ and we denote it by $\delta{I}$. Also one can denote 
\begin{equation}
	\sum_{i=1}\left(\frac{\partial{f}}{\partial{q_i}}\epsilon\eta_i+\frac{\partial{f}}{\partial{\dot{q_i}}}\epsilon\dot{\eta_i}\right)dt\nonumber
\end{equation}
by $\delta{f}$ i.e., 
\begin{equation}
	\delta{I}=\delta\int_{t_0}^{t_1}fdt=\int_{t_0}^{t_1}\delta{f}dt\nonumber
\end{equation}
i.e., variation of the integral =integral of the variation of the integrand.

\underline{Note } : As $\delta{q_i=\epsilon\eta_i(t)}$ so $\delta\dot{q_i}=\epsilon\dot{\eta_i}(t)=\frac{d}{dt}(\epsilon\eta_i)=\frac{d}{dt}(\delta{q_i})$.

$\bullet$ {\textbf{Variation of a function with		
		variation of path with respect to time}} :
\begin{figure}
	\centering
	\includegraphics[width=0.4\textwidth]{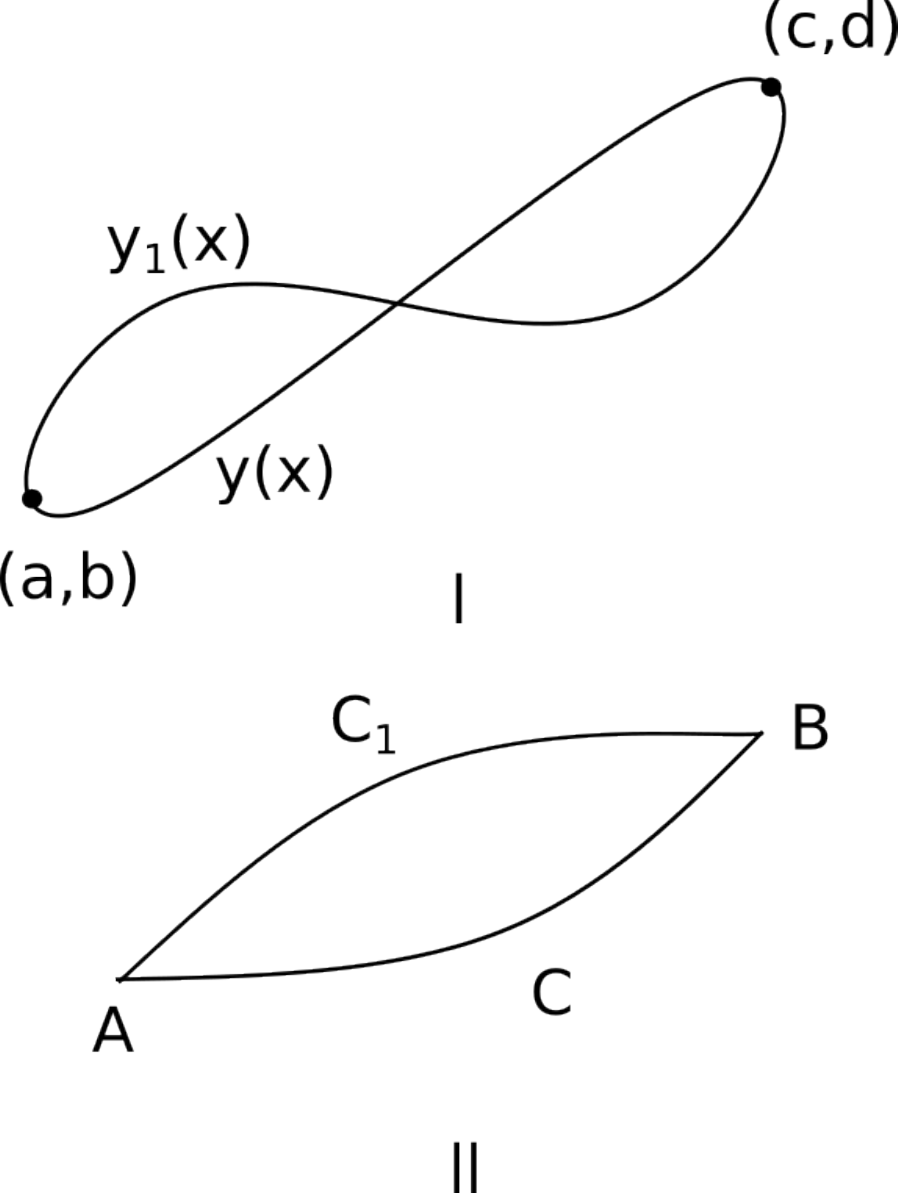}\\
	\label{fig1}
\end{figure}

Let us consider two points $A$ and $B$ in the configuration space and let us join these two points by two neighbouring curves $C$ and $C_1$. Let $q(t)$ denote the actual motion of a certain system along $C$ for the time $t_0\le{t}\le{t_1}$. Further, let $q_1(t)$denote the arbitrary varying motion of the same system along $C_1$ for $t'_0\le{t}\le{t'_1}$. Let us consider the difference 
\begin{eqnarray}
	q_1(t+\Delta)-q_1(t)&\simeq&{q_1(t)}+\Delta{t}\dot{q_1}-q(t)\nonumber\\
	&=&\delta{q}+\Delta{t}\frac{d}{dt}(q+\delta{q})\nonumber\\
	&=&\delta{q}+\Delta{t}\dot{q_1}~\mbox{upto 1st order}\nonumber
\end{eqnarray} 
We denote this difference by $\Delta{q}$ and call it the 1st variation of $q(t)$ with variation of time. 
\begin{subequations}
	\begin{equation}
		\therefore\Delta{q}=\delta{q}+\Delta{t}\dot{q_1}\label{g13}
	\end{equation}
\end{subequations}
Now differentiating equation (\ref{g13}) with respect to '$t$' we get 
\begin{subequations}
	\begin{eqnarray}
		\frac{d}{dt}(\Delta{q})&=&\frac{d}{dt}(\delta{q})+\Delta{t}\frac{d}{dt}(\dot{q})+\dot{q}\frac{d}{dt}(\Delta{t})\nonumber\\
		&=&\delta{\dot{q}}+\Delta{t}\frac{d}{dt}(\dot{q})+\dot{q}\frac{d}{dt}(\Delta{t})\nonumber\\
		&=&\Delta{\dot{q}}+\dot{q}\frac{d}{dt}(\Delta{t})\nonumber
	\end{eqnarray}
	\begin{equation}
		\mbox{i.e.,}~\frac{d}{dt}(\Delta{q})=\Delta{\dot{q}}+\dot{q}\frac{d}{dt}(\Delta{t})\label{g14}
	\end{equation}
\end{subequations}
Hence $\frac{d}{dt}(\Delta{q})\neq\Delta{\dot{q}}$.

Analogously, the variation of a function on $f(q,\dot{q},t)$ is defined as 
\begin{subequations}
	\begin{eqnarray}
		\therefore\Delta{f}&=&\delta{f}+\Delta{t}\dot{f}\nonumber\\
		&=&\frac{\partial{f}}{\partial{q}}\delta{q}+\frac{\partial{f}}{\partial{\dot{q}}}\delta{\dot{q}}+\Delta{t}\dot{f}\nonumber\\
		&=&\frac{\partial{f}}{\partial{q}}\delta{q}+\frac{\partial{f}}{\partial{\dot{q}}}\delta{\dot{q}}+\Delta{t}\left(\frac{\partial{f}}{\partial{q}}\dot{q}+\frac{\partial{f}}{\partial{\dot{q}}}{\ddot{q}}+\frac{\partial{f}}{\partial{t}}\right)\nonumber\\
		&=&\frac{\partial{f}}{\partial{q}}(\delta{q}+\dot{q}\Delta{t})+\frac{\partial{f}}{\partial{\dot{q}}}(\delta{\dot{q}}+\ddot{q}\Delta{t})+\Delta{t}\frac{\partial{f}}{\partial{t}}\nonumber\\
		&=&\frac{\partial{f}}{\partial{q}}\Delta{q}+\frac{\partial{f}}{\partial{\dot{q}}}\Delta{\dot{q}}+\Delta{t}\frac{\partial{f}}{\partial{t}}\nonumber
	\end{eqnarray}
	\begin{equation}
		\mbox{i.e.,}~\Delta{f}=\frac{\partial{f}}{\partial{q}}\Delta{q}+\frac{\partial{f}}{\partial{\dot{q}}}\Delta{\dot{q}}+\Delta{t}\frac{\partial{f}}{\partial{t}}\label{g15}
	\end{equation}
\end{subequations}

$\bullet$ {\textbf{Variation of integral with variation of time}} :

Let us consider the integral $I=\int_{t_0}^{t_1}fdt$ for the actual motion $C$ and $I=\int_{t_0+\Delta{t_0}}^{t_1+\Delta{t_1}}f_1dt$, for the varied path of motion $C_1$ with $f_1=f+\delta{f}$. Now, 
\begin{subequations}
	\begin{equation}
		I_1=\int_{t_0}^{t_1}f_1dt+\int_{t_1}^{t_1+\Delta{t_1}}f_1dt-\int_{t_0}^{t_0+\Delta{t_0}}f_1dt\nonumber
	\end{equation}
	\begin{eqnarray}
		\therefore{I_1-I}&=&\int_{t_0}^{t_1}(f_1-f)dt+(f_1)_{t_1}\Delta{t_1}-(f_1)_{t_0}\Delta{t_0}\nonumber\\
		&&\left(\mbox{By mean value theorem }\int_{a}^{b}f(x)\phi(x)dx=f(\xi)\int_{a}^{b}\phi(x)dx\right)\nonumber\\
		&=&\int_{t_0}^{t_1}\delta{f}dt+f_1\Delta{t_1}|_{t_0}^{t_1}~~(\mbox{assuming continuity of the function} f_1(t))\nonumber\\
		&=&\int_{t_0}^{t_1}\left[\delta{f}+\frac{d}{dt}(f\Delta{t})\right]dt~~(\mbox{upto 1st order})\nonumber
	\end{eqnarray}
\end{subequations}
Thus 
\begin{subequations}
	\begin{equation}
		\Delta{I}=\int_{t_0}^{t_1}\left[\delta{f}+\frac{d}{dt}(f\Delta{t})\right]dt\label{g16}
	\end{equation}
\end{subequations}

 \section{Appendix III: Uniqueness and other properties of Lagrangian function}
 
For a system of $n$ second order differential equations of the form $F_i(t,x_j,\dot{x}_j,\ddot{x}_j)=0,~ i,j=1,~2,~3~...~n$, a natural question arises whether these differential equations may be the Euler-Lagrange equations corresponding to a Lagrangian $L=L(t,x_j,\dot{x}_j)$. From the point of view of dynamics, a physical system with $n$ d.f can be described by $n$ number of second order differential equations.

In the literature, one has Helmholtz conditions which are necessary and sufficient conditions for the above 2nd order diff. eqs to be the Euler-Lagrange eqs. corresponding to a Lagrangian. The conditions are

$(i)~\frac{\partial F_i}{\partial \ddot{x}_j}=\frac{\partial F_j}{\partial \ddot{x}_i}$, $(ii)~ \frac{\partial F_i}{\partial {x}_j}-\frac{\partial F_j}{\partial {x}_i}=\frac{1}{2}\frac{d}{dt}\left(\frac{\partial F_i}{\partial \dot{x}_j}-\frac{\partial F_j}{\partial \dot{x}_i}\right)$ and $(iii)~ \frac{\partial F_i}{\partial \dot{x}_j}+\frac{\partial F_j}{\partial \dot{x}_i}=2\frac{d}{dt}\left(\frac{\partial F_j}{\partial \ddot{x}_i}\right)$, $\forall ~i,j=1,~2,~3~...~ n$.

For a single variable the conditions $(i)$ and $(ii)$ are identically satisfied and relation $(iii)$ simplifies to $\frac{\partial F}{\partial \dot{x}_i}=\frac{d}{dt}\left( \frac{dF}{d\ddot{x}}\right)$. Note that Helmholtz conditions do not yield the Lagrangian of the system, only thing one can say is that the differential equations may be the E-L equations for some Lagrangian. A natural question that immediately arises is that whether Helmholtz conditions are unique for Euler-Lagrange equations.

  We shall illustrate it by two examples: \\
  1. The Euler-Lagrange equation is   $\ddot{q}f(\dot{q},q) + g(\dot{q},q)=0$.
  From H's condition\\
  $~~~~~~~~~~~~~\ddot{q}\frac{\partial f}{\partial{\dot{q}}} + \frac{\partial g}{\partial{\dot{q}}}=\frac{d}{dt}{\{f(q,\dot{q})} \}$\\
  $~~~~~~~~\Rightarrow \ddot{q}\frac{\partial f}{\partial{\dot{q}}} + \frac{\partial g}{\partial{\dot{q}}}=\frac{\partial f}{\partial q} \dot{q}+ \frac{\partial f}{\partial \dot{q}} \ddot{q}$\\
  $~~~~~~~~\Rightarrow  \frac{\partial g}{\partial{\dot{q}}}=\frac{\partial f}{\partial q} \dot{q}$
    
$\hspace{-0.7 cm}$ 2. The E-L equation is  $\ddot{q}f(q)+g(\dot{q},q)=0$. From the H's condition: $\frac{\partial g}{\partial \dot{q}}=\frac{d (f(q))}{dt}=\dot{q}\frac{\partial f}{\partial q}$\\
$i.e., \frac{\partial g}{\partial{\dot{q}}}=\frac{\partial f}{\partial q} \dot{q}$.

Thus we see that though the E-L equations are distinct in the above two examples still the H's condition is identical for the above diff. equations to be Euler-Lagrange equations. Hence Helmholtz conditions are not unique to identify the Euler-Lagrange equations.

We shall now try to determine the possible Lagrangian for examples 1 and 2.

  \textbf{Example 1:} suppose $L=\frac{\dot{q}^2}{2}h(q) +\chi(\dot{q},q)$ \\
$\frac{\partial L}{\partial{\dot{q}}}=\dot{q}h(q)+ \frac{\partial \chi}{\partial \dot{q}}~,~\frac{\partial L}{\partial q}=\frac{\dot{q}^2}{2}h'(q)+\frac{\partial \chi}{\partial q}$

$\therefore$ The E-L equation is \\
$\frac{d}{dt}\left(\frac{\partial L}{\partial \dot{q}}\right)-\frac{\partial L}{\partial q}=0 ~~\Rightarrow \ddot{q}h(q)+\dot{q}^2h'(q) +\ddot{q} \frac{\partial^2\chi}{\partial \dot{q}^2}+\dot{q}\frac{\partial^2\chi}{\partial q \partial \dot{q} }- \frac{\dot{q}^2}{2}h'(q)-\frac{\partial \chi}{\partial q}=0$ \\
$\Rightarrow \ddot{q} \{ h(q)+\frac{\partial^2 \chi }{\partial {\ddot{q}^2}}\} +\frac{\dot{q}^2}{2}h'(q)+\dot{q}\frac{\partial^2 \chi}{\partial q \partial \dot{q}} -\frac{\partial \chi}{\partial q}=0$

Now comparing with E-L equation\\
$f(q,\dot{q})=h(q)+\frac{\partial^2 \chi}{\partial \dot{q}^2}~~....~(a)$ \\ $g(q,\dot{q})=\frac{\dot{q}^2}{2}h'(q)+\dot{q}\frac{\partial^2 \chi}{\partial q \partial \dot{q}}-\frac{\partial \chi}{\partial q}~~....~(b)$

\vspace{1 cm}
 \hspace{-0.6 cm } From (b),\\     $\frac{\partial g}{\partial \dot{q}}=\dot{q}h'(q)+\frac{\partial^2 \chi}{\partial q\partial \dot{q}}+\dot{q}\frac{\partial^3 \chi}{\partial q\partial \dot{q}^2}-\frac{\partial^2 \chi}{\partial q\partial \dot{q}}$\\
$~\hspace{0.5 cm}=\dot{q}h'(q)+\dot{q}\frac{\partial^3 \chi}{\partial q\partial \dot{q}^2}$\\

 \hspace{-0.6 cm } Also    $~~~\dot{q}\frac{\partial f}{\partial q}=\dot{q}h'(q)+\dot{q}\frac{\partial^3 \chi}{\partial q\partial \dot{q}^2}$\\
 $\bullet$ Thus, $\frac{\partial g}{\partial \dot{q}}=\dot{q} \frac{\partial f}{\partial q}~~\rightarrow$ The Helmholtz condition is identically satisfied.
 
\hspace{-0.65 cm} $\bullet$ One may note that eqs. (a) and (b) can not determine $h$ and $\chi$.

\hspace{-0.65 cm} $\bullet$ If one assumes that $\frac{\partial \chi}{\partial q}$ is a hom. fn. of $\dot{q}$ of degree $n$, then\\
$\dot{q}\frac{\partial}{\partial \dot{q}}\left( \frac{\partial \chi}{\partial q}\right)=n\frac{\partial \chi}{\partial q}$\\
 \hspace{-0.6 cm } From (b), $g=\frac{\dot{q}^2}{2}h'(q)+(n-1)\frac{\partial \chi}{\partial q}$\\
  \hspace{-0.6 cm } So for $n=1, ~~g=\frac{\dot{q}^2}{2}h'(q),~~~~h(q)=\int\frac{2g}{\dot{q}^2} dq$. Then $\chi$ can be determined from (a).
  
  \textbf{Example 2:} Suppose $L=\frac{\dot{q}^2}{2}f(q)+h(q)$\\
  Then, $\frac{\partial L}{\partial \dot{q}}=\dot{q}f(q)$ , $\frac{\partial L}{\partial q}=\frac{\dot{q}^2}{2}f'(q)+h'(q)$\\
  So, the Euler-Lagrange eq. is \\
  
  $\frac{d}{dt}\left(\frac{\partial L}{\partial \dot{q}}\right)-\frac{\partial L}{\partial q}=0 \Rightarrow \ddot{q}f(q)+\dot q^2f'(q)-\frac{1}{2}\dot{q}^2f'(q)-h'(q)=0$\\
 $~~~\hspace{3.2 cm} \Rightarrow~\ddot{q}f(q)+\frac{1}{2}\dot{q}^2f'(q)-h'(q)=0 $ \\
 Using H's condition $\Rightarrow \ddot{q}f(q)+\frac{1}{2} \dot{q}\frac{\partial g}{\partial q}-h'(q)=0$\\
 Comparing with the E-L eq. in the example, $g(q, \dot{q})=\frac{1}{2}\dot{q}\frac{\partial g}{\partial \dot{q}}-h'(q)$\\
 $\Rightarrow h(q)=\bigintss\left[g(q, \dot{q})- \frac{1}{2}\dot{q}\frac{\partial g}{\partial \dot{q}}  \right]dq=-\left(\frac{n-2}{2}\right)\bigintsss g(q, \dot{q}) dq,~$   if $g$ is a hom. fn. of $\dot{q}$ of degree $n$.  
  In particular, if $n=2$, then $h(q)=0$.
  
  We shall now discuss some typical form of the Lagrangian and the corresponding E-L eqs.

  \textbf{Example 3:}  $L(x,\dot{x})=\frac{1}{2}\dot{x}^2f(x)+g(x)$, then the E-L eq. is: $f(x)\ddot{x}+\frac{1}{2}\dot{x}^2f'(x)-g'(x)=0$.
  
  \textbf{Example 4:} $L(x,\dot{x})=\frac{1}{2}\dot{x}^2f(x)+\dot{x}h(x)+g(x)$, here the E-L eq. is: $f(x)\ddot{x}+\frac{1}{2}\dot{x}^2f'(x)-g'(x)=0$ \\
  Note: for both the Lagrangians in Ex.-3 and Ex.-4 the E-L eq. is identical. So, Lagrangian for a physical system is note unique.
  
  \textbf{Example 5:} $L=\dot{x}^nf(x)+g(x)$\\
  Here, $\frac{\partial L}{\partial \dot{x}}=n\dot{x}^{n-1}f(x),~\frac{\partial L}{\partial x}=\dot{x}^{n}f'(x)+g'(x)$\\
  So, the E-L eq. is $\frac{d }{dt}(n\dot{x}^{n-1}f(x))-\dot{x}^{n}f'(x)-g'(x)=0$\\
  $\Rightarrow n(n-1)\dot{x}^{n-2}\ddot{x}f(x)+n\dot{x}^nf'(x)-\dot{x}^nf'(x)-g'(x)=0$

    \textbf{Example 6:} $L=\dot{x}^nf(x)+\dot{x}^mh(x)+g(x)$\\
    $\frac{\partial L}{\partial \dot{x}}=n\dot{x}^{n-1}f(x)+m\dot{x}^{m-1}h(x), \frac{\partial L}{\partial x}=\dot{x}^{n}f'(x)+\dot{x}^{m}h'(x)+g'(x) $\\
    The Euler-Lagrange eq. is:\\ $n(n-1)\dot{x}^{n-2}\ddot{x}f(x)+m(m-1)\dot{x}^{m-2}h(x)\ddot{x}+n\dot{x}^{n}f'(x)+m\dot{x}^{m}h'(x)-\dot{x}^{n}f'(x)-\dot{x}^{m}h'(x)-g'(x)=0 $\\
    $\Rightarrow \{n(n-1)\dot{x}^{n-2} f(x)+m(m-1)\dot{x}^{m-2} h(x)\}\ddot{x}+(n-1)\dot{x}^nf'(x)+(m-1)\dot{x}^mh(x)-g'(x)=0$
    
    \textbf{Note:} (a) A second order differential equation is a Euler-Lagrange equation corresponding to a Lagrangian if it contains second order differential in linear form.\\
    (b) If the first order derivative  is of degree $n$ in the diff. eq. then the corresponding Lagrangian has $\dot{q}$ of at most degree $(n+1)$.\\
    Conversely, if the Lagrangian has $\dot{q}$ of degree $n$, then the corresponding E-L equation contains $\dot{q}$ at most of degree $n$ \\
    (c) The Lagrangian $L=L(q^{\alpha},\dot{q}^{\alpha},t)$ and the E-L eqs. are: $\frac{d}{dt}\left(\frac{\partial L}{\partial \dot{q}_{\alpha}}\right)-\frac{\partial L}{\partial q_{\alpha}}=0$,\\
    Then, $dL=\frac{\partial L}{\partial q^{\alpha}}dq^{\alpha}+\frac{\partial L}{\partial \dot{q}^{\alpha}}d\dot{q}^{\alpha}=\frac{d}{dt}(\frac{\partial L}{\partial \dot{q}^{\alpha}})dq^{\alpha}+\frac{\partial L}{\partial \dot{q}^{\alpha}} \frac{d}{dt}(dq^{\alpha})=\frac{d}{dt}(\frac{\partial L}{\partial \dot{q}^{\alpha}}dq^{\alpha})$\\
    Now from the E-L equation:\\
    $\lambda^{\alpha}\frac{d}{dt}(\frac{\partial L}{\partial \dot{q}^{\alpha}})-\lambda^{\alpha}\frac{\partial L}{\partial q^{\alpha}}=0$\\
    $\Rightarrow \frac{d}{dt}(\lambda^{\alpha}\frac{\partial L}{\partial \dot{q}^{\alpha}})-\lambda^{\alpha}\frac{\partial L}{\partial q^{\alpha}}=\left(\frac{d \lambda^{\alpha}}{dt~}\right)\frac{\partial L}{\partial \dot{q}^{\alpha}}-\lambda^{\alpha}\frac{\partial L}{\partial q^{\alpha}}+\frac{d}{dt}(\lambda^{\alpha}\frac{\partial L}{\partial \dot{q}^{\alpha}})=0$\\
    $\Rightarrow \lambda^{\alpha}\frac{\partial L}{\partial q^{\alpha} } +\left(\frac{d \lambda^{\alpha}}{dt~}\right)\frac{\partial L}{\partial \dot{q}^{\alpha}}=\frac{d}{dt}(\lambda^{\alpha}\frac{\partial L}{\partial \dot{q}^{\alpha}})$\\
    $\Rightarrow \vec{X}L=\frac{d}{dt}(\lambda^{\alpha}\frac{\partial L}{\partial \dot{q}^{\alpha}})$, where $\vec{X}=\lambda^{\alpha}\frac{\partial}{\partial q^{\alpha}}+ \left(\frac{d \lambda^{\alpha}}{dt~}\right)\frac{\partial }{\partial \dot{q}^{\alpha}}$. 
    
    If $Q=\lambda^{\alpha}\frac{\partial L}{\partial \dot{q}^{\alpha}}$ is conserved i.e., $\frac{dQ}{dt}=0$, then, $\vec{X}$ is a Noether symmetry vector of the physical system and Q is termed as Noether charge.

{\Large{\textbf{{\underline{Non-Uniqueness of the Lagrangian :}}}}}
 
The Lagrangian of a given system is not unique. A Lagrangian $L$ can be multiplied by a non-zero constant $'a'$ and shift by an arbitrary constant $'b'$ and the new Lagrangian $L'=aL+b$ will describe the same motion as $L$. If one restricts the trajectories $\overrightarrow{q}$ over given time interval $[t_{in},t_{fin}]$ and fixed end points $P_{in}=q(t_{st})$ and $P_{fin}=q(t_{fin}),$ then two Lagrangians describing the same system can differ by total time derivative of a function $f(\overrightarrow{q},t)$ i.e.,
\begin{equation}
	L'(\overrightarrow{q},\dot{\overrightarrow{q}},t)=L+\frac{d}{dt}f(\overrightarrow{q},t)\nonumber
\end{equation}
As 
\begin{equation}
	\frac{df}{dt}=\frac{\partial{f}}{\partial{t}}+\sum\frac{\partial{f}}{\partial{q_i}}\dot{q_i}~~,\nonumber
\end{equation}
so
\begin{eqnarray}
	S'[\overrightarrow{q}]&=&\int_{t_{in}}^{t_{fin}}L'(\overrightarrow{q},\dot{\overrightarrow{q}},t)dt=\int_{t_{in}}^{t_{fin}}L(\overrightarrow{q},\dot{\overrightarrow{q}},t)dt+\int_{t_{in}}^{t_{fin}}\frac{df}{dt}dt\nonumber\\
	&=&S[\overrightarrow{q}]+f(P_{fin},t_{fin})-f(P_{in},t_{in})\nonumber
\end{eqnarray}
As the 2nd and 3rd terms are independent of $q$ so both the actions $S$ and $S'$  have same equations of motion i.e., both the Lagrangian have the same equations of motion.

{\Large{\textbf{\underline{Invariance under point transformation :}}}}

Let us consider a point transformation $q_{\alpha}=q_{\alpha}(s_{\beta})$ in the configuration space. $\dot{q_{\alpha}}=\frac{\partial{q_{\alpha}}}{\partial{s_{\beta}}}\dot{s_{\beta}}$.
Then 
\begin{eqnarray}
	\frac{\partial{L}}{\partial{s_{\beta}}}&=&\frac{\partial{L}}{\partial{q_{\alpha}}}\frac{\partial{q_{\alpha}}}{\partial{s_{\beta}}},~~~\frac{\partial{L}}{\partial{\dot{s_{\beta}}}}=\frac{\partial{L}}{\partial{\dot{q_{\alpha}}}}\frac{\partial{\dot{q_{\alpha}}}}{\partial{\dot{s_{\beta}}}}=\frac{\partial{L}}{\partial{\dot{q_{\alpha}}}}\frac{\partial{q_{\alpha}}}{\partial{s_{\beta}}}\nonumber\\
	&\implies&\frac{d}{dt}\left(\frac{\partial{L}}{\partial{\dot{s_{\beta}}}}\right)-\frac{\partial{L}}{\partial{{s_{\beta}}}}=0\nonumber\\
	&\implies&\frac{d}{dt}\left(\frac{\partial{L}}{\partial{\dot{q_{\alpha}}}}\frac{\partial{q_{\alpha}}}{\partial{s_{\beta}}}\right)-\frac{\partial{L}}{\partial{{q_{\alpha}}}}\frac{\partial{q_{\alpha}}}{\partial{s_{\beta}}}=0\nonumber\\
	&\implies&\frac{\partial{q_{\alpha}}}{\partial{s_{\beta}}}\frac{d}{dt}\left(\frac{\partial{L}}{\partial{\dot{q_{\alpha}}}}\right)-\frac{\partial{q_{\alpha}}}{\partial{s_{\beta}}}\frac{\partial{L}}{\partial{{q_{\alpha}}}}=0\nonumber\\
	&\implies&\frac{d}{dt}\left(\frac{\partial{L}}{\partial{\dot{q_{\alpha}}}}\right)-\frac{\partial{L}}{\partial{{q_{\alpha}}}}=0\nonumber\\
	&\implies&\mbox{Lagrangian's equations of motion is invariant under point transformation}\nonumber\\ &&\mbox{in configuration space.}\nonumber
\end{eqnarray}

{\Large{\textbf{{A comparison between Newtonian Mechanics and Lagrangian mechanics :}}}}

In Newtonian mechanics, the time varying constraint forces which keep the particle in the constrained motion are solved by Newton's law of motion. In particular, If the size and shape of a massive object is negligible then it can be treated as a point particle. Thus for a system  of $N$ such point particles having masses $m_1,~m_2,...,m_N$ and position vectors $\overrightarrow{r_1},~\overrightarrow{r_1},\overrightarrow{r_2},...,\overrightarrow{r_N}$ (with $\overrightarrow{r_k}=(x_k,y_k,z_k)$) then the Newtonian equations of motion i.e., Newton's second law are given by
\begin{equation}
	\Sigma{m_i}\frac{d^2\overrightarrow{r_i}}{dt^2}=\Sigma\overrightarrow{F_i}
\end{equation}

These are $3N$ number of second order differential equations. In Lagrangian formulation, instead of forces the energies of the system is used. The primary object is the Lagrangian function which characterized the state of a physical system or the dynamics of the system. The Lagrangian function has the dimension of energy and in mechanics it is just the difference between the K.E.(energy of motion) and the potential energy (energy of position) i.e.,$L=T-V$.

In Lagrangian mechanics, a convenient set of independent variables (known as generalised coordinates) are chosen to characterised the possible motion of the system (i.e., particles). In Lagrangian formulation, the constraint forces do not appear into the system of equations and the number of equations will be less as the influence of the constraints on the particles is not taken into account. The equations of motion in Lagrangian formulation are obtained from the action principle which states that the action functional of the system derivative from $L$ must remain at a stationary point (namely a maxima/a minimum/a saddle) throughout the time evolution of the system.

As the K.E. is the energy of the system's motion, so it is a function only of the velocities $\overrightarrow{v_k}$ but not of the positions $\overrightarrow{r_k}$ and time $t$ i.e.,$T=T(\overrightarrow{v_1},\overrightarrow{v_2},...,\overrightarrow{v_N}).$ For example for a system of point particles $T=\frac{1}{2}\sum_{k=1}^{N}m_kv_k^2.$ The P.E. of a physical system measures the amount of energy to one particle due to all the others and others external influences. However, for conservative system the P.E. is a function vectors of the particles only i.e.,$V=V(\overrightarrow{r_1},\overrightarrow{r_2},...,\overrightarrow{r_N})$ while for non-conservative forces (which can be derived from an appropriate potential (e.g. electromagnetic potential)) the P.E. is a function of both position and velocity i.e.,$V=V(\overrightarrow{r_1},\overrightarrow{r_2},...,\overrightarrow{r_N},\overrightarrow{V_1},\overrightarrow{V_2},...,\overrightarrow{V_N})$. Further, due to presence of external forces the above potential function also depends on time i.e., $V=V(\overrightarrow{r_k},\overrightarrow{V_k},t),k=1,2,...,N$.

\vspace{0.5 cm}
{\Large{\textbf{{\underline{Comments :}}}}}

\begin{itemize}
	\item For dissipative forces another function must be introduced along side $L$.
	
	\item The above form of $L$ does not hold in relativistic lagragian mechanics and it must be replaced by a function consistent with special or general relativity.
	
	\item For holonomic constraint system, the constrained equations determine the allowed path the particle can move along but where they are or how fast they go at every instant of time. Lagrangian mechanics is applicable to systems whose constraints are holonmic in nature.
	
	\item Lagrange formulation is not applicable to system having non-holonomic constraints which depend on the particle velocities, accelerations or higher derivatives of position.
	
	\item If both $T$ and $V$ have explicit time dependence then $L$ will have explicit time dependence. On the other hand, if both of them do not have explicit time dependence then,  $L$  will also have no explicit time dependence but has implicit dependence on time through the generalized co-ordinates.
	
	\item The Lagrangian equation of motion :
	\begin{equation}
		\frac{d}{dt}\left(\frac{\partial{L}}{\partial{\dot{q_j}}}\right)-\frac{\partial{L}}{\partial{{q_j}}}\nonumber
	\end{equation} 
	are second order differential equations in the generalized co-ordinates. These equations do not include constraint forces at all, only non-constraint forces are taken into account in Lagrangian formulation.
	
\end{itemize}

{\Large{\textbf{{Lagrange's equation of motion in curved space-time :}}
{\Large{\textbf{{{A path from Newtonian to Lagrangian mechanics }}}}}}}

For a point particle Newton's law of motion is $\overrightarrow{F}=m\overrightarrow{a}$ or in component form : $F^{\mu}=ma^{\mu}$. In Euclidean geometry, cartesian co-ordinate system is commonly used and Newton's law is very suitable in cartesian co-ordinate system. For any curvilinear co-ordinate system or in curved space time one can extend the idea as follows : in curved space time the acceleration vector is defined as 

\begin{eqnarray}
	a^{\mu}&=&\frac{\delta{v^{\mu}}}{d\tau}=(\nabla_{\alpha}v^{\mu})\frac{d{x^{\alpha}}}{d\tau}=\left(\frac{\partial{v^{\mu}}}{\partial{x^{\alpha}}}+\Gamma^{\mu}_{\alpha\beta} v^{\beta}\right)\frac{d{x^{\alpha}}}{d\tau}\nonumber\\
	&=&\frac{dv^{\mu}}{d\tau}+\Gamma^{\mu}_{\alpha\beta}v^{\alpha}v^{\beta}=\frac{d^2{x^{\alpha}}}{d\tau^2}+\Gamma^{\mu}_{\alpha\beta}\frac{d{x^{\alpha}}}{d\tau}\frac{d{x^{\beta}}}{d\tau}\nonumber
\end{eqnarray}

So the Newton's laws of motion becomes 
\begin{eqnarray}
	F^{\mu}=m\left(\frac{d^2x^{\mu}}{d\tau^2}+\Gamma^{\mu}_{\alpha\beta}\frac{dx^{\alpha}}{d\tau}\frac{dx^{\beta}}{d\tau}\right)\nonumber
\end{eqnarray}
Here $\Gamma^{\mu}_{\alpha\beta}$ is the christoffel symbol of second kind in curved space-time.

Now we shall show that the above laws of motion can be derived from the point-like Lagrangian  

\begin{equation}
	L=T-V=\frac{1}{2}mg_{bc}\frac{dx^b}{d\tau}\frac{dx^c}{d\tau}-V(x)\nonumber
\end{equation}
Now, 
\begin{equation}
	\frac{\partial L}{\partial\dot{x}^l}=\frac{\partial{T}}{\partial\dot{x}^l}=mg_{lc}\dot{x}^c\nonumber
\end{equation}
\begin{eqnarray}
	\frac{d}{dt}\left(\frac{\partial L}{\partial\dot{x}^l}\right)&=&m\frac{\partial{g_{lc}}}{\partial{x^k}}\dot{x^k}\dot{x^c}+mg_{lc}\ddot{x}^c\nonumber\\
	&=&\frac{m}{2}\left\{\frac{\partial{g_{lc}}}{\partial{x^k}}\dot{x^k}\dot{x^c}+\frac{\partial{g_{lk}}}{\partial{x^c}}\dot{x^c}\dot{x^k}\right\}+mg_{lc}\ddot{x}^c\nonumber\\
	\frac{\partial L}{\partial{x}^l}&=&\frac{m}{2}\frac{\partial{g_{lc}}}{\partial{x^l}}\dot{x^l}\dot{x^c}\nonumber
\end{eqnarray}
From Euler-Lagrange equations
\begin{eqnarray}
	g^{al}\left\{\frac{d}{dt}\left(\frac{\partial{T}}{\partial{x^l}}\right)-\frac{\partial{T}}{\partial{x^l}}\right\}=g^{al}\left(-\frac{\partial{V}}{\partial{x^l}}\right)&&\nonumber\\
	\implies{g^{al}}\left\{mg_{lc}\ddot{x}+\frac{m}{2}\left[\frac{\partial{g_{lc}}}{\partial{x^k}}\dot{x^k}\dot{x^c}+\frac{\partial{g_{lk}}}{\partial{x^c}}\dot{x^c}\dot{x^k}\right]\right\}-\frac{\partial{g_{bc}}}{\partial{x^l}}\dot{x}^b\dot{x}^c&=&g^{al}F_{l}\nonumber\\
	\implies{m}\delta_{c}^{a}\ddot{x}^c+mg^{al}\Gamma_{kcl}\dot{x}^k\dot{x}^c&=&F^a\nonumber\\
	\implies{m}(\ddot{x}^a+\Gamma^{a}_{kc}\dot{x}^k\dot{x}^c)&=&F^a\nonumber
\end{eqnarray}
Thus in curved space-time the general form of Lagrangian as 
\begin{equation}
	L=\frac{m}{2}g_{\alpha\beta}\frac{dx^{\alpha}}{d\tau}\frac{dx^{\beta}}{d\tau}-V(x)\nonumber
\end{equation}
and the generalized form of the Newton's law of motion is 
\begin{equation}
	m\left(\frac{d^2x^{\alpha}}{d\tau^2}+\Gamma_{\beta\delta}^{\alpha}\frac{dx^{\alpha}}{d\tau}\frac{dx^{\beta}}{d\tau}\right)=F^{\alpha}\nonumber
\end{equation}
It is to be noted that the above equation of motion is nothing but the Euler-Lagrange equation corresponding to the above Lagrangian.

 \section{Appendix IV: Symplectic structures on Manifolds}
 Let, $M^{2n}$ be an even dimensional differentiable manifold. A symplectic structure on $M^{2n}$ is a closed non degenerate differential 2-form $\omega^{2}$ such that $d\omega^{2}=0$ and $\forall \xi\neq0$, there exists $\eta:\omega^{2}(\xi,\eta)\neq 0, \xi,\eta\in TM_{x}$. The pair $(M^{2n}, \omega^{2})$ is called a symplectic manifold.
 \begin{itemize}
 	\item A symplectic structure on a manifold is a closed non-degenerate differential 2-form.
 	\item On a symplectic manifold (on Riemannian manifold), there is a natural isomorphism between vector fields and 1-forms. 
 	\item A vector field on a symplectic manifold corresponding to the differential of a function is called a Hamiltonian vector field.
 	\item A vector field on a manifold determines a phase flow, a one-parameter group of diffeomorphisms.
 	\item The phase flow of a Hamiltonian vector field (hvf) on a symplectic manifold preserves the symplectic structure of phase space.
 	\item The vector fields on a manifold form a Lie algebra. The hvf on a symplectic manifold also form a Lie algebra. The operation in this algebra is called the Poisson bracket.
 \end{itemize}
\textbf{Note:} In $R^{2n}$ with coordinates $(q^{i},p^{i})$, $\omega^{2}=\sum_{i}dp_{i}\wedge dq_{i}$. In $R^{2}$ the pair $(R^{2},\omega^{2})$ is the pair: (the plane, area).\\ \\
\textbf{Cotangent Bundle}
Let $V$ be an $n$-dimensional differentiable manifold. A $1$-form on the tangent space to $V$ at a point $x$ is called a cotangent vector to $V$ at $x$. The set of all cotangent vectors to $V$ at $x$ form an $n$-D vector space, dual to the tangent space $TV_{x}$. This vector space (VS) of cotangent vectors is denoted by $T^{*}V_{x}$ and call it the cotangent space to V at $x$.

	The union of the cotangent spaces to the manifold at all of its points is called the cotangent bundle of V and is denoted by $T^{*}V$. It has a natural structure of a differentiable manifold of dimension $2n$.
	
	An element of $T^{*}V$ is a $1$-form on the tangent space to $V$ at some point of $V$. If $q$ is a choice of $n$ local coordinates for points in $V$, then such a form is given by its $n$ components $p$. Together, the $2n$ numbers $(p,q)$ form a collection of local coordinates for points in $T^{*}V$. There is a natural projection $f:T^{*}V\rightarrow V$ (i.e, sending every one-form on $T^{*}V$ to the point $x$). This projection $f$ is differentiable and surjective. The pre-image of a point $x\in V$ under $f$ is the cotangent space $T^{*}V_{x}$.\\
	\textbf{Note:} In Lagrangian mechanics if $V$ be the configuration manifold and $L$ be the Lagrangian function then the Lagrangian generalized velocity $\dot{q}$ is a tangent vector to $V$ and the generalized momentum $p=\dfrac{\partial L}{\partial \dot{q}}$ is a cotangent vector. Then the $``p,q"$ phase space of the Lagrangian system is the cotangent bundle of the configuration manifold. The phase space has a natural symplectic manifold structure. \\
	\textbf{Note:} A Riemannian structure on a manifold establishes an isomorphism between the spaces of tangent vectors and $1$-forms. A symplectic structure establishes a similar isomorphism.\\
	\textbf{Note} To each vectors $\xi$, tangent to a symplectic manifold $(M^{2n}, \omega^{2})$ at the point $x$, one can associate a $1$-form $\omega_{\xi}^{1}$ on $TM_{x}$ by the relation: $\omega_{\xi}^{1}(\eta)=\omega^{2}(\eta,\xi),$ for all $\eta \in TM_{x}$. This isomorphism $T^{*}M_{x}\rightarrow TM_{x}$ is denoted by $I$. Suppose, $H$ be a function on a symplectic manifold $M^{2n}$. Then $dH$ is a differential $1$-form on $M$. At every point there is a tangent vector to $M$ associated to it. We denote this vector field as $IdH$ on $M$. This vector field $IdH$ is called a Hamiltonian vector field and $H$ is called the Hamiltonian function.\\
	Example- If $M^{2n}=R^{2n}=\{(p,q)\}$ then the velocity vector field of Hamilton's canonical equations:\\
	$\dot{x}=IdH(x)\implies \dot{p}=-\dfrac{\partial H}{\partial q}$ and $\dot{q}=\dfrac{\partial H}{\partial p}$.\\
	Definition: A Lie algebra is a vector space $L$, together with a bi-linear skew-symmetric operation $L\times L\rightarrow L$ which satisfies the Jacobi identity.\\
	Vector field and differential operators: Let $M$ be a smooth manifold and $\vec{A}$ is a smooth vector field on $M$; at every point $x\in M$ we are given a tangent vector $\vec{A}(x)\in TM_{x}$. Then with such vector field we have:\\
	One parameter group of diffeomorphisms or flow: Let $A^{t}: M\rightarrow M$ for which $\vec{A}$ is the velocity vector field i.e, $\dfrac{d}{dt}|_{t=0}A^{t}x=\vec{A}(x)$. Then $A^{t}$ is called a one parameter group of diffeomorphisms or flow $M\rightarrow M$. Let $L_{A}$ denotes a first order differential operator denoting  differentiation along the direction of the field $\vec{A}$ then for any function $\phi:M\rightarrow R$, the derivative of $\phi$ in the direction of $\vec{A}$ is a new function $L_{A}\phi$ whose value at a point $x$ is $(L_{A}\phi)(x)=\dfrac{d}{dt}|_{t=0} \phi(A^{t}x)$.\\
	The Poisson bracket of vector fields: Let $\vec{A}$ and $\vec{B}$ are two given vector fields on a manifold $M$. The corresponding flows $\vec{A^{t}}$ and $\vec{B^{t}}$ do not commute in general i.e, $\vec{A^{t}}B^{s}\neq B^{s}\vec{A^{t}}$. The Poisson bracket or commutator of two vector fields $\vec{A}$ and $\vec{B}$ on a manifold $M$ is the vector field $\vec{C}$ for which $L_{c}=L_{B}L_{A}-L_{A}L_{B}$ i.e, $\vec{C}=[\vec{A},\vec{B}]$.\\
	The Poisson/ commutator bracket makes the vector space of vector fields on a manifold $M$ into a Lie algebra as they satisfy the Jacobi identity.\\
	\textbf{Note:} To measure the degree of non-commutativity of the two flows $A^{t}$ and $B^{s}$ we consider the points $A^{t}B^{s}x$ and $B^{s}A^{t}x$. To estimate the difference between these points we compare the value of some smooth function $\phi$ on the manifold $M$ at those two points. The difference $\Delta (t;s;x)=\phi(A^{t}B^{s}x)-\phi(B^{s}A^{t}x)$ is a differentiable function which is zero for $s=0$ and for $t=0$. Note that $\Delta(t;s;x)|_{t=0}=0$, $\Delta(t;s;x)|_{s=0}=0$. So the Taylor series in $s$ and $t$ at $(0,0)$ contains $st$ and the other terms of second order vanish. This principal bi-linear term of $\Delta$ at $(0,0)$ gives the commutator $[L_{A},L_{B}](x)|_{(0,0)}$. \\
	\textbf{Note:} The Hamiltonian vector fields on a symplectic manifold form a sub algebra of the Lie algebra of al fields. The Hamiltonian functions also form a Lie algebra. The operation in this algebra is called the Poisson bracket of functions.
	\section{Appendix V:  Application of differential geometry to General Mechanics}
	The space of all generalized co-ordinates is termed as configuration space. The fiber bundle with the configuration space as the base manifold and fibers along the velocity field is called a tangent bundle or tangent space. Similarly, the fiber bundle with base manifold as the configuration space and fibers along the momenta co-vectors is termed as co-tangent bundle or cotangent space. If there are $n$ independent generalized co-ordinates then the configuration manifold is of dimension $n$ while both tangent and co-tangent spaces are of dimension $2n$.
	
	Let $\mathcal{L}=\mathcal{L}(q,\dot{q})$ is defined over tangent bundle while $\mathcal{H}=\mathcal{H}(q,p)$ is defined over co-tangent space or phase space. In phase space one can define a simplectic 2-form as
	\begin{equation}
		 \underline{\omega}=\underbar{d}q\wedge \underbar{d}p=\underline{d}q\bigotimes\underline{p}-\underline{d}p\bigotimes\underline{d}q\nonumber
		\end{equation}
\textbf{Result}: In phase-space, there exists a vector field $\vec{V}$ in the phase-space for which $\mathcal{L}_{\vec{V}}\underline{\omega}=0$. This vector field identifies the dynamical path.\\
	\textbf{Proof}: As, $\mathcal{L}_{\vec{V}}~\underline{\omega}=0$ i.e, $\underline{d}~[\underline{\omega}(\vec{V})]+(\underline{d}~\underline{w})~(\vec{V})=0$. From the very definition $\underline{d}~\underline{\omega}=0$ and hence $\mathcal{L}_{\vec{V}}~\underline{\omega}=0\implies \underline{d}~[\underline{\omega}(\vec{V})]=0$. Now as $\vec{V}=\dfrac{d}{dt}=\dot{q}\dfrac{\partial}{\partial q}+\dot{p}\dfrac{\partial}{\partial p}$, so $\underline{\omega}(\vec{V})=\dot{q}\underline{d}p-\dot{p}\underline{d}q=\dfrac{\partial H}{\partial p}\underline{d}p+\dfrac{\partial H}{\partial q}\underline{d}q\implies \underline{d}[\underline{\omega}(\vec{V})]=\dfrac{\partial^{2}H}{\partial q \partial p}\underline{d}q\wedge\underline{d}p+\dfrac{\partial^{2}H}{\partial p\partial q}\underline{d}p\wedge \underline{d}q=0$. On the other hand, if $\underline{d}[\underline{\omega}(\vec{V})]=0$, then $\underline{\omega}(\vec{V})=\underline {d}H$ for some function $H=H(q,p)$ in the phase space. Thus, $\dot{q}\underline{d}p-\dot{p}\underline{d}q=\dfrac{\partial H}{\partial q}\underline{q}+\dfrac{\partial H}{\partial p}\underline{p}$ i.e, $\dot{q}=\dfrac{\partial H}{\partial p}$ and $\dot{p}=-\dfrac{\partial H}{\partial q}$, the Hamilton's equation of motion. Thus a vector field $\vec{V}$ in phase-space (i.e, co-tangent space) which satisfies $\mathcal{L}_{\vec{V}}\underline{\omega}=0$ is called a Hamiltonian vector field. Now, a natural question arises: can we have a Hamiltonian vector field corresponding to a function $f=f(q,p)$ in the phase-space. 	The answer is as follows:\\
	If $\vec{X_{f}}$ be the Hamiltonian vector field for the function $f$ then it is defined as $\underline{\omega}(\vec{X_{f}})=\underline{d}f$. Suppose, $\vec{X_{f}}=a\dfrac{\partial}{\partial q}+b\dfrac{\partial}{\partial p}$, then the above relation gives $a~\underline{d}p-b~\underline{d}q=\dfrac{\partial f}{\partial p}\underline{d}p+\dfrac{\partial f}{\partial q}\underline{d}q$. Thus, comparing the coefficients of $\underline{d}p$ and $\underline{d}q$ we have, $a=\dfrac{\partial f}{\partial p}$ and $b=-\dfrac{\partial f}{\partial q}$. Hence, the Hamiltonian vector field corresponding to the function $f$ is $\vec{X_{f}}=\dfrac{\partial f}{\partial p}\dfrac{\partial}{\partial q}-\dfrac{\partial f}{\partial q}\dfrac{\partial}{\partial p}$. 
		Note: If $\vec{v}$ be a Hamiltonian vector field, then $\vec{v}$ is tangent to the solution curves in phase-space and $\mathcal{L}_{\vec{v}}H=0$. Also, we can say that the system is conservative in nature.\\ \\
	\textbf{Canonical Transformation:} A transformation from  $(q,p)\rightarrow(Q,P)$ in phase space is said to be canonical if it leaves the Hamiltonian's equation of motion to be invariant. In differential geometric notion, a canonical transformation is defined as that transformation in phase space which leaves the $2$-form $\underline{\omega}$ to be invariant. Thus $\underline{\omega}=\underline{d}q\wedge\underline{d}p=\underline{d}Q\wedge\underline{d}P$ with $Q=Q(q,p)$ and $P=P(q,p)$. So, $\underline{d}Q=\dfrac{\partial Q}{\partial q}\underline{d}q+\dfrac{\partial Q}{\partial p}\underline{d}p$ and $\underline{d}P=\dfrac{\partial P}{\partial q}\underline{d}q+\dfrac{\partial P}{\partial p}\underline{d}p$. So, $\underline{d}Q\wedge\underline{d}P=\left(\dfrac{\partial Q}{\partial q}\dfrac{\partial P}{\partial p}-\dfrac{\partial Q}{\partial p}\dfrac{\partial P}{\partial q}\right)\underline{d}q\wedge \underline{d}p$. Hence, $\dfrac{\partial Q}{\partial q}\dfrac{\partial P}{\partial p}-\dfrac{\partial Q}{\partial p}\dfrac{\partial P}{\partial q}=1$, is the necessary and sufficient condition for canonical transformation. A trivial choice is $Q=p$ and $P=-q$. Now suppose, $p=p(q,Q)$ and $P=P(q,Q)$ then $\underline{d}p=\dfrac{\partial p}{\partial q}\underline{d}q+\dfrac{\partial P}{\partial Q}\underline{d}Q$ and $\underline{d}P=\dfrac{\partial P}{\partial q}\underline{d}q+\dfrac{\partial P}{\partial Q}\underline{d}Q$. So, $\underline{d}q\wedge \underline{d}p=\dfrac{\partial p}{\partial Q}\underline{d}q\wedge \underline{d}Q$ and $\underline{d}Q\wedge \underline{d}P=-\dfrac{\partial P}{\partial q}\underline{d}q\wedge\underline{d}Q$. Thus, for Canonical Transformation $\dfrac{\partial F_{1}}{\partial Q}=-P$ and $\dfrac{\partial F_{1}}{\partial q}=p$ then the above condition for Canonical Transformation is automatically satisfied. Hence $F_{1}$ is called a generating function for Canonical Transformation. There are three other types of generating functions namely $F_{2}(q,P)$, $F_{3}(p,Q)$ and $F_{4}(p,P)$. Thus, four types of generating functions are possible for Canonical Transformation.\\ \\
	\textbf{Poisson Bracket:} Let $f=f(q,p)$ and $g=g(q,p)$ are functions in phase-space. Suppose, $\vec{X_{f}}$ and $\vec{X_{g}}$ be the corresponding Hamiltonian vector fields. Then Poisson bracket (PB) between two functions is defined as 
	\begin{eqnarray}
		\{f,g\}=\underline{\omega}(\vec{X_{f}},\vec{X_{g}})=<\underline{d}f,\vec{X_{g}}>
	=<\dfrac{\partial f}{\partial q}\underline{d}q+\dfrac{\partial f}{\partial p}\underline{d}p,~\dfrac{\partial g}{\partial p}\dfrac{\partial}{\partial q}-\dfrac{\partial g}{\partial q}\dfrac{\partial}{\partial p}>\nonumber\\
		=\dfrac{\partial f}{\partial q}\dfrac{\partial g}{\partial p}-\dfrac{\partial f}{\partial p}\dfrac{\partial g}{\partial q}\nonumber
	\end{eqnarray}
Note: Poisson Bracket is independent of the choice of co-ordinates in phase-space, it depends only on $\underline{\omega}$.\\
Properties:
\begin{enumerate}
	\item The Poisson Bracket satisfies the Jacobi identity:\\
	$\{f,\{g,h\}\}+\{g,\{h,f\}\}+\{h,\{f,g\}\}=0$, for any $C^{2}$ functions $f,~g$ and $h$. For proof see page 175.
	\item $\vec{X_{f}}(g)=-\dfrac{\partial f}{\partial q}\dfrac{\partial g}{\partial p}+\dfrac{\partial f}{\partial p}\dfrac{\partial g}{\partial q}=\{g,f\}$. If $f=H$, then $\vec{X_{H}}(f)=\dfrac{df}{dt}$.
	\item $[\vec{X_{f}},\vec{X_{g}}]=-\vec{X}_{\{f,g\}}$.\\
	Proof: As $\{f,h\}=-\vec{X_{f}}(h)$, so from Jacobi's identity \\
	$\{f,\{g,h\}\}+\{g,\{h,f\}\}=\{f,-\vec{X_{g}}(h)\}+\{g,\vec{X_{f}}(h)\}=\vec{X_{f}}\vec{X_{g}}(h)-\vec{X_{g}}\vec{X_{f}}(h)=[\vec{X_{f}},\vec{X_{g}}](h)$. So, $\{h,\{f,g\}\}=\vec{X}_{\{f,g\}}(h)$. Thus, $\vec{X}_{\{f,g\}}(h)=-[\vec{X_{f}},\vec{X_{g}}](h)$, for arbitrary $h$ i.e, $\vec{X}_{\{f,g\}}=-[\vec{X_{f}},\vec{X_{g}}]$.
\end{enumerate}
\textbf{Note}: The above result shows that if $\vec{X_{f}}$ and $\vec{X_{g}}$ are Hamiltonian vector fields then $\vec{X}_{\{f,g\}}$ is also a Hamiltonian vector field. So, the set of all Hamiltonian vector fields form a Lie algebra.
	\section{Appendix VI: Rigid body motion}
	\textbf{d'Alembert's principle: Virtual displacement and virtual work}
	In a mechanical system, virtual work arises as an application to the principle of least action. As an application of a force on a particle, the work done changes with the change in displacement. Now, among all possible displacements, the one which minimize the action is called the virtual displacement and the corresponding work on the particle is termed as virtual work.
	
	An important concept in mechanics is the principle of least constraint (by Gauss). It is essentially a variational formulation of classical mechanics. It states that the acceleration of a constrained physical system should be similar as possible to that of the corresponding unconstrained system.
	
	The principle of virtual work basically corresponds to equilibrium configuration (i.e, static situation) and it states that the total virtual work done by the applied forces in a physical system is zero. Due to Newton's second law of motion, in equilibrium state the algebraic sum of the applied forces is equal and opposite to the algebraic sum of the reactions/ constraint forces of the system and hence the algebraic sum of the virtual works done by the constraint forces should be zero. 
	
	d'Alembert's principle is a generalization of the principle of virtual work from static to dynamical system. By introducing the notion of forces of inertia to the applied forces in a physical system result in dynamic equilibrium.
	
	d'Alembert's principle also known as Lagrange-d'Alembert principle is a statement of the fundamental classical laws of motion. It generalizes the principle of virtual work from static to dynamical systems by introducing forces of inertia which when added to the applied forces in a system result in dynamic equilibrium.
	
\textbf{Note:} This principle can be applied to kinetic constraints which depend on velocities. However, this principle is not applicable for irreversible displacements.

Mathematically, Newton's second law of motion can be written as 
\begin{equation}
	\sum \vec{F}^{I}-\sum \dot{\vec{p_{i}}}=0
\end{equation}
where $F_{i}^{I}$ denotes the total forces acting on the i-th particle, $\vec{p_{i}}$ is the momentum of the $i$-th particle and $\dot{\vec{p_{i}}}$ is termed as forces of inertial or the effective force on the $i$-th particle. Thus, Newton's second law of motion states that the algebraic sum of the total forces on the particles and the algebraic sum of the reverse effective forces (or the reverse forces of inertia) makes the system in a dynamic equilibrium. Consequently, the mathematical statement of the d'Alembert's principle can be written as
\begin{equation}
	\sum (F^{I}_{i}-\dot{\vec{p_{i}}})~\delta\vec{r_{i}}=0
\end{equation}
Here $\delta\vec{r_{i}}$ is the virtual displacement of the i-th particle, consistent with the constraints. As by definition $\vec{p_{i}}=m_{i}\vec{v_{i}}$ so the above statement can be written as
\begin{equation}
	\sum \left(F_{i}^{I}-m_{i}\dot{\vec{v_{i}}}-\dot{m_{i}}\vec{v_{i}}\right)\delta\vec{r_{i}}=0\label{eq7.9}
\end{equation} where $m_{i}$ is the mass of the i-th partcile, $\vec{v_{i}}$ is the velocity vector of the i-th particle and $\vec{a_{i}}=\dot{\vec{v_{i}}}$ measures the acceleration of the i-th particle, Now, splitting the total forces as the algebraic sum of the applied forces and the constraint forces i.e,
\begin{equation}
	\vec{F_{i}^{I}}=\vec{F_{i}}+\vec{c_{i}}
\end{equation}
then equation (\ref{eq7.9}) has the explicit form as
\begin{equation}
	\sum \vec{F_{i}}\delta\vec{r_{i}}+\sum \vec{c_{i}}\vec{\delta r_{i}}-\sum m_{i}(\vec{a_{i}}\delta \vec{r_{i}})=0
\end{equation}
where masses of the system of particle are assumed to be constants. Further, if the arbitrary displacements are restricted to orthogonal to the corresponding constraint forces i.e, $\sum \vec{c_{i}}\delta\vec{r_{i}}=0$ then the above equation simplifies to
\begin{equation}
	\sum(\vec{F_{i}}-m_{i}\vec{a_{i}})\delta \vec{r_{i}}=0\label{eq7.99}
\end{equation}
The above special displacements are termed as constraint consistent displacements and the relation (\ref{eq7.99}) is termed as d' Alembert's principle. Also, it is known as principle of virtual work for applied forces. \\
\textbf{Rigid body motion:}
Now due to rotation if $\vec{\omega}$ be the angular velocity then the linear velocity can be written as 
\begin{equation}
	\vec{v}=\vec{\omega}\times \vec{r}
\end{equation}
So for a particle of mass $m$, the angular momentum is given by
\begin{equation}
	\vec{L}=\vec{r}\times m\vec{v}=m\left(\vec{r}\times (\vec{\omega}\times \vec{r})\right)
\end{equation}
Hence for a rotating rigid body about an axis through a point O of the rigid body, the total angular momentum is given by
\begin{equation}
	\vec{M}=\sum_{	i}m_{i}\left(\vec{r_{i}}\times (\vec{\Omega}\times \vec{r_{i}})\right)
\end{equation}
with $\vec{\Omega}$, the angular velocity about the axis through O. The above angular momentum equation can be written as an operator equation of the form 
\begin{equation}
	A\vec{\Omega}=\vec{M}
\end{equation}
i.e, there exists a linear operator $A$ which operating on $\vec{\Omega}$ gives the angular momentum vector. Now for any two vectors $\vec{X}$ and $\vec{Y}$ one has the inner product
\begin{eqnarray}
	(A\vec{X},\vec{Y})=
	=\sum_{	i}m_{i}[(\vec{r_{i}}\times (\vec{X}\times \vec{r_{i}})).\vec{Y}]\nonumber\\
	=-\sum_{i}m_{i}[((\vec{X}\times \vec{r_{i}})\times \vec{r_{i}}).\vec{Y}]\nonumber\\
	=-\sum_{i}m_{i}(\vec{X}\times \vec{r_{i}}).(\vec{r_{i}}\times \vec{Y})\nonumber\\
	=\sum_{	i}m_{i}(\vec{X}\times \vec{r_{i}}).(\vec{Y}\times \vec{r_{i}})
\end{eqnarray}
The last line shows that the above inner product is symmetric in $\vec{X}$ and $\vec{Y}$ and hence $A$ is a symmetric linear operator. Now choosing, $\vec{X}=\vec{Y}=\vec{\Omega}$,
\begin{eqnarray}
	\left(A\vec{\Omega}, \vec{\Omega}\right)=\sum_{	i}m_{i}(\vec{\Omega}\times \vec{r_{i}}).(\vec{\Omega}\times \vec{r_{i}})\nonumber\\
	=\sum_{	i}m_{i}(\vec{\Omega}\times \vec{r_{i}})^{2}=\sum_{	i}m_{i}v_{i}^{2}
\end{eqnarray}
Thus, the K.E of the rigid body can be written as
\begin{equation}
	T=\dfrac{1}{2}\sum m_{i}v_{i}^{2}=\dfrac{1}{2}\sum m_{i}(\vec{\Omega}\times \vec{r_{i}})^{2}=\dfrac{1}{2}	\left(A\vec{\Omega}, \vec{\Omega}\right)=\dfrac{1}{2}\left(\vec{M}, \vec{\Omega}\right)
\end{equation}
The above symmetric operator $A$ is called the \underline{inertia operation} of the rigid body.
Now due to symmetric nature of the linear operator $A$ there are three mutually orthogonal characteristic directions $(\vec{e_{1}},\vec{e_{2}},\vec{e_{3}})$ along which the eigen values are $I_{i}~(i=1,2,3)$ i.e, 
\begin{eqnarray}
	A\vec{e_{i}}=I_{i}\vec{e_{i}}, ~i=1,2,3\nonumber\\
	A\vec{\Omega}.\vec{e_{i}}=I_{i}\vec{\Omega}.\vec{e_{i}}\nonumber\\
	\vec{M}\vec{e_{i}}=M_{i}(\hat{\Omega}.\vec{e_{i}}),~M_{i}=I_{i}|\vec{\Omega}|
\end{eqnarray}
So, the K.E $T$ has the expression 
\begin{equation}
	T=\dfrac{1}{2}(I_{1}\Omega_1^{2}+I_{2}\Omega_2^{2}+I_{3}\Omega_{3}^{2})
\end{equation}
\textbf{Note:}
\begin{enumerate}
	\item The eigen directions $\vec{e_{i}}$ are the principal axes of the rigid body about O.
	\item The eigen values $I_{i}$ of the inertial operator $A$ are the moments of inertia of the rigid body w.r.t the principal axes $\vec{e_{i}}$.
\end{enumerate}
In general, suppose the rigid body rotates about an axis $\vec{e}$, the unit vector along the axis of rotation. Then, $\vec{\Omega}=\Omega_{e}\vec{e}$ and $\vec{v_{i}}=\vec{\Omega}\times \vec{r_{ie}}=\Omega_{e}.(\vec{e}\times \vec{r_{ie}})=\Omega_{e}|\vec{r_{ie}}|\tilde{e_{l}}$ where $\tilde{e_{l}}$ is the unit vector perpendicular to $\vec{e}$ and $\vec{r_{ie}}$. Now, the Moment of inertia of the rigid body about $\vec{e}$ axis is $I_{e}=\sum m_{i}|\vec{r_{ie}}|^{2}$. Hence the total angular momentum of the rigid body is 
\begin{eqnarray}
	\vec{M}=\sum_{	i}m_{i}\left(\vec{r_{ie}}\times (\vec{\Omega}\times \vec{r_{ie}})\right)=\sum_{	i}m_{i}\left(\vec{r_{ie}}\times (\Omega_{e}\vec{e}\times \vec{r_{ie}})\right)\nonumber\\
	=\sum_{	i}m_{i}\left(|\vec{r_{i}}|^{2}\Omega_{e}.\vec{e}\right)\nonumber\\
	=\left(\sum m_{i}|\vec{r_{ie}}|^{2}\right)\Omega_{e}\vec{e}=I_{e}.\Omega_{e}\vec{e}
\end{eqnarray}
Suppose, $\vec{\Omega_{e}}=\dfrac{\vec{e}}{\sqrt{I_{e}}}$ then $\vec{M}=\sqrt{I_{e}}\vec{e}$ and $T=\dfrac{1}{2}I_{e}\Omega_{e}^{2}=\dfrac{1}{2}$. But, $T=\dfrac{1}{2}(A\vec{\Omega},\vec{\omega})=\dfrac{1}{2}\implies(A\vec{\Omega},\vec{\Omega})=1$ i.e, $I_{1}\Omega_1^{2}+I_{2}\Omega_2^{2}+I_{3}\Omega_{3}^{2}=1$, an ellipsoid. This ellipsoid consists of those angular velocity vector $\vec{\Omega}$ whose K.E  is $\dfrac{1}{2}$. In particular, this ellipsoid: $\{\vec{\Omega}: (A\vec{\Omega},\vec{\Omega})=1\}$ is called the inertia  ellipsoid of the rigid body about O. Here, the principal axes of the inertia ellipsoid are directed along the principal axes of the rigid body about O and their lengths are inversely proportional to $\pm\sqrt{I_{i}}$. 

\textbf{Note:} Suppose a rigid body is stretched out along a direction, then MI w.r.t this axis is small and consequently the inertia ellipsoid is also stretched out along this axis. So the inertia may resemble the shape of the body.

\textbf{Euler's dynamical equations:}
Let, $\vec{M}=A\vec{\Omega}$ be the angular momentum vector of the rigid body. Suppose, the motion of the rigid body is considered around a stationary point O and $\vec{M}$ be the angular momentum of the body relative to O in the body. Then,
\begin{equation}
	\dfrac{d\vec{M}}{dt}=\dfrac{\partial \vec{M}}{\partial t}+\vec{\Omega}\times \vec{M}=0
\end{equation}
i.e, $\dfrac{\partial \vec{M}}{\partial t}=\vec{M}\times \vec{\Omega}$. This is known as Euler's equations. Suppose, $\vec{M}=M_{1}\vec{e_{1}}+M_{2}\vec{e_{2}}+M_{3}\vec{e_{3}}$ and $\vec{\Omega}=\Omega_1\vec{e_{1}}+\Omega_2\vec{e_{2}}+\Omega_{3}\vec{e_{3}}$ be the decomposition of $\vec{M}$ and $\vec{\Omega}$ about the principal axes at O. The components of these two vectors are related as $M_{i}=I_{i}\Omega_{i}$, $i=1,2,3$ where $I_{i}$ is the moment of inertia about the principal axis $\vec{e_{i}}$. Now, in component form the Euler's equations are
\begin{eqnarray}
	\dfrac{\partial M_{1}}{\partial t}=M_{2}\Omega_{3}-M_{3}\Omega_2=M_{2}\dfrac{M_{3}}{I_{3}}-M_{3}\dfrac{M_{2}}{I_{2}}=\dfrac{(I_{2}-I_{3})}{I_{2}I_{3}}M_{2}M_{3}\label{eq7.24}
\end{eqnarray}
$\dfrac{\partial M_{1}}{\partial t}=a_{1}M_{2}M_{3}$, $a_{1}=\dfrac{(I_{2}-I_{3})}{I_{2}I_{3}}$. The above Euler's equation can also be written as 
\begin{equation}
	I_{1}\dfrac{\partial \Omega_1}{\partial t}=(I_{2}-I_{3})\Omega_2\Omega_{3}
\end{equation}
From equation (\ref{eq7.24}), 
\begin{equation}
	M_{1}\dfrac{\partial M_{1}}{\partial t}+M_{2}\dfrac{\partial M_{2}}{\partial t}+M_{3}\dfrac{\partial M_{3}}{\partial t}=0
\end{equation}
i.e, $M_{1}^{2}+M_{2}^{2}+M_{3}^{2}=M^{2}$, a constant. Thus, total angular momentum is conserved. Similarly, 
\begin{eqnarray}
	\dfrac{M_{1}}{I_{1}}\dfrac{\partial M_{1}}{\partial t}+\dfrac{M_{2}}{\partial t}+\dfrac{M_{3}}{I_{3}}\dfrac{\partial M_{3}}{\partial t}=0\nonumber\\
	\implies \dfrac{M_{1}^{2}}{I_{1}}+\dfrac{M_{2}^{2}}{I_{2}}+\dfrac{M_{3}^{2}}{I_{3}}=2E
\end{eqnarray}
This equation is nothing but the conservation of energy. Note that the above equations are nothing but the first integrals of the Euler's equations.  From the above conservation of angular momentum and conservation of energy it is clear that the angular momentum vector $\vec{M}$ lies in the intersection of an ellipsoid and a sphere. To study, the structure of the curves of intersection, at first fix the ellipsoid for a given $E>0$ and change the radius $M$ of the sphere. Suppose, $I_{1}>I_{2}>I_{3}$. Then semi-axes of the ellipsoid will be $\sqrt{2EI_{1}}>\sqrt{2EI_{2}}>\sqrt{2EI_{3}}$. Now, the following cases will arise:
\begin{enumerate}
	\item $M<\sqrt{2EI_{3}}$ or $M>\sqrt{2EI_{1}}$. In this case, the sphere and the ellipsoid do not intersect. So, no rigid body motion is possible corresponding to such values of $M$ and $E$.
	\item $M=\sqrt{2EI_{3}}/\sqrt{2EI_{1}}$. Here the sphere and the ellipsoid intersect at two points at the end of the smallest/ largest semi axes.
	\item $\sqrt{2EI_{3}}<M<\sqrt{2EI_{2}}$. There are two curves around the ends of the smallest semi axes along which the motion is possible.
	\item $M=\sqrt{2EI_{2}}$. The intersection consists of two circles with centre at the two ends of the semi-axes of the middle principal axes.
	\item $\sqrt{2EI_{2}}<M<\sqrt{2EI_{1}}$. This case is same as 3.
	\item $M=\sqrt{2EI_{1}}$. This cone is same as case 2. 
\end{enumerate}
\textbf{Observations:}
\begin{enumerate}
	\item Each of the six end points of the semi-axes of the ellipsoid is a separate trajectory of the Euler equations. These trajectories correspond to stationary position of $\vec{M}$. So, they correspond to fixed values of the angular velocity vector directed along one of the principal axes $\vec{e_{i}}$. So, along these motions $\vec{\Omega}$ is collinear with $\vec{M}$. The body simply rotates with fixed angular velocity around the principal axis of inertia $\vec{e_{i}}$, which is stationary in space. (The motion of a rigid body under which its angular velocity remains constant is called a stationary rotation.)
	\item A rigid body fixed at a point O admits a stationary rotation around any of the three principal axes $\vec{e_{1}},~\vec{e_{2}}$ and $\vec{e_{3}}$.
	\item If, $I_{1}>I_{2}>I_{3}$, then the r.h.s of the Euler equation do not become zero anywhere else. So, there are no other stationary rotations.
	\item The stationary solutions $\vec{M_{1}}=M_{1}\vec{e_{1}},~\vec{M_{3}}=M_{3}\vec{e_{3}}$ of the Euler equations corresponding to the largest and smallest principal axes are stable, while the solution with $\vec{M_{2}}=M_{2}\vec{e_{2}}$ is unstable.
	\item The motion of the angular momentum and angular velocity vectors in a rigid body ($\vec{M}$ and $\vec{\Omega}$) will be periodic if $M\neq \sqrt{2EI_{i}},~i=1,2,3.$
\end{enumerate}

\end{document}